\documentclass[12pt,notitlepage,letterpaper]{report}             %
\usepackage{amsfonts}
\usepackage{tabularx}
\usepackage{amssymb}
\usepackage{amstext}
\usepackage{amsmath}
\usepackage{bm}
\usepackage{enumerate}
\usepackage{float}

\usepackage{setspace}
\usepackage{xspace}
\usepackage{ntheorem}
\usepackage{graphicx}
\usepackage{url}
\usepackage{graphics}
\usepackage{colordvi}
\usepackage{xcolor}
\definecolor{ForestGreen}{rgb}{0.1333,0.5451,0.1333}
\definecolor{DarkRed}{rgb}{0.8,0,0}
\definecolor{Red}{rgb}{1,0,0}
\usepackage[
  linktocpage=true,
  pagebackref=true,
  colorlinks,
  linkcolor=DarkRed,
  citecolor=ForestGreen,
  bookmarks,bookmarksopen,bookmarksnumbered
  ]{hyperref}
\PassOptionsToPackage{unicode}{hyperref}

\usepackage{caption}
\usepackage{framed}
\usepackage{subfigure}
\usepackage{cleveref}
\usepackage[ruled, linesnumbered]{algorithm2e}
\usepackage{bookmark}
\usepackage{tabularx}
\usepackage[page,toc,titletoc,title]{appendix}
\usepackage{thmtools}
\usepackage{thm-restate}
\usepackage{etoolbox}
\usepackage{multirow}
\usepackage[preset=reset]{phfnote}

\newcolumntype{L}{>{\raggedright\arraybackslash}X}

\advance \oddsidemargin by -0.8in
\newcommand{\myparskip}{3pt}
\parskip \myparskip

\newenvironment{proofof}[1]{\noindent{\bf Proof of #1.}}%
        {\hspace*{\fill}$\Box$\par\vspace{4mm}}

\newcommand{\yes}{\mathsf{yes}}

\newcommand{\yi}{{\sc Yes-Instance}\xspace}
\renewcommand{\ni}{{\sc No-Instance}\xspace}
\newcommand{\YI}{\yi}
\newcommand{\NI}{\ni}
\newcommand{\yis}{{\sc Yes-Instances}\xspace}
\newcommand{\nis}{{\sc No-Instances}\xspace}

\newcommand{\bad}{\mathsf{bad}}

\newcommand{\bfH}{\mathbf{H}}

\newcommand{\ceil}[1]{\ensuremath{\left\lceil#1\right\rceil}}
\newcommand{\floor}[1]{\ensuremath{\left\lfloor#1\right\rfloor}}

\newcommand{\event}{{\cal{E}}}

\newcommand{\NP}{\mbox{\sf NP}}
\newcommand{\NPHard}{\mbox{\sf NP-Hard}\xspace}

\newcommand{\polylog}[1]{\mathrm{polylog(#1)}}
\newcommand{\DTIME}{\mbox{\sf DTIME}}
\newcommand{\BPTIME}{\mbox{\sf BPTIME}}
\newcommand{\ZPTIME}{\mbox{\sf ZPTIME}}
\newcommand{\RTIME}{\mbox{\sf RTIME}}
\newcommand{\APX}{\mbox{\sf APX}}

\newcommand{\opt}{\mathsf{OPT}}

\newcommand{\set}[1]{\left\{ #1 \right\}}
\newcommand{\sse}{\subseteq}

\newcommand{\tset}{{\mathcal T}}
\newcommand{\iset}{{\mathcal{I}}}
\newcommand{\isetodd}{\iset_{\text{odd}}}
\newcommand{\iseteven}{\iset_{\text{even}}}
\newcommand{\hiset}{\hat \iset}

\newcommand{\pset}{{\mathcal{P}}}
\newcommand{\hpset}{\hat{\pset}}
\newcommand{\qset}{{\mathcal{Q}}}
\newcommand{\hqset}{\hat{\qset}}
\newcommand{\gset}{{\mathcal{G}}}
\newcommand{\lset}{{\mathcal{L}}}
\newcommand{\bset}{{\mathcal{B}}}
\newcommand{\aset}{{\mathcal{A}}}
\newcommand{\cset}{{\mathcal{C}}}
\newcommand{\fset}{{\mathcal{F}}}
\newcommand{\eset}{{\mathcal{E}}}
\newcommand{\dset}{{\mathcal D}}
\newcommand{\nset}{{\mathcal N}}
\newcommand{\mset}{{\mathcal M}}
\newcommand{\tmset}{\tilde \mset}
\newcommand{\hmset}{\hat \mset}

\newcommand{\jset}{{\mathcal{J}}}

\newcommand{\wset}{{\mathcal{W}}}
\newcommand{\uset}{{\mathcal{U}}}

\newcommand{\yset}{{\mathcal{Y}}}
\newcommand{\rset}{{\mathcal{R}}}

\newcommand{\hset}{{\mathcal{H}}}
\newcommand{\thset}{\tilde \hset}
\newcommand{\sset}{{\mathcal{S}}}

\newtheorem{theorem}{Theorem}[section]
\newtheorem{lemma}[theorem]{Lemma}
\newtheorem{observation}[theorem]{Observation}
\newtheorem{corollary}[theorem]{Corollary}
\newtheorem{claim}[theorem]{Claim}

\newtheorem*{definition}{Definition.}

\newtheorem{assumption}[theorem]{Assumption}

\newtheorem{fact}[theorem]{Fact}
\newenvironment{proof}{\par \smallskip{\bf Proof:}}{\hfill\stopproof}
\def\stopproof{\square}
\def\square{\vbox{\hrule height.2pt\hbox{\vrule width.2pt height5pt \kern5pt
\vrule width.2pt} \hrule height.2pt}}

\crefname{observation}{Observation}{Observations}
\crefname{claim}{Claim}{Claims}
\crefname{assumption}{Assumption}{Assumptions}
\crefname{table}{Table}{Tables}

\crefalias{constraint}{equation}
\crefname{constraint}{Constraint}{Constraints}
\creflabelformat{constraint}{#2\textup{(#1)}#3}

\crefname{subsubsubappendix}{Appendix}{Appendices}

\crefalias{invariant}{item}
\crefname{invariant}{Invariant}{Invariants}
\creflabelformat{invariant}{#2\textup{(#1)}#3}

\crefalias{step}{item}
\crefname{step}{Step}{Steps}
\creflabelformat{step}{#2\textup{(#1)}#3}

\crefalias{property}{enumitem}
\crefname{property}{Property}{Properties}
\creflabelformat{property}{#2\textup{P(#1)}#3}

\crefalias{guarantee}{enumitem}
\crefname{guarantee}{Guarantee}{Guarantees}
\creflabelformat{guarantee}{#2\textup{G(#1-4)}#3}

\crefname{algocf}{alg.}{algs.}
\Crefname{algocf}{Algorithm}{Algorithms}
\crefalias{AlgoLine}{line}%
\makeatletter
\let\cref@old@stepcounter\stepcounter
\def\stepcounter#1{%
  \cref@old@stepcounter{#1}%
  \cref@constructprefix{#1}{\cref@result}%
  \@ifundefined{cref@#1@alias}%
    {\def\@tempa{#1}}%
    {\def\@tempa{\csname cref@#1@alias\endcsname}}%
  \protected@edef\cref@currentlabel{%
    [\@tempa][\arabic{#1}][\cref@result]%
    \csname p@#1\endcsname\csname the#1\endcsname}}
\makeatother

\makeatletter
\let\save@mathaccent\mathaccent
\newcommand*\if@single[3]{%
  \setbox0\hbox{${\mathaccent"0362{#1}}^H$}%
  \setbox2\hbox{${\mathaccent"0362{\kern0pt#1}}^H$}%
  \ifdim\ht0=\ht2 #3\else #2\fi
  }
\newcommand*\rel@kern[1]{\kern#1\dimexpr\macc@kerna}
\newcommand*\widebar[1]{\@ifnextchar^{{\wide@bar{#1}{0}}}{\wide@bar{#1}{1}}}
\newcommand*\wide@bar[2]{\if@single{#1}{\wide@bar@{#1}{#2}{1}}{\wide@bar@{#1}{#2}{2}}}
\newcommand*\wide@bar@[3]{%
  \begingroup
  \def\mathaccent##1##2{%
    \let\mathaccent\save@mathaccent
    \if#32 \let\macc@nucleus\first@char \fi
    \setbox\z@\hbox{$\macc@style{\macc@nucleus}_{}$}%
    \setbox\tw@\hbox{$\macc@style{\macc@nucleus}{}_{}$}%
    \dimen@\wd\tw@
    \advance\dimen@-\wd\z@
    \divide\dimen@ 3
    \@tempdima\wd\tw@
    \advance\@tempdima-\scriptspace
    \divide\@tempdima 10
    \advance\dimen@-\@tempdima
    \ifdim\dimen@>\z@ \dimen@0pt\fi
    \rel@kern{0.6}\kern-\dimen@
    \if#31
      \overline{\rel@kern{-0.6}\kern\dimen@\macc@nucleus\rel@kern{0.4}\kern\dimen@}%
      \advance\dimen@0.4\dimexpr\macc@kerna
      \let\final@kern#2%
      \ifdim\dimen@<\z@ \let\final@kern1\fi
      \if\final@kern1 \kern-\dimen@\fi
    \else
      \overline{\rel@kern{-0.6}\kern\dimen@#1}%
    \fi
  }%
  \macc@depth\@ne
  \let\math@bgroup\@empty \let\math@egroup\macc@set@skewchar
  \mathsurround\z@ \frozen@everymath{\mathgroup\macc@group\relax}%
  \macc@set@skewchar\relax
  \let\mathaccentV\macc@nested@a
  \if#31
    \macc@nested@a\relax111{#1}%
  \else
    \def\gobble@till@marker##1\endmarker{}%
    \futurelet\first@char\gobble@till@marker#1\endmarker
    \ifcat\noexpand\first@char A\else
      \def\first@char{}%
    \fi
    \macc@nested@a\relax111{\first@char}%
  \fi
  \endgroup
}
\makeatother

\newcommand{\LP}{\mbox{\sf LP}\xspace}

\newenvironment{prog}[1]{
\begin{minipage}{5.8 in}
\begin{center}
{\sc #1}
\end{center}
}
{
\end{minipage}}

\newcommand{\program}[2]{\fbox{\vspace{2mm}\begin{prog}{#1} #2 \end{prog}\vspace{2mm}}}

\renewcommand{\phi}{\varphi}
\newcommand{\eps}{\epsilon}
\newcommand{\Y}{\Upsilon}

\newcommand{\half}{\ensuremath{\frac{1}{2}}}

\newcommand{\poly}{\operatorname{poly}}

\newcommand{\reals}{{\mathbb R}}
\newcommand{\naturals}{{\mathbb N}}

\newcommand{\vectZ}{\bm {Z}}
\newcommand{\vectDelta}{\bm {\Delta}}
\newcommand{\vectMu}{\bm {\mu}}

\newcommand{\expect}[2][]{\text{\bf E}_{#1}\left [#2\right]}
\newcommand{\prob}[2][]{\text{\bf Pr}_{#1}\left [#2\right]}

\newenvironment{properties}[2][0]
{
\begin{enumerate} \setcounter{enumi}{#1}}{\end{enumerate}}

\newcommand{\NDP}{\mbox{\sf NDP}\xspace}
\newcommand{\EDP}{\mbox{\sf EDP}\xspace}
\newcommand{\NDPgrid}{{\sf NDP-Grid}\xspace}

\newcommand{\NDPplanar}{{\sf NDP-Planar}\xspace}

\newcommand{\EDPwall}{{\sf EDP-Wall}\xspace}
\newcommand{\EDPplanar}{{\sf EDP-Planar}\xspace}
\newcommand{\NDPwall}{{\sf NDP-Wall}\xspace}
\newcommand{\NDPwC}{{\sf NDPwC}\xspace}
\newcommand{\EDPwC}{{\sf EDPwC}\xspace}

\newcommand{\threecol}{\mbox{\sf 3COL(5)}\xspace}
\newcommand{\WGP}{{\sf (r,h)-GP}\xspace}
\newcommand{\WGPwB}{{\sf (r,h)-GPwB}\xspace}
\newcommand{\WGPfull}{{\sf (r,h)}-Graph Partitioning Problem\xspace}
\newcommand{\WGPwBfull}{{\sf (r,h)}-Graph Partitioning Problem with Bundles\xspace}
\newcommand{\DkS}{{\sf DkS}\xspace}

\newcommand{\cro}{\operatorname{cr}}

\newcommand{\NCM}{\mbox{\sf NCM}\xspace}

\newcommand{\LIS}{\mbox{\sf LIS}\xspace}

\newcommand{\bestbound}[1]{\frac{n}{{#1}\log n}\cdot \left(\frac{\alpha}{d}\right )^{#1}}
\newcommand{\simplebound}[1]{\frac{n}{#1 \log^2 n} \cdot \frac{\alpha^3}{d^5}}

\newcommand{\poefull}{Path-of-Expanders }
\newcommand{\doefull}{Duo-of-Expanders }

\newcommand{\posfull}{Strong Path-of-Sets }
\newcommand{\posexp}{Expanding Path-of-Sets }

\newcommand{\alg}{\ensuremath{\mathsf{Alg}}\xspace}
\newcommand{\optncm}{\mathsf{OPT}_{\mbox{\textup{\scriptsize{NCM}}}}}

\newcommand{\optlis}{\mathsf{OPT}_{\mbox{\textup{\scriptsize{LIS}}}}}

\newcommand{\alglis}{\alg_{\mbox{\textup{\scriptsize{LIS}}}}}
\newcommand{\algncm}{\alg_{\mbox{\textup{\scriptsize{NCM}}}}}
\newcommand{\alphancm}{\alpha_{\mbox{\textup{\scriptsize{NCM}}}}}

\newcommand{\optNDP}{\mathsf{OPT}_{\mathsf{NDP}}}

\newcommand{\approxfactorNDPAlgo}{2^{O \left(\sqrt{\log n} \cdot \log \log n \right)}}
\newcommand{\constantForSizeOfBlocks}{1024}
\newcommand{\twiceConstantForSizeOfBlocks}{2048}
\newcommand{\cyi}{c_{\mbox{\textup{\tiny{YI}}}}}

\newcommand{\algcheckblock}{\mathsf{AlgCheckBlock}}

\newcommand{\jump}{\mathsf{jump}}
\newcommand{\cover}{\mathsf{cover}}

\newcommand{\ncminssizeexp}{9}
\newcommand{\ncmdboundasG}{|G|^{10^{-\ncminssizeexp}}}
\newcommand{\ncmgammaboundasd}{(d(G))^{10^{-3}}}

\newcommand{\tG}{\tilde G}
\newcommand{\hG}{\hat G}
\newcommand{\row}{\operatorname{row}}
\newcommand{\col}{\operatorname{col}}
\newcommand{\block}{K}
\newcommand{\badevent}{\mathcal{\event}_{\mathsf{bad}}}
\newcommand{\vbl}{\operatorname{vbl}}

\newcommand{\advice}{\mathsf{advice}}
\newcommand{\leftover}{\mathsf{leftover}}
\newcommand{\new}{\mathsf{new}}

\newcommand{\ourwork}{our work}
\renewcommand{\ourwork}{this thesis}

\begin{document}
\newtoggle{ncm}
\newtoggle{exp}
\newtoggle{ndp}
\newtoggle{intro}

    \begin{titlepage}
    \begin{center}
        \vspace*{0.8cm}
            
        \Huge
        \textbf{Graph Theory and Its Uses in Graph Algorithms and Beyond}
            
        \vspace{0.8cm}
        \LARGE \textbf{Rachit Nimavat}
        \vspace{0.8cm}
        
        \large

        A thesis submitted\\
        in partial fulfillment of the requirements for\\
        the degree of\\
        \vspace{0.8cm}
        Doctor of Philosophy in Computer Science\\
        \vspace{0.8cm}
        at the\\

        \vspace{0.8cm}
        TOYOTA TECHNOLOGICAL INSTITUTE AT CHICAGO\\
        Chicago, Illinois\\

        \vspace{0.8cm}
        August 2023\\
            
        \vspace{0.8cm}
        \vspace{0.8cm}

        Thesis Committee:\\
        \vspace{0.4cm}
        Julia Chuzhoy (Thesis Advisor)\\
        \vspace{0.4cm}
        Sanjeev Khanna\\
        \vspace{0.4cm}
        Yury Makarychev\\

    \end{center}
\end{titlepage}

\thispagestyle{empty}
\pagenumbering{roman}
\begin{abstract}

Graphs are fundamental objects that find widespread applications across computer science and beyond.
Graph Theory has yielded deep insights about structural properties of various families of graphs, which have been leveraged in the design and analysis of algorithms for graph optimization problems and other computational optimization problems. 
These insights have also proved especially helpful in understanding the limits of efficient computation by providing constructions of hard problem instances.
At the same time, algorithmic tools and techniques provide a fresh perspective on graph theoretic problems, often leading to novel discoveries.
In this thesis, we exploit this symbiotic
relationship between graph theory and algorithms for graph optimization problems and beyond.
This thesis consists of three parts.

\vspace{0.4cm}

In the first part, we study a classical graph routing problem called the Node-Disjoint Paths (\NDP) problem.
Given an undirected graph and a set of source-destination pairs of its vertices, the goal in this problem is to route the maximum number of pairs via node-disjoint paths.
We come close to resolving the approximability of \NDP
by showing that 
\NDP is $n^{\Omega(1/\poly \log \log n)}$-hard to approximate, even on grid graphs, where $n$ is the number of grid vertices.
In the second part of this thesis, we use graph decomposition techniques developed for efficient algorithms and tools from the analysis of random processes to derive a graph theoretic result.
Specifically, we show that for every $n$-vertex expander graph $G$, if $H$ is any graph containing at most $O(n/\log n)$ vertices and edges, then $H$ is a minor of $G$.
We also present an algorithm that, given a target graph computes its model in the input host graph.
These two parts highlight the intimate relationship between graph theory and graph algorithms.

\vspace{0.4cm}

In the last part of this thesis, we show that the graph theoretic tools and graph algorithmic techniques can shed light on problems seemingly unrelated to graphs. 
Specifically, we demonstrate that the randomized space complexity of the Longest Increasing Subsequence (\LIS) problem in the streaming model is intrinsically tied to the query-complexity of the Non-Crossing Matching problem on graphs in a new model of computation that we define. 
Leveraging the insights obtained  from this connection, we also design a better randomized algorithm for the \LIS problem in the streaming model.
\end{abstract}

\newpage
\section*{Acknowledgements}

I would like to start by expressing my gratitude to my advisor Julia Chuzhoy for her immense patience, guidance, support, and unwavering belief in me.
She graciously explained research ideas, taught me techniques, and meticulously reviewed my writing to bring it (a bit) closer to her standards.
I am forever grateful for her generosity with her time and insights about both: research and life.
Thank you, Julia, for the delicious flourless chocolate cakes and for introducing me to the joy of snacking on coffee beans!
I am thankful to Sanjeev Khanna for his valuable advice, and am grateful that I got a chance to learn from him.
His humility and wisdom continue to inspire me.
I would also like to thank Yury Makarychev for being a part of my thesis committee and for his encouragement and support.
I am thankful to Madhur Tulsiani for his time, support, understanding, and for being a constant source of inspiration.
I am also thankful to my co-authors and collaborators Sanjeev Khanna and David Kim.
It was a pleasure to discuss ideas and work with them.

From the bottom of my heart, I am grateful to TTIC for providing a friendly and vibrant environment for students.
With an abundance of talks and food, student outings, and close interactions with professors, visitors, and fellow students, TTIC has been the highlight of my experience and my home in the US.
A huge thanks to the admin staff, especially Amy and Chrissy, for making everything happen seamlessly.
I am also eternally grateful to Mary for her kindness and pampering of students, treating us like her own children.
Many thanks to friends at TTIC and UChicago for making my time here fun and interesting: 
Anmol, Behnam, Chris, David, Diego, Fernando, Goutham,
Hao, Jerry, Kshitij, Marziyeh, Naren, Omar,
Riddhima, Robby, Shane, Srinadh, Subhasmita, Suriya,
Willie, Yuval, Zihan.
Shubham, my first roommate in Chicago, made my life here so much more happening with his moral support, quick wit, and (mis)adventures.
Haris and Mrinal were the other theory students at TTIC when I joined, and I thank them both for patiently answering my questions without making me feel stupid.
A warm thank you to Somaye for showing me the secret access to the rooftop - I've lost count of the number of rainbows, snowfights, and sunsets I've enjoyed there.
I had a great time bantering with office-mates Pedro and Qingming - thank you for watching my back whenever I took naps.
I thank Akash and Shubhendu for their semi-permanent presence at TTIC and lake respectively.
Thank you Ankita for delicious food and sharing my frustration of not understanding how seasons work.
Special thanks to Pushkar and Sudarshan for fighting with their landlady and allowing me to stay at their place.
I had many `insightful' discussions with Shashank while exploring Chicago.
Thank you for putting up with me and laughing at my jokes when no one else would.
I hope to convince you for a late night Red Line ride someday.

When I arrived in Chicago, I was unsure what to expect.
I am thankful to Raut uncle and his family for easing my transition.
As I began to explore the city, I found myself pleasantly surprised by its charm and warmth.
Like taking a plunge into the cool waters of Lake Michigan, once you get past the initial shock, you'll find that the city is surprisingly open and embraces you with open arms.
With its rich history and diverse population, Chicago offered me the opportunity to interact with people from all walks of life.
As I immersed myself in its vibrant culture, I realized that Chicago is a place that I could be proud to be a part of.

I thank my cousin Pinakin for choosing to come to Chicago for his studies and for always cheering me up.
Huge thanks to my friends Bhagwat and Lokesh for frequently stopping by, checking on me, and taking me on long road-trips.
I have had the pleasure of going on memorable trips with Arnab, Deep, Dharmil, Divya, Eeshit, Naman, Pinakin, Rohit, Shubham, Yajat, and Yuval.
These trips were filled with laughter and adventure and allowed me to discover the vast and picturesque landscapes of the US, which I fell in love with.
Along the way, I met many kind-hearted strangers who helped me on countless occasions.
Their warmth and generosity never cease to amaze me.

My gratitude goes out to all my friends from IIT Kanpur, especially those from my wing and department, who made me feel right at home.
I am deeply grateful to Prof. Surender Baswana, without whom I would never have considered grad school.
It is an honor to be his student.
During my school years, Ankur Sir and Deepak Sir sparked my curiosity in maths, that I hope will stay with me for a lifetime.
A special thanks to Aakash and Samarth for sharing soda with me whenever we meet.
Even though it may have been months since we last spoke, talking with them feels as if we had just met yesterday.

I cannot express enough gratitude to my parents, Anil and Arti, for their unconditional support and trust.
From early childhood, they have given me the freedom to explore and learn, always encouraging me with no questions asked.
You have always found happiness in my happiness.
I am forever grateful for your love.
I am also grateful to my \textit{masi}, Asha, for her love and encouragement since as far back as I can remember.
You are a second mother to me.
To Divya, I am deeply grateful for your understanding, encouragement, and love.
I admire your curiosity and perseverance and I hope to emulate your kindness, even if it is only a constant factor approximation.
I would like to thank my late grandparents, my extended family in Gujarat, and my neighbors in Bhavnagar.
Their cheers and encouragement have always lifted my spirits, and I feel privileged to have received their love.

I consider myself extremely fortunate to be surrounded by amazing people in life: family, friends, neighbors, classmates, teachers, and all other fellow life-travelers.
Thank you all for being there for me.
Your love and support mean the world to me.

\tableofcontents
\newpage

    \thispagestyle{empty}
    \listoffigures
    \listoftables

    \chapter{Introduction}  \label{chap: intro}
    \pagenumbering{arabic}
    \toggletrue{intro}

Graphs are fundamental objects that find widespread applications across computer science and beyond.
They are a powerful tool for modeling relationships between objects, such as web pages, molecules, and cities, among others.
These relationships give rise to rich graph structures that are both interesting in their own right and have direct real-world applications. 
The field of Graph Theory is dedicated to studying structural properties of various families of graphs.
This area has a long-standing symbiotic relationship with the field of Graph Algorithms, where the focus is on efficiently solving optimization problems involving graphs.

Graph theory has yielded deep insights that have been leveraged in the design and analysis of algorithms for graph optimization problems and other computational optimization problems. 
Apart from Theoretical Computer Science, a wide variety of fields, including basic sciences, Machine Learning, Data Science, Social Science, Operation Research, Medicines and beyond have benefited from these insights and algorithms.
The insights drawn from graph theory have also proved especially helpful in understanding the limits of efficient computation by hinting at possible hard problem instances.
At the same time, algorithmic tools and techniques provide a fresh perspective on graph theoretic problems, often leading to novel discoveries.
The quest for designing better algorithms for the real world problems
often point us to previously undiscovered graph structures and the interplay between various such structures.

In this thesis, we exploit this intimate relationship between graph theory and graph algorithms in the context of graph optimization problems and beyond.
This thesis consists of three parts.
In the first part, we utilize the tools developed in both graph theory and graph algorithms to study the hardness of approximation of the Node-Disjoint Paths problem, one of the fundamental graph routing problems.
Then in the second part of this thesis, we apply techniques developed for efficient graph algorithms and the analysis of random processes to derive a graph theoretic result.
Specifically, we show that for every $n$-vertex expander graph $G$, if $H$ is any graph containing at most $O(n/\log n)$ vertices and edges, then $H$ is a minor of $G$.
We also present an algorithm that computes the model of such large target graphs in the host graph.
The first two parts of this thesis highlight the intimate back-and-forth relationship between graph theory and graph algorithms.
In the last part, we demonstrate that the complexity of the Longest Increasing Subsequence problem in the streaming model is intrinsically tied to that of the Non-Crossing Matching problem on graphs.
Leveraging the insights obtained from this matching problem, we design a novel randomized streaming algorithm for the Longest Increasing Subsequence problem.
This relationship affirms that right graph theoretic tools and graph algorithmic techniques can shed light on problems seemingly unrelated to graphs.

In the remainder of this chapter, we briefly introduce the problems explored in each of the three parts of this thesis in turn.
A more detailed introduction to these problems can be found in the subsequent sections of the respective chapters.

\section{Node-Disjoint Paths}
Graph routing problems are a class of graph optimization problems that involves finding paths or routes between nodes in a graph.
These problems arise in various applications, including transportation networks, communication networks, and chip design.
In \Cref{chap: ndp} we study a classical routing problem called the Node-Disjoint Paths (\NDP) problem.
The problem takes as input an undirected graph $G$ with $n$ vertices and a set of $k$ pair of its vertices, called \emph{source-destination pairs} or \emph{demand pairs}. 
The objective is to find the largest collection of paths,
each of which connects a distinct source vertex to its corresponding destination vertex, such that no two paths share a common node.

The original motivation for this problem came from its applications in VLSI design, where layout strategies relied on heuristics for finding node-disjoint paths solution connecting all demand pairs.
This is a classic case of a fundamental graph routing problem originating in the real world and subsequently extensively studied in both graph theory and theoretical computer science communities.
Robertson and Seymour~\cite{RobertsonS,flat-wall-RS} explored the problem in their seminal Graph Minor series, providing an efficient algorithm for \NDP when  the number $k$ of the demand pairs is a constant.
However, when $k$ is a part of input, it is a classical \NPHard problem~\cite{Karp-NDP-hardness,EDP-hardness}.
Consequently, research focus naturally shifted to designing approximation algorithms: efficient algorithms that compute approximate solutions, minimizing the \emph{approximation factor}, the ratio between the size of the optimal solution and the size of the solution returned by the algorithm.

The best current approximation factor of $O(\sqrt{n})$ for \NDP is achieved by a simple greedy algorithm \cite{KolliopoulosS}.
Surprisingly, until recently, this was the state-of-the-art even for the graphs containing seemingly exploitable structures, such as grid graphs.
The holy grail in the field of approximation algorithms is to achieve a tight understanding of the approximability of a problem through a two prongs approach: demonstrate an efficient algorithm that achieves some factor of approximation; and prove that improving this factor any further is \NPHard.
Until our works, only factor $\Omega \left( \log^{1/2 - \eps}{n} \right)$-hardness of approximation, for any constant $\eps > 0$, was known for the \NDP problem~\cite{AZ-undir-EDP,ACGKTZ}.
The approximability status of \NDP thus remained wide open.

The focus then shifted to understanding families of instances with more structure, both in the underlying graph and the placement of the demand pairs in it.
In the special case where the underlying graph is a grid graph, called \NDPgrid, one might expect that good approximation algorithms can be designed for this problem or, at the very least, that the problem should be easy to understand.
On the algorithmic side, Chuzhoy and Kim~\cite{NDP-grids} designed an $\tilde{O}(n^{1/4})$-approximation for \NDPgrid.
They also showed that \NDPgrid is \APX-Hard.
However, despite many efforts, various approximation techniques to obtain sub-polynomial approximation factors failed for \NDPgrid and establishing its approximability remained elusive.
A natural question then arises: is the \NDPgrid problem (and hence, \NDP problem) genuinely very hard to approximate within a reasonable factor of approximation?
Is it possible to achieve better, say sub-polynomial, factor of approximation, at least when the source and destination vertices of the demand pairs in \NDPgrid obey some structure?
In~\cite{NDP-hard-old} the \NDP problem was shown to be hard to approximate within a factor of $2^{\Omega(\sqrt{\log n})}$ under the standard complexity theoretic assumptions, even if the underlying graph is a subgraph of a grid graph, with all sources placed on the grid boundary.
Later,~\cite{NDP-algo} showed an efficient algorithm for \NDPgrid achieving roughly the same $2^{\tilde O(\sqrt{\log n})}$ factor of approximation, where the input graph has $n$ vertices, provided all source-vertices of the demand pairs appear `close' to the grid boundary.
Together, these result seem to suggest that sub-polynomial approximation algorithms may be achievable for \NDPgrid, and we might also have gotten a handle on the `correct' approximation factor achievable.

However, in~\cite{NDP-hard-new}, that we discuss in this thesis (see, \Cref{chap: ndp} for more details), we show that this is unlikely to be the case, and come close to resolving the approximability status of \NDPgrid, and of \NDP problem in general.
We show that \NDPgrid is $2^{\Omega \left( \log^{1 - \eps}{n} \right)}$-hard to approximate for any constant $\eps > 0$ under standard complexity theoretic assumptions and a factor $n^{\Omega \left(1/(\log \log n)^2 \right)}$ hardness of approximation under a stronger complexity theoretic assumption.
To establish this hardness result, we perform a Cook reduction from a known \NP-Complete problem \threecol, using a newly defined graph partitioning problem as a proxy.
We show that the \NDPgrid problem is almost as hard as this graph partitioning problem.
This is a Karp reduction exploiting graph theoretic insights of viewing routing in grids as a form of graph drawing, and that graphs with low crossing numbers have small balanced separators.
In contrast, the reduction between \threecol and the graph partitioning problem is a Cook reduction that uses the celebrated technique of parallel repetition of two-prover games.

\section{Large Minors in Expanders}
Graph minors have been used to prove many fundamental results in graph theory and computer science. 
A graph $H$ is a minor of a host graph $G$ if $H$ can be obtained by deleting some edges and vertices from $G$ and contracting some of the remaining edges.
Perhaps, the most well known result in the study of graph minors is Wagner's theorem~\cite{Wagner1937}, that provides a complete characterization of planar graphs as a family of graphs with \emph{forbidden minors}.
The seminal graph minor theorem by Robertson and Seymour~\cite{RS} proved as a part of their graph minor series is a natural extension of this result, demonstrating that any minor-closed family of graphs can be characterized as a family of graphs with forbidden minors.
An important ingredient in this result is the excluded grid theorem~\cite{gmt_5}, that is a fundamental and widely used graph theoretic result in its own right.
Among other things, this theorem plays a crucial role in their efficient algorithm for the \NDP problem when the number of demand pairs is a constant~\cite{RobertsonS,flat-wall-RS}.
Informally, this theorem states that any graph $G$ that has large \emph{treewidth} (which is a measure of how far is the graph from being tree-like) contains a grid graph of size dependent on this treewidth, as a minor.
In a long line of works~\cite{gmt_5, RST_exclude_planar, KK_gmt, leaf_gmt, CC_gmt, C_gmt, gmt_julia_arxiv,CT18}, progress has been made in one direction of this relationship, showing the existence of larger and larger grid minors, as a function of the treewidth of the host graph.
The best current lower bound~\cite{CT18} states that every graph with treewidth $t$ has a grid minor of size at least $\Omega(t^{1/9}/\poly \log t) \times \Omega(t^{1/9}/\poly \log t)$.
But the other direction of providing an upper bound on the size of the largest possible grid minor as a function of the treewidth remains elusive.
The family of host graphs achieving the best known upper bound is \emph{expander graphs}.
It is well known that constant-degree expanders have treewidth $\Theta(n)$ and there exists a family of expander graphs with girth $\Omega(\log n)$, where $n$ is the number of vertices in the graph.
Robertson et al.~\cite{RST_exclude_planar} observe that for this family of expander graph, the size of the largest grid minors does not exceed $O(n/ \log n)$.
They suggest that this bound is tight, that is, every graph with treewidth $t$ has a grid minor of size $\Omega(t/\log t)$.

Expanders are ubiquitous in discrete mathematics, theoretical computer science and beyond, arising in a wide variety of fields ranging from computational complexity to designing robust computer networks (see, ~\cite{avi_survey} for a survey).
A graph $G$ is an $\alpha$-expander, if, for every partition $(A,B)$ of its vertices into non-empty subsets, the number of edges connecting vertices of $A$ to vertices of $B$ is at least $\alpha \cdot \left(\min\set{|A|,|B|} \right)$.
We say that $G$ is an expander, its \emph{expansion} $\alpha$ is at least a constant.
This concept has been extensively studied in the literature, and there are multiple known explicit families of expanders with constant vertex-degree (see, for example,~\cite{alon1996explicit}).
In this section, we will refer to these bounded vertex-degree expanders as simply expanders.
The seemingly contradictory properties of being extremely well connected while being sparse make expanders fascinating objects to study in graph theory, in addition to their algorithmic applications. 
It is therefore important to understand the size of the largest grid minor in expander graphs, as it  not only sheds light on the structure of these objects  but also deepens our understanding of the excluded grid theorem.

The problem of finding large minors in expanders was first studied by Kleinberg and Rubinfield~\cite{KR}.
Building on the random walk-based techniques of Broder et al. \cite{BFU}, they showed that every expander $G$ on $n$ vertices contains every graph with $O(n/\log^{\kappa} n)$ vertices and edges as a minor.
The exponent $\kappa$ in this result depends on the expansion of $G$ and its maximum vertex-degree, and could be substantially larger than $1$.
In~\cite{large-minors-in-expanders}, that we discuss in this thesis (see, \Cref{chap: exp} for more details), we improve upon this result by showing that every expander $G$ on $n$ vertices with expansion $\alpha$ and maximum vertex-degree $d$ contain every other graph with $O\left( \frac{n}{\log n} \cdot \left( \frac{\alpha}{d} \right)^c \right)$ vertices and edges as minor, for some absolute constant $c$.
Notably, exponent of $\log n$ in our bound is $1$ and does not depend on the expansion $\alpha$ or maximum vertex degree of the host graph $G$.
As demonstrated by~\cite{RST_exclude_planar} this bound  achieves optimal dependence on $n$, since there exist expanders with $n$ vertices that do not contain grid graphs of size greater than $O(n/\log n)$ as minors.
Our result employs a known combinatorial object called paths-of-set system and its generalization paths-of-expanders system that we propose, among others.
These objects provide a `large enough canvas' in the host expander graph in which any target graph can be embedded.
The computation of such embedding involves finding a routing of a collection of carefully chosen demand pairs via internally node-disjoint paths.
Our algorithm for computing such a routing is inspired by the algorithm of Frieze~\cite{journal-Frieze} that efficiently routes a large set of demand pairs in an expander graph via edge-disjoint paths.
We also present an algorithm that finds such a minor in the host graph with runtime polynomial in $n$, but has super-polynomial dependency on the expansion and maximum vertex-degree of $G$.
Specifically, the runtime of our algorithm is $O \left( \poly(n) \cdot 2^{\log^2{(d/\alpha)}} \right)$, where $\alpha$ is the expansion of the host graph and $d$ is its maximum vertex-degree.

We also show a simpler algorithm whose running time is truly polynomial in the size of the host graph $G$, but can only find graphs containing fewer than 
$O \left( \frac{n}{\log^2 n} \cdot \frac{\alpha^3}{d^5}\right)$ vertices and edges as minors in $G$, where $\alpha$ is the expansion of $G$ and $d$ is its maximum vertex-degree.
Unlike the previous result, it does not employ paths-of-expanders system that inflates the running time of the algorithm as well as the dependence on $\alpha$ and $d$ in the guaranteed minor size.
Instead, we use the graph routing algorithm of~\cite{LeightonRao} and the constructive version of Lovász Local Lemma~\cite{Moser-Tardos} directly on the host expander graph.
While this simplifies the analysis, it also leads to a weaker dependence of the minor size on $n$ than our first result.

Additionally, we show that expanders are the `most minor-rich' family of graph in the following sense.
Using a simple counting technique, we show that for every graph $G$ with $O(n)$ vertices and edges, there exists a graph $H$ with $\Theta(n/\log n)$ vertices and edges, such that $H$ is not a minor of $G$.
In other words, expanders contain all graphs as minors with size up to this fundamental limit for sparse graphs.

Independently from our work, Krivelevich and Nenadov~\cite{expander-minor} provide an elegant proof of a similar but stronger result.
Their result achieves optimal dependence on all three relevant parameters: $n$, the expansion of $G$, and its maximum vertex-degree.
They also provide an efficient algorithm that finds such a minor in the host graph.

\section{Longest Increasing Subsequence and Non-Crossing Matchings}
Streaming algorithms have grown in importance in recent years due to the massive amounts of data generated by modern applications such as network monitoring, data mining, and machine learning. 
In these settings, traditional `efficient' algorithms are no longer efficient because storing this massive amount of data in memory for processing is impractical.
In contrast, streaming algorithms process data in a single pass and with limited memory, making them a useful tool for handling such large-scale data sets.
Over the last few decades, there has been a significant body of work on the study of various optimization problems in the streaming model, especially discrete optimization problems and graph problems~\cite{muthukrishnan2005data,Bar-Yossefstream}.
One such problem is the classical {\em Longest Increasing Subsequence} (\LIS) problem  which naturally fits into the streaming setting.
In this problem, we are given a sequence $S = (a_1, \ldots, a_N)$ of $N$ elements with the objective of estimating the size of its longest increasing subsequence.
Formally, we say that a subsequence $S' = (a_{i_1}, a_{i_2}, \ldots, a_{i_k})$ of length $k$ is an increasing subsequence of $S$ iff $1 \leq i_1 < i_2 < \ldots < i_k \leq N$ and $a_{i_1} < a_{i_2} < \ldots < a_{i_k}$.
In this section, we assume that $S$ is a permutation of the \emph{range} $\set{1, \ldots, N}$ and each of its elements can be stored in unit memory.

The \LIS problem and its variants have been extensively studied in the framework of streaming models and other related models. 
In the classical model of computation, there is a textbook dynamic programming algorithm that solves exactly computes \LIS in $O(n^2)$ time.
Fredman~\cite{FREDMAN197529} improved this algorithm to $O(n \log n)$ time using a technique that is now widely known and used as `Patience Sorting.'
In the streaming model, the problem of computing the exact length of \LIS is very well understood: $\Theta(N)$ space is both necessary and sufficient~\cite{Liben-NowellVZ06,sqrt-n-det-lis-soda, SunW07}.
Thus, to achieve sublinear space complexity, one must settle for an \emph{approximate solution}, where the approximation factor, the ratio of the optimal length of \LIS and the estimate returned by an algorithm, serves as the measure of quality.

The approximability of deterministic streaming algorithms for the \LIS problem is equally well understood: $\Theta \left( \sqrt{\frac{{N}}{\alpha - 1}} \right)$ space is both necessary and sufficient for achieving factor-$\alpha$ approximation to \LIS length for any $\alpha > 1$~\cite{sqrt-n-det-lis-soda,GalG07,ErgunJ08}.
However, approximability of randomized streaming algorithms remains wide open.
No known algorithm achieves strictly superior space complexity than deterministic ones, while our understanding of lower bounds is elementary in this setting.
The only known lower bound result is due to Sun and Woodruff~\cite{SunW07}, who established a lower bound of $\Omega(1/\eps)$ for the space complexity of randomized algorithms that achieve a $(1+\eps)$-factor approximation succeeding with probability at least $2/3$.
In other words, even a factor-$1.01$ approximation to \LIS length in $O(1)$ space cannot be ruled out.

Motivated by this wide gap in our understanding of randomized streaming algorithms for the \LIS problem, we initiate the study of bipartite {\em Non-Crossing Matching} (\NCM) problem in the framework that we refer to as \emph{hybrid model}.
In the \NCM problem, we are given a bipartite graph $G = (L, R, E)$ with ordered vertex sets $L$ and $R$.
We say that a matching $M$ in $G$ is \emph{non-crossing} if for every pair of edges $(u,v) , (u', v') \in M$, if $u < u'$ then $v < v'$ holds.
The \NCM problem is a generalization of \LIS problem in a sense, since a \LIS problem instance $S$ can be thought of as a \NCM problem instance where all vertex-degrees are $1$.
In the hybrid model, we assume that the algorithm has the access to the graph $G = (L, R, E)$ as follows.
We are given access to vertices $L$ and $R$ beforehand, while the edges of $G$ are revealed over the course of $|L|$ rounds.
In the $i^{th}$ round, the algorithm is given a superset of edges incident on the $i^{th}$ vertex of $L$.
It then selects a subset of these `advice-edges' to \emph{query}, and then receives the set of `real-edges' among its queried edge-slots.
At the end of processing the last vertex of $L$, the algorithm reports its estimate on the cardinality of the maximum non-crossing matching in $G$.
The aim in this model is to minimize the number of queries per vertex as a fraction of the number of advice-edges incident on it, while disregarding the space complexity.
In \Cref{chap: ncm} of this thesis, we establish a connection between the query complexity of the \NCM problem in the hybrid model and the randomized space complexity of the \LIS problem in the streaming model.
Our choice of the hybrid model is motivated by the fact that understanding one problem is closely linked to understanding the other.

There is a trivial algorithm in the hybrid model that queries every advice-edge and exactly computes maximum non-crossing matching.
We show that improving on the number of edges queried by this trivial algorithm, by even a small factor, has non-trivial consequences for the \LIS problem.
Specifically, it would imply the existence of a randomized algorithm that achieves factor-$N^{o(1)}$-approximation to \LIS length in space $N^{1/2 - \eps}$, for some absolute constant $\eps$, which depends solely on the assumed algorithm for the \NCM problem in the hybrid model.
We show a low space complexity randomized \LIS algorithm in the streaming model, that, without reduction to the \NCM problem, achieves factor $N^{o(1)}$-approximation with low space complexity, except for input instances that satisfy some stringent technical properties. 
For the latter case, we provide a structural result that allows us to reduce \LIS instances satisfying this technical condition to the \NCM problem in the hybrid model.

The intrinsic `self-reducibility' or `self-summarizing' property of the \LIS problem is the key ingredient in both, the explicit \LIS algorithm and the reduction to the \NCM problem. 
This property enables us to obtain a hierarchical decomposition of the input \LIS instance $S$ and process the instance differently at various `scales'.
It facilitates a recursive sampling based \LIS streaming algorithm, that achieves the desired guarantee, except for cases that meet the technical condition.
Fortunately, the same technical condition enables us to obtain a reduction to a collection of \NCM problem instances, each with a size of roughly $O(\log \log N)$, where $N$ is the length of $S$.
The constraints of the hybrid model allows us to simulate the supposed \NCM algorithm while processing $S$ in the streaming model, and the hierarchical decomposition ensures that we perform a small number of parallel executions of the \NCM algorithm.
Finally, the small \NCM instance size enables us to perform this simulation with an overhead of $N^{o(1)}$ in space complexity.
As a consequence, we reduce the problem of obtaining low space complexity algorithm for the \LIS problem in the streaming model to the problem of obtaining low query complexity algorithm for the \NCM problem in the hybrid model.

We also establish a converse connection, that requires a technical restriction of the streaming \LIS algorithm.
Specifically, we require that the algorithm is \emph{comparison-based}, that is, it may only compare the values of the input stream elements but not use their absolute values.
We show that the existence of factor-$(1+\eps)$ approximate comparison-based \LIS streaming randomized algorithm, for any sufficiently small constant $\eps > 0$, implies a non-trivial algorithm for the \NCM problem in the hybrid model that significantly outperforms the above-mentioned trivial algorithm.
Our results thus demonstrate that the space complexity of randomized streaming algorithms for \LIS problem is intrinsically linked to the query complexity of the \NCM problem in the hybrid model.
This relationship between the \LIS and \NCM problems highlights how graph-theoretic tools and graph-algorithmic techniques can provide insights into the complexity of seemingly unrelated problems, even in unconventional models of computation.

Exploiting the techniques developed in showing the equivalence of \NCM and \LIS problems, we additionally show a randomized factor $\alpha$-approximation algorithm for the $\LIS$ problem in the streaming model with space-complexity $\tilde O\left(\frac{\sqrt{N}}{\alpha-1} \right)$ for any $\alpha > 1$.
This is in contrast to the deterministic algorithms, which require space $\Omega\left( \sqrt{\frac{N}{\alpha - 1}} \right)$ achieving for $\alpha$-approximation~\cite{GalG07}.
To the best of our knowledge, this is the first known proof of separation of space complexity between deterministic and randomized algorithms for the \LIS problem in the streaming model.
    
    \chapter{Node-Disjoint Paths}  \label{chap: ndp}
    \toggletrue{ndp}
\newtoggle{ndp-algo}
\newtoggle{ndp-hard}

\section{Introduction}  \label{sec: ndp-intro}

In this chapter, we explore the classical Node-Disjoint Paths (\NDP) problem from the perspective of approximation algorithms and the hardness of approximation.
The problem's input consists of an undirected graph $G$ with $n$ vertices and a collection $\mset=\set{(s_1,t_1),\ldots,(s_k,t_k)}$ of $k$ vertex-pairs, called \emph{source-destination} or \emph{demand} pairs.
We refer to the vertices in  set $S=\set{s_1,\ldots,s_k}$ as \emph{source vertices};  to the vertices in set $T=\set{t_1,\ldots,t_k}$ as   \emph{destination vertices}, and to the vertices in set $S\cup T$ collectively as \emph{terminals}.
We say that a path $P$ \emph{routes} a demand pair $(s_i,t_i)$ iff the endpoints of $P$ are $s_i$ and $t_i$.
The goal of this problem is to compute a maximum-cardinality set $\pset$ of node-disjoint paths, where each path $P\in \pset$ routes a distinct demand pair in $\mset$.
The optimal solution size is denoted as $\optNDP(G, \mset)$ or simply $\optNDP$ when the instance is clear from the context.

While the original motivation for this problem came from its applications in VLSI design, being a fundamental graph routing problem, it has been extensively studied in both graph theory and theoretical computer science communities.
Robertson and Seymour~\cite{RobertsonS,flat-wall-RS} explored the problem in their Graph Minor series, providing an efficient algorithm for \NDP when  the number $k$ of the demand pairs is bounded by a constant.
However, when $k$ is a part of input, the problem becomes $\NP$-hard~\cite{Karp-NDP-hardness,EDP-hardness}.
Consequently, the focus of the research then naturally shifted to studying the \NDP problem when the underlying graph $G$ is `simpler'.
Two special cases of \NDP that have received a lot of attention are when the underlying graph $G$ is a planar graph and when it is a square grid.
We denote the former problem as \NDPplanar and the latter as \NDPgrid.
\footnote{We use the standard convention of denoting $n=|V(G)|$, and so the underlying graph in the \NDPgrid instance is a grid graph $G$ with $(\sqrt{n}\times \sqrt{n})$; we assume that $\sqrt{n}$ is an integer.}.
It is worth noting that both of these special cases, \NDPplanar and \NDPgrid, are \NP-hard~\cite{npc_planar, npc_grid}.
We summarize the following known results in \Cref{table: ndp-ndp-bounds}.

On the side of approximation algorithms, the best current approximation factor of $O(\sqrt{n})$ for \NDP is achieved by a simple greedy algorithm \cite{KolliopoulosS}.
Until recently, this was also the best approximation algorithm for the above mentioned special cases: \NDPplanar and \NDPgrid.
From the direction of hardness of approximation, before our work, only an $\Omega(\log^{1/2-\eps}n)$-hardness of approximation was known for the \NDP problem, for any constant $\eps$, unless $\NP \subseteq \ZPTIME(n^{\poly \log n})$~\cite{AZ-undir-EDP,ACGKTZ}.
In the special cases of \NDPplanar and \NDPgrid, only \APX-hardness was known~\cite{NDP-grids}.
The approximability status of \NDP remained wide open, and the designing better approximation algorithms, especially in the conceivably simple setting of \NDPgrid remained a tantalizing open question.
Grid graphs are extremely well-structured, and one might expect that good approximation algorithms can be designed for them or, at the very least, that they should be easy to understand.
However, despite many efforts, various approximation techniques to obtain sub-polynomial approximation factors failed for \NDPgrid and establishing its approximability has remained elusive so far.

One natural technique of attacking \NDP problem to design approximation algorithms is via the standard multicommodity flow relaxation.
Instead of connecting (routed) demand pairs with a path, this technique involves sending the maximum possible amount of (possibly fractional) flow between demand pairs such that no vertex harbors more than one unit of flow. 
Such an optimal fractional flow can be computed by writing a textbook linear program (\LP) and solving it optimally by standard methods.
The hope is to `round' this optimal fractional solution and obtain a good feasible solution to the original \NDP problem.
The approximation factor that we get is the ratio between the optimal flow value and the size of rounded feasible solution to the \NDP problem.
The $O(\sqrt{n})$-approximation algorithm of~\cite{KolliopoulosS} can be cast as an \LP-rounding algorithm of precisely this multicommodity flow relaxation.
Unfortunately, it is well-known that the  integrality gap of this relaxation is $\Omega(\sqrt{n})$, even when the underlying graph is a grid, even with all terminals lying on its boundary.
In other words, there are \NDPgrid instances, with terminals lying on the grid boundary, such that the optimal multicommodity flow relaxation has value $\Omega(\sqrt{n} \cdot \optNDP)$ and we cannot hope to achieve a $o(\sqrt{n})$-factor approximation with this approach.

\vspace*{1em}
\begin{table}[ht]
    \centering
     \begin{tabular}{|m{4cm} | c | m{5.5cm} |} 
        \hline
        Special Case & Approximation factor & References \\ [0.2em]
        \hline\hline
        {Constant $k$} & $1$ & \cite{RobertsonS,flat-wall-RS}\\ \hline \hline
        \multirow{4.8}{*}{General Case} & $O(n^{1/2})$ & \cite{KolliopoulosS}\\ [0.2em]
        & $\Omega(\log^{1/2-\eps}n)$ & \cite{AZ-undir-EDP,ACGKTZ}\\[0.2em]
        & $2^{\Omega(\sqrt{\log n})}$ & \cite{NDP-hard-old}\\[0.2em]
        & $n^{\Omega \left(1/(\log \log n)^2 \right)}$ & \cite{NDP-hard-new}\\
        \hline \hline
        \multirow{2.9}{*}{\NDPgrid} & $\tilde O(n^{1/4})$ & \cite{NDP-grids}\\ [0.2em]
        & $n^{\Omega \left(1/(\log \log n)^2 \right)}$ & this thesis; see also \cite{NDP-hard-new}\\\hline
        \NDPgrid with sources on boundary & $2^{\tilde O(\sqrt{\log n})}$ & \cite{NDP-algo}\\
        \hline \hline
        \multirow{3.9}{*}{\NDPplanar} & $\tilde O(n^{9/19})$ &\cite{NDP-planar}\\ [0.2em]
        & $2^{\Omega(\sqrt{\log n})}$ & \cite{NDP-hard-old}\\[0.2em]
        & $n^{\Omega \left(1/(\log \log n)^2 \right)}$ & this thesis; see also \cite{NDP-hard-new}\\
        \hline
     \end{tabular}
     \caption{Approximation tradeoffs for efficient algorithms for the \NDP problem in the general case and some special cases.}
     \label{table: ndp-ndp-bounds}
\end{table}
\vspace*{1em}

Bypassing this multicommodity flow integrality gap barrier, Chuzhoy and Kim~\cite{NDP-grids} designed an $\tilde{O}(n^{1/4})$-approximation for \NDPgrid; and along with Li, generalized it to obtain an $\tilde O(n^{9/19})$-approximation algorithm for \NDPplanar~\cite{NDP-planar}.
Their key observation is that, if all terminals lie close to the grid boundary, say within distance $O(n^{1/4})$ of the grid boundary, then a simple dynamic programming-based algorithm yields an  $O(n^{1/4})$-approximation.
On the other hand, if, for every demand pair, either the source or the destination lies at a distance at least $\Omega(n^{1/4})$ from the grid boundary, then the integrality gap of the multicommodity flow relaxation improves, and one can obtain an $\tilde{O}(n^{1/4})$-approximation via \LP-rounding.
A natural question is whether the integrality gap for \NDPgrid improves even further, if all terminals lie further away from the grid boundary.
Unfortunately, the authors show in~\cite{NDP-grids} that the integrality gap remains at least $\Omega(n^{1/8})$, even if all terminals lie within distance $\Omega(\sqrt n)$ from the grid boundary.
A natural question then arises: is it possible to achieve better, say sub-polynomial, factor of approximation when the terminals lie close to the grid boundary?
Or more specifically, can we improve over the simple dynamic programming approach of \cite{NDP-grids} in this special case?

In our work~\cite{NDP-algo}, we answer this question in affirmative and show an $\approxfactorNDPAlgo$-approximation algorithm for \NDPgrid provided all sources appear on the grid boundary.
Note that we do not require any restriction on the placement of the destination vertices: they may appear anywhere on the underlying grid.
As mentioned earlier, Chuzhoy and Kim~\cite{NDP-grids} show that the integrality gap of the multicommodity flow relaxation problem remains $\Omega(n^{1/8})$ in this regime.
To get around this barrier, we define a new set of sufficient conditions.
We show that we can efficiently route each subset $\mset' \subseteq \mset$ of demand pairs that satisfies these conditions.
We also show an efficient algorithm, that given an \NDPgrid instance with sources lying on the grid boundary, computes a subset $\mset' \subseteq \mset$ of demand pairs of cardinality $\optNDP/\approxfactorNDPAlgo$, that satisfy these sufficient conditions.
We then generalize our result to instances where the sources lie within a prescribed distance from the grid boundary, at the expense of larger factor of approximation.

In our earlier work~\cite{NDP-hard-old}, we show that \NDPplanar (and hence, \NDP) is hard to approximate to within a $2^{\Omega(\sqrt{\log n})}$ factor  unless $\NP\subseteq \DTIME(n^{O(\log n)})$, even if the underlying graph is a sub-cubic planar graph with all the sources lying on the boundary of a single face.
We also show that the same hardness of approximation holds, even if the underlying graph is a subgraph of a grid graph, with sources lying on the grid boundary.
Equivalently, the underlying graph of these hard instances is a grid graph with certain carefully chosen vertices removed; and the demand pairs of these hard instances have sources lying on the (original) grid boundary.
Thus, our work~\cite{NDP-algo} can be seen as complementing this result.
Together, they seem to suggest that sub-polynomial approximation algorithms may be achievable for \NDPgrid and $2^{\tilde \Theta(\sqrt{\log n})}$ might even be the `correct' approximation factor achievable.

In~\cite{NDP-hard-new}, we show that this is unlikely to be the case, and come close to resolving the approximability status of \NDPgrid, and of \NDP problem in general.
We show that \NDPgrid is $2^{\Omega(\log^{1-\eps}n)}$-hard to approximate for any constant $\eps > 0$ assuming ${\NP\nsubseteq \BPTIME(n^{\poly\log n})}$.
We further show that it is $n^{\Omega\left( 1/(\log \log n)^2 \right)}$-hard to approximate, assuming that for some constant $\delta > 0$, $\NP \not \subseteq \RTIME(2^{n^\delta})$.
To show this hardness, we perform a Cook reduction: if we have a good approximation efficient algorithm for the \NDPgrid problem, we use it repeatedly to solve a known \NP-Complete \threecol problem.
By choosing different tradeoffs between the \NDPgrid instance sizes and the approximation factor, we obtain our hardness results.
We note that unlike the hard instances presented in~\cite{NDP-hard-old}, the hard instances we construct in this work require all terminals to be situated far away from the boundary of the grid.
This is in line with our earlier work in~\cite{NDP-algo}, where we demonstrated an efficient algorithm with an improved approximation factor, specifically designed for instances where the source vertices are in close proximity to the grid boundary.

The \NDP problem has been studied in a variety of restricted settings.
Cutler and Shiloach~\cite{Cutler-Shiloach} studied a more restricted version of \NDPgrid, where all source vertices lie on the top row $R^*$ of the grid, and all destination vertices lie on a single row  $R'$ of the grid, far enough from its top and bottom boundaries.
We note that \cite{NDP-grids} extended their algorithm to a more general setting, where the destination vertices may appear anywhere in the grid, as long as the distance between any pair of the destination vertices, and any destination vertex and the boundary of the grid, is large enough. 
Robertson and Seymour~\cite{NDP-surface} provided sufficient conditions for the existence of node-disjoint routing of a given set of demand pairs in the more general setting of graphs drawn on surfaces, and they designed an algorithm whose running time
scales at least exponentially in $k$, the number of demand pairs.
Their result implies the existence of the routing in grids, when the destination vertices are sufficiently far from each other and from the grid boundaries, but it does not provide an efficient algorithm to compute such a routing.
Aggarwal, Kleinberg, and Williamson~\cite{AKW} considered another special case of the  \NDPgrid problem, where the set of  the demand pairs is a permutation: that is, every vertex of the grid participates in exactly one demand pair.
They show that $\Omega(\sqrt{n}/\log n)$ demand pairs are routable in this case via node-disjoint paths.

Another basic routing problem that is closely related to \NDP is the Edge-Disjoint Paths (\EDP) problem.
The input to this problem is the same as before: an undirected graph $G=(V,E)$ and a set $\mset=\set{(s_1,t_1),\ldots,(s_k,t_k)}$ of demand pairs.
The goal, as before, is to route the largest number of the demand pairs via paths.
However, we now allow the paths to share vertices, and only require that they are mutually edge-disjoint.

In general, it is easy to see that  \EDP is a special case of \NDP.
Indeed, given an \EDP instance $(G,\mset)$, computing the line graph of the input graph $G$ transforms it into an equivalent instance of \NDP.
However, this transformation may potentially inflate the number of the graph vertices by a quadratic factor, and so approximation factors that depend on $|V(G)|$ may no longer be preserved.
Moreover, this transformation does not preserve planarity, and no relationship is known between \NDP and \EDP when the underlying graph is constrained to be a  planar graph.

\begin{figure}[h]
    \centering
    \scalebox{0.4}{\includegraphics{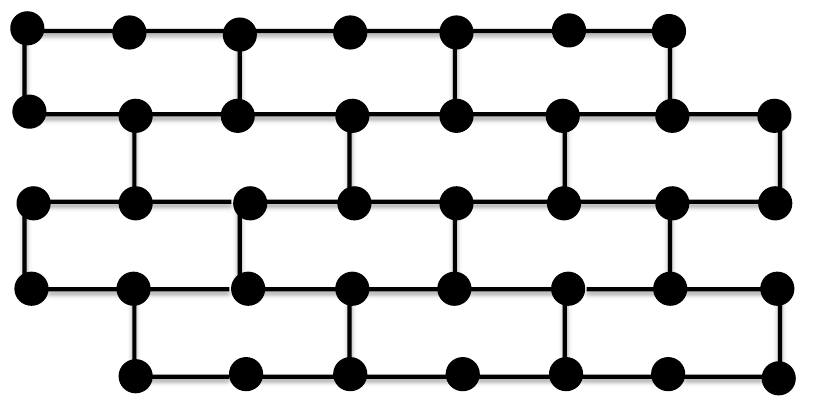}}
    \caption{A wall graph.\label{fig: ndp-hard-wall}}
\end{figure}

The approximability status of \EDP is very similar to that of \NDP: the best current approximation algorithm achieves an $O(\sqrt{n})$-approximation factor~\cite{EDP-alg}.
On the other side, until recently, our result~\cite{NDP-hard-old} showed the best known $2^{\Omega(\sqrt{\log n})}$-hardness of approximation for \EDP on planar graphs, under the same complexity assumption.
Interestingly, \EDP appears to be relatively easy on grid graphs, and has a constant-factor approximation for this special case~\cite{grids1, grids3,grids4}. 
The analogue of \NDPgrid in the setting of \EDP seems to be the wall graph (see Figure~\ref{fig: ndp-hard-wall}): the approximability status of \EDP on wall graphs (that we call \EDPwall) is similar to that of \NDPgrid.
Until recently, the best current upper bound was $\tilde O(n^{1/4})$-approximation algorithm and the best lower bound was of \APX-hardness~\cite{NDP-grids}. 

Both of our results also extend to the \EDPwall problem.
In~\cite{NDP-algo}, we extend our approximation algorithm for \NDPgrid to \EDPwall when the sources are located on the wall boundary.
In~\cite{NDP-hard-new}, we show that the hardness of approximation factor for \NDPgrid extends to \EDPwall under the same complexity assumptions.
We note that for \EDP on planar graphs (\EDPplanar), the best upper bound is still $O(\sqrt{n})$, and it remains an open question whether the techniques of \cite{NDP-planar} that achieve an $\tilde O(n^{9/19})$-approximation for \NDPplanar can be extended to \EDPplanar.

\vspace*{1em}
\begin{table}[ht]
    \centering
     \begin{tabular}{|m{14em} |c | m{5.4cm} |} 
        \hline
        Special Case & Approximation factor & References \\ [0.2em]
        \hline\hline
        {Constant $k$} & $1$ & \cite{RobertsonS,flat-wall-RS}\\ \hline \hline
        \multirow{3.5}{*}{General Case} & $O(n^{1/2})$ & \cite{EDP-alg}\\ [0.2em]
        & $2^{\Omega(\sqrt{\log n})}$ & \cite{NDP-hard-old}\\[0.2em]
        & $n^{\Omega \left(1/(\log \log n)^2 \right)}$ & this thesis; see also \cite{NDP-hard-new}\\
        \hline \hline
        \multirow{3}{14em}{\EDP on grids, grid-like graphs and trees} & \multirow{3}{*}{$O(1)$} & \multirow{3}{5cm}{\cite{grids1, AwerbuchGLR94, grids3,grids4,apx_tree,ChekuriMS07}}\\ 
        &&\\
        &&\\
        \hline \hline
        \multirow{3}{14em}{\EDP on bounded-degree expanders and graphs with large minimum cut} & \multirow{3}{*}{$\tilde O(1)$} & \multirow{3}{5cm}{\cite{LeightonRao,BFU92,BroderFSU94,KR,journal-Frieze,RaoZhou}}\\ 
        &&\\
        &&\\
        \hline \hline
        \multirow{3}{14em}{\EDP on Eulerian planar graphs and $4$-edge-connected planar graphs} & \multirow{3}{*}{$\tilde O(1)$} & \multirow{3}{5cm}{\cite{KK-planar,Kleinberg-planar}}\\ 
        &&\\
        &&\\
        \hline \hline
        \multirow{2.3}{*}{\EDPwall} & $\tilde O(n^{1/4})$ & \cite{NDP-grids}\\ [0.2em]
        & $n^{\Omega \left(1/(\log \log n)^2 \right)}$ & this thesis; see also  \cite{NDP-hard-new}\\
        \hline \hline
        \multirow{3.5}{*}{\EDPplanar} & $\tilde O(n^{9/19})$ &\cite{NDP-planar}\\ [0.2em]
        & $2^{\Omega(\sqrt{\log n})}$ & \cite{NDP-hard-old}\\[0.2em]
        & $n^{\Omega \left(1/(\log \log n)^2 \right)}$ & this thesis; see also \cite{NDP-hard-new}\\
        \hline
     \end{tabular}
     \caption{Approximation tradeoffs for efficient algorithms for the \EDP problem in the general case and some special cases.}
     \label{table: ndp-edp-bounds}
\end{table}
\vspace*{1em}

As alluded earlier, unlike \NDP, several other special cases of \EDP are known to have reasonably good approximation algorithms.
We summarize some of the following results in \Cref{table: ndp-edp-bounds}.
For example, for the special case of Eulerian planar graphs, Kleinberg~\cite{Kleinberg-planar} showed an $O(\log^2 n)$-approximation algorithm,
while Kawarabayashi and Kobayashi~\cite{KK-planar} provide an improved $O(\log n)$-approximation for both Eulerian and $4$-connected planar graphs.
Bounded-degree expander graphs are also known to admit polylogarithmic approximation algorithms~\cite{LeightonRao,BFU,BroderFSU94,KR,journal-Frieze}.
Constant-factor approximation algorithms are known for trees~\cite{apx_tree,ChekuriMS07}, as well as grids and grid-like graphs~\cite{grids1,AwerbuchGLR94,grids3,grids4}.
Rao and Zhou~\cite{RaoZhou} showed an efficient randomized polylogarithmic approximation algorithm for the special case of \EDP where underlying graphs have a large global minimum cut.
Recently, Fleszar et al.~\cite{fleszar_et_al} designed an $\tilde O(\sqrt{r})$-approximation algorithm for \EDP, where $r$ is the size of the feedback vertex set of $G$: the smallest number of vertices that need to be deleted from $G$ in order to turn it into a forest.
Since $r \leq n$, their bound matches the result of~\cite{EDP-alg} up to logarithmic factors and strengthens it when $r = o(n)$.
In a recent work, Huang~et~al.~\cite{edp-fully-planar-constant-approx} showed a constant factor approximation algorithm provided the underlying graph, together with the demand pairs visualized as regular edges, forms a planar graph.

Another natural variant of \NDP and \EDP relaxes the the disjointedness constraint by allowing a small vertex-congestion or edge-congestion has been a subject of extensive study. 
In the \NDP with Congestion (\NDPwC) problem, the input consists of an undirected graph and a set of demand pairs as before, and additionally a non-negative integer $c$.
The goal is to route a maximum number of the demand pairs with congestion $c$, that is, each vertex may participate in at most $c$ paths in the solution.
The \EDP with Congestion problem (\EDPwC) is defined similarly, except that now the congestion is measured on the graph edges and not vertices.
The famous result of Raghavan and Thompson~\cite{RaghavanT}, that introduced the randomized \LP-rounding technique, obtained a constant-factor approximation for \NDPwC and \EDPwC, for a congestion value $c=\Theta(\log n/\log\log n)$.
A long sequence of work~\cite{CKS,Raecke,Andrews,RaoZhou,Chuzhoy11,ChuzhoyL12,ChekuriE13,NDPwC2} has led to an $O(\poly\log k)$-approximation for \NDPwC and \EDPwC with congestion bound $c=2$.
This result is essentially optimal, since it is known that for every constant $\eps$, and for  every congestion value $c=o(\log\log n/\log\log\log n)$, both problems are hard to approximate to within a factor $\Omega((\log n)^{\frac{1-\eps}{c+1}})$, unless $\NP \subseteq \ZPTIME(n^{\poly \log n})$~\cite{ACGKTZ}.
When the input graph is planar, Seguin-Charbonneau and Shepherd~\cite{EDP-planar-c2}, improving on the result of Chekuri, Khanna and Shepherd~\cite{EDP-planar-constant-cong}, have shown a constant-factor approximation for \EDPwC with congestion 2.
Combined with these results, our work shows a dramatic contrast in the approximability of \EDP in the presence of tiniest non-trivial congestion.

\subsection{Our Results and Informal Overview of Techniques}

Our main result is summarized in the following theorem:

\begin{theorem} \label{thm: ndp-hard-master NDP}
For every constant $\eps > 0$, there is no efficient $2^{O(\log^{1-\eps} n)}$-approximation algorithm for the \NDP problem, assuming that $\NP\nsubseteq\RTIME(n^{\poly\log n})$. Moreover, there is no efficient $n^{O(1/(\log\log n)^2)}$-approximation algorithm for the \NDP problem, assuming that  for some constant $\delta>0$, $\NP\nsubseteq \RTIME(2^{n^\delta})$. These results hold even when the input graph is a grid graph or a wall graph.
\end{theorem}

We also extend this hardness of approximation result to the closely related \EDP problem under the same complexity assumptions.
As mentioned earlier, \EDPwall can be thought of as an analogue to the \NDPgrid problem.

\begin{theorem} \label{thm: ndp-hard-master EDP}
For every constant $\eps > 0$, there is no efficient $2^{O(\log^{1-\eps} n)}$-approximation algorithm for the \EDP problem, assuming that $\NP\nsubseteq\RTIME(n^{\poly\log n})$. Moreover, there is no efficient $n^{O(1/(\log\log n)^2)}$-approximation algorithm for the \EDP problem, assuming that  for some constant $\delta>0$, $\NP\nsubseteq \RTIME(2^{n^\delta})$. These results hold even when the input graph is a wall graph.
\end{theorem}

We now provide a high-level overview of our techniques behind the proof of \Cref{thm: ndp-hard-master NDP}.
We define a new graph partitioning problem, that we refer to as \WGPfull, and denote it by \WGP for short.
In this problem, we are given a bipartite graph $\tilde G = (V_1,V_2,E)$ along with two integral parameters $r,h > 0$.
A solution to this problem is a partition $(W_1,W_2,\ldots,W_r)$ of $V_1\cup V_2$ into $r$ subsets, and for each $1\le i\le r$, a subset $E_i\subseteq E(W_i)$ of edges, so that $|E_i|\le h$ holds, and the goal is to maximize $\sum_{i=1}^r |E_i|$.
We say that a solution to \WGP is a \emph{perfect solution} if it has maximum possible value: $r \cdot h$.
An intuitive way to think about this problem is that we would like to partition $\tilde G$ into a large number of subgraphs, in a roughly balanced way (with respect to the number of edges), so as to preserve as many of the edges as possible.

Our goal is to show that \NDPgrid is at least as hard as \WGP for some choice of parameters $r$ and $h$, and that \WGP itself is \NP-hard to approximate for the same choice of parameters.
Unfortunately, despite our feeling that \WGP is somewhat similar to Densest $k$-Subgraph problem (\DkS), we were unable to extend its hardness results to \WGP.
In the \DkS problem, we are given a graph $G=(V,E)$ and a parameter $k$, and the goal is to find a subset $U\subseteq V$ of $k$ vertices, that maximizes the number of edges in the induced graph $G[U]$. Intuitively, in the \WGP problem, the goal is to partition the graph into many dense subgraphs, and so in order to prove that \WGP is hard to approximate, it is natural to employ techniques used in hardness of approximation proofs for \DkS. 
The best current approximation algorithm for \DkS achieves a $n^{1/4+\eps}$-approximation for any constant $\eps$~\cite{dks10}. Even though the problem appears to be very hard to approximate, its hardness of approximation proof has been elusive until recently: only constant-factor hardness results were known for \DkS under various worst-case complexity assumptions, and $2^{\Omega(\log^{2/3}n)}$-hardness under average-case assumptions~\cite{Feige02,dks_average_hardness,Khot04,RaghavendraSteurer10}. 
In a recent breakthrough, Manurangsi~\cite{Manurangsi16} has shown that for some constant $c$, \DkS is hard to approximate to within a factor $n^{1/(\log\log n)^c}$,  under the Exponential Time Hypothesis.
Even though \WGP problem itself appears similar in flavor to \DkS, we were unable to extend the techniques of~\cite{Manurangsi16} to this problem, or to prove its hardness of approximation via other techniques.

We overcome this difficulty as follows.
First, we define a graph partitioning problem that we call \WGPwBfull (or \WGPwB for short), which is more general than \WGP.
The definition of this problem is somewhat technical and is deferred to \Cref{subsec: ndp-hard-WGP}.
Intuitively, we want this problem to be hard enough so that we can prove that approximating it is \NP-hard, but on the other hand, we also want it to be `easier' than \NDPgrid, so that, we can show that good approximation algorithm for \NDPgrid implies good approximation algorithm for \WGPwB.
To this end, we define subsets of edges in the underlying \WGPwB instance, that we call bundles, and require that in each one of the $r$ subgraph, at most single edge from a bundle can be present. 
As before, we say that a solution to \WGPwB is a \emph{perfect solution} if it has maximum possible value: $r \cdot h$.

\paragraph{Relating \NDPgrid and \WGPwB.}
Assume that we have an efficient $\alpha$-approximation algorithm $\aset$ for \NDPgrid for some parameter $\alpha$.
We show that, using $\aset$, we can come up with an efficient algorithm $\aset'$ for \WGPwB with roughly the same approximation factor.
To this end, assume that we are given an instance $\iset'$ of \WGPwB problem and our goal is to come up with an instance $\iset$ of \NDPgrid such that $\iset$ has large solution iff $\iset'$ has a large solution.
We can then use the assumed efficient algorithm $\aset$ on the \NDPgrid instance $\iset$.

Intuitively, our reduction proceeds as follows.
We start with a large enough grid $G$ that will serve as the underlying graph for the \NDPgrid instance $\iset$.
The demand pairs $\mset$ of $\iset$ corresponds to the edges of \WGPwB instance $\iset'$: the sources are placed in a single row far away from the grid boundary and similarly, the destinations are placed in a single row far away from the grid boundary and from the sources.
We place the terminals in their respective rows in a carefully chosen suitable random ordering.
We now want to show that
(i) if the original \WGPwB instance $\iset'$ has a solution of value $\beta$, then with constant probability, the \NDPgrid instance $\iset$ has a solution routing roughly $\beta$ demand pairs; and
(ii) if we are given some feasible routing of $\beta$ demand pairs in the \NDPgrid instance $\iset$, we can efficiently come up with a feasible solution of value roughly $\beta$ in the original \WGPwB instance $\iset'$.

For the first part, we define a notion of \emph{distance property}.
Intuitively, we say that a set of demand pairs $\mset$ has distance property if, for every two demand pairs $(s,t), (s', t') \in \mset$, the distance between $t$ and $t'$ is much larger than the same between $s$ and $s'$.
This separatedness allows us to show a combinatorial algorithm, that given any set $\mset'$ of demand pairs with distance property, routes them via node-disjoint paths.
We then show that if the optimum solution of \WGPwB instance $\iset'$ has value $\beta$, then there is a subset of roughly $\beta$ demand pairs in the \NDPgrid instance $\iset$ with this distance property; and hence, $\iset$ has a solution of value at least roughly $\beta$.

For the second part, we are given a node-disjoint routing of a subset $\mset' \subseteq \mset$ of $\beta$ demand pairs of $\iset$.
Our goal is to show that the original \WGPwB instance $\iset'$ has a solution of value roughly $\min{(\beta, r h)}$.
We consider the underlying graph $G'$ of $\iset'$ and let $E'$ be the subset of its edges corresponding to the demand pairs in $\mset'$.
We discard all but the edges $E'$ from $G'$.
Exploiting the fact that we have a node-disjoint routing of demand pairs in $\mset'$ on a grid, and from our careful placement of terminals on rows of the grid, we show that we can draw the graph $G'$ with relatively few crossings.
We show that given such a drawing of $G'$ with few crossings, we can efficiently compute a good balanced cut of $G'$, where we divide $G'$ into two subgraphs $G'_1$ and $G'_2$, such that there are only a small number of cut edges and the number of edges in $G'_1$ and $G'_2$ are roughly $|E'|/2 = \beta/2$.
We repeat this process recursively until we are left with $r$ subgraphs, each consisting of roughly $\beta/r$ edges.
We also show that we can perform this step while ensuring that each subgraph has at most single edge from each bundle.
Thus, we have computed a feasible solution to the \WGPwB instance $\iset'$ of value roughly $\min{(\beta, r h)}$.

Combining both these parts together, we show that if there is an efficient $\alpha$-approximation algorithm for \NDPgrid, there is an efficient $\tilde \Omega(\alpha)$-approximation algorithm for \WGPwB.

\paragraph{Hardness of \WGPwB.}

The high level overview of the reduction to show that \WGPwB itself is hard, is as follows.
We perform a reduction from a known \NP-Complete problem \threecol, which is a special case of the $3$-coloring problem, where the underlying graph is $5$-regular.
It is well known that there is some constant $\eps$, such that it is \NP-Hard to distinguish between its \yi (the input graph is vertex-colorable by three colors such that no edge is monochromatic) and \ni (for every three-coloring, at least an $\eps$-fraction of the edges are monochromatic)~\cite{Feige3COL5}.
We start with a \threecol instance $G^*$ and a parameter $\ell > 0$, that depends on the approximation factor $\alpha$ that we are considering.
We construct a bipartite constraint graph $H$ with the following properties.
The vertices of $H$ on one side correspond to $\ell$-tuples of edges of $G^*$; and the vertices on its other side correspond to $\ell$-tuples of vertices of $G^*$.
There is an edge between two vertices if the other vertex corresponds to a tuple of endpoints of the edges corresponding to the first vertex.
We then consider a `cover-graph' $L(H)$, where each vertex of $H$ is replaced by a cloud of vertices taking into account potential feasible colorings of the original graph $G^*$.
We also analogously define cover-graph $L(H')$ for each subgraph $H'$ of $H$.
For each subgraph $H'$ of $H$, we think of this cover-graph $L(H')$ as an instance of \WGPwB for some suitable choice of parameters $r$ and $h$ depending upon $\ell$ and the size of the graph $H'$.

We show that if $G^*$ is a \yi, the corresponding \WGPwB instance $L(H)$ has a perfect solution.
In fact, we can also show that if $G^*$ is a \yi, for every subgraph $H'$ of $H$, the corresponding \WGPwB instance $L(H')$ has a perfect solution.
Ideally we would like to show that if $G^*$ is a \ni, the \WGPwB instance has small optimal solution.
Indeed if that is the case, we can then use the assumed efficient approximation algorithm $\aset'$ for \WGPwB to distinguish between the \yi and \ni of \threecol problem, proving that approximating \WGPwB is \NP-Hard.
Unfortunately, it might still be possible that $L(H)$ has a large feasible solution even though $G^*$ is a \ni.

To overcome this issue, we employ an iterative Cook-type reduction.
In every iteration $j$, we reduce the \threecol instance $G^*$ to a collection $\iset_j$ of instances of \WGPwB, corresponding to a collection $\hset_j$ of the subgraphs of $H$; and apply the algorithm $\aset'$ to each of them.
If $G^*$ is a \YI, then we show that each resulting instance of \WGPwB has perfect solution, and so all solutions returned by $\aset'$ are large.
If on the other hand, $G^*$ is a \NI, it is still possible that the resulting instances of \WGPwB will have large solutions, or even perfect solutions.
However, we can use these resulting solutions in order to further refine our reduction, and construct a new collection $\iset_{j+1}$ of instances of \WGPwB, corresponding to a fresh collection $\hset_{j+1}$ of the subgraphs of $H$.
While in the \YI case we will continue to obtain large solutions to all \WGPwB instances that we construct, we can show that in the \NI case, in some iteration of the algorithm, we will fail to find such a large solution.
Our reduction is crucially  sequential, and we exploit the solutions returned by algorithm $\aset'$ in previous iterations in order to construct new instances of \WGPwB for the subsequent iterations.

We show that the running time of our reduction is polynomial in the size of $H$: which itself scales as $N^{\ell + O(1)}$, where $N$ is the number of vertices in the original \threecol instance $G^*$.
Recall that the choice of $\ell$ depends upon the approximation factor $\alpha$.
Performing tradeoffs between the different values for $\alpha$ and the resulting running time, we obtain the guarantees of \Cref{thm: ndp-hard-master NDP}.

\subsection{Organization}
We start with Preliminaries in \Cref{subsec: ndp-hard-prelims}, and introduce the new graph partitioning problems in \Cref{subsec: ndp-hard-WGP}.
The hardness of approximation proof for \NDPgrid appears in \Cref{subsec: ndp-hard-the hardness proof}, with the reduction from the graph partitioning problem to \NDPgrid deferred to \Cref{subsec: ndp-hard-from WGP to NDP}.
Finally, we extend our hardness results to \NDPwall and \EDPwall in \Cref{subsec: ndp-hard-from NDP to EDP}.

\toggletrue{ndp-hard}

    \section{Preliminaries}\label{subsec: ndp-hard-prelims}

We use standard graph-theoretic notation. Given a graph $G$ and a subset $W\subseteq V(G)$ of its vertices, $E(W)$ denotes the set of all edges of $G$ that have both their endpoints in $W$. Given a path $P$ and a subset $U$ of vertices of $G$, we say that $P$ is \emph{internally disjoint} from $U$ iff every vertex in $P\cap U$ is an endpoint of $P$. Similarly, $P$ is internally disjoint from a subgraph $G'$ of $G$ iff $P$ is internally disjoint from $V(G')$. Given a subset $\mset'\subseteq\mset$ of the demand pairs in $G$, we denote by $S(\mset')$ and $T(\mset')$ the sets of the source and the destination vertices of the demand pairs in $\mset'$, respectively. We let $\tset(\mset')=S(\mset')\cup T(\mset')$ denote the set of all terminals participating as a source or a destination in $\mset'$. All logarithms in this paper are to the base of 2.

\paragraph{Grid Graphs.}
For a pair  $h,\ell> 0$ of integers, we let $G^{h,\ell}$ denote the grid of height $h$ and length $\ell$.
The set of its vertices is $V(G^{h,\ell})=\set{v(i,j)\mid 1\leq i\leq h, 1\leq j\leq \ell}$, and the set of its edges is the union of two subsets: the set $E^H=\set{(v_{i,j},v_{i,j+1})\mid 1\leq i\leq h, 1\leq j<\ell}$ of horizontal edges  and the set $E^{V}=\set{(v_{i,j},v_{i+1,j})\mid 1\leq i< h, 1\leq j\leq\ell}$ of vertical edges. The subgraph of $G^{h,\ell}$ induced by the edges of $E^H$ consists of $h$ paths, that we call the \emph{rows} of the grid; for $1\leq i\leq h$, the $i$th row $R_i$ is the row containing the vertex $v(i,1)$. Similarly, the subgraph induced by the edges of $E^V$ consists of $\ell$ paths that we call the \emph{columns} of the grid, and for $1\leq j\leq \ell$, the $j$th column $W_j$ is the column containing $v(1,j)$. We think of the rows  as ordered from top to bottom and the columns as ordered from left to right. Given a vertex $v=v(i,j)$ of the grid, we denote by $\row(v)$ and $\col(v)$ the row and the column of the grid, respectively, that contain $v$. We say that $G^{h,\ell}$ is a \emph{square grid} iff $h=\ell$. The \emph{boundary of the grid} is $R_1\cup R_h\cup W_1\cup W_{\ell}$. We sometimes refer to $R_1$ and $R_h$ as the top and the bottom boundary edges of the grid respectively, and to $W_1$ and $W_{\ell}$ as the left and the right boundary edges of the grid. 

Given a subset $\rset'$ of consecutive rows of $G$ and a subset $\wset'$ of consecutive columns of $G$, the \emph{sub-grid of $G$ spanned by the rows in $\rset'$ and the columns in $\wset'$} is the sub-graph of $G$ induced by the set $\set{v \mid \row(v)\in \rset', \col(v)\in \wset'}$ of its vertices.

Given two vertices $u=v(i,j)$ and $u'=v(i',j')$ of a grid $G$, the shortest-path distance between them is denoted by $d(u,u')$. Given two vertex subsets $X,Y\subseteq V(G)$, the distance between them is $d(X,Y)=\min_{u\in X,u'\in Y}\set{d(u,u')}$. When $H, H'$ are subgraphs of $G$, we use $d(H,H')$ to denote $d(V(H),V(H'))$.

\paragraph{Wall Graphs.}
Let $G=G^{\ell,h}$ be a grid of length $\ell$ and height $h$. Assume that $\ell>0$ is an even integer, and that $h>0$.
For every column $W_j$ of the grid, let $e^j_1,\ldots,e^j_{h-1}$ be the edges of $W_j$ indexed in their top-to-bottom order. Let $E^*(G)\subseteq E(G)$ contain all edges $e^j_z$, where $z\neq j \mod 2$, and let $\hat G$ be the graph obtained from $G\setminus E^*(G)$, by deleting all degree-$1$ vertices from it.
Graph $\hat G$ is called a \emph{wall of length $\ell/2$ and height $h$} (see Figure~\ref{fig: ndp-hard-wall}).
Consider the subgraph of $\hat G$ induced by all horizontal edges of the grid $G$ that belong to $\hat G$. This graph is a collection of $h$ node-disjoint paths, that we refer to as the \emph{rows} of $\hat G$, and denote them by $R_1,\ldots,R_h$ in this top-to-bottom order; notice that $R_j$ is also the $j$th row of the grid $G$ for all $j$. Graph $\hat G$ contains a unique collection $\wset$ of $\ell/2$ node-disjoint paths that connect vertices of $R_1$ to vertices of $R_h$ and are internally disjoint from $R_1$ and $R_h$. We refer to the paths in $\wset$ as the \emph{columns} of $\hat G$, and denote them by $W_1,\ldots,W_{\ell/2}$ in this left-to-right order. Paths $W_1, W_{\ell/2},R_1$ and $R_h$ are called the left, right, top and bottom boundary edges of $\hat G$, respectively, and their union is the boundary of $\hat G$.%

\paragraph{The 3COL(5) problem.}
The starting point of our reduction is the 3COL(5) problem. In this problem, we are given a $5$-regular graph $G=(V,E)$. Note that, if $n=|V|$ and $m=|E|$, then $m=5n/2$.  We are also given a set $\cset=\set{r,b,g}$ of $3$ colors. A \emph{coloring} $\chi: V\rightarrow \cset$ is an assignment of a color in $\cset$ to every vertex  in $V$. 
We say that an edge $e=(u,v)$ is \emph{satisfied} by the coloring $\chi$ iff $\chi(u)\neq \chi(v)$. 
The coloring $\chi$ is \emph{valid} iff it satisfies every edge. We say that $G$ is a \yi iff there is a valid coloring $\chi: V\rightarrow \cset$. We say that it is a \ni with respect to some given parameter $\eps$, iff for every coloring $\chi: V\rightarrow \cset$, at most a $(1-\eps)$-fraction of the edges are satisfied by $\chi$. We use the following theorem of Feige et al.~\cite{Feige3COL5}:

\begin{theorem} \label{thm: ndp-hard-GAPCOL is hard}[Proposition 15  in \cite{Feige3COL5}]
There is some constant $\eps$, such that distinguishing between the \yis and the  \nis (with respect to $\eps$) of 3COL(5) is NP-hard.
\end{theorem}

\paragraph{A Two-Prover Protocol.}
We use the following two-prover protocol. The two provers are called an edge-prover and a vertex-prover. Given a $5$-regular graph $G$, the verifier selects an edge $e=(u,v)$ of $G$ uniformly at random, and then selects a random endpoint (say $v$) of this edge. It then sends $e$ to the edge-prover and $v$ to the vertex-prover. The edge-prover must return an assignment of colors from $\cset$ to $u$ and $v$, such that the two colors are distinct; the vertex-prover must return an assignment of a color from $\cset$ to $v$. The verifier accepts iff both provers assign the same color to $v$. Given a $2$-prover game $\gset$, its \emph{value} is the maximum acceptance probability of the verifier over all possible strategies of the provers.  

Notice that, if $G$ is a \yi, then there is a strategy for both provers that guarantees acceptance with probability $1$: the provers fix a valid coloring $\chi: V\rightarrow \cset$ of $G$ and respond to the queries according to this coloring. 

We claim that if $G$ is a \ni, then for any strategy of the two provers, the verifier accepts with probability at most $(1-\eps/2)$. Note first that we can assume without loss of generality that the strategies of the provers are deterministic. Indeed,  if the provers have a probability distribution over the answers to each query $Q$, then the edge-prover, given a query $Q'$, can return an answer that maximizes the acceptance probability of the verifier under the random strategy of the vertex-prover. This defines a deterministic strategy for the edge-prover that does not decrease the acceptance probability of the verifier. The vertex-prover in turn, given any query $Q$, can return an answer that maximizes the acceptance probability of the verifier, under the new deterministic strategy of the edge-prover. The acceptance probability of this final deterministic strategy of the two provers is at least as high as that of the original randomized strategy. The deterministic strategy of  the vertex-prover defines a coloring of the vertices of $G$. This coloring must dissatisfy at least $\eps m$ edges. The probability that the verifier chooses one of these edge is at least $\eps$. The response of the edge-prover on such an edge must differ from the response of the vertex-prover on at least one endpoint of the edge. The verifier chooses this endpoint with probability at least $\half$, and so overall the verifier rejects with probability at least $\eps/2$. 
Therefore, if $G$ is a \yi, then the value of the corresponding game is $1$, and if it is a \ni, then the value of the game is at most $(1-\eps/2)$.

\paragraph{Parallel Repetition.}
We perform $\ell$ rounds of parallel repetition of the above protocol, for some integer $\ell>0$, that may depend on $n=|V(G)|$. Specifically, the verifier chooses a sequence $(e_1,\ldots,e_{\ell})$ of $\ell$ edges, where each edge $e_i$ is selected independently uniformly at random from $E(G)$. For each chosen edge $e_i$, one of its endpoints $v_i$ is then chosen independently at random. The verifier sends $(e_1,\ldots,e_{\ell})$ to the edge-prover, and $(v_1,\ldots,v_{\ell})$ to the vertex-prover. The edge-prover returns a coloring of both endpoints of each edge $e_i$. This coloring must satisfy the edge (so the two endpoints must be assigned different colors), but it need not be consistent across different edges. In other words, if two edges $e_i$ and $e_j$ share the same endpoint $v$, the edge-prover may assign different colors to each occurrence of $v$. The vertex-prover returns a coloring of the vertices in $(v_1,\ldots,v_{\ell})$. Again, if some vertex $v_i$ repeats twice, the coloring of the two occurrences need not be consistent. The verifier accepts iff for each $1\leq i\leq \ell$, the coloring of the vertex $v_i$ returned by both provers is consistent. (No verification of consistency is performed across different $i$'s. So, for example, if $v_i$ is an endpoint of $e_j$ for $i\neq j$, then it is possible that the two colorings do not agree and the verifier still accepts). We say that a pair $(A,A')$ of answers to the two queries $(e_1,\ldots,e_{\ell})$ and $(v_1,\ldots,v_{\ell})$  is \emph{matching}, or \emph{consistent}, iff it causes the verifier to accept. We let $\gset^\ell$ denote this 2-prover protocol with $\ell$ repetitions.

\begin{theorem}[Parallel Repetition]\cite{RazParallelRep,HolensteinParallelRep,RaoParallelRep}
There is some constant $0<\gamma'<1$, such that for each $2$-prover game $\tilde \gset$, if the value of $\tilde \gset$ is $x$, then the value of the game $\tilde\gset^\ell$, obtained from $\ell>0$ parallel repetitions of $\tilde\gset$, is $x^{\gamma'\ell}$. 
\end{theorem}

\begin{corollary}\label{cor: ndp-hard-parallel repetition}
There is some constant $0<\gamma<1$, such that, if $G$ is a \yi, then $\gset^\ell$ has value $1$, and if $G$ is a \ni, then $\gset^\ell$ has value at most  $2^{-\gamma\ell}$.
\end{corollary}

We now summarize the parameters and introduce some basic notation:

\begin{itemize}
\item Let $\qset^E$ denote the set of all possible queries to the edge-prover, so each query is an $\ell$-tuple of edges. Then $|\qset^E|=m^{\ell}=(5n/2)^{\ell}$. Each query has $6^{\ell}$ possible answers -- $6$ colorings per edge. The set of feasible answers is the same for each edge-query, and we denote it by $\aset^E$.

\item Let $\qset^V$ denote the set of all possible queries to the vertex-prover, so each query is an $\ell$-tuple of vertices. Then $|\qset^V|=n^{\ell}$. Each query has $3^{\ell}$ feasible answers -- $3$ colorings per vertex. The set of feasible answers is the same for each vertex-query, and we denote it by $\aset^V$.

\item We think about the verifier as choosing a number of random bits, that determine the choices of the queries $Q\in \qset^E$ and $Q'\in \qset^V$ that it sends to the provers. We sometimes call each such random choice a ``random string''. The set of all such choices is denoted by $\rset$, where for each $R\in \rset$, we denote $R=(Q,Q')$, with $Q\in \qset^E$, $Q'\in \qset^V$ --- the two queries sent to the two provers when the verifier chooses $R$. Then $|\rset|=(2m)^{\ell}=(5n)^{\ell}$, and each random string $R\in \rset$ is chosen with the same probability.

\item It is important to note that each query $Q=(e_1,\ldots,e_{\ell})\in \qset^E$ of the edge-prover participates in exactly $2^{\ell}$ random strings (one random string for each choice of one endpoint per edge of $\set{e_1,\ldots,e_{\ell}}$), while each query $Q'=(v_1,\ldots,v_{\ell})$ of the vertex-prover participates in exactly $5^{\ell}$ random strings (one random string for each choice of an edge incident to each of $v_1,\ldots,v_{\ell}$).
\end{itemize}

A function $f: \qset^E\cup \qset^V\rightarrow \aset^E\cup \aset^V$ is called a \emph{global assignment of answers to queries} iff for every query $Q\in \qset^E$ to the edge-prover, $f(Q)\in \aset^E$, and for every query $Q'\in\qset^V$ to the vertex-prover, $f(Q')\in \aset^V$.  We say that $f$ is a \emph{perfect global assignment} iff for every random string $R=(Q^E,Q^V)$, $(f(Q^E),f(Q^V))$ is a matching pair of answers. The following simple theorem, whose proof appears in \Cref{appdx: ndp-hard-proof of yi-partition-of-answers thm} of the Appendix, shows that in the \yi, there are many perfect global assignments, that neatly partition all answers to the edge-queries.

\begin{theorem} \label{thm: ndp-hard-yi-partition-of-answers}
Assume that $G$ is a \yi. Then there are $6^{\ell}$ perfect global assignments $f_1,\ldots,f_{6^{\ell}}$ of answers to queries, such  that:

\begin{itemize}
\item for each query $Q\in \qset^E$ to the edge-prover, for each possible answer $A\in \aset^E$, there is exactly one index $1\leq i\leq 6^{\ell}$ with $f_i(Q)=A$; and

\item for each query $Q'\in \qset^V$ to the vertex-prover, for each possible answer $A'\in \aset^V$ to $Q'$, there are exactly $2^{\ell}$ indices $1\leq i\leq 6^{\ell}$, for which $f_i(Q')=A'$.
\end{itemize}
\end{theorem}

\paragraph{Two Graphs.}
Given a 3COL(5) instance $G$ with $|V(G)|=n$, and an integer $\ell>0$, we associate a graph $H$, that we call the \emph{constraint graph}, with it. For every query $Q\in \qset^E\cup \qset^V$, there is a vertex $v(Q)$ in $H$, while for each random string $R=(Q,Q')$, there is an edge $e(R)=(v(Q),v(Q'))$. Notice that $H$ is a bipartite graph. We denote by $U^E$ the set of its vertices corresponding to the edge-queries, and by $U^V$ the set of its vertices corresponding to the vertex-queries. Recall that $|U^E|=(5n/2)^{\ell}$, $|U^V|=n^{\ell}$; the degree of every vertex in $U^E$ is $2^{\ell}$; the degree of every vertex in $U^V$ is $5^{\ell}$, and $|E(H)|=|\rset|=(5n)^{\ell}$.

Assume now that we are given some subgraph $H'\subseteq H$ of the constraint graph. We build a bipartite graph $L(H')$ associated with it (this graph takes into account the answers to the queries; it may be convenient  for now  to think that $H'=H$, but later we will use smaller sub-graphs of $H$). The vertices of $L(H')$ are partitioned into two subsets:

\begin{itemize}
\item For each edge-query $Q\in \qset^E$ with $v(Q)\in H'$, for each possible answer $A\in \aset^E$ to $Q$, we introduce a vertex $v(Q,A)$. We denote by $S(Q)$ the set of these $6^{\ell}$ vertices corresponding to $Q$, and we call them a \emph{group representing $Q$}. We denote by $\hat U^E$ the resulting set of vertices:

\[\hat U^E=\set{v(Q,A)\mid (Q\in \qset^E\mbox{ and } v(Q)\in H'), A\in \aset^E}.\]

\item For each vertex-query $Q'\in \qset^V$ with $v(Q')\in H'$, for each possible answer $A'\in \aset^V$ to $Q'$, we introduce $2^{\ell}$ vertices $v_1(Q',A'),\ldots,v_{2^{\ell}}(Q',A')$. We call all these vertices \emph{the copies of answer $A'$ to query $Q'$}. We denote by $S(Q')$ the set of all vertices corresponding to $Q'$:

\[S(Q')=\set{v_i(Q',A')\mid A'\in \aset^V, 1\leq i\leq 2^{\ell}},\]
 
 so $|S(Q')|=6^{\ell}$. We call $S(Q')$ \emph{the group representing $Q'$}. We denote by $\hat U^V$ the resulting set of vertices:
 
 \[\hat U^V=\set{v_i(Q',A')\mid  (Q'\in \qset^V\mbox{ and } v(Q')\in H'), A'\in \aset^V, 1\leq i\leq 2^{\ell}}.\]
 
\end{itemize}

The final set of vertices of $L(H')$ is $\hat U^E\cup \hat U^V$. We define the set of edges of $L(H')$ as follows. For each random string $R=(Q^E,Q^V)$ whose corresponding edge $e(R)$ belongs to $H'$, for every answer $A\in \aset^E$ to $Q^E$, let $A'\in \aset^V$ be the unique answer to $Q^V$ consistent with $A$. For each copy $v_i(Q^V,A')$ of answer $A'$ to query $Q^V$, we add an edge $(v(Q^E,A),v_i(Q^V,A'))$. Let 

\[E(R)=\set{(v(Q^E,A),v_i(Q^V,A'))\mid A\in \aset^E, A'\in \aset^V, \mbox{ $A$ and $A'$ are consistent answers to $R$}, 1\leq i\leq 2^{\ell}}\]

be the set of the resulting edges, so $|E(R)|=6^{\ell}\cdot 2^{\ell}=12^{\ell}$. We denote by $\hat E$ the set of all edges of $L(H')$ --- the union of the sets $E(R)$ for all random strings $R$ with $e(R)\in H'$.

Recall that we have defined a partition of the set $\hat U^E$ of vertices into groups $S(Q)$ --- one group for each query $Q\in \qset^E$ with $v(Q)\in H'$. We denote this partition by $\uset_1$. Similarly, we have defined a partition of $\hat U^V$ into groups, that we denote by $\uset_2=\set{S(Q')\mid Q'\in \qset^V\mbox{ and } v(Q')\in H'}$.
Recall that for each group $U\in \uset_1\cup\uset_2$, $|U|=6^{\ell}$.

Finally, we need to define bundles of edges in graph $L(H')$. 
For every vertex $v\in \hat U^E\cup \hat U^V$, we define a partition $\bset(v)$ of the set of all edges incident to $v$ in $L(H')$ into bundles, as follows. Fix some group $U\in \uset_1\cup \uset_2$ that we have defined. If there is at least one edge of $L(H')$ connecting $v$ to the vertices of $U$, then we define a bundle containing all edges connecting $v$ to the vertices of $U$, and add this bundle to $\bset(v)$. Therefore, if $v\in S(Q)$, then for each random string $R$ in which $Q$ participates, with $e(R)\in H'$, we have defined one bundle of edges in $\bset(v)$. For each vertex $v\in \hat U^E\cup \hat U^V$, the set of all edges incident to $v$ is thus partitioned into a collection of bundles, that we denote by $\bset(v)$, and we denote $\beta(v)=|\bset(v)|$. Note that, if $v\in S(Q)$ for some query $Q\in \qset^E\cup \qset^V$, then $\beta(v)$ is exactly the degree of the vertex $v(Q)$ in graph $H'$. Note also that $\bigcup_{v\in V(H')}\bset(v)$ does not define a partition of the edges of $\hat E$, as each such edge belongs to two bundles. However, each of $\bigcup_{v\in \hat U^E}\bset(v)$ and $\bigcup_{v\in \hat U^V}\bset(v)$ does define a partition of $\hat E$.
It is easy to verify that every bundle that we have defined contains exactly $2^{\ell}$ edges.

    \section{The \WGPwBfull}\label{subsec: ndp-hard-WGP}
    We will use a graph partitioning problem as a proxy in order to reduce the \threecol problem to \NDPgrid. The specific graph partitioning problem is somewhat complex. We first define a simpler variant of this problem, and then provide the intuition and the motivation for the more complex variant that we eventually use. 

In the basic \WGPfull, that we denote by \WGP, we are given a bipartite graph $\tG=(V_1,V_2,E)$ and two integral parameters $h,r>0$. A solution consists of a partition $(W_1,\ldots,W_r)$ of $V_1\cup V_2$ into $r$ subsets, and for each $1\leq i\leq r$, a subset $E_i\subseteq E(W_i)$ of edges, such that $|E_i|\leq h$. The goal is to maximize $\sum_i|E_i|$.

One intuitive way to think about the \WGP problem is that we would like to partition the vertices of $\tG$ into $r$ clusters, that are roughly balanced (in terms of the number of edges in each cluster). However, unlike the standard balanced partitioning problems, that attempt to minimize the number of edges connecting the different clusters, our goal is to maximize the total number of edges that remain in the clusters. We suspect that the \WGP problem is very hard to approximate; in particular it appears to be somewhat similar to the Densest $k$-Subgraph problem (\DkS). Like in the \DkS problem, we are looking for dense subgraphs of $\tG$ (the subgraphs $\tG[W_i]$), but unlike \DkS, where we only need to find one such dense subgraph, we would like to partition all vertices of $\tG$ into a prescribed number of dense subgraphs. We can prove that \NDPgrid is at least as hard as \WGP (to within polylogarithmic factors; see below), but unfortunately we could not prove strong hardness of approximation results for \WGP. In particular, known hardness proofs for \DkS do not seem to generalize to this problem. To overcome this difficulty,  we define a slightly more general problem, and then use it as a proxy in our reduction.
Before defining the more general problem, we start with intuition.

\paragraph{Intuition:} Given a \threecol instance $G$, we can construct the graph  $H$, and the graph $L(H)$, as described above. We  can then view $L(H)$ as an instance of \WGP, with $r=6^{\ell}$ and $h=|\rset|$. Assume that $G$ is a \yi. Then we can use the perfect global assignments $f_1,\ldots,f_r$ of answers to the queries, given by \Cref{thm: ndp-hard-yi-partition-of-answers}, in order to partition the vertices of $L(H)$ into $r=6^{\ell}$ clusters $W_1,\ldots,W_r$, as follows. Fix some $1\leq i\leq r$. For each query $Q\in \qset^E$ to the edge-prover, set $W_i$ contains a single vertex $v(Q,A)\in S(Q)$, where $A=f_i(Q)$. For each query $Q'\in \qset^V$ to the vertex-prover, set $W_i$ contains a single vertex $v_j(Q',A')$, where $A'=f_i(Q')$, and the indices $j$ are chosen so that every vertex $v_j(Q',A')$ participates in exactly one cluster $W_i$. From the construction of the graph $L(H)$ and the properties of the assignments $f_i$ guaranteed by \Cref{thm: ndp-hard-yi-partition-of-answers}, we indeed obtain a partition $W_1,\ldots,W_r$ of the vertices of $L(H)$. For each $1\leq i\leq r$, we then set $E_i=E(W_i)$. Notice that for every query $Q\in \qset^E\cup \qset^V$, exactly one vertex of $S(Q)$ participates in each cluster $W_i$. Therefore, for each group $U\in \uset_1\cup \uset_2$, each cluster $W_i$ contains exactly one vertex from this group. It is easy to verify that for each $1\leq i\leq r$, for each random string $R\in \rset$, set $E_i$ contains exactly one edge of $E(R)$, and so $|E_i|=|\rset|=h$, and the solution value is $h\cdot r$. Unfortunately, in the \ni, we may still obtain a solution of a high value, as follows: instead of distributing, for each query $Q\in \qset^E\cup \qset^V$, the vertices of $S(Q)$ to different clusters $W_i$, we may put all vertices of $S(Q)$ into a single cluster. While in our intended solution to the \WGP problem instance each cluster can be interpreted as an assignment of answers to the queries, and the number of edges in each cluster is bounded by the number of random strings satisfied by this assignment, we may no longer use this interpretation with this new type of solutions\footnote{We note that a similar problem arises if one attempts to design naïve hardness of approximation proofs for \DkS.}. Moreover, unlike in the \yi solutions,
if we now consider some cluster $W_i$, and some random string $R\in \rset$, we may add several edges of $E(R)$ to $E_i$, which will further allow us to accumulate a high solution value. One way to get around this problem is to impose additional restrictions on the feasible solutions to the \WGP problem, which are consistent with our \yi solution, and thereby obtain a more general (and hopefully more difficult) problem. But while doing so we still need to ensure that we can prove that \NDPgrid remains at least as hard as the newly defined problem. Recall the definition of bundles in graph $L(H)$. It is easy to verify that in our intended solution to the \yi, every bundle contributes at most one edge to the solution. This motivates our definition of a slight generalization of the \WGP problem, that we call \WGPwBfull, or \WGPwB.

The input to \WGPwB problem is almost the same as before: we are given a bipartite graph $\tG=(V_1,V_2,E)$, and two integral parameters $h,r>0$. Additionally, we are given a partition $\uset_1$ of $V_1$ into groups, and a partition $\uset_2$ of $V_2$ into groups, so that for each $U\in \uset_1\cup \uset_2$, $|U|=r$.
Using these groups, we define bundles of edges as follows: for every vertex $v\in V_1$, for each group $U\in \uset_2$, such that some edge of $E$ connects $v$ to a vertex of $U$, the set of all edges that connect $v$ to the vertices of $U$ defines a single bundle. Similarly, for every vertex $v\in V_2$, for each group $U\in \uset_1$, all edges that connect $v$ to the vertices of $U$ define a bundle. We denote, for each vertex $v\in V_1\cup V_2$, by $\bset(v)$ the set of all bundles into which the edges incident to $v$ are partitioned, and we denote by $\beta(v)=|\bset(v)|$ the number of such  bundles. We also denote by $\bset=\bigcup_{v\in V_1\cup V_2}\bset(v)$ -- the set of all bundles. Note that as before, $\bset$ is not a partition of $E$, but every edge of $E$ belongs to exactly two bundles: one bundle in $\bigcup_{v\in V_1}\bset(v)$, and one bundle in $\bigcup_{v\in V_2}\bset(v)$.
As before, we need to compute a partition $(W_1,\ldots,W_r)$ of $V_1\cup V_2$ into $r$ subsets, and for each $1\leq i\leq r$, select a subset $E_i\subseteq E(W_i)$ of edges, such that $|E_i|\leq h$.  But now there is an additional restriction: we require that for each $1\leq i\leq r$, for every bundle $B\in \bset$, $E_i$ contains at most one edge $e\in B$. As before, the goal is to maximize $\sum_i|E_i|$.

\paragraph{Valid Instances and Perfect Solutions.}

Given an instance $\iset=(\tG=(V_1,V_2,E), \uset_1,\uset_2,h,r)$ of \WGPwB, let $\beta^*(\iset)=\sum_{v\in V_1}\beta(v)$. Note that for any solution to $\iset$, the 
solution value must be bounded by $\beta^*(\iset)$, since for every vertex $v\in V_1$, for every bundle $B\in \bset(v)$, at most one edge from the bundle may contribute to the solution value. In all instances of \WGPwB that we consider, we always set $h=\beta^*(\iset)/r$. Next, we define valid instances; they are defined so that the instances that we obtain when reducing from \threecol are always valid, as we show later.

\begin{definition}
We say that instance $\iset$ of \WGPwB is \emph{valid} iff $h=\beta^*(\iset)/r$ and $h\geq \max_{v\in V_1\cup V_2}\set{\beta(v)}$.
\end{definition}

Recall that for every group $U\in \uset_1\cup \uset_2$, $|U|=r$. We now define perfect solutions to the \WGPwB problem. We will ensure that our intended solutions in the \yi are always perfect, as we show later.

\begin{definition}
We say that a solution $((W_1,\ldots,W_r),(E_1,\ldots,E_r))$ to a valid \WGPwB instance $\iset$ is \emph{perfect} iff:

\begin{itemize}
\item For each group $U\in \uset_1\cup \uset_2$, exactly one vertex of $U$ belongs to each cluster $W_i$;  and
\item For each $1\leq i\leq r$, $|E_i|=h$.
\end{itemize}
\end{definition}

Note that the value of a perfect solution to a valid instance $\iset$ is $h\cdot r=\beta^*(\iset)$, and this is the largest value that any solution can achieve.

\subsection{From \threecol to \WGPwB}

 Suppose we are given an instance $G$ of the  \threecol problem, and an integral parameter $\ell>0$ (the number of repetitions). Consider the corresponding constraint graph $H$, and suppose we are given some subgraph $H'\subseteq H$. We define an instance $\iset(H')$ of \WGPwB, as follows.

\begin{itemize}
\item The underlying graph is $L(H')=(\hat U^E,\hat U^V,\hat E)$;

\item The parameters are $r=6^{\ell}$ and $h=|E(H')|$;

\item The partition $\uset_1$ of $\hat U^E$ is the same as before: the vertices of $\hat U^E$ are partitioned into groups $S(Q)$ --- one group for each query $Q\in \qset^E$ with $v(Q)\in V(H')$. Similarly, the partition $\uset_2$ of $\hat U^V$ into groups is also defined exactly as before, and contains, for each query $Q'\in \qset^V$ with $v(Q')\in V(H')$, a group $S(Q')$. (Recall that for all $Q\in \qset^E\cup \qset^V$ with $v(Q)\in H'$, $|S(Q)|=6^{\ell}$).
\end{itemize}

\begin{claim}\label{clm: ndp-hard-yi of 3col gives perfect solution to WGP}
Let $G$ be an instance of the \threecol problem, $\ell>0$ an integral parameter, and $H'\subseteq H$ a subgraph of the corresponding constraint graph. Consider the corresponding instance $\iset(H')$ of \WGPwB.  Then $\iset(H')$ is a valid instance, and moreover, if $G$ is a \yi, then there is a perfect solution to $\iset(H')$.
\end{claim}

\begin{proof}
We first verify that $\iset(H')$ is a valid instance of \WGPwB. Recall that for a query $Q\in \qset^E$ to the edge-prover and an answer $A\in \aset^E$, the number of bundles incident to vertex $v(Q,A)$ in $L(H')$ is exactly the degree of the vertex $v(Q)$ in graph $H'$. The total number of bundles incident to the vertices of $S(Q)$ is then the degree of $v(Q)$ in $H'$ times $|\aset^E|$. Therefore, $\beta^*(\iset)=\sum_{v(Q,A)\in \hat U^E}|\beta(v)|=|E(H')|\cdot |\aset^E|=h\cdot r$. It is now immediate to verify that $h=\beta^*(\iset)/r$. Similarly, for a vertex $v=v_j(Q',A')\in U^V$, the number of bundles incident to $v$ is exactly the degree of $v$ in $H'$. Since $h=|E(H')|$, we get that $h\geq \max_{v\in V_1\cup V_2}\set{\beta(v)}$, and so $\iset(H')$ is a valid instance. 

Assume now that $G$ is a \yi. We  define a perfect solution $((W_1,\ldots,W_r),(E_1,\ldots,E_r))$ to this instance.
Let $\set{f_1,f_2,\ldots,f_{6^\ell}}$ be the collection of perfect global assignments of answers to the queries, given by \Cref{thm: ndp-hard-yi-partition-of-answers}. Recall that $\uset_1=\set{S(Q)\mid Q\in\qset^E, v(Q)\in H'}$ and $\uset_2=\set{S(Q')\mid Q'\in\qset^V, v(Q')\in H'}$, where each group in $\uset_1\cup \uset_2$ has cardinality $r=6^{\ell}$. We now fix some $1\leq i\leq r$, and define the set $W_i$ of vertices. For each query $Q\in \qset^E$ to the edge-prover, if $A=f_i(Q)$, then we add the vertex $v(Q,A)$ to $W_i$. For each query $Q'\in \qset^V$ to the vertex-prover, if $A'=f_i(Q')$, then we select some index $1\leq j\leq 2^{\ell}$, and add the vertex $v_j(Q',A')$ to $W_i$. The indices $j$ are chosen so that every vertex $v_j(Q',A')$ participates in at most one cluster $W_i$. From the construction of the graph $L(H')$ and the properties of the assignments $f_i$ guaranteed by \Cref{thm: ndp-hard-yi-partition-of-answers}, it is easy to verify that $W_1,\ldots,W_r$ partition the vertices of $L(H')$, and moreover, for each group $S(Q)\in \uset_1\cup \uset_2$, each set $W_i$ contains exactly one vertex of $S(Q)$. 

Finally, for each $1\leq i\leq r$, we set $E_i=E(W_i)$. We claim that for each bundle $B\in \bset$, set $E_i$ may contain at most one edge of $B$. Indeed, let $v\in W_i$ be some vertex, let $U\in \uset_1\cup \uset_2$ be some group, and let $B$ be the bundle containing all edges that connect $v$ to the vertices of $U$. Since $W_i$ contains 
 exactly one vertex of $U$, at most one edge of $B$ may belong to $E_i$.

It now remains to show that $|E_i|=h$ for all $i$. Fix some $1\leq i\leq r$. It is easy to verify that for each random string $R=(Q,Q')$ with $e(R)\in H'$, set $W_i$ contains a pair of vertices $v(Q,A)$, $v_j(Q',A')$, where $A$ and $A'$ are matching answers to $Q$ and $Q'$ respectively, and so the corresponding edge connecting this pair of vertices in $L(H')$ belongs to $E_i$. Therefore, $|E_i|=|E(H')|=h$.
\end{proof}

\subsection{From \WGPwB to \NDPgrid} 

The following definition will be useful for us later, when we extend our results to \NDP and \EDP on wall graphs.

\begin{definition}
Let $\pset$ be a set of paths in a grid $\hat G$.
We say that $\pset$ is a \emph{spaced-out} set iff for each pair $P,P' \in \pset$ of paths, $d(V(P),V( P')) \geq 2$, and all paths in $\pset$ are internally disjoint from the boundaries of the grid $\hat G$.    
\end{definition}

Note that if $\pset$ is a set of paths that is spaced-out, then all paths in $\pset$ are mutually node-disjoint.
The following theorem is central to our hardness of approximation proof.

\begin{theorem}\label{thm: ndp-hard-from WGP to NDP}
There  is a constant $c^*>0$, and there is an efficient randomized algorithm, that, given a valid instance $\iset=(\tilde G, \uset_1,\uset_2,h,r)$ of \WGPwB with $|E(\tilde G)|=M$, constructs an instance $\hat{\iset}=(\hat G,\mset)$ of \NDPgrid with $|V(\hat G)|=O(M^4\log^2M)$, such that the following hold:

\begin{itemize}
\item If $\iset$ has a perfect solution (of value $\beta^*=\beta^*(\iset)$), then with probability at least $\half$ over the construction of $\hat \iset$, instance $\hat\iset$ has a solution $\pset$ that routes at least $\frac{\beta^*}{c^*\log^3M}$ demand pairs, such that the paths in $\pset$ are spaced-out; and

\item There is a deterministic efficient algorithm, that, given a solution $\pset^*$ to the \NDPgrid problem instance $\hat{\iset}$, constructs a solution to the \WGPwB instance $\iset$, of value at least $\frac{|\pset^*|}{c^*\cdot \log^3M}$.
\end{itemize}
\end{theorem}

We note that the theorem is slightly stronger than what is needed in order to prove hardness of approximation of \NDPgrid: if $\iset$ has a perfect solution, then it is sufficient to ensure that the set $\pset$ of paths in the corresponding \NDPgrid instance $\hat \iset$ is node-disjoint. But we will use the stronger guarantee that it is spaced-out when extending our results to \NDP and \EDP in wall graphs. Note that in the second assertion we are only guaranteed that the paths in $\pset^*$ are node-disjoint.
The proof of the theorem is somewhat technical and is deferred to \Cref{subsec: ndp-hard-from WGP to NDP}.

Assume now that we are given an instance $G$ of \threecol, an integral parameter $\ell>0$, and a subgraph $H'\subseteq H$ of the corresponding constraint graph. Recall that we have constructed a corresponding instance $\iset(H')$ of \WGPwB. We can then use \Cref{thm: ndp-hard-from WGP to NDP} to construct a (random) instance of \NDPgrid, that we denote by $\hat{\iset}(H')$.
Note that $|E(L(H'))|\leq 2^{O(\ell)}\cdot |E(H)|\leq n^{O(\ell)}$. Let $\hat c$ be a constant, such that $|E(L(H'))|\leq n^{\hat c\ell}$ for all $H'\subseteq H$; we can assume w.l.o.g. that $\hat c>1$. We can also assume w.l.o.g. that $c^*\geq 1$, where $c^*$ is the constant from \Cref{thm: ndp-hard-from WGP to NDP}, and we denote $\cyi=(\hat c\cdot c^*)^3$.
We obtain the following immediate corollary of \Cref{thm: ndp-hard-from WGP to NDP}:

\begin{corollary}\label{cor: ndp-hard-YI for 3COL has good routing}
Suppose we are given \threecol instance $G$ that is a \yi, an integer $\ell>0$, and a subgraph $H'\subseteq H$ of the corresponding constraint graph. Then with probability at least $\half$, instance $\hat{\iset}(H')$ of \NDPgrid
has a solution of value at least $\frac{|E(H')|\cdot 6^{\ell}}{\cyi\ell^3\log^3n}$, where $n=|V(G)|$. (The probability is over the random construction of $\hat\iset(H')$).
\end{corollary}

\begin{proof}
From \Cref{clm: ndp-hard-yi of 3col gives perfect solution to WGP}, instance $\iset(H')=(L(H'), \uset_1,\uset_2,r,h)$ of \WGPwB is a valid instance, and it has a perfect solution, whose value must be $\beta^*=\beta^*(\iset(H'))=h\cdot r=|E(H')|\cdot 6^{\ell}$. From \Cref{thm: ndp-hard-from WGP to NDP}, with probability at least $1/2$, instance $\hat \iset(H')$ of \NDPgrid has a solution of value at least 
$\frac{|E(H')|\cdot 6^{\ell}}{c^*\log^3M}$, where $M=|E(L(H'))|$. Since $\log M\leq \hat c \ell\log n$, and the corollary follows.
\end{proof}

    \section{The Hardness Proof}\label{subsec: ndp-hard-the hardness proof}
    Let $G$ be an input instance of 3COL(5). 
Recall that $\gamma$ is the absolute constant from the Parallel Repetition Theorem (\Cref{cor: ndp-hard-parallel repetition}).
We will set the value of the parameter $\ell$ later, ensuring that $\ell>\log^2 n$, where $n=|V(G)|$.
Let $\alpha^*=2^{\Theta(\ell/\log n)}$ be the hardness of approximation factor that we are trying to achieve.

Given the tools developed in the previous sections, a standard way to prove hardness of \NDPgrid would work as follows. Given an instance $G$ of 3COL(5) and the chosen parameter $\ell$, construct the corresponding graph $H$ (the constraint graph), together with the graph $L(H)$. We then construct an  instance $\iset(H)$ of \WGPwB as described in the previous section, and convert it into an instance $\hat{\iset}(H)$ of \NDPgrid. 

We note that, if $G$ is a \yi, then from \Cref{cor: ndp-hard-YI for 3COL has good routing}, with constant probability there is a solution to $\hat \iset(H)$ of value $\frac{|\rset|\cdot 6^{\ell}}{\cyi\ell^3\log^3n}$. 
Assume now that $G$ is a \ni. If we could show that any solution to the corresponding \WGPwB instance $\iset(H)$ has value less than $\frac{|\rset|\cdot 6^{\ell}}{\cyi^2\cdot \alpha^*\ell^6\log^6n}$, we would be done. Indeed, in such a case,  from \Cref{thm: ndp-hard-from WGP to NDP}, every solution to the \NDPgrid instance $\hat\iset(H)$ routes fewer than $\frac{ |\rset|\cdot 6^{\ell}}{\cyi\alpha^*\ell^3\log^3n}$ demand pairs. If we assume for contradiction that an $\alpha^*$-approximation algorithm exists for \NDPgrid, then, if $G$ is a \yi, the algorithm would have to return a solution to $\hat\iset(H)$ routing at least $\frac{|\rset|\cdot 6^{\ell}}{\cyi\alpha^*\ell^3\log^3n}$ demand pairs, while, if $G$ is a \ni, no such solution would exist. Therefore, we could use the $\alpha^*$-approximation algorithm for \NDPgrid to distinguish between the \yis and the \nis of 3COL(5).

Unfortunately, we are unable to prove this directly. Our intended solution to the \WGPwB instance $\iset(H)$, defined over the graph $L(H)$, for each query $Q\in \qset^E\cup\qset^V$, places every vertex of $S(Q)$ into a distinct cluster. Any such solution will indeed have a low value in the \ni. But a cheating solution may place many vertices from the same set $S(Q)$ into some cluster $W_j$. Such a solution may end up having a high value, but it may not translate into a good strategy for the two provers, that satisfies a large fraction of the random strings. In an extreme case, for each query $Q$, we may place all vertices of $S(Q)$ into a single cluster $W_i$. The main idea in our reduction is to overcome this difficulty by noting that such a cheating solution can be used to compute a partition of the constraint graph $H$. The graph is partitioned into as many as $6^{\ell}$ pieces, each of which is significantly smaller than the original graph $H$. At the same time, a large fraction of the edges of $H$ will survive the partitioning procedure. Intuitively, if we now restrict ourselves to only those random strings $R\in \rset$, whose corresponding edges have survived the partitioning procedure, then the problem does not become significantly easier, and we can recursively apply the same argument to the resulting subgraphs of $H$. We make a significant progress in each such iteration, since the sizes of the resulting sub-graphs of $H$ decrease very fast.
The main tool that allows us to execute this plan is the following theorem.

\begin{theorem}\label{thm: ndp-hard-good strategy or partition}
Suppose we are given an instance $G$ of the 3COL(5) problem with $|V(G)|=n$, and an integral parameter $\ell>\log^2 n$, together with some subgraph $H'\subseteq H$ of the corresponding constraint graph $H$, and a parameter $P>1$. Consider the corresponding instance $\iset(H')$ of \WGPwB, and assume that we are given a solution to this instance of value at least $|E(H')|\cdot 6^{\ell}/\alpha$, where $\alpha=\cyi^2\cdot \alpha^*\cdot \ell^6\log^6n$. Then there is a  randomized algorithm whose running time is $O\left (n^{O(\ell)}\cdot  \log P\right )$, that returns one of the following:

\begin{itemize}
\item Either a randomized strategy for the two provers that satisfies, in expectation, more than a $2^{-\gamma \ell/2}$-fraction of the constraints $R\in \rset$ with $e(R)\in E(H')$; or

\item A collection $\hset$ of disjoint sub-graphs of $H'$, such that for each $H''\in \hset$, $|E(H'')|\leq |E(H')|/2^{\gamma\ell/16}$, and with probability at least $(1-1/P)$, $\sum_{H''\in \hset}|E(H'')|\geq \frac{c'|E(H')|}{\ell^2 \alpha^2}$, for some universal constant $c'$.
\end{itemize}
\end{theorem}

We postpone the proof of the theorem to the following subsection, after we complete the hardness proof for \NDPgrid.
We assume for contradiction that we are given a factor-$\alpha^*$ approximation algorithm $\aset$ for \NDPgrid (recall that $\alpha^*=2^{\Theta(\ell/\log n)}$). We will use this algorithm to distinguish between the \yis and the \nis of 3COL(5). Suppose we are given an instance $G$ of 3COL(5).

For an integral parameter $\ell>\log^2 n$, let $H$ be the constraint graph corresponding to $G$ and $\ell$. We next show a randomized algorithm, that uses $\aset$ as a subroutine, in order to determine whether $G$ is a \yi or a \ni. The running time of the algorithm is $n^{O(\ell)}$. %

Throughout the algorithm, we maintain a collection $\hset$ of sub-graphs of $H$, that we sometimes call clusters. Set $\hset$ is in turn partitioned into two subsets: set $\hset_1$ of active clusters, and set $\hset_2$ of inactive clusters. Consider now some inactive cluster $H'\in \hset_2$. This cluster defines a $2$-prover game $\gset(H')$, where the queries to the two provers are $\set{Q^E\in \qset^E \mid v(Q^E)\in V(H')}$, and $\set{Q^V\in \qset^V \mid v(Q^V)\in V(H')}$ respectively, and the constraints of the verifier are $\rset(H')=\set{R\in \rset\mid e(R)\in E(H')}$. For each inactive cluster $H'\in \hset_2$, we will store a (possibly randomized) strategy of the two provers for game $\gset(H')$, that satisfies at least a  $2^{-\gamma\ell/2}$-fraction of the constraints in $\rset(H')$.

\begin{figure}[h]
  \center
  \includegraphics[width=17cm]{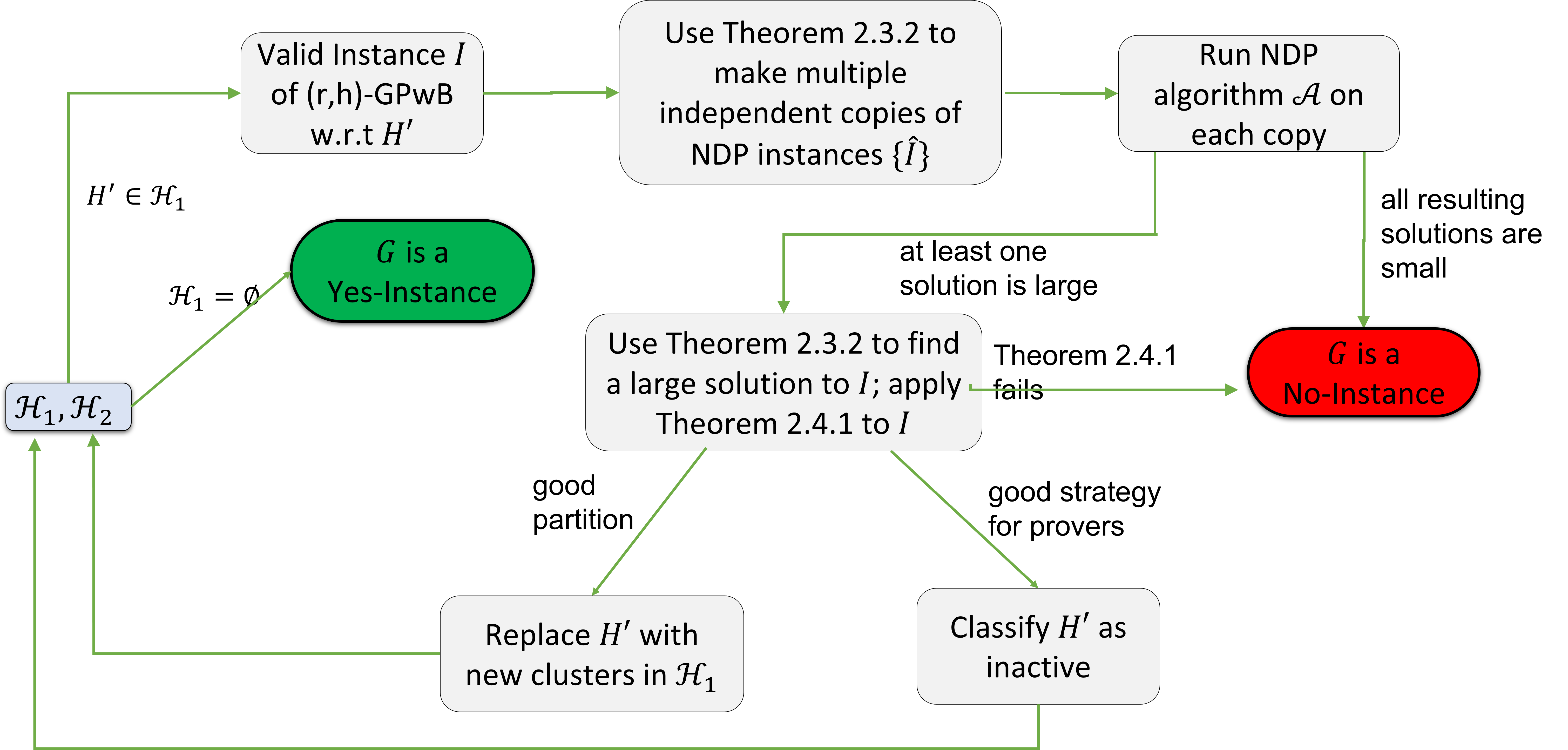}

  \caption{Flowchart for an iteration execution of the algorithm of \Cref{thm: ndp-hard-master NDP}} \label{fig: ndp-hard-flowchart}
\end{figure}

At the beginning, $\hset$ contains a single cluster -- the graph $H$, which is active. The algorithm is executed while $\hset_1\neq \emptyset$, and its execution is partitioned into phases. In every phase, we process each of the clusters that belongs to $\hset_1$ at the beginning of the phase. Each phase is then in turn is partitioned into iterations, where in every iteration we process a distinct active cluster $H'\in \hset_1$. We describe an iteration when an active cluster $H'\in \hset_1$ is processed in \Cref{fig: ndp-hard-iteration} (see also the flowchart in \Cref{fig: ndp-hard-flowchart}).

If the algorithm terminates with $\hset$ containing only inactive clusters, then we return ``$G$ is a \yi''.

\paragraph{Correctness.}
 We establish the correctness of the algorithm in the following two lemmas.
 
\begin{lemma}\label{lem: ndp-hard-correctness soundness}
  If $G$ is a \yi, then with high probability, the algorithm returns ``$G$ is a \yi''.
\end{lemma}

\begin{proof}
  Consider an iteration of the algorithm when an active cluster $H'$ is processed. Notice that the algorithm may only determine that $G$ is a \ni in \Cref{ndp-hard-alg returns NI 1} or in \Cref{ndp-hard-alg returns NI2}. We now analyze these two steps.
    
  Consider first \Cref{ndp-hard-alg returns NI 1}.
  From \Cref{cor: ndp-hard-YI for 3COL has good routing}, with probability at least $1/2$, a random graph $\hat \iset (H')$ has a solution of value at least $\frac{|E(H')|\cdot 6^{\ell}}{\cyi\ell^3\log^3n}$, and our $\alpha^*$-approximation algorithm to \NDPgrid must then return a solution of value at least $\frac{|E(H')|\cdot 6^{\ell}}{\cyi \alpha^*\ell^3\log^3n}$ . Since we use $n^{4\ell}$ independent random constructions of $\hat \iset (H')$, with high probability, for at least one of them, we will obtain a solution of value at least $\frac{|E(H')|6^{\ell}}{\cyi \alpha^*\ell^3\log^3n}$.  Therefore, with high probability our algorithm will not return ``$G$ is a \ni'' due to \Cref{ndp-hard-alg returns NI 1} in this iteration.
  
  Consider now \Cref{ndp-hard-alg returns NI2}. The algorithm can classify $G$ as a \ni in this step only if $\sum_{H''\in \hset}|E(H'')|< \frac{c'|E(H')|}{\ell^2 \alpha^2}$. From \Cref{thm: ndp-hard-good strategy or partition}, this happens with probability at most $1/P$, and from our setting of the parameter $P$ to be $n^{c\ell}$ for a large enough constant $c$, with high probability our algorithm will not return ``$G$ is a \ni'' due to \Cref{ndp-hard-alg returns NI2} in this iteration.
  
  It is not hard to see that our algorithm performs $n^{O(\ell)}$ iterations, and so, using the union bound, with high probability, it will classify $G$ as a \yi.
 \end{proof}

\begin{figure}[H]
  \program{Iteration for Processing a Cluster $H'\in \hset_1$}{
    \begin{enumerate}
      \item Construct an instance $\iset(H')$ of \WGPwB. 

      \item Use \Cref{thm: ndp-hard-from WGP to NDP} to independently construct $n^{4\ell}$ instances $\hat{\iset}(H')$ of the \NDPgrid problem.
      \item Run the $\alpha^*$-approximation algorithm $\aset$ on each such instance $\hat{\iset}(H')$. If the resulting solution, for each of these instances, routes fewer than $\frac{|E(H')|6^{\ell}}{\cyi \alpha^*\ell^3\log^3n}$ demand pairs, halt and return ``$G$ is a \ni''. \label[step]{ndp-hard-alg returns NI 1}

      \item Otherwise, fix any instance $\hat\iset(H')$ for which the algorithm returned a solution routing at least $\frac{|E(H')|6^{\ell}}{\cyi \alpha^*\ell^3\log^3n}$ demand pairs. Denote $|E(L(H'))|=M$, and recall that $M\leq n^{\hat c \ell}$. Use \Cref{thm: ndp-hard-from WGP to NDP} to compute a solution $((W_1,\ldots,W_r),(E_1,\ldots,E_r))$ to the instance $\iset(H')$ of the \WGPwB problem, of value at least:

      \[\frac{|E(H')|\cdot 6^{\ell}}{(\cyi  \alpha^*\ell^3\log^3n)(c^*\log^3M)}\geq \frac{|E(H')|\cdot 6^{\ell}}{\cyi c^* \hat c^3 \alpha^*\ell^6\log^6n}\geq  \frac{|E(H')|\cdot 6^{\ell}}{\cyi^2  \alpha^*\ell^6\log^6n}= \frac{|E(H')|\cdot 6^{\ell}}{\alpha}.\]

      \item Apply the algorithm from \Cref{thm: ndp-hard-good strategy or partition} to this solution, with the parameter $P=n^{c\ell}$, for a sufficiently large constant $c$.
      
      \begin{enumerate}
        \item If the outcome is a strategy for the provers satisfying more than a $2^{-\gamma\ell/2}$-fraction of constraints $R\in \rset$ with $e(R)\in E(H')$, then declare cluster $H'$  inactive and move it from $\hset_1$ to $\hset_2$. Store the resulting strategy of the provers. 
        
        \item Otherwise, let $\tilde \hset$ be the collection of sub-graphs of $H'$ returned by the algorithm. If $\sum_{H''\in \hset}|E(H'')|< \frac{c'|E(H')|}{\ell^2 \alpha^2}$, then  return ``$G$ is a \ni''. Otherwise, remove $H'$ from $\hset_1$ and add all graphs of $\tilde \hset$ to $\hset_1$. \label[step]{ndp-hard-alg returns NI2}
      \end{enumerate}

    \end{enumerate}
  }

  \caption{An iteration description of the algorithm of \Cref{thm: ndp-hard-master NDP} \label{fig: ndp-hard-iteration}}
\end{figure}

\begin{lemma}\label{lem: ndp-hard-correctness completeness}
 If $G$ is a \ni, then the algorithm always returns ``$G$ is a \ni''.
 
\end{lemma}
\begin{proof}
  From \Cref{cor: ndp-hard-parallel repetition}, it is enough to show that, whenever the algorithm classifies $G$ as a \yi, there is a strategy for the two provers, that satisfies more than a fraction-$2^{-\gamma\ell}$ of the constraints in $\rset$. 
  
  Note that the original graph $H$ has  at most $n^{\hat c \ell}$ edges. In every phase, the number of edges in each active graph decreases by a factor of at least $2^{\gamma \ell/16}$. Therefore, the number of phases is bounded by $O(\log n)$. If the algorithm classifies $G$ as a \yi, then it must terminate when no active clusters remain. In every phase, the number of edges in $\bigcup_{H'\in \hset}E(H')$ goes down by at most a factor $\ell^2\alpha^2/c'$. Therefore, at the end of the algorithm:
  
  \[\sum_{H'\in \hset_2}|E(H')|\geq \frac{|\rset|}{(\ell^2\alpha^2/c')^{O(\log n)}}=\frac{|\rset|}{(\ell^{14}\cdot (\alpha^*)^2\cdot \log ^{12}n)^{O(\log n)}}=\frac{|\rset|}{(\alpha^*)^{O(\log n)}}.\] 
  
  By appropriately setting $\alpha^*=2^{\Theta( \ell/\log n)}$, we will ensure that the number of edges remaining in the inactive clusters $H'\in \hset_2$ is at least $|\rset|/2^{\gamma\ell/4}$. Each such edge corresponds to a distinct random string $R\in \rset$. Recall that for each inactive cluster $H'$, there is a strategy for the provers in the corresponding game $\gset(H')$ that satisfies at least $|E(H')|/2^{\gamma\ell/2}$ of its constraints. Taking the union of all these strategies, we can satisfy more than $|\rset|/2^{\gamma\ell}$ constraints of $\rset$, contradicting the fact that $G$ is a \ni.
\end{proof}

\paragraph{Running Time and the Hardness Factor.}
As observed above, our algorithm has at most $n^{O(\ell)}$ iterations, where in every iteration it processes a distinct active cluster $H'\subseteq H$.  The corresponding graph $L(H')$ has at most $n^{O(\ell)}$ edges, and so each of the $n^{O(\ell)}$ resulting instances of \NDPgrid contains at most $n^{O(\ell)}$ vertices. Therefore, the overall running time of the algorithm is $n^{O(\ell)}$. From the above analysis, if $G$ is a \ni, then the algorithm always classifies it as such, and if $G$ is a \yi, then the algorithm classifies it as a \yi with high probability. The hardness factor that we obtain is $\alpha^* = 2^{\Theta(\ell/\log n)}$, while we only apply our approximation algorithm to instances of \NDPgrid containing at most $N=n^{O(\ell)}$ vertices. The running time of the algorithm is $n^{O(\ell)}$, and it is a randomized algorithm with a one-sided error.

Setting $\ell=\log^pn$ for a large enough integer $p$, we obtain $\alpha^*=2^{\Theta((\log N)^{1-2/(p+1)})}$, giving us a $2^{(\log n)^{(1-\eps)}}$-hardness of approximation for \NDPgrid for any constant $\eps$, assuming $\NP \not \subseteq \RTIME (n^{\poly \log n})$.

Setting $\ell=n^{\delta}$ for some constant $\delta$, we get that $N=2^{O(n^{\delta}\log n)}$ and $\alpha^*=2^{\Theta(n^{\delta}/\log n)}$, giving us a $n^{\Omega (1/(\log \log n)^2)}$-hardness of approximation for \NDPgrid, assuming that $\NP \not \subseteq \RTIME(2^{n^{\delta}})$ for some constant $\delta>0$.

\subsection{Prover Strategy or Partition --- Proof of Theorem \ref{thm: ndp-hard-good strategy or partition}}

Recall that each edge of graph $H'$ corresponds to some constraint $R\in \rset$. Let $\rset'\subseteq\rset$ be the set of all constraints $R$ with $e(R)\in E(H')$. Denote the solution to the \WGPwB instance $\iset(H')$ by $((W_1,\ldots,W_r),(E_1,\ldots,E_r))$, and let $E'=\bigcup_{i=1}^r E_i$. Recall that for each random string $R\in \rset'$, there is a set $E(R)$ of $12^{\ell}$ edges in graph $L(H')$ representing $R$. Due to the way these edges are partitioned into bundles, at most $6^{\ell}$ edges of $E(R)$ may belong to $E'$. We say that a random string $R\in \rset'$ is \emph{good} iff $E'$ contains at least $6^{\ell}/(2\alpha)$ edges of $E(R)$, and we say that it is bad otherwise. 

\begin{observation}
At least $|\rset'|/(2\alpha)$ random strings of $\rset'$ are good.
\end{observation}
\begin{proof}
Let $x$ denote the fraction of good random strings in $\rset'$. A good random string contributes at most $6^{\ell}$ edges to $E'$, while a bad random string contributes at most $6^{\ell}/(2\alpha)$. If $x<1/(2\alpha)$, then a simple accounting shows that $|E'|<|\rset'|\cdot 6^{\ell}/\alpha=|E(H')|\cdot 6^{\ell}/\alpha$, a contradiction.
\end{proof}

Consider some random string $R\in \rset'$, and assume that $R=(Q^E,Q^V)$. We denote by $E'(R)=E(R)\cap E'$. Intuitively, say that a cluster $W_i$ is a \emph{terrible cluster} for $R$ if the number of edges of $E(R)$ that lie in $E_i$ is much smaller than $|Q^E\cap W_i|$ or $|Q^V\cap W_i|$. We now give a formal definition of a terrible cluster.

\begin{definition}
Given a random string $R\in \rset$ and an index $1\leq i\leq 6^{\ell}$, we say that a cluster $W_i$ is a \emph{terrible cluster}  for $R$, if: 

\begin{itemize}
\item Either $|E'(R)\cap E_i|< |W_i\cap S(Q^V)|/(8\alpha)$; or
\item $|E'(R)\cap E_i|< |W_i\cap S(Q^E)|/(8\alpha)$.
\end{itemize}

We say that an edge $e\in E'(R)$ is a \emph{terrible edge} if it belongs to the set $E_i$, where $W_i$ is a terrible cluster for $R$. 
\end{definition}

\begin{observation}
For each good random string $R\in \rset'$, at most $6^{\ell}/(4\alpha)$ edges of $E'(R)$ are terrible.
\end{observation}

\begin{proof}
Assume for contradiction that more than $6^\ell/(4\alpha)$ edges of $E'(R)$ are terrible. Denote $R=(Q^E,Q^V)$.  Consider some such terrible edge $e\in E'(R)$, and assume that $e\in E_i$ for some cluster $W_i$, that is terrible for $R$. We say that $e$ is a type-$1$ terrible edge iff $|E'(R)\cap E_i|< |W_i\cap S(Q^V)|/(8\alpha)$, and it is a type-2 terrible edge otherwise, in which case $|E'(R)\cap E_i|< |W_i\cap S(Q^E)|/(8\alpha)$ must hold. Let $E^1(R)$ and $E^2(R)$ be the sets of all terrible edges of $E'(R)$ of types $1$ and $2$, respectively. Then either $|E^1(R)|> 6^{\ell}/(8\alpha)$, or $|E^2(R)|> 6^{\ell}/(8\alpha)$ must hold. %

Assume first that $|E^1(R)|> 6^{\ell}/(8\alpha)$. Fix some index $1\leq i\leq 6^{\ell}$, such that $W_i$ is a cluster that is terrible for $R$, and  $|E(R)\cap E_i|< |W_i\cap S(Q^V)|/(8\alpha)$. We assign, to each edge $e\in E_i\cap E^1(R)$, a set of $8\alpha$ vertices of $W_i\cap S(Q^V)$ arbitrarily, so that every vertex is assigned to at most one edge; we say that the corresponding edge is \emph{responsible} for the vertex. Every edge of $E_i\cap E^1(R)$ is now responsible for $8\alpha$ distinct vertices of  $W_i\cap S(Q^V)$. Once we finish processing all such clusters $W_i$, we will have assigned, to each edge of $E^1(R)$, a set of $8\alpha$ distinct vertices of $S(Q^V)$. We conclude that $|S(Q^V)|\geq 8\alpha |E^1(R)|> 6^{\ell}$. But $|S(Q^V)|=6^{\ell}$, a contradiction.

The proof for the second case, where $|E^2(R)|> 6^{\ell}/(16\alpha)$ is identical, and relies on the fact that $|S(Q^E)|=6^{\ell}$.
\end{proof}

We will use the following simple observation.

\begin{observation}\label{obs: ndp-hard-not terrible}
Let $R\in \rset'$ be a good random string, with $R=(Q^E,Q^V)$, and let $1\leq i\leq 6^{\ell}$ be an index, such that $W_i$ is not terrible for $R$. Then $|W_i\cap S(Q^V)|\geq |W_i\cap S(Q^E)|/(8\alpha)$ and $|W_i\cap S(Q^E)|\geq |W_i\cap S(Q^V)|/(8\alpha)$.
\end{observation}

\begin{proof}
Assume first for contradiction that $|W_i\cap S(Q^V)|< |W_i\cap S(Q^E)|/(8\alpha)$. Consider the edges of $E(R)\cap E_i$. Each such edge must be incident to a distinct vertex of $S(Q^V)$. Indeed, if two edges $(e,e')\in E(R)\cap E_i$ are incident to the same vertex $v_j(Q^V,A)\in S(Q^V)$, then, since the other endpoint of each such edge lies in $S(Q^E)$, the two edges belong to the same bundle, a contradiction. Therefore, $|E_i\cap E(R)|\leq |W_i\cap S(Q^V)|< |W_i\cap S(Q^E)|/(8\alpha)$, contradicting the fact that $W_i$ is not a terrible cluster for $R$. 

The proof for the second case, where  $|W_i\cap S(Q^E)|< |W_i\cap S(Q^V)|/(8\alpha)$ is identical. As before, each edge of  $E(R)\cap E_i$ must be incident to a distinct vertex of $S(Q^E)$, as otherwise, a pair $e,e'\in E(R)$ of edges that are incident on the same vertex $v(Q^E,A)\in S(Q^E)$ belong the same bundle. Therefore, $|E_i\cap E(R)|\leq |W_i\cap S(Q^E)|< |W_i\cap S(Q^V)|/(8\alpha)$, contradicting the fact that $W_i$ is not a terrible cluster for $R$. 
\end{proof}

For each good random string $R\in \rset'$, we discard the terrible edges from set $E'(R)$, so $|E'(R)|\geq 6^{\ell}/(4\alpha)$ still holds. 

Let $z=2^{\gamma\ell/8}$. We say that cluster $W_i$ is \emph{heavy} for a random string $R=(Q^E,Q^V)\in \rset'$ iff $|W_i\cap S(Q^E)|,|W_i\cap S(Q^V)|>z$. We say that an edge $e\in E'(R)$ is heavy iff it belongs to set $E_i$, where $W_i$ is a heavy cluster for $R$. Finally, we say that a  random string $R\in \rset'$ is \emph{heavy} iff at least half of the edges in $E'(R)$ are heavy. Random strings and edges that are not heavy are called light.
We now consider two cases. The first case happens if at least half of the good random strings are light. In this case, we compute a randomized strategy for the provers to choose assignment to the queries, so that at least a $2^{-\gamma\ell/2}$-fraction of the constraints in $\rset'$ are satisfied in expectation. In the second case, at least half of the good random strings are heavy. We then compute a partition $\hset$ of $H'$ as desired.
We now analyze the two cases. Note that if $|E(H')|<z/(8\alpha)$, then Case 2 cannot happen. This is since $h=|E(H')|<z/(8\alpha)$ in this case, and so no random strings may be heavy. Therefore, if $H'$ is small enough, we will return  a strategy of the provers that satisfies a large fraction of the constraints in $\rset'$.

\paragraph{Case 1.} This case happens if at least half of the good random strings are light. Let $\lset\subseteq \rset'$ be the set of the good light random strings, so $|\lset|\geq |\rset'|/(4\alpha)$. For each such random string $R\in \lset$, we let $E^L(R)\subseteq E'(R)$ be the set of all light edges corresponding to $R$, so $|E^L(R)|\geq 6^{\ell}/(8\alpha)$. We now define a randomized algorithm to choose an answer to every query $Q\in \qset^E\cup \qset^V$ with $v(Q)\in H'$. Our algorithm chooses a random index $1\leq i\leq r$. For every query $Q\in \qset^E\cup \qset^V$ with $v(Q)\in H'$, we consider the set $\aset(Q)$ of all answers $A$, such that some vertex $v(Q,A)$ belongs to $W_i$ (for the case where $Q\in \qset^V$, the vertex is of the form $v_j(Q,A)$). We then choose one of the answers from $\aset(Q)$ uniformly at random, and assign it to $Q$. If $\aset(Q)=\emptyset$, then we choose an arbitrary answer to $Q$.

We claim that the expected number of satisfied constraints of $\rset'$ is at least $|\rset'|/2^{\gamma\ell/2}$. Since $\lset\geq |\rset'|/(4\alpha)$, it is enough to show that the expected fraction the good light constraints that are satisfied is at least $4\alpha |\lset|/2^{\gamma\ell/2}$, and for that it is sufficient to show that each light constraint $R\in \lset$ is satisfied with probability at least $4\alpha/2^{\gamma\ell/2}$. 

Fix one such constraint $R=(Q^E,Q^V)\in \lset$, and consider an edge $e\in E^L(R)$. Assume that $e$ connects a vertex $v(Q^E,A)$ to a vertex $v_j(Q^V,A')$, and that $e\in E_i$. We say that edge $e$ is happy iff our algorithm chose the index $i$, the answer $A$ to query $Q^E$, and the answer $A'$ to query $Q^V$. Notice that due to our construction of bundles, at most one edge $e\in E^L(R)$ may be happy with any choice of the algorithm; moreover, if any edge $e\in E^L(R)$ is happy, then the constraint $R$ is satisfied. The probability that a fixed edge $e$ is happy is at least $1/(8\cdot 6^{\ell}z^2\alpha)$. Indeed, we choose the correct index $i$ with probability $1/6^{\ell}$. Since $e$ belongs to $E_i$, $W_i$ is a light cluster for $R$, and so either $|S(Q^E)|\leq z$, or  $|S(Q^V)|\leq z$. Assume without loss of generality that it is the former; the other case is symmetric. Then, since $e$ is not terrible, from \Cref{obs: ndp-hard-not terrible}, $|S(Q^V)|\leq 8 \alpha z$, and so $|\aset(Q^V)|\leq 8\alpha z$, while $|\aset(Q^E)|\leq z$. Therefore, the probability that we choose answer $A$ to $Q^E$ and answer $A'$ to $A^V$ is at least $1/(8\alpha z^2)$, and overall,  the probability that a fixed constraint $R\in \lset$ is satisfied is at least $|E^L(R)|/(8\cdot 6^{\ell}z^2\alpha)\geq 1/(64z^2\alpha^2)\geq 4\alpha/ 2^{\gamma\ell/2}$, since $z=2^{\gamma\ell/8}$, and $\alpha<2^{\gamma\ell/32}$.  

\paragraph{Case 2.} This case happens if at least half of the good random strings are heavy. Let $\rset''\subseteq \rset'$ be the set of the heavy random strings, so $|\rset''|\geq |\rset'|/(4\alpha)$. For each such random string $R\in \rset''$, we let $E^H(R)\subseteq E'(R)$ be the set of all heavy edges corresponding to $R$. Recall that $|E^H(R)|\geq 6^{\ell}/(8\alpha)$.

Fix some heavy random string $R\in \rset''$ and assume that $R=(Q^E,Q^V)$. For each $1\leq i\leq r$, let $E_i(R)=E^H(R)\cap E_i$. Recall that, if $E_i(R)\neq\emptyset$, then 
$|W_i\cap S(Q^E)|, |W_i\cap S(Q^V)|\geq z$ must hold, and, from the definition of terrible clusters, $|E_i(R)|\geq z/(8\alpha)$. It is also immediate that $|E_i(R)|\leq |E'(R)|\leq 6^{\ell}$.

We partition the set $\set{1,\ldots,6^{\ell}}$ of indices into at most $\log(|E^H(R)|)\leq \log(6^{\ell})$ classes, where index $1\leq y\leq 6^{\ell}$ belongs to class $\cset_j(R)$ iff $2^{j-1}<|E^H(R)\cap E_y|\leq 2^j$.  Then there is some index $j_R$, so that $\sum_{y\in \cset_{j_R}(R)}|E^H(R)\cap E_y|\geq |E^H(R)|/\log (6^{\ell})$. We say that $R$ \emph{chooses} the index $j_R$. Notice that:

\[ \sum_{y\in \cset_{j_R}(R)}|E^H(R)\cap E_y|\geq  \frac{|E^H(R)|}{\log(6^{\ell})}\geq \frac{ 6^{\ell}}{8\ell \alpha\log 6}.\] 

Moreover,

\begin{equation}
|\cset_{j_R}(R)|\geq\frac{|E^H(R)|}{\log(6^{\ell})\cdot 2^{j}}\geq \frac{6^{\ell}}{8\cdot 2^j\cdot \ell\alpha\log 6}. \label{eq: bound on C}\end{equation}

Let $j^*$ be the index that was chosen by at least $|\rset''|/\log(6^{\ell})$ random strings, and let $\rset^*\subseteq \rset''$ be the set of all random strings that chose $j^*$. 
We are now ready to define a collection $\hset=\set{H_1,\ldots,H_{6^{\ell}}}$ of sub-graphs of $H'$. 
We first define the sets of vertices in these subgraphs, and then the sets of edges. Choose a random ordering of the clusters $W_1,\ldots,W_{6^{\ell}}$; re-index the clusters according to this ordering. For each query $Q\in \qset^E\cup \qset^V$ with $v(Q)\in H'$, add the vertex $v(Q)$ to set $V(H_i)$, where $i$ is the smallest index for which $W_i$ contains at least $2^{j^*-1}$ vertices of $S(Q)$; if no such index $i$ exists, then we do not add $v(Q)$ to any set.

In order to define the edges of each graph $H_i$, for every random string $R=(Q^E,Q^V)\in \rset^*$, if $i\in \cset_{j^*}(R)$, and both $v(Q^E)$ and $v(Q^V)$ belong to $V(H_i)$, then we add the corresponding edge $e(R)$ to $E(H_i)$. 
This completes the definition of the family $\hset=\set{H_1,\ldots,H_{6^{\ell}}}$ of subgraphs of $H'$. We now show that the family $\hset$ of graphs has the desired properties.
It is immediate to verify that the graphs in $\hset$ are disjoint.

\begin{claim}
For each $1\leq i\leq 6^{\ell}$, $|E(H_i)|\leq |E(H')|/2^{\gamma\ell/16}$.
\end{claim}

\begin{proof}
Fix some index $1\leq i\leq 6^{\ell}$. An edge $e(R)$ may belong to $H_i$ only if $R\in \rset^*$, and $i\in \cset_{j^*}(R)$. In that case, $E_i$ contained at least $z/(8\alpha)$ edges of $E(R)$ (since $W_i$ must be heavy for $R$ and it is not terrible for $R$). Therefore, the number of edges in $H_i$ is bounded by $|E_i|\cdot 8\alpha /z\leq 8\alpha h/z=8\alpha |E(H')|/2^{\gamma\ell/8}\leq |E(H')|/2^{\gamma\ell/16}$, since $\alpha \leq 2^{\gamma\ell/32}$.
\end{proof}

\begin{claim}
$\expect{\sum_{i=1}^r|E(H_i)|}\geq \frac{|E(H')|}{128\ell^2\alpha^2\log^26}$.
\end{claim}
\begin{proof}
Recall that $|\rset^*|\geq |\rset''|/\log(6^{\ell})\geq |\rset'|/(4\alpha\log(6^{\ell}))$. We now fix $R\in \rset^*$ and analyze the probability that $e(R)\in \bigcup_{i=1}^rE(H_i)$. Assume that $R=(Q^E,Q^V)$. 
Let $J$ be the set of indices $1\leq y\leq 6^{\ell}$, such that $|W_i\cap S(Q^V)|\geq 2^{j^*-1}$. Clearly, $|J|\leq  6^{\ell}/2^{j^*-1}$, and $v(Q^V)$ may only belong to graph $H_i$ if $i\in J$. Similarly, let $J'$ be the set of indices $1\leq y\leq 6^{\ell}$, such that $|W_i\cap S(Q^E)|\geq 2^{j^*-1}$. As before, $|J'|\leq  6^{\ell}/2^{j^*-1}$, and $v(Q^E)$ may only belong to graph $H_i$ if $i\in J$. %
Observe that every index $y\in \cset_{j^*}(R)$ must belong to $J\cap J'$, and, since $j^*=j_R$, from \Cref{eq: bound on C}, $|\cset_{j^*}(R)|\geq \frac{6^{\ell}}{8\cdot 2^{j^*}\cdot \ell\alpha\log 6}$.

 Let $y\in J\cup J'$ be the first index that occurs in our random ordering. If $y\in \cset_{j^*}(R)$, then edge $e(R)$ is added to $H_y$. The probability of this happening is at least:
 
 \[\frac{|\cset_{j^*}(R)|}{|J\cup J'|}\geq \frac{6^{\ell}/(8\cdot 2^{j^*}\cdot \ell\alpha\log 6)}{2\cdot 6^{\ell}/2^{j^*-1}}=\frac{1}{32\ell\alpha \log 6}.\]
 
  Overall, the expectation of $\sum_{i=1}^r|E(H_i)|$ is at least:

\[\frac{|\rset^*|}{32\ell\alpha\log 6}\geq \frac{|\rset'|}{128\ell^2\alpha^2\log^26}= \frac{|E(H')|}{128\ell^2\alpha^2\log^26}.\]
\end{proof}
 
Denote the expectation  of $\sum_{i=1}^r|E(H_i)|$  by $\mu$, and let $c=128\log^26$, so that $\mu=|E(H')|/(c \ell^2\alpha^2)$. 
Let $\event$ be the event that $\sum_{i=1}^r|E(H_i)|\geq |E(H')|/(2c \ell^2\alpha^2)=\mu/2$.
We claim that $\event$ happens with probability at least $1/(2c\ell^2\alpha^2)$. Indeed, assume that it happens with probability $p<1/(2c\ell^2\alpha^2)$. If $\event$ does not happen, then  $\sum_{i=1}^r|E(H_i)|\leq \mu/2$, and if it happens, then $\sum_{i=1}^r|E(H_i)|\leq |E(H')|$. Overall, this gives us that $\expect{\sum_{i=1}^r|E(H_i)|} \leq (1-p)\mu/2+p|E(H')|<\mu$, a contradiction.
We repeat the algorithm for constructing $\hset$ $O(\ell^2\alpha^2\poly\log n\log P)$ times. We are then guaranteed that with probability at least $(1-1/P)$, event $\event$ happens in at least one run of the algorithm. It is easy to verify that the running time of the algorithm is bounded by $O(n^{O(\ell)}\cdot \log P)$, since $|V(L(H'))|\leq n^{O(\ell)}$.

    \section{From \WGPwBfull to \NDP on Grid Graphs}\label{subsec: ndp-hard-from WGP to NDP}
    In this section we prove \Cref{thm: ndp-hard-from WGP to NDP}, by providing a reduction from \WGPwB to \NDPgrid.
We assume that we are given an instance $\iset=(\tG=(V_1\cup V_2,E), \uset_1,\uset_2,h, r)$ of \WGPwB.  Let $|V_1|=N_1,|V_2|=N_2$, $|E|=M$, and $N=N_1+N_2$.
We assume that  $\iset$ is a valid instance, so, if we denote by $\beta^*= \beta^*(\iset) = \sum_{v\in V_1}\beta(v)$, then $h=\beta^*/r$, and $h\geq \max_{v\in V_1\cup V_2}\set{\beta(v)}$.

We start by describing a randomized construction of the instance $\hiset=(\hG,\mset)$ of \NDPgrid.

\subsection{The Construction}\label{subsubsec: ndp-hard-the construction}

Fix an arbitrary ordering $\rho$ of the groups in $\uset_1$. Using $\rho$, we define an ordering $\sigma$ of the vertices of $V_1$, as follows. 
The vertices that belong to the same  group $U\in \uset_1$ are placed consecutively in the ordering $\sigma$, in an arbitrary order. The ordering between the groups in $\uset_1$ is the same as their ordering in $\rho$. We assume that $V_1=\set{v_1,v_2,\ldots,v_{N_1}}$, where the vertices are indexed according to their order in $\sigma$. Next, we select a {\bf random} ordering $\rho'$ of the groups in $\uset_2$. We then define an ordering $\sigma'$ of the vertices of $V_2$ exactly as before, using the ordering $\rho'$ of $\uset_2$. We assume that $V_2=\set{v'_1,v'_2,\ldots,v'_{N_2}}$, where the vertices are indexed according to their ordering in $\sigma'$. We note that the choice of the ordering $\rho'$ is the only randomized part of our construction.

Consider some vertex $v\in V_1$. Recall that   $\bset(v)$ denotes the partition of the edges incident to $v$ into bundles, where every bundle is a  non-empty subsets of edges, and that $\beta(v)=|\bset(v)|$. Each such bundle $B\in \bset(v)$ corresponds to a single group $U(B)\in \uset_2$, and contains all edges that connect $v$ to the vertices of $U(B)$. The ordering $\rho'$ of the groups in $\uset_2$ naturally induces an ordering of the bundles in $\bset(v)$, where $B$ appears before $B'$ in the ordering iff $U(B)$ appears before $U(B')$ in $\rho'$. We denote $\bset(v) = \set{B_1(v), B_2(v), \ldots, B_{\beta(v)}(v)}$, where the bundles are indexed according to this ordering.

Similarly, for a vertex $v' \in V_2$, every bundle $B\in \bset(v')$ corresponds to a group $U(B)\in \uset_1$,  and contains all edges that connect $v'$ to the vertices of $U(B)$. As before, the ordering $\rho$ of the groups in $\uset_1$ naturally defines an ordering of the bundles in $\bset(v')$. We denote $\bset(v') = \set{B_1(v'), B_2(v'), \ldots, B_{\beta(v')}(v')}$, and we assume that the bundles are indexed according to this ordering.

We are now ready to define the instance $\hiset=(\hG,\mset)$ of \NDPgrid, from the input instance $(\tilde G=(V_1,V_2,E),\uset_1,\uset_2,h,r)$ of \WGPwB.
Let $\ell= \twiceConstantForSizeOfBlocks \cdot \ceil{M^2 \cdot \log M}$.
The graph $\hG$ is simply the $(\ell\times \ell)$-grid, so $V(\hG)=O(M^4\log^2M)$ as required. We now turn to define the set $\mset$ of the demand pairs. We first define the set $\mset$ itself, without specifying the locations of the corresponding vertices in $\hat G$, and later specify a mapping of all vertices participating in the demand pairs to $V(\hG)$.

Consider the underlying graph $\tilde G=(V_1,V_2,E)$ of the \WGPwB problem instance. Initially, for every edge $e=(u,v)\in E$, with $u\in V_1,v\in V_2$, we define a demand pair $(s(e),t(e))$ representing $e$, and add it to $\mset$, so that the vertices participating in the demand pairs are  all distinct. Next, we process the vertices $v\in V_1\cup V_2$ one-by-one. Consider first some vertex $v\in V_1$, and some bundle $B\in \bset(v)$. Assume that $B=\set{e_1,\ldots,e_z}$. Recall that for each $1\leq i\leq z$, set $\mset$ currently contains a demand pair $(s(e_i),t(e_i))$ representing $e_i$. We unify all vertices $s(e_1),\ldots,s(e_z)$ into a single vertex $s_B$. We then replace the demand pairs $(s(e_1),t(e_1)),\ldots,(s(e_z),t(e_z))$ with the demand pairs $(s_{B},t(e_1)),\ldots,(s_{B},t(e_z))$.  Once we finish processing all vertices in $V_1$, we perform the same procedure for every vertex of $V_2$: given a vertex $v'\in V_2$, 
for every bundle $B' \in \bset(v')$, we unify all destination vertices $t(e)$ with $e\in B'$ into a single destination vertex, that we denote by $t_{B'}$, and we update $\mset$ accordingly. This completes the definition of the set $\mset$ of the demand pairs.

Observe that each edge of $e\in E$ still corresponds to a unique demand pair in $\mset$, that we will denote by $(s_{B(e)}, t_{B'(e)} )$, where $B(e)$ and $B'(e)$ are the two corresponding bundles containing $e$. Given a subset $E'\subseteq E$ of edges of $\tG$, we denote by $\mset(E')=\set{(s_{B(e)},t_{B'(e)})\mid e\in E'}$ the set of all demand pairs corresponding to the edges of $E'$.

In order to complete the reduction, we need to show a mapping of all source and all destination vertices of $\mset$ to the vertices of $\hG$.
Let $R'$ and $R''$ be two rows of the grid $\hG$, lying at a distance at least $\xi/4$ from each other and from the top and the bottom boundaries of the grid. We will map all vertices of $S(\mset)$ to $R'$, and all vertices of $T(\mset)$ to $R''$.

\paragraph{Locations of the sources.}
Let $\block_1,\block_2,\ldots,\block_{N_1}$ be a collection of $N_1$ disjoint sub-paths of $R'$, where each sub-path contains {$\constantForSizeOfBlocks \cdot\ceil{ h\cdot \log M}$} vertices; the sub-paths are indexed according to their left-to-right ordering on $R'$, and every consecutive pair of the paths is separated by at least $10M$ vertices from each other and from the left and the right boundaries of $\hG$. Observe that the width $\ell$ of the grid is large enough to allow this, as $h\leq M$ must hold.
For all $1\leq i\leq N_1$,
we call $\block_i$ the \emph{block representing the vertex $v_i\in V_1$}.
We now fix some $1 \leq i \leq N_1$ and consider the block $\block_i$ representing the vertex $v_i$. We map the source vertices $s_{B_1(v_i)},s_{B_2(v_i)},\ldots,s_{B_{\beta(v_i)}(v_i)}$ to vertices of $\block_i$, so that they appear on $\block_i$ in this order, so that every consecutive pair of sources is separated by exactly {$512 \cdot \ceil{h\cdot \log M/\beta(v_i)}$} vertices.

\paragraph{Locations of the destinations.}
 Similarly, we let $\block'_1,\block'_2,\ldots,\block'_{N_2}$ be a collection of $N_2$ disjoint sub-paths of $R''$, each of which contains {$\constantForSizeOfBlocks \cdot \ceil{h\cdot \log M}$}  vertices, so that the sub-paths are indexed according to their left-to-right ordering on $R''$, and every consecutive pair of the paths is separated by at least $10M$ vertices from each other and from the left and the right boundaries of $\hG$. We call $\block'_i$ the \emph{block representing the vertex $v'_i\in V_2$}.
 We now fix some $1 \leq i \leq N_2$ and consider the block $\block'_i$ representing the vertex $v'_i$. We map the destination vertices $t_{B_1(v'_i)},t_{B_2(v'_i)},\ldots,t_{B_{\beta(v'_i)}(v'_i)}$ to vertices of $\block'_i$, so that they appear on $\block'_i$ in this order, and every consecutive pair of destinations is separated by exactly {$512 \cdot \ceil{h\cdot \log M/\beta(v'_i)}$} vertices.

 This concludes the definition of the instance $\hiset=(\hG,\mset)$ of \NDPgrid. In the following subsections we analyze its properties. 
 The following immediate observation will be useful to us.
 
 \begin{observation}\label{obs: ndp-hard-ordering of sources and destinations}
 Consider a vertex $v_j\in V_1$, and let $\nset_j\subseteq\mset$ be any subset of demand pairs, whose sources are all distinct and lie on $\block_j$. Assume that $\nset_j=\set{(s_1,t_1),\ldots,(s_{y},t_y)}$, where the demand pairs are indexed according to the left-to-right ordering of their source vertices on $\block_j$. Then $t_1,\ldots,t_y$ appear in this left-to-right order on $R''$.
 \end{observation}

 We will also use the following two auxiliary lemmas, whose proofs are straightforward and are deferred to \Cref{appdx-subsec: ndp-hard-auxiliarly lemmas}.

\paragraph{Auxiliary Lemmas}  Assume that we are given a set $U$ of $n$ items, such that  $P$ of the items are pink, and $Y=n-P$ items are yellow. Consider a random permutation $\pi$ of these items. 

\begin{lemma} \label{lem: ndp-hard-random ordering}
For any $\log n\leq \mu\leq Y$, the probability that there is a sequence of $\ceil{4n\mu/P}$ consecutive items in $\pi$ that are all yellow, is at most $n/e^{\mu}$.
\end{lemma}

\begin{lemma} \label{lem: ndp-hard-random ordering2}
For any $\log n\leq \mu\leq P$, the probability that there is a set $S$ of $\floor{\frac{n\mu}{P}}$ consecutive items in $\pi$, such that more than $4\mu$ of the items are pink, is at most $n/4^{\mu}$.
\end{lemma}

\subsection{From Partitioning to Routing}\label{subsubsec: ndp-hard-YI from partitioning to routing}

The goal of this subsection is to prove the following theorem.

\begin{theorem}\label{thm: ndp-hard-yi}
Suppose we are given a valid instance $\iset=(\tG=(V_1,V_2,E),\uset_1,\uset_2,h,r)$ of \WGPwB, such that $\iset$ has a perfect solution. Then with probability at least $1/2$ over the random choices made in the construction of the corresponding instance $\hiset$ of \NDPgrid, there is a solution to $\hiset$, routing $\Omega(\beta^*(\iset)/\log^3M)$ demand pairs via a set of spaced-out paths.
\end{theorem}

The remainder of this subsection is devoted to the proof of the theorem. We assume w.l.o.g. that $|E|=M > 2^{50}$, as otherwise, since $\beta^*(\iset)\leq |E|$, routing a single demand pair is sufficient.

Let  $((W_1,\ldots,W_r),(E_1,\ldots,E_r))$ be a perfect solution to $\iset$.
Recall that by the definition of a perfect solution, for each group $U\in \uset_1 \cup \uset_2$, every set $W_ i$ contains exactly one vertex of $U$, and moreover, for each $1\leq i\leq r$, $|E_i|=h=\beta^*(\iset)/r$.

We let $E^0=\bigcup_{i=1}^rE_i$, so $|E^0|=\beta^*(\iset) = hr$. Let $\mset^0\subseteq \mset$ be the set of all demand pairs corresponding to the edges of $E^0$. Note that we are guaranteed that no two demand pairs in $\mset^0$ share a source or a destination, since no two edges of $E^0$ belong to the same bundle.

Next we define a property of subsets of the demand pairs, called a \emph{distance property}. We later show that every subset $\mset'\subseteq \mset^0$ of demand pairs that has this property can be routed via spaced-out paths, and that there is a large subset $\mset'\subseteq \mset^0$ of the demand pairs with this property.

Given a subset $\mset'\subseteq \mset^0$ of the demand pairs, we start by defining an ordering $\sigma_{\mset'}$ of the destination vertices in $T(\mset')$. This ordering is somewhat different from the ordering of the vertices of $T(\mset')$ on row $R''$. We first provide a motivation and an intuition for this new ordering $\sigma_{\mset'}$. Recall that the rows $R'$ and $R''$ of $\hat G$, where all source and all destination vertices lie, respectively, are located at a distance at least $\ell/4$ from each other and from the grid boundaries. Let $R$ be any row of $\hat G$, lying between $R'$ and $R''$, at a distance at least $\ell/16$ from both $R'$ and $R''$. Let $X$ be some subset of $|\mset'|$ vertices of $R$. If we index the vertices of $T(\mset')$ as $\set{t_1,t_2,\ldots,t_{|\mset'|}}$ according to their order in the new ordering $\sigma_{\mset'}$, then we view the $i$th vertex of $X$, that we denote by $x_i$, as representing the terminal $t_i$. For each $1\leq i\leq |\mset'|$,  we denote the source vertex corresponding to $t_i$ by $s_i$, that is, $(s_i,t_i)\in \mset'$. Note that the ordering of the vertices of $S(\mset')$ on $R'$ may be completely different from the one induced by these indices.  Similarly, the ordering of the vertices of $T(\mset')$ on $\rset''$ may be inconsistent with this indexing. Eventually, we will construct a set $\pset$ of spaced-out paths routing the demand pairs in $\mset'$, so that the path $P_i\in \pset$, connecting $s_i$ to $t_i$, intersects the row $R$ exactly once -- at the vertex $x_i$. In this way, we will use the ordering $\sigma_{\mset'}$ of the destination vertices in $T(\mset')$ to determine the order in which the path of $\pset$ intersect $R$.

Assume now that we are given some subset $\mset'\subseteq \mset^0$ of demand pairs. Recall that the sources and the destinations of all demand pairs in $\mset'$ are distinct. We are now ready to define the ordering $\sigma_{\mset'}$ of $T(\mset')$. We partition the vertices of $T(\mset')$ into subsets $J_1,J_2,\ldots,J_r$, as follows. 
Consider some vertex $v'_j\in V_2$ of $\tilde G$, and assume that it lies in the cluster $W_i$. Then all destination vertices of $T(\mset')$ that belong to the corresponding block $K'_j$ are added to the set $J_i$. 
 To obtain the final ordering $\sigma_{\mset'}$, we place the vertices of  $J_1,J_2,\ldots,J_r$ in this order, where within each set $J_i$, the vertices are ordered  according to their ordering along the row $R''$. Notice that a selection of a subset $\mset'\subseteq \mset_0$ completely determines the ordering $\sigma_{\mset'}$. Given two demand pairs $(s,t),(s',t')\in \mset'$, we let $N_{\mset'}(s,s')$ denote the number of destination vertices that lie between $t$ and $t'$ in the ordering $\sigma_{\mset'}$ (note that this is well-defined as the demand pairs in $\mset'$ do not share their sources or destinations). Recall that $d(s,s')$ is the distance between $s$ and $s'$ in graph $\hat G$.

\begin{definition} Suppose we are given a subset $\mset'\subseteq \mset^0$ of the demand pairs. 
We say that two distinct vertices $s,s'\in S(\mset')$ are \emph{consecutive} with respect to $\mset'$, iff no other vertex of $S(\mset')$ lies between $s$ and $s'$ on $R'$. We say that $\mset'$ has the \emph{distance property} iff for every pair $s,s'\in S(\mset')$ of vertices that are consecutive with respect to $\mset'$, $N_{\mset'}(s,s')<d(s,s')/4$.
\end{definition}

We first show that there is a large subset of the demand pairs in $\mset^0$ with the distance property in the following lemma, whose proof appears in the next subsection.

\begin{lemma}\label{lem: ndp-hard-large set with distance property}
With probability at least $1/2$ over the construction of $\hat \iset$, there is a subset $\mset'\subseteq \mset^0$ of demand pairs that has the distance property, and $|\mset'|=\Omega(|\mset^0|/\log^3M)$.%
\end{lemma}

Finally, we show that every set $\mset'$ of demand pairs with the distance property can be routed via spaced-out paths.

\begin{lemma}\label{lem: ndp-hard-can find routing}
Assume that $\mset'\subseteq \mset^0$ is a subset of demand pairs that has the distance property. %
Then there is a spaced-out set $\pset$ of paths routing all pairs of $\mset'$ in graph $\hat G$.
\end{lemma}

The above two lemmas finish the proof of \Cref{thm: ndp-hard-yi}, since $|\mset^0|=\beta^*(\iset)$. We prove these lemmas in the following two subsections.

\proofof{\Cref{lem: ndp-hard-large set with distance property}}

We assume that $|\mset^0|>c\log^3M$ for some large enough constant $c$, since otherwise we can return a set $\mset'$ containing a single demand pair. 
We gradually modify the set $\mset^0$ of the demand pairs, by selecting smaller and smaller subsets $\mset^1,\mset^2$, $\mset^3$, and $\mset'$.
For each vertex vertex $v \in V_1 \cup V_2$ of the \WGPwB instance $\tG$, let $\delta(v)$ denote the set of all edges of $E(\tG)$ incident to $v$.

We start by performing two ``regularization'' steps on the vertices of $V_2$ and $V_1$ respectively. Intuitively, we will select two integers $p$ and $q$, and a large enough subset $\mset^2\subseteq \mset^0$ of demand pairs, so that for every vertex $v_i\in V_1$, either no demand pair in $\mset^2$ has its source on $\block_i$, or roughly $2^p$ of them do. Similarly, for every vertex $v'_j\in V_2$, either no demand pairs in $\mset^2$ has its destination on $K'_j$, or roughly $2^q$ of them do. We will not quite achieve this, but we will come close enough.

\paragraph{Step 1 [Regularizing the degrees in $V_2$].} In this step we select a large subset $\mset^1\subseteq \mset^0$ of the demand pairs, and an integer $q$, such that, for each vertex $v\in V_2$, the number of edges of $E(\tG)$ incident to $v$, whose corresponding demand pair lies in $\mset^1$, is either $0$, or roughly $2^q$. In order to do this,
we partition the vertices of $V_2$ into classes $Z_1,\ldots,Z_{\ceil{\log M}}$, where a vertex $v' \in V_2$ belongs to class $Z_y$ iff $2^{y-1}\leq |E^0 \cap \delta(v')|< 2^y$. If $v\in Z_y$, then we say that all edges in $\delta(v)\cap E^0$ belong to the class $Z_y$.
Therefore, each edge of $E^0$ belongs to exactly one class, and there is some index $1\leq q\leq \ceil{\log M}$, such that at least $\Omega(|\mset^0|/\log M)$ edges of $E^0$ belong to class $Z_{q}$. We let $E^1\subseteq E^0$ be the set of all edges that belong to the class $Z_q$, and we let $\mset^1 = \mset(E^1) \subseteq \mset^0$ be the corresponding subset of the demand pairs.

\paragraph{Step 2 [Regularizing the degrees in $V_1$].} This step is similar to the previous step, except that it is now performed on the vertices of $V_1$.
We partition the vertices of $V_1$ into classes $Y_1,\ldots,Y_{\ceil{\log M}}$, where a vertex $v \in V_1$ belongs to class $Y_z$ iff $2^{z-1}\leq |E^1 \cap \delta(v)|<2^z$. If $v\in Y_z$, then we say that all edges in $\delta(v)\cap E^1$ belong to the class $Y_z$.
As before, every edge of $E^1$ belongs to exactly one class, and there is some index $1\leq p\leq \ceil{\log M}$, such that at least $\Omega(|E^1|/\log M)\geq \Omega(|\mset^0|/\log^2 M)$ edges of $E^1$ belong to the class $Y_{p}$. We let $E^2\subseteq E^1$ denote the set of all edges that belong to class $Y_p$, and $\mset^2 = \mset(E^2) \subseteq \mset^1$ denote the corresponding subset of the demand pairs, so that $|\mset^2|=\Omega(|\mset^0|/\log^2M)$.

Notice that so far, for every vertex $v\in V_1$, if $\delta(v)\cap E^2\neq \emptyset$, then $2^{p-1}\leq |\delta(v)\cap E^2|<2^{p}$. However, for a vertex $v\in V_2$ with $\delta(v)\cap E^2\neq \emptyset$, we are only guaranteed that $|\delta(v)\cap E^2|<2^{q}$, since we may have discarded some edges that were incident to $v$ from $E^1$. Moreover, the subset $\mset^2$ of the demand pairs is completely determined by the solution $((W_1,\ldots,W_r),(E_1,\ldots,E_r))$ to the \WGPwB problem, and is independent of the random choices made in our construction of the \NDPgrid problem instance.
The following simple observation, whose proof is deferred to \Cref{prf-obs: ndp-hard-h is large} will be useful for us later.

\begin{observation}\label{obs: ndp-hard-h is large}
$h\geq 2^{p}\cdot 2^{q}/4$.
\end{observation}

For each vertex $v_i\in V_1$, let $X_i\subseteq V(\block_i)$ be the set of all vertices that serve as the sources of the demand pairs in $\mset$, so $X_i=S(\mset)\cap V(\block_i)$. Recall that $|X_i|=\beta(v_i)$, and every pair of vertices in $X_i$ is separated by at least $512\ceil{\frac{h\log M}{\beta(v_i)}}$ vertices of $\block_i$. We let $X'_i\subseteq X_i$ denote the subset of vertices that serve as sources of the demand pairs in $\mset^2$. We say that a sub-path $Q\subseteq \block_i$ is \emph{heavy} iff $|V(Q)|=\floor{\frac{512 h\log^2M}{2^p}}$, and $|V(Q)\cap X'_i|>16\log M$.

\begin{observation}\label{obs: ndp-hard-no heavy path exist}
With probability at least $0.99$ over the choice of the random permutation $\rho'$, for all $v_i\in V_1$, no heavy sub-path $Q\subseteq \block_i$ exists.
\end{observation}

We prove \Cref{obs: ndp-hard-no heavy path exist} in \Cref{prf-obs: ndp-hard-no heavy path exist}.
Let $\badevent$ be the bad event that for some $v_i\in V_1$, block $\block_i$ contains a heavy path. From \Cref{obs: ndp-hard-no heavy path exist}, the probability of $\badevent$ is at most $0.01$.

The following claim will be used to bound the values $N_{\mset^2}(s,s')$.

\begin{claim}\label{clm: ndp-hard-dist prop}
Consider some vertex $v_j\in V_1$ in graph $\tilde G$, and the block $\block_j$ representing it.
Then with probability at least $(1-1/M^3)$, for every pair $s,s'\in S(\mset^2)\cap V(\block_j)$ of source vertices that are consecutive with respect to $\mset^2$, $N_{\mset^2}(s,s')\leq 128h\log M/2^p$.
\end{claim}

\begin{proof}
Fix some vertex $v_j\in V_1$ and consider the block $\block_j$ representing it. Assume that $v_j$ belongs to the cluster $W_i$ in our solution to the \WGPwB problem.
Let $s,s'\in S(\mset^2)\cap V(\block_j)$ be a pair of source vertices  that are consecutive with respect to $\mset^2$.  Recall that we have defined a subset $J_i\subseteq T(\mset^2)$ of destination vertices, that appear consecutively in the ordering $\sigma_{\mset^2}$, and contain all vertices of $T(\mset^2)$, that lie in blocks $\block'_{j'}$, whose corresponding vertices $v_{j'}\in V_2\cap W_i$.

Let $A= W_i\cap V_2$; let $A'\subseteq A$ contain all vertices that have an edge of $E^2\cap E_i$ incident to them; and let $A''\subseteq A'$ contain all vertices that have an edge of $E^2\cap E_i$ connecting them to $v_j$. Since the solution to the \WGPwB problem instance is perfect, every vertex of $A$ (and hence $A'$ and $A''$) belongs to a distinct group of $U\in \uset_2$. We denote by $\uset'\subseteq \uset_2$ the set of all groups to which the vertices of $A'$ belong, and we define $\uset''\subseteq \uset'$ similarly for $A''$. Consider now some group $U\in \uset'$, and let $v'_a$ be the unique vertex of $U$ that belongs to $A'$. We denote by $C(U)$ the set of all vertices of the corresponding block $\block'_a$ that belong to $T(\mset^2)$. Therefore, we now obtain a partition $\set{C(U)}_{U\in \uset'}$ of all vertices of $J_i$ into subsets, where each subset contains at most $2^q$ vertices. Moreover, in the ordering $\sigma_{\mset^2}$, the vertices of each such set $C(U)$ appear consecutively, in the order of their appearance on $R''$, while the ordering between the different sets $C(U)$ is determined by the ordering of the corresponding groups $U$ in $\rho'$.  Let $\rho''$ be the ordering of the groups in $\uset'$ induced by $\rho'$, so that $\rho''$ is a random ordering of $\uset'$. Observe that, since the choice of the set $\mset^2$ is independent of the ordering $\rho'$ (and only depends on the solution to the \WGPwB problem instance), so is the choice of the sets $\uset'$ and $\uset''$.

Let $t$ and $t'$ be the destination vertices that correspond to $s$ and $s'$, respectively, that is, $(s,t),(s',t')\in \mset^2$. Assume that $t\in \block'_z$ and $t'\in \block'_{z'}$, where $v'_z$ and $v'_{z'}$ are vertices of $V_2$. From our definition, both $v'_z$ and $v'_{z'}$ must belong to the set $A''$. Assume that $v'_z$ belongs to the group $U'$ in $\uset_2$, while $v'_{z'}$ belongs to group $U''$. Again, from our definitions, both $U',U''\in \uset''$. From the above discussion, if the number of groups $U\in \uset'$ that fall between $U'$ and $U''$ is $\gamma$, then the number of destination vertices lying between $t$ and $t'$ in $\sigma_{\mset^2}$ is at most $2^q\cdot (\gamma+2)$. Therefore, it is now enough to bound the value of $\gamma$. In order to do so, we think of the groups of $\uset''$ as pink, and the remaining groups of $\uset'$ as yellow.
Let $P$ denote the total number of all pink groups, and let $n^*=|\uset'|$. From the construction of $\mset^2$, $P=|S(\mset^2)\cap V(\block_j)|\geq 2^{p-1}$. We use the following simple observation, whose proof is deferred to \Cref{prf-obs: ndp-hard-nset is small} to upper-bound $n^*$.

\begin{observation} \label{obs: ndp-hard-nset is small}
$n^* \leq h/2^{q-1}$.
\end{observation}

Let $\mu=4\log M$, and let $\event_Y$ be the event that there are at least  $\ceil{4n^*\mu/P}$ consecutive yellow groups in the ordering $\rho''$ of $\uset'$. From \Cref{lem: ndp-hard-random ordering}, the probability of $\event_Y$ is at most  $n^*/e^{\mu}\leq M/e^{\mu}\leq 1/M^3$.  If event $\event_Y$ does not happen, then  the length of the longest consecutive sub-sequence of $\rho''$ containing only yellow groups is bounded by:

\[\ceil{\frac{4n^*\mu}{P}}\leq \frac{64h\log M}{2^{q} 2^p}.\]

Assume now that $\event_Y$ does not happen, and consider any two vertices $s,s'\in  S(\mset^2)\cap V(\block_j)$ that are consecutive with respect to  $\mset^2$. Assume that their corresponding destination vertices are $t$ and $t'$ respectively, and that $t$ and $t'$ belong to the groups $U$ and $U'$, respectively. Then $U$ and $U'$ are pink groups. Moreover, since the ordering of the vertices of $S(\mset)\cap V(\block_j)$ on $R'$ is identical to the ordering of the groups of $\uset''$ to which their destinations belong in $\rho'$, no other pink group appears between $U$ and $U'$ in $\rho''$. Therefore, at most $\frac{64 h\log M}{2^{q} 2^p}$ groups of $\uset'$ lie between $U$ and $U'$ in $\rho''$. Recall that for each group $U''\in \uset'$, at most one vertex $v'_z\in U''$ belongs to $W_i$, and that the corresponding block $\block'_z$ may contribute at most $2^q$ destination vertices to $T(\mset^2)$. We conclude that the number of vertices separating $t$ from $t'$ in $\sigma_{\mset_3}$ is bounded by:

\[\left(\frac{64h\log M}{2^{q} 2^p}+2\right)\cdot 2^q\leq \frac{128h\log M}{2^p}.\]

(We have used \Cref{obs: ndp-hard-h is large}).
Therefore, if event $\event_Y$ does not happen, then  for every pair $s,s'\in  S(\mset^2)\cap V(\block_j)$ of vertices that are consecutive with respect to $\mset^2$, $N_{\mset^2}(s,s')\leq  \frac{128h\log M}{2^p}$.
\end{proof}

Let $\badevent'$ be the bad event that for some vertex $v_j\in V_1$, for some pair $s,s'\in S(\mset^2)\cap V(\block_j)$ of source vertices that are consecutive with respect to $\mset^2$, $N_{\mset^2}(s,s')> 128h\log M/2^p$. By applying the Union Bound to the result of \Cref{clm: ndp-hard-dist prop}, we get that the probability of $\badevent'$ is at most $1/M^2$. 
Notice that the probability that neither $\badevent$ nor $\badevent'$ happen is at least $1/2$. We assume from now on that this is indeed the case, and show how to compute a large subset $\mset'\subseteq \mset^2$ of the demand pairs that has the distance property. This is done in the following two steps.

\paragraph{Step 3 [Sparsifying the Sources.]}

Assume that $\mset^2=\set{(s_1,t_1),\ldots,(s_{|\mset^2|},t_{|\mset^2|})}$, where the demand pairs are indexed according to the left-to-right ordering of their sources on $R'$, that is $s_1,s_2,\ldots,s_{|\mset^2|}$ appear on $R'$ in this order. We now define:

\[\mset^3=\set{(s_i,t_i)\mid i=1\mod \ceil{32\log M}}.\]

Let $E^3\subseteq E^2$ be the set of edges of $\tG$ whose corresponding demand pairs belong to $\mset^3$.
It is easy to verify that $|\mset^3|\geq \Omega(|\mset^2|/\log M)=\Omega(|\mset^0|/\log^3 M)$.

 We also obtain the following claim.
 
\begin{claim}\label{clm: ndp-hard-almost distance property}
Assume that events $\badevent$ and $\badevent'$ did not happen. Then
for each $1\leq j\leq N_1$, for every pair  $s,s'\in S(\mset^3)\cap V(\block_j)$ of source vertices, $N_{\mset^3}(s,s')\leq 128 d(s,s')$.
\end{claim}
\begin{proof}
Fix some $1\leq j\leq N_1$, and some pair $s,s'\in S(\mset^3)\cap V(\block_j)$ of vertices. Let $S'=\set{s_1,s_2,\ldots,s_z}$ be the set of all vertices of $S(\mset^2)$ that appear between $s$ and $s'$ on $R'$. Assume w.l.o.g. that $s$ lies to the left of $s'$ on $R'$, and denote $s_0=s$ and $s_{z+1}=s'$. Assume further that the vertices of $S'$ are indexed according to their left-to-right ordering on $R'$. Note that, from the definition of $\mset^3$, $z\geq \ceil{32\log M}$ must hold. 

Let $I\subseteq \block_j$ be the sub-path of $\block_j$ between $s$ and $s'$. We partition $I$ into paths containing $\floor{512 h\log^2M/2^p}$ vertices each, except for the last path that may contain fewer vertices. Since no such path may be heavy, we obtain at least $n'=\floor{\frac{z}{16\log M}}$ disjoint sub-paths of $I$, each of which contains $\floor{512 h\log^2M/2^p}$ vertices. We conclude that:

\[d(s,s')\geq n'\cdot \floor{512 h\log^2M/2^p}-1\geq \floor{\frac{z}{16\log M}}\cdot \floor{512 h\log^2M/2^p}-1\geq \frac{2zh\log M}{2^p}.\]

On the other hand, since $\sigma_{\mset^3}$ is the same as the ordering of $T(\mset^3)$ induced by $\sigma_{\mset^2}$, we get that:

\[N_{\mset^3}(s,s')\leq N_{\mset^2}(s,s')\leq \sum_{i=0}^zN_{\mset^2}(s_i,s_{i+1})\leq (z+1)\cdot \frac{128 h\log M}{2^p}\leq 128 d(s,s').\]
\end{proof}

\paragraph{Step 4 [Decreasing the Values $N_{\mset^3}(s,s')$].}
We are now ready to define the final set $\mset'\subseteq \mset^3$ of demand pairs. 

Assume that $\mset^3=\set{(s_1,t_1),\ldots,(s_{|\mset^3|},t_{|\mset^3|})}$, where the demand pairs are indexed according to the ordering of their destinations in $\sigma_{\mset^3}$ (notice that this is the same as the ordering of $T(\mset^3)$ induced by $\sigma_{\mset^2}$). We now define:

\[\mset'=\set{(s_i,t_i)\mid i=1\mod 512}.\]

We claim that $\mset'$ has the distance property. Indeed, consider any two pairs $(s,t),(s',t')\in \mset'$, such that $s$ and $s'$ are consecutive with respect to $\mset'$. If $s$ and $s'$ lie in different blocks $\block_j$, then, since the distance between any such pair of blocks is at least $10M$, while $|\mset'|\leq M$, we get that $N_{\mset'}(s,s')\leq d(s,s')/4$. Otherwise, both $s$ and $s'$ belong to the same block $\block_j$. But then it is easy to verify that $N_{\mset'}(s,s')\leq N_{\mset^3}(s,s')/512\leq d(s,s')/4$ from \Cref{clm: ndp-hard-almost distance property}.
We conclude that with probability at least $1/2$, neither of the events $\badevent$, $\badevent'$ happens, and in this case, $\mset'$ has the distance property.

\proofof{\Cref{lem: ndp-hard-can find routing}}
Recall that $R'$ and $R''$ are the rows of $\hat G$ containing the vertices of $S(\mset)$ and $T(\mset)$ respectively, and that $R'$ and $R''$ lie at distance at least $\ell/4$ form each other and from the top and the bottom boundaries of $\hat G$, where $\ell>M^2$ is the dimension of the grid. Let $R$ be any row lying between $R'$ and $R''$, within distance at least $\ell/16$ from each of them.

We denote $M'=|\mset'|$, and $\mset'=\set{(s_1,t_1),\ldots,(s_{M'},t_{M'})}$, where the pairs are indexed according to the ordering of their destination vertices in $\sigma_{\mset'}$. To recap, the ordering $\sigma_{\mset'}$ of $T(\mset')$ was defined as follows. We have defined a partition $(J_1,\ldots,J_r)$ of the vertices of $T(\mset')$ into subsets, where each set $J_i$ represents a cluster $W_i$ in our solution to the \WGPwB problem instance, and contains all destination vertices $t\in T(\mset')$ that lie in blocks $\block'_j$, for which the corresponding vertex $v_j\in V_2$ belongs to $W_i$. The ordering of the destination vertices inside each set $J_i$ is the same as their ordering on $R''$, and the different sets $J_1,\ldots,J_r$ are ordered in the order of their indices.

Let $X=\set{x_i\mid 1\leq i\leq M'}$ be a set of vertices of $R$, where $x_i$ is the $(2i)$th vertex of $R$ from the left. We will construct a set $\pset^*$ of spaced-out paths routing all demand pairs in $\mset'$, so that the path $P_i$ routing $(s_i,t_i)$ intersects the row $R$ at exactly one vertex --- the vertex $x_i$.
Notice that row $R$ partitions the grid $\hat G$ into two sub-grids: a top sub-grid $G^t$ spanned by all rows that appear above $R$ (including $R$), and a bottom sub-grid $G^b$ spanned by all rows that appear below $R$ (including $R$).

It is now enough to show that there are two sets of spaced-out paths: set $\pset^1$ routing all pairs in $\set{(s_i,x_i)\mid 1\leq i\leq M'}$ in the top grid $G^t$, and  set $\pset^2$ routing all pairs in $\set{(t_i,x_i)\mid 1\leq i\leq M'}$ in the bottom grid $G^b$, so that both sets of paths are internally disjoint from $R$.

\paragraph{Routing in the Top Grid.}Consider some vertex $v_j\in V_1$, and the corresponding block $\block_j$. We construct a sub-grid $\hat \block_j$ of $\hat G$, containing $\block_j$,  that we call a box, as follows. Let $\cset_j$ be the set of all columns of $\hat G$ intersecting $\block_j$. We augment $\cset_j$ by adding $2M$ columns lying immediately to the left and $2M$ columns lying immediately to the right of $\cset_j$, obtaining a set $\hat{\cset}_j$ of columns. Let $\hat \rset$ contain three rows: row $R'$; the row lying immediately above $R'$; and the row lying immediately below $R'$. Then box $\hat \block_j$ is the sub-grid of $\hat G$ spanned by the rows in $\hat \rset$ and the columns in $\hat {\cset}_j$. Since every block is separated by at least $10M$ columns from every other block, as well as the left and the right boundaries of $\hat G$, the resulting boxes are all disjoint, and every box is separated by at least $2M$ columns of $\hat G$  from every other box, and from the left and the right boundaries of $\hat G$. 

We will initially construct a set $\pset^1$ of spaced-out paths in $G^t$, such that each path $P_i\in \pset^1$, for $1\leq i\leq M'$, originates from the vertex $x_i$, and visits the boxes $\hat \block_1,\hat \block_2,\ldots,\hat \block_{N_1}$ in turn. We will ensure that each such path $P_i$ contains the corresponding source vertex $s_i$. Eventually, by suitably truncating each such path $P_i$, we will ensure that it connects $x_i$ to $s_i$.

\begin{claim}\label{clm: ndp-hard-selecting columns}
Consider some vertex $v_j\in V_1$, and the corresponding box $\hat \block_j$. Denote $Y_j=S(\mset')\cap V(\block_j)$, $M_j=|Y_j|$, and assume that $Y_j=\set{s_{i_1},s_{i_2},\ldots,s_{i_{M_j}}}$, where the indexing of the vertices of $Y_j$ is consistent with the indexing of the vertices of $S(\mset')$ that we have defined above, and $i_1<i_2<\ldots<i_{M_j}$. 
Then there is a set $\wset_j$ of $M'$  columns of the box $\hat \block_j$, such that:

\begin{itemize}
\item set $\wset_j$ does not contain a pair of consecutive columns; and
\item for each $1\leq z\leq M_j$, the $i_z$th column of $\wset_j$ from the left contains the source vertex $s_{i_z}$.
\end{itemize}
\end{claim}

\begin{proof}
Observe that from \Cref{obs: ndp-hard-ordering of sources and destinations}, the vertices $s_{i_1},s_{i_2},\ldots,s_{i_{M_j}}$ must appear in this left-to-right order on $R'$, while the vertices $x_{i_1},x_{i_2},\ldots,x_{i_{M_j}}$ appear in this left-to-right order on $R$. Moreover, for all $1\leq z<M_j$, $i_{z+1}-i_z-1=N_{\mset'}(s_{i_z},s_{i_{z+1}})\leq d(s_{i_z},s_{i_{z+1}})/4$. We add to $\wset_j$ all columns of $\hat \block_j$ where the vertices of $Y_j$ lie. For each $1\leq z<M_j$, we also add to $\wset_j$ an arbitrary set of $(i_{z+1}-i_z-1)$ columns lying between the column of $s_{i_z}$ and the column of $s_{i_{z+1}}$, so that no pair of columns in $\wset_j$ is consecutive. Finally, we add to $\wset_j$ $(i_1-1)$ columns that lie to the left of the column of $s_{i_1}$, and $(M'-i_{M_j})$ columns that lie to the right of the column of $s_{i_{M_j}}$. We make sure that no pair of columns in $\wset_j$ is consecutive -- it is easy to see that there are enough columns to ensure that. 
\end{proof}

\begin{claim}\label{clm: ndp-hard-the paths}
There is a set $\pset^1=\set{P_1,\ldots,P_{M'}}$ of spaced-out paths in $G^t$, that are internally disjoint from $R$, such that for each $1\leq i\leq M'$, path $P_i$ originates from vertex $x_i$, and for all  $1\leq j\leq N_1$, it contains the $i$th column of $\wset_j$; in particular, it contains $s_i$.
\end{claim}

\begin{figure}[H]
\centering
\includegraphics[width = 12cm]{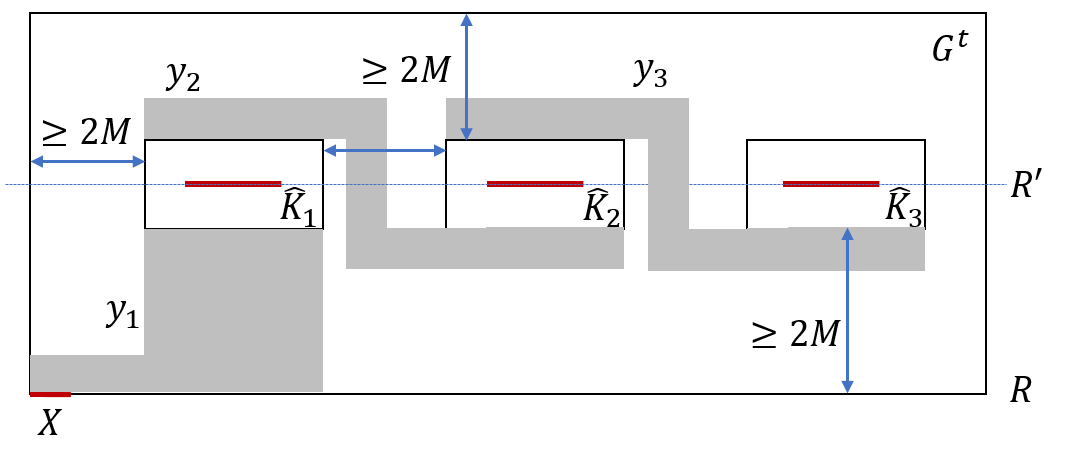} \label{fig: ndp-hard-global-routing}
\caption{Routing in the top grid.}
\label{fig: ndp-hard-routing}
\end{figure}

We defer the proof of the claim to \Cref{appdx: ndp-hard-routing in yi}; see \Cref{fig: ndp-hard-routing} for an illustration of the routing.

\paragraph{Routing in the Bottom Grid.}

Consider some vertex $v_j'\in V_2$, and the corresponding block $\block'_j$. We construct a box $\hat \block'_j$ containing $\block'_j$ exactly as before. As before, the resulting boxes are all disjoint, and every box is separated by at least $2M$ columns of $\hat G$  from every other box, and from the left and the right boundaries of $\hat G$. 

As before, we will initially construct a set $\pset^2$ of spaced-out paths in $G^b$, such that each path $P'_i\in \pset^2$, for $1\leq i\leq M'$, originates from the vertex $x_i$, and visits the boxes $\hat \block'_1,\hat \block'_2,\ldots,\hat \block'_{N_2}$ in turn. We will ensure that each such path $P_i'$ contains the corresponding destination vertex $t_i$. Eventually, by suitably truncating each such path $P'_i$, we will ensure that it connects $x_i$ to $t_i$.

\begin{claim}\label{clm: ndp-hard-selecting columns2}
Consider some vertex $v_j'\in V_2$, and the corresponding box $\hat \block_j'$. Denote $Y_j'=T(\mset')\cap V(\block'_j)$, $M'_j=|Y'_j|$, and assume that $Y'_j=\set{t_{i_1},t_{i_2},\ldots,t_{i_{M_j}}}$, where the indexing of the vertices of $Y'_j$ is consistent with the indexing of the vertices of $T(\mset')$ that we have defined above, and $i_1<i_2<\ldots<i_{M'_j}$. 
Then there is a set $\wset'_j$ of $M'$  columns of the box $\hat \block_j'$, such that:

\begin{itemize}
\item set $\wset_j'$ does not contain a pair of consecutive columns; and
\item for each $1\leq z\leq M_j'$, the $i_z$th column of $\wset'_j$ from the left contains the destination vertex $t_{i_z}$.
\end{itemize}
\end{claim}

\begin{proof}
From our construction of $\sigma_{\mset'}$, the vertices $t_{i_1},t_{i_2},\ldots,t_{i_{M_j}}$ appear consecutively in this order in $\sigma_{\mset'}$. Moreover, from the construction of $\mset$, every pair of these destination vertices is separated by at least one vertex. We add to $\wset_j'$ all columns in which the vertices $t_{i_1},t_{i_2},\ldots,t_{i_{M_j}}$ lie. We also add to $\wset_j'$ $i_1-1$ columns that lie to the left of the column of $t_{i_1}$, and $M'-i_{M_j'}$ columns that lie to the right of the column of $t_{i_{M_j}}$ in $\hat \block_j'$. We make sure that no pair of columns in $\wset_j'$ is consecutive -- it is easy to see that there are enough columns to ensure that. 
\end{proof}

The proof of the following claim is identical to the proof of \Cref{clm: ndp-hard-the paths} and is omitted here.
\begin{claim}\label{clm: ndp-hard-the paths2}
There is a set $\pset^2=\set{P_1',\ldots,P_{M'}'}$ of spaced-out paths in $G^b$, that are internally disjoint from $R$, such that for each $1\leq i\leq M'$, path $P_i'$ originates from vertex $x_i$, and for all  $1\leq j\leq N_2$, it contains the $i$th column of $\wset_j'$; in particular it contains $t_i$
\end{claim}

By combining the paths in sets $\pset^1$ and $\pset^2$, we obtain a new set $\pset^*=\set{P_1^*,\ldots,P_{M'}^*}$ of spaced-out paths, such that for all $1\leq i\leq M'$, path $P_i^*$ contains $s_i,x_i$ and $t_i$. By suitably truncating each such path, we obtain a collection of spaced-out paths routing all demand pairs in $\mset'$.

\endproofof
\subsection{From Routing to Partitioning}\label{subsubsec: ndp-hard-NI from routing to partitioning}

The goal of this subsection is to prove the following theorem, that will complete the proof of \Cref{thm: ndp-hard-from WGP to NDP}.

\begin{theorem}\label{thm: ndp-hard-can get good partition from routing}
There is a deterministic  efficient algorithm, that,  given a valid instance $\iset=(\tilde G=(V_1,V_2,E),\uset_1,\uset_2,h,r)$ of the \WGPwB problem with $|E|=M$, the corresponding (random) instance $\hat {\iset}$ of \NDPgrid, and 
 a solution $\pset^*$ to $\hat{\iset}$, computes a solution to the \WGPwB instance $\iset$ of value at least $\Omega(|\pset^*|/\log^3M)$.
\end{theorem}

Let $\mset^*\subseteq \mset$ be the set of the demand pairs routed by the solution $\pset^*$, and let $E^*\subseteq E$ be the set of all edges $e$, whose corresponding demand pair belongs to $\mset^*$. Let $\tG'\subseteq \tG$ be the sub-graph of $\tG$ induced by the edges in $E^*$. Notice that whenever two edges of $\tG$ belong to the same bundle, their corresponding demand pairs share a source or a destination. Since all paths in $\pset^*$ are node-disjoint, all demand pairs in $\mset^*$ have distinct sources and destinations, and so no two edges in $E^*$ belong to the same bundle.

Note that, if $|\pset^*| \leq 2^{64} h \log^3 M$, then we can return the solution $((W_1,\ldots,W_r), (E_1,\ldots,E_r))$, where $W_1=V(\tG)$ and $W_2=W_3=\cdots=W_r=\emptyset$; set $E_1$ contains an arbitrary subset of $\ceil{\frac{|\pset^*|}{2^{64}\log^3M}} \leq h$ edges of $E^*$, and all other sets $E_i$ are empty. Since no two edges of $E^*$ belong to the same bundle, we obtain a feasible solution to the \WGPwB problem instance of value  $\Omega(|\pset^*|/\log^3M)$.
Therefore, from now on, we assume that $|\pset^*|>2^{64}h\log^3M$.

Our algorithm computes a solution to the \WGPwB instance $\iset$ by repeatedly partitioning $\tG'$ into smaller and smaller sub-graphs, by employing suitably defined balanced cuts.

Recall that, given a graph $\bfH$, a \emph{cut} in $\bfH$ is a bi-partition $(A,B)$ of its vertices. We denote by $E_{\bfH}(A,B)$ the set of all edges with one endpoint in $A$ and another in $B$, and by $E_{\bfH}(A)$ and $E_{\bfH}(B)$ the sets of all edges with both endpoints in $A$ and in $B$, respectively.
Given a cut $(A,B)$ of $\bfH$, the \emph{value} of the cut is $|E_{\bfH}(A,B)|$. We will omit the subscript $\bfH$ when clear from context. 

\begin{definition}
    Given a graph $\bfH$ and a parameter $0<\rho<1$, a cut $(A,B)$ of $\bfH$ is called a $\rho$-edge-balanced cut iff $|E(A)|,|E(B)| \geq \rho \cdot |E(\bfH)|$. 
\end{definition}

The following theorem is central to the proof of \Cref{thm: ndp-hard-can get good partition from routing}.

\begin{theorem}\label{thm: ndp-hard-balanced partition}
There is an efficient algorithm, that, given a vertex-induced subgraph $\bfH$ of $\tG'$ with $|E(\bfH)| >  2^{64}h\log^3M$, computes a $1/32$-edge-balanced cut of $\bfH$, of value at most $\frac{|E(\bfH)|}{64 \log M}$.
\end{theorem}

We prove \Cref{thm: ndp-hard-balanced partition} below, after we complete the proof of \Cref{thm: ndp-hard-can get good partition from routing} using it.
Our algorithm maintains a collection $\gset$ of disjoint vertex-induced sub-graphs of $\tG'$, and consists of a number of phases. The input to the first phase is the collection $\gset$ containing a single graph - the graph $\tG'$. The algorithm continues as long as $\gset$ contains a graph $\bfH\in \gset$ with $|E(\bfH)| > 2^{64} \cdot h \log^3 M$; if no such graph $\bfH$ exists, the algorithm terminates. Each phase is executed as follows. We process every graph $\bfH\in \gset$ with $|E(\bfH)| > 2^{64} \cdot h \log^3 M$ one-by-one. When graph $\bfH$ is processed, we apply \Cref{thm: ndp-hard-balanced partition} to it, obtaining a $1/32$-edge-balanced cut $(A,B)$ of $\bfH$, of value at most $\frac{|E(\bfH)|}{64 \log M}$. We then remove $\bfH$ from $\gset$, and add $\bfH[A]$ and $\bfH[B]$ to $\gset$ instead. This completes the description of the algorithm.
 We use the following claim to analyze it.

\begin{claim} \label{clm: ndp-hard-gset is good solution2}
 Let $\gset'$ be the final set of disjoint sub-graphs of $\tG'$ obtained at the end of the algorithm. Then    $\sum_{\bfH \in \gset'} |E(\bfH)| \geq \Omega(|E(\tG')|)$, and   $|\gset'| \leq r$.
\end{claim}
\begin{proof}
We construct a binary partitioning tree $\tau$ of graph $\tG'$, that simulates the graph partitions computed by the algorithm. 
For every sub-graph $\bfH\subseteq \tG'$ that belonged to $\gset$ over the course of the algorithm, tree $\tau$ contains a vertex $v(\bfH)$. The root of the tree is the vertex $v(\tG')$. If, over the course of our algorithm, we have partitioned the graph $\bfH$ into two disjoint vertex-induced sub-graphs $\bfH'$ and $\bfH''$, then we add an edge from $v(\bfH)$ to $v(\bfH')$ and to $v(\bfH'')$, that become the children of $v(\bfH)$ in $\tau$. 

The \emph{level} of a vertex $v(\bfH)$ in the tree is the length of the path connecting $v(\bfH)$ to the root of the tree; so the root of the tree is at level $0$. The \emph{depth} of the tree, that we denote by $\Delta$, is the length of the longest leaf-to-root path in the tree. 
Since the cuts computed over the course of the algorithm are $1/32-$edge-balanced, $\Delta\leq \frac{\log M}{\log{(32/31)}}$. 
Consider now some  level $0\leq i\leq \Delta$, and let $V_i$ be the set of all vertices of the tree $\tau$ lying at level $i$. Let $\hat E_i=\bigcup_{v(\bfH)\in V_i}E(\bfH)$ be the set of all edges contained in all sub-graphs of $\tG'$, whose corresponding vertex belongs to level $i$. Finally, let $m_i=|E_i|$. Then $m_0=|E(\tG')|$, and
for all $1\leq i\leq \Delta$, the number of edges discarded over the course of phase $i$ is $m_{i-1}-m_{i}\leq m_{i-1}/(64\log M)\leq m_0/(64\log M)$ --- this is since, whenever we partition a graph $\bfH$ into two subgraphs, we lose at most $|E(\bfH)|/(64 \log M)$ of its edges. Overall, we get that:

\[m_0-m_{\Delta}\leq \frac{\Delta m_0}{64\log M}  \leq \frac{\log M}{\log{32/31}}\cdot \frac{m_0}{64\log M}\leq \frac{m_0} 2,\]

and so $\sum_{\bfH \in \gset'} |E(\bfH)|=m_{\Delta}\geq m_0/2$. This finishes the proof of the first assertion. We now turn to prove the second assertion. Recall that no two edges of $E^*$ may belong to the same bundle. Since $h=\beta^*(\iset)/r=\left(\sum_{v\in V_1}\beta(v)\right )/r$, for any subset $E'\subseteq E^*$ of edges, $|E'|\leq \sum_{v\in V_1}\beta(v)\leq hr$ must hold, and in particular, $|E^*|\leq hr$. It is now enough to prove that for every leaf vertex $v(\bfH)$ of $\tau$, $|E(\bfH)|\geq h$ --- since all graphs in $\gset'$ are mutually disjoint, and each such graph corresponds to a distinct leaf of $\tau$, this would imply that $|\gset'|\leq r$.

Consider now some leaf vertex $v(\bfH)$ of $\tau$, and let $v(\bfH')$ be its parent. The $|E(\bfH')|\geq 2^{64}h\log^3M$, and, since the partition of $\bfH'$ that we have computed was $1/32$-balanced, $|E(\bfH)|\geq |E(\bfH')|/32\geq h$. We conclude that $|\gset'|\leq r$.
\end{proof}

We are now ready to define the solution $((W_1,\ldots,W_r),(E_1,\ldots,E_r))$ to the \WGPwB problem instance $\iset$. Let $\gset'$ be the set of the sub-graphs of $\tG'$ obtained at the end of our algorithm, and denote $\gset'=\set{\bfH_1,\bfH_2,\ldots,\bfH_z}$. Recall that from \Cref{clm: ndp-hard-gset is good solution2}, $z\leq r$. For $1\leq i\leq z$, we let $W_i=V(\bfH_i)$. If $|E(\bfH_i)|\leq h$, then we let $E_i=E(\bfH_i)$; otherwise, we let $E_i$ contain any subset of $h$ edges of $E(\bfH_i)$. Since $|E(\bfH_i)|\leq 2^{64}h\log^3M$, in either case, $|E_i|\geq \Omega(|E(\bfH_i)|/\log^3M)$. For $i>z$, we set $W_i=\emptyset$ and $E_i=\emptyset$. Since, as observed before, no pair of edges of $E^*$ belongs to the same bundle, it is immediate to verify that we obtain a feasible solution to the \WGPwB problem instance. The value of the solution is:

\[\sum_{i=1}^r|E_i|\geq \sum_{i=1}^r\Omega(|E(\bfH_i)|/\log^3M)=\Omega(|E(\tG')|/\log^3M)= \Omega(|\pset^*|/\log^3M),\]

from \Cref{clm: ndp-hard-gset is good solution2}. In order to complete the proof of \Cref{thm: ndp-hard-can get good partition from routing}, it now remains to prove \Cref{thm: ndp-hard-balanced partition}.

\begin{proofof}{\Cref{thm: ndp-hard-balanced partition}}
Let $\bfH$ be a vertex-induced subgraph of $\tG'$ with $|E(\bfH)| >  2^{64}h\log^3M$. 
Our proof consists of two parts. First, we show an efficient algorithm to compute a drawing of $\bfH$ with relatively few crossings. Next, we show how to exploit this drawing in order to compute a $1/32$-edge-balanced cut of $\bfH$ of small value. We start by defining a drawing of a given graph in the plane and the crossings in this drawing.

\begin{definition}
A \emph{drawing} of a given graph $\bfH'$ in the plane is a mapping, in which every vertex of $\bfH$ is mapped to a point in the plane, and every edge to a continuous curve connecting the images of its endpoints, such that no three curves meet at the same point; no curve intersects itself; and no curve contains an image of any vertex other than its endpoints.
A \emph{crossing} in such a drawing is a point contained in the images of two edges.
\end{definition}

\begin{lemma}\label{lem: ndp-hard-drawing of H}
There is an efficient algorithm that, given a solution $\pset^*$ to the \NDPgrid problem instance $\hat \iset$, and a vertex-induced subgraph $\bfH\subseteq \tG'$, computes a drawing of $\bfH$ with at most  $\twiceConstantForSizeOfBlocks \cdot |E(\bfH)| \ceil{h \log M}$ crossings.%
\end{lemma}

\begin{proof}
We think of the grid $\hat G$ underlying the \NDPgrid instance $\hat \iset$ as the drawing board, and map the vertices of $\bfH$ to points inside some of its carefully selected cells. Consider a vertex $v_i \in V(\bfH) \cap V_1$, and let $\block_i$ be the block representing this vertex.
Let $\kappa_i$ be any cell of the grid $\hat G$ that has a vertex of $\block_i$ on its boundary. We map the vertex $v_i$ to a point $p_i$ lying in the middle of the cell $\kappa_i$  (it is sufficient that $p_i$ is far enough from the boundaries of the cell).
For every vertex $v'_j\in V(\bfH)\cap V_2$, we select a cell $\kappa'_j$ whose boundary contains a vertex of the corresponding block $\block'_j$, and map $v'_j$ to a point $p'_j$ lying in the middle of $\kappa'_j$ similarly. 

Next, we define the drawings of the edges of $E(\bfH)$. 
Consider any such edge $e = (v_i,v'_j) \in E(\bfH)$, with $v_i\in V_1$ and $v'_j\in V_2$, and let $(s,t)\in \mset^*$ be its corresponding demand pair.
Let $\block_i$ and $\block'_j$ be the blocks containing $s$ and $t$ respectively, and let $P\in \pset^*$ be the path routing the demand pair $(s,t)$ in our solution to the \NDPgrid problem instance $\hat \iset$.
The drawing of the edge $e$ is a concatenation of the following three segments: (i) the image of the path $P$, that we refer to as a type-1 segment; (ii) a straight line connecting $s$ to the image of $v_i$, that we refer to as a type-2 segment; and (iii) a straight line connecting $t$ to the image of $v'_i$, that we refer to as a type-3 segment. If the resulting curve has any self-loops, then we delete them.

We now bound the number of crossings in the resulting drawing of $\bfH$. Since the paths in $\pset^*$ are node-disjoint, whenever the images of two edges $e$ and $e'$ cross, the crossing must be between the type-1 segment of $e$, and either the type-2 or the type-3 segment of $e'$ (or the other way around). 

Consider now some edge $e=(v_i,v'_j) \in E(\bfH)$ with $v_i\in V_1$ and $v'_j\in V_2$, and let $\block_i$ and $\block'_j$ be the blocks representing $v_i$ and $v'_j$ respectively. Let $(s,t)\in \mset^*$ be the demand pair corresponding to $e$. Assume that a type-1 segment of some edge $e'$ crosses a type-$2$ segment of $e$. This can only happen if 
the path $P'\in \pset^*$ routing the demand pair $(s',t')$ corresponding to the edge $e'$ contains a vertex of $\block_i$. Since $|V(\block_i)| \leq \constantForSizeOfBlocks \cdot \ceil{h \log M}$, at most $ \constantForSizeOfBlocks \cdot \ceil{h \log M}$ type-1 segments of other edges may cross the type-2 segment of $e$. The same accounting applies to the type-3 segment of $e$. 
Overall, the number of crossings in the above drawing is bounded by: 

\[\sum_{e \in E(\bfH)} 2 \cdot \constantForSizeOfBlocks \cdot \ceil{h \log M} = \twiceConstantForSizeOfBlocks \cdot |E(\bfH)| \ceil{h \log M}.\]%
\end{proof}

Next, we show an efficient algorithm that computes a small balanced partition of any given graph $\bfH'$, as long as its maximum vertex degree is suitably bounded, and we are given a drawing of $\bfH'$ with a small number of crossings.

\begin{lemma}\label{lem: ndp-hard-good balanced cut in low crossing number}
There is an efficient algorithm that, given any graph $\bfH$ with $|E(\bfH)|=m$ and maximum vertex degree at most $d$, and a drawing $\phi$ of $\bfH$ with at most $\cro\leq md\alpha$ crossings for some $\alpha>1$, such that $m>2^{20}d\alpha$, computes a $1/32$-edge-balanced cut $(A,B)$ of $\bfH$ of value $|E(A,B)|\leq 64\sqrt{8md\alpha}$.
\end{lemma}

Before we complete the proof of the lemma, we show that the proof of \Cref{thm: ndp-hard-balanced partition} follows from it. Let $m=|E(\bfH)|$. Note that the maximum vertex degree in graph $\bfH$ is bounded by $\max_{v\in V_1\cup V_2}\set{\beta(v)}$, as $E(\bfH)$ cannot contain two edges that belong to the same bundle. From the definition of valid instances, $h\geq \beta(v)$, and so the maximum vertex degree in $\bfH$ is bounded by $d=h$. From \Cref{lem: ndp-hard-drawing of H}, the drawing $\phi$ of $\bfH$ has at most $2048 |E(\bfH)| \ceil{h \log M}\leq 2^{12}m d\log M$ crossings. Setting $\alpha=2^{12}\log M$, the number of crossings $\cro$ in $\phi$ is bounded by $md\alpha$. Moreover, since $m>2^{64}h\log^3M$, we get that $m>2^{20}d\alpha$. We can now apply \Cref{lem: ndp-hard-good balanced cut in low crossing number} to graph $\bfH$ to obtain a $1/32$-edge-balanced cut $(A,B)$ with $|E(A,B)|\leq 64\sqrt{8md\alpha}\leq 64\sqrt{2^{15}mh\log M}\leq 64\sqrt{m^2/(2^{49}\log^2M)}\leq \frac{|E(\bfH)|}{64\log M}$, since we have assumed that $|E(\bfH)|=m>2^{64}h\log^3M$. It now remains to prove \Cref{lem: ndp-hard-good balanced cut in low crossing number}.

\begin{proofof}{\Cref{lem: ndp-hard-good balanced cut in low crossing number}}
For each vertex $v \in V(\bfH)$, we denote the degree of $v$ in $\bfH$ by $d_v$. We assume without loss of generality that for all $v\in V(\bfH)$, $d_v\geq 1$: otherwise, we can remove all isolated vertices from $\bfH$ and then apply our algorithm to compute a $1/32$-edge-balanced cut $(A,B)$ in the remaining graph. At the end we can add the isolated vertices to $A$ or $B$, while ensuring that the cut remains $1/32$-balanced, and without increasing its value.

We construct a new graph $\hat \bfH$ from graph $\bfH$ as follows. For every vertex $v\in V(\bfH)$, we add a  $(d_v \times d_v)$-grid $Q_v$ to $\hat \bfH$, so that the resulting grids are mutually disjoint. We call the edges of the resulting grids \emph{regular edges}. 
Let $e_1(v), \ldots, e_{d_v}(v)$ be the edges of $\bfH$ incident to $v$, indexed in the clockwise order of their entering the vertex $v$ in the drawing $\phi$ of $\bfH$. We denote by $\Pi(v)=\set{p_1(v),\ldots,p_{d_v}(v)}$ the set of vertices on the top boundary of $Q_v$, where the vertices are indexed in the clock-wise order of their appearance on the boundary of $Q_v$. We refer to the vertices of $\Pi(v)$ as the \emph{portals} of $Q_v$ (see \Cref{fig: ndp-hard-v_to_gv}). Let $\Pi=\bigcup_{v\in V(\bfH)}\Pi(v)$ be the set of all portals. For every edge $e=(u,v)\in E(\bfH)$, we add a new \emph{special edge} to graph $\hat \bfH$, as follows. Assume that $e=e_i(v)=e_j(u)$. Then we add an edge $(p_i(v),p_j(u))$ to $\hat \bfH$. We think of this edge as the special edge representing $e$. This finishes the definition of the graph $\hat \bfH$. It is immediate to see that the drawing $\phi$ of $\bfH$ can be extended to a drawing $\phi'$ of $\hat \bfH$ without introducing any new crossings, that is, the number of crossings in $\phi'$ remains at most $\cro$. Note that every portal vertex is incident to exactly one special edge, and the maximum vertex degree in $\hat \bfH$ is $4$. We will use the following bound on $|V(\hat \bfH)|$:

\begin{figure}[H]
\center
\includegraphics[width=8cm]{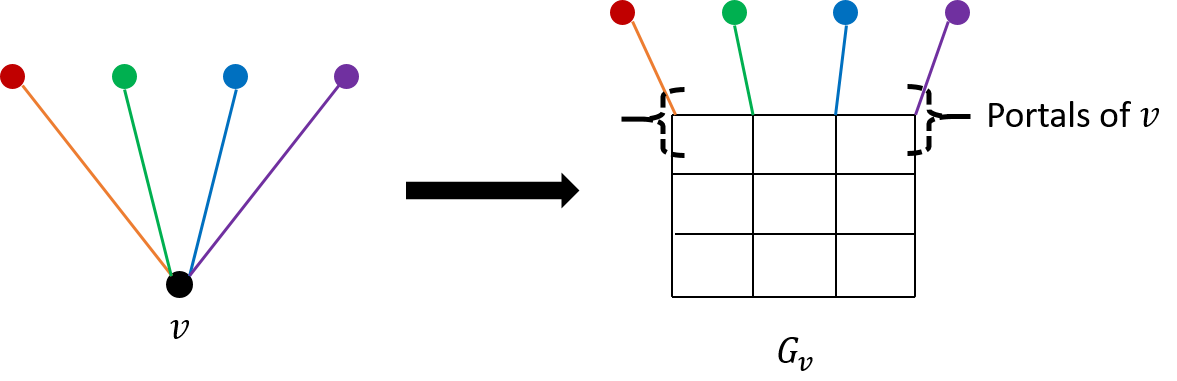}
\caption{Grid $Q_v$ obtained from $v$} \label{fig: ndp-hard-v_to_gv}
\end{figure}

\begin{observation}\label{obs: ndp-hard-number of vertices in hat H}
    $|V(\hat \bfH)|\leq (2m+d)d$.
\end{observation}

The proof of \Cref{obs: ndp-hard-number of vertices in hat H} is present in \Cref{prf-obs: ndp-hard-number of vertices in hat H}.
Let $\hat \bfH'$ be the graph obtained from $\hat \bfH$ by replacing every intersection point in the drawing $\phi'$ of $\hat \bfH$ with a vertex.
Then $\hat \bfH'$ is a planar graph with,

\[ |V(\hat \bfH')|\leq |V(\hat \bfH)|+\cro\leq  (2m+d)d+md\alpha\leq 4md\alpha,\]

as $\alpha\geq 1$. We assign weights to the vertices of $\hat \bfH'$ as follows: every vertex of $\Pi$ is assigned the weight $1$, and every other vertex is assigned the weight $0$. Note that the weight of a vertex is exactly the number of special edges incident to it, and the total weight of all vertices is $W=|\Pi|=2m$.
We will use the following version of the planar separator theorem~\cite{planar-separator-theorem2,planar-separator-theorem1,planar-separator-theorem3}.

\begin{theorem}[\cite{planar-separator-theorem1}]\label{thm: ndp-hard-balanced separator}
    There is an efficient algorithm, that, given a planar graph $G=(V,E)$  with $n$ vertices, and an assignment $w:V\rightarrow R^+$ of non-negative weights to the vertices of $G$, with $\sum_{v\in V}w(v)=W$, computes a partition $(A,X,B)$ of $V(G)$, such that:

    \begin{itemize}
        \item no edge connecting a vertex of $A$ to a vertex of $B$ exists in $G$;
        \item $\sum_{v\in A}w(v),\sum_{v\in B}w(v)\leq 2W/3$; and
        \item $|X|\leq 2\sqrt{2n}$.
    \end{itemize}
\end{theorem}

We apply \Cref{thm: ndp-hard-balanced separator} to graph $\hat \bfH'$, to obtain a partition $(A,X,B)$ of $V(\hat \bfH')$, with $|X|\leq 2\sqrt{2|V(\hat \bfH')|}\leq 2\sqrt{8md\alpha}$. Since $W=\sum_{v\in V(\hat \bfH')}w(v)=2m$, we get that $|A\cap \Pi|= \sum_{v\in A}w(v)\leq 2W/3\leq 4m/3$, and similarly $|B\cap \Pi|\leq 4m/3$.
Assume without loss of generality that $|A\cap \Pi|\leq |B\cap \Pi|$. We obtain a bi-partition $(A',B')$ of $V(\hat \bfH')$ by setting $A'=A\cup X$ and $B'=B$. Since $|X|\leq 2\sqrt{8md\alpha}\leq m/3$ (as $m>2^{20}d\alpha$), we are guaranteed that $|A'\cap \Pi|,|B'\cap \Pi|\leq 4m/3$ holds. Moreover, as all vertex degrees in $\hat \bfH'$ are at most $4$, $|E(A',B')|\leq 4|X|\leq 8\sqrt{8md\alpha}$.

Unfortunately, the cut $(A',B')$ of $\hat H'$ does not directly translate into a balanced cut in $\bfH$, since for some vertices $v\in V(\bfH)$, the corresponding grid $Q_v$ may be split between $A'$ and $B'$. We now show how to overcome this difficulty, by moving each such grid entirely to one of the two sides.
Before we proceed, we state a simple fact about grid graphs, whose proof is present in \Cref{prf-obs: ndp-hard-cuts in grids} for the sake of completeness.

\begin{observation}\label{obs: ndp-hard-cuts in grids}
    Let $z>1$ be an integer, and let $Q$ be the $(z\times z)$-grid. Let $U$ be the set of vertices lying on the top row of $Q$, and let $(X,Y)$ be a bi-partition of $V(Q)$. Then $|E_Q(X,Y)|\geq \min\set{|U\cap X|,|U\cap Y|}$.
\end{observation}

We say that a vertex $v \in V(\bfH)$ is \textit{split} by the cut $(A',B')$ iff $V(Q_v)\cap A'$ and $V(Q_v)\cap B'\neq \emptyset$. We say that it is \emph{split evenly} iff $|\Pi(v)\cap A'|,|\Pi(v)\cap B'|\geq d_v/8$; otherwise we say that it is \emph{split unevenly}.
We modify the cut $(A',B')$ in the following two steps, to ensure that no vertex of $V(\bfH)$ remains split.

\paragraph{Step 1 [Unevenly split vertices].}
We process each vertex $v\in V(\bfH)$ that is unevenly split one-by-one. Consider any such vertex $v$. If $|\Pi(v)\cap A'|>|\Pi(v) \cap B'|$, then we move all vertices of 
$Q_v$ to $A'$; otherwise we move all vertices of $Q_v$ to $B'$. Assume without loss of generality that the former happens.
Notice that from \Cref{obs: ndp-hard-cuts in grids}, $|E(A',B')|$ does not increase, since $E(Q_v)$ contributed at least $|\Pi(v)\cap B'|$ regular edges to the cut before the current iterations.  Moreover, $|\Pi(v)\cap A'|$ increases by the factor of at most $8/7$. Therefore, at the end of this procedure, once all unevenly split vertices of $\bfH$ are processed, $|A'\cap \Pi|,|B'\cap \Pi|\leq \frac 8 7 \cdot \frac 4 3 m=\frac{32}{21}m$ and $|E(A',B')|\leq 8\sqrt{8md\alpha}$.

\paragraph{Step 2 [Evenly split vertices].}
In this step, we process each vertex $v\in V(\bfH)$ that is evenly split one-by-one. Consider an iteration where some such vertex $v\in V(\bfH)$ is processed. If $|A'\cap \Pi|\leq |B'\cap \Pi|$, then we move all vertices of $Q_v$ to $A'$; otherwise we move them to $B'$. Assume without loss of generality that the former happened. Then before the current iteration $|A'\cap \Pi|\leq |\Pi|/2\leq m$, and,  since $|\Gamma(v)|\leq d<m/21$,  $|A'\cap \Pi|\leq \frac{32}{21}m$, while $|B'\cap \Pi|\leq \frac{32}{21}m$ as before. Moreover, from \Cref{obs: ndp-hard-cuts in grids}, before the current iteration, the regular edges of $Q_v$ contributed at least $d(v)/8$ edges to $E(A',B')$, and after the current iteration, no regular edges of $Q_v$ contribute to the cut, but we may have added up to $d(v)$ new special edges to it. Therefore, after all vertices of $\bfH$ that are evenly split are processed, $|E(A',B')|$ grows by the factor of at most $8$, and remains at most $64\sqrt{8md\alpha}$.

We are now ready to define the final cut $(A^*,B^*)$ in graph $\bfH$. We let $A^*$ contain all vertices $v\in V(\bfH)$ with $V(Q_v)\subseteq A'$, and we let $B^*$ contain all remaining vertices of $V(\bfH)$. Clearly, $|E_{\bfH}(A^*,B^*)|\leq |E_{\hat \bfH'}(A',B')|\leq 64\sqrt{8md\alpha}$. It remains to show that $|E_{\bfH}(A^*)|, |E_{\bfH}(B^*)| \geq |E(\bfH)|/32$. We show that $|E_{\bfH}(A^*)| \geq |E(\bfH)|/32$; the proof that $|E_{\bfH}(B^*)| \geq |E(\bfH)|/32$ is symmetric.
Observe that $\sum_{v\in B^*}d_v=|B'\cap \Pi|\leq \frac{32m}{21}$, while $\sum_{v\in V(\bfH)}d_v=2m$. Therefore, $\sum_{v\in A^*}d_v\geq 2m-\frac{32m}{21}=\frac{10m}{21}$. But $|E_{\bfH}(A^*,B^*)|\leq 64\sqrt{8md\alpha}\leq 64\sqrt{m^2/2^{17}}<m/4$ (since $m>2^{20}d\alpha$). Therefore,

\[|E_{\bfH}(A^*)|=\frac{\sum_{v\in A^*}d_v-|E_{\bfH}(A^*,B^*)|}{2}\geq \frac{5m}{21}-\frac m 8\geq\frac m {32}.\]

\end{proofof} \end{proofof}

    \section{Hardness of \NDP and \EDP on Wall Graphs}\label{subsec: ndp-hard-from NDP to EDP}
    
In this subsection we extend our results to \NDP and \EDP on wall graphs, completing the proofs of \Cref{thm: ndp-hard-master NDP} and \Cref{thm: ndp-hard-master EDP}. We first prove hardness of \NDPwall, and show later how to extend it to \EDPwall.
Let $\hat G=G^{\ell,h}$ be a grid of length $\ell$ and height $h$, where $\ell>0$ is an even integer, and $h>0$.
We denote by $\hat G'$ the wall corresponding to $\hat G$, as defined in \Cref{subsec: ndp-hard-prelims}.
We prove the following analogue of \Cref{thm: ndp-hard-from WGP to NDP}.

\begin{theorem}\label{thm: ndp-hard-from WGP to NDP in walls}
    There  is a constant $c^*>0$, and there is an efficient randomized algorithm, that, given a valid instance $\iset=(\tilde G, \uset_1,\uset_2,h,r)$ of \WGPwB with $|E(\tilde G)|=M$, constructs an instance $\hat{\iset}'=(\hat G',\mset)$ of \NDPwall with $|V(\hat G')|=O(M^4\log^2M)$, such that the following hold:

    \begin{itemize}
        \item If $\iset$ has a perfect solution (of value $\beta^*=\beta^*(\iset)$), then with probability at least $\half$ over the construction of $\hat \iset'$, instance $\hat\iset'$ has a solution $\pset'$ that routes at least $\frac{\beta^*}{c^*\log^3M}$ demand pairs via node-disjoint paths; and

        \item There is a deterministic efficient algorithm, that, given a solution $\pset^*$ to the \NDPwall problem instance $\hat{\iset}'$, constructs a solution to the \WGPwB instance $\iset$, of value at least $\frac{|\pset^*|}{c^*\cdot \log^3M}$.
    \end{itemize}
\end{theorem}

Notice that plugging \Cref{thm: ndp-hard-from WGP to NDP in walls} into the hardness of approximation proof instead of \Cref{thm: ndp-hard-from WGP to NDP}, we extend the hardness result to the \NDP problem on wall graphs and complete the proof of \Cref{thm: ndp-hard-master NDP}.

\begin{proofof}{\Cref{thm: ndp-hard-from WGP to NDP in walls}}
Let $\hat \iset=(\hat G,\mset)$ be the instance of \NDPgrid constructed in \Cref{thm: ndp-hard-from WGP to NDP}. In order to obtain an instance $\hat \iset'$ of \NDPwall, we replace the grid $\hat G$ with the corresponding wall $\hat G'$ as described above; the set of the demand pairs remains unchanged. We now prove the two assertions about the resulting instance $\hat{\iset}'$, starting from the second one.

Suppose we are given a solution $\pset^*$ to the \NDPwall problem instance $\hat{\iset}'$. Since $\hat G'\subseteq \hat G$, and the set of demand pairs in instances $\hat \iset$ and $\hat \iset'$ is the same, $\pset^*$ is also a feasible solution to the \NDPgrid problem instance $\hat \iset$, and so we can use the deterministic efficient algorithm from \Cref{thm: ndp-hard-from WGP to NDP} to construct  a solution to the \WGPwB instance $\iset$, of value at least $\frac{|\pset^*|}{c^*\cdot \log^3M}$.

It now remains to prove the first assertion. Assume that $\iset$  has a perfect solution. Let $\event$ be the good event that the instance $\hat \iset$ of \NDPgrid has a solution $\pset$  that routes at least $\frac{\beta^*}{c^*\log^3M}$ demand pairs via  paths that are spaced-out. From \Cref{thm: ndp-hard-from WGP to NDP}, event $\event$ happens with probability at least $\half$. It is now enough to show that whenever event $\event$ happens, there is a solution of value $\frac{\beta^*}{c^*\log^3M}$ to the corresponding instance $\hat \iset'$ of \NDPwall.

Consider the spaced-out set $\pset$ of paths in $\hat G$. Recall that for every pair $P,P'$ of paths, $d(V(P),V(P'))\geq 2$, and all paths in $\pset$ are internally disjoint from the boundaries of the grid $\hat G$. For each path $P\in \pset$, we will slightly modify $P$ to obtain a new path $P'$ contained in the wall $\hat G'$, so that the resulting set $\pset'=\set{P'\mid P\in \pset}$ of paths is node-disjoint.

For all $1\leq i<\ell$, $1\leq j\leq \ell$, let $e_i^j$ denote the $i$th edge from the top lying in column $W_j$ of the grid $\hat G$, so that $e^i_j=(v(i,j),v(i+1,j))$.
Let $E^*=E(\hat G)\setminus E(\hat G')$ be the set of edges that were deleted from  the grid $\hat G$ when constructing the wall $\hat G'$. We call the edges of $E^*$ \emph{bad edges}.
Notice that only vertical edges may be bad, and, if $e^j_i\in E(W_j)$ is a bad edge, for $1<j<\ell$, then $e^{j+1}_i$ is a good edge. Consider some bad edge $e^j_i=(v(i,j),v(i+1,j))$, such that $1<j<\ell$, so $e^j_i$ does not lie on the boundary of $\hat G$. Let $Q^j_i$ be the path $(v(i,j),v(i,j+1),v(i+1,j+1),v(i+1,j))$. Clearly, path $Q^j_i$ is contained in the wall $\hat G'$. For every path $P\in \pset$, we obtain the new path $P'$ by replacing every bad edge $e^j_i\in P$ with the corresponding path $Q^j_i$. It is easy to verify that $P'$ is a path with the same endpoints as $P$, and that it is contained in the wall $\hat G'$. Moreover, since the paths in $\pset$ are spaced-out, the paths in the resulting set $\pset'=\set{P'\mid P\in \pset}$ are node-disjoint.
\end{proofof}

This completes the proof of \Cref{thm: ndp-hard-master NDP}. In order to prove \Cref{thm: ndp-hard-master EDP}, we show an approximation-preserving reduction from \NDPwall to \EDPwall.

\begin{claim} \label{clm: ndp-hard-NDPwall to EDP wall}
Let $\iset=(G,\mset)$ be an instance of \NDPwall, and let $\iset'$ be the instance of \EDPwall consisting of the same graph $G$ and the same set $\mset$ of demand pairs.
Let $\opt$ and $\opt'$ be the optimal solution values for $\iset$ and $\iset'$, respectively. Then $\opt'\geq \opt$, and there is an efficient algorithm, that, given any solution $\pset'$ to instance $\iset'$ of \EDPwall, computes a solution $\pset$ to instance $\iset$ of \NDPwall of value $\Omega(|\pset'|)$.
\end{claim}

The proof of \Cref{clm: ndp-hard-NDPwall to EDP wall} follows standard techniques and is deferred to \Cref{appdx: ndp-hard-from NDP to EDP}.
The following corollary immediately follows from \Cref{clm: ndp-hard-NDPwall to EDP wall} and completes the proof of \Cref{thm: ndp-hard-master EDP}.

\begin{corollary}
    If there is an $\alpha$-approximation algorithm for \EDPwall with running time $f(n)$, for $\alpha>1$ that may be a function of the graph size $n$, then there is an $O(\alpha)$-approximation algorithm for \NDPwall with running time $f(n)+\poly(n)$.
\end{corollary}

    \addtocontents{toc}{\protect\newpage}
    \chapter{Large Minors in Expanders}  \label{chap: exp}
    \toggletrue{exp}

\section{Introduction} \label{sec: exp-intro}
In this chapter we study about large minors of expander graphs and describe results that were published in \cite{large-minors-in-expanders}.
We start by informally defining the notions of expander graphs and graph minors.
A graph $G$ is an \emph{expander}, if, for every partition $(A,B)$ of its vertices into non-empty subsets, the number of edges connecting vertices of $A$ to vertices of $B$ is at least $\Omega(\min\set{|A|,|B|})$.
More generally, we say that $G$ is an \emph{$\alpha$-expander}, if, for every such partition $(A,B)$, the number of edges connecting vertices of $A$ to those of $B$ is at least $\alpha\cdot \min\set{|A|,|B|}$.
A graph $H$ is a \emph{minor} of a given graph $G$, if one can obtain a graph isomorphic to $H$ from $G$, via a sequence of edge- and vertex-deletions and edge-contractions.

Bounded-degree expanders are graphs that are simultaneously extremely well connected, while being sparse. Expanders are ubiquitous in discrete mathematics, theoretical computer science and beyond, arising in a wide variety of fields ranging from computational complexity to designing robust computer networks (see \cite{avi_survey} for a survey on expanders and their applications).
In this chapter we study an extremal problem about expanders: what is the largest function $f(n,\alpha,d)$, such that every $n$-vertex $\alpha$-expander with maximum vertex degree at most $d$ contains \emph{every} graph with at most $f(n,\alpha,d)$ vertices and edges as a minor?

Our main result is that there is an absolute constant $c$, such that
$f(n,\alpha,d)\geq \bestbound{c}$.
We note that this result achieves an optimal dependence on $n$.
We also provide a randomized algorithm that, given an $n$-vertex $\alpha$-expander with maximum vertex degree at most $d$, and 
another graph $H$ containing at most $\bestbound{c}$ edges and vertices, with high probability finds a model of $H$ in $G$, in time $\poly(n)\cdot (d/\alpha)^{\log(d/\alpha)}$.
Additionally, we show a simple randomized algorithm with running time $\poly(n,d/\alpha)$, that achieves a bound that has a slightly worse dependence on $n$ but a better dependence on $d$ and $\alpha$:
if $G$ is an $n$-vertex $\alpha$-expander with maximum vertex degree at most $d$, and $H$ is any graph  with at most $\simplebound{{c'}}$ edges and vertices, for some universal constant $c'$, the algorithm finds a model of $H$ in $G$ with high probability.

Independently from our work, Krivelevich and Nenadov (see Theorem 8.1 in~\cite{expander-minor}) provide an elegant proof of a similar but stronger result: namely, they show that $f(n,\alpha,d)=\Omega(\frac{n\alpha^2}{d^2\log n})$, and provide an efficient algorithm, that, given an $n$-vertex $\alpha$-expander of maximum vertex degree at most $d$, and a graph $H$ with $O(\frac{n\alpha^2}{d^2\log n})$ vertices and edges, finds a model of $H$ in $G$.

One of our main motivations for studying this question is the Excluded Grid Theorem of Robertson and Seymour.
This is a fundamental result in graph theory,  that was proved by Robertson and Seymour~\cite{gmt_5} as part of their Graph Minors series.
The theorem states that there is a function $t: \mathbb{Z}^+ \to \mathbb{Z}^+$, such that for every integer $g>0$, every graph of treewidth at least $t(g)$ contains the $(g\times g)$-grid as a minor.
The theorem has found many applications in graph theory and algorithms, including routing problems~\cite{flat-wall-RS},
fixed-parameter tractability~\cite{bidimensionality,DemaineH07}, and 
Erdős-Pósa-type results
\cite{gmt_5,thomassen1988presence, Reed-chapter, FominST11}. 
For an integer $g>0$, let $t(g)$ be the smallest value, such that  every graph of treewidth at least $t(g)$ contains the $(g \times g)$-grid as a minor.
An important open question is establishing tight bounds on the function $t$.
Besides being a fundamental graph-theoretic question in its own right, improved upper bounds on $t$ directly affect the running times of numerous algorithms that rely on the theorem, as well as parameters in various graph-theoretic results, such as, for example,
Erdős-Pósa-type results.

In a series of works \cite{gmt_5, RST_exclude_planar, KK_gmt, leaf_gmt, CC_gmt, C_gmt, gmt_julia_arxiv,CT18}, it was shown that $t(g) = \tilde O(g^{9})$ holds. The best currently known negative result, due to Robertson et al. \cite{RST_exclude_planar} is that $t(g) = \Omega(g^2 \log g)$. This is shown by employing a family bounded-degree expander graphs of large girth. Specifically, consider an $n$-vertex expander $G$ whose maximum vertex degree is bounded by a constant independent of $n$, and whose girth is $\Omega(\log n)$. It is not hard to show that the treewidth of $G$ is $\Omega(n)$. Assume now that $G$ contains the $(g\times g)$-grid as a minor, for some value $g$. Such a grid contains $\Omega(g^2)$ disjoint cycles, each of which must consume $\Omega(\log n)$ vertices of $G$, and so $g\leq O(\sqrt{n/\log n})$. This simple argument is the best negative result that is currently known for the Excluded Grid Theorem. In fact, Robertson and Seymour conjecture that this bound is tight, that is, $t(g)=\Theta(g^2\log g)$ must hold. A natural question therefore is whether this analysis is tight, and in particular, whether every $n$-vertex bounded-degree expander must contain a $(g\times g)$-grid as a minor, for $g=O(\sqrt {n/\log n})$. In this paper we answer this question in the affirmative, and moreover, we show that \emph{every} graph with at most $O(n/\log n)$ vertices and edges is a minor of such an expander.

The problem of finding large minors in bounded-degree expanders was first considered by Kleinberg and Rubinfield \cite{KR}.
Building on the random walk-based techniques of Broder et al. \cite{BFU}, they showed that every expander $G$ on $n$ vertices contains every graph with $O(n/\log^{\kappa} n)$ vertices and edges as a minor. The exponent $\kappa$ depends on the expansion $\alpha$ and the maximum degree $d$ of the expander; we estimate it to be at least $\Theta(\log^2d/\log^2(1/\alpha))$.
They also show an efficient algorithm for finding a model of  such a graph in $G$.
We summarize the known results in \Cref{table: exp-bounds}. 

\vspace*{1em}
\begin{table}[ht]
    \centering
     \begin{tabular}{| c | c | c |} 
        \hline
        Size & Runtime & References \\ [0.2em]
        \hline\hline
        $O \left( \frac{n}{\log^{\kappa(\alpha, d)}{(n)}} \right)$ & $\poly(n)$ & \cite{KR}  \\ \hline
        $\frac{n}{\log n} \cdot \left( \frac{\alpha}{d} \right)^{O(1)}$  & ${\poly(n)\cdot(d/\alpha)^{O(\log(d/\alpha))}}$ & \cite{large-minors-in-expanders}\\\hline
        $O \left(\frac{n}{\log^2 n} \cdot \frac{\alpha^3}{d^5} \right)$ & $\poly(n)$ & \cite{large-minors-in-expanders}\\\hline
        $O(\frac{n}{\log n} \cdot \frac{\alpha^2}{d^2})$ & $\poly(n)$ & \cite{expander-minor}\\
        \hline
     \end{tabular}
     \caption{Maximum size of arbitrary graphs as minors in an expander $G$ with $n$ vertices, maximum vertex-degree $d$ and expansion $\alpha$.}
     \label{table: exp-bounds}
\end{table}
\vspace*{1em}

Another related direction of research is the existence of large clique minors in graphs.
The study of the size of the largest clique minor in a graph is motivated by Hadwiger's conjecture from 1943 \cite{hadwiger-original}.
This conjecture states that, if the chromatic number (the minimal number of colors required to color the vertices such that no edge is monochromatic) of a graph is at least $k$, then it contains a clique with $k$ vertices as a minor (see, \cite{hadwiger-survey} for a recent survey).
One well-known result in this area, due to Kawarbayashi and Reed \cite{clique_or_sep_opt}, shows that every $\alpha$-expander $G$ with $n$ vertices and maximum vertex-degree bounded by $d$ contains a clique with $\Omega(\alpha\sqrt{n}/d)$ vertices as a minor.
Recently, Krivelevich and Nenadov \cite{KN2018} improved the dependence on the expansion $\alpha$ and the maximum vertex degree $d$ under a somewhat stronger definition of expansion.
We note that both these bounds have tight dependence on $n$, since $G$ contains only $O(n)$ edges.
Our results imply a weaker bound of $\Omega \left( \left(\frac{\alpha}{d} \right)^{c'} \sqrt{n/\log n} \right)$ on the size of the clique minor, for some absolute constant $c'$.

The existence of large clique minors is also studied in the context of random graphs.
Recall that $G \sim \gset(n,p)$ is a (Erdős–Rényi) random graph on $n$ vertices, whose edges are added independently with probability $p$ each.
Bollob\'{a}s, Catlin and Erd\H{o}s \cite{hadwiger_erdos} showed that Hadwiger's conjecture is true for almost all graphs $\gset(n,p)$ for every constant $p > 0$.
Fountoulakis et al. \cite{ccl_random} later showed that for every constant $\epsilon > 0$, the graph $G \sim \gset \left(n, \frac{1+\epsilon}{n} \right)$ contains a clique minor on $\Omega(\sqrt{n})$ vertices with probability $1-o(1)$.
Using a theorem from \cite{exp_random}, our results imply a slightly weaker bound of $\Omega(\sqrt{n/\log n})$ on the clique minor size.

\subsection{Our Results and Informal Overview of Techniques} \label{sec: exp-overview}

All graphs that we consider are finite; they do not have loops or parallel edges.
Given a graph $H$, we define its \emph{size} to be $|V(H)|+|E(H)|$.
Our main result is summarized in the following theorem:
\begin{theorem}\label{thm: exp-general main}
	There is a constant $c^*$,
	such that for all $0<\alpha<1$ and $d\geq 1$, if $G$ is an $n$-vertex $\alpha$-expander with maximum vertex degree at most $d$,
	and $H$ is a graph of size at most ${\bestbound{c^*}}$, then $H$ is a minor of $G$.
	Moreover, there is a randomized algorithm, whose running time is ${\poly(n)\cdot(d/\alpha)^{O(\log(d/\alpha))}}$, that, given $G$ and $H$ as above, with high probability, finds a model of $H$ in $G$.
\end{theorem}

As discussed above, the theorem implies that we cannot get stronger negative results for the Excluded Grid Theorem using bounded-degree expanders.
Our next result provides a simpler algorithm, with better running time and a better dependence on $d$ and $\alpha$, at the cost of slightly weaker dependence on $n$ in the minor size.

\begin{theorem}\label{thm: exp-constructive main}
	There is a constant $\tilde c^*$ and an efficient randomized algorithm, that,
	given an $n$-vertex $\alpha$-expander $G$ with maximum vertex degree at most $d$, where $0<\alpha<1$, and
	another graph $H$ of size at most $\simplebound{\tilde c^*}$,
	with high probability computes a model of $H$ in $G$.
\end{theorem}

The following corollary easily follows from \Cref{thm: exp-general main} and a result of \cite{exp_random}.

\begin{corollary}\label{cor: exp-random}
	For every $\epsilon > 0$, there is a constant $c_{\epsilon}$ depending only on $\epsilon$, such that a random graph $G \sim \gset \left(n, \frac{1+\epsilon}{n} \right)$  with high probability contains every graph of size at most $c_{\epsilon} n/\log n$ as a minor.
\end{corollary}

As mentioned earlier, similar but somewhat stronger results were obtained independently by Krivelevich and Nenadov (see Theorem 8.1 in~\cite{expander-minor}).
As a final comment, we show in \Cref{appn-subsec: exp-lower bound} that expanders are the `most minor-rich' family of graphs in the following sense:

\begin{observation}	\label{obs: exp-lower bound}
	For every graph $G$ of size $s \geq 2$, there is a graph $H_G$ of size at most $20 s/ \log s$ such that $G$ does not contain $H_G$ as a minor.
\end{observation}

We now turn to describe our techniques, starting with the simpler result: \Cref{thm: exp-constructive main}. Given an $n$-vertex $\alpha$-expander $G$ with maximum vertex degree at most $d$, we compute a partition of $G$ into two disjoint subgraphs, $G_1$ and $G_2$, such that $G_1$ is a connected graph; $G_2$ is an $\alpha'$-expander for a  somewhat weaker parameter $\alpha'$, and a large matching $\mset$ connecting vertices of $G_1$ to vertices of $G_2$. We refer to the edges of $\mset$, and to their endpoints, as \emph{terminals}. %
Assume now that we are given a graph $H$, containing at most $\simplebound{\tilde c^*}$ vertices and edges.
Using the transitivity of the minor relation, we can assume w.l.o.g. that the maximum vertex degree in $H$ is at most $3$,
and we denote $|V(H)|=n'$.
Using the standard grouping technique, we partition the graph $G_1$ into connected subgraphs $S_1,\ldots,S_{n'}$, each of which contains at least $\Theta(d^2\log^2n/\alpha^2)$ terminals. Assume that $H=\set{v_1,\ldots,v_{n'}}$. We map the vertex $v_i$ of $H$ to the graph $S_i$. Let $E_i\subseteq \mset$ be the set of edges of $\mset$ incident to the vertices of $S_i$. Every edge $(v_i,v_j)\in E(H)$ is embedded into a path in the expander $G_2$, that connects some edge of $E_i$ to some edge of $E_j$. The paths are found using standard techniques: we use the classical result of Leighton and Rao~\cite{LeightonRao} to show that for every edge $e=(v_i,v_j)$ of $H$, there is a large set $\pset_e$ of paths in $G_2$, connecting edges of $E_i$ to edges of $E_j$, such that all resulting paths in $\pset=\bigcup_{e\in E(H)}\pset_e$ are short, and cause a small vertex-congestion in $G_2$. We then use the constructive proof of the Lovász Local Lemma by Moser and Tardos~\cite{Moser-Tardos} to select a single path $P_e$ from each such set $\pset_e$, so that the resulting paths are disjoint in their vertices.

\begin{figure}[h]
    \center
    \includegraphics[width=0.6\linewidth]{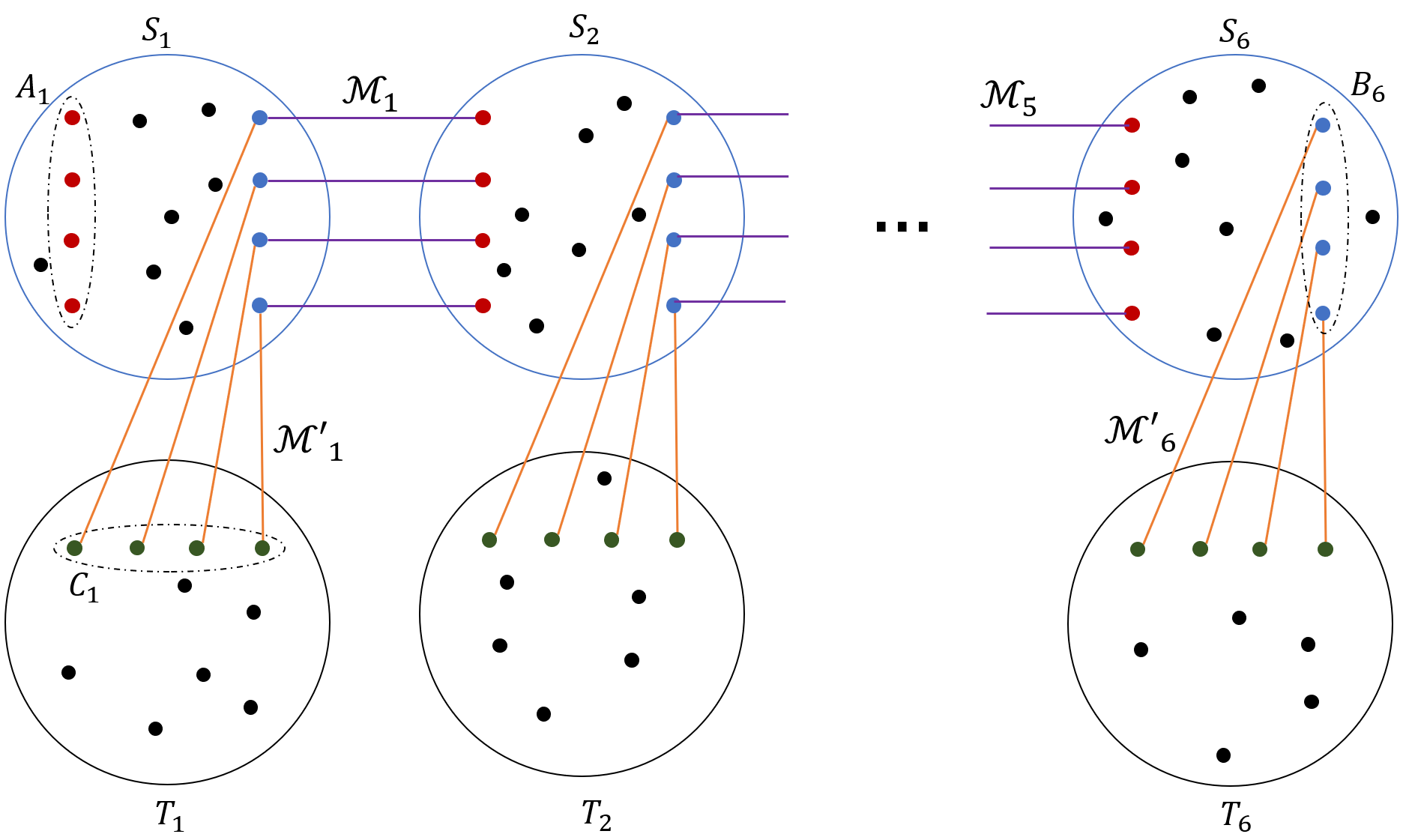}
	\caption{An illustration of the \poefull System $\Pi = (\sset, \mset, A_1, B_{6},\tset, \mset')$.
	For each $1\leq i\leq 6$, the vertices of $A_i$, $B_i$ and $C_i$ are shown in red, blue and green, respectively.
	}
    \label{fig: exp-poe-intro}
\end{figure}

The proof of \Cref{thm: exp-general main} is somewhat more complex. As before, we assume w.l.o.g. that maximum vertex degree in the graph $H$ is at most $3$. We define a new combinatorial object called a \poefull System (see \Cref{fig: exp-poe-intro}). At a high level, a \poefull System of width $w$ and expansion $\alpha'$ consists of 12 graphs: graphs $T_1,\ldots,T_6$ that are $\alpha'$-expanders, and graphs $S_1,\ldots,S_6$ that are connected graphs. For each $1\leq i\leq 6$, we are also given a matching $\mset'_i$ of cardinality $w$ connecting vertices of $S_i$ to vertices of $T_i$; the endpoints of the edges of $\mset'_i$ in $S_i$ and $T_i$ are denoted by $B_i$ and $C_i$, respectively. For each $1\leq i< 6$, we are given a matching $\mset_i$ connecting every vertex of $B_i$ to some vertex of $S_{i+1}$; the endpoints of the edges of $\mset_i$ that lie in $S_{i+1}$ are denoted by $A_{i+1}$. We show that an $n$-vertex $\alpha$-expander with maximum vertex degree at most $d$ must contain a \poefull System of width $w\geq n(\alpha/d)^c$ and expansion $\alpha'=(\alpha/d)^{c'}$ for some constants $c$ and $c'$, and provide an algorithm with running time $\poly(n)\cdot (d/\alpha)^{O(\log(d/\alpha))}$ to compute it. Next, we split the \poefull System into three parts. The first part is the union of the graphs $S_2,T_2$ and the matching $\mset'_2$. We view the vertices of $B_2$ as terminals, and we use the graph $T_2$ and the matching $\mset'_2$ in order to partition them into large enough groups, and to define a connected sub-graph of $T_2\cup \mset'_2$ spanning each such group, like in the proof of \Cref{thm: exp-constructive main}. We ensure that the number of groups is equal to the number of vertices in the graph $H$ that we are trying to embed into $G$. Every vertex of $H$ is then embedded into a separate group, together with the corresponding connected sub-graph of $T_2\cup \mset'_2$ spanning the group.

We use the graphs $S_3,\ldots,S_6,T_3\ldots,T_6$ in order to route all but a small fraction of the edges of $H$. The algorithm in this part is inspired by the algorithm of Frieze~\cite{journal-Frieze} for routing a large set of demand pairs in an expander graph via edge-disjoint paths. Lastly, the remaining edges of $H$ are routed in graph $S_1\cup T_1\cup \mset'_1$, using essentially the same algorithm as the one in the proof of \Cref{thm: exp-constructive main}.

\subsection{Organization}
We start with preliminaries in \Cref{sec: exp-prelims}.
The proof of \Cref{thm: exp-general main} is provided in \Cref{sec: exp-new main}, with technical details deferred to \Cref{sec: exp-exp to poe,sec: exp-embedding in poe}.
The proof of \Cref{thm: exp-constructive main} appears in \Cref{sec: exp-constructive proof}, and the proofs of \Cref{cor: exp-random} and \Cref{obs: exp-lower bound} appear in \Cref{appn-sec: exp-proofs of exp-intro}.

    \section{Preliminaries} \label{sec: exp-prelims}
Throughout the chapter, for an integer $\ell\geq 1$, we denote $[\ell]=\set{1,\ldots,\ell}$. All logarithms in the chapter are to the base of $2$.
All graphs that we consider are finite; they do not have loops or parallel edges.
We will use the following simple observation, whose proof is deferred to \Cref{appn-subsec: exp-simple partition proof}.

\begin{observation}\label{obs: exp-simple partition} 
	There is an efficient algorithm, that, given a set
	$\set{x_1,\ldots,x_r}$ of non-negative integers, with
	$\sum_{i}x_i=N$, and $x_i\leq 3N/4$ for all $i$, computes a
	partition $(A,B)$ of $\set{1,\ldots,r}$, such that $\sum_{i\in A}x_i
	\geq N/4$ and $\sum_{i\in B}x_i\geq N/4$.
\end{observation}

Given a graph $G=(V,E)$ and a subset $V' \subseteq V$ of its vertices, we denote by $\delta_G(V')$ the set of all edges that have exactly one  endpoint in $V'$, and by $E_G[V']$ the set of all edges with both endpoints in $V'$.
For readability, we write $\delta_G(v)$ instead of $\delta_G(\set{v})$. Given a pair $V', V'' \subseteq V$  of disjoint subsets of vertices, we denote by $E_G(V', V'')$ the set of all the edges with one endpoint in $V'$ and another in $V''$.
We will omit the subscript $G$ when the underlying graph is clear from context.
For a subset $V'\subseteq V$ of vertices of $G$, 
we denote by $G[V']$ the subgraph of $G$ induced by $V'$.

Given a path $P$ in a graph $G$, we denote by $V_P$ and $E_P$ the sets of all its vertices and edges, respectively. 
Given a path $P$ and a subset $V' \subseteq V$ of vertices of $G$, we say that $P$ is \emph{disjoint} from $V'$ iff $V_P\cap V'=\emptyset$. We say that $P$ is \emph{internally disjoint} from $V'$ iff every vertex of $V'\cap V_P$ is an endpoint of $P$.
Similarly, suppose we are given two paths $P,P'$ in a graph $G$. We say that the two paths are  \emph{disjoint} iff $V_P \cap V_{P'} = \emptyset$, and we say that they are  \emph{internally disjoint} iff all vertices in $V_P \cap V_{P'}$ serve as endpoints of both these paths.

Let $\pset$ be any set of paths in a graph $G$. We say that $\pset$ is a set of \emph{disjoint} paths iff every pair  $P,P' \in \pset$ of distinct paths are disjoint.
We say that $\pset$ is a set of \emph{internally disjoint} paths iff every pair  $P,P' \in \pset$ of distinct paths are internally disjoint.
We denote by $V(\pset) = \bigcup_{P \in \pset} V_P$ the set of all vertices participating in the paths of $\pset$.
Given a pair $V',V''$ of subsets of vertices of $V$ (that are not necessarily disjoint), we say that a path $P\in \pset$  \emph{connects} $V'$ to $V''$ iff one of its endpoints is in $V'$ and the other endpoint is in $V''$. We use a shorthand $\pset : V' \rightsquigarrow V''$ to indicate that $\pset$ is a collection of disjoint paths, where each path $P \in \pset$ connects $V'$ to $V''$.
Notice that each path in $\pset$ must originate at a distinct vertex of $V'$ and terminate at a distinct vertex of $V''$.

Finally, assume that we are given a (partial) matching $\mset$ over the vertices of $G$, and a set $\pset$ of $|\mset|$ paths. We say that $\pset$ \emph{routes $\mset$} iff for every pair of vertices $(v',v'')\in \mset$, there is a path $P \in \pset$, whose endpoints are $v'$ and $v''$.

We now turn to the definitions pertaining expansion of a graph.
A \emph{cut}  in $G$ is a bipartition $(S, S')$ of its vertices, that is, $S \cup S' = V$, $S \cap S' = \emptyset$ and $S, S' \neq \emptyset$.
The \emph{sparsity} of the cut $(S, S')$ is $|E(S,S')|/\min{ \set{|S|, |S'|}}$. The \emph{expansion} of a graph $G$, denoted by $\phi(G)$, is the minimum sparsity of any cut in $G$. 

\begin{definition}[Expanders]
Given a parameter $\alpha>0$, we say that a graph $G$ is an $\alpha$-expander iff $\phi(G)\geq \alpha$. Equivalently, for every subset $S$ of at most $|V(G)|/2$ vertices of $G$, $|\delta_G(S)|\geq \alpha |S|$.
\end{definition}

Cheeger's inequality, one of the most fundamental result in the spectral graph theory, shows that, for any graph $G$, whose maximum vertex degree is bounded by $d$, $\frac{ \lambda(G)}{2} \leq \varphi(G) \leq  \sqrt{2d \lambda(G)}$, where $\lambda(G)$ is the second smallest eigenvalue of the Laplacian of $G$.
Combining it with the algorithm of~\cite{cheeger-alg}, we obtain the following theorem (see also~\cite{Cheeger1,Cheeger2,Alon}).

\begin{theorem}\label{thm: exp-spectral}
There is an efficient algorithm, that, given an $n$-vertex graph $G$ with maximum vertex degree at most $d$, computes a cut $(A,B)$ in $G$ of sparsity $O(\sqrt{d\phi(G)})$.
\end{theorem}

Finally, we use the following simple claim several times; the claim allows one to ``fix'' an expander, after a small number of edges were deleted from it.

\begin{claim}\label{clm: exp-large expanding subgraph-edges}
     Let $T$ be an $\alpha$-expander, and let $E'$ be any subset of edges of $T$. Then there is an $\alpha/4$-expander $T'\subseteq T\setminus E'$, with $|V(T')|\geq |V(T)|-\frac{4|E'|}{\alpha}$.
\end{claim}

The proof of \Cref{clm: exp-large expanding subgraph-edges} is deferred to \Cref{appn-subsec: exp-large expanding subgraph}.
We now turn to the definition of graph minors.
Given a graph $G$, all graphs $H$ that can be obtained by vertex-deletions, edge-deletions, and edge-contractions from $G$ are \emph{minors} of $G$.
It is a well known fact that graph minors can also be defined in the following equivalent way.

\begin{definition}[Graph Minors]\label{def: exp-minor}
	We say that a graph $H = (U, F)$ is a \emph{minor} of a graph $G = (V, E)$ iff there is a map $f$, called a \emph{model of $H$ in $G$}, mapping every vertex $u \in U$ to a subset $X_u\subseteq V$ of vertices, and mapping every edge $e \in F$ to a path $P_e$ in $G$, such that:
	\begin{itemize}
		\item For every vertex $u \in U$, $G[X_u]$ is connected;
		\item For every edge $e = (u,v) \in F$, the path $P_e$ connects $X_u$ to $X_v$;
		\item For every pair $u,v \in U$ of distinct vertices, $X_u\cap X_v=\emptyset$; and
		\item Paths $\set{P_e \mid e \in F}$ are internally disjoint from each other and they are internally disjoint from the set $\bigcup_{u \in U}X_u$ of vertices. 
	\end{itemize}
For a vertex $u\in U$ we sometimes call $G[X_u]$ the \emph{embedding of $u$ into $G$}, and for an edge $e\in F$, we sometimes refer to $P_e$ as the \emph{embedding of $e$ into $G$}.
\end{definition}

We now turn to a slight variation of the standard notion of well-linkedness that plays an important role in several decomposition results.

\begin{definition}[Well-Linkedness]
We say that a set $A$ of vertices in a graph $G$ is \emph{well-linked} iff for every pair $A',A''$ of disjoint equal-cardinality subsets of $A$, there is a set  $\pset : A' \rightsquigarrow A''$ of $|A'|$ paths in $G$, that are internally disjoint from $A$. (Note that the paths in $\pset$ must be disjoint).
\end{definition}

Next, we define a Path-of-Sets system, that was first introduced in~\cite{CC_gmt} (a somewhat similar object called \emph{grill} was introduced by \cite{leaf_gmt}), and was used since then in a number of graph theoretic results.

\begin{figure}
    \centering
    \begin{minipage}{0.45\linewidth}
			\includegraphics[width=\linewidth]{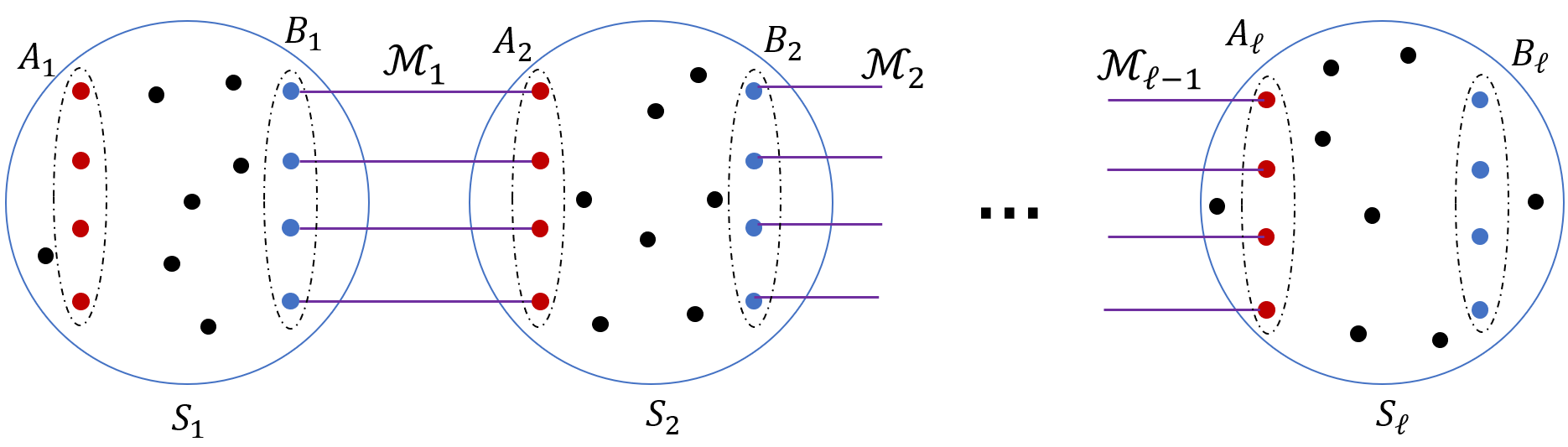}
			\caption{An illustration of a Path-of-Sets System $(\sset, \mset, A_1, B_{\ell})$.
				For each $i \in [\ell]$, the vertices of $A_i$ and $B_i$ are shown in red and blue respectively.
			}
    		\label{fig: exp-pos}
    \end{minipage}
    \hspace{0.04\linewidth}
    \begin{minipage}{0.45\linewidth}
            \includegraphics[width=\linewidth]{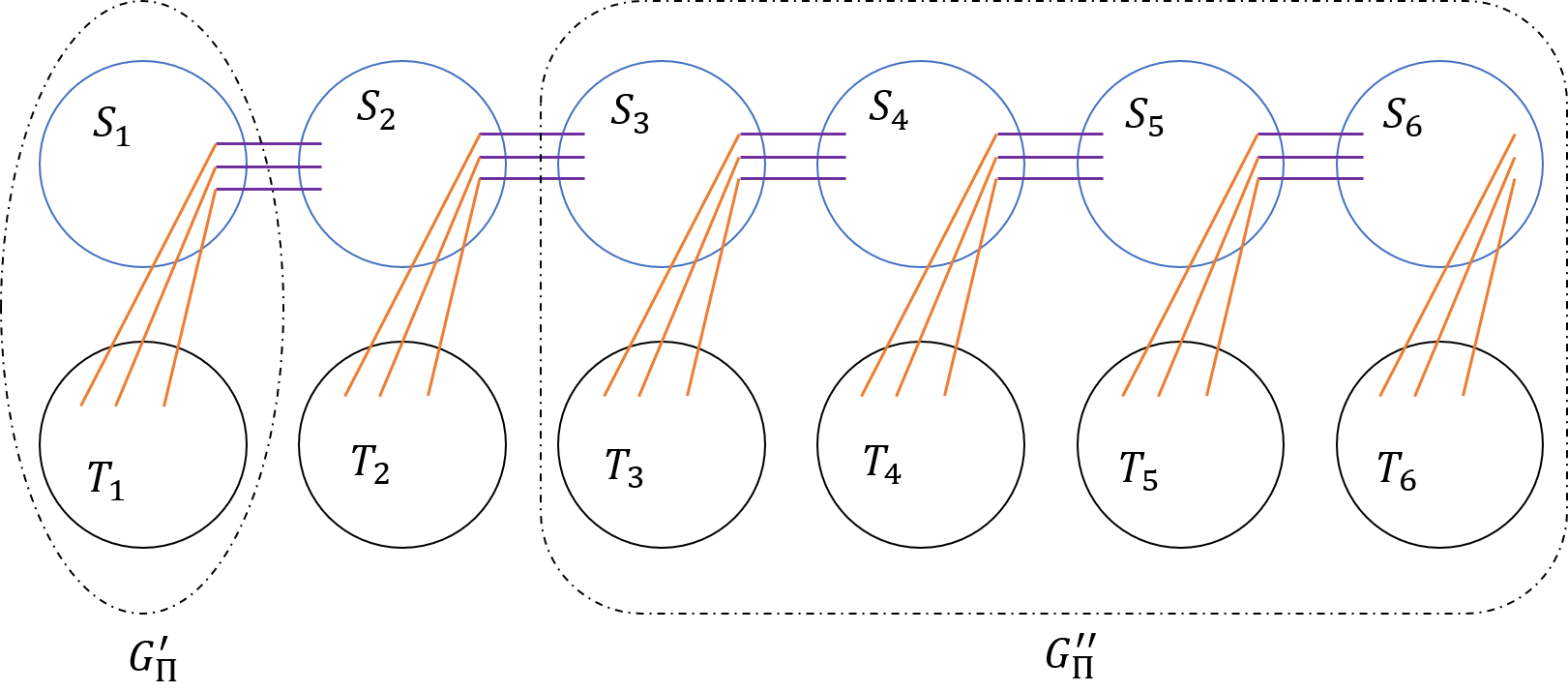}
			\caption{An illustration of the subgraphs $G'_{\Pi}$ and $G''_{\Pi}$ of $G_{\Pi}$.}
    		\label{fig: exp-g_prime_and_double_prime}
    \end{minipage}
\end{figure}

\begin{definition}[Path-of-Sets System] \label{def: exp-pos}
	Given integers $w,\ell>0$ a Path-of-Sets System of width $w$ and length $\ell$  (see \Cref{fig: exp-pos}) consists of:
	\begin{itemize}
		\item a sequence $\sset = (S_1, \ldots, S_{\ell})$ of $\ell$ disjoint connected graphs, that we refer to as \emph{clusters};

		\item for each $1\leq i \leq \ell$, two disjoint subsets, $A_i, B_i \subseteq V(S_i)$ of $w$ vertices each; and

		\item For each $1 \leq i < \ell$, a collection $\mset_i$ of edges, connecting every vertex of $B_i$ to a distinct vertex of $A_{i+1}$.
	\end{itemize}
	We denote the Path-of-Sets  System by $\Sigma = (\sset, \mset, A_1, B_{\ell})$, where $\mset = \bigcup_i \mset_i$. We also denote by $G_{\Sigma}
	$ the graph defined by the Path-of-Sets System, that is, $G_{\Sigma}=\left(\bigcup_{i=1}^{\ell}S_i\right )\cup \mset$.
	
	We say that a given Path-of-Sets System is \emph{a \posfull System} iff all $1\leq i\leq \ell$, the vertices of $A_i\cup B_i$ are well-linked in $S_i$. We say that it is $\alpha$-expanding, iff for all $1\leq i\leq \ell$, graph $S_i$ is an $\alpha$-expander. Note that a \posfull System is not necessarily $\alpha$-expanding and vice versa.
\end{definition}

Finally, we turn to \poefull System, the main structural object that we introduce.

\begin{definition}[\poefull System]\label{def: exp-poe}
	Given an integer $w>0$ and a parameter $0<\alpha<1$, a  \poefull System of width $w$ and expansion $\alpha$  (see \Cref{fig: exp-poe-intro}) consists of:
	\begin{itemize}
		\item a \posfull System $\Sigma = (\sset, \mset, A_1, B_6)$ of width $w$ and length $6$;

		\item a sequence $\tset = (T_1, \ldots, T_{6})$ of $6$ disjoint connected graphs, such that for each $1\leq i\leq 6$, $T_i$ is disjoint from $S_1,\ldots,S_6$, and it is an $\alpha$-expander; and

		\item for each $1\leq i\leq 6$, a perfect matching $\mset'_i$ between $B_i$ and some subset $C_i$ of $w$ vertices of $T_i$.
	\end{itemize}
	We denote the \poefull System by $\Pi = (\sset, \mset,  A_1,B_6,\tset, \mset')$, where $\mset' = \bigcup_i \mset'_i$.
	For convenience, for each $1\leq i \leq 6$, we denote by $W_i$ be the graph obtained from the union of the graphs $S_i$ and $T_i$, and the matching $\mset'_i$.
\end{definition}

Similarly to the Path-of-Sets System, we associate with the \poefull System $\Pi$ a graph $G_{\Pi}$, obtained by taking the union of the graphs $S_1,\ldots,S_6$, $T_1,\ldots,T_6$ and the sets $\mset,\mset'$ of edges.
We will be interested in three subgraphs of $G_\Pi$ (see \Cref{fig: exp-g_prime_and_double_prime}): 
(i) Graph $W_1$, that we denote by $G'_{\Pi}$;
(ii) Graph $W_2$; and
(iii) Graph $G''_{\Pi}$, obtained by taking the union of $W_3 \cup W_4 \cup W_5 \cup W_6$ and the edges of $\mset_3\cup\mset_4\cup \mset_5$.

We say that a graph $G$ contains a Path-of-Sets System of width $w$ and length $\ell$ as a minor iff there is a Path-of-Sets System  $\Sigma$ of width $w$ and length $\ell$, such that its corresponding graph $G_{\Sigma}$ is a minor of $G$. Similarly, 
we say that  a graph $G$ contains a \poefull  System of width $w$ and expansion $\alpha$ as a minor iff there is a \poefull  System $\Pi$ of width $w$ and expansion $\alpha$, such that its corresponding graph $G_{\Pi}$ is a minor of $G$.
As one of our main technical contribution, we show that, an expander must contain a \poefull System of large enough expansion and width as a minor, and provide an algorithm to compute its model in the expander.

    \section{Expanders Contain all Not-So-Large Graphs as Minors}
\label{sec: exp-new main}
The goal of this section is to prove \Cref{thm: exp-general main}.
As alluded earlier, we will prove it in two steps.
In the first step, we show that every expander contains a \poefull System with large enough size and expansion as a minor.
Moreover, there is an efficient algorithm, that given a graph $G$, either finds such a \poefull System $\Pi$ or finds a cut in the original graph with small sparsity.
In the second step, we show that we can embed every not-so-large graph in such a \poefull System.
Unfortunately, this algorithm is efficient only if $d/\alpha \leq 2^{O(\sqrt{\log n})}$, where $d$ and $\alpha$ are the maximum vertex-degree and expansion of $G$ respectively.
This is precisely the reason why we cannot efficiently find a model of $H$ in $G$ if $d/\alpha$ is large.
We formalize this idea in the following two theorems, whose proofs are deferred to \Cref{sec: exp-embedding in poe,sec: exp-exp to poe} respectively.

\begin{theorem} \label{thm: exp-embedding algo}
    There is constant $c_0$ and a randomized algorithm, that, 
    given a \poefull System $\Pi$ with expansion $\alpha$ and width $w$,
    such that the maximum vertex degree in $G_{\Pi}$ is at most $d$ and $|V(G_{\Pi})|\leq n$ for some $n>c_0$,
    together with a graph $H$ of maximum vertex degree at most $3$ and $|V(H)|\leq \frac{w^2\alpha^2}{2^{19}d^4n \log n}$,
    with high probability,
    in time $\poly(n)$,
    finds a model of $H$ in $G_{\Pi}$. 
\end{theorem}

\begin{theorem} \label{thm: exp-poe}
    There are constants $\hat c_1,\hat c_2$,  and an algorithm, that, given an $\alpha$-expander $G$ with $|V(G)|=n$, whose maximum vertex degree is at most $d$, and $0<\alpha<1$, constructs a \poefull System $\Pi$ of expansion $\tilde \alpha \geq \left(\frac{\alpha}{d}\right )^{\hat c_1}$ and width $w \geq n \cdot \left(\frac{\alpha}{d}\right )^{\hat c_2} $, such that the corresponding graph $G_{\Pi}$ has maximum vertex degree at most $d+1$ and is a minor of $G$. Moreover, the algorithm computes a model of $G_{\Pi}$ in $G$. The running time of the algorithm is $\poly(n)\cdot (d/\alpha)^{O(\log(d/\alpha))}$.
\end{theorem}

Before proving \Cref{thm: exp-poe,thm: exp-embedding algo}, we complete the proof of \Cref{thm: exp-general main} using them.
Let $G$ be the given $\alpha$-expander with $|V(G)|=n$, and maximum vertex degree at most $d$. 
Recall that $0<\alpha < 1$.
By letting $c^*$ be a sufficiently large constant, we can assume that $n$ is sufficiently large,
so that, for example, $n>c_0$, where $c_0$ is the constant from \Cref{thm: exp-embedding algo}.
Indeed, otherwise, it is enough to show that the graph with $1$ vertex is a minor of $G$, which is trivially true.
Therefore, we assume from now on that $n$ is sufficiently large.

From \Cref{thm: exp-poe}, $G$ contains as a minor a \poefull System $\Pi$
of width $w \geq n \cdot \left(\frac{\alpha}{d}\right )^{\hat c_2} $ and
expansion  $\tilde \alpha \geq \left(\frac{\alpha}{d}\right )^{\hat c_1}$,
such that the maximum vertex degree in $G_{\Pi}$ is at most $d+1$. %
We claim that, if $H'$ is a graph with maximum vertex degree at most $3$, and
$|V(H')|\leq \frac{3n}{c^*\log n}\cdot \left(\frac{\alpha}{d} \right)^{c^*}$,
then, $G$ contains $H'$ as a minor.
Indeed, using the above bounds, we get that:

\[\begin{split}
\frac{w^2 \tilde \alpha^2}{2^{19}(d+1)^4n\log n}&\geq  n^2 \cdot \left(\frac{\alpha}{d}\right )^{2\hat c_2} \cdot  \left(\frac{\alpha}{d}\right )^{2\hat c_1}\cdot \frac{1}{2^{23}d^4 n\log n}\\
&=\frac{n \alpha^{2(\hat c_1+\hat c_2)}}{2^{23}d^{4+2(\hat c_1+\hat c_2)}\log n}\\
&\geq \frac{3n}{c^*\log n}\cdot \left(\frac{\alpha}{d} \right)^{c^*},
\end{split}\]
for $c^* \geq \max{\set{4 + 2 (\hat c_1 + \hat c_2), c_0, 2^{25}}}$.
Therefore, if $H'$ is a graph with maximum vertex degree at most $3$, and
$|V(H')|\leq \frac{3n}{c^*\log n}\cdot \left(\frac{\alpha}{d} \right)^{c^*}$,
then, from \Cref{thm: exp-embedding algo},
$G$ contains $H'$ as a minor, and from \Cref{thm: exp-poe,thm: exp-embedding algo}, its model in $G$ can be computed with high probability by a randomized algorithm, in time $\poly(n)\cdot (d/\alpha)^{O(\log(d/\alpha))}$.

Consider now any graph $H=(U,F)$ of size at most
$\frac{n}{c^*\log n}\cdot \left(\frac{\alpha}{d} \right)^{c^*}$.
Let $n'=|U|$ and $m'=|F|$, so
$n'+m'\leq \frac{n}{c^*\log n}\cdot \left(\frac{\alpha}{d} \right)^{c^*}$.
We construct another graph $H'$, whose maximum vertex degree is at most $3$ and $|V(H')|\leq n'+2m' \leq  \frac{3n}{c^*\log n}\cdot \left(\frac{\alpha}{d} \right)^{c^*} $, such that $H$ is a minor of $H'$. Since $H'$ must be a minor of $G$, it follows that $H$ is a minor of $G$. In order to construct graph $H'$ from $H$, we consider every vertex $u\in U$ of degree $d_u>3$ in turn, and replace it with a cycle $C_u$ on $d_u$ vertices, such that every edge incident to $u$ in $H$ is incident to a distinct vertex of $C_u$. It is easy to verify that the resulting graph $H'$ has maximum vertex degree at most $3$, that $H$ is a minor of $H'$, and that $|V(H')|\leq 2m'+n'$, completing the proof of \Cref{thm: exp-general main}. Notice that this proof is constructive, that is, there is a randomized algorithm that constructs a model of $H$ in $G$ in time $\poly(n)\cdot (d/\alpha)^{O(\log(d/\alpha))}$.
This completes the proof of \Cref{thm: exp-general main}, assuming \Cref{thm: exp-poe,thm: exp-embedding algo} that we prove in \Cref{sec: exp-embedding in poe,sec: exp-exp to poe} respectively.

    \section{Embedding Target-Graph in a \poefull System} \label{sec: exp-embedding in poe}

This section is devoted to the proof \Cref{thm: exp-embedding algo}.
We assume that we are given a  \poefull System $\Pi = (\sset, \mset, A_1, B_{6},\tset, \mset')$ of width $w$ and expansion $\alpha$, whose corresponding graph $G_{\Pi}$ contains at most $n$ vertices, where $n>c_0$ for some large enough constant $c_0$, and its maximum vertex degree is bounded by $d$. 
In order to simplify the notation, we denote $G_{\Pi}$ by $G$. 
We also use the following parameter: $\rho = 2^{16}\floor{\frac{d^3n\log n}{\alpha^2w}}$. 

We are also given a graph $H = (U,F)$ of maximum degree $3$, with
$ |U| \leq \frac{w^2\alpha^2}{2^{19}d^4n\log n} \leq \frac{w}{8d \rho}$.
Our goal is to find a model of $H$ in $G$.
Our algorithm consists of three steps. In the first step, we associate with each vertex $u\in U$, a subset $X_u$ of vertices of $W_2$, such that $W_2[X_u]$ is a connected graph. This defines the embeddings of the vertices of $H$ into $G$ for the model of $H$ that we are computing.
In the second step, we embed all but a small fraction of the edges of $H$ into $G''_{\Pi}$, and in the last step, we embed the remaining edges of $H$ into $G'_{\Pi}$. 
We now describe each step in detail.

\paragraph{Step 1: Embedding the Vertices of $H$.}
In this step we compute an embedding of every vertex of $H$ into a connected subgraph of $W_2$. Recall that graph $W_2$ is the union of the graphs $S_2$ and $T_2$, and the matching $\mset'_2$, connecting the vertices of $B_2\subseteq V(S_2)$ to the vertices of $C_2\subseteq V(T_2)$, where $|B_2|=|C_2|=w$.
We use the following simple observation, that was used extensively in the literature (often under the name of ``grouping technique'') (see e.g. ~\cite{CKS, RaoZhou, Andrews,Chuzhoy11}). The proof is deferred to \Cref{appn-subsec: exp-proof of decompose by spanning tree} of the Appendix.

\begin{observation} \label{obs: exp-decompose by spanning tree}
    There is an efficient algorithm that, given a connected graph $\hG$ with maximum vertex degree at most $d$, an integer $r\geq 1$, and a subset  $R \subseteq V(\hG)$ of vertices of $\hG$ with $|R|\geq r$, computes a collection $\set{V_1, \ldots, V_r}$ of $r$ mutually disjoint subsets of $V(\hat G)$, such that:
    \begin{itemize}
        \item For each $i \in [r]$, the induced graph $\hG[V_i]$ is connected; and
        \item For each $i \in [r]$, $|V_i \cap R| \geq \floor{|R|/(d r)}$.
    \end{itemize}
\end{observation}

\begin{figure}[h]
    \center
    \includegraphics[width=0.45\linewidth]{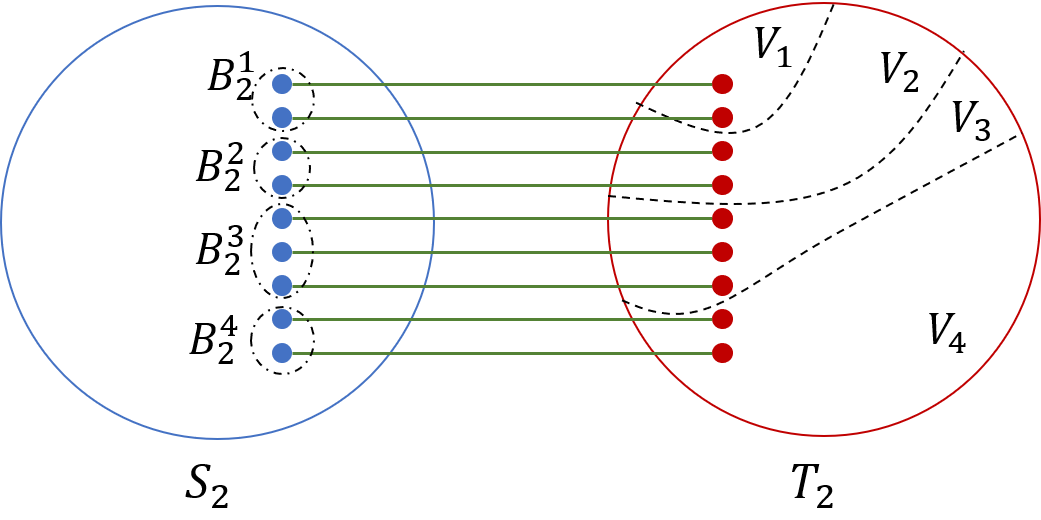}
    \caption{A sketch of the partition of $T_2$ and $B_2$ with $|U|=4$. Vertices of $B_2$ and $C_2$ are shown in blue and red respectively. }
    \label{fig: exp-embedding in w2}
\end{figure}

We apply the above observation to the graph $T_2$, together with vertex set $R=C_2$ and parameter $r=\floor{\frac{w}{8d\rho}}$.
Let $\uset$ be the resulting collection of $r$ subsets of vertices of $T_2$. Recall that for each set $V_i\in \uset$, $|V_i\cap C_2|\geq \floor{\frac{|C_2|}{dr}}\geq \floor{\frac{w}{d\floor{w/8d\rho}}} \geq 3\rho$.
Since $|U|\leq \frac{w}{8d\rho}$, we can choose $|U|$ distinct sets $V_1,\ldots,V_{|U|}\in \uset$ (see, \Cref{fig: exp-embedding in w2}).
We also denote $U=\set{u_1,\ldots,u_{|U|}}$.
Finally, for each $1\leq i\leq |U|$, we let $E^i\subseteq \mset'_2$ be the subset of edges that have an endpoint in $V_i$, and we let $B^i_2$ be the subset of vertices of $B_2$ that serve as endpoints of the edges in $E^i$.
Since  $|V_i\cap C_2|\geq 3\rho$, we have $|B^i_2|\geq 3\rho$ for each $1 \leq i \leq |U|$.

We are now ready to define the embeddings of the vertices of $H$ into $G$. For each $1\leq i\leq |U|$, we let $f(u_i)=G[B_2^i\cup V_i]$. Notice that for all $1\leq i\leq |U|$, $f(u_i)$ is a connected graph, and for all $1\leq i<j\leq | U|$, $f(u_i)\cap f(u_j)=\emptyset$. 
In the remaining steps, we focus on embedding  the edges of $H$ into $G$, such that the resulting paths are internally disjoint from $B_2 \cup T_2$.

\paragraph{Step 2: Routing all but a small fraction of edges of $H$ in $G''_{\Pi}$.}
Consider some vertex $u_i \in U$, its corresponding graph $f(u_i)$, and the set $B_2^i\subseteq B_2$ of vertices that lie in $f(u_i)$.
Recall that the maximum vertex degree in $H$ is at most $3$ and for each $1 \leq i \leq |U|$ we have $|B_2^i|\geq 3\rho$.
For every edge $e\in \delta_H(u_i)$, we now select an arbitrary subset $B_2^i(e)\subseteq B_2^i$ of $\rho$ vertices, so that all resulting sets $\set{B_2^i(e)}_{e\in \delta_H(u_i)}$ are mutually disjoint.

Recall that graph $G_{\Pi}$ contains a perfect matching $\mset_2$ between the vertices of $B_2$ and the vertices of $A_3$. We let $\hat E^i\subseteq \mset_2$ be the subset of edges whose endpoints lie in $B_2^i$, and denote by $A_3^i\subseteq A_3$ the set of endpoints of the edges of $E^i$ that lie in $A_3$.
For every edge $e\in \delta(v_i)$, we let $A_3^i(e)\subseteq A_3^i$ be the set of $\rho$ vertices that are connected to the vertices of $B_2^i(e)$ with an edge of $\mset_2$. Clearly, all resulting vertex sets $\set{A_3^i(e)}_{e\in \delta_H(u_i)}$ are mutually disjoint.
Let $A_3' = \bigcup_{u_i \in U} \bigcup_{e \in \delta_H(u_i)} A_3^i(e)$, and notice that
\[ |A_3'| \leq 3 \rho \cdot |U| \leq 3 \rho \cdot \frac{w}{8d\rho}  = \frac{3w}{8d} \leq \frac{w}{2}.\]

The following lemma, whose proof is deferred to \Cref{subsec: exp-routing in toe}, allows us to embed a large number of edges of $H$ in $G''_{\Pi}$.

\begin{lemma}\label{lem: exp-routing in toe}
    There is an efficient algorithm, that, given a \poefull System $\Pi = (\sset, \mset,  A_1, B_{6}, \tset, \mset')$ of expansion $\alpha$ and width $w$, where $0<\alpha<1$ and $w$ is an integral multiple of $4$, whose corresponding graph $G_{\Pi}$ contains at most $n$ vertices and has  maximum vertex degree at most $d$, together with a subset $A'_3 \subseteq A_3$ of at most $w/2$ vertices, and a collection $\set{A_3^1, \ldots, A_3^{2 r}}$ of mutually disjoint subsets of $A'_3$ of cardinality $\rho= 2^{16}\floor{\frac{d^3n\log n}{\alpha^2w}}$ each, where $r> \frac{w \alpha^2 (\log \log n)^2}{d^3 \log^3 n}$, returns 
 a partition $\iset',\iset''$ of $\set{1,\ldots,r}$, and a set $\pset^* = \set{P^*_j \mid j \in \iset'}$ of disjoint paths in $G''_{\Pi}$, such that for each $j \in \iset'$, path $P^*_j$ connects $A_3^j$ to $A_3^{j+r}$, and $|\iset''|\leq  \frac{w \alpha^2 (\log \log n)^2}{d^3 \log^3 n}$.
 \end{lemma}

We obtain the following immediate corollary of the lemma.

\begin{corollary}\label{cor: exp-step 2}
There is an efficient algorithm to compute a partition $(F_1,F_2)$ of the set $F$ of edges of $H$, and for each edge $e=(u_i,u_j)\in F_1$, a path $P^*_e$ in graph $G''_{\Pi}$, connecting a vertex of $A_3^i(e)$ to a vertex of $A_3^j(e)$, such that all paths in set $\pset^*_1=\set{P^*(e)\mid e\in F_1}$ are disjoint, and $|F_2|\leq \frac{w \alpha^2 (\log \log n)^2}{d^3 \log^3 n}$.
\end{corollary}
\begin{proof}
By appropriately ordering the collection $\set{A_3^i(e)\mid u_i\in U,e\in \delta_H(e)}$ of vertex subsets, and applying \Cref{lem: exp-routing in toe} to the resulting sequence of subsets of $A_3'$, we obtain a set $F_1\subseteq F$ of edges of $H$, and for each edge $e=(u_i,u_j)\in F_1$, a path $P^*_e$, connecting a vertex of $A_3^i(e)$ to a vertex of $A_3^j(e)$ in graph $G''_{\Pi}$, such that all paths in set $\pset^*_1=\set{P^*(e)\mid e\in F_1}$ are disjoint. Let $F_2=F\setminus F_1$. From \Cref{lem: exp-routing in toe}, $|F_2| \leq  \frac{w \alpha^2 (\log \log n)^2}{d^3 \log^3 n}$. 
\end{proof}

\begin{figure}
    \centering
    \begin{minipage}{0.45\linewidth}
            \includegraphics[width=\linewidth]{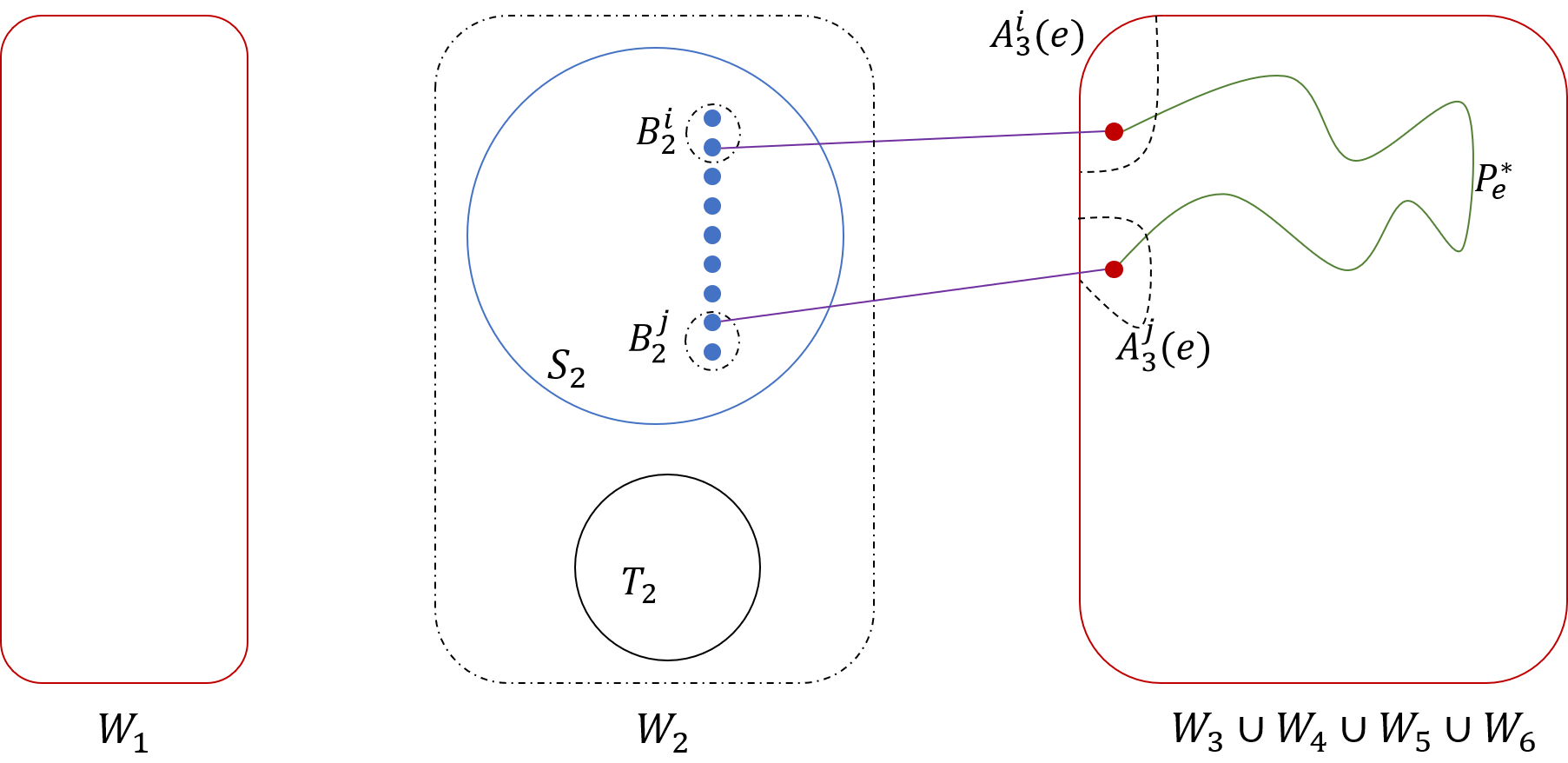}
            \caption{An illustration of a path $P^*_e \in \pset^*_1$ routing an edge $e=(u_i, u_j) \in F_1$. Dashed boundaries represent the labeled subsets.}
            \label{fig: exp-pset 1}
    \end{minipage}
    \hspace{0.04\linewidth}
    \begin{minipage}{0.45\linewidth}
            \includegraphics[width=\linewidth]{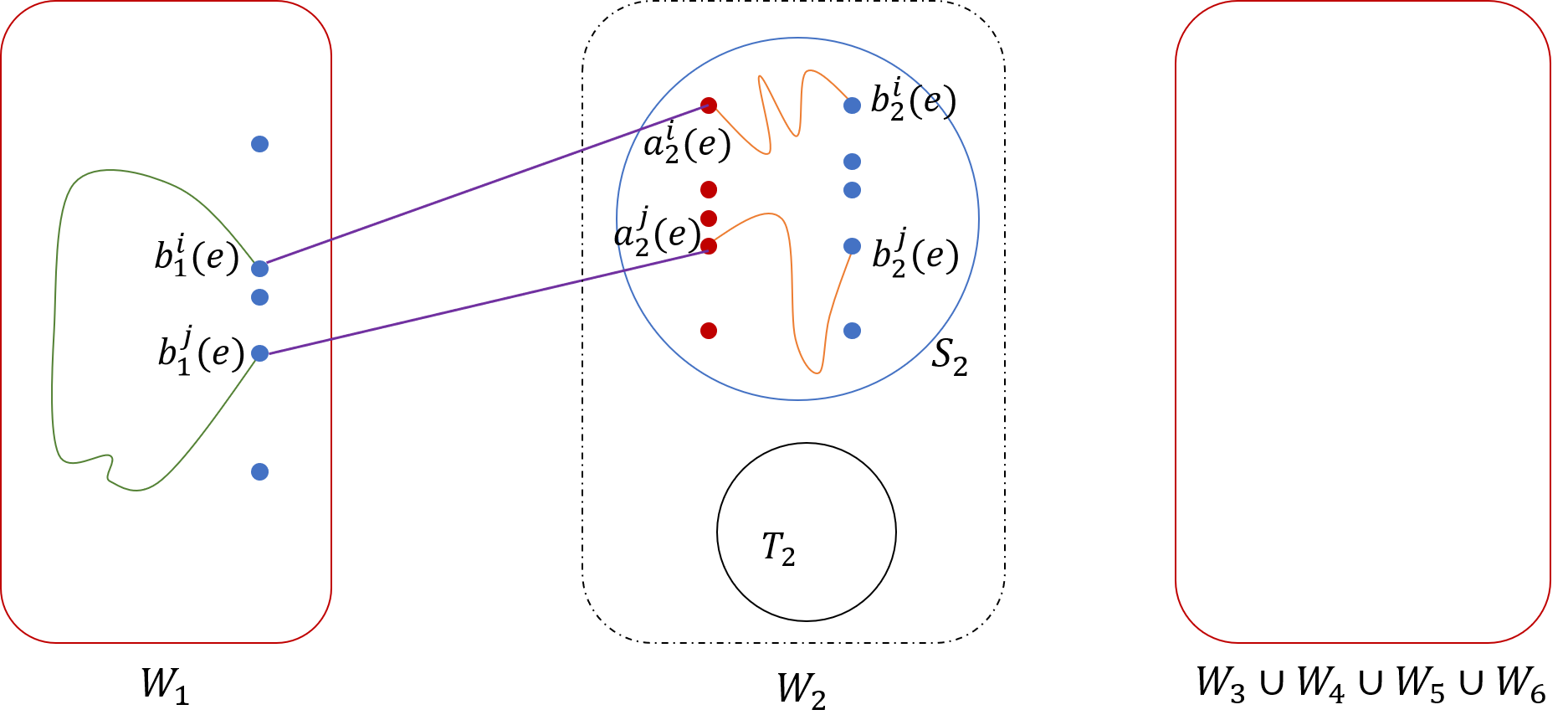}
            \caption{An illustration of the path $P^*_e \in \pset^*_2$ connecting $e=(u_i, u_j) \in F_2$.}
            \label{fig: exp-pset 2}
    \end{minipage}
\end{figure}

For each edge $e=(u_i,u_j)\in F_1$, we extend the path $P^*_e$ to include the two edges of $\mset_2$ incident to its endpoints, so that $P^*_e$ now connects a vertex of $B_2^i$ to a vertex of $B_2^j$. Path $P^*_e$ becomes the embedding $f(e)$ of $e$ in the model $f$ of $H$ that we are constructing. For convenience, the resulting set of paths $\set{P^*_e\mid e\in F_1}$ is still denoted by $\pset^*_1$. The paths in $\pset^*_1$ remain disjoint from each other; they are internally disjoint from $W_2$, and completely disjoint from  $W_1$ (see \Cref{fig: exp-pset 1}).

\paragraph{Step 3: Routing remaining edges of $H$ in $G'_{\Pi}$.}
In this last step we complete the construction of a minor of $H$ in $G$, by embedding the edges of $F_2$.
The main tool that we use is the following lemma, whose proof is deferred to \Cref{subsec: exp-routing in doe}.

\begin{lemma}\label{lem: exp-routable b}
    There is a universal constant $c$, and an efficient algorithm that, given a \poefull System $\Pi = (\sset, \mset,A_1, B_{6}, \tset, \mset')$ of expansion $\alpha$ and width $w$, such that the corresponding graph $G_{\Pi}$ contains at most $n$ vertices and has  maximum vertex degree at most $d$, computes a subset $B'_1 \subseteq B_1$ of at least $ \frac{cw \alpha^2}{d^3 \log^2 n}$ vertices, such that the following holds.
    There is an efficient randomized algorithm, that given any matching $\mset^*$ over the vertices of $B'_1$, with high probability returns a set $\pset$ of disjoint paths in $W_1$, routing $\mset^*$.
\end{lemma}

We now complete the embedding using the above lemma.
Let $B'_1 \subseteq B_1$ be the subset of at least $\frac{c w \alpha^2}{d^3 \log^2 n}$ vertices, computed by algorithm from \Cref{lem: exp-routable b}.
Let $A'_2 \subseteq A_2$ be the set of all the vertices connected to the vertices of $B'_1$ by the edges of the matching $\mset_1$.
Observe that $|A'_2|\geq 2|F_2|$, since:

\[ 2|F_2|\leq \frac{2 w \alpha^2 (\log \log n)^2}{d^3 \log^3 n} \leq \frac{c w \alpha^2}{d^3 \log^2 n} = |B'_1| = |A'_2|.\]

Here, the last inequality follows since we have assumed that $n$ is sufficiently large. 

We let $A''_2$ be an arbitrary subset of $2|F_2|$ vertices of $A'_2$.
Recall that for every vertex $u_i \in U$ and every edge $e\in \delta_H(u_i)$ incident on it, we have defined a subset $B_2^i(e)\subseteq B_2^i$ of vertices. 
We select an arbitrary representative vertex $b_2^i(e)\in B_2^i(e)$, and we let $B'_2=\set{b_2^i(e)\mid u_i\in U, e\in \delta(u_i)\cap F_2}$ be the resulting set of representative vertices, so that $|B'_2|=2|F|$. 

Since $(A_2\cup B_2)$ are well-linked in $S_2$, there is a set $\qset_2$ of $2|F_2|$ disjoint paths in $S_2$, connecting every vertex of $B'_2$ to some vertex of $A''_2$, such that the paths in $\qset_2$ are internally disjoint from $A_2 \cup B_2$.
For each vertex $b_2^i(e) \in B'_2$, let $a_2^i(e) \in A''_2$ be the corresponding endpoint of the path of $\qset_2$ that originates at $b_2^i(e)$ (see \Cref{fig: exp-pset 2}).
Let $b_1^i(e) \in B'_1$ be the vertex of $B_1$ that is connected to $a_2^i(e)$ with an edge from $\mset_1$. We can now naturally define a matching $\mset^*$ over the vertices of $B'_1$, where for every edge $e=(u_i,u_j)\in F_2$, we add the pair $(b_1^i(e),b_1^j(e))$ of vertices to the matching.
From \Cref{lem: exp-routable b}, with high probability we obtain a collection $\pset^*_2 = \set{P^*_e \mid e  \in F_2}$ of disjoint paths in $W_1$, such that, for every edge $e = (u_i, u_j) \in F_2$, the  corresponding path $P^*_e$ connects $b_1^i(e)$ to $b_1^j(e)$. We extend this path to connect the vertex $b_2^i(e)$ to the vertex $b_2^j(e)$, by using the edges of $\mset_1$ that are incident to $b_1^i(e)$ to $b_1^j(e)$, and the paths of $\qset_2$ that are incident to $a_2^i(e)$ to $a_2^j(e)$. 
Notice that  the resulting extended paths are internally disjoint from $B_2$, and are completely disjoint from $T_2\cup G''_{\Pi}$.
We now embed each edge $e\in F_2$ into the path $P^*_e$, that is, we set $f(e)=P^*_e$.
This completes the construction of the model of $H$ in $G_{\Pi}$, except for the proofs of \Cref{lem: exp-routing in toe,lem: exp-routable b}, that are provided in \Cref{subsec: exp-routing in toe,subsec: exp-routing in doe}, respectively.

    \subsection{Embedding Almost All Edges of the Target-Graph} \label{subsec: exp-routing in toe}
This section is dedicated to the proof of \Cref{lem: exp-routing in toe}.
We define a new combinatorial object, called a \doefull System.

\begin{figure}[h]
    \center
    \includegraphics[width=0.45\linewidth]{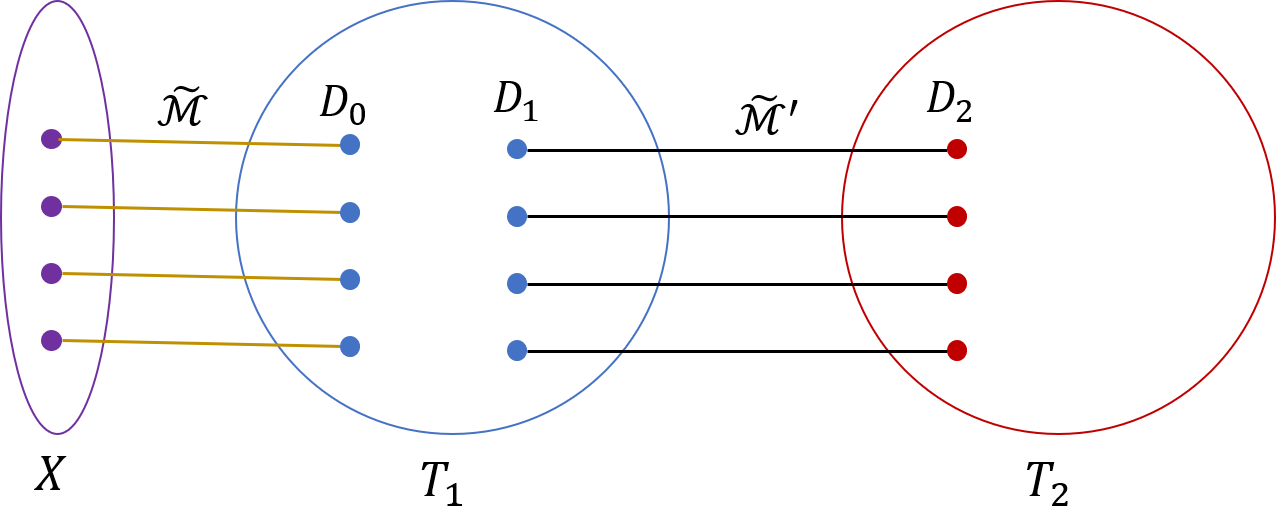}
    \caption{An illustration of the \doefull System.}
    \label{fig: exp-new doe}
\end{figure}

\begin{definition}
    A \doefull System of width $w$, expansion $\alpha$ (see \Cref{fig: exp-new doe}) consists of:
    \begin{itemize}
        \item two disjoint graphs $T_1,T_2$, each of which is an $\alpha$-expander;
        \item a set $X$ of $w$ vertices that are disjoint from $T_1\cup T_2$, and three subsets $D_0,D_1\subseteq V(T_1)$ and $D_2\subseteq V(T_2)$ of $w$ vertices each, where all three subsets are disjoint; and
        \item a complete matching $\tmset$ between the vertices of $X$ and the vertices of $D_0$, and a complete matching $\tmset'$ between the vertices of $D_1$ and the vertices of $D_2$, so $|\tmset|=|\tmset'|=w$.
    \end{itemize}

    We denote the \doefull System by $\dset = (T_1, T_2, X, \tmset, \tmset')$. The set $X$ of vertices is called the \emph{backbone} of $\dset$.
Let $G_{\dset}$ be the graph corresponding to the \doefull System $\dset$, so $G_{\dset}$ is the union of graphs $T_1,T_2$, the set $X$ of vertices, and the set $\tmset\cup \tmset'$ of edges.
\end{definition}

Similarly to \poefull System, given a graph $G$, we say that it contains a \doefull System $\dset$ as a minor iff $G_{\dset}$ is a minor of $G$.

The following lemma is central to the proof of \Cref{lem: exp-routing in toe}.

\begin{lemma}\label{lem: exp-routing in new doe}
    There is an efficient algorithm that, given a \doefull System $\dset$ of width $w/4$ and expansion $\alpha$, for some $0<\alpha<1$, such that the corresponding graph $G_{\dset}$ contains at most $n$ vertices and has maximum vertex degree at most $d$, together with a collection $\set{X_1, \ldots, X_{2 r}}$ of  mutually disjoint subsets of the backbone $X$ of cardinality $\sigma = 2^{15}\floor{\frac{d^3n\log n}{\alpha^2w}}$ each, where $r > \frac{w \alpha^2 (\log \log n)^2}{d^3 \log^3 n}$, returns a partition $\iset',\iset''$ of $\set{1,\ldots,r}$, and for each $j\in \iset'$, a path $P_j$ connecting a vertex of $X_j$ to a vertex of $X_{j+r}$ in $G_{\dset}$, such that the paths in set $\pset = \set{P_j \mid j \in \iset'}$ are disjoint, and $|\iset''|\leq r \cdot \frac{\log \log n}{\log n}$.
    \end{lemma}

We prove \Cref{lem: exp-routing in new doe} after we complete the proof of \Cref{lem: exp-routing in toe} using it.
Recall that we are given a \poefull System $\Pi = (\sset, \mset, A_1, B_6,\tset, \mset')$, together with its corresponding graph $G_{\Pi}$. 
We are also given a subset $A'_3\subseteq A_3$ of at most $w/2$ vertices, and a partition of $A'_3$ into $2r$ disjoint subsets $A^1_3,\ldots,A^{2r}_3$, of cardinality $\rho=2^{16}\floor{\frac{d^3n\log n}{\alpha^2w}}$ each, where $r\geq \frac{w\alpha^2(\log\log n)^2}{d^3\log^3n}$.

For each $1\leq i\leq 2r$, we arbitrarily partition $A_3^i$ into two subsets, $W^i_1,W^i_2$, of cardinality $\rho/2$ each (note that $\rho$ is an even integer). Let $W_1=\bigcup_{i=1}^{2r}W_1^i$ and let $W_2=\bigcup_{i=1}^{2r}W^i_2$. Note that $|W_1|,|W_2|\leq  |A'_3|/2\leq w/4$. We add arbitrary vertices of $A_3\setminus A_3'$ to $W_1$ and $W_2$, until each of them contains $w/4$ vertices (recall that $w/4$ is an integer), while keeping them disjoint. The vertices of $A_3\setminus (W_1\cup W_2)$ are then arbitrarily partitioned into two subsets, $Y_1$ and $Y_2$, of cardinality $w/4$ each.

Next, we show that graph $G''_{\Pi}$ contains two disjoint \doefull Systems as minors. We will then use \Cref{lem: exp-routing in new doe} in each of the two \doefull Systems in turn in order to obtain the desired routing.

\begin{claim}\label{clm: exp-two does}
There is an efficient algorithm to compute two disjoint subgraphs, $G^{(1)}$ and $G^{(2)}$ of $G''_{\Pi}$, and for each $z\in \set{1,2}$, to compute a model $f^{(z)}$ of a \doefull System $\dset^{(z)}=(T_1^{(z)},T_2^{(z)},X^{(z)},\tmset^{(z)},(\tmset')^{(z)})$ of width $w/4$ and expansion $\alpha$ in $G^{(z)}$,  such that the corresponding graph $G_{\dset^{(z)}}$ has maximum vertex degree at most $d$, and for every vertex $w\in W_z$, there is a distinct vertex $v(w)$ in the backbone $X^{(z)}$, such that $w\in f^{(z)}(v(w))$.
\end{claim}

The proof of this claim is deferred to \Cref{appn-subsec: clm: exp-two does}.
We are now ready to complete the proof of \Cref{lem: exp-routing in toe}.
We apply \Cref{lem: exp-routing in new doe} to $\dset^{(1)}$, together with vertex sets $W_1^1,\ldots,W_1^{2r}$, each of which now contains $\rho/2=2^{15}\floor{\frac{d^3n\log n}{\alpha^2w}}$ vertices obtaining a partition $(\iset',\iset'')$ of $\set{1,\ldots,r}$, together with a set $\pset_1= \set{P_j \mid j \in \iset'}$ of disjoint paths in $G_{\dset^{(1)}}$, such that for all $j \in \iset'$ path $P_j $ connects a vertex of $W_1^j$ to a vertex of $W_1^{r+j}$, and $|\iset''|\leq r \cdot \frac{\log \log n}{\log n}$. Since $G_{\dset^{(1)}}$ is a minor of $G^{(1)}$, it is immediate to obtain a collection $ 
\pset_1'= \set{P'_j \mid j \in \iset'}$ of disjoint paths in $G^{(1)}$, such that for all $j \in \iset'$ path $P'_j $ connects a vertex of $A_3^j$ to a vertex of $A_3^{j+r}$.

If $|\iset''|\leq \frac{w \alpha^2 (\log \log n)^2}{d^3 \log^3 n}$, then we terminate the algorithm, and return the set $\pset'$ of paths, together with the partition $(\iset',\iset'')$ of $\iset$.
Thus, we assume from now on that $|\iset''| > \frac{w \alpha^2 (\log \log n)^2}{d^3 \log^3 n}$.

For convenience, we denote $r' := |\iset''|$.
We apply \Cref{lem: exp-routing in new doe} to $\dset^{(2)}$, together with vertex sets $\set{W^j_2,W^{j+r}_2 \mid j\in \iset''}$, that are appropriately ordered.
We then obtain a partition $\iset_1,\iset_2$ of $\iset''$,  and a set $\pset_2= \set{P_j \mid j \in \iset_1}$ of disjoint paths in $G_{\dset^{(2)}}$, such that for each $j \in \iset_2$ path $P_j $ connects a vertex of $W^j_2$ to a vertex of $W_2^{j+r}$, and  $|\iset_2|\leq r' \cdot \frac{\log \log n}{\log n}\leq r \cdot \frac{(\log \log n)^2}{\log^2 n}$. As before, since $G_{\dset^{(2)}}$ is a minor of $G^{(2)}$, it is immediate to obtain a collection $\pset_2'= \set{P'_j \mid j \in \iset_1}$ of disjoint paths in $G^{(2)}$, such that for all $j \in \iset_1$ path $P'_j $ connects a vertex of $A_3^j$ to a vertex of $A_3^{j+r}$.
We return the partition $(\iset'\cup \iset_1,\iset_2)$ of $\set{1,\ldots,r}$, together with the set $\pset_1'\cup \pset_2'$ of paths.
Since the graphs $G^{(1)}$ and $G^{(2)}$ are disjoint, all paths in $\pset_1'\cup \pset_2'$ are disjoint.
It now remains to show that $|\iset_2|\leq \frac{w \alpha^2 (\log \log n)^2}{d^3 \log^3 n}$.

Recall that the set $A_3'$, consisting of at most $w/2$ vertices, is partitioned into $2r$ subsets of cardinality $\rho=2^{16}\floor{\frac{d^3n\log n}{\alpha^2w}}$ each. Therefore:

\[r \leq \frac{w}{4\rho}  =  \frac{w}{2^{18}\floor{d^3n\log n/(\alpha^2w)}} \leq \frac{w^2\alpha^2}{2^{17}d^3n\log n} \leq \frac{w\alpha^2}{2^{17}d^3\log n}.\]

We now conclude that $|\iset_2|\leq r \cdot \frac{(\log \log n)^2}{\log^2 n}\leq \frac{w\alpha^2(\log \log n)^2}{d^3\log^3 n}$, as required.
This completes the proof of \Cref{lem: exp-routing in toe} assuming \Cref{lem: exp-routing in new doe} that we complete in the remainder of this subsection.

\subsubsection{Routing in \doefull | Proof of \Cref{lem: exp-routing in new doe}} \label{subsubsec: exp-routing in new doe}
Our proof is inspired by the algorithm of Frieze~\cite{journal-Frieze} for routing a large set of demand pairs in an expander graph via edge-disjoint paths.
Recall that we are given a \doefull System $\dset$ of width $w/4$ and expansion $\alpha$, for some $0<\alpha<1$, such that the maximum vertex degree in the corresponding graph $G_{\dset}=(V,E)$ is at most $d$, and $|V|\leq n$. We are also given   mutually disjoint subsets $\set{X_1, \ldots, X_{2 r}}$ of the backbone $X$, of cardinality $\sigma=2^{15}\floor{\frac{d^3n\log n}{w\alpha^2}}$ each, where $r>\frac{w \alpha^2 (\log \log n)^2}{d^3 \log^3 n}$. In particular, since $|X|=w/4$, we get that $2r\sigma\leq w/4$, and so $r\leq \frac{w}{8\sigma}\leq  \frac{w}{8\cdot 2^{15}\floor{d^3n\log n/(w\alpha^2)}}\leq \frac{w^2\alpha^2}{2^{17}d^3n\log n}$.
 Therefore, we obtain the following bounds on $r$ that we will use throughout the proof:
 
 \begin{equation}\label{eqn: exp-bounds on r}
\frac{w \alpha^2}{d^3}\cdot\frac{ (\log \log n)^2}{\log^3 n}<r\leq \frac{w^2\alpha^2}{2^{17}d^3n\log n}.
\end{equation}

For convenience, we will denote $G_{\dset}$ by $G$ for the rest of this subsection.
We will iteratively construct the set $\pset$ of disjoint paths in $G$, where for each path $P\in \pset$, there is some index $j\in \set{1,\ldots,r}$, such that $P$ connects $X_j$ to $X_{j+r}$. Whenever a path $P$ is added to $\pset$, we delete all vertices of $P$ from $G$.
Throughout the algorithm, we say that an index $j \in [r]$ is \emph{settled} iff there is a path $P_j \in \pset$ connecting $X_j$ to $X_{j+r}$, and otherwise we say that it is \emph{not settled}.
We use a parameter $\gamma = 512 n d^2/w\alpha$.
We say that a path $P$ in $G$ is \emph{permissible} iff $P$ contains at most $\gamma \log \log n$ nodes of $T_1$ and at most $\gamma \log n$ nodes of $T_2$.

\begin{minipage}{\linewidth}
    \begin{framed}
    \paragraph{The Algorithm.}
    Start with $\pset = \emptyset$.
    While there is an index $j \in [r]$ and a permissible path $P^*_j$ in the current graph $G := G_{\dset}$ such that:
    \begin{itemize}
        \item $j$ is not settled; \vspace{-0.2em}
        \item $P^*_j$ connects $X_j$ to $X_{j+r}$; and \vspace{-0.2em}
        \item $P^*_j$ is internally disjoint from $X$:
    \end{itemize}
    add $P^*_j$ to $\pset$ and delete all vertices of $P^*_j$ from $G$.
    \end{framed}
\end{minipage}

In order to complete the proof of \Cref{lem: exp-routing in new doe}, it is enough to show that, when the algorithm terminates, at most $\frac{r \log \log n}{\log n}$ indices $j \in [r]$ are not settled. Assume for contradiction that this is not true.
Let $\pset$ be the path set obtained at the end of the algorithm, and let $\tilde V = V(\pset)$ be the set of vertices participating in the paths of $\pset$. We further partition $\tilde V$ into three subsets: $\tilde V_1=\tilde V\cap V(T_1)$; $\tilde V_2=\tilde V\cap V(T_2)$; and $\tilde X=\tilde V\cap X$. Note that, since $|\pset|\leq r$, we are guaranteed that $|\tilde V_1|\leq \gamma  r\log\log n$; $|\tilde V_2|\leq \gamma r\log n$, and, since we have assumed that $|\pset|\leq r(1-\log \log n/\log n)$, and all paths in $\pset$ are internally disjoint from $X$, we get that $|\tilde X|\leq 2r-\frac{2r\log \log n}{\log n}$.

We now proceed as follows. First, we show that $T_1\setminus \tilde V_1$ and $T_2\setminus \tilde V_2$ both contain very large $\alpha/4$-expanders. We also show that there is a large number of edges in $\tmset'$ that connect these two expanders. This will be used to show that there must still be a permissible path $P^*_j$, connecting two sets $X_j$ and $X_{j+r}$ for some index $j$ that is not settled yet, leading to a contradiction. We start with the following claim that allows us to find large expanders in $T_1\setminus \tilde V_1$ and $T_2\setminus \tilde V_2$. 

\begin{claim}\label{clm: exp-large expanding subgraph}
     Let $T$ be an $\alpha$-expander with maximum vertex degree at most $d$, and let $Z$ be any subset of vertices of $T$. Then there is an $\alpha/4$-expander $T'\subseteq T\setminus Z$, with $|V(T')|\geq |V(T)|-\frac{4d|Z|}{\alpha}$.
\end{claim}

The proof of \Cref{clm: exp-large expanding subgraph} follows immediately from \Cref{clm: exp-large expanding subgraph-edges}, by letting $E'$ be the set of all edges incident to the vertices of $Z$.
The following corollary follows immediately from \Cref{clm: exp-large expanding subgraph}

\begin{corollary}\label{clm: exp-surviving step 1}
There is a subgraph $T'_1\subseteq T_1\setminus \tilde V_1$ that is an $\alpha/4$-expander, and $|V(T_1)\setminus V(T'_1)|\leq 4dr\gamma\log\log n/\alpha$. Similarly, there is a subgraph $T'_2\subseteq T_2\setminus \tilde V_2$ that is an $\alpha/4$-expander, and $|V(T_2)\setminus V(T'_2)|\leq 4dr\gamma\log n/\alpha$.
\end{corollary}

Let $R_1=V(T_1)\setminus V(T_1')$ and let $R_2=V(T_2)\setminus V(T_2')$. We refer to the vertices of $R_1$ and $R_2$ as the vertices that were \emph{discarded} from $T_1$ and $T_2$, respectively. The vertices that belong to $T_1'$ and $T_2'$ are called \emph{surviving} vertices. It is easy to verify that $|R_1|,|R_2|\leq w/64$. Indeed, observe that $|R_1|,|R_2|\leq 4dr\gamma\log n/\alpha$. Since, from \Cref{eqn: exp-bounds on r}, $r\leq \frac{w^2\alpha^2}{2^{17}d^3n\log n}$, we get that altogether:

\[|R_1|,|R_2|\leq \frac{4dr\gamma\log n}{\alpha}\leq \frac{\gamma w^2\alpha}{2^{15} d^2 n}\leq \frac{w}{64},\]

since $\gamma=512nd^2/w\alpha$.

Recall that the \doefull $\dset$ contains a matching $\tmset'$ between the set $D_1\subseteq V(T_1)$ of $w/4$ vertices and the set $D_2\subseteq V(T_2)$ of $w/4$ vertices. Next, we show that there are large subsets $D'_1\subseteq D_1$ and $D'_2\subseteq D_2$ of surviving vertices, such that a subset of $\tmset'$ defines a complete matching between them.

\begin{observation}
There are two sets $D'_1\subseteq D_1$ and $D_2'\subseteq D_2$ containing at least $w/16$ vertices each, and a subset $\hmset\subseteq \tmset'$ of edges, such that $\hmset$ is a complete matching between $D_1'$ and $D_2'$.
\end{observation}
\begin{proof}
Let $\hat D_1=D_1\setminus R_1$. Since $|R_1|\leq w/64$, $|\hat D_1|\geq w/8$. Let $\hmset'\subseteq \mset'$ be the set of edges whose endpoints lie in $\hat D_1$, and let $\hat D_2\subseteq D_2$ be the set of vertices that serve as endpoints for the edges in $\hmset'$, so $|\hat D_2|\geq w/8$. Finally, let $D_2'=D_2\setminus R_2$, so $|D_2'|\geq w/8-|R_2|\geq w/16$. We let $\hmset\subseteq \hmset'$ be the set of all edges incident to the vertices of $D_2'$, and we let $D_1'$ be the set of endpoints of these edges.
\end{proof}

Our second main tool is the following claim, that shows that for any pair of large enough sets of vertices in an expander, there is a short path connecting them. The proof uses standard methods and is deferred to \Cref{appn-subsec: exp-proof of decompose by spanning tree}.

\begin{claim}\label{clm: exp-short paths in expanders}
Let $T$ be an $\alpha'$-expander for some $0<\alpha'<1$, such that $|V(T)|\leq n$, and the maximum vertex degree in $T$ is at most $d$. Let $Z,Z'\subseteq V(T)$ be two vertex subsets, with $|Z|=z$ and $|Z'|=z'$. Then there is a path in $T$, connecting a vertex of $Z$ to a vertex of $Z'$, whose length is at most $\frac{8d}{\alpha'}(\log (n/z)+\log(n/z'))$. In particular, for every pair $v,v'$ of vertices in $T$, there is a path of length at most $16d\log n/\alpha'$ connecting $v$ to $v'$ in $T$.
\end{claim}

Let $J\subseteq \set{1,\ldots,r}$ be the set of indices that are not settled yet. From our assumption, $|J|\geq \frac{r\log\log n}{\log n}$. For every index $j\in J$, consider the corresponding sets $X_j,X_{j+r}$ of vertices of $X$, and let $Y_j,Y_{j+r}$ be the sets of vertices of $D_0$, that are connected to $X_j$ and $X_{j+r}$ via the matching $\tmset$. Let $Y_j'=Y_j\setminus R_1$ and let $Y_{j+r}'=Y_{j+r}\setminus R_1$ be the subsets  of surviving vertices in $Y_j$ and $Y_{j+r}$ respectively. We say that index $j$ is \emph{bad} iff $|Y'_j|<\sigma/2$ or $|Y'_{j+r}|<\sigma/2$; otherwise we say that it is a \emph{good index}. Recall that $|R_1|\leq 4dr\gamma\log\log n/\alpha$. Therefore, the total number of bad indices is at most:

\[\begin{split}
\frac{2|R_1|}{\sigma}&\leq \frac{8dr\gamma\log\log n}{\alpha\cdot 2^{15}\floor{d^3n\log n/(w\alpha^2)}}\\
&\leq \frac{w\alpha r\gamma\log\log n}{2^{11}d^2 n\log n}\\
&\leq \frac{r\log\log n}{4\log n}\cdot \frac{w\alpha \gamma}{512 d^2n}\\
&\leq \frac{r\log\log n}{4\log n},\end{split}\]

since $\gamma=512 n d^2/w\alpha$.

Let $J'\subseteq J$ be the set of all good indices, so $|J'|\geq \frac{r\log\log n}{2\log n}$.
We say that an index $j\in J'$ is \emph{happy} iff there is a path $P_1(j)$ in $T'_1$, of length at most $(\gamma\log\log n)/4$, connecting a vertex of $Y'_j$ to a vertex of $D'_1$, and there is a path $P_2(j)$ in $T'_1$,  of length at most $(\gamma\log\log n)/4$, connecting a vertex of $Y'_{j+r}$ to a vertex of $D'_1$. The following claim, whose proof is deferred to \Cref{prf-clm: exp-happy index} will complete the proof of \Cref{lem: exp-routing in toe}.

\begin{claim}\label{clm: exp-happy index}
At least one index of $J'$ is happy.
\end{claim}

Consider the paths $P_1(j)$ and $P_2(j)$ in $T'$, given by \Cref{clm: exp-happy index}, and assume that path $P_1(j)$ connects a vertex $v\in Y'_j$ to a vertex $v'\in D'_1$. Let $v''\in D'_2$ be the vertex connected to $v'$ by an edge of $\hat \mset$, that we denote by $e_v$. Similarly, assume that path $P_2(j)$  connects a vertex $u\in Y'_j$ to a vertex $u'\in D'_1$. Let $u''\in D'_2$ be the vertex connected to $u'$ by an edge of $\hat \mset$, that we denote by $e_u$.  From \Cref{clm: exp-short paths in expanders}, there is a path $P$ in $T_2'$, of length at most $64d\log n/\alpha<\gamma\log n$, connecting $v''$ to $u''$. By combining $P_1(j),e_v,P',e_u,P_2(j)$, together with the edges of $\tmset$ incident to $u$ and $v$, we obtain an admissible path, connecting a vertex of $X_j$ to a vertex of $X_{j+r}$, a contradiction.
This completes the proof of \Cref{lem: exp-routing in toe}.

    \subsection{Embedding Remaining Edges of the Target-Graph}\label{subsec: exp-routing in doe}

The goal of this subsection is to prove \Cref{lem: exp-routable b}.
We use the following lemma, whose proof uses standard techniques and is deferred to \Cref{appn-subsec: exp-routing along short paths}.

\begin{restatable}{lemma}{routeInExpanders}
    \label{lem: exp-can route large disjoint subsets in expanders}
    There is a universal constant $c$, and an efficient randomized algorithm, that,
    given graph $G = (V,E)$ with $|V|\leq n$,
    such that the maximum vertex degree in $G$ is at most $d$ and
    a parameter $0<\alpha<1$,
    together with a collection $\set{C_1, \ldots, C_{2 r}}$ of mutually disjoint subsets of $V$ of cardinality $q =\ceil{ cd^2 \log^2n/\alpha^2}$ each,
    computes one of the following:
    \begin{itemize}
        \item either a collection $\qset = \set{Q_1, \ldots, Q_{r}}$ of paths in $G$,
        where for each $1\leq j\leq r$, path $Q_j$ connects a vertex of $C_{j}$ to a vertex of $C_{r + j}$, and
        with high probability the paths in $\qset$ are disjoint; or
        \item a cut $(S,S')$ in $G$ of sparsity less than $\alpha$.
    \end{itemize}
\end{restatable}

Consider the subgraph $W'$ of $G_{\Pi}$; recall that it consists of two graphs, $S_1$ and $T_1$, where $S_1$ is a connected graph and $T_1$ is an $\alpha$-expander. Recall that $S_1$ contains a set $B_1$ of $w$ vertices; $T_1$ contains a set $C_1$ of $w$ vertices, and $\mset'_1$ is a perfect matching between these two sets.

We let $q=\ceil{cd \log^2n/\alpha^2}$, where $c$ is the constant from \Cref{lem: exp-can route large disjoint subsets in expanders}, and we let $r=\floor{w/dq}=\Omega(w\alpha^2/d^3\log^2n)$.
Observe that $q\leq \floor{w/dr}$.
We use \Cref{obs: exp-decompose by spanning tree} to compute $r$ connected subgraphs $S^1,\ldots,S^{r}$ of $S_1$, each of which contains at least $\floor{w/dr}\geq q$ vertices of $B_1$.
For $1\leq i\leq r$, we denote $B^i=B_1 \cap V(S^i)$.
We also let $\mset^i\subseteq \mset'_1$ be the set of edges incident to the vertices of $B^i$ in $\mset'_1$, and
we let $C^i\subseteq C_1$ be the set of the endpoints of the edges of 
$\mset^i$
that lie in $C_1$.
Observe that for all $1\leq i\leq 2r$, $|C^i|\geq q$.
For each $1\leq i\leq 2r$, we select an arbitrary vertex $b_i\in B^i$, and we let $B'=\set{b_i\mid 1\leq i\leq 2r}$,
so that $|B'|=2r=\Omega(w\alpha^2/d^3\log^2n)$, as required.

Assume now that we are given an arbitrary matching $\mset^*$ over the vertices of $B'$. By appropriately re-indexing the sets $B^i$, we can assume w.l.o.g. that $\mset^*=\set{(b_i, b_{r+i})}_{i=1}^r$.
Since $T_1$ is an $\alpha$-expander, the algorithm of \Cref{lem: exp-can route large disjoint subsets in expanders} computes a collection $\qset = \set{Q_1, \ldots, Q_{r}}$ of paths in
$T_1$,
where for each $1\leq j\leq r$, path $Q_j$ connects some vertex $c^*_j\in C^j$ to some vertex $c^*_{j+r}\in C^{j+r}$, and with high probability the paths in $\qset$ are disjoint. 

Consider now some index $1\leq j\leq 2r$.
We let $e_j$ be the unique edge of the matching $\mset'_1$ incident to $c^*_j$, and we let $b^*_j\in B^j$ be the other endpoint of this edge.
Since graph $S^j$ is connected, and it contains both $b_j$ and $b^*_j$, we can find a path $P_j$ in $S^j$, connecting $b_j$ to $b^*_j$. 
For each $1\leq j\leq r$, let $P^*_j$ be the path obtained by concatenating $P_j,e_j,Q_j,e_{j+r},P_{j+r}$, and let $\pset^*=\set{P^*_j\mid 1\leq j\leq r}$.
It is immediate to verify that, if the paths in $\qset$ are disjoint from each other, then so are the paths in $\pset^*$, since all graphs in $\set{S^j\mid 1\leq j\leq 2r}$ are disjoint from each other and from $T_1$.
Moreover, for each $1\leq j\leq r$, path $P^*_j$ connects $b_j$ to $b_{j+r}$.
Thus, we have obtained a set $\pset^*$ of disjoint paths in $W_1$ routing $\mset^*$ as required.
This completes the proof of \Cref{lem: exp-routable b}.

    \section{Large \poefull System in Expanders}\label{sec: exp-exp to poe}
The goal of this section is to prove \Cref{thm: exp-poe}.
Recall that we are given an $\alpha$-expander $G$ and our goal is to construct a \poefull System $\Pi$ with large enough expansion and width, such that the corresponding graph $G_\Pi$ is low degree and is a minor of $G$.
Our algorithm consists of three parts.
In the first part, we construct an $\alpha'$-expanding Path-of-Sets System of length $24$ in $G$, for some $\alpha'$ to be fixed later.
In the second part, we transform it into a \posfull System of the same length.
In the third and the final part, we turn the \posfull System into a \poefull System with the guarantees claimed in \Cref{thm: exp-poe}.

\subsection{Expanding Path-of-Sets System} \label{subsec: exp-exp to pos}

The main technical result of this section is the following theorem.

\begin{theorem}\label{thm: exp-split an expander}
There is a constant $c_x>3$, and a deterministic algorithm, that,  given an $n$-vertex $\alpha$-expander $G$ with maximum vertex degree at most $d$, where $0<\alpha<1$, computes, in time $\poly(n)\cdot \left ( \frac{d}{\alpha}\right)^{O(\log(d/\alpha))}$ a partition $(V',V'')$ of $V(G)$, such that $|V'|,|V''|\geq  \frac{\alpha |V(G)|}{256d}$, and each graph $G[V'],G[V'']$ is an $\alpha^*$-expander, for $\alpha^*\geq \left(\frac{\alpha}{d}\right )^{c_x}$. 
\end{theorem}

The main tool that we use in the proof of the theorem is the following lemma, that we prove first.

\begin{lemma}\label{lem: exp-first cluster}
There is a constant $c'_x$, and deterministic algorithm, that,  given an $n$-vertex $\alpha$-expander $G$ with maximum vertex degree at most $d$, where $0<\alpha<1$, computes, in time $\poly(n)\cdot \left ( \frac{d}{\alpha}\right)^{O(\log(d/\alpha))}$,  a subset $V'\subseteq V(G)$ of vertices, such that $\frac{\alpha |V(G)|}{256d}\leq |V'|\leq \frac{\alpha |V(G)|}{8d}$, and $G[V']$ is an $\hat \alpha^*$-expander, for $\hat \alpha^*\geq \left(\frac{\alpha}{d}\right )^{c'_x}$.
\end{lemma}

\begin{proof}
Given a graph $G$, we say that a partition $(U',U'')$ of $V(G)$ is a \emph{balanced cut} iff $|U'|,|U''|\geq |V(G)|/4$. 
Our starting point is the following claim whose proof by Cheeger's inequality (\Cref{thm: exp-spectral}) is present in \Cref{appn-subsec: clm: exp-cheeger toy}.

\begin{claim} \label{clm: exp-cheeger toy}
	There is an efficient algorithm that, given an $n$-vertex graph $G=(V,E)$, and a parameter $\beta$, returns one of the following:
	\begin{itemize}
		\item either a subset $V'\subseteq V$ of vertices, such that $n/2\leq |V'|\leq 3n/4$ and $G[V']$ is an $\Omega(\frac{\beta^2}{d})$-expander; 
		\item or a partition $(S, T)$ of $V$ with $|E_G(S,T)| < \beta \cdot \min\set{|S|,|T|}$.
	\end{itemize} 
\end{claim}

By combining \Cref{clm: exp-cheeger toy} with \Cref{obs: exp-simple partition}, we obtain the following simple corollary.

\begin{corollary}\label{cor: exp-expander or balanced}
	There is an efficient algorithm that, given an $n$-vertex graph $G=(V,E)$ with maximum vertex degree at most $d$, and a parameter $\beta$, returns one of the following:
	\begin{itemize}
		\item either a subset $V'\subseteq V$ of vertices, such that $n/4\leq |V'|\leq 3n/4$ and $G[V']$ is an $\Omega(\frac{\beta^2}{d})$-expander; 
		\item or a {\bf balanced} partition $(S, T)$ of $V$ with $|E_G(S,T)| < \beta \cdot \min\set{|S|,|T|}$.
	\end{itemize} 
\end{corollary}

The proof of \Cref{cor: exp-expander or balanced} is present in \Cref{appn-subsec: cor: exp-expander or balanced}.
We now turn to complete the proof of \Cref{lem: exp-first cluster}.
We denote $|V(G)|=n$, and let ${n^*=\alpha |V(G)|/(8d)}$.
Our goal now is to compute a subset $V'\subseteq V(G)$ of vertices, with $n^*/32\leq |V'|\leq n^*$, such that $G[V']$ is an $\hat \alpha^*$-expander, where $\hat\alpha^*\geq \left(\frac{\alpha}{d}\right )^{c'_x}$ for some constant $c'_x$.
Our algorithm is recursive. Over the course of the algorithm, we will consider smaller and smaller sub-graphs of $G$, containing at least $n^*/4$ vertices each. For each such subgraph $G'\subseteq G$, we define its \emph{level} $L(G')$ as follows. Let $n'=|V(G')|$. If $n'\leq 4n^*/3$, then $L(G')=0$; otherwise, $L(G')=\ceil{\log_{4/3}(n'/n^*)}$. Intuitively, $L(G')$ is the number of recursive levels that we will use for processing $G'$.
Notice that, from the definition of $n^*$, $L(G)\leq O(\log (d/\alpha))$.
 We use the following claim.

\begin{claim}\label{clm: exp-recursion}
There is a deterministic algorithm, that, given a subgraph
 $G'\subseteq G$, such that $|V(G')|\geq n^*/4$, and a parameter $0<\beta<1$, returns one of the following:
\begin{itemize}
\item Either a balanced cut $(S,T)$ in $G'$ with $|E_{G'}(S,T)|<\beta\cdot\min\set{|S|,|T|}$; or
\item A subset $V'\subseteq V(G')$ of vertices of $G'$, such that $n^*/32\leq |V'|\leq n^*$, and $G'[V']$ is an $\hat \beta$-expander, for $\hat \beta\geq \Omega\left(\frac{\beta^2}{d\cdot 2^{10L(G')}}\right )$.
\end{itemize}
The running time of the algorithm is $\poly(n)\cdot \left (\frac{256 d}{\hat \beta}\right)^{L(G')}$.
\end{claim}

We prove the claim below, after we complete the proof of \Cref{lem: exp-first cluster} using it.
We apply \Cref{clm: exp-recursion} to the input graph $G$ and the parameter $\alpha$. Since $G$ is an $\alpha$-expander, we cannot obtain a cut $(S,T)$ in $G$ with $|E(S,T)|<\alpha\min\set{|S|,|T|}$. Therefore, the outcome of the algorithm is a subset $V'\subseteq V$ of vertices of $G$, with $n^*/32\leq V'\leq n^*$, such that $G[V']$ is a $\hat\alpha$-expander, for $\hat \alpha=\Omega\left(\frac{\alpha^2}{d\cdot 2^{10 L(G)}}\right )$, in time $\poly(n)\cdot \left (\frac{256 d}{\hat \alpha}\right)^{L(G)}$. Recall that $L(G)\leq O(\log (d/\alpha))$. Therefore, we get that $\hat \alpha=\Omega\left(\frac{\alpha^2}{d\cdot 2^{O(\log(d/\alpha))}}\right )\geq (\alpha/d)^{c'_x}$ for some constant $c'_x$, and the running time of the algorithm is $\poly(n)\cdot \left ( \frac{d}{\alpha}\right)^{O(\log(d/\alpha))}$.
This completes the proof of \Cref{lem: exp-first cluster} assuming \Cref{clm: exp-recursion} that we show next.
\end{proof}

\begin{proofof}{\Cref{clm: exp-recursion}}
	We denote $|V(G')|=n'$.
	We let $c$ be a large enough constant. We prove by induction on $L(G')$ that the claim is true, with the running time of the algorithm bounded by $n^c\cdot  \left (256d/\beta \right)^{L(G')}$. The base of the recursion is when $L(G')=0$, and so $n^*/4\leq n'\leq 4n^*/3$. We apply \Cref{cor: exp-expander or balanced} to graph $G'$ with the parameter $\beta$. If the outcome of the corollary is a subset $V'\subseteq V(G')$ of vertices with $n'/4\leq |V'|\leq 3n'/4$, such that $G'[V']$ is an $\Omega(\beta^2/d)$-expander, then we terminate the algorithm and return $V'$. Notice that in this case, we are
	guaranteed that $n^*/16\leq |V'|\leq n^*$. Otherwise, the algorithm returns a balanced cut $(S,T)$ in $G'$, with $|E_{G'}(S,T)|<\beta\cdot\min\set{|S|,|T|}$. We then return this cut. The running time of the algorithm is $\poly(n)$.
	
	We now assume that the theorem holds for all graphs $G'$ with $L(G')<i$, for some integer $i>0$, and prove it for a given graph $G'$ with $L(G')=i$. Let $n'=|V(G')|$. The proof is somewhat similar to the proof of \Cref{cor: exp-expander or balanced}. Throughout the algorithm, we maintain a balanced cut $(U',U'')$ of $G'$, with $|U'|\geq |U''|$. Initially, we start with an arbitrary such balanced cut. Notice that $|E(U',U'')|\leq |E(G')|\leq n'd$. While $|E(U',U'')|\geq \beta n'/4$, we perform iterations (that we call phases for convenience, since each of them consists of a number of iterations). At the end of every phase, we either compute a subset $V'\subseteq V(G')$ of vertices of $G'$, such that $n^*/32 \leq |V'|\leq n^*$, and $G'[V']$ is an $\hat \beta$-expander, in which case we terminate the algorithm and return $V'$; or we compute a new balanced cut $(J',J'')$ in $G'$, such that $|E(J',J'')|\leq |E(U',U'')|-\frac{\beta n'}{32}$.  If $|E(J',J'')|<\beta n'/4$, then we return this cut; it is easy to verify that $|E(J',J'')|<\beta\cdot\min\set{|J'|,|J''|}$. Otherwise, we replace $(U',U'')$ with the new cut $(J',J'')$, and continue to the next iteration. Since initially $|E(U',U'')|\leq n'd$, and since $|E(U',U'')|$ decreases by at least $\frac{\beta n'}{32}$ in every phase, the number of phases is bounded by $\frac{32d}{\beta}$. We now proceed to describe a single phase.

	\paragraph{An execution of a phase.}
	We assume that we are given a balanced cut $(U',U'')$ in $G'$, with $|U'|\geq |U''|$, and $|E(U',U'')|\geq \beta n'/4$. Our goal is to either compute a subset $V'$ of vertices of $G'$ such that $n^*/32\leq |V'|\leq n^*$ and $G'[V']$ is an $\hat \beta$-expander, or return another balanced cut $(J',J'')$ in $G'$, with $|E(J',J'')|\leq |E(U',U'')|-\frac{\beta n'}{32}$. Let $\beta'=\beta/32$. Over the course of the algorithm, we will maintain a set $E'$ of edges that we remove from the graph, starting with $E'=\emptyset$, and a collection $\gset$ of subgraphs of $G[U']$ (that will contain at most $4$ such subgraphs). As each graph $H\in \gset$ is a subgraph of $G[U']$, we are guaranteed that $|V(H)|\leq 3n'/4$, and so $L(H)\leq L(G')-1$. We start with $\hset$ containing a single graph, the graph $G'[U']$. We then iterate, while there is a graph $H\in \hset$ with $|V(H)|>|U'|/2$.

	In every iteration, we let $H\in \hset$ be the unique graph with $|V(H)|> |U'|/2$. Notice that $|V(H)|\geq n'/4 \geq n^*/3$, since we have assumed that $L(G')>0$ and so $n'\geq 4n^*/3$.
	We apply the algorithm from the induction hypothesis to $H$, with the parameter $\beta'=\beta/32$. If the outcome is a subset $V'\subseteq V(H)$ of vertices of $G'$, such that  $n^*/32 \leq |V'|\leq n^*$ and $H[V']$ is a $\hat \beta'$-expander, for $\hat \beta'\geq \Omega\left(\frac{(\beta')^2}{d\cdot 2^{10L(H)}}\right )$ then we terminate the algorithm and return $V'$. Notice that, since $L(H)\leq L(G)-1$, and $\beta'=\beta/32$, we get that $\frac{(\beta')^2}{d\cdot 2^{10L(H)}}\geq \frac{\beta^2}{d\cdot 2^{10L(G')}}$, so $G'[V']$ is a $\hat \beta$-expander. Otherwise, the algorithm returns a balanced cut $(S,T)$ of $V(H)$, such that $|E(S,T)|<\beta'\cdot\min\set{|S|,|T|}$. We add the edges of $E(S,T)$ to $E'$, remove $H$ from $\hset$, and add $H[S]$ and $H[T]$ to $\hset$. The algorithm terminates once for every graph $H\in \hset$, $|V(H)|\leq |U'|/2$. Let $r=|\hset|$ at the end of the algorithm. Since the cuts $(S,T)$ that we compute in every iteration are balanced, it is easy to verify that we run the algorithm from the induction hypothesis at most $3$ times, and that $r\leq 4$, since in every iteration the size of the largest graph in $\hset$ decreases by at least factor $3/4$, and $(3/4)^3<1/2$. Denote $\hset=\set{H_1,\ldots,H_r}$, and for each $1\leq j\leq r$, let $V_j=V(H_j)$, and let $m_j=|E(V_j,U'')|$. Since $|E(U',U'')|\geq \beta n'/4$, there is some index $1\leq j\leq r$, such that $|E(V_j,U'')|\geq \beta n'/16$. We define a new balanced cut $(J',J'')$, by setting $J'=U'\setminus V_j$ and $J''=U''\cup V_j$. Since $|V_j|\leq |U'|/2$, it is immediate to verify that it is a balanced cut. Moreover, it is immediate to verify that $|E'|\leq \beta'|U'|\leq 3\beta'n'/4\leq \beta n'/32$, and so:

	\[|E(J',J'')|\leq |E(U',U'')|-|E(V_j,U'')|+|E'|\leq |E(U',U'')|-\frac{\beta n'}{16}+\frac{\beta n'}{32}\leq |E(U',U'')|-\frac{\beta n'}{32}.\]

	Finally, we bound the running time of the algorithm. The running time is at most $\poly(n)$ plus the time required for the recursive calls to the same procedure.
	Recall that the number of phases in the algorithm is at most $32d/\beta$, and every phase requires up to $3$ recursive calls. Therefore, the total number of recursive calls is bounded by $100d/\beta$. Each recursive call is to a graph $H$ that has $L(H)<L(G)$. From the induction hypothesis, the running time of each recursive call is bounded by $n^c\cdot  \left (256d/\hat \beta' \right)^{L(G)-1}\leq n^c\cdot  \left (256d/\hat \beta \right)^{L(G)-1}$, and so the total running time of the algorithm is bounded by:

	\[n^c+\frac{100 d}{\beta}\cdot n^c\cdot  \left (\frac{256d}{\hat \beta} \right)^{L(G)-1}\leq n^c\cdot  \left (\frac{256d}{\hat \beta }\right)^{L(G)},\]

	since $\beta>\hat \beta$.
\end{proofof}

We are now ready to complete the proof of \Cref{thm: exp-split an expander}.

\begin{proofof}{\Cref{thm: exp-split an expander}}
We start with the input $n$-vertex $\alpha$-expander $G$ and apply \Cref{lem: exp-first cluster} to it, obtaining a subset $V_1\subseteq V(G)$ of vertices, such that $G[V_1]$ is a $\hat \alpha^*$-expander and $\frac{\alpha n}{256d}\leq |V_1|\leq \frac{\alpha n}{8d}$. Let $E'=\delta_G(V_1)$. Since the maximum vertex degree in $G$ is at most $d$, $|E'|\leq \frac{\alpha n}{8}$. 

We use the following claim, which is similar to \Cref{clm: exp-large expanding subgraph-edges}, except that it provides an efficient algorithm instead of the existential result of \Cref{clm: exp-large expanding subgraph-edges}, at the expense of obtaining  somewhat weaker parameters. The proof appears in \Cref{appn-subsec: exp-efficient expander fixing}.

\begin{claim} \label{clm: exp-large expanding subgraph cheeger}
There is an efficient algorithm, that given an $\alpha$-expander $G = (V,E)$ with maximum vertex degree at most $d$ and a subset $E' \subseteq E$ of its edges, computes a subgraph $H\subseteq G\setminus E'$ that is an $\Omega\left(\frac{\alpha^2}{d}\right)$-expander, and $|V(H)|\geq |V| - \frac{4|E'|}{\alpha}$.
\end{claim}

We apply \Cref{clm: exp-large expanding subgraph cheeger} to graph $G$ and the set $E'$ of edges computed above. Let $H\subseteq G\setminus E'$ be the resulting graph, and let $V_2=V(H)$. From \Cref{clm: exp-large expanding subgraph cheeger},  $|V_2|\geq n-\frac{4|E'|}{\alpha}\geq n/2$. Since $|V_1|<n/2$ and the set $E'$ of edges disconnects the vertices of $V_1$ from the rest of the graph, while $H$ is an $\Omega\left(\frac{\alpha^2}{d}\right)$-expander and therefore a connected graph, $V_1\cap V_2=\emptyset$. 

We are now ready to define the final partition $(V',V'')$ of $V(G)$, by letting it be the minimum cut separating the vertices of $V_1$ from the vertices of $V_2$ in $G$: that is, we require that $V_1\subseteq V'$, $V_2\subseteq V''$, and among all such partitions $(V',V'')$ of $V(G)$, we select the one minimizing $|E(V',V'')|$. The partition $(V',V'')$ can be computed efficiently using standard techniques: we construct a new graph $\hat G$ by starting with $G$, contracting all vertices of $V_1$ into a source $s$, contracting all vertices of $V_2$ into a destination $t$, and computing a minimum $s$-$t$ cut in the resulting graph. The resulting cut naturally defines the partition $(V',V'')$ of $V(G)$. Let $E''=E(V',V'')$, and denote $|E''|=z$. From Menger's theorem, there is a set $\pset$ of $z$ edge-disjoint paths in $G$, connecting $V_1$ to $V_2$. Therefore, there is a set $\pset_1$ of $z$ edge-disjoint paths in $G[V']\cup E''$, where each path in $\pset_1$ connects a distinct edge of $E''$ to a vertex of $V_1$, and similarly, there is a set $\pset_2$ of $z$ edge-disjoint paths in $G[V'']\cup E''$, where each path in $\pset_2$ connects a distinct edge of $E''$ to a vertex of $V_2$.

We claim that each of the graphs $G[V'],G[V'']$ is an $\alpha^*$-expander, for $\alpha^*=\frac{\alpha\hat \alpha^*}{512d}$. We prove this for $G[V']$; the proof for $G[V'']$ is similar. Assume for contradiction that $G[V']$ is not an $\alpha^*$-expander. Then there is a cut $(X,Y)$ in $G[V']$, such that $|E(X,Y)|<\alpha^*\cdot \min\set{|X|,|Y|}$.
Assume w.l.o.g. that $|X\cap V_1|\leq |Y\cap V_1|$. We now consider two cases.

The first case happens when $|X\cap V_1|\geq \frac{\alpha |X|}{512d}$. In that case, since $G[V_1]$ is an $\hat \alpha^*$-expander, there are at least $\hat \alpha^*\cdot |X\cap V_1|\geq \frac{\hat \alpha^*\cdot \alpha |X|}{512d}\geq \alpha^*|X|$ edges connecting $X\cap V_1$ to $Y\cap V_1$, and so $|E(X,Y)|>\alpha^*\cdot \min\set{|X|,|Y|}$, a contradiction. Therefore, we assume now that $|X\cap V_1|<\frac{\alpha |X|}{512d}$.

\begin{figure}[h]
        \center
        \scalebox{0.30}{\includegraphics{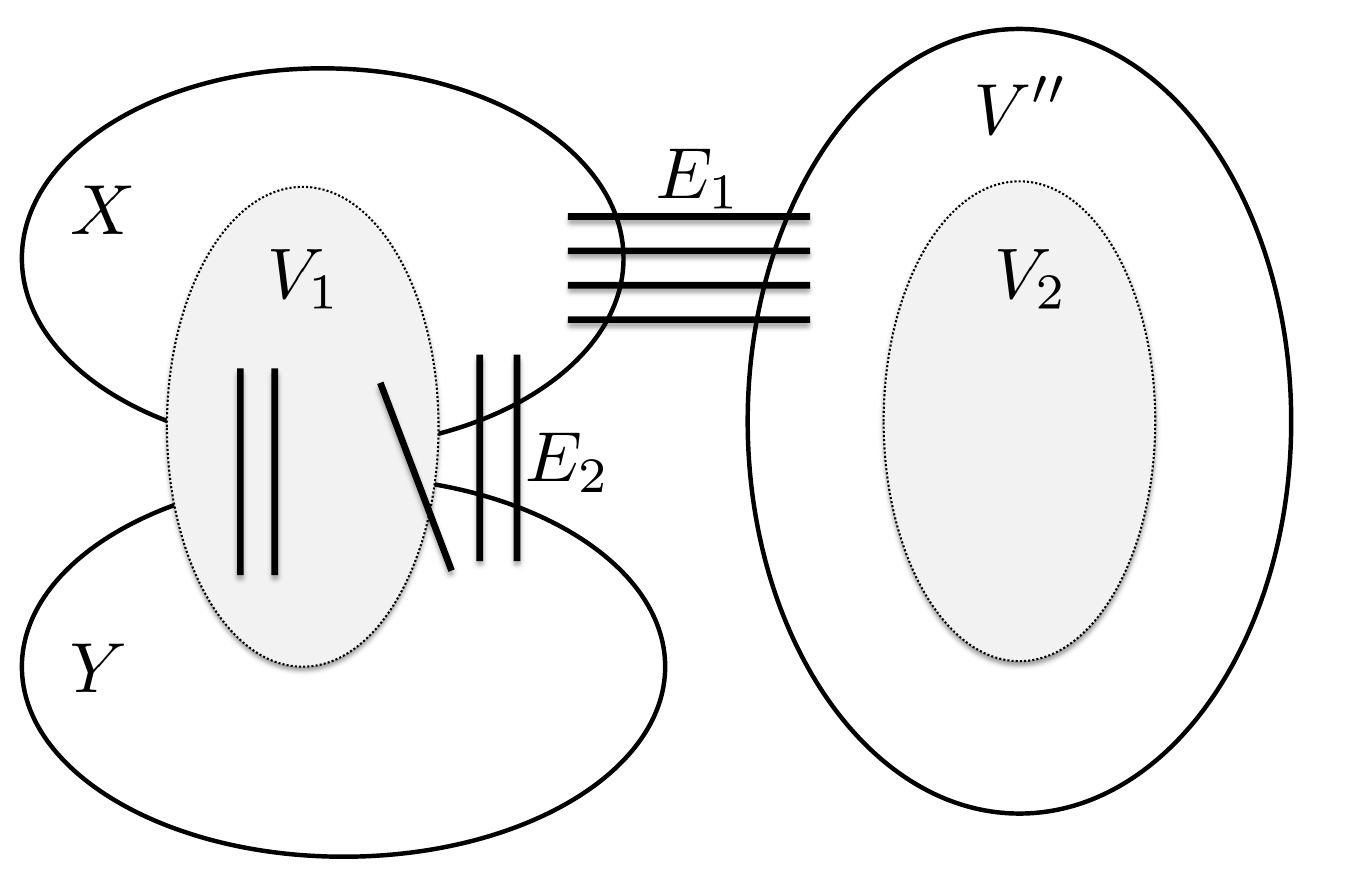}}
        \caption{An illustration for the proof of \Cref{thm: exp-split an expander}}
        \label{fig: exp-expansion proof}
    \end{figure}

 We partition the edges of $\delta_G(X)$ into two subsets: set $E_1$ contains all edges that lie in $E(V',V'')$, and set $E_2$ contains all remaining edges, so $E_2=E(X,Y)$ (see \Cref{fig: exp-expansion proof}). Note that from the definition of the cut $(X,Y)$, $|E_2|<\alpha^*|X|$.
 Recall that for every edge $e\in E(V',V'')$, there is a path $P_e\in \pset_1$ contained in $G[V']\cup E(V',V'')$, connecting $e$ to a vertex of $V_1$, such that all paths in $\pset_1$ are edge-disjoint. Let $\tilde \pset\subseteq \pset_1$ be the set of paths originating at the edges of $E_1$. We further partition $\tilde \pset$ into two subsets: set $\tilde \pset'$ contains all paths $P_e$ that contain an edge of $E_2$, and $\tilde \pset''$ contains all remaining paths. Notice that $|\tilde \pset'|\leq |E_2|< \alpha^*|X|$. On the other hand, every path $P_e\in \tilde \pset''$ is contained in $G[X]\cup E_1$, and contains a vertex of $V_1\cap X$ -- the endpoint of $P_e$. Since we have assumed that $|V_1\cap X|<\frac{\alpha |X|}{512d}$, and since the maximum vertex degree in $G$ is at most $d$, while the paths in $\tilde \pset''$ are edge-disjoint, we get that $|\tilde \pset''|<\frac{\alpha |X|}{512}$. Altogether, we get that $|E_1|=|\tilde \pset|\leq \alpha^*|X|+\frac{\alpha |X|}{512}$, and $|\delta_G(X)|=|E_1|+|E_2|\leq 2\alpha^*|X|+\frac{\alpha |X|}{512}\leq \frac{\alpha |X|}{256}<\alpha\cdot \min\set{|X|,n/256}\leq \alpha\cdot\min\set{|X|,|V(G)\setminus X|}$, since $|V(G)\setminus X|\geq n/256$, as $V_2\cap X=\emptyset$. This contradicts the fact that $G$ is an $\alpha$-expander.
\end{proofof}

\begin{corollary} \label{cor: exp-exp to pos}
    There is an algorithm, that, given, an $n$-vertex $\alpha$-expander $G$ with maximum vertex degree at most $d$ and an integer $\ell\geq 1$, where $0<\alpha<1/3$, 
    computes an $\alpha_{\ell}$-expanding Path-of-Sets system $\Sigma$ of length $\ell$ and width $w_{\ell}=\ceil{ \alpha_{\ell}n}$, together with a subgraph $G_{\Sigma}$ of $G$, where $\alpha_{\ell}=\alpha^{{c_x}^{\ell-1}}/d^{c_x^{2\ell-2}}$, and $c_x\geq 3$ is the constant from \Cref{thm: exp-split an expander}. The running time of the algorithm is $\poly(n)\cdot\left(\frac{d}{\alpha_{\ell}}\right )^{O(\log(d/\alpha_{\ell}))}$.
\end{corollary}

We note that we will use the corollary for with $\ell=48$, and so the resulting Path-of-Sets System will have expansion $(\alpha/d)^{O(1)}$, and the running time of the algorithm from \Cref{cor: exp-exp to pos} is $\poly(n)\cdot \left(\frac{d}{\alpha}\right)^{O(\log(d/\alpha))}$.

\begin{proof}
The proof is by induction on $\ell$. The base case is when $\ell=1$. We choose two arbitrary disjoint subsets $A_1,B_1$ of $\ceil{w_1}<n/2$ of vertices, and we let $S_1=G$. This defines an $\alpha$-expanding Path-of-Sets System of length $1$ and width $w_1$.

We now assume that we are given an integer $\ell>1$, and an $\alpha_{\ell-1}$-expanding Path-of-Sets System $\Sigma=(\sset,\mset,A_1,B_{\ell-1})$ of length $\ell-1$ and width $w_{\ell-1}$, where $G_{\Sigma}\subseteq G$. We assume that $\sset=(S_1,\ldots,S_{\ell-1})$.  We compute an $\alpha_{\ell}$-expanding Path-of-Sets System $\Sigma'=(\sset',\mset',A'_1,B'_{\ell})$ of length $\ell$ and width $w_{\ell}$. 
We will denote $\sset'=(S_1',\ldots,S_{\ell}')$, and for each $1\leq i\leq \ell'$, the corresponding vertex sets $A_i$ and $B_i$ in $S'_i$ are denoted by $A'_i$ and $B'_i$, respectively.

For all $1\leq i<\ell-1$, we set $S'_i=S_i$. We also let $A'_1\subseteq A_1$ be any subset of $w_{\ell}$ vertices, and for $1\leq i<\ell-2$, we let $\mset'_i\subseteq \mset_i$ be any subset of $w_{\ell}$ edges; the endpoints of these edges lying in $B_i$ and $A_{i+1}$ are denoted by $B'_i$ and $A'_{i+1}$ respectively. It remains to define $S'_{\ell-1},S'_{\ell}$, the matchings $\mset'_{\ell-2}$ and $\mset'_{\ell-1}$ (that implicitly define the sets $B_{\ell-2}',A_{\ell-1}',B_{\ell-1}',A_{\ell}'$ of vertices), and the set $B'_{\ell}$ of vertices.

We apply \Cref{thm: exp-split an expander} to graph $S_{\ell-1}$, and compute, in time $\poly(n)\cdot \left ( \frac{d}{\alpha_{\ell-1}}\right)^{O(\log(d/\alpha_{\ell-1}))}$ a partition $(V',V'')$ of $V(S_{\ell-1})$, such that $|V'|,|V''|\geq  \frac{\alpha_{\ell-1} |V(S_{\ell-1})|}{256d}$, and each graph $G[V'],G[V'']$ is an $\alpha^*$-expander, for $\alpha^*\geq \left(\frac{\alpha_{\ell-1}}{d}\right )^{c_x}$. 

One of the two subsets, say $V'$, must contain at least half of the vertices of $A_{\ell-1}$. We set $S'_{\ell-1}=S_{\ell-1}[V']$ and $S'_{\ell}=S_{\ell-1}[V'']$. Recall that:
$|V'|,|V''|\geq \frac{\alpha_{\ell-1} |V(S_{\ell-1})|}{256d}\geq   \frac{\alpha_{\ell-1} w_{\ell-1}}{128d}$. Since graph $S_{\ell-1}$ is an $\alpha_{\ell-1}$-expander, there are at least $\frac{\alpha_{\ell-1}^2 w_{\ell-1}}{128d}$ edges connecting $V'$ to $V''$. Since maximum vertex degree in $G$ is at most $d$, there is a matching $\mset$, between vertices of $V'$ and vertices of $V''$, with $|\mset|\geq \frac{\alpha_{\ell-1}^2 w_{\ell-1}}{128d^2}$. We claim that $|\mset|\geq w_{\ell}$. In order to see this, it is enough to prove that $w_{\ell}\leq \frac{\alpha_{\ell-1}^2 w_{\ell-1}}{128d^2}$. Since $w_{\ell}=\ceil{\alpha_{\ell}n}$, this is equivalent to proving that:

\[\alpha_{\ell}\leq \frac{\alpha_{\ell-1}^3}{256 d^2}.\] 

This is easy to verify from the definition of $\alpha_{\ell}$ and the fact that $c_x\geq 3$.
We let $\mset'_{\ell-1}$ be any subset of $\mset$ containing $w_{\ell}$ edges. The endpoints of the edges of $\mset'_{\ell-1}$ lying in $V'$ and $V''$ are denoted by $B'_{\ell-1}$ and $A'_{\ell}$ respectively. We let $B_{\ell}'$ be any subset of $w_{\ell}$ vertices of $V''\setminus A_{\ell}'$. Finally, we let $A_{\ell-1}'$ any subset of $w_{\ell}$ vertices of $(V'\cap A_{\ell-1})\setminus B'_{\ell-1}$; $\mset'_{\ell-2}\subseteq \mset_{\ell-2}$ the subset of edges whose endpoints lie in $A'_{\ell-1}$; and $B'_{\ell-2}$ the set of endpoints of the edges of $\mset'_{\ell-2}$ lying in $B_{\ell-2}$. This completes the construction of the Path-of-Sets System $\Sigma'$. It is immediate to verify that it has length $\ell$, width $w_{\ell}$, and that $G_{\Sigma'}\subseteq G$. It remains to prove that it is $\alpha_{\ell}$-expanding, or equivalently, that $S'_{\ell-1}$ and $S'_{\ell}$ are $\alpha_{\ell}$-expanders. Recall that \Cref{thm: exp-split an expander} guarantees that both these graphs are $\alpha^*$-expanders, where $\alpha^*\geq \left(\frac{\alpha_{\ell-1}}{d}\right )^{c_x}$. It is now enough to verify that $\alpha^*\geq \alpha_{\ell}$, which is immediate to do from the definition of $\alpha_{\ell}$:

\[
\alpha^*\geq \left(\frac{\alpha_{\ell-1}}{d}\right )^{c_x}=\frac{\left(\alpha^{{c_x}^{\ell-2}}/d^{c_x^{2\ell-4}}\right)^{c_x}}{d^{c_x}}
=\frac{\alpha^{{c_x}^{\ell-1}}}{d^{c_x^{2\ell-3}}\cdot d^{c_x}}\geq \frac{\alpha^{{c_x}^{\ell-1}}}{d^{c_x^{2\ell-2}}}=\alpha_{\ell}
\]

Lastly, the running time of the algorithm is dominated by partitioning $S_{\ell-1}$, and is bounded by $\poly(n)\cdot \left ( \frac{d}{\alpha_{\ell-1}}\right)^{O(\log(d/\alpha_{\ell-1}))}\leq \poly(n)\cdot \left ( \frac{d}{\alpha_{\ell}}\right)^{O(\log(d/\alpha_{\ell}))}$, as required.
\end{proof}

We apply \Cref{cor: exp-exp to pos} to the input graph $G$, with the parameter $\ell=48$, obtaining a sub-graph $G_{\Sigma}\subseteq G$, and an $\alpha'$-expanding Path-of-Sets System $\Sigma$ of length $48$ and width $w'=\ceil{\alpha' n}$, where $\alpha'=(\alpha/d)^{O(1)}$. The running time of the algorithm is $\poly(n)\cdot \left(\frac{d}{\alpha}\right)^{O(\log(d/\alpha))}$.

\subsection{From Expanding Path-of-Sets System to Strong Path-of-Sets System} \label{subsec: exp-exp to strong}

The goal of this subsection is to prove the following theorem:

\begin{theorem}\label{thm: exp-exp pos to strong pos}
 There is an efficient algorithm, that, given a parameter $\ell>0$, and an $\alpha$-\posexp System $\Sigma$ of
    width $w$ and length $4 \ell$,
    where $0<\alpha <1$,
    such that the corresponding graph $G_{\Sigma}$ has
     maximum vertex-degree at most $d$,
    computes a \posfull System $\Sigma'$, of
    width $w' =\Omega(\alpha^3w/d^4)$ and length $\ell$, such that the maximum vertex degree in the corresponding graph $G_{\Sigma'}$ is at most $d$, and $G_{\Sigma'}$ is a minor of $G_{\Sigma}$. Moreover, the algorithm computes a model of $G_{\Sigma'}$ in $G_{\Sigma}$.
\end{theorem}

We use the following simple claim, whose proof is deferred to \Cref{appn-subsec: exp-flow in expander}.

 \begin{claim} \label{clm: exp-flow in expander}
There is an efficient algorithm, that, given an $\alpha$-expander $G$, whose maximum vertex degree is at most $d$, where $0<\alpha<1$, together with two disjoint subsets $A,B$ of its vertices of cardinality $z$ each, computes a collection $\pset$ of $\ceil{\alpha z/d}$ disjoint paths, connecting vertices of $A$ to vertices of $B$ in $G$.
\end{claim}

We will also use the following theorem, whose proof is similar to some arguments that appeared in~\cite{CC_gmt}, and is deferred to \Cref{appn-subsec: exp-exp to wl}.

\begin{theorem} \label{thm: exp-exp to wl}
        There is an efficient algorithm, that, given an $\alpha$-\posexp System $\Sigma = (\sset, \mset, A_1, B_3)$ of width $w$ and length $3$, where $0<\alpha <1$, and the corresponding graph $G_{\Sigma}$ has maximum vertex degree at most $d$, computes subsets $\hat A_1\subseteq A_1,\hat B_3\subseteq B_3$ of $\Omega(\alpha^2 w/d^3)$ vertices each, such that $\hat A_1\cup \hat B_3$ is well-linked in $G_{\Sigma}$.
\end{theorem}

We are now ready to complete the proof of \Cref{thm: exp-exp pos to strong pos}.

\begin{proofof}{\Cref{thm: exp-exp pos to strong pos}}
We construct a \posfull System $\Sigma'=(\sset',\mset',A_1',B_{\ell}')$ of length $\ell$ and width $w'$, denoting $\sset'=(S_1',\ldots,S_{\ell}')$. For all $1\leq i\leq\ell$, the corresponding vertex sets $A_i$ and $B_i$ are denoted by $A'_i$ and $B'_i$, respectively. 

For all $1\leq i\leq \ell$, we let $\Sigma_i$ be the $\alpha$-expanding Path-of-Sets System of width $w$ and length $3$ obtained by using the clusters $S_{4i-3},S_{4i-2}$, $S_{4i-1}$, and the matchings $\mset_{4i-3}$ and $\mset_{4i-2}$. In order to define the new Path-of-Sets System, for each $1\leq i\leq \ell$, we set $S'_i=G_{\Sigma_i}$. We apply \Cref{thm: exp-exp to wl} to $\Sigma_i$, to obtain subsets $\hat A_i\subseteq A_{4i-3}$, $\hat B_i\subseteq B_{4i-1}$ of $\Omega(\alpha^2w/d^3)$ vertices each, such that $\hat A_i\cup \hat B_i$ are well-linked in $S'_i$.

In order to complete the construction of the Path-of-Sets System $\Sigma'$, we let $A_1'\subseteq \hat A_1$ be any subset of $w'$ vertices, and we define $B'_{\ell}\subseteq \hat B_{\ell}$ similarly. It remains to define, for each $1\leq i<\ell$, the matching $\mset'_i$. We will ensure that the endpoints of the resulting matching are contained in $\hat B_i$ and $\hat A_{i+1}$, respectively, ensuring that the resulting Path-of-Sets System is strong.

Consider some index $1\leq i<\ell$. Recall that we have computed the sets $\hat B_i\subseteq B_{4i-1}$, $\hat A_{i+1}\subseteq A_{4i+1}$ of vertices. We let $E'_i\subseteq \mset_{4i-1}$ be the set of edges incident to the vertices of $\hat B_i$, and we denote by $\tilde A_{4i}\subseteq A_{4i}$ the set of vertices in $A_{4i}$ that serve as their endpoints. Similarly, we let $E''_i\subseteq \mset_{4i}$ be the set of edges incident to the vertices of $\hat A_{i+1}$, and we denote by $\tilde B_{4i}\subseteq B_{4i}$ the set of vertices in $B_{4i}$ that serve as their endpoints. From \Cref{clm: exp-flow in expander}, there is a set $\qset_i$ of disjoint paths in $S_{4i}$, connecting vertices of $\tilde A_{4i}$ to vertices of $\tilde B_{4i}$, of cardinality $w'=\Omega(\alpha^3w/d^4)$. By extending the paths in $\qset_i$ to include the edges of $E'_i\cup E''_i$ incident to them, we obtain a collection $\qset'_i$ of $w'$ disjoint paths in $S_{4i}\cup \mset_{4i-1}\cup\mset_{4i}$, connecting vertices of $\hat B_i$ to vertices of $\hat A_{i+1}$. We denote the endpoints of the paths in $\qset'_i$ lying in $\hat B_i$ by $B_i'$, and the endpoints of the paths in $\qset'_i$ lying in $\hat A_{i+1}$ by $A'_{i+1}$. The paths in $\qset'_i$ naturally define the matching $\mset'_i$ between the vertices of $B_i'$ and the vertices of $A_{i+1}'$. This concludes the definition of the Path-of-Sets System $\Sigma'$. It is immediate to verify that it is a strong Path-of-Sets System of length $\ell $ and width $w'$, and to obtain a model of $G_{\Sigma'}$ in $G_{\Sigma}$. Note that graph $G_{\Sigma'}$ has maximum vertex degree at most $d$.
\end{proofof}

Recall that in Part 1 of the algorithm, we have obtained  a sub-graph $G_{\Sigma}\subseteq G$, and an $\alpha'$-expanding Path-of-Sets System $\Sigma$ of length $48$ and width $w'=\ceil{\alpha' n}$, where $\alpha'=(\alpha/d)^{O(1)}$. 
Applying \Cref{thm: exp-exp pos to strong pos} to $\Sigma$, we obtain a \posfull System $\Sigma'$ of length $12$ and width
\[ w''=\Omega\left(\frac{(\alpha')^3w'}{d^4}\right )=\Omega\left(\frac{(\alpha')^4}{d^4}n\right )=\left(\frac{\alpha}{d}\right )^{O(1)}\cdot n.\]

We have also computed a model of $G_{\Sigma'}$ in $G$, and established that the maximum vertex degree in $G_{\Sigma'}$ is at most $d$.
For convenience, we let $c'$ be a constant, such that $w''\geq \frac{\alpha^{c'}}{d^{c'}}n$.
\subsection{From \posfull System to \poefull System} \label{subsec: exp-pos to poe}

The goal of this subsection is to prove the following theorem:
\begin{theorem} \label{thm: exp-pos to poe}
    There is an efficient algorithm, that, given a  \posfull System $\Sigma$ of width $w$ and length $12$, such that the corresponding graph $G_{\Sigma}$ has at most $n$ vertices and has maximum vertex degree at most $d$,  computes a \poefull System $\Pi$ of width $\hat w =\Omega\left(\frac{w^4}{d^2n^3}\right )$ and expansion $\hat \alpha \geq \Omega\left (\frac{w^2}{n^2d}\right )$, whose corresponding graph $G_{\Pi}$ has maximum vertex degree at most $d+1$ and is a minor of $G_{\Sigma}$. Moreover, the algorithm computes a model of $G_{\Pi}$ in $G_{\Sigma}$.
\end{theorem}

Before we prove \Cref{thm: exp-pos to poe}, we complete the proof of \Cref{thm: exp-poe} using it.

\proofof{\Cref{thm: exp-poe}}
    Recall that our input is an $\alpha$-expander $G$, for some $0<\alpha<1$, with $|V(G)|=n$, such that the maximum vertex degree in $G$ is at most $d$. Our goal is to provide an algorithm that computes a \poefull System $\Pi$ of expansion $\tilde \alpha \geq \left(\frac{\alpha}{d}\right )^{\hat c_1}$ and width $\tilde w \geq n \cdot \left(\frac{\alpha}{d}\right )^{\hat c_2}$, such that the maximum vertex degree in $G_{\Pi}$ is at most $d+1$, and to compute a minor of $G_{\Pi}$ in $G$.

    Recall that in Step 2 we have constructed  a \posfull System $\Sigma'$ of length $12$ and width $w''\geq \frac{\alpha^{c'}}{d^{c'}}n$, for some constant $c'$, such that $G_{\Sigma'}$ has maximum vertex degree at most $d$. We have also computed a model of $G_{\Sigma'}$ in $G$.
    Our last step is to apply \Cref{thm: exp-pos to poe} to $\Sigma'$.
    As a result, we obtain a  \poefull System $\Pi$ of width $\hat w =\Omega\left(\frac{(w'')^4}{d^2n^3}\right )$ and expansion $\hat \alpha \geq \Omega\left (\frac{(w'')^2}{n^2d}\right )$, whose corresponding graph $G_{\Pi}$ has maximum vertex degree at most $d+1$. We also obtain a model of $G_{\Pi}$ in $G_{\Sigma}$. 
    Substituting the value $w''\geq \frac{\alpha^{c'}}{d^{c'}}n$, we get that the width of the \poefull System $\Pi$ is $\Omega\left(\frac{\alpha^{4c'}}{d^{2+4c'}}\right )\cdot n$, and that its expansion is $\Omega \left(\frac{\alpha^{2c'}}{d^{2c'+1}}\right )$. By appropriately setting the constants $\hat c_1$ and $\hat c_2$, we ensure that the width of the \poefull System is at least $n \cdot \left(\frac{\alpha}{d}\right )^{\hat c_2}$ and its expansion is at least  $\left(\frac{\alpha}{d}\right )^{\hat c_1}$.
    This completes the proof of \Cref{thm: exp-poe}.
\endproofof

In the remainder of this section, we prove \Cref{thm: exp-pos to poe}.
We can assume w.l.o.g. that  $w^4\geq 2^{14}n^3d^2$, since otherwise it is sufficient to produce a \poefull System of width $1$, which is trivial to do.
 We denote the input \posfull System by $\Sigma = (\sset, \mset, A_1, B_{12})$, where $\sset=(S_1,\ldots,S_{12})$, and we let $G_{\Sigma}$ be its corresponding graph.
For convenience, we denote by $\iseteven$ and $\isetodd$ the sets of all even and all odd indices in $\set{1,\ldots,12}$, respectively.

Our algorithm consists of three steps. In the first step, for every index $i\in \iseteven$, we find a large set $\pset_i$ of disjoint paths connecting $A_i$ to $B_i$ in $S_i$, and a subgraph $T_i\subseteq S_i$ that is an $\hat \alpha$-expander, such that the paths in $\pset_i$ are disjoint from $T_i$. In the second step, for each such index $i\in\iseteven$, we compute another set $\qset_i$ of disjoint paths in $S_i$, and a large enough subset $\pset'_i\subseteq \pset_i$ of paths, such that every path in $\qset_i$ connects a vertex on a distinct path of $\pset'_i$ to a distinct vertex of $T_i$. In the third and the final step we compute the \poefull System $\Pi$ and a model of $G_{\Pi}$ in $G_{\Sigma}$.

\paragraph{Step 1.} In this step, we prove the following lemma.

\begin{lemma}\label{lem: exp-step 1}
There is an efficient algorithm, that, given an index $i\in \iseteven$, computes a set $\pset_i$ of $\floor{\frac{w^2}{16nd}}$ paths in $S_i$, and a subgraph $T_i\subseteq S_i$, such that:

\begin{itemize}
\item graph $T_i$ is an $\hat \alpha$-expander, and it contains at least $w/2$ vertices of $A_i$;
\item the paths in $\pset_i$ are disjoint from each other; they are also disjoint from $T_i$ and internally disjoint from $A_i\cup B_i$; 
\item every path in $\pset_i$ connects a vertex of $A_i$ to a vertex of $B_i$; and
\item every path in $\pset_i$  has length at most $2n/w$.
\end{itemize}
\end{lemma}

\begin{proof}
For convenience, we omit the subscript $i$ in this proof. We are given a graph $S$ that contains at most $n$ vertices and has maximum vertex degree at most $d$, and two disjoint subsets $A,B$ of $V(S)$ of cardinality $w$ each, such that each of $A\cup B$ is well-linked in $S$. Therefore, there is a set $\pset$ of $w$ disjoint paths in $S$, connecting vertices of $A$ to vertices of $B$, such that the paths in $\pset$ are internally disjoint from $A\cup B$. We say that a path in $\pset$ is \emph{short} if it contains at most $2n/w$ vertices, and otherwise it is long. Since $|V(S)|\leq n$, at most $w/2$ paths in $\pset$ can be long, and the remaining paths must be short. Let $\pset'\subseteq \pset$ be any subset of $\floor{\frac{w^2}{16nd}}$ paths in $\pset$. It is now sufficient to show an algorithm that computes an $\hat \alpha$-expander $T\subseteq S$, such that $T$ is disjoint from the paths in $\pset'$. In order to do so, we let $E'$ be the set of all edges lying on the paths in $\pset'$, so $|E'|\leq |\pset'|\cdot \frac{2n}{w}\leq \floor{\frac{w^2}{16nd}}\cdot \frac{2n}{w}\leq \frac{w}{8}$. 
 
We start with $T=S\setminus E'$, and then iteratively remove edges from $T$, until we obtain a connected component of the resulting graph that is an $\hat \alpha$-expander, containing at least $w/2$ vertices of $A$. Notice that the original graph $T$ is not necessarily connected. We also maintain a set $E''$ of edges that we remove from $T$, initialized to $E''=\emptyset$. 
Our algorithm is iterative. In every iteration, we apply \Cref{thm: exp-spectral} to the current graph $T$, to obtain a cut $(Z,Z')$ in $T$. If the sparsity of the cut is at least $\frac{w}{16n}$, that is, $|E_T(Z,Z')|\geq \frac{w}{16n}\min\set{|Z|,|Z'|}$, then we terminate the algorithm. \Cref{thm: exp-spectral} then guarantees that the expansion of $T$ is $\Omega\left (\frac{w^2}{n^2d}\right )$, that is, $T$ is a $\hat \alpha$-expander. Otherwise, $|E_T(Z,Z')|< \frac{w}{16n}\min\set{|Z|,|Z'|}$. Assume w.l.o.g. that $|Z\cap A|\geq |Z'\cap A|$. We then add the edges of $E_{T}(Z,Z')$ to $E''$, set $T=T[Z]$, and continue to the next iteration. Note that the number of edges added to $E''$ during this iteration is at most $\frac{|Z'|w}{16n}$.

Clearly, the graph $T$ we obtain at the end of the algorithm is an $\hat \alpha$-expander, and it is disjoint from all paths in $\pset'$. It now only remans to show that $T$ contains at least $w/2$ vertices of $A$. Assume for contradiction that this is false.

Assume that the algorithm performs $r$ iterations, and for each $1\leq j\leq r$, let $(Z_j,Z'_j)$ be the cut computed by the algorithm in iteration $j$, where $|Z_j\cap A|\geq |Z'_j\cap A|$. But then for all $1\leq j\leq r$, $|Z'_j\cap A|\leq w/2$ must hold. 
Let $n_j=|Z'_j\cap A|$. Since the vertices of $A$ are well-linked in $S$, $\delta_S(Z'_j)\geq n_j$. Therefore:

\[\sum_{j=1}^r|\delta_S(Z'_j)|\geq \sum_{j=1}^rn_j\geq w/2,\]

since we have assumed that the final graph $T$ has fewer than $w/2$ vertices of $A$. On the other hand, all edges in $\bigcup_{j=1}^r\delta_S(Z'_j)$ are contained in $E'\cup E''$, and so:

\[\sum_{j=1}^r|\delta_S(Z'_j)|\leq 2|E'\cup E''|.\]

Recall that $|E'|\leq \frac{w}{8}$, and it is easy to verify that $|E''|\leq \frac{w}{16n}\cdot n=\frac{w}{16}$. Therefore, $\sum_{j=1}^r|\delta_S(Z'_j)|<\frac w 2$, a contradiction.
\end{proof}

\paragraph{Step 2.} For every index $i\in \iseteven$, let $A'_i\subseteq A_i$ be the subset of vertices that serve as endpoints for the paths in $\pset_i$. The goal of this step is to prove the following lemma.

\begin{lemma}\label{lem: exp-Q-paths}
There is an efficient algorithm, that, given an index $i\in \iseteven$, computes a subset $\pset'_i\subseteq \pset_i$ of $\hat w$ paths, and, for each path $P\in \pset'_i$, a path $Q_P$ in $S_i$, that connects a vertex of $P$ to a vertex of $T_i$, such that the paths in set $\qset_i=\set{Q_P\mid P\in \pset'_i}$ are disjoint from each other, internally disjoint from $T_i$, and internally disjoint from the paths in $\pset'_i$.
\end{lemma}

\begin{proof}
We fix an index $i\in \iseteven$, and for convenience omit the subscript $i$ for the remainder of the proof. Recall that we are given a set $A'\subseteq A$ of $\floor{\frac{w^2}{16nd}}$ vertices, that serve as endpoints of the paths in $\pset$. Recall that $T$ contains at least $w/2$ vertices of $A$. We let $A''\subseteq A$ be any set of $\floor{\frac{w^2}{16nd}}$ vertices of $A$ lying in $T$. Since the set $A$ of vertices is well-linked in $S$, there is a set $\qset$ of $\floor{\frac{w^2}{16nd}}$ node-disjoint paths, connecting the vertices of $A'$ to the vertices of $A''$ in $S$. We say that a path in $\qset$ is \emph{short} if it contains fewer than $\frac{64n^2d}{w^2}$ vertices, and otherwise we say that it is \emph{long}. Since $S$ contains at most $n$ vertices, and the paths in $\qset$ are disjoint, at most $\frac{w^2}{64nd}$ paths of $\qset$ are long. We let $\hqset\subseteq \qset$ be the set of all short paths, so $|\hqset|\geq \frac{w^2}{64nd}$, and we let $\hat A\subseteq A'$ be the set of vertices that serve as endpoints of the paths in $\hqset$. We also let $\hpset\subseteq \pset$ the set of paths originating from the vertices in $\hat A$. We are now ready to compute the set $\pset'$ of paths, and the corresponding paths $Q_P$ for all $P\in \pset'$.

We start with $\pset'=\emptyset$, and then iterate. While $\hpset\neq \emptyset$, let $P$ be any path in $\hpset$, and let $a\in \hat A$ be the vertex from which it originates. Let $Q$ be the path of $\hqset$ originating at $a$. We prune the path $Q$ as needed, so that it connects a vertex of $P$ to a vertex of $T$, but is internally disjoint from $P$ and $T$. Let $Q'$ be the resulting path. We then add $P$ to $\pset'$, and we let $Q_P=Q'$. Next, we delete from $\hpset$ all paths that intersect $Q'$ (since the length of $Q'$ is at most $\frac{64n^2d}{w^2}$, we delete at most $\frac{64n^2d}{w^2}$ paths from $\hpset$), and for every path $P^*$ that we delete from $\hpset$, we delete from $\hqset$ the path sharing an endpoint with $P^*$ (so at most $\frac{64n^2d}{w^2}$ paths are deleted from $\hqset$). Similarly, we delete from $\hqset$ every path that intersects $P$ (since the length of $P$ is at most $2n/w$, we delete at most $\frac{2n}{w}\leq \frac{64n^2d}{w^2}$ paths from $\hqset$), and for every path $Q^*$ that we delete from $\hqset$, we delete from $\hpset$ the path sharing an endpoint with $Q^*$ (again, at most $\frac{64n^2d}{w^2}$ paths are deleted from $\hpset$). Overall, we delete at most $\frac{128n^2d}{w^2}$ paths from $\hpset$, and at most $\frac{128n^2d}{w^2}$ paths from $\hqset$. The paths that remain in both sets form pairs -- that is, for every path $P^*\in \hpset$, there is a path $Q^*\in \hqset$ originating at the same vertex of $A$, and vice versa. Furthermore, and all paths in $\hpset\cup \hqset$ are disjoint from the paths in $\pset'\cup \set{Q_P\mid P\in \pset'}$. 

At the end of the algorithm, we obtain a subset $\pset'\subseteq \pset$ of paths, and for each path $P\in \pset'$, a path $Q_P$ in $S$, connecting a vertex of $P$ to a vertex of $T$, such that  the paths in set $\qset'=\set{Q_P\mid P\in \pset'}$ are disjoint from each other, internally disjoint from $T$, and internally disjoint from the paths in $\pset'$. It now only remains to show that $|\pset'|\geq \hat w$. 

Recall that we start with $|\hpset|\geq \frac{w^2}{64nd}$. In every iteration, we add one path to $\pset'$, and delete at most $\frac{128n^2d}{w^2}$ paths from $\hpset$. Since we have assumed that $w^4\geq 2^{14}n^3d^2$, we get that $\frac{256 n^2d}{w^2}\leq \frac{w^2}{64nd}$. 
It is then easy to verify that at the end of the algorithm, $|\pset'|\geq \floor{\frac{|\hpset|}{256n^2d/w^2}}\geq \Omega\left(\frac{w^4}{n^3d^2}\right )=\hat w$.
\end{proof}

\paragraph{Step 3.} In this step we complete the construction of the \poefull System $\Pi$. We will also define a minor $G'$ of $G_{\Sigma}$ and compute a model of $G_{\Pi}$ in $G'$; it is then easy to obtain a model of $G_\Pi$ in $G_{\Sigma}$.

\begin{figure}[h]
\centering
\subfigure[Paths $P$ (shown in blue) and $Q_P$ (shown in red) before edge contractions]{\scalebox{0.3}{\includegraphics{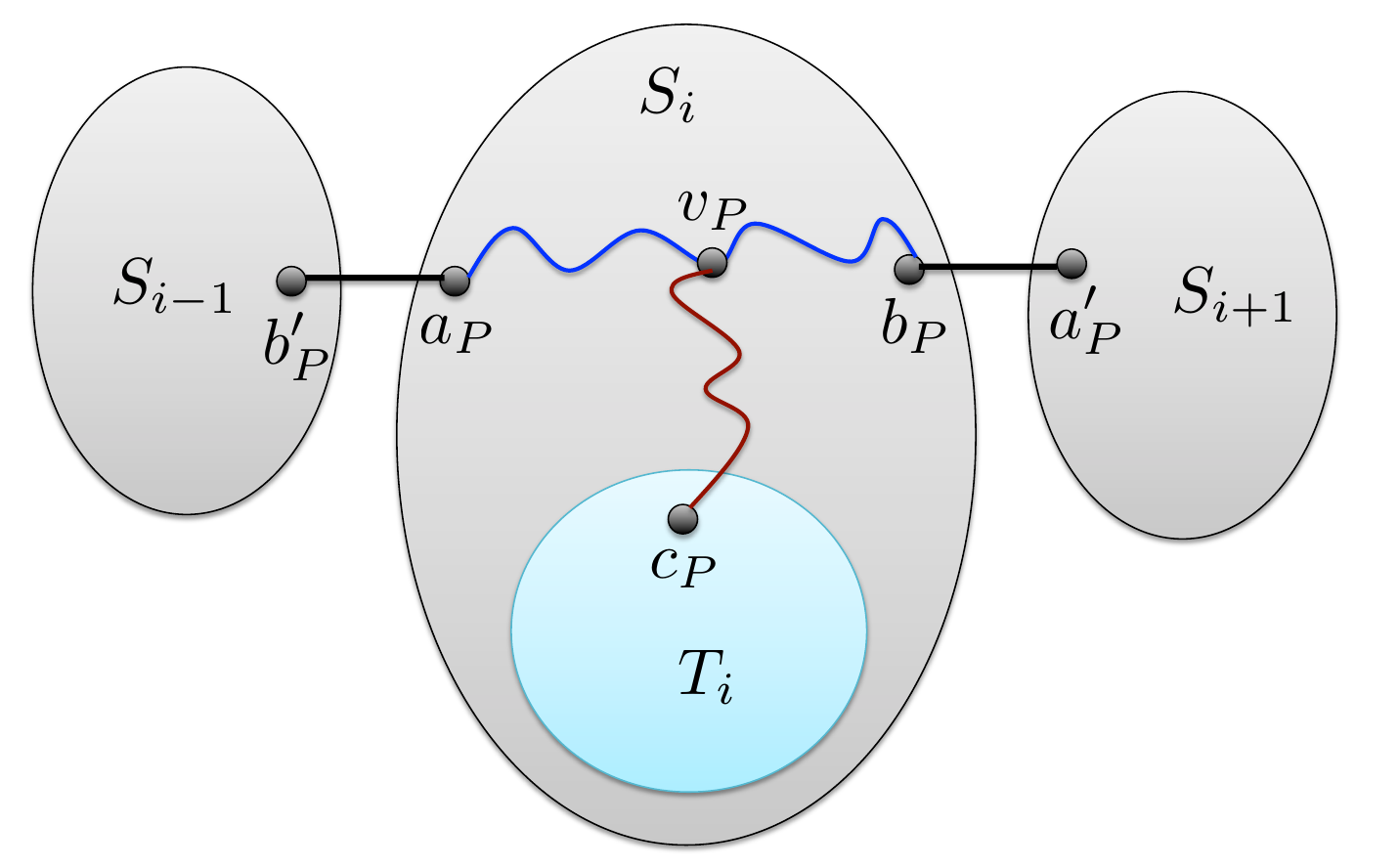}}\label{fig: exp-before contraction}}
\hspace{1cm}
\subfigure[After edge contractions]{
\scalebox{0.3}{\includegraphics{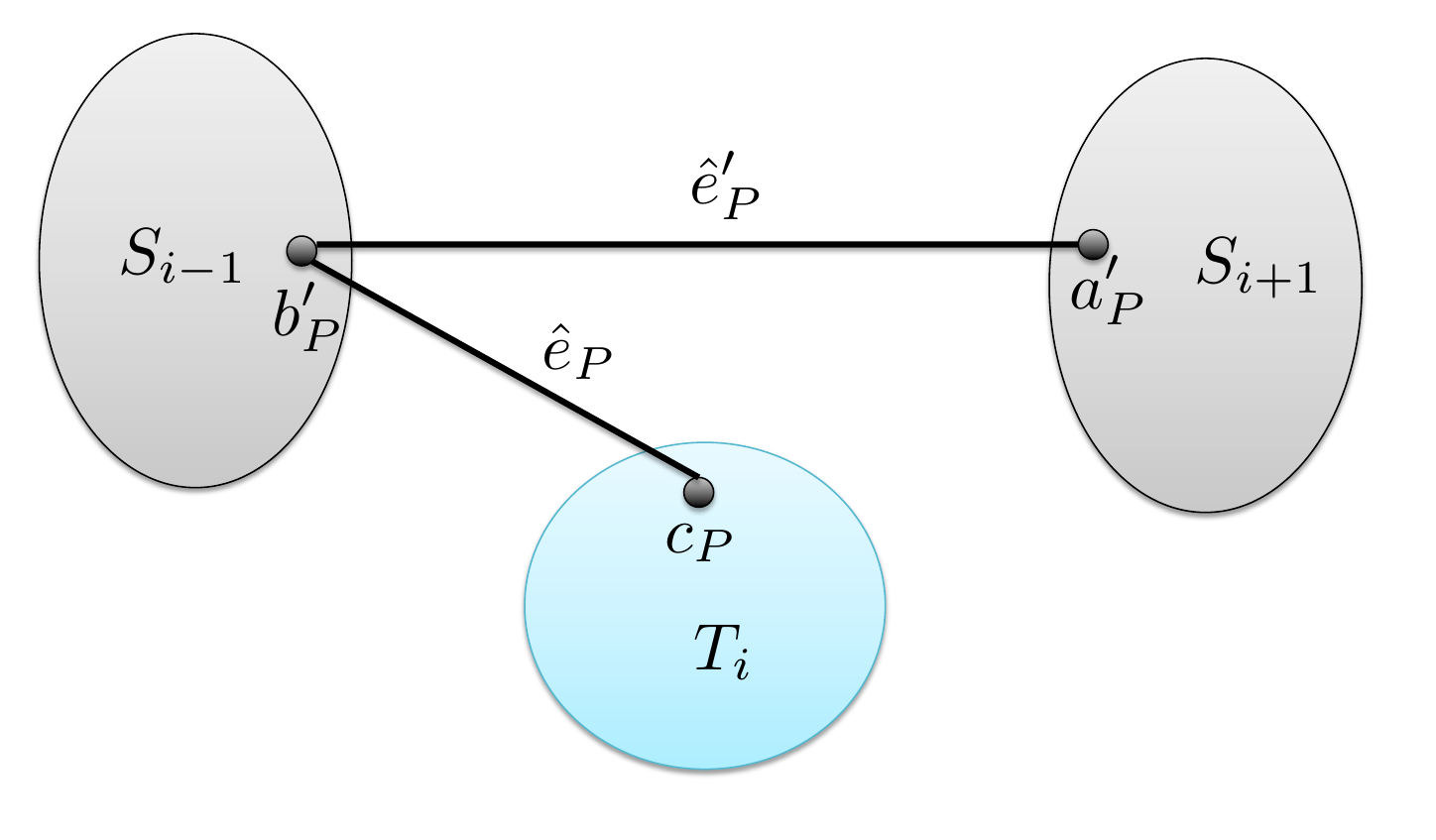}}\label{fig: exp-after contraction}}
\caption{The contractions of the edges on paths $P$ and $Q_P$. \label{fig: exp-edge contractions}}
\end{figure}

Consider some index $i\in \iseteven$, and the sets $\pset'_i,\qset_i$ of paths computed in Step 2. Let $P\in \pset'_i$ be any such path, and assume that it connects a vertex $a_P\in A_i$ to a vertex $b_P\in B_i$. Let $v_P\in P$ be the endpoint of $Q_P$ lying on $P$, and let $c_P$ be its other endpoint. Finally, let $e_P$ be the edge of $\mset_{i-1}$ incident to $a_P$ and let $b'_P\in B_{i-1}$ be its other endpoint. Similarly, if $i\neq 12$, let $e'_P$ be the edge of $\mset_i$ incident to $b_P$, and let $a'_P\in A_{i+1}$ be its other endpoint (see \Cref{fig: exp-before contraction}).

We contract the edge $e_P$ and all edges lying on the sub-path of $P$ between $a_P$ and $v_P$, so that $v_P$ and $b'_P$ merge. The resulting vertex is denoted by $b'_P$. We also suppress all inner vertices on the path $Q_P$, obtaining an edge $\hat e_P$, connecting $b'_P$ to $c_P$. Finally, if $i\neq 12$, then we contract all edges on the sub-path of $P$ between $v_P$ and $b_P$, obtaining an edge $\hat e'_P=(b_P,a'_P)$. We let $\hat E_i=\set{\hat e_P\mid P\in \pset'_i}$ and we let $\hat E_i'=\set{\hat e_P'\mid P\in \pset'_i}$ be the sets of these newly defined edges. 
Notice that the edges of $\hat E_i$ connect a subset of $\hat w$ vertices of $B_{i-1}$ (that we denote by $\hat B_{i-1}$) to a subset of $\hat w$ vertices of $T_{i}$ (that we denote by $\hat C_{i})$, and for $i\neq 12$,  the edges of $\hat E_i'$  connect every vertex of $\hat B_{i-1}$ to some vertex of $A_{i+1}$; we denote the set of endpoints of these edges that lie in $A_{i+1}$ by $\hat A_{i+1}$.

Once we perform this procedure for every path $P\in \pset'_i$, for all $i\in \iseteven$, we delete from the resulting graph all edges and vertices except those lying in graphs $S_i$ for $i\in \isetodd$, graphs $T_i$ for $i\in \iseteven$, and the edges in $\hat E_i\cup \hat E_i'$ for $i\in \iseteven$. The resulting graph, denoted by $G'$, is a minor of $G$, and it is easy to verify that its maximum vertex degree is at most $d+1$. 

We now define a \poefull System
$\Pi=(\tilde \Sigma,\tilde \mset, \tilde A_1,\tilde B_6, \tilde \tset,\tilde \mset')$, where the clusters of $\tilde \Sigma $ are denoted by $\tilde S_1,\ldots,\tilde S_6$; for each $1\leq i\leq 6$ the corresponding sets $A_i,B_i,C_i$ of vertices are denoted by $\tilde A_i$, $\tilde B_i$ and $\tilde C_i$ respectively; the matching $\mset'_i$ is denoted by $\tmset'_i$ and the expander $T_i$ is denoted by $\tilde T_i$. For all $1\leq i<6$, we also denote the matching $\mset_i$ by $\tmset_i$.
    
For each $1\leq i\leq 6$, we let the cluster $\tilde S_i$ of $\tilde \Sigma$ be $S_{2i-1}$, and we let the expander $\tilde T_i$ be $T_{2i}$. We also set $C_i=\hat C_{2i}$, and $\tmset'_i=\hat E_{2i}$.
If $i>1$, then we let $\tilde A_i=\hat A_{2i-1}$, and we let $\tilde A_1$ be any subset of $\hat w$ vertices of $A_1$. Similarly, if  $i<6$, then we let $\tilde B_i=\hat B_{2i-1}$, and we let $\tilde B_6$ be any subset of $\hat w$ vertices of $B_6$. Finally, for $i<6$, we let $\tilde \mset_i=\hat E_{2i}'$.
 It is immediate to verify that we have obtained a \poefull  System of width $\hat w$ and expansion $\hat \alpha$, and a model of $G_{\Pi}$ in $G'$. It is now immediate  to obtain a model of $G_{\Pi}$ in $G_{\Sigma}$.
This completes the proof of \Cref{thm: exp-pos to poe}.

    \section{Efficiently Finding Models of all Not-So-Large Graphs in Expanders} \label{sec: exp-constructive proof}

The goal of this section is to provide the proof of \Cref{thm: exp-constructive main}. Notice that \Cref{thm: exp-constructive main} provides slightly weaker dependence on $n$ in the minor size than \Cref{thm: exp-general main}, but it has several advantages: its proof is much simpler, the algorithm's running time is polynomial in $n,d$ and $\alpha$, and it provides a better dependence on $\alpha$ and $d$ in the bound on the minor size.
Our algorithm also has an additional useful property:  if it fails to find the required model, then with high probability it certifies that the input graph is not an $\alpha$-expander by exhibiting a cut of sparsity less than $\alpha$.

\newcommand{\consbound}{\floor{\frac{n}{\tilde c^*\log^2 n}\cdot \frac{\alpha^{3}}{d^{5}}}}

Let $G=(V,E)$ be the given $n$-vertex $\alpha$-expander with maximum vertex degree at most $d$.
As in the proof of \Cref{thm: exp-general main}, given a graph $H$ with $n'$ vertices and $m'$ edges,
we can construct another graph $H'$, whose maximum vertex-degree is at most $3$ and $|V(H')| \leq n' + 2m' \leq 2 \consbound$,
such that $H$ is a minor of $H'$.
It is now enough to provide an efficient algorithm that computes a model of $H'$ in $G$.
For convenience of notation, we denote $
H'$ by $H=(U,F)$, and we denote $U= \set{u_1, \ldots, u_{|U|}}$.
We can assume that $n>c_0$ for a large enough constant $c_0$ by appropriately setting the constant $\tilde c^*$, as otherwise it is enough to show that every graph of size $1$ is a minor of $G$, which is trivial.

Our algorithm consists of a number of iterations.
We say that a partition $(V', V'')$ of $V$ is \emph{good} iff $|V'|, |V''| \geq n/(4d)$; and $G[V'], G[V'']$ are both connected graphs.
We start with an arbitrary good partition $(V_1, V_2)$ of $V$, obtained by using the algorithm from \Cref{obs: exp-decompose by spanning tree} with $r=2$.
Assume without loss of generality that $|V_1| \geq |V_2|$.
We now try to compute a model of $H$ in $G$, by first embedding the vertices of $H$ into connected sub-graphs of $G[V_2]$, and then routing the edges of $H$ in $G[V_1]$.
We show an efficient algorithm, that with high probability returns one of the following:
\begin{itemize}
    \item either a good partition $(V'_1, V'_2)$ such that $|E(V'_1, V'_2)| < |E(V_1, V_2)|$
            (in this case, we proceed to the next iteration); or
    \item a model of $H$ in $G$ (in this case, we terminate the algorithm and return the model).
\end{itemize}
Clearly, we terminate after $|E|$ iterations, succeeding with high probability.
We now describe a single iteration in detail.
Recall that we are given a good partition $(V_1, V_2)$ of $V$ with $|V_1|\geq |V_2|$.
Since $G$ is an $\alpha$-expander, we have $|E(V_1, V_2)| \geq \alpha n/(4d)$ (note that, if this is not the case, we have found a cut $(V_1, V_2)$ of sparsity less than $\alpha$).
Since the maximum vertex-degree in $G$ is bounded by $d$, we can efficiently find a matching $\mset\subseteq E(V_1,V_2)$ of cardinality at least $\alpha n/(8d^2)$.
We denote the endpoints of the edges in $\mset$ lying in $V_1$ and $V_2$ by $Z$ and $Z'$, respectively.
Let $\rho := 3 \cdot \ceil{4 c d^2 \log^2 n/\alpha^2}$,
where $c$ is the constant from \Cref{lem: exp-can route large disjoint subsets in expanders}.

    Recall that $U$ is the set of vertices in the graph $H$.
We apply \Cref{obs: exp-decompose by spanning tree} to the graph $G[V_2]$, together with $R = Z'$ and parameter $r = |U|$, 
to obtain a collection $\wset=\set{W_1,\ldots,W_{|U|}}$ of disjoint connected subgraphs of $G[V_2]$,
such that for all $1\leq i\leq |U|$,
\[ |V(W_i)\cap Z'|\geq \floor{\frac{|Z'|}{d|U|}}
    \geq \floor{\frac{\alpha n}{8d^3|U|}}
    \geq \floor{\frac{\alpha n}{8d^3} \cdot \frac{\tilde c^* d^5\log^2n}{2n\alpha^3}}
    = \floor{\frac{ \tilde c^* d^2\log^2n}{16\alpha^2}}
\]

Here, we have used the fact that $|U|\leq 2 \consbound$.
By appropriately setting the constant $\tilde c^* $ in the bound on $|U|$, we can ensure that for all $1\leq i\leq |U|$, $|V(W_i)\cap Z'|\geq 3\rho$.

Recall that we are given a graph $H = (U,F)$ with maximum vertex-degree $3$ and that we have denoted $U=\set{u_1,\ldots,u_{|U|}}$.
For $1\leq i\leq |U|$, we think of the graph $W_i$ as representing the vertex $u_i$ of $H$.
For each $1\leq i\leq |U|$, and for each edge $e\in \delta_{H}(u_i)$,
we select an arbitrary subset $Z'_i(e) \subseteq V(W_i) \cap Z'$ of $\rho$ vertices,
such that all resulting sets $\set{Z'_i(e)\mid e\in \delta_H(u_i)}$ of vertices are mutually disjoint.
Let $E_i(e)\subseteq \mset$ be the subset of edges of $\mset$ that have an endpoint in $Z'_i(e)$, so $|E_i(e)|=\rho$.
We let $Z_i(e)$ be the set of vertices of $Z$ that serve as endpoints of the edges in $E_i(e)$.
Notice that all resulting sets $\set{Z_i(e)\mid 1\leq i\leq |U|, e\in \delta_H(u_i)}$ are mutually disjoint, and each of them contains $\rho'$ vertices. 

We apply the algorithm of \Cref{lem: exp-can route large disjoint subsets in expanders} to the graph $G[V_1]$,
together with the parameter $\alpha/2$ and the family
$\set{Z_i(e)\mid 1\leq i\leq |U|, e\in \delta_H(u_i)}$ of vertex subsets, that we order appropriately.

\paragraph{Case 1. The algorithm returns a cut.}
In this case, we obtain a cut $(X, Y)$ in $G[V_1]$ of sparsity less than $\alpha/2$ by computing a good partition $(V'_1, V'_2)$ of $V$ with $|E(V'_1, V'_2)|~<~|E(V_1, V_2)|$.
We need the following simple observation, that we prove after completing the proof of \Cref{thm: exp-constructive main} assuming it.

\begin{observation} \label{obs: exp-connected sparsest cut}
    There is an efficient algorithm, that given a connected graph $G = (V,E)$ and a cut $(X,Y)$ in $G$, produces a cut $(X^*, Y^*)$, whose sparsity is less than or equal to that of $(X,Y)$,
    such that both $G[X^*]$ and  $G[Y^*]$ are connected.
\end{observation}

We apply \Cref{obs: exp-connected sparsest cut} to graph $G[V_1]$ and cut $(X,Y)$, obtaining a new cut $(X^*,Y^*)$ of sparsity less than $\alpha/2$,
such that both $G[X^*]$ and $G[Y^*]$ are connected.
For convenience, we denote the cut $(X^*,Y^*)$ by $(X,Y)$, and we assume without loss of generality that $|Y|\leq |X|$. 
Notice that $|Y| \leq |V_1|/2 \leq |V|/2$.
Since $G$ is an $\alpha$-expander, $|\delta_G(Y)| \geq \alpha |Y|$ (note that, if this is not the case, then have found a cut $(Y, V \backslash Y)$ of sparsity less than $\alpha$.).

Since $\delta_G(Y)=E(X,Y)\cup E(Y,V_2)$, we get that $|E(Y,V_2)|\geq \alpha |Y|/2$, and $|E(X,Y)|<|E(Y,V_2)|$.
In particular, $E(Y,V_2)\neq \emptyset$.
We now define a new cut $(V'_1,V'_2)$ of $G$, where $V'_2 = V_2 \cup Y$ and $V'_1 = X$.
We claim that $(V'_1,V'_2)$ is a good partition of $V(G)$.
It is immediate to verify that $|V'_1|,|V'_2|\geq n/(4d)$, and that $G[V'_1]=G[X]$ is connected.
Moreover, since $G[Y]$ is connected and $E(Y,V_2)\neq \emptyset$, $G[V'_2]=G[V_2 \cup Y]$ is also connected.
Lastly, we claim that $|E(V'_1,V'_2)| < |E(V_1, V_2)|$.
Indeed, since $|E(X, Y)| < |E(V_2, Y)|$:

\[|E(V'_1, V'_2)| = |E(V_1, V_2)| - |E(V_2, Y)| + |E(Y,X)| < |E(V_1,V_2)|.\]

Therefore, we have computed a good partition $(V'_1, V'_2)$ of $V(G)$, with $|E(V'_1, V'_2)| < |E(V_1, V_2)|$ as required.

\paragraph{Case 2. The algorithm returns paths.}
In this case, we have obtained, for every edge $e=(u_i,u_j)\in F$, 
a path $Q(e)$ in $G[V_1]$, connecting a vertex of $Z_i(e)$ to a vertex of $Z_j(e)$,
such that, with high probability, the paths in $\set{Q(e)\mid e\in F}$ are mutually disjoint.
If the paths in $\set{Q(e)\mid e\in F}$ are not mutually disjoint, the algorithm fails.
We assume from now on that the paths in $\set{Q(e)\mid e\in F}$ are mutually disjoint.
We extend each path $Q(e)$ to include the two edges of $\mset$ that are incident to its endpoints,
so that $Q(e)$ now connects a vertex of $Z'_i(e)$ to a vertex of $Z'_j(e)$. 
 
We are now ready to define the model of $H$ in $G$.
For every $1 \leq i \leq |U|$, we let $f(u_i)=W_i$, and for every edge $e\in F$, we let $f(e)=Q(e)$.
It is immediate to verify that this mapping indeed defines a valid model of $H$ in $G$.
This completes the proof of \Cref{thm: exp-constructive main} assuming \Cref{obs: exp-connected sparsest cut} that we prove next.

\proofof{\Cref{obs: exp-connected sparsest cut}}
    We start with the cut $(X,Y)$ and perform a number of iterations. In every iteration, we modify the cut $(X,Y)$ so that the number of connected components in $G\setminus E(X,Y)$ strictly decreases, while ensuring that the cut sparsity does not increase.
    We now describe the execution of an iteration. Let $(X,Y)$ be the current cut. Let $\cset_X$ and $\cset_Y$ be the sets of all connected components of $G[X]$ and $G[Y]$ respectively. If $|\cset_X| = |\cset_Y| = 1$, then we return the cut $(X,Y)$, and terminate the algorithm. We assume from now on that this is not the case.

    Assume w.l.o.g. that $|X|\leq |Y|$. 
    Let $\rho_X := \frac{|E(X,Y)|}{|X|}$ and $\rho_Y := \frac{|E(X,Y)|}{|Y|}$.
    We consider the following two cases.

    \paragraph{Case 1: } The first case happens when $|\cset_X| > 1$.
        Recall that $|E(X,Y)| = \rho_X |X|$.
        Thus, there is a connected component $C \in \cset_X$ such that $|E(C, Y)| \geq \rho_X |C|$.
        Consider a new partition $(X', Y')$, obtained by setting $X' = X \backslash C$ and $Y' = Y \cup C$.
        Notice that the number of connected components in $G\setminus E(X',Y')$ decreases by at least one.
        The sparsity of the new cut is:
            \[ \frac{|E(X', Y')|}{\min\set{|X'|, |Y'|}} = \frac{|E(X', Y')|}{|X'|} = \frac{|E(X,Y)| - |E(C,Y)|}{|X| - |C|} \leq \frac{\rho_X |X| - \rho_X |C|}{|X| - |C|} = \rho_X.\]

    \paragraph{Case 2:} If Case 1 does not happen, then $|\cset_Y| > 1$ must hold.
        As before, there is a connected component $C \in \cset_Y$ such that $|E(C,Y)| \geq \rho_Y |C|$.
        Consider the new partition $(X', Y')$ by setting $X' = X \cup C$ and $Y' = Y \backslash C$.
        Notice that the number of connected components in $G\setminus E(X',Y')$ decreases by at least one.
        In order to bound the sparsity of the new cut, we consider two cases.
        If $|X'| \geq |Y'|$, then the sparsity of the new cut is 
            \[  \frac{|E(X', Y')|}{|Y'|} = \frac{|E(X,Y)| - |E(C,Y)|}{|Y| - |C|} \leq \frac{\rho_Y|Y| - \rho_Y|C|}{|Y| - |C|} = \rho_Y \leq \rho_X.\]
        Otherwise, the sparsity of the new cut is 
            \[ \frac{|E(X', Y')|}{|X'|} = \frac{|E(X,Y)| - |E(C,Y)|}{|X|+|C|} < \frac{|E(X,Y)|}{|X|} = \rho_X.\]
            
    It is immediate to verify that the algorithm is efficient, and that it produces the cut $(X^*,Y^*)$ with the required properties.
\endproofof

    \chapter{Longest Increasing Subsequence and Non-Crossing Matching}  \label{chap: ncm}

\toggletrue{ncm}
\newtoggle{ncm-algo-to-lis}
\newtoggle{partitioning}
\newtoggle{alpha-approx-lis}
\newcommand{\LISNCMgammabound}{2-3\eps}

\section{Introduction}
In this work, we initiate the study of bipartite {\em non-crossing matching} (\NCM) problem in the framework that we refer to as `hybrid model'.
In the \NCM problem, we are given a bipartite graph $G = (L, R, E)$ where $L$ and $R$ are ordered sets of vertices and $E \subseteq L \times R$ is the set of edges.
A matching $M$ in $G$ is \emph{non-crossing}, if, for every pair of edges $(u,v), (u', v') \in M$, if $u < u'$, then  $v < v'$ holds.
Our choice of hybrid model is motivated by its close relationship with the well-studied {\em longest increasing subsequence} (\LIS) problem in the streaming model that we establish.
In the hybrid model, we assume that the access to the graph $G = (L, R, E)$ is provided to an algorithm as follows.
We are given (offline) access to vertices $L$ and $R$, while the edges of $G$ are revealed over the course of $|L|$ rounds.
In the $i^{th}$ round, the algorithm is revealed a superset of edges incident on the $i^{th}$ vertex of $L$.
It then selects a subset of these `advice-edges' to \emph{query}, and then receives the set of `real-edges' among its queried edge-slots.
At the end of processing the last vertex of $L$, the algorithm reports its estimate on the cardinality of the maximum non-crossing matching in $G$.
The aim in this model is to minimize the number of queries per vertex, as a fraction of the number of advice-edges incident on it, while disregarding the space complexity.
Our choice of this access model for the \NCM problem, is motivated precisely by the fact that it is the query complexity in this model that sheds light on space complexity of the \LIS problem in the streaming model.

We note that the initial study of the \NCM problem in the classical setting was motivated by its applications in VLSI design~\cite{ncm-first}.
Variants of the \NCM problem have also been studied in computational geometry, where the (potentially non-bipartite) graph is embedded into a plane, and the goal is to find its non-crossing matching maximizing certain objective functions.

A closely related model to the hybrid model is the query model, where the algorithm may directly query the adjacency matrix or adjacency list of the input graph.
The problem of estimating the size of a (standard) matching has been well-studied in the query model, and it is known that $\tilde{O}(n)$ queries suffice to achieve an $O(1)$-approximation of maximum matching size whenever the maximum matching size is $\tilde{\Omega}(n)$, where $n$ is the number of vertices in $G$~\cite{onak2012near,kapralov2020space,Behnezhad21}.
These algorithms implicitly rely on the {\em robustness} of matchings: the decision to include any  edge in a matching solution can only lower the optimum solution value by at most one.
In fact, even $O(1)$-approximation to the size of maximal matching leads to $O(1)$-approximation of the maximum matching size.
In contrast, inclusion of a single edge can potentially reduce the optimal non-crossing matching solution value by a factor of $\Omega(n)$, showing a factor $\Omega(n)$ gap between optimal and maximal non-crossing matching sizes.
This lack of robustness makes the task of designing a sublinear-query algorithm for \NCM size estimation much harder, even in query model.

Before describing the relation between the space complexity of the \LIS problem in the streaming model and the query complexity of the \NCM problem in the hybrid model, we summarize known result for the former.
Liben-Nowell, Van, and Zhu~\cite{Liben-NowellVZ06} showed that an exact computation of the \LIS length in the streaming model, even by a randomized algorithm, requires $\Omega(\sqrt{N})$ space where $N$ is the stream length.
This was soon afterwards strengthened to an $\Omega(N)$ space lower bound for an exact computation of \LIS length even by a randomized algorithm in independent works by Gopalan {\em et al.}~\cite{sqrt-n-det-lis-soda}, and by Sun and Woodruff~\cite{SunW07}.
We note that the $\Omega(N)$ space lower bound holds even for instances where the input is a permutation of $[1..N]$.

The focus then shifted to the design of space-efficient algorithms for approximating \LIS length.
The first non-trivial result in this direction was obtained by Gopalan {\em et al.}~\cite{sqrt-n-det-lis-soda} who gave an $O(\sqrt{N/\eps})$ space one-pass deterministic streaming algorithm for approximating the \LIS length to within a $(1+\eps)$-factor for any $\eps > 0$.
Moreover, they conjectured that the space-complexity of their algorithm is optimal.
This conjecture was shown to be true independently by G\'{a}l and Gopalan~\cite{GalG07}, and by Erg\"{u}n and Jowhari~\cite{ErgunJ08} who established an essentially matching space lower bound for one-pass deterministic streaming algorithms.
We note that the result of G\'{a}l and Gopalan~\cite{GalG07} also implies a deterministic space lower bound of $\Omega \left( \sqrt{N/(\alpha-1)} \right)$ for any $\alpha > 1$.
This bound is tight, since the deterministic algorithm of \cite{sqrt-n-det-lis-soda} can be easily adapted to provide $\alpha$-approximation to \LIS length for any $\alpha > 1$ in space $O \left( \sqrt{N/(\alpha-1)} \right)$.
Together, these results give a complete understanding of the tradeoff between space complexity and approximation factor for deterministic one-pass streaming algorithms for estimating \LIS length.

The corresponding picture for randomized algorithms is far from being resolved though.
Saks and Seshadhri~\cite{SaksS13} gave an additive approximation error randomized one-pass streaming algorithm, that, for any $\delta > 0$, achieves a $(\frac{\delta N}{1 + \delta})$ additive approximation to \LIS length in $O\left( (\log^2 N)/\delta \right)$ space.
In the regime when the \LIS length is $\omega\left( \sqrt{N} \log^2 N \right)$, this bound implies a multiplicative approximation algorithm that has better space complexity 
than~\cite{sqrt-n-det-lis-soda}.
In particular, for any sequence $S$, setting $\delta = \eps \frac{\optlis(S)}{N}$ gives a $(1+\eps)$-approximation to \LIS length in ${O}\left (\frac{N \log^2 N}{\eps \optlis(S)}\right )$ space.
On the other hand, \cite{Liben-NowellVZ06} showed that \LIS length can be computed exactly in ${O}\left( \optlis(S) \right)$ space.
However, the regime where $\optlis(S) \approx \sqrt{N}$ remains a bottleneck for the current state-of-the-art for streaming algorithms for \LIS\ -- all known algorithms require ${\Omega}\left( N^{1/2-o(1)} \right)$ space in this regime for even an $N^{o(1)}$-approximation to the \LIS length. 
At the same time, known lower bound results do not even rule out the possibility that a poly-logarithmic space randomized algorithm may achieve an $\tilde{O}(1)$-approximation to \LIS length~\cite{SunW07}.
It is worth emphasizing that the ${\Omega}\left( N^{1/2-o(1)} \right)$ space requirements for current known algorithms holds even when the input sequence is a permutation of $[1..N]$, which is the setting of the \LIS problem that we focus on.

We show that the goals of understanding the randomized space complexity of \LIS and \NCM problems are closely intertwined, and under some natural conditions, progress on the current state-of-the-art for one problem will have significant implications for the other. 
Note that a trivial exact algorithm for the \NCM problem in the hybrid model is to query all the advice-edges. 
We show that an \NCM algorithm in the hybrid model with a slightly better query complexity than this naive algorithm and even a moderately high factor of approximation implies an $N^{o(1)}$-approximation streaming algorithm for \LIS with space complexity bounded by $N^{1/2-\epsilon}$, where $N$ is the \LIS stream length and $\eps$ is some absolute constant.
Specifically, we show a low space complexity randomized \LIS algorithm in the streaming model, that, without reduction to the \NCM problem, achieves factor $N^{o(1)}$-approximation with low space complexity, except for input instances that satisfy some stringent technical properties.
For the latter case, we reduce \LIS instances satisfying these technical properties to the \NCM problem in the hybrid model, where the input instance $G = (L, R, E_\advice, E)$ satisfies $|L| = |R|$ and $\optncm(G)$ is relatively large.

We also show a converse connection, namely, that the existence of a $(1+\eps)$-approximation randomized algorithm for the \LIS problem in streaming model with space complexity $N^{1/2-\delta}$ would imply a randomized algorithm for the \NCM problem in the hybrid model with per-vertex query complexity of $(d(G) |L|)^{1/2-\delta}$ that achieves ${O}(1)$-approximation in the regime where $|L| = |R|$ and the optimal solution value is close to $|L|$, where the input instance is $G~=~(L, R, E_\advice, E)$ and $d(G)$ is the maximum number of edge-slots of $E_{\advice}$ that are incident to a single vertex of $L$.
This converse connection, however, requires a technical restriction, namely, that the streaming \LIS algorithm is {\em comparison-based}.
To our knowledge, all known \LIS streaming algorithms are comparison-based.
Thus, if one can rule out the existence of an algorithm for \NCM in the hybrid model with this bound on its per-vertex query complexity and approximation guarantee, even in the above-mentioned regime, it would also eliminate the possibility of the existence of an algorithm for \LIS problem in the streaming model with space complexity significantly lower than $\sqrt{N}$.
As noted earlier, our current state of knowledge for \LIS neither rules out such a result (for a randomized algorithm) nor gives an algorithm to achieve this.
Any progress in either direction would fundamentally improve our understanding of the \LIS problem.

In the next subsection, we formally define our models and then give a technical overview of our main results.

\subsection{Our Results and Informal Overview of Techniques}
All the algorithms that we consider in this work are randomized algorithms succeeding with probability at least $2/3$ unless stated otherwise.
Before presenting our results, we define the Longest Increasing Subsequence (\LIS) and Non-Crossing Matching (\NCM) problems, and various models of computation that we analyze.

\paragraph{Longest Increasing Subsequence.}
The Longest Increasing Subsequence (\LIS) problem takes as input a sequence $S = (a_1, \ldots, a_N)$ of $N$ elements from an ordered universe $\uset = \set{1, \ldots, M}$.
We say that a subsequence $S' = (a_{i_1}, \ldots, a_{i_k})$ is an \emph{increasing subsequence} of $S$ iff $1 \leq i_1 < \ldots < i_k \leq N$ and $a_{i_1} < \ldots < a_{i_k}$.
Let $\optlis(S)$ be the length of the longest increasing subsequence of $S$, and denote it by $\optlis$ or $\opt$ when the sequence $S$ is clear from the context.
Given such a sequence $S$, the objective of the \LIS problem is to produce an estimate $k'$ of $\optlis(S)$.
We say that a randomized algorithm $\alg$ is an $\alpha$-approximation algorithm for \LIS, if, given as input a sequence $S$,  with probability at least $2/3$ it produces an estimate $k'$ such that $\optlis(S)/\alpha \leq k' \leq \optlis(S)$.
For some $\alpha' \geq \beta'$, we say that $\alg$ solves \emph{$(\alpha', \beta')$-gap \LIS} problem iff it distinguishes between the two cases by reporting \emph{yes} with probability at least $2/3$ when $\optlis(S) \geq \alpha'$ and reporting \emph{no} with probability at least $2/3$ when $\optlis(S) < \beta'$.
In this work, unless stated otherwise, we will restrict ourselves to the case where the input sequence $S$ is a permutation of the universe $\uset = \set{1, \ldots, N}$.

\paragraph{\LIS in the Streaming Model.}
We assume that the universe $\uset = \set{1, \ldots, M}$ is known to us in advance, and each of its element can be stored in a single unit of space.
In the standard one-pass streaming model, an algorithm with limited memory is presented the elements of $S = (a_1, \ldots, a_N)$ one by one, in this sequential order.
At the end of the stream, it has to output an estimate $k'$ of $\optlis(S)$.
We do not place any constraints on the operations it is allowed to perform.

\paragraph{Non-Crossing Matching Problem.}
The input to the Non-Crossing Matching (\NCM) problem is a bipartite graph $G=(L, R, E)$ where $L$ and $R$ are ordered sets of vertices.
We say that a matching $M$ in $G$ is \emph{non-crossing}, if, for every pair of edges $(u,v), (u', v') \in M$, where $u < u'$, it holds that $v < v'$.
Given such an instance $G$, the goal in the \NCM problem is to produce an estimate $k'$ of the size of the largest non-crossing matching, denoted by $\optncm(G)$ or $\opt$ when $G$ is clear from context. 
We say that a randomized algorithm $\alg$ is an $\alpha$-approximation algorithm  for \NCM if with probability at least $2/3$ it produces an estimate $k'$ on the size of the largest non-crossing matching in $G$ such that $\optncm(G)/\alpha\leq k' \leq \optncm(G)$.
For some $\alpha' \geq \beta'$,  we say that $\alg$ solves \emph{$(\alpha', \beta')$-gap \NCM} problem iff it distinguishes between the two cases by reporting \emph{yes} with probability at least $2/3$ when $\optncm(G) \geq \alpha'$ and reporting \emph{no} with probability at least $2/3$ when $\optncm(G) < \beta'$.

\paragraph{\NCM Problem in the Hybrid Model.}
In the hybrid model, an algorithm $\alg$ is given access to an \NCM problem instance $G = (L, R, E)$ along with a superset $E_{\advice} \supseteq E$ of edge-slots, that we call `advice' edges.
We will denote the resulting instance by $G = (L, R, E_\advice, E)$ and say that its \emph{size} is $|G| := |L| \cdot |R|$.
We denote by $d(G)$, the maximum number of edge-slots of $E_{\advice}$ incident to a single vertex of $L$.
We assume that the vertex-sets $L$, $R$ along with $d(G)$ are known to $\alg$ beforehand, but the sets $E$ and $E_\advice$ of edge-slots are unknown.
The algorithm $\alg$ is revealed the sets $E$ and $E_\advice$ over the course of $|L|$ rounds.
In the $i^{th}$ round, the algorithm $\alg$ receives the set of the advice edges $E_\advice(u_i)$ incident on $u_i$, where $u_i$ is the $i^{th}$ vertex of $L$.
It then selects a subset $E'_\advice(u_i)$ of these edge-slots to query and receives the set $E'_\advice(u_i) \cap E$ of `real edges' among its queried edge-slots.
At the end of processing the last vertex of $L$, the algorithm outputs its estimate $k'$ of $\optncm(G)$.
We say that the algorithm $\alg$ has a per-vertex query complexity $q$ iff for all $u \in L$, $|E'_\advice(u)| \leq q$.
In the hybrid model, the goal is to optimize the per-vertex query complexity, and we ignore both, the space complexity and the time complexity of our algorithm.

\paragraph{From Query-Efficient \NCM to Space-Efficient \LIS.}
We start by discussing our algorithm for the \LIS problem in the Streaming model, which uses as a blackbox, the algorithm for the \NCM problem in the hybrid model.
To establish this relation between \NCM and \LIS problems, we prove the following theorem:

\begin{theorem}\label{thm: LIS to NCM final}
    Suppose there exists a constant $\delta > 0$ and an algorithm $\algncm$ for the $(\gamma |L|, \gamma |L|/|G|^{o(1)})$-gap \NCM problem in the hybrid model
    for the special case where
    the input instance $G~=~(L, R, E_\advice, E)$ has $|L| = |R|$ and $d(G) \geq \ncmdboundasG$, for all $\ncmgammaboundasd \leq \gamma \leq 1$,
    with per-vertex query complexity $\left(d(G)\right)^{1 - \delta}$.
    Then there is a constant $\eps  = \eps(\delta) > 0$ and a factor $N^{o(1)}$-approximation algorithm $\alglis$ for the \LIS problem in the streaming model with space complexity $N^{1/2 - \eps}$, where $N$ is the size of the input stream.
\end{theorem}

Before we give a brief overview of our techniques behind \Cref{thm: LIS to NCM final}, we discuss some of its implications.
For the ease of notation, assume in this section that $N$ is a large enough integral power of $2$.
Consider an instance $G = (L, R, E_\advice, E)$ for the \NCM Problem in the hybrid model, and let $d = d(G)$ be the maximum number of edge-slots of $E_{\advice}$ incident to a single vertex of $L$.
Note that there is a trivial $1$-approximation algorithm that performs $d$ queries per vertex.
Indeed, we can query all the advice-edges $E_\advice$ and incur the per-vertex query complexity of $d$.
Suppose there exists an algorithm that improves this bound by a relatively modest factor of $d^{\delta}$, for some constant $\delta > 0$, even at the cost of a `large' factor of $(|G|)^{o(1)}$ in the approximation guarantee in the special case where $|L| = |R|$ and $d \geq |G|^{10^{-9}}$.
Consider a special case where the input instance $G = (L, R, E_\advice, E)$ for the \NCM Problem in the hybrid model is guaranteed to have $\optncm(G) \geq d^{10^{-3}} \geq |G|^{10^{-12}}$.
Assume that there is an algorithm $\algncm$ in the hybrid model that solves $(\gamma |L|, \gamma |L|/|G|^{o(1)})$-gap \NCM problem with per-vertex query complexity $(d(G))^{1\delta}$, for all $\ncmgammaboundasd \leq \gamma \leq 1$ and some constant $\delta > 0$, where the input instance is $G$.
We then show that there are non-trivial consequences for the \LIS problem in the streaming model.
Specifically, we demonstrate that the long-standing space complexity upper bound barrier 
of $N^{1/2 - o(1)}$ for achieving $N^{o(1)}$-approximation algorithm
for the \LIS problem in the streaming model, where $N$ is the input stream length,
can be broken in this case.
\footnote{
    We note that for such an algorithm $\algncm$ to exist, $\delta \leq 10^{-3}$ must hold.
    Indeed, consider an input instance $G = (L, R, E_\advice, E)$ of the \NCM problem in the hybrid model where $|L| = |R|$ and $d(G) = |G|^{10^{-9}}$.
    It is immediate to verify that to distinguish between the cases where $E$ is a non-crossing matching of size $d^{10^{-3}}$ from the case where $E = \emptyset$, one must query at least $\Omega \left(|L| d / d^{10^{-3}} \right) = \Omega \left(|L| \cdot d^{(1 - 10^{-3})} \right)$ edge-slots of $E_\advice$, which translates to per-vertex query complexity of $\Omega \left( d^{(1 - 10^{-3})} \right)$.
}
As mentioned earlier, we do not put any restriction on the space complexity of $\algncm$ in the hybrid model.
To obtain this `space-agnostic' reduction, we use an intrinsic `self-reducibility' property of the \LIS problem that we outline below.

\paragraph{Self-reducibility Property and Hierarchical Decomposition.}
Consider an instance $S$ of the \LIS problem and a parameter $Z$ that is an integral power of $2$.
We assume that $S$ is a permutation of the range $H^* = (1, \ldots, N)$, where $N$ is known to us in advance, and is an integral power of $2$.
We prove the existence of two parameters, $X(Z)$ and $Y(Z)$, also integral powers of $2$, and an increasing subsequence $S^*$ of $S$ with size comparable to $\optlis(S)$, such that the following property holds. 
Notice that $S^*$ can also be naturally thought of as an increasing subsequence of the range $H^*$.
Consider the partition of $S$ into \emph{stream-blocks}, where each stream-block is a contiguous subsequence in $S$ of length exactly $X(Z)$.
Similarly, consider the partition of the range $H^*$ into \emph{range-blocks}, where each range-block is a contiguous subsequence in $H^*$ of length exactly $Y(Z)$.
Then each block of these partitions either contributes exactly $Z$ elements to $S^*$ or none at all.
Consider a pair $(B, B')$ of blocks, where $B$ is a stream-block and $B'$ is a range-block.
From the above discussion, for each such pair $(B, B')$ of blocks, either $S^*$ contains exactly $Z$ elements of $B$ with values in $B'$, or there are zero such elements.
Consider a pair $(B, B')$ that contributes elements to $S^*$.
We refer to such pairs as \emph{yes-pairs}.
Our algorithm ensures that each yes-pair $(B, B')$ is statistically similar to every other yes-pair in terms of the number of their common elements ($B \cap B'$) and the `distance' at which the elements of $S^*$ appear in the resulting common subsequence $B \cap B'$.
This `regularity property' will prove advantageous in subsequent analysis.

We can think of these partitions as naturally giving rise to a \NCM instance $G = (L, R, E_\advice, E)$ in the hybrid model.
For each stream-block of the partition of the stream $S$, there is a unique vertex in $L$.
Similarly, for each range-block of the partition of the range $H^*$, there is a unique vertex in $R$.
For each pair of blocks that are potential candidates for contributing elements to $S^*$, by the virtue of having the `correct' number of shared elements; we add the corresponding pair of vertices to the set $E_\advice$ of advice-edges.
If these blocks indeed share an increasing subsequence of size comparable to $Z$ (which may contain elements that do not lie in $S^*$), we add the corresponding pair of vertices to the set $E$ of edges.
Here, we are glossing over the simulation of this reduction while accessing $S$ in the streaming model with limited memory.

Consider a pair $(B, B')$ of a stream-block and a range-block.
We view each such pair $(B, B')$ as a fresh \LIS problem instance, where the input stream is the sequence $B \cap B'$ of elements of $B$ with their values in the range-block $B'$.
The range of this new \LIS problem instance is $B'$, which is in turn, a subset of the range $H^*$ of the original input stream $S$.
Moreover, if $(B, B')$ is a yes-pair, we have $\optlis(B \cap B') \geq Z$.
On the other hand, if it is a no-pair, we do not have any constraint on $\optlis(B \cap B')$.
Notice that we can recursively apply our partitioning process to these resulting instances.
We refer to this property as the `self-reducibility property,' and the resulting hierarchical family of partitions as a `hierarchical decomposition.'

Naively, this is an existential result: that there exists a large increasing subsequence $S^*$ of $S$ and a hierarchical decomposition that obeys the regularity properties outlined above w.r.t. the subsequence $S^*$.
As mentioned earlier, we show that these regularity properties, and hence, the relevant parameters of the hierarchical decomposition, depend solely on the `level' at which they appear, and it suffices to consider a relatively small set of such parameters.
We show that we can efficiently `guess' such a hierarchical decomposition (but not the increasing subsequence $S^*$), even before processing the input sequence $S$, with a small enough overhead in space complexity.
We can thus assume from now on, that the algorithm is given a hierarchical decomposition for which there is a large enough increasing subsequence $S^*$ of $S$ that satisfies the above-mentioned regularity properties.

\paragraph*{Space-efficient \LIS Algorithm.}
We note that existing algorithms can be combined in order to achieve a constant-factor approximation for \LIS in space ${N^{1/2-\eps + o(1)}}$, except for the case where $N^{1/2-\eps}\leq \optlis(S)\leq N^{1/2+\eps}$.
We thus focus here on the interesting regime where such a result is not known, namely, 
where $N^{1/2-\eps}\leq \optlis(S)\leq N^{1/2+\eps}$.
We define four technical conditions, that describe the interplay between the parameters of our hierarchical decomposition, and show a structural theorem that if $N^{1/2-\eps}\leq \optlis(S)\leq N^{1/2+\eps}$, at least one of them must hold.
Thus, it suffices to consider four special cases of the problem, each restricted to instances satisfying one of these conditions.
Intuitively, three our of these four special cases can be dealt with directly, without exploiting the algorithm $\algncm$ for the \NCM problem in the hybrid model.
We reduce the last special case to the \NCM problem in the hybrid model.
We proceed to provide a brief overview of these conditions, and the efficient \LIS algorithm in the streaming model for each of these resulting special cases.

The first special case occurs when the size of the stream-blocks is relatively small at some level of our hierarchical decomposition.
In this case, we present a simple algorithm that samples elements of $S$ as they arrive in the streaming model while gradually narrowing down on the pairs of stream-blocks and range-blocks that share a prescribed number of elements and hence can possibly contribute elements to $S^*$.
For each of these promising pairs, we employ one of the existing \LIS algorithms to compute the size of their largest shared increasing subsequence.
Using the self reducibility property, we show that these computations suffice to compute an estimate to the size of $S^*$.
In this special case, we obtain $O(1)$-approximation to $\optlis(S)$, directly, without exploiting the algorithm $\algncm$.

The second special case occurs when there is some level in our hierarchical decomposition, such that the contribution of the yes-pairs to $S^*$ is miniscule compared to the number of elements they share. 
Our algorithm for this case, that also obtains $O(1)$-approximation to $\optlis(S)$, is similar to that for the first special case, but with minor technical differences.
As in first special case, this algorithm does not exploit $\algncm$.

The first two special cases deal with situations where the relevant condition occurs at a unique level.
However, the third special case is more complex and involves an interplay of parameters across multiple levels.
Unlike the previous cases, this condition may occur independently across a large number of levels, necessitating the need to apply the following approach recursively on all such occurrences.
Informally, this case arises when a significant proportion of block-pairs at a single level seem promising, making it challenging to determine the ones to pursue.
Exploiting the statistical properties of our hierarchical decomposition, we show that for this challenge to arise, there must be a large enough increasing subsequence (which may be unrelated to both, $S^*$ and the optimum longest increasing subsequence of $S$) that uses elements from these promising pairs.
Our algorithm then pursues as many of these pairs as possible; when this number exceeds a carefully crafted threshold, we prune some of these searches by falling back on this existential guarantee.
Unfortunately, the resulting increasing subsequence may not adhere to our hierarchical decomposition.
Nevertheless, we can still obtain an $N^{o(1)}$-approximation to $\optlis(S)$ in this special case directly, without exploiting the algorithm $\algncm$.

Recall that we can view each level of partition in our hierarchical decomposition as a \NCM problem instance.
The fourth and final special case arises when a large fraction these instances are so tiny, that we can execute the algorithm $\algncm$ on them with a negligible space overhead.
This is precisely the point where the self-reducibility property is most useful: it allows us to create \NCM problem instances $G$, that are tiny, of size roughly $\poly \log \log N$, so that even if $\algncm$ uses space $2^{\poly(|G|)}$, this space overhead can be bounded by $N^{o(1)}$.
Similar to the third special case, this phenomenon occurs on a large number of levels, and we need to execute multiple instances of $\algncm$ recursively across multiple levels.
Assuming that each execution of $\algncm$ on the input instance $G$ achieves factor $|G|^{o(1)}$-approximation, we achieve factor $N^{o(1)}$-approximation to $\optlis(S)$ in this special case.

\paragraph{From Space-Efficient \LIS to Query-Efficient \NCM.}
Ideally, we would like a similar relation in the opposite direction: that is, an $N^{o(1)}$-approximation \LIS algorithm in the streaming model with space complexity $N^{1/2 - \eps}$, for some small absolute constant $\eps > 0$, should imply an algorithm for the $\left( \gamma |L|, \gamma |L|/|G|^{o(1)} \right)$-gap \NCM problem in the hybrid model with query complexity $(d(G))^{1 - \delta}$, for a constant $\delta$ that depends solely on $\eps$, when the input instance $G = (L, R, E_\advice, E)$ satisfies $|L| = |R|$ and $d(G) \geq \ncmdboundasG$ for all $\ncmgammaboundasd \leq \gamma \leq 1$.
Unfortunately, we can only show the reduction from a special case of the \NCM problem in the hybrid model, where the problem instance $G = (L, R, E_\advice, E)$ satisfies even stricter restrictions to the \LIS problem in the comparison-based streaming model that we define next.

\paragraph{\LIS in the Comparison-Based Streaming Model.}
In contrast to the standard streaming model, the comparison-based streaming model only allows an algorithm to gain information about the elements in the stream by \emph{comparing} them, without using any information about their \emph{absolute values}.
Consider the input \LIS instance $S$ of $N$ elements in the streaming model.
A (randomized) $s$-space comparison-based streaming algorithm maintains a `memory-state' $\Gamma$ along with an array $I$ consisting of at most $s$ elements from the stream observed so far.
While processing the $i$-th element $a_i$ from the stream $S$, the algorithm is allowed to compare elements of $I$ with each other and the element $a_i$ with elements of $I$.
Based solely on the current memory state $\Gamma$ and these comparisons, the algorithm updates its memory state.
It may then choose to add $a_i$ to $I$ and/or discard some elements from $I$, while ensuring that $|I| \leq s$ holds.
Note that are no restrictions on the algorithm's time complexity or the size of its memory state $\Gamma$, except for the number of elements $I$ that it may save.
To our knowledge, all the previously known \LIS streaming algorithms are comparison-based.
We provide a formal description of this model in \Cref{sec: ncm-lis-to-ncm-hybrid}.

We are now ready to state the relation between the \LIS problem and \NCM problem in the reverse direction.

\begin{restatable}{theorem}{lisToHybrid}
    \label{thm: comparison lis to hybrid ncm}
    Suppose there are constants $0 < \epsilon < 1/3$ and $\delta > 0$, such that there exists a factor $(1+\eps)$-approximation algorithm for the $\LIS$ problem in the comparison-based streaming model with space-complexity $N^{1/2 - \delta}$, where $N$ is the length of the input stream $S$.
    Then there is an algorithm for the $(\gamma |L|, \eps \gamma |L|)$-gap \NCM problem in the hybrid model 
    for the special case where the input instance $G~=~(L, R, E_\advice, E)$ has $|L| = |R|$ and $\gamma > 1/(\LISNCMgammabound)$,
    that achieves  per-vertex query complexity of $\left(2d(G) |L| \right)^{1/2 - \delta}$.
\end{restatable}

Note that in this result, we are assuming the existence of an algorithm for the \LIS problem with much stronger approximation guarantee in a more stringent model than in \Cref{thm: LIS to NCM final}.
Despite these stronger requirements, we are only able to get a non-trivial algorithm for the \NCM problem in the hybrid model where there is a significant separation between the cardinality of the non-crossing matching in the yes- and no-instances.
To obtain this reduction, we use a simulation-based argument.

Consider an input instance $G = (L, R, E_\advice, E)$ with $|L| = |R|$ of the \NCM problem in the hybrid model and a parameter $\gamma > 1/(\LISNCMgammabound)$.
Our goal is to distinguish the case where $\optncm(G) \geq \gamma |L|$ from the case where $\optncm(G) < \eps \gamma |L|$ with low per-vertex query complexity.
A natural way to reduce this decision problem to the \LIS problem is the following.
We would like to construct a stream $S_G$ from $G$ of length $N$ such that we can use the $(1+\eps)$-approximation algorithm $\alglis$ on $S_G$ to distinguish between these two cases.
We also need to ensure that such a stream $S_G$ can be constructed while processing $G$ in the hybrid model by querying small number of advice edges incident on each vertex of $L$.
Notice that once we explicitly construct such a stream $S_G$, we do not need to query more edge-slots to execute the run of $\alglis$ on $S_G$.

\paragraph{The \LIS Instance $S_G$.}
We show that there is such a stream $S_G$ of length $N = 2dn$, where $n = |L|$ and $d = d(G)$ is the maximum number of edge-slots of $E_{\advice}$ incident to a single vertex of $L$, that is obtained as follows.
For each vertex $u \in L$, there are two stream-blocks denoted by $B^1(u)$ and $B^2(u)$.
These blocks appear in $S_G$ in their natural order $\left(B^1(u_1), B^2(u_1), B^1(u_2), \ldots, B^{2}(u_{n})\right)$, where $u_1,\ldots, u_{n}$ are the vertices of $L$ in that order. 
Consider now a vertex $u \in L$.
The first stream-block $B^1(u)$ consists of a sequence of elements, each corresponding to a unique advice-edge incident on $u$.
The second stream-block $B^2(u)$ contains the same number of elements as $B^1(u)$, each corresponding to a unique advice-edge incident on $u$, with the following additional property.
For each edge $e \in E$ incident on $u$, the corresponding elements in these two stream-blocks form an increasing subsequence.
Conversely, for each edge-slot $e \in E_\advice \backslash E$, the corresponding elements in these two stream-blocks form a decreasing subsequence.
We also ensure that for each non-crossing matching $M$ in $G$, the corresponding $2|M|$ elements form an increasing subsequence of $S_G$. 
Using these conditions, we prove that $2\optncm(G) \leq \optlis(S_G) \leq n + \optncm(G)$.
Therefore, in the regime where $\gamma > 1/(\LISNCMgammabound)$, the execution of the $(1+\eps)$-approximation algorithm $\alglis$ on the input sequence $S_G$ can indeed distinguish the case where $\optncm(G) \geq \gamma |L|$ from the case where $\optncm(G) < \eps \gamma |L|$.
However, the naive approach to constructing $S_G$ as described above requires querying all the advice-edges $E_\advice$ of $G$, which yields a per-vertex query-complexity of $d$ and is unwieldy for a low query-complexity algorithm.

\paragraph*{Simulating the Execution of $\alglis$ on $S_G$.}
To address the high query-complexity of this approach, we construct a stream $\hat S_G$ with the guarantee that $\alglis$ behaves identically on both the streams.
To construct the stream $\hat S_G$ while processing $G$ in hybrid model, we proceed as follows.
As before, we partition $\hat S_G$ into $2n$ stream-blocks, where for each vertex $u \in L$ there are two blocks $\hat B^1(u)$ and $\hat B^2(u)$.
Intuitively, consider some vertex $u \in L$ and assume that the state of algorithm $\alglis$ on $S_G$ before processing the first element of $B^1(u)$ is identical to that  of algorithm $\alglis$ on $\hat S_G$ before processing the first element of $\hat B^1(u)$.
When we receive advice-edges incident on $u$ in the hybrid model, we generate the elements of $\hat B^1(u)$ that are identical to those of $B^1(u)$.
By coupling the randomness used by $\alglis$ on both executions, we ensure that their states remain identical at the end of processing these blocks.
Since $\alglis$ is comparison based, it must have saved a subset of elements of the stream it has processed so far.
We show that by querying advice-edges incident on $u$ corresponding to these saved elements, we can generate the block $\hat B^2(u)$ such that the states of both algorithms remain identical after the blocks $B^2(u)$ and $\hat B^2(u)$ are processed.
In doing so, the number of edges that we query remains bounded by the number of saved elements of $\alglis$, which is bounded by $|N|^{1/2 - \delta} = (2d|L|)^{1/2 - \delta}$.
It is worthwhile to note that the values of $\optlis(S_G)$ and $\optlis(\hat S_G)$ are incomparable to each other: we can only guarantee that $\alglis$ cannot distinguish between these two streams and hence, must behave identically in their processing.
Our algorithm now processed $G$ in the hybrid model while generating the \LIS input sequence $\hat S_G$ and in parallel, executing the algorithm $\alglis$ on $\hat S_G$.
At the end of processing the last vertex of $L$, it reports whether $\alglis$ reported yes or no.

\paragraph{Improved $\alpha$-approximation \LIS Streaming Algorithm.}
Consider some approximation parameter $\alpha > 1$ and the problem of achieving $\alpha$-approximation to the \LIS-length in the streaming model.
The result of G\'{a}l and Gopalan~\cite{GalG07} implies a space lower bound of $\Omega \left( \sqrt{N/(\alpha-1)} \right)$ for this problem for deterministic algorithms, where the input stream has length $N$ and range $\set{1, \ldots, \alpha \cdot N} \subseteq \set{1, \ldots, N^2}$.
Using the techniques discussed above, we show that there is a regime for the parameter $\alpha$ where the space-complexity of randomized algorithms is strictly better than that for the deterministic ones.
This result is summarized in the following theorem.

\begin{theorem} \label{thm: sqrt n by alpha randomized lis algo}
    For every $1 < \alpha \leq N^{1/4}$, there exists a factor $\alpha$-approximation randomized algorithm $\alglis$ for the $\LIS$ problem in the streaming model with space-complexity $\tilde O\left( \sqrt{N}/(\alpha-1) \right)$ where the input stream has length $N$ and range $\set{1, \ldots, \poly(N)}$.
\end{theorem}

We note that the related work of Erg\"{u}n and Jowhari \cite{ErgunJ08} shows  the space lower bound of $\Omega(\sqrt{N})$ for deterministic streaming algorithms for this problem in the special case where $1 < \alpha < 2$ is some constant and the input stream has length $N$.
As a result, their lower bound does not apply to our setting where we require the approximation factor $\alpha = \omega(1)$ is a part of the input.
Our result (\Cref{thm: sqrt n by alpha randomized lis algo}) shows that in the regime where $\polylog{N} \leq \alpha \leq N^{1/4}$, there is an $\alpha$-approximation \LIS streaming algorithm with space complexity $\tilde O(\sqrt{N}/\alpha)$.
Thus, in this regime, randomized algorithms achieve strictly superior space-complexity than their deterministic counterparts, which must use $\Omega(\sqrt{N/\alpha})$ space.
The key tool behind this algorithm is a hierarchical decomposition curtailed to two levels.

\paragraph*{Hierarchical Decomposition.}
We note that existing deterministic algorithms already provide factor-$\alpha$ approximation to the \LIS-length in space $\tilde O\left(\sqrt{N / (\alpha - 1)} \right)$.
We can assume from now on that $\alpha \geq \poly \log N$, since otherwise we can use the existing deterministic algorithm.
We also assume w.l.o.g. that $\alpha$ is an integral power of $2$.
Let $S$ be the input instance of the \LIS problem $S$, of length $N$, where $N$ is an integral power of $2$.
We consider parameters $Z_1 = \alpha Z_2$ that are also integral powers of $2$, whose explicit values will be chosen later.
We show that there exists a partition of $S$ into level-$1$ blocks and a partition of these level-$1$ blocks into level-$2$ subblocks with the following guarantee.
There is an increasing subsequence $S^*$ of $S$, of size $\tilde \Omega(\optlis(S))$, such that each level-$2$ subblock contributes either $Z_2$ elements to it or none at all.
Similarly, each level-$1$ block contributes either $Z_1$ elements to it or none at all.
Note that if a level-$1$ block contributes $Z_1$ elements, it must contain exactly $Z_1/Z_2 = \alpha$ level-$2$ subblock that contribute elements to $S^*$.
As before, we can ensure that the block-sizes of all level-$1$ blocks are identical and so are the block-sizes of all level-$2$ sub-blocks.
This allows us to guess the sizes, say, $X_1$ and $X_2$ respectively, with a low space overhead.
Let us assume that we have the correct level-$1$ and level-$2$ partitions of $S$, which we refer to as a hierarchical decomposition into two levels.

We observe that the existing algorithms can be combined to achieve a factor-$2$ approximation for \LIS-length in space $\tilde O \left(\sqrt{N} / \alpha \right)$, except for the case where $\sqrt{N} / \alpha \leq \optlis(S)\leq \alpha \sqrt{N}$.
We thus focus on this regime, namely, where $\sqrt{N} / \alpha \leq \optlis(S)\leq \alpha \sqrt{N}$.
Using standard techniques, we can assume w.l.o.g. that we are given a parameter $\sqrt{N} / \alpha \leq \tau^* \leq \alpha \sqrt{N}$ and our goal is to distinguish the case where the above-mentioned well-behaved increasing subsequence $S^*$ has cardinality exactly $\tau^*$ from the case where $\optlis(S) < \tau^*/\alpha$.
In the former case, note that there are (exactly) $|S^*|/Z_1 = \tilde \Omega\left(\tau^* / Z_1 \right)$ level-$1$ blocks that contribute (exactly) $Z_1$ elements to $S^*$.
We exploit this structure of $S^*$ with the help of the following dynamic program.

\paragraph*{Dynamic Program.}
We maintain a dynamic programming table $T$ containing $\tilde O \left(\tau^* / Z_1 \right)$ entries while processing the elements of $S$ in the streaming model.
After each level-$1$ block $B$ is processed, the $i^{\text{th}}$ entry stores the smallest range-element $e$, such that there is an increasing subsequence of size at least $iZ_2$ among the elements seen so far, with values at most $e$.
If no such element exist, the entry is undefined.
We report yes at the end of stream $S$ if the last entry of $T$ is defined.
Otherwise, we report no.
By appropriately choosing the ratio $Z_1/Z_2$, we can ensure that if the algorithm reports yes, there must be an increasing subsequence of $S$ of cardinality at least $\tilde \Omega \left(\tau^* / Z_1 \right) \cdot Z_2 > \tau^*/\alpha$.
On the other hand, if we report no, we show that there is no increasing subsequence of size at least $\tau^*$, that adheres to our $2$-level hierarchical decomposition.
Thus, we successfully distinguish the case where $S^*$ has cardinality exactly $\tau^*$ from the case where $\optlis(S) < \tau^*/\alpha$.

We analyze the space complexity of the algorithm, which can be divided into two parts.
The first part is the space required to store the dynamic programming table $T$, which is trivially bounded by the number of entries: $\tilde O\left(\tau^* / Z_1 \right)$.
In the second part, we consider the space required to update its entries. 
Consider a level-$1$ block $B$.
We observe that a naive way to update its entries while processing the elements of $B$ in the streaming model would use $\Omega(Z_2)$ space for each entry.
Hence, the overall space used is $\tilde \Omega(\tau^*) = \tilde \Omega (\alpha \sqrt{N})$, which is not permissible.
We show that by sub-sampling level-$2$ subblocks of $B$ at an appropriate rate, we can `amortize' the space required and achieve a bound of $\tilde O(Z_2)$ for simultaneously updating all the entries of $T$.
The space complexity of our algorithm is then bounded by $\tilde O\left(\tau^* / Z_1 \right) + \tilde O(Z_2)$.
By appropriately choosing parameters $Z_1 = \tilde \Omega \left(\tau^*/\alpha \right)$ and $Z_2 = \tilde O \left(\tau^* / \alpha^2 \right)$, we can further bound this space complexity by $\tilde O\left( \tau^*/\alpha^2 + \alpha \right) = \tilde O \left( \sqrt{N} / \alpha \right)$, since $\tau^* \leq \alpha \sqrt{N}$ and $\alpha \leq N^{1/4}$.

As mentioned earlier, in the regime where $\alpha \geq \poly \log N$, the randomized space complexity $\tilde O(\sqrt{N} / \alpha)$ for achieving $\alpha$-approximation to the \LIS length is strictly better than the lower bound of $\Omega(\sqrt{N/\alpha})$ for deterministic algorithms.

    \subsection{Previous Work}
    \paragraph*{Longest Increasing Subsequence.}
The problem of finding the longest increasing subsequence (\LIS) and its variants have been extensively studied in the framework of streaming models and other related models. 
In the classical model of computation, there is a textbook dynamic programming algorithm that exactly computes \LIS in $O(n^2)$ time.
Fredman~\cite{FREDMAN197529} improved this algorithm to $O(n \log n)$ time using a technique that is now known as `Patience Sorting.'
In what follows, we denote by $N$ the length of the input sequence $S$ and by $\optlis$ the cardinality of its longest increasing subsequence.

Liben-Nowell {\em et al.}~\cite{Liben-NowellVZ06} initiated the study of the \LIS problem in the streaming model and showed that the $O(N \log N)$ time deterministic algorithm of~\cite{FREDMAN197529} can be implemented in the streaming model using $O(\optlis)$ space.
Since $\optlis$ can be as large as $\Omega(N)$, the worst-case space complexity of this algorithm is $O(N)$.
They also provided the first non-trivial lower bound for computing \LIS length in the streaming model by proving that exact computation of $\optlis$, even by randomized algorithms, requires space $\Omega(\sqrt{N})$ in the streaming model.
As mentioned earlier, this lower bound was strengthened to $\Omega(N)$ in independent works by Gopalan {\em et al.}~\cite{sqrt-n-det-lis-soda}, and by Sun and Woodruff~\cite{SunW07}, even for randomized algorithms even when the input stream $S$ is a permutation of the range $\set{1, \ldots, N}$.
This shows that the patience sorting is indeed optimal for exact computation of \LIS-length and the focus then naturally shifted to the design of sublinear space algorithms for \emph{approximating} \LIS length.

Gopalan {\em et al.}~\cite{sqrt-n-det-lis-soda} gave the first sublinear $O(\sqrt{N/\eps})$ space one-pass deterministic streaming algorithm for approximating the \LIS length to within a $(1+\eps)$-factor for any $\eps > 0$.
Moreover, they conjectured that their algorithm is the optimal deterministic one, in the sense that deterministic single pass streaming algorithms indeed require $\Omega(\sqrt{N})$ space to achieve factor-$(1+\eps)$ approximation to $\optlis$.
As a first step, they proved this conjecture for a restricted class of algorithms that they called `natural algorithms', which is precisely the class of algorithms that we refer to as `comparison-based' algorithms.
\footnote{See \Cref{sec: ncm-prelims} for the formal definition of comparison-based streaming algorithms.}
This restriction was lifted very soon and their conjecture was proven independently by G\'{a}l and Gopalan~\cite{GalG07}, and by Erg\"{u}n and Jowhari~\cite{ErgunJ08} who established a matching space lower bound for one-pass deterministic streaming algorithms.
It is worth noting that the result of G\'{a}l and Gopalan~\cite{GalG07} is more general in the sense that it also implies a space lower bound of $\Omega \left(\sqrt{\frac{N}{\alpha-1}}\right)$ for $\alpha$-approximation of \LIS-length by deterministic one-pass streaming algorithms for any approximation factor $\alpha > 1$, provided that the range of the input stream consists of at least $\Omega\left( (\alpha-1) N \right)$ elements.
Note that this bound is tight and matches the deterministic algorithm of \cite{sqrt-n-det-lis-soda} by choosing $\eps = \alpha - 1$.
Together, these results provide a complete understanding of the space-approximation tradeoff for deterministic one-pass streaming algorithms for estimating the \LIS length.
To our knowledge, no multipass algorithm has been proposed that improves upon these one-pass algorithms.
In the multipass algorithms framework, the only known lower bound in this multipass setting is the result of G\'{a}l and Gopalan~\cite{GalG07} that shows that $R$-pass deterministic streaming algorithm must use at least $\Omega \left(\frac{1}{R} \cdot \sqrt{\frac{N}{\eps}}\right)$ space to achieve $(1+\eps)$-approximation of \LIS-length for any $\eps > 0$, provided that the range of the input stream consists of at least $\Omega(\eps N)$ elements.

The understanding of the tradeoff between space complexity and approximation factor for randomized streaming algorithms for \LIS length is not clear, even for one-pass comparison-based algorithms.
The only known negative result in this setting is due to Sun and Woodruff~\cite{SunW07}, who established a lower bound of $\Omega(1/\eps)$ for the space complexity of randomized algorithms that achieve a $(1+\eps)$-factor approximation with probability at least $2/3$, provided that the range of the input stream consists of at least $\Omega(1/\eps)$ elements. 
On the algorithmic side, no improvement over the deterministic algorithms in the worst case is currently known, except in the special case where $\optlis \gg \sqrt{N}$.
Saks and Seshadhri~\cite{SaksS13} proposed a randomized single-pass algorithm that achieves a factor-$(1+\eps)$ approximation to the \LIS length in $\tilde O\left( \frac{N}{\eps \optlis} \right)$ space, that improves on the longstanding deterministic bound of $O(\sqrt{N})$~\cite{sqrt-n-det-lis-soda} in this regime.
As a consequence, there is a huge gap in our understanding of the \LIS problem in the framework of randomized algorithms: no known method exploits randomness to obtain better streaming algorithms than their deterministic counterparts; at the same time, the presence of randomness breaks the lower bound arguments of ~\cite{GalG07,ErgunJ08}.
Both these results rely on reductions from certain $2$-player communication complexity problems for which randomized communication protocols with communication complexity $O(\poly \log N)$ were shown by Chakrabarti~\cite{Chakrabarti-polylog-communication}.
We summarize these results in \Cref{table: ncm-lis-bounds}.

\vspace*{1em}
\begin{table}[ht]
    \centering
     \begin{tabular}{|c | c | c | c |} 
        \hline
        Setting & Approximation factor & Space & References \\ [0.2em]
        \hline\hline
        \multirow{3.5}{*}{Deterministic $1$-pass} & $1$ & $\Theta(N)$ & \cite{Liben-NowellVZ06,sqrt-n-det-lis-soda,SunW07}  \\ [0.2em]
        & $1+\eps$, for any $\eps > 0$ & $\Theta(\sqrt{N})$ & \cite{sqrt-n-det-lis-soda,GalG07,ErgunJ08}\\ [0.2em]
        & $\alpha$, for any $\alpha \geq 2$ & $\Theta(\sqrt{N/\alpha})$ & \cite{sqrt-n-det-lis-soda,GalG07}\\
        \hline \hline
        {Deterministic $R$-pass} & $1+\eps$, for any $\eps > 0$ & $\Omega(\sqrt{N}/R)$ & \cite{GalG07}  \\
        \hline \hline
        \multirow{4.6}{*}{Randomized $1$-pass} & $1$ & $\Theta(N)$ & \cite{sqrt-n-det-lis-soda,SunW07}  \\  [0.2em]
        & $1+\eps$, for any $\eps > 0$  &$\tilde O\left(N / \left({\eps \optlis}\right) \right)$ & \cite{SaksS13}\\ [0.2em]
        & $1+\eps$, for any $\eps > 0$  &$\Omega({1/\eps})$ &\cite{SunW07}\\ [0.2em]
        & $\alpha$, for any $\alpha \geq 2$ & $\tilde O(\sqrt{N}/\alpha)$ & [\ourwork] \\
        \hline
     \end{tabular}
     \caption{Space-approximation tradeoffs for the \LIS problem in the streaming model.}
     \label{table: ncm-lis-bounds}
\end{table}
\vspace*{1em}

The \LIS problem is also studied in the related query, sublinear-space, and sublinear-time models of computation, where there has been a recent flurry of activity.
In these models, the input sequence $S$ is provided on a read-only tape via query access.
The objective of the query model is to estimate the length of the \LIS while minimizing the number of indices at which the values of $S$ are queried.
Note that these queries may either be adaptive (where the subsequent queries depend on the responses to previous queries) or non-adaptive.
It is worth noting that a non-adaptive algorithm with $q$ queries can be translated to a streaming algorithm with space complexity $q$ with identical correctness guarantees.
\footnote{Here we note that the convention in the streaming algorithms community is to optimize the space used \emph{while} processing the input, disregarding the space used preprocessing and post-processing the input.}
It is important to note that no such reduction applies to an algorithm performing adaptive queries. 
In the sublinear-time model, the objective is to optimize for running time while disregarding the space used and the number of queries performed.
Similarly, in the sublinear-space model, the aim is to optimize for working space, disregarding the runtime and the number of queries performed.
Note that a sublinear-time algorithm with running time $T$ directly implies a sublinear-space and a query algorithm with the space-complexity $T$ and query-complexity $T$ respectively.

The \LIS problem has been studied in two different regimes in these models.
The earlier works were motivated by testing monotonicity: whether the input sequence is sorted ($\optlis = N$) or is it \emph{far} from being monotone ($\optlis < (1 - \eps)N$).
The other regime where ${\optlis \ll N}$ had seen much less progress until recently when \cite{SS10-estimate-lis-polylog} showed an algorithm for additive $\eps N$-approximation to \LIS length in time $(1/\eps)^{O(1/\eps)} \cdot \polylog{N}$ for any $\eps > 0$.
This algorithm is truly sublinear and achieves meaningful approximation in the regime where $\optlis \geq \Omega(N/\log N)$.
In the regime where $\optlis \leq o(N/ \log N)$, \cite{RSSS19} provided an $O(\lambda^3)$-approximation of \LIS length in time $\tilde O(\sqrt{N} \cdot \lambda^7)$ for any $\lambda \geq N/\optlis$.
In \cite{MS21}, the approximation guarantee was improved to $O(\lambda^\eps)$ for any constant $\eps > 0$, but at the cost of higher runtime of $O\left(N^{1 - \Omega(\eps)} \cdot \left(\lambda \cdot \log N \right)^{O(1/\eps)} \right)$.
It is worth noting that all of these algorithms employ adaptive queries.
In the framework of non-adaptive query algorithms, \cite{NV21} obtained an $O(\lambda)$-approximation algorithm that perform $\tilde O(\sqrt{r} \lambda^2)$ non-adaptive queries, for any $\lambda \geq N/\optlis$, where $r$ is the number of elements in the range of the input sequence.
Very recently \cite{ANSS22} improved this algorithm and obtained $N^{o(1)}$-approximation algorithm in time $N^{1 + o(1)}/\optlis$, that performs non-adaptive queries.
In particular, their result offers an alternate streaming algorithm that almost matches the space complexity of \cite{SaksS13}, albeit at the cost of much higher approximation factor.
\footnote{We note that the algorithm of~\cite{SaksS13} requires iterating over the input and does not offer improvement over~\cite{ANSS22} in the framework of query or sublinear-time algorithms.}

\paragraph{Matching problems and hybrid model.}
Our proposed computation model for the \NCM problem, known as the hybrid model, sits between two well-established models, namely the semi-streaming and query models. 
While the \NCM problem is not studied in these models, the (standard) matching problem, both on general and bipartite graphs, has been studied in both models, yet there still exist large gaps in our understanding of its complexity.

The semi-streaming model is one of the standard sublinear algorithm frameworks for processing graphs: the algorithm with only $\tilde O(n)$ memory is allowed access to a sequence of edges of the underlying graph, in potentially adversarial order, where $n$ is the number of vertices in the graph.
The (standard) matching problem admits a very simple deterministic greedy algorithm that maintains a maximal matching (which is a factor-$2$ approximation to maximum matching) using $O(n)$ space.
However, when the algorithm is only required to estimate the \emph{size} of the maximum matching, the current lower bounds do not rule out such algorithms with $o(n)$ space complexity.
On the side of lower bounds, the result of~\cite{matching-streaming-det-lowerbound} shows that any deterministic algorithm that achieves factor-$3/2$ approximation must use $\Omega(n)$ space, even on bipartite graphs that obtained by taking disjoint union of paths of length up to $3$.
Their  guarantee drops to the space lower bound of $\Omega(\sqrt{n})$ for randomized algorithms on the same input family, but does not rule out $o(n)$-space randomized algorithms.
To the best of our knowledge, there is no randomized algorithm, that achieves $n^{o(1)}$-approximation to the size of the maximum matching in space $n^{1-\delta}$ for any constant $\delta > 0$, even for bipartite graphs.
In the sublinear-time model, and consequently, in the query model, a useful primitive is the randomized greedy maximal matching algorithm that randomly permutes the edges and then iterates over them, greedily adding the edges to compute a maximal matching.
Although explicitly computing such a maximal matching takes linear time, one may locally estimate if a specific edge is part of the resulting maximal matching.
This approach was pioneered by~\cite{NguyenK08} who showed a randomized algorithm achieving additive $\eps n$-approximation to the size of the maximal matching that performs $2^{O(d)}/\eps^2$ queries to the graph presented as adjacency-list, where $d$ is the maximum vertex-degree of the input graph.
This bound was later improved to $O(d^4/\eps^2)$ by~\cite{YoshidaYI09} and recently to $O(\bar{d} \log n)$~\cite{Behnezhad21}, where $\bar d$ is the average vertex-degree in the input graph.
In the case of adjacency-matrix model,~\cite{Behnezhad21} also shows an additive $\eps n$-approximation to maximal matching by a randomized algorithm that queries $\tilde O(n/\eps^3)$ entries of the adjacency-matrix of the input graph.
This bound is essentially tight, since distinguishing between a graph that does not contain any edges from the graph that contains a single matching of size $\Omega(n)$ requires $\Omega(n)$ queries to its adjacency-matrix representation.

    \subsection{Organization}
    The rest of the sections are organized as follows.
In \Cref{sec: ncm-lis-to-ncm-hybrid} we prove \Cref{thm: comparison lis to hybrid ncm}.
In \Cref{sec: ncm-partition-lemma}, we introduce our key technique, called partition lemma.
This is a `self reducibility property' that allows us to perform a recursive partition of the \LIS problem instance into a hierarchical decomposition.
Using this partition lemma, we give the proof of \Cref{thm: LIS to NCM final} in \Cref{sec: ncm-ncm-hybrid-to-lis}.
Finally, in \Cref{sec: ncm-alpha-approx}, we prove \Cref{thm: sqrt n by alpha randomized lis algo} using the tools developed in \Cref{{sec: ncm-ncm-hybrid-to-lis},{sec: ncm-partition-lemma}}.

    \section{Preliminaries} \label{sec: ncm-prelims}
    We denote by $S$ the input \LIS problem instance that is a sequence of length $N$.
The values of $S$ lie in the range $H^* = \set{1, \ldots, N}$, unless specified otherwise.
We also assume that each element of $H^*$ can be stored in single unit of space.
For a subsequence $S'$ of $S$, we denote by $\optlis(S')$ the length of the largest increasing subsequence of $S'$.

The $\tilde O (\cdot)$ notation hides multiplicative $\poly \log N$ factors and $o(1)$ denotes a function of the length $N$ that approaches $0$ as $N$ approaches infinity.
All the graphs that we consider are undirected simple bipartite graphs.
The success probability of all the randomized algorithms is at least $2/3$ unless specified otherwise.

We will use the algorithm of Liben-Nowell {\em et al.}~\cite{Liben-NowellVZ06} as a primitive in a variety of settings.
As mentioned earlier, this algorithm is essentially a streaming implementation of the $O(N \log N)$ time persistence sorting based deterministic algorithm of Fredman~\cite{FREDMAN197529}.
At a high level, this algorithm maintains a dynamic programming table, where at each time, the $i^{th}$ entry, if defined, stores the smallest range-element such that there exists an increasing sequence of length $i$, using only the elements seen so far, with values up to this range-element.
This algorithm is summarized in the following lemma.

\begin{lemma}\cite{Liben-NowellVZ06} \label{obs: ncm-det-opt-space-sound}
    There is a deterministic algorithm $\alg_1$ for the \LIS problem in streaming model, that, given input sequence $S$ of length $N$ and range $H^* = \set{1, \ldots, M}$, along with a parameter $\tau$, solves the $(\tau, \tau)$-gap \LIS problem with space complexity $O(\tau)$.
    Moreover, if it reports $\optlis(S) \geq \tau$, it also reports the smallest range-element $r^* \in H^*$ such that $\optlis(S_{ \leq r^*}) = \tau$, where $S_{\leq r^*}$ is the subsequence of $S$ consisting of its elements with value at most $r^*$.
\end{lemma}

We will also use the algorithm of Saks and Seshadhri~\cite{SaksS13} that achieves factor-$(1+\eps)$ approximation to the \LIS length in space $\tilde O\left( \frac{N}{\eps \optlis} \right)$ for any $\eps > 0$, where $\optlis$ is the length of the longest increasing subsequence in the input sequence.
In our setting, we are interested in a large factor of approximation and restate their result as the following lemma.

\begin{lemma}\cite{SaksS13}\label{obs: ncm-det-n-by-opt-space-sound}
    There is a randomized algorithm $\alg_2$ for the \LIS problem in streaming model, that, given input sequence $S$ of length $N$ and range $H^* = \set{1, \ldots, N}$, along with a parameter $\tau$, solves the $(\tau, \tau/2)$-gap \LIS problem with space complexity $\tilde O(N/\tau)$.
\end{lemma}

We also use the deterministic approximation algorithm of Gopalan {\em et al.}~\cite{sqrt-n-det-lis-soda} that is summarized in the following lemma.

\begin{lemma}\cite{sqrt-n-det-lis-soda}\label{obs: ncm-det-sqrt-n-space-sound}
    There is a deterministic algorithm $\alg_3$ for the \LIS problem in streaming model, that, given input sequence $S$ of length $N$ and range $H^* = \set{1, \ldots, N}$, reports a value $\optlis(S)/2 \leq \tau^* \leq \optlis(S)$ with space complexity $O(\sqrt{N})$.
\end{lemma}

It will be convenient for us to use the following weaker, but simpler version of the standard Chernoff bound.

\begin{fact}[Chernoff bound]\cite{measure-concentration}\label{fact: ncm-chernoff-version}
    Let $X_1, \ldots, X_n$ be independent boolean random variables and let $\mu := \expect{\sum_i X_i}$
    Then for any $\mu^* \geq {2 \mu}$, $\prob{\sum_i X_i \geq \mu^*} \leq e^{\left( - \frac{\mu^*}{6} \right)}$.
    Moreover, for any $\mu^{**} \leq {\mu/3}$, $\prob{\sum_i X_i \leq \mu^{**}} \leq e^{\left( - \frac{2\mu}{9} \right)}$.
\end{fact}

Finally, we will also need the following two simple facts.

\begin{fact} \label{fact: ncm-pigeonhole}
    For all $x_1, \ldots, x_n \in \reals$, there is some $i \in \set{1, \ldots, n}$ with $x_i \geq (\sum_j x_j)/(2i^2)$.
\end{fact}
\begin{proof}
    Let $X = \sum_j x_j$.
    Assume for contradiction that for each $i \in \set{1, \ldots, n}$, we have $x_i < X/(2i^2)$.
    But then,
        \[X = \sum_{j=1}^n x_j < \sum_{j=1}^{n} \frac{X}{2j^2} < \frac{\pi^2}{12}X < X,\]
    a contradiction.
    Here, we have used the well known inequality that $\sum_{j=1}^{\infty} 1/j^2 = \pi^2/6$.
\end{proof}

    \begin{fact} \label{fact: ncm-log-concave}
        For all $x_1, \ldots, x_n \in \reals$ such that $x_1, \ldots, x_n \geq 1$, $\left( \prod_i \log{x_i} \right)^{1/n} \leq \log{\left( \left( \prod_i x_i \right)^{1/n} \right)}$.
    \end{fact}
    \begin{proof}
        If there exists $1 \leq j \leq n$ such that $x_j = 1$, then $\Pi_i \log{x_i} = 0$ and there is nothing to show.
        We assume from now on that for each $1 \leq j \leq n$, $x_j > 1$ and hence, $\log \log x_j$ is well-defined.
        Since the $\log$ function is monotonically increasing, it suffices to show that $\left( \sum_i \log \log x_i \right)/n \leq \log \log{\left( \left( \prod_i x_i \right)^{1/n}\right)}$.
        But indeed, $\log$ is a concave function and we have,

        \begin{align*}
            \frac{\sum_i \log \log x_i}{n} &\leq \log{\left( \frac{\sum_i \log x_i}{n} \right)}\\
            &= \log{\left( \frac{\log{\prod_i x_i}}{n} \right)}\\
            &= \log{\left( \log{ \left( \left( \prod_i x_i \right)^{1/n} \right)} \right)}.
        \end{align*}
    \end{proof}

    We note that \Cref{fact: ncm-log-concave} holds for any base of the logarithm, in particular for the natural logarithm with base $e$.

    \section{From \LIS in Streaming Model to \NCM in Hybrid Model}\label{sec: ncm-lis-to-ncm-hybrid}
    The goal of this section is to prove \Cref{thm: comparison lis to hybrid ncm}.
Before we proceed, we give a formal definition of the \LIS problem in the comparison-based streaming model.

\paragraph{\LIS in comparison-based streaming model.} %
Consider an instance $S = (a_1, \ldots, a_N)$ of the \LIS problem, where the elements $a_i \in S$ belong to an ordered universe $\uset$.
In the streaming model, we assume that $S$ is revealed over the course of $N$ iterations, where in iteration $i$, the element $a_i$ is revealed.
We assume that the length $N$ of the sequence $S$ and the universe $\uset$ is known to the algorithm beforehand.
We further assume that each element of $\uset$ can be stored in unit space.
The algorithm needs to produce an estimate $k'$ on $\optlis(S)$ after the arrival of the last element of the stream $S$.

We say that a (possibly randomized) streaming algorithm $\alg$ is a \emph{$s$-space comparison-based streaming algorithm}
\footnote{We note that comparison-based algorithms is a natural class of algorithms to study in the streaming model and was recently used by \cite{comparison-based-cite} to give lower-bounds for approximating quantiles in data streams.
}
iff the following holds.
We assume that $\alg$ has access to a sequence $\wset \in \set{0,1}^*$ of random bits.
We think of the memory $M$ of $\alg$ as being divided into two parts: $M = (I, \Gamma)$, where $I$ denotes the \emph{item array} storing some items from $S$ and the \emph{general memory} $\Gamma$.
We assume that $\alg$ can store a single element $a_i$ of stream in each memory cell of $I$.
For $\alg$ to be $s$-space, we require $|I| \leq s$ throughout the algorithm, but do not put any restriction on the size of $\Gamma$.
We also assume that $\alg$ can only perform comparison tests on elements of $I$.

We now formalize this definition.
Let $P(I) \in \set{-1, 0, 1}^{|I| \times |I|}$ be the comparison matrix, where the $(i,j)^{\text{th}}$ entry is defined as:
\begin{equation}
        P_{ij}(I) =
          \begin{cases}
            -1 & \text{if $I[i] < I[j]$}\\
            0 & \text{if $I[i] = I[j]$}\\
            1 & \text{if $I[i] > I[j]$}
          \end{cases}       
\end{equation}

For convenience, we pad extra entries of $P(I)$ by $0$ to ensure that $P(I) \in \set{-1, 0, 1}^{s \times s}$.
The algorithm $\alg$ must be completely determined by computable functions $f : \set{0,1}^{*} \mapsto \set{1, \ldots, s}$, ${g : \set{-1, 0, 1}^{s \times s} \times \set{0,1}^{*} \times \set{0,1}^{*} \mapsto \set{0,1}^*}$ and $h : \set{0,1}^* \mapsto \naturals$ used as follows.
Before seeing any element of the stream $S = (a_1, \ldots, a_N)$, $\alg$ initializes $I \gets \emptyset$ and $\Gamma = \emptyset$.
For each successive element $a_t$ observed, $\alg$ updates $I[f(\Gamma)] \gets a_t$ and $\Gamma \gets g(P(I), \Gamma, \wset)$.
At the end of the stream, $\alg$ outputs $h(\Gamma)$ as the estimate of $\optlis(S)$.
We denote the output of the algorithm on input $S$ along with a sequence $\wset \in \set{0,1}^*$ of random bits by $\alg(S, \wset)$.
Intuitively, the general memory of the algorithm depends on the item array at that time only via the underlying comparison matrix.

We say that $\alg$ is an $\alpha$-approximation algorithm succeeding with probability $p$ iff the probability of the event 
$\optlis(S)/\alpha \leq \alg(S, \wset) \leq \optlis(S)$ is at least $p$.
Unless mentioned otherwise, we say that a randomized algorithm $\alg$ is an $\alpha$-approximation algorithm iff it succeeds with probability $p \geq 2/3$.
If $\alg$ is a deterministic algorithm, it does not depend on $\wset$, and we say that its success probability is $p = 1$.

In the remainder of this section, we prove \Cref{thm: comparison lis to hybrid ncm}.
Our reduction is based on a simulation argument.
We fix the constants $0 < \epsilon < 1/3$ and $\delta > 0$.
We assume that we are given a $(1+\eps)$-approximation algorithm $\alglis$ for the $\LIS$ problem in the comparison-based streaming model with space complexity $N^{1/2 - \delta}$, where $N$ is the input stream length.
Consider an instance $G = (L, R, E_\advice, E)$ of the \NCM problem in the streaming model with $|L| = |R|$ and a parameter $\gamma \geq 1/(\LISNCMgammabound)$.
We first show that there is an instance $S_G$ of \LIS, such that $2\optncm(G) \leq \optlis(S_G) \leq |R| + \optncm(G)$.
We then show that we can use $\alglis$ on $S_G$ to distinguish the case where $\optncm(G) \geq \gamma |L|$ from the case where $\optncm(G) \leq \eps \gamma |L|$.
Naively, to construct such an \LIS instance $S_G$, we would need to learn all the edges of $G$ leading to querying \emph{all} advice-edges.
To circumvent this issue, we construct another stream $\hat S_G$ while performing only a small number of queries to $G$, but with the guarantee that the behavior of $\alglis$ on $\hat S_G$ is identical to that on $S_G$.
Moreover, we ensure that we can construct the stream $\hat S_G$ while processing $G$ in hybrid model.
Here, we critically use the fact that $\alglis$ is a comparison-based algorithm.
We now formally describe the reduction.

As mentioned earlier, we are given an \NCM instance $G = (L, R, E_\advice, E)$ in the hybrid model along with the maximum vertex-degree $d$ of the advice-edges $E_\advice$.
We are also given a parameter $\gamma \geq 1/(\LISNCMgammabound)$ and our goal is to distinguish the case where $\optncm(G) \geq \gamma |L|$ (\yi) from the case where $\optncm(G) < \eps \gamma |L|$ (\ni).
We have assumed that $|L| = |R|$ and let $n := |L|$.
By adding sufficiently many extra edges to $E_\advice$, we can assume w.l.o.g. that each vertex of $L$ has exactly $d$ advice edges incident to it.
For each vertex $u \in L$, we let $E_\advice(u) = (e_1(u), \ldots, e_d(u))$ be the set of advice edges incident on $u$ in their natural increasing order of the corresponding endpoints in $R$.
To reduce this gap-problem in the hybrid model to the \LIS problem in the streaming model, we consider a stream $S_G$ of integers defined as follows.

\begin{figure}[ht]
    \centering
    \includegraphics[width = 12cm]{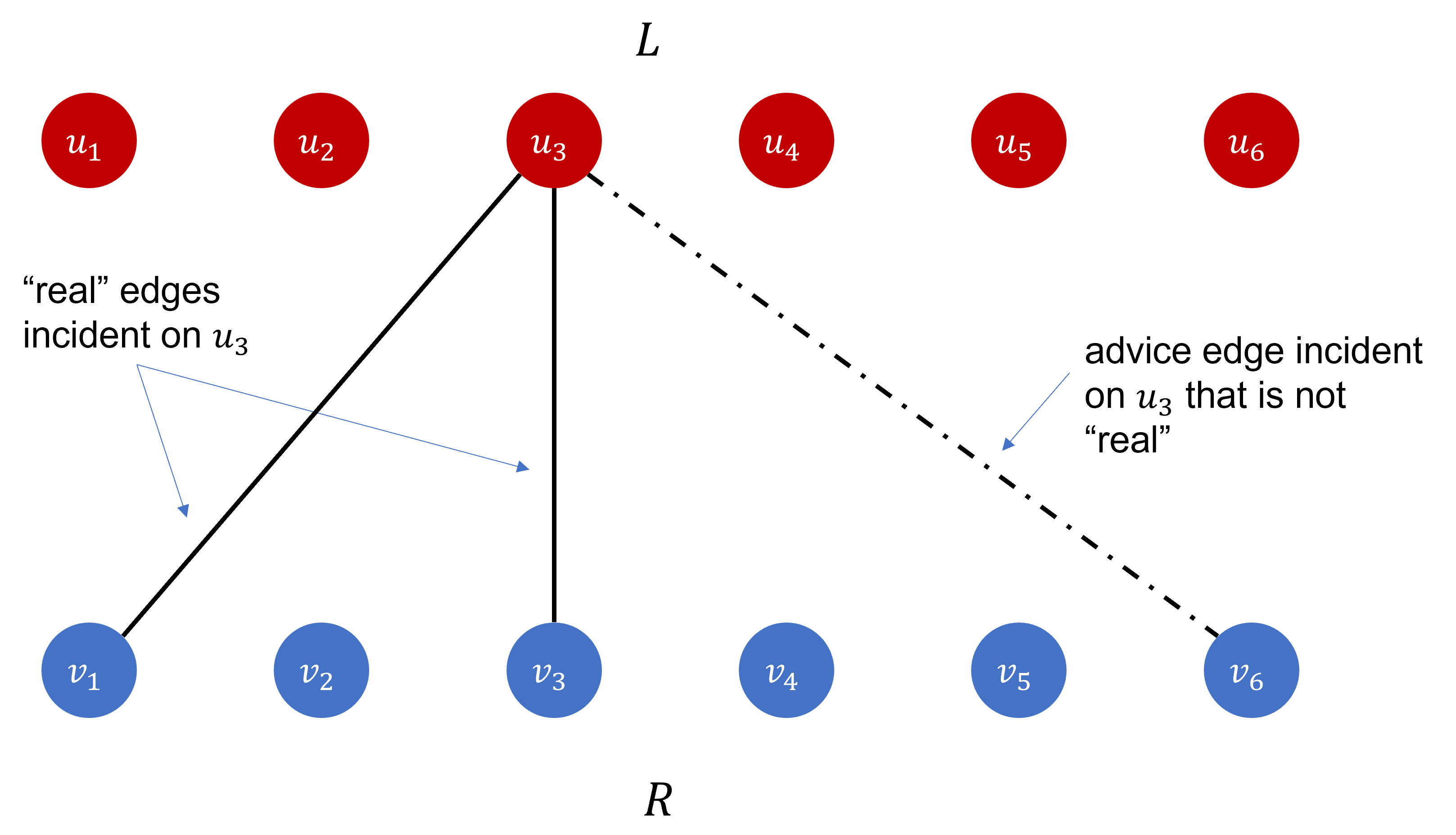}
    \caption{Input \NCM instance $G = (L, R, E_\advice, E)$, where, $L$ and $R$ contain $6$ vertices each.
    Only the edges and advice-edges incident on the vertex $u_3$ are shown. 
    The advice-edges incident on $u_3$ are $e_1(u_3) = (u_3, v_1)$, $e_2(u_3) = (u_3, v_3)$, and $e_3(u_3) = (u_3, v_6)$; where, $e_1(u_3), e_2(u_3) \in E$ and $e_3(u_3) \in E_\advice \backslash E$.}
    \label{fig: ncm-ncm-algo-ncm-instance-G}
\end{figure}

\begin{figure}[h!]
    \centering
    \includegraphics[width = 16cm]{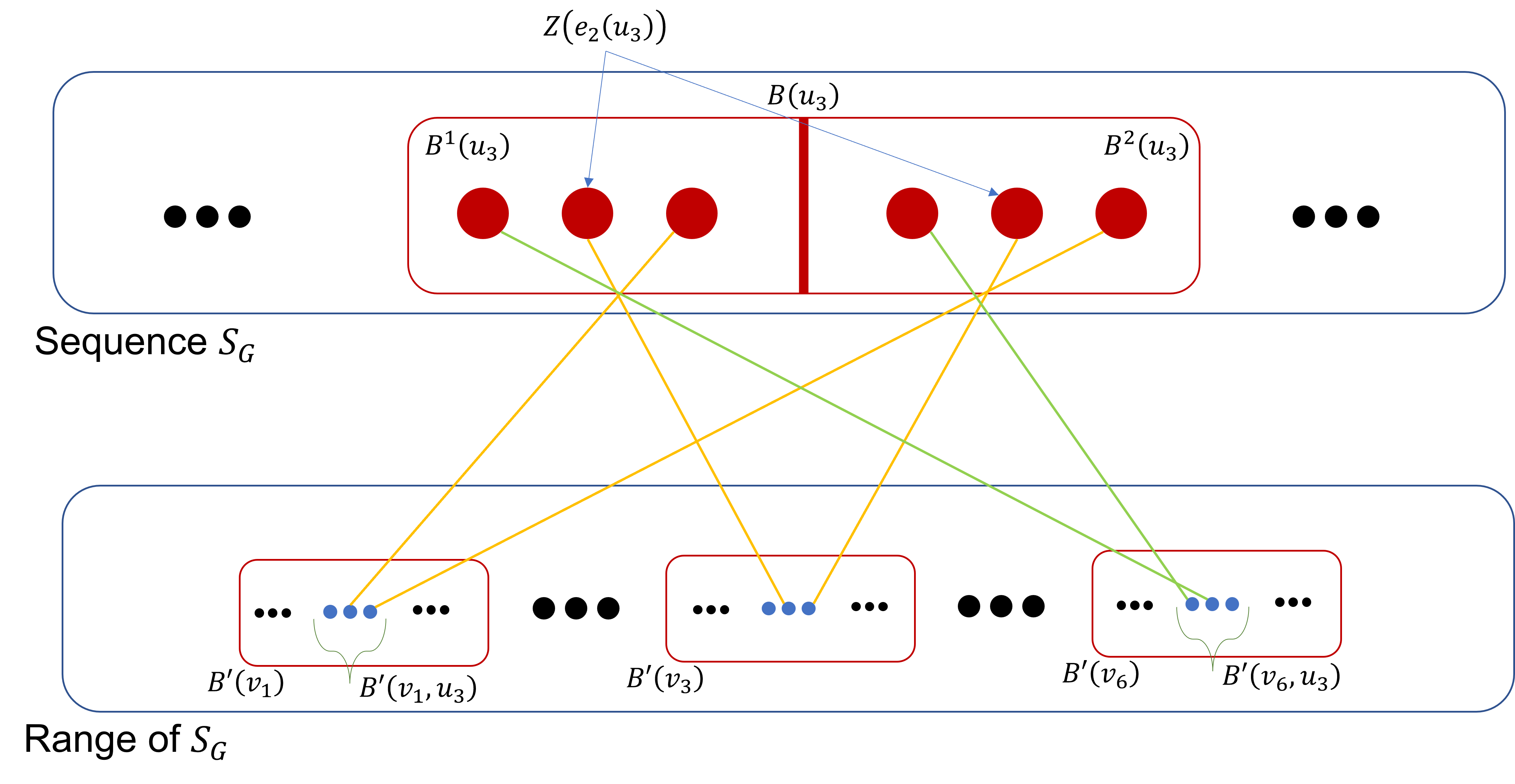}
    \caption{\LIS instance $S_G$ corresponding to the \NCM instance $G$ of \Cref{fig: ncm-ncm-algo-ncm-instance-G}.
    Only the elements of subsequence $B(u_3)$ corresponding to the vertex $u_3$ are shown.
    Notice that the subsequences $Z(e_1(u_3))$ and $Z(e_2(u_3))$ corresponding to `real' edges $e_1(u_3)$ and $e_2(u_3)$ are increasing; while the sequence $Z(e_3(u_3))$ corresponding to the edge $e_3(u_3)$ is decreasing.}
    \label{fig: ncm-ncm-algo-lis-instance-s-g}
\end{figure}

The stream $S_G$ has length $N = 2dn$ and is partitioned into blocks and sub-blocks as follows (see, \Cref{fig: ncm-ncm-algo-ncm-instance-G,fig: ncm-ncm-algo-lis-instance-s-g}). 
We first partition $S_G$ into $n$ equal length blocks $B(u_1), \ldots, B(u_n)$, that represent the vertices of $L$.
Each block $B(u)$ is further partitioned into two length $d$ sub-blocks $B^1(u)$ and $B^2(u)$.
For $r \in \set{1,2}$ and $1 \leq j \leq d$, we identify the $j^{\text{th}}$ from last element of $B^r(u)$ with the corresponding edge $e_j(u)$ of $E_\advice(u)$.
We denote by $Z(e_j(u))$ the sub-sequence comprising of exactly $2$ elements corresponding to the edge-slot $e_j(u)$.
Note that $Z(e_j(u))$ contains precisely $2$ elements: the first element is the $j^{\text{th}}$ from last element of $B^1(u)$ and the second element is $j^{\text{th}}$ from last element of $B^2(u)$.
We assign values to the elements of $S_G$ from $\set{1, \ldots, 3|L||R|}$, that we call the \emph{range} of $S_G$.
We partition the range of $S_G$ into blocks and sub-blocks as follows.
We first partition the range into $n$ equal size blocks $B'(v_1), \ldots, B'(v_n)$, representing vertices of $R$.
The $j^{\text{th}}$ block $B'(v_j)$ is partitioned into $|L|$ equal sub-blocks representing vertices of $L$ in the reverse order: $B'(v_j, u_n), \ldots, B'(v_j, u_1)$, each consisting of $3$ consecutive numbers.
We say that the sub-block corresponding to edge-slot $e = (u_i, v_j)$ with $u_i \in L$ and $v_j \in R$ is $B'(e) := B'(v_j, u_i)$.
We assign the values to the stream $S_G$ ensuring that the values assigned to elements of $Z(e)$ for each advice-edge $e \in E_\advice$ are from the corresponding range sub-block $B'(e)$.
For each advice-edge $e \in E$  we arbitrarily assign values to $Z(e)$ from $B'(e)$ such that $Z(e)$ forms an increasing sub-sequence.
For the remainder of the advice-edges $e \not \in E$, we do the same, but we ensure that the sub-sequence is decreasing.
We denote the resulting stream by $S_G = (a_1, \ldots, a_{N})$.

\begin{figure}[ht]
    \centering
    \includegraphics[width = 16cm]{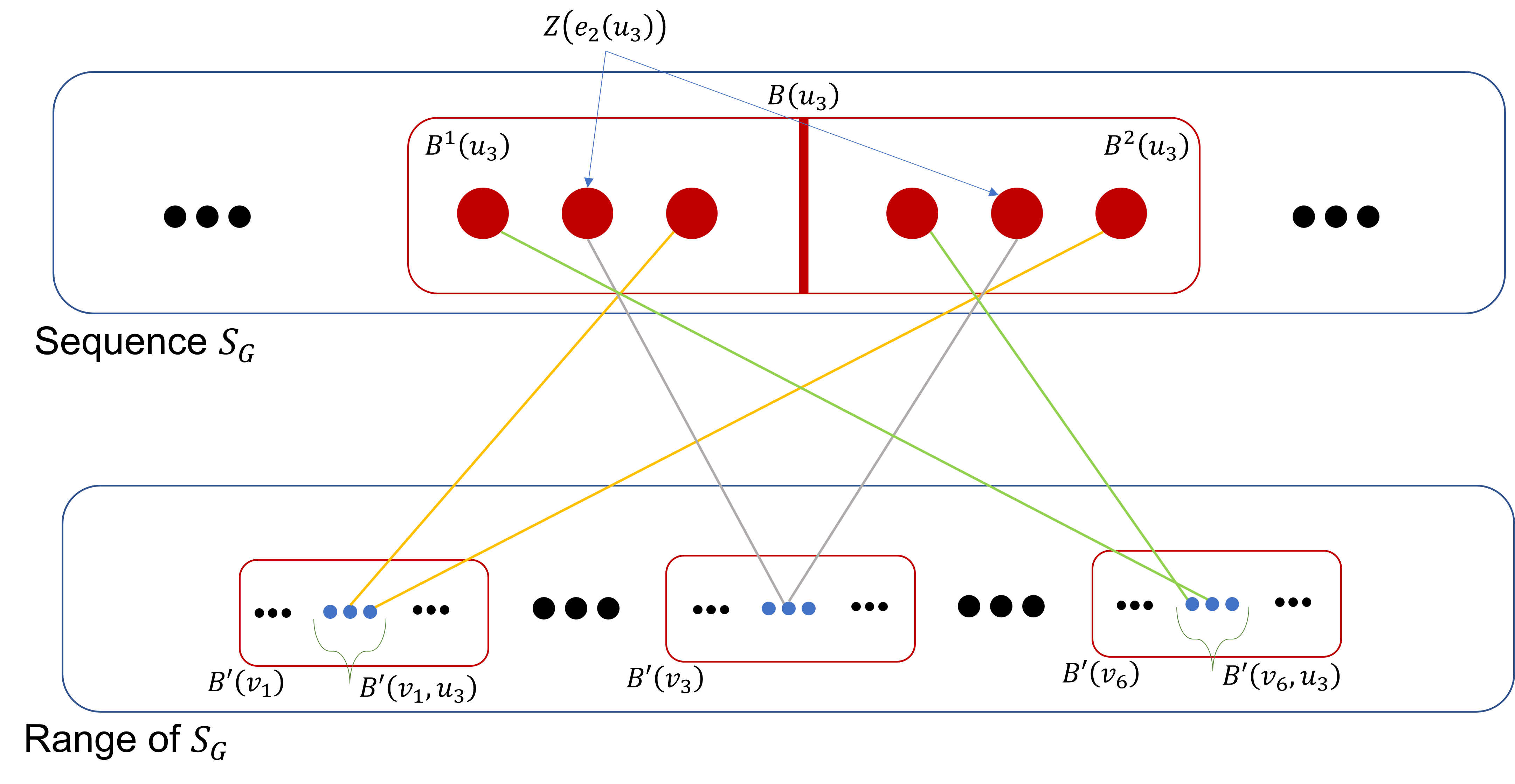}
    \caption{\LIS instance $\hat S_G$ corresponding to the \LIS and \NCM instances of \Cref{{fig: ncm-ncm-algo-ncm-instance-G},{fig: ncm-ncm-algo-lis-instance-s-g}}.
    Only the elements of subsequence $B(u_3)$ corresponding to the vertex $u_3$ are shown.
    We have assumed that the first elements of $Z(e_1(u_3))$ and $Z(e_3(u_3))$ are present in the item array of our algorithm while processing the second elements of $Z(e_1(u_3))$ and $Z(e_3(u_3))$ respectively; and the first element of $Z(e_2(u_3))$ is absent in the item array of our algorithm while processing the second element of $Z(e_2(u_3))$.
    Notice that the subsequence $Z(e_1(u_3))$ is increasing; while the sequences $Z(e_2(u_3))$ and $Z(e_3(u_3))$ are \emph{not} increasing.}
    \label{fig: ncm-ncm-algo-lis-instance-hat-s-g}
\end{figure}

Notice that constructing the stream $S_G$ as described above requires full knowledge of both, $E$ and $E_\advice$.
We start with the following claim whose proof by standard arguments is completed after we complete the proof of \Cref{thm: comparison lis to hybrid ncm} assuming it.

\begin{claim} \label{clm: optlis between optncm}
    $2\optncm(G) \leq \optlis(S_G) \leq |R| + \optncm(G)$.
\end{claim}

From the above claim, if $G$ is a \yi, we have $\optlis(S_G) \geq 2\optncm(G) \geq 2 \gamma n$.
On the other hand, if $G$ is a \ni, we have $\optlis(S_G) \leq n + \optncm(G) < n + \epsilon \gamma n \leq \frac{1}{1+\eps} \cdot 2\gamma n$,
where the last inequality holds for all $\gamma \geq \frac{1}{\LISNCMgammabound} > \frac{1+\eps}{2 - \eps(1+\eps)}$.
Thus, in order to distinguish between \yi and \ni of $G$, it suffices to distinguish between the case where $\optlis(S_G) \geq 2 \gamma n$ from the case where $\optlis(S_G) < \frac{2 \gamma n}{1+\eps}$.
Since the algorithm $\alglis$ is an $(1+\eps)$-approximation algorithm, it can indeed distinguish between these two cases.
Thus, if we could construct the stream $S_G$ while processing $G$ in the streaming model, we would be done.

Unfortunately, constructing $S_G$ naively requires complete knowledge of both, $E$ and $E_\advice$.
As a workaround, we run $\alglis$ on a carefully constructed stream $\hat S_G$
such that the output of $\alglis$ on input $\hat S_G$ is identical to its output on input $S_G$.
We note that $\optlis(S_G)$ might be incomparable to $\optlis(\hat S_G)$ -- since we only need to guarantee that the final output of $\alglis$ is same on both the streams.
We will also ensure that we can construct the stream $\hat S_G$ while processing $G$ in the hybrid model, while performing a small number of queries.
We now formally define our algorithm to construct $\hat S_G$.

We start by drawing a set $\wset^* \in_R \set{0,1}^*$ of random bits.
We consider the execution of $\alglis$ on the stream $S_G = (a_1, \ldots, a_N)$ of length $N$ where $\wset^*$ is the set of the random bits used.
Recall that $\alglis$ maintains a subset $I$ of the elements it has seen so far along with its internal memory content $\Gamma$.
For each $1 \leq t \leq N$, we let $M^{t} = (I^{t}, \Gamma^{t})$ the memory content of $\alglis$ running on $S_G$ just before processing item $a_t$.
We also let $M^{N+1} = (I^{N+1}, \Gamma^{N+1})$ the final memory content of $\alglis$.
Initially, we have $M^1 = (I^1, \Gamma^1) = (\emptyset, \emptyset)$.
Notice that we do not explicitly know the stream $S_G$ and hence, $M^t$ is unknown for $t > 1$.

We construct another stream $\hat S_G = (b_1, \ldots, b_N)$ as follows and consider the execution of $\alglis$ on the stream $\hat S_G$ along with the sequence $\wset^*$ of random bits.
For each $1 \leq t \leq N$, we let $\hat M^{t} = (\hat I^{t}, \hat \Gamma^{t})$ its memory content just before processing the element $b_t$.
We also let $\hat M^{N+1} = (\hat I^{N+1}, \hat \Gamma^{N+1})$ its final memory content.
For $1 \leq t \leq N$, we compute the element $b_t$ as follows.
Let $e = (u,v) \in E_\advice$ be the edge-slot corresponding to the element $a_t$.
Let $r \in \set{1,2}$ be such that $a_t$ is the $r^{th}$ element of $Z(e)$.
Recall that $B'(e)$ contains $3$ consecutive numbers of the range $\set{1, \ldots, 3|L||R|}$.
If $r = 1$, we assign to $b_t$ the value of the middle element of $B'(e)$.
Otherwise, if $r = 2$, we distinguish between the following two cases.
Let ${t'} < t$ be such that $a_{t'}$ is the $1^{st}$ element of $Z(e)$.
If $b_{t'} \not \in \hat I^t$, we assign to $b_t$ the value of the middle element of $B'(e)$.
Otherwise, we query the advice-edge $e$.
If $e \in E$, we assign to $b_t$ the value of the last element of $B'(e)$.
Otherwise, if $e \not \in E$, we assign to $b_t$ the value of the first element of $B'(e)$.
This completes the description of the value assigned to $b_t$, and as a result, the description of the stream $\hat S_G$ (see, \Cref{fig: ncm-ncm-algo-lis-instance-hat-s-g}).
At the end of the stream $\hat S_G$, we report the outcome of $\alglis$ on $\hat S_G$.

In the following claim, we show that the index-array and the memory contents of the execution of $\alglis$ on $S_G$ and $\hat S_G$ are always identical, and hence, the output of $\alglis$ with input $S_G$ is identical to that with input $\hat S_G$.

\begin{claim} \label{clm: correctness of ncm from lis}
    For each $1 \leq t \leq N+1$, $M^t = \hat M^t$.
\end{claim}
\begin{proof}
    Recall that for an index-array $I$, we defined a comparison matrix $P(I) \in \set{-1, 0, 1}^{|I| \times |I|}$, where the $(i,j)^{\text{th}}$ entry records the comparison between $i^{\text{th}}$ and $j^{\text{th}}$ elements of $I$.
    We let $J(I) \in [N]^{|I|}$ be the array where$i^{\text{th}}$ entry records the index of the $i^{\text{th}}$ element of $I$ in $S$.
    To prove \Cref{clm: correctness of ncm from lis}, it now suffices show that for each $1 \leq t \leq N+1$, $J(\hat I^t) = J(I^t)$ and $P(\hat I^t) = P(I^t)$.
    We proceed by induction.
    The base case is when $t = 1$, and the assertion trivially holds since $I^1 = \hat I^1 = \emptyset$.
    We now fix some $2 \leq t \leq N+1$ and assume that $J(\hat I^{t-1}) = J(I^{t-1})$ and $P(\hat I^{t-1}) = P(I^{t-1})$.
    Our goal is to show that $J(\hat I^{t}) = J(I^{t})$ and $P(\hat I^t) = P(I^t)$, which implies, $\hat M^{t} = M^t$.

    \paragraph{Showing that $J(\hat I^{t}) = J(I^{t})$.}
    Recall that $I^{t}$ is obtained by setting the $f(\Gamma^{t-1})^{th}$ entry of $I^{t-1}$ to $a_t$.
    Similarly, $\hat I^{t}$ is obtained by setting $f(\hat \Gamma^{t-1})^{th}$ entry of $\hat I^{t-1}$ to $b_t$.
    The assertion now follows, since $J(\hat I^{t-1}) = J(I^{t-1})$ and $\Gamma^{t-1} = \hat \Gamma^{t-1}$ from our hypothesis.

    \paragraph{Showing that $P(\hat I^t) = P(I^t)$.}
    From our hypothesis, we have $P(\hat I^{t-1}) = P(I^{t-1})$ and $J(\hat I^{t}) = J(I^{t})$.
    Consider now index $\tau \in J(I^t)$ of some saved element $a_\tau \in I^t$.
    Let $b_\tau \in \hat S_G$ be the corresponding element of $\hat S_G$ at index $\tau$.
    It now suffices to show that, for each such $\tau$, if $a_\tau < a_t$ holds then $b_\tau < b_t$ holds, if $a_\tau = a_t$ holds then $b_\tau = b_t$ holds, and if $a_\tau > a_t$ holds then $b_\tau > b_t$ holds.
    But from our construction of $S_G$ and $\hat S_G$, it suffices to show the above relation in the special case where there is some advice-edge $e \in E_\advice$, such that $a_\tau$ and $b_\tau$ correspond to the first element of $Z(e)$, while $a_t$ and $b_t$ correspond to the second element of $Z(e)$.
    In this case, the elements $a_\tau$ and $b_\tau$ are assigned the value of the middle element of $B'(e)$.
    If $e \in E$, both $a_t$ and $b_t$ are assigned the last element of $B'(e)$.
    On the other hand, if $e \not \in E$, both $a_t$ and $b_t$ are assigned the first element of $B'(e)$.
    This completes the proof of \Cref{clm: correctness of ncm from lis}.
\end{proof}

We are now ready to describe and analyze our algorithm $\algncm$ for the \NCM problem in the hybrid model.
Our algorithm $\algncm$ for processing the instance $G = (L, R, E_\advice, E)$ now reports the outcome of $\alglis$ on the stream $\hat S_G$.
The correctness of $\algncm$ now follows from \Cref{clm: correctness of ncm from lis}.
It is immediate to see that we can report this outcome while processing $G$ in the hybrid model.
Indeed, consider a vertex $u \in L$ and the corresponding block $B(u)$ of $\hat S_G$.
Also consider the state of our algorithm $\algncm$ just before processing $u$.
Let $E_\advice(u)$ be the set of advice-edges incident on $u$.
We can compute the elements of the stream subblock $B^1(u)$ from $E_\advice(u)$.
At the end of the stream $B^1(u)$, let $\hat M^{u} = (\hat I^{u}, \hat \Gamma^{u})$ be memory content of $\alglis$ running on $\hat S_G$ just after processing the last element of $B^1(u)$.
Let $E'_\advice(u) \subseteq E_\advice(u)$ be the set of advice-edges, whose corresponding elements of $Z(e)$ are present in $\hat I^{u}$.
We query this subset $E'_\advice(u)$ of advice-edges and let $E^*_\advice(u) = E'_\advice(u) \cap E$ be the set of reported edges.
It is now immediate to assign the values to the subblock $B^2(u)$ from $E^*_\advice(u)$.
Thus, we can indeed simulate the stream $\hat S_G$ while processing $G$ in the hybrid model.

We now analyze the query-complexity of $\algncm$.
Consider the execution of $\alglis$ on input stream $\hat S_G$ as defined above.
Consider a vertex $u \in L$ and the memory-state of algorithm $\hat M^t = (\hat I^t, \hat \Gamma^t)$ just before processing the first vertex of $B^{2}(u)$.
Note that we query at most $|\hat I^t| = |I^t|$ edges of $E_\advice$ incident on $u$.
Thus, the query complexity of $\algncm$ is at most $\max_{1 \leq t \leq 2 dn} |\hat I^{t}| \leq \left(2dn \right)^{1/2 - \delta}$.

This completes the proof of \Cref{thm: comparison lis to hybrid ncm} except for \Cref{clm: optlis between optncm} that we show next.

\proofof{\Cref{clm: optlis between optncm}}
    We first show that $\optlis(S_G) \geq 2\optncm(G)$.
    Fix an optimal non-crossing matching $M$ of size $\optncm(G)$ in $G$.
    Let $M = (e_1, \ldots, e_{|M|})$ in their natural increasing order.
    From our construction, for each edge $e \in E$, $Z(e)$ is an increasing sequence of length $2$ with values in the range sub-block $B'(e)$.
    Since $M$ forms a non-crossing matching, the corresponding range sub-blocks $B'(e_1), \ldots, B'(e_{|M|})$ are disjoint and appear in this order.
    It is now immediate to verify that $\bigcup_{e \in M} Z(e)$ is an increasing subsequence of size $2|M|$ in $S_G$.

    Next, we show that $\optlis(S_G) \leq |R| + \optncm(G)$.
    Fix an optimum increasing subsequence $S'$ of size $\opt(S_G)$ in $S_G$.
    We first claim that for each edge-slot $e \in E_\advice \backslash E$, $S'$ contains at most one element with value in the range sub-block $B'(e)$.
    Indeed, the only elements of $S_G$ with values in $B'(e)$ lie in the subsequence $Z(e)$ of $S_G$.
    But since $e \not \in E$, we are guaranteed that $Z(e)$ has maximum increasing subsequence of size $1$ and the claim follows.
    We further claim that for each vertex $v \in R$, there is at most $1$ advice-edge $e$ incident to it that contributes elements to $S'$.
    Assume otherwise for contradiction that there are $2$ edges $e_1$ and $e_2$ incident on $v$ that contributes elements to $S'$.
    Let $u, u' \in L$ be the corresponding endpoints with $u < u'$ such that $e_1 = (u, v)$ and $e_2 = (u, v')$.
    But then all the elements of $B(u)$ appear before those of $B(u')$ in the stream $S_G$ and have their values after those of $B(u')$, a contradiction.
    Let $R' \subseteq R$ be the vertices that contribute more than $1$ elements to $S'$.
    Since the remainder of the vertices contribute at most $1$ element, we have $\optlis(S) \leq 2|R'| + \left(|R| - |R'|\right) = |R| + |R'|$.
    From the above claim, each such vertex $v \in R'$ contributes exactly $2$ elements to $S'$ and the corresponding edge $e_v \in E$.
    Furthermore, from our construction, such edges $\set{e_v \> | \> v \in R'}$ form a non-crossing matching in $G$.
    Hence, $\optncm(G) \geq |R'| \geq \optlis(S) - |R|$, or in other words, $\optlis(S) \leq |R| + \optncm(G)$.
    This completes the proof of \Cref{clm: optlis between optncm}.
\endproofof

    \section{Hierarchical Decomposition} \label{sec: ncm-partition-lemma}
    \toggletrue{partitioning}
\newcommand{\ysup}[1]{\yset^{(#1)}}

In this section, we consider an input sequence $S=(a_1,\ldots,a_N)$ of $N$ elements for the \LIS problem, where $N$ is an integral power of $2$, and $S$ is a permutation of the range $H^*=(1,\ldots,N)$. 
Our aim is to recursively partition the stream $S$ and the range $H^*$ into a hierarchical structure of blocks, such that there exists a well-behaved increasing subsequence $S^*$ of $S$ of length comparable to $\optlis(S)$ with respect to this partition.
The precise definition of the required `hierarchical decomposition' and the `well-behavedness' is technical and is formally stated later in this section. 
Intuitively, we seek to partition $S$ and $H^*$ into blocks at each level, where each block of this partition should behave identically to every other block at the same level.
We are interested in several statistics about this hierarchical decomposition, such as the size and number of these blocks, the number of shared elements among various pairs of blocks, and the contribution to $S^*$ due to various blocks.
It is also desirable to ensure that the number of levels in our hierarchical decomposition is not too large and that the block sizes do not differ drastically across levels.
Before presenting the technical details, we provide a formal definition of blocks and partitions.

\paragraph{Blocks and subblocks.} 
Given a sequence $A = (a_1, \ldots, a_n)$, we say that a subsequence $B$ of $A$ is a \emph{block} iff $B = (a_i, a_{i+1}, \ldots, a_{i+\ell-1})$ is a contiguous subsequence of $A$.
In this case, we say that $B$ is a block of length $\ell$.
To emphasize that $B$ is contained in $A$, we sometimes refer to it as a \emph{subblock} of $A$.
In the special case where $A$ is the input sequence of the \LIS problem, we say that $B$ is a \emph{stream-block}.
Similarly, if $A$ is the range-sequence of the \LIS problem, we say that $B$ is a \emph{range-block}.

\paragraph*{Partition into blocks.}
Consider a sequence $A = (a_1, \ldots, a_n)$, where $n$ is an integral power of $2$.
For a positive integer $\eta$ that is also an integral power of $2$, we denote by $\bset_\eta(A)$ the unique partition of $A$ into exactly $\eta$ blocks, each containing exactly $n/\eta$ elements.
In other words, for $1 \leq i \leq \eta$, the $i^{th}$ block $B_i \in \bset_\eta(A)$ contains exactly $n/\eta$ contiguous elements of $A$: $\set{a_{(i-1)(n/\eta) + 1}, \ldots, a_{i (n/\eta)}}$.
We say that $\bset_\eta(S)$ is the \emph{partition of the stream $S$ into stream-blocks} and $\bset_\eta(H^*)$ is the \emph{partition of the range $H^*$ into range-blocks}.

As alluded earlier, our goal is to obtain a hierarchical decomposition of $S$ and $H^*$ into stream-blocks and range-blocks each, such that for a large enough increasing subsequence of $S$, the contribution due to different blocks is roughly balanced.
Before describing this hierarchical decomposition, we start with its vanilla version, that shows a single level of partition.

\begin{lemma} \label{lem: ncm-vanilla-partition}
    Suppose we are given a sequence $S = (a_1, \ldots, a_N)$ of $N$ elements from the range $H^* = (1, \ldots, N)$ where $N$ is an integral power of $2$.
    For each subsequence $S'$ of $S$ and each integer $1 \leq Z \leq |S'|$ that is an integral power of $2$, there is an integer $\eta$, that is also an integral power of $2$, with the following property.
    There is a subsequence $S'' \subseteq S'$ of size at least $|S'|/16  \log^2{(2\eta)}$ such that for all $B \in \bset_\eta(S)$, either $|B \cap S''| = 0$ or $Z \leq |B \cap S''| \leq 16Z \log^2{(2\eta)}$.
\end{lemma}

The proof of \Cref{lem: ncm-vanilla-partition} is deferred to \Cref{prf-lem: ncm-vanilla-partition}.
This lemma ensures that, there is a large enough increasing subsequence of $S$, such that each stream-block either contributes nothing or roughly the same amount to it.
We further aim to ensure that if a block contributes elements to such an subsequence, it contributes exactly the same amount as others.

\paragraph{$Z$-canonical subsequence w.r.t. partition $\bset_\eta(S)$.}
    Suppose we are given a sequence $S$ consisting of $N$ elements, where $N$ is an integral power of $2$.
    For a positive integer $Z$ we say that a subsequence $S'$ of $S$ is \emph{$Z$-canonical} w.r.t. the partition $\bset_\eta(S)$ iff every stream-block $B$ of $\bset_\eta(S)$ contributes either exactly $Z$ elements to $S'$ or none at all.
    We say that a block $B \in \bset_\eta(S)$ is a yes-block of such a sequence $S'$ iff it contributes $Z$ elements to it and a \emph{no-block} otherwise.

\begin{corollary} \label{cor: ncm-vanilla-threshold-partition}
    Consider a sequence $S$ consisting of $N$ elements, where $N$ is an integral power of $2$.
    For each subsequence $S'$ of $S$ and each integer $1 \leq Z \leq |S'|$ that is an integral power of $2$, there is an integer $\eta$, that is also an integral power of $2$, with the following property.
    There is a subsequence $S'' \subseteq S'$ of size at least $|S'|/256  \log^4{(2\eta)}$ that is $Z$-canonical w.r.t $\bset_\eta(S)$.
\end{corollary}
\begin{proof}
    From \Cref{lem: ncm-vanilla-partition}, there is an integer $\eta$, that is an integral power of $2$, and a subsequence $S''$ of $S'$ of size at least $\frac{|S'|}{16 \log^2{(2\eta)}}$ such that for each block $B \in \bset_\eta(S)$, either $|B \cap S''| = 0$ or $Z \leq |B \cap S''| \leq 16 Z \log^2{2\eta}$.
    For each block $B \in \bset_\eta(S)$ with $|B \cap S''| \geq Z$, we discard all but $Z$ elements of $S''$.
    It is easy to see that at least 
    \[ \frac{|S''|}{16 \log^2{(2\eta)}} \geq \frac{|S'|}{16 \log^2{(2\eta)}} \cdot \frac{Z}{16 Z \log^2{(2\eta)}} = \frac{|S'|}{256 \log^4{(2\eta)}}\] 
    elements of $S''$ survive and the corollary now follows.
\end{proof}

The above corollary implies that for any arbitrary $Z$ that is an integral power of $2$, there exists a near-optimal increasing subsequence of $S$ that is $Z$-canonical with respect to some partition $\bset_\eta(S)$ of $S$ into stream-blocks.
In our specific application, we require a similar guarantee that holds across multiple levels in our hierarchical decomposition, say, consisting of $\ell$ levels. We then want a sequence $\vectZ = (Z_1, \ldots, Z_\ell)$ of integral powers of $2$, and a near-optimal increasing subsequence of $S$ that is $Z_i$-canonical for every level $i$ of this partition.
We will also want to ensure that the number of child-blocks for each stream-block in our partition is relatively small.

\subsection{Partition Lemma} \label{subsec: ncm-partition-lemma}
We are now ready to define our hierarchical partition formally and establish the necessary parameters.

\paragraph{Hierarchical partition of stream-blocks.}
Consider a sequence $S$ of $N$ elements, where $N$ is an integral power of $2$.
Also consider a sequence $\Psi = (\eta_1, \ldots, \eta_r)$ of $r$ positive integers, each an integral power of $2$, such that $\Pi_{i=1}^r \eta_i \leq N$.
The \emph{hierarchical partition} $\bset_\psi(S)$ of $S$ into stream-blocks is obtained as follows.
We view $S$ as a single block and call it a level-$0$ stream-block.
For each $1 \leq i \leq r$, the level-$i$ stream-blocks are obtained by further partitioning each level-$(i-1)$ block into exactly $\eta_i$ equal length blocks.
Formally, we let $\bset_{\Psi}(S) = \left(\bset^0_\Psi(S), \bset^1_{\Psi}(S), \ldots, \bset^r_{\Psi}(S) \right)$, where $\bset^0_{\Psi}(S) = \set{S}$ and for each $1 \leq i \leq r$, $\bset^i_{\Psi}(S) = \bigcup_{B \in \bset^{i-1}_{\Psi}(S)}\bset_{\eta_i}(B)$.
For each level $0 \leq i \leq r$, we say that the blocks of $\bset^i_\Psi(S)$ constitute a \emph{level-$i$ partition} of $S$ and refer to the stream-blocks in $\bset^i_\Psi(S)$ as \emph{level-$i$ stream-blocks}.
For every pair $1 \leq i < j \leq r$ of levels, we say that a block $B^* \in \bset^i_\Psi(S)$ is an \emph{ancestor} of a block $B \in \bset^j_{\Psi}(S)$ iff $B$ is a contained in $B^*$.
Additionally, if $i = j-1$, we say that $B^*$ is the \emph{parent-block} of $B$ and $B$ is a \emph{child-block} of $B^*$.
For each $1 \leq \eta \leq N$ that is an integral power of $2$, we denote by $\Psi^*(N, \eta) := (\eta, \ldots, \eta)$ a sequence of exactly $r(N, \eta) := \floor{\frac{\log N}{\log \eta}}$ integers, all with value $\eta$.
Note that $\bset_{\Psi^*}(S)$ is a hierarchical partition of $S$ with $1 + r(N, \eta)$ levels, and we will use it throughout this subsection.

We will now extend the definition of $Z$-canonical subsequences to the multiple levels of a hierarchical partition as follows.

\paragraph{$\vectZ$-canonical subsequence w.r.t. hierarchical partition of stream-blocks.}
    Consider a sequence $S$ of $N$ elements where $N$ is an integral power of $2$ and a hierarchical partition $\bset_\Psi(S)$ of $S$, where $\Psi = (\eta_1, \ldots, \eta_r)$.
    For a sequence $\vectZ = (Z_0, \ldots, Z_r)$ of integers, we say that a subsequence $S'$ of $S$ is \emph{$\vectZ$-canonical w.r.t $\bset_{\Psi}(S)$} iff
        (i) for each level $0 \leq i \leq r$, $S'$ is $Z_i$-canonical w.r.t the level-$i$ partition $\bset^i_{\Psi}(S)$; and
        (ii) for each level $0 < i \leq r$ and a level-$i$ yes-block $B \in \bset^i_{\Psi}(S)$ of $S'$, its parent-block $B^* \in \bset^{i-1}_{\Psi}$ is also a level-$(i-1)$ yes-block of $S'$.

We also naturally extend this definition of the hierarchical partition of $S$ to the hierarchical partition of the range $H^*$.

\paragraph{Hierarchical partition of range-blocks.}
Consider a sequence $S$ of $N$ elements from the range $H^* = (1, \ldots, N)$, where $N$ is an integral power of $2$.
Also consider a sequence $\Psi' = (\eta'_1, \ldots, \eta'_r)$ of $r$ positive integers, each an integral power of $2$, such that $\Pi_{i=1}^r \eta'_i \leq N$.
As before, we let $\bset_{\Psi'}(H^*) = \left(\bset^0_{\Psi'}(H^*), \bset^1_{\Psi'}(H^*), \ldots, \bset^r_{\Psi'}(H^*) \right)$ be a \emph{hierarchical partition} of $H^*$, where $\bset^0_{\Psi}(H^*) = \set{H^*}$ and for each $1 \leq i \leq r$, $\bset^i_{\Psi'}(H^*) = \bigcup_{B \in \bset^{i-1}_{\Psi'}(H^*)}\bset_{\eta'_i}(B)$.
For each level $0 \leq i \leq r$, we say that the blocks of $\bset^i_{\Psi'}(H^*)$ constitute a \emph{level-$i$ partition} of $H^*$ and refer to the range-blocks in $\bset^i_{\Psi'}(H^*)$ as \emph{level-$i$ range-blocks}.
In other words, the hierarchical partition $\bset_{\psi'}(H^*)$ is obtained as follows.
We view the range $H^*$ as a single block and call it a level-$0$ range-block.
For each level $1 \leq i \leq r$, the level-$i$ range-blocks are obtained by further partitioning each level-$(i-1)$ range-block into exactly $\eta'_i$ equal length child-blocks.
As in the case of stream-blocks, we have the natural ancestor-descendant relationship between various blocks of $\bset_{\Psi'}(H^*)$.
Assume now that we are also given a hierarchical partition $\bset_\Psi(S)$ of $S$ into stream-blocks consisting of $1+r$ levels.
For a level $0 \leq i \leq r$, we refer to a pair of a level-$i$ stream-block and a level-$i$ range-block $(B, B') \in \bset^i_\Psi(S) \times \bset^i_{\Psi'}(H^*)$ as a \emph{pair}.
We also naturally extend ancestor-descendant and parent-child relationships between such pairs at different levels.

\paragraph{$\vectZ$-canonical subsequence w.r.t. hierarchical partitions $\bset_\Psi(S)$ and $\bset_{\Psi'}(H^*)$.}
Consider a sequence $S$ of $N$ elements from the range $H^* = (1, \ldots, N)$, where $N$ is an integral power of $2$.
Also consider a pair of hierarchical partitions $\bset_\Psi(S)$ and $\bset_{\Psi'}(H^*)$ of $S$ and $H^*$ into stream-blocks and range-blocks respectively, consisting of $1+r$ levels each, for some non-negative integer $r$.
Suppose we are given an increasing subsequence $S'$ of $S$.
Notice that the values of $S'$ form a subsequence of $H^*$.
Assume now that we are given another sequence $\vectZ = (Z_0, \ldots, Z_r)$ of integers such that the \textbf{values of $S'$} form a $\vectZ$-canonical subsequence of $H^*$ w.r.t $\bset_{\Psi'}(H^*)$.
In this case, we say that $S'$ is \emph{$\vectZ$-canonical} w.r.t $\bset_{\Psi'}(H^*)$.
If $S'$ is $\vectZ$-canonical w.r.t both, $\bset_\Psi(S)$ and $\bset_{\Psi'}(H^*)$, then for each level $0 \leq i \leq r$, each level-$i$ pair $(B, B') \in \bset^i_{\Psi}(S) \times \bset^i_{\Psi'}(H^*)$ of blocks contribute either exactly $Z_i$ elements to $S'$ (in which case, we say it is a \emph{yes-pair}) or none at all (in which case, we say it is a  \emph{no-pair}).
We refer to the constituent blocks of the yes-pairs as \emph{yes-blocks} and the constituent blocks of the no-pairs as \emph{no-blocks}

\phantomsection
\paragraph{Ensemble $\Upsilon$ and $\Upsilon$-canonical subsequence.} \label{para: ncm-ensemble}
Consider a sequence $S$ that is a permutation of the range $H^* = (1, \ldots, N)$ where $N$ is an integral power of $2$.
Let $\Psi = (\psi_1, \ldots, \psi_r)$ be a sequence of integral powers of $2$ such that $\prod_{0 \leq i \leq r} \psi_r \leq N$.
Similarly, let $\Psi' = (\psi'_1, \ldots, \psi'_r)$ be a sequence of integral powers of $2$ such that $\prod_{0 \leq i \leq r} \psi'_r \leq N$.
In our partition lemma, the goal is to construct a hierarchical partition $\bset_\Psi(S)$ of $S$ into stream-blocks and a hierarchical partition $\bset_{\Psi'}(H^*)$ of $H^*$ into range-blocks, consisting of $1+r$ levels each.
We require that there is a large increasing subsequence $S^*$ of $S$, of length comparable to $\optlis(S)$, that satisfies the following four properties.

\begin{properties}[0]{P}
    \item \label[property]{prop: ncm-p1} There is a sequence $\vectZ = (Z_0, \ldots, Z_r)$ of integral powers of $2$, such that $S^*$ is $\vectZ$-canonical w.r.t. both $\bset_\Psi(S)$ and $\bset_{\Psi'}(H^*)$.
    
    \item \label[property]{prop: ncm-p2} There is a sequence $\vectDelta = (\Delta_0, \ldots, \Delta_r)$ of integral powers of $2$, such that for each level $i$, each level-$i$ yes-pair $(B, B') \in \bset^i_\Psi(S) \times \bset^i_{\Psi'}(H^*)$ has $Z_i \Delta_i \leq |B \cap B'| < 2 Z_i \Delta_i$.
    In other words, for each level $i$, we require that each level-$i$ yes-pair contributes elements to $S^*$ at roughly an equal `rate'.

    \item \label[property]{prop: ncm-p3} There is a sequence $\vectMu = (\mu_0, \ldots, \mu_r)$ of integral powers of $2$, such that the following holds.
    Consider some level $0 \leq i < r$ and a level-$i$ yes-pair $(B, B')$.
    For each child yes-pair $(\hat B, \hat B')$, $Z_{i+1} \mu_i \leq |\hat B \cap B'| < 2 Z_{i+1} \mu_i$.
    \footnote{Note that this property is not relevant for level $r$. We let $\mu_r$ to be a part of $\vectMu$ to ensure that it is a sequence of $1+r$ integers just as $\vectZ$ and $\vectMu$.}
    This property will allow us to argue that the elements of $S^*$ are roughly `evenly spaced-out' in $S$.

    \item \label[property]{prop: ncm-p4}
    For each level $0 \leq i < r$, exactly one of the following must hold:
    (i)$\frac{Z_i}{Z_{i+1}} \geq \log^3{\left(2\eta \right)}$;  or
    (ii) $Z_{i} = Z_{i+1}$ and $\psi'_{i+1} = 1$.
    This property can be thought of as follows.
    Ideally, we would like to ensure that the sequence $(Z_0, \ldots, Z_r)$ is a `rapidly' decreasing sequence.
    Unfortunately, it might be impossible to achieve this.
    To compensate for that, we require that for each level $i$ where $Z_i = Z_{i+1}$, the level-$(i+1)$ partition $\bset^{(i+1)}_{\Psi'}(H^*)$ of $H^*$ remains identical to the level-$i$ partition $\bset^{(i)}_{\Psi'}(H^*)$ of $H^*$.
\end{properties}

We refer to hierarchical partitions $\bset_\Psi(S)$ and $\bset_{\Psi'}(H^*)$ of $S$ and $H^*$ into stream-blocks and range-blocks respectively along with collection of sequences of integers $\vectZ$, $\vectDelta$, and $\vectMu$ as an \emph{ensemble}, that we denote by $\Upsilon = (\Psi, \Psi', \vectZ, \vectDelta, \vectMu)$.
We refer to the number of levels in the underlining hierarchical partitions $\bset_\Psi(S)$ and $\bset_{\Psi'}(H^*)$ as the \emph{length} of the ensemble $\Upsilon$.
If a subsequence $S^*$ of $S$ satisfies Properties \ref{prop: ncm-p1}-\ref{prop: ncm-p4}, we call $S^*$ a \emph{$\Upsilon$-canonical subsequence}.

Recall that $\Psi^*(N, \eta) = (\eta, \ldots, \eta)$ is a sequence of exactly $r(N, \eta) := \floor{\frac{\log N}{\log \eta}}$ integers, all with value $\eta$.
We are now ready to state the main result of this subsection.

\begin{lemma}[Partition Lemma]\label{lem: ncm-partition-lemma}
    Consider a sequence $S$ that is a permutation of the range $H^* = (1, \ldots, N)$ where $N$ is an integral power of $2$.
    For each $1 < \eta < N$ that is an integral power of $2$, there is an ensemble $\Upsilon = \left(\Psi^*({N, \eta}), \Psi', \vectZ, \vectDelta, \vectMu \right)$ of length $1 + r(N, \eta)$, and an $\Upsilon$-canonical increasing subsequence $S^*$ of $S$ of length at least ${\optlis(S)} / \left({\eta^3 N^{O \left( \frac{\log \log \eta}{\log \eta}\right)}}\right)$.
\end{lemma}
\begin{proof}
    For convenience, we denote $r := r({N,\eta}) = \floor{\frac{\log N}{\log \eta}}$ and $\Psi := \Psi^*({N, \eta}) = (\eta, \ldots, \eta)$ the sequence of $r$ integers, all with value $\eta$. 
    We consider the hierarchical partition $\bset_\Psi(S)$ of $S$ into stream-blocks with $1+r$ levels and fix an optimal increasing subsequence $S'$ of $S$ with length $\optlis(S)$.
    We will construct a hierarchical partition $\bset_{\Psi'}(H^*)$ of the range $H^*$ into range-blocks with $1+r$ levels and the sequences $\vectZ$, $\vectDelta$, and $\vectMu$ of integral powers of $2$.
    We will ensure that for the resulting ensemble $\Upsilon = (\Psi, \Psi', \vectZ, \vectDelta, \vectMu)$, there is a large $\Upsilon$-canonical subsequence $S^*$ of $S'$.
    In the rest of the section, we denote for readability, $\Psi' = (\psi_1, \ldots, \psi_r)$.
    
    For convenience, for each level $0 \leq i \leq r$, we denote by $\bset_i = \bset^i_\Psi(S)$ the partition of $S$ into level-$i$ stream-blocks.
    We will start from level $r$ and gradually proceed to level $0$.
    At each level $i$, we will compute a partition $\tilde \bset_i$ of the range $H^*$ into range-blocks, that will serve as level-$i$ partition the hierarchical partition $\bset_{\Psi'}(H^*)$.
    For level $0$, we have $\tilde \bset_0 = \set{H^*}$.
    For each level $0 < i \leq r$, we will define $\psi_i$, an integral power of $2$ and define $\tilde \bset_{i} = \bigcup_{B' \in \tilde \bset_{i-1}} \bset_{\psi_i}(B')$.
    In other words, partitioning range-blocks of $\tilde \bset_{i-1}$ into exactly $\psi_i$ equal length blocks each, yields the partition $\bset_{i+1}(H^*)$.
    We will select $\set{\psi_i}_{1 \leq i \leq r}$ such that the partition $\tilde \bset_r$ has exactly $\eta^r$ equal length range-blocks.
    If we can ensure this, then indeed for $\Psi' = (\psi_1, \ldots, \psi_r)$, we have a hierarchical partition $\bset_{\Psi'}(H^*)$ of $H^*$ into range-blocks with $1+r$ levels and for each level $0 \leq i \leq r$, the partition $\tilde \bset_i$ of $H^*$ into range-blocks constitute the level-$i$ partition of $\bset_{\Psi'}(H^*)$.

    For each level $i$, starting from level $r$ and proceeding up to level $0$, we will construct a series of subsequences $S^{'} \supseteq S_r \supseteq \ldots \supseteq S_0$ of $S$ with $S^* = S_0$ being the desired subsequence of $S$ that we report.
    For the base case of level $r$, we will compute a subsequence $S_r$ of $S'$, and parameters $Z_r$, $\Delta_r$ and $\mu_r$, that satisfies Properties \ref{prop: ncm-p1}-\ref{prop: ncm-p4}, albeit only for level $r$.
    For each level $0 \leq i < r$, to obtain $S_i$ from $S_{i+1}$, we will proceed as follows.
    We consider the level-$(i+1)$ yes-pairs of $\bset_{i+1} \times \tilde \bset_{i+1}$ for $S_{i+1}$.
    The subsequence $S_i$ of $S_{i+1}$ will be obtained by either choosing all elements of $S_{i+1}$ participating in such a yes-pair or by completely discarding all elements of $S_{i+1}$ from the said yes-pair.
    We will choose the parameters $Z_i$, $\Delta_i$ and $\mu_i$ such that $S_i$ satisfies Properties \ref{prop: ncm-p1}-\ref{prop: ncm-p4} for levels $\set{i, \ldots, r}$.
    We will continue this process until we reach level $0$ and obtain the desired subsequence $S^* = S_0$ of $S'$ with the claimed properties.

    \paragraph{Level $r$.}
    For the base case, we consider level $r$.
    As mentioned earlier, we fix the partition $\tilde \bset_r = \bset_{\eta^r}(H^*)$ of range-elements.
    Since the size of blocks in both, $\bset_r$ and $\tilde \bset_{r}$, is exactly $N/\eta^r < \eta$, there is a subsequence $S'_r \subseteq S^{'}$ of size at least $|S'|/\eta^2$ such that $S_r$ is $1$-canonical w.r.t $\bset_r$ and $\tilde \bset_r$.
    Indeed, such a subsequence $S'_r$ can be obtained from $S'$ by choosing exactly one element belonging to each stream-block of $\bset_i$ that contributes elements to $S'$, and further discarding all but exactly one element, if any, belonging to each range-block of $\tilde \bset_i$.
    
    We now set $Z_r = 1$ and let $\yset'_r \subseteq \bset_r \times \tilde \bset_r$ be the set of $|S'_r|$ yes-pairs for $S'_r$.
    Notice that for each such yes-pair $(B, B') \in \yset'_r$, we have $1 \leq |B \cap B'| < \eta$.
    From the pigeonhole principle, there is some $1 \leq \Delta_r < \eta$, an integral power of $2$, such that at least $|S'_r|/\log \eta$ yes-pairs $(B, B')$ have $\Delta_r \leq |B \cap B'| < 2 \Delta_r$.
    We let $\yset_r \subseteq \yset'_r$ be the set of all such yes-pairs and let $S_r \subseteq S'_r$ the subsequence of $S'_r$ consisting of elements participating in these yes-pairs.
    Since $Z_r = 1$, we indeed have $Z_r \Delta_r \leq |B \cap B'| < 2 Z_r \Delta_r$ for each such yes-pair $(B, B') \in \yset_r$.
    We have thus computed the subsequence $S_r$ of $S'$ with length,
    
    \begin{equation}\label{eqn: ncm-partition-s-r-bound}
        |S_r| \geq \frac{|S'|}{\eta^2 \log \eta} = \frac{\optlis(S)}{\eta^2 \log \eta}
    \end{equation}
    
    that satisfies Properties \ref{prop: ncm-p1}-\ref{prop: ncm-p3}, albeit only for level $r$.
    Note that Property \ref{prop: ncm-p4} is not relevant for level $r$.

    \paragraph{Level $i$.}
    We are given an integer $0 \leq i < r$, and we consider level $i$.
    We are given a subsequence $S_{i+1}$ of $S'$ that is $Z_{i+1}$-canonical w.r.t. the partitions $\bset_{i+1}$ of $S$ and $\tilde \bset_{i+1}$ of $H^*$.
    Furthermore, we assume that $S_{i+1}$ satisfies Properties \ref{prop: ncm-p1}-\ref{prop: ncm-p4}, albeit only for the levels of $\set{i+1, \ldots, r}$.
    Let $\yset_{i+1} \subseteq \bset_{i+1} \times \tilde \bset_{i+1}$ be the yes-pairs for $S_{i+1}$.
    We are also given $\Delta_{i+1}$ and $\mu_{i+1}$, both integral powers of $2$, such that for each yes-pair $(\hat B, \hat B') \in \yset_{i+1}$ the following holds:
    (i) $Z_{i+1} \Delta_{i+1} \leq |\hat B \cap \hat B'| < 2 Z_{i+1} \Delta_{i+1}$; and
    (ii) for each level-$(i+2)$ child-block $\tilde B$ of $\hat B$, if any, that contributes elements to $S_{i+1}$, we have $Z_{i+2} \mu_{i+1} \leq |\tilde B \cap \hat B'| < 2 Z_{i+2} \mu_{i+1}$.
    Our goal is to compute $\psi_{i+1}$, an integral power of $2$ and a partition $\tilde \bset_i$ of $H^*$, such that $\bigcup_{B' \in \tilde \bset_{i}} \bset_{\psi_{i+1}}(B') = \tilde \bset_{i+1}$.
    In other words, partitioning each range-block of $\tilde \bset_i$ into $\psi_{i+1}$ blocks each, yields the partition $\tilde \bset_{i+1}$ of $H^*$.
    Additionally, our goal is to obtain a subsequence $S_i$ of $S_{i+1}$ with the following guarantees:

    \begin{properties}[0]{G}
        \item \label{prop: ncm-g1} For each yes-pair of $\yset_{i+1}$, $S_i$ contains either all elements of $S_{i+1}$ participating in it or none at all.
        
        \item \label{prop: ncm-g2} There is $Z_i$, an integral power of $2$, such that $S_i$ is $Z_i$-canonical w.r.t. the partitions $\bset_i$ and $\tilde \bset_i$.
    
        \item \label{prop: ncm-g3} If $\psi_{i+1} \neq 1$, then $Z_i \geq Z_{i+1} \log^3{\left(2\eta\right)}$. Otherwise, if $\psi_{i+1} = 1$, then $Z_i = Z_{i+1}$.
    
        \item \label{prop: ncm-g4} Let $\yset_i \subseteq \bset_i \times \tilde \bset_i$ be the set of all yes-pairs of $S_i$. Then for each such yes-pair $(B, B')$, $Z_i \Delta_i \leq |B \cap B'| < 2 Z_i \Delta_i$.
        
        \item \label{prop: ncm-g5} There is $\mu_i$, an integral power of $2$, such that for each pair $(\hat B, B') \in \bset_{i+1} \times \tilde \bset'_i$ that contributes elements to $S_i$, $Z_{i+1} \mu_i \leq |\hat B \cap B'| < 2 Z_{i+1} \mu_i$.
    \end{properties}

    It is easy to verify that if we achieve these guarantees, $S_{i}$ indeed satisfies Properties \ref{prop: ncm-p1}-\ref{prop: ncm-p4}, for the levels $\set{i, \ldots, r}$.
    We will first fix the partition $\tilde \bset_i$ of $H^*$ and then compute a large enough subsequence $S_i$ of $S$ that satisfies the above-mentioned five properties.

    \newcommand{\jBound}{3 \log \log {(2\eta)}}

    \paragraph{Computing $\tilde \bset_i$.}
    We start by analyzing the structure of the sequence $S_{i+1}$.
    Recall that we are given a level-$i$ partition $\bset_i$ of $S$ into stream-blocks.
    We say that a yes-pair $(\hat B, \hat B') \in \yset_{i+1}$ \emph{belongs} to a stream-block $B \in \bset_i$ iff $B$ is the parent-block of $\hat B$.
    We partition level-$i$ stream-blocks $\bset_i$ into $1 + \log{\eta}$ classes $\cset_0, \ldots, \cset_{\log \eta}$, where a block $B \in \bset_i$ lies in a class $\cset_j$ iff at least $2^j$ and less than $2^{j+1}$ yes-pairs of $\yset_{i+1}$ belong to $B$.
    Recall that each block $B \in \bset_i$ has exactly $\eta$ child-blocks in $\bset_{i+1}$ and a block $\hat B \in \bset_{i+1}$ appears in at most $1$ yes-pair of $\yset_{i+1}$.
    By the pigeonhole principle, there is a class $\cset_j$ such that at least $|\yset_{i+1}|/\log{(2\eta)}$ yes-pairs of $\yset_{i+1}$ belong to the blocks present in the class $\cset_j$.
    We fix some such $j$ and let $\yset^{(1)}_{i+1}$ be the set of at least 
    
    \begin{equation} \label{eqn: ncm-yset-1-to-yset}
        |\ysup{1}_{i+1}| \geq \frac{|\yset_{i+1}|}{\log{(2\eta)}}
    \end{equation}
    
    yes-pairs of $\yset_{i+1}$ that belong to the blocks present in the class $\cset_j$.
    If $j < \jBound$, we set $\psi_{i+1} = 1$ and $\tilde B_{i} = \tilde B_{i+1}$.
    Thus, assume from now on that $j \geq \jBound$.

    We view the range-blocks of $\tilde \bset_{i+1}$ as a sequence $A_i = (\hat B'_1, \ldots, \hat B'_\ell)$ of its constituent range-blocks in the natural order of their respective range-elements, where $\ell = |\tilde \bset_{i+1}|$, is an integral power of $2$.
    Recall that $S_{i+1}$ is an increasing subsequence of $S$.
    Thus, values of $S_{i+1}$ form an increasing subsequence of $H^*$, or in other words, the yes-blocks of $S_{i+1}$ in $A_i$ form a subsequence.
    We let $A'_i \subseteq A_i$ a subsequence consisting of range-blocks participating in $\yset^{(1)}_{i+1}$.

    From \Cref{lem: ncm-vanilla-partition}, there is $\psi_{i+1}$, an integral power of $2$, and a subsequence $A''_i \subseteq A'_i$ of size at least $|A'_i|/(16 \log^2{(2 \psi_{i+1})})$ such that for all $\bm{B} \in \bset_{\psi_{i+1}}(A_i)$,
    either $2^{j+1} \leq |\bm{B} \cap A''_i| \leq 2^{j + 5} \log^2{(2\psi_{i+1})}$ or  $\bm{B} \cap A''_i = \emptyset$.
    We now choose $\tilde \bset_i$ to be the unique partition of $H^*$ such that $\bigcup_{B' \in \tilde \bset_i} \bset_{\psi_{i+1}}(B') = \tilde \bset_{i+1}$.
    Indeed, such a partition exists, since both, $\ell = |\tilde \bset_{i+1}|$ and $\psi_{i+1}$ are integral powers of $2$.

    This completes the description of our choice of $\psi_{i+1}$ and the partition $\tilde \bset_i$ of the range $H^*$ such that $\bigcup_{B' \in \tilde \bset_i} \bset_{\psi_{i+1}}(B') = \tilde \bset_{i+1}$.
    We now focus on choosing the subsequence $S_i$ of $S_{i+1}$ that satisfies guarantees \ref{prop: ncm-g1} to \ref{prop: ncm-g5}.

    \paragraph*{Computing $S_i$.}
    As before, we say that a yes-pair $(\hat B, \hat B') \in \yset_{i+1}$ \emph{belongs to a range-block $B'$} iff $\hat B'$ is contained in $B'$.
    So far we have computed a partition $\tilde \bset_i$ of the range $H^*$ and some $j$, such that, each stream-block $B \in \bset_i$ has either roughly $2^j$ yes-pairs of $\yset^{(1)}_{i+1}$ belonging to it or none at all.
    Our goal is to show the following claim.

    \begin{claim}\label{clm: ncm-partition-si-guarantees-and-size}
        There are $Z_i, \Delta_i, \mu_i$, all integral powers of $2$, such that there is a subsequence $S_i$ of $S_{i+1}$ that satisfies guarantees \ref{prop: ncm-g1} to \ref{prop: ncm-g5}.
        Moreover, the cardinality of $S_i$ is, $|S_i| \geq \frac{|S_{i+1}|}{2^{30} \log^2{(2\eta)}\log^{10}{(2\psi_{i+1})} \log^3{\left(\frac{2 Z_i \Delta_i}{Z_{i+1} \Delta_{i+1}} \right)}  }$.
    \end{claim}

    To prove the above claim, we distinguish the case where $2^j < \jBound$ from the case where  $2^j \geq \jBound$, beginning with the former.

    \proofof{\Cref{clm: ncm-partition-si-guarantees-and-size} when $2^j < \jBound$}
        Recall that in this case we have set $\psi_{i+1} = 1$ and $\tilde \bset_i = \tilde \bset_{i+1}$.
        In this case, we set $Z_i = Z_{i+1}$ to satisfy guarantee \ref{prop: ncm-g3}.
        We let $\yset'_i \subseteq \bset_i \times \tilde \bset_i$ be the set of pairs obtained as follows.
        For each level-$i$ stream-block $B \in \bset_i$, if there is some pair $(\hat B, \hat B') \in \yset^{(1)}_{i+1}$ belonging to it, we add exactly one such pair $(B, \hat B')$ to $\yset'_i$.
        We let $S'_i \subseteq S_{i+1}$ be the subsequence obtained by choosing all elements of $S_{i+1}$ belonging to the pairs in $\yset'_i$.
        It is immediate to verify that $S'_i$ is $Z_i = Z_{i+1}$-canonical w.r.t. the partitions $\bset_i$ and $\tilde \bset_i$ and has size 
        
        \begin{equation} \label{eqn: ncm-case1-s-prime-i-s-i-plus-1}
            |S'_i| = |\yset'_i| Z_i > \frac{|\yset^{(1)}_{i+1}| Z_{i+1}}{2^{j+1}} \geq \frac{|\yset_{i+1}| Z_{i+1}}{\log^4{(2\eta)}} = \frac{|S_{i+1}|}{\log^4{(2\eta)}}
        \end{equation}

        Here, the inequalities follows from the fact that fewer than $2^{j+1}$ pairs of $\yset_{i+1}$ belong to a single stream-block of $\bset_i$ and $|\yset^{(1)}_{i+1}| \geq |\yset_{i+1}|/\log{(2\eta)}$.
        Note that $S'_i$ already satisfies the guarantees \ref{prop: ncm-g1} to \ref{prop: ncm-g3}.
        Further, notice that any subsequence $S''_i \subseteq S'_i$ that is obtained by either choosing all elements of $S'_i$ from a pair of $\yset'_i$ or discarding all elements from the said pair, also satisfies the guarantees \ref{prop: ncm-g1} to \ref{prop: ncm-g3}.
        It now remains to show that there is large enough subsequence $S_i$ of $S'_i$, obtained by either choosing all elements of $S'_i$ from a pair of $\yset'_i$ or discarding all elements from the said pair, that  additionally satisfies guarantees \ref{prop: ncm-g4} and \ref{prop: ncm-g5}.
    
        Consider some pair $(B, B') \in \yset'_i$.
        From our choice of $\yset'_i$, there is at least one pair $(\hat B, B')$ in $\yset_{i+1}$, where $\hat B$ is a child-block of $B$.
        Thus, $|B \cap B'| \geq Z_{i+1} \Delta_{i+1} = Z_i \Delta_{i+1}$.
        We now partition pairs of $\yset'_i$ into a number of classes, where a pair $(B, B')$ belongs to a class $\cset'_{j'}$ iff $Z_i \Delta_{i+1} \cdot 2^{j'-1} \leq |B \cap B'| < Z_i \Delta_{i+1} \cdot 2^{j'}$.
        For each positive integer $j'$, we let $x_{j'} = |\cset'_{j'}|$, the number of pairs assigned to the class $\cset'_{j'}$.
        Since $\sum_{j'} x_{j'} = |\yset'_i|$, from \Cref{fact: ncm-pigeonhole}, there is an index $j'$ such that $x_{j'} \geq |\yset'_i|/2(j')^2$.
        In other words, there is a class $\cset'_{j'}$ containing at least $\frac{|\yset'_i|}{2(j')^2}$ pairs of $\yset'_i$.
        Let $\yset_i \subseteq \yset'_i$ be the set of at least $\frac{|\yset'_i|}{2(j')^2}$ such pairs belonging to the class $\cset'_{j'}$.
        We let $S_i \subseteq S'_i$ be the subsequence of $S'_i$ consisting of elements of these pairs.

        We now choose $\Delta_i := 2^{j'-1}\Delta_{i+1}$ and $\mu_i := \Delta_{i+1}$.
        We claim that $S_i$ satisfies guarantees \ref{prop: ncm-g4} and \ref{prop: ncm-g5}.
        Indeed, $\yset_i$ is the set of yes-pairs of $\bset_i \times \tilde \bset_i$ for $S_i$.
        Consider now some such yes-pair $(B, B') \in \yset_i$.
        Since $(B, B')$ belongs to the class $\cset'_{j'}$, we have $Z_i \Delta_i \leq |B \cap B'| < 2 Z_i \Delta_i$, satisfying the guarantee \ref{prop: ncm-g4}.
        Consider now some pair $(\hat B, B') \in \bset_{i+1} \times \tilde \bset_{i}$ that contributes elements to $S_i$.
        But $(\hat B, B')$ is also a yes-pair in $\yset_{i+1}$ and has $Z_{i+1} \Delta_{i+1} \leq  |\hat B \cap B'| < 2 Z_{i+1} \Delta_{i+1} $.
        Since $\mu_i = \Delta_{i+1}$, we now have $Z_{i+1} \mu_i \leq  |\hat B \cap B'| < 2 Z_{i+1} \mu_i$, satisfying the guarantee \ref{prop: ncm-g5}.

        Finally, from \Cref{eqn: ncm-case1-s-prime-i-s-i-plus-1} and the fact that $j' = \log{(2\Delta_i/\Delta_{i+1})}$, the cardinality of $S_i$ is,

        \begin{equation} \label{eqn: ncm-case-1-s-i-bound}
            \begin{split}
                |S_i| = |\yset_i| Z_i \geq \frac{|\yset'_i| Z_i}{2 \log^2{(2\Delta_i / \Delta_{i+1})}} = \frac{|S'_i|}{2 \log^2{(2\Delta_i / \Delta_{i+1})}}\\
                \indent \geq \frac{|S_{i+1}|}{2\log^4{(2\eta)} \log^2{ \left( \frac{2\Delta_i}{\Delta_{i+1}} \right)}}\\
                \indent = \frac{|S_{i+1}|}{2\log^4{(2\eta)} \log^{10}{(2\psi_{i+1})} \log^2{ \left( \frac{2 Z_i\Delta_i}{Z_{i+1} \Delta_{i+1}} \right)}}
            \end{split}
        \end{equation}
    
        Here, the last equality follows from the facts that $\psi_{i+1} = 1$ and $Z_i = Z_{i+1}$.
        This completes the proof of \Cref{clm: ncm-partition-si-guarantees-and-size} for the case where $j < \jBound$.
    \endproofof

    We are now ready to complete the proof of \Cref{clm: ncm-partition-si-guarantees-and-size} in the remaining case.

    \proofof{\Cref{clm: ncm-partition-si-guarantees-and-size} for the case where $j \geq \jBound$}

        Recall that in this case, we are given a set $A''_i$ of $|\yset^{(1)}_{i+1}|/(16 \log^2{(2\psi_{i+1})})$ range-blocks participating in $\yset^{(1)}_{i+1}$ such that each range-block of $\tilde \bset_i$ contains either between $2^{j+1}$ and $2^{j+5} \log^2{2 \psi_{i+1}}$ range-blocks of $A''_i$, or none at all.
        We let $\yset^{(2)}_{i+1} \subseteq \yset^{(1)}_{i+1}$ the set of $|A''_i|$ yes-pairs whose corresponding range-blocks are present in $A''_i$.
        From our construction,

        \begin{equation}\label{eqn: ncm-yset-2-to-yset-1}
            |\yset^{(2)}_{i+1}| = |A''_i| \geq \frac{|A'_i|}{16 \log^2{(2\psi_{i+1})}} = \frac{|\yset^{(1)}_{i+1}|}{16 \log^2{(2\psi_{i+1})}}
        \end{equation}

        We say that a pair $(\hat B, \hat B') \in \yset^{(2)}_{i+1}$ \emph{belongs} to a pair $(B, B') \in \bset_i \times \tilde \bset_i$ iff it belongs to both $B$ and $B'$.
        In other words, $B$ is the parent of $\hat B$ and $B'$ is the parent of $\hat B'$.
        In this case, we also say that the pair $(B, B')$ \emph{contributes} the pair $(\hat B, \hat B')$ to $\yset^{(2)}_{i+1}$.
        Notice that less than $2^{j+1}$ pairs of $\yset^{(2)}_{i+1}$ belong to a stream-block $B \in \bset_i$.
        It might still happen that some stream-block $B$ may contain a very small non-zero number of such pairs.
        We will first show that this, in a sense, does not happen too often and show the existence of a large subset, such that each pair $(B, B') \in \bset_i \times \tilde \bset_i$ either contributes roughly an equal number of pairs to it or none at all.

        \begin{claim}\label{clm: ncm-yset-3-all-or-none}
            There is a subset $\yset^{(3)}_{i+1} \subseteq \yset^{(2)}_{i+1}$ of at least $\frac{|\yset^{(2)}_{i+1}|}{2^{24} \log^{8}{(2\psi_{i+1})}}$ pairs such that each pair $(B, B') \in \bset_i \cup \tilde \bset_i$, either contributes exactly $\ceil{\frac{2^j}{2^6 \log^2{(2\psi_{i+1})}}}$ pairs to $\yset^{(3)}_{i+1}$ or none at all.
        \end{claim}

        We defer the proof of this claim to \Cref{prf-clm: ncm-yset-3-all-or-none}.
        We will now refine the pairs of $\yset^{(3)}_{i+1}$ to obtain a large enough subset of pairs that will serve as yes-pairs for a large enough subsequence $S_i$ of $S_{i+1}$ that we will compute next.
        Our goal is to ensure that $S_i$ satisfies all the guarantees \ref{prop: ncm-g1} to \ref{prop: ncm-g5}.
        Consider some pair $(\hat B, \hat B') \in \yset^{(3)}_{i+1}$ and let $(B, B') \in \bset_i \times \tilde \bset_i$ be its parent-pair.
        In other words, $B \in \bset_i$ is the unique stream-block containing $\hat B$ and $B' \in \tilde \bset_i$ is the unique range-block containing $\hat B'$.
        We say that a pair $(\hat B, \hat B')$ is \emph{$\sigma$-friendly} iff $\sigma \leq |B \cap B'| < 2\sigma$.
        We defer the proof of the following claim to \Cref{prf-clm: ncm-sigma-i}.
        
        \begin{claim}\label{clm: ncm-sigma-i}
            There is a subset $\yset^{(4)}_{i+1} \subseteq \yset^{(3)}_{i+1}$ of at least $\frac{|\yset^{(3)}_{i+1}|}{2\log^2{\left(\frac{2 \sigma_i}{Z_{i+1} \Delta_{i+1}} \right)}}$ pairs that are $\sigma_i$-friendly, where $\sigma_i$ is an integral power of $2$. 
        \end{claim}

        Consider some pair $(\hat B, \hat B') \in \ysup{4}_{i+1}$ and let $(B, B')$ be its parent-pair.
        It is immediate to see that all pairs contributed by $(B, B')$ to $\ysup{3}_{i+1}$ are also present in $\ysup{4}_{i+1}$.
        Thus, each pair $(B, B') \in \bset_i \times \tilde \bset_i$ either contributes exactly $\ceil{\frac{2^{j}}{2^6 \log^2{(2\psi_i)}}}$ pairs to $\ysup{4}_{i+1}$ or none at all.
        Moreover, for each stream-block $B \in \bset_i$ that contributes pairs to $\ysup{4}_{i+1}$, there is a unique range-block $B' \in \tilde \bset_i$ containing the respective range-blocks of all such pairs.
    
        Consider some pair $(\hat B, \hat B') \in \ysup{4}_{i+1}$ and let $(B, B')$ be its parent-pair.
        We say that $(\hat B, \hat B')$ is \emph{$\sigma'$-heavy} iff $\sigma' \leq |\hat B \cap B'| < 2\sigma'$.
        Proceeding exactly as in \Cref{clm: ncm-sigma-i} we show the following claim whose proof is present in \Cref{prf-clm: ncm-sigma-prime-heavy}.
    
        \begin{claim}\label{clm: ncm-sigma-prime-heavy}
            There is a subset $\yset^{(5)}_{i+1} \subseteq \yset^{(4)}_{i+1}$ of at least $\frac{|\ysup{4}_{i+1}|}{\log{ \left( \frac{2\sigma_i}{Z_{i+1} \Delta_{i+1}} \right)}}$  pairs that are $\sigma'_i$-heavy, where $\sigma'_i$ is an integral power of $2$.
        \end{claim}

        As before, we partition pairs of $\yset^{(5)}_{i+1}$ into $1 + \log \eta$ classes as follows.
        Consider some pair $(\hat B, \hat B') \in \yset^{(5)}_{i+1}$ and let $B$ and $B'$ be parent-blocks of $\hat B$ and $\hat B'$ in $\bset_i$ and $\tilde \bset_i$ respectively.
        Recall that $B$ has exactly $\eta$ child-blocks in $\bset_{i+1}$ and hence, the pair $(B, B')$ contributes at most $\eta$ pairs to $\yset^{(5)}_{i+1}$.
        Moreover, if such a pair $(B, B')$ contributes pairs to $\ysup{5}_{i+1}$, there is no other range-block $B'' \in \tilde \bset_i$ such that $(B, B'')$ contributes pairs to $\ysup{5}_{i+1}$.
        We say that $(\hat B, \hat B')$ belongs to class $\cset'_{j'}$ for some $0 \leq j' \leq \log \eta$ iff the pair $(B, B')$ contributes at least $2^{j'}$ and less than $2^{j'+1}$ pairs to $\yset^{(5)}_{i+1}$.
        From the pigeonhole principle, there is some class $\cset'_{j^*}$  containing at least $\frac{|\yset^{(5)}_{i+1}|}{1 + \log \eta} = \frac{|\yset^{(5)}_{i+1}|}{\log{(2\eta)}}$ pairs of $\yset^{(5)}_{i+1}$.
        We let $\yset^*_{i+1}$ be the set of such pairs and let $\yset_i \subseteq \bset_i \times \tilde \bset_i$ be the set of pairs contributing pairs to $\yset^*_{i+1}$.
        As mentioned earlier, each stream-block $B \in \bset_i$ appears in at most one pair of $\yset_i$ and similarly, each range-block $B' \in \tilde \bset_i$ appears in at most one pair of $\yset_i$.
        For each pair $(B, B') \in \yset_i$, we discard all but exactly $2^{j^*}$ pairs of $\yset^*_{i+1}$ belonging to it and let $\yset^{**}_{i+1} \subseteq \yset^*_{i+1}$ be the surviving set of pairs.
        Notice that the number of surviving pairs is at least 
        
        \begin{equation} \label{eqn: ncm-yset-star-star-to-yset-5}
            |\yset^{**}_{i+1}| > \frac{|\yset^*_{i+1}|}{2} \geq \frac{|\ysup{5}_{i+1}|}{2\log{(2\eta)}}    
        \end{equation}

        We now set $Z_i := 2^{j^*}$, $\Delta_i := \sigma_i/Z_i$, and $\mu_i := \sigma'_i/Z_{i+1}$.
        We let $S_i$ be the subsequence of $S_{i+1}$ comprising of all of its elements participating in the pairs of $\yset^{**}_{i+1}$.
        From our choice of $Z_i$, the subsequence $S_i$ is indeed a $Z_i$-canonical subsequence of $S$ w.r.t. the partitions $\bset_i$ and $\tilde \bset_i$, and $\yset_i$ is the set of its yes-pairs.
        Consider some such yes-pair $(B, B') \in \yset_i$.
        From \Cref{clm: ncm-sigma-i}, $Z_i \Delta_i \leq |B \cap B'| < 2 Z_{i} \Delta_i$.
        Finally, for each child-block $\hat B$ of $B$, that contributes elements to $S_i$, from \Cref{clm: ncm-sigma-prime-heavy}, $Z_{i+1} \mu_i \leq |\hat B \cap B'| < 2 Z_{i+1} \mu_i$.
        Thus, $S_i$ is indeed the desired subsequence of $S_{i+1}$ satisfying all the guarantees \ref{prop: ncm-g1} to \ref{prop: ncm-g5}.
        To prove \Cref{clm: ncm-partition-si-guarantees-and-size} for the case where $j \geq \jBound$, it suffices to show that the cardinality of $S_i$ is sufficiently large, which we show now.
        Since the sequence $S_i$ contains all the elements participating in the pairs of $\yset^{**}_{i+1}$, we obtain,

        \begin{equation} \label{eqn: ncm-case-2-s-i-bound-one}
            \begin{split}
                |S_i| = |\yset^{**}_{i+1}| Z_{i+1} = \frac{|\yset^{**}_{i+1}|}{\yset_{i+1}} \cdot |\yset_{i+1}| Z_{i+1} = \frac{|\yset^{**}_{i+1}|}{\yset_{i+1}} \cdot  |S_{i+1}|.
            \end{split}
        \end{equation}

        Thus, to obtain a lower bound on the cardinality of $S_i$, it suffices to bound $|\yset^{**}_{i+1}|$.
        From \Cref{{eqn: ncm-yset-star-star-to-yset-5},{clm: ncm-sigma-prime-heavy},{clm: ncm-sigma-i},{clm: ncm-yset-3-all-or-none},{eqn: ncm-yset-2-to-yset-1},{eqn: ncm-yset-1-to-yset}},
        
        \begin{equation} \label{eqn: ncm-case-2-s-i-bound-two}
            \begin{split}
            |\yset^{**}_{i+1}| \geq \frac{|\ysup{5}_{i+1}|}{2\log{(2\eta)}} &\geq \frac{|\ysup{4}_{i+1}|}{{2\log{(2\eta)}  \left( \frac{2\sigma_i}{Z_{i+1} \Delta_{i+1}} \right)}}\\
            &\geq \frac{|\yset^{(3)}_{i+1}|}{{2^2\log{(2\eta)}}  \log^3{\left(\frac{2 \sigma_i}{Z_{i+1} \Delta_{i+1}} \right)}}\\
            &\geq \frac{|\yset^{(2)}_{i+1}|}{{2^{26}\log{(2\eta)}} \log^{8}{(2\psi_{i+1})} \log^3{\left(\frac{2 \sigma_i}{Z_{i+1} \Delta_{i+1}} \right)}}\\
            &\geq \frac{|\yset^{(1)}_{i+1}|}{{2^{30}\log{(2\eta)}} \log^{10}{(2\psi_{i+1})} \log^3{\left(\frac{2 \sigma_i}{Z_{i+1} \Delta_{i+1}} \right)}}\\
            &\geq \frac{|\yset_{i+1}|}{{2^{30}\log^2{(2\eta)}} \log^{10}{(2\psi_{i+1})} \log^3{\left(\frac{2 \sigma_i}{Z_{i+1} \Delta_{i+1}} \right)}}.
            \end{split}
        \end{equation}

        Plugging \Cref{eqn: ncm-case-2-s-i-bound-two} into \Cref{eqn: ncm-case-2-s-i-bound-one}, we can bound,

        \begin{equation} \label{eqn: ncm-case-2-s-i-bound}
            \begin{split}
                |S_i| = \frac{|\yset^{**}_{i+1}|}{\yset_{i+1}} \cdot  |S_{i+1}| &\geq \frac{|S_{i+1}|}{{2^{30}\log^2{(2\eta)}} \log^{10}{(2\psi_{i+1})} \log^3{\left(\frac{2 \sigma_i}{Z_{i+1} \Delta_{i+1}} \right)}}.
            \end{split}
        \end{equation}

        This completes the proof of \Cref{clm: ncm-partition-si-guarantees-and-size} for the case where $j \geq \jBound$.
    \endproofof

    \paragraph{Wrapping up the analysis of level $i$.}
    So far, we have computed a partition $\tilde \bset_i$ of the range $H^*$ and $\psi_{i+1}$, an integral power of $2$, such that $\bigcup_{B' \in \tilde \bset_i} \bset_{\psi_{i+1}}(B') = \tilde \bset_{i+1}$.
    We have also computed $Z_i$, $\Delta_i$, and $\mu_i$, all integral powers of $2$ along with a large enough subsequence $S_i$ of $S_{i+1}$ that satisfies all the guarantees \ref{prop: ncm-g1} to \ref{prop: ncm-g5}, and hence, Properties \ref{prop: ncm-p1}-\ref{prop: ncm-p4} for the levels $\set{i, \ldots, r}$.
    This completes the analysis of level $i$.

    At the end of processing level $0$, we have computed a hierarchical partition $\bset_{\Psi'}(H^*)$ of the range $H^*$, a subsequence $S^* := S_0$ of $S'$, along with the sequences $\vectZ = (Z_0, \ldots, Z_r)$, $\vectDelta = (\Delta_0, \ldots, \Delta_r)$, and $\vectMu = (\mu_0, \ldots, \mu_r)$ of integral powers of $2$, that satisfies Properties \ref{prop: ncm-p1}-\ref{prop: ncm-p4}.
    Thus, for $\Upsilon = (\Psi, \Psi', \vectZ, \vectDelta, \vectMu)$, $S^*$ is indeed a $\Upsilon$-canonical increasing subsequence of $S$.

    \paragraph{Analyzing the cardinality of $S^*$.}
    From \Cref{clm: ncm-partition-si-guarantees-and-size}, the cardinality of $S^* = S_0$ is,

    \begin{equation} \label{eqn: ncm-s-star-with-log-prod}
        \begin{split}
            |S^*| = |S_0| \geq \frac{|S_r|}{\prod_{i=0}^{r-1}{ \left( 2^{30} \log^4{(2\eta)}\log^{10}{(2\psi_i)} \log^3{\left(\frac{2 Z_i \Delta_i}{Z_{i+1} \Delta_{i+1}} \right)} \right) }} \\
            \indent = \frac{|S_r|}{2^{30r} \log^{4r}{(2\eta)} \prod_{i=0}^{r-1} \left(  \log^{10}{(2\psi_i)} \log^3{\left(\frac{2 Z_i \Delta_i}{Z_{i+1} \Delta_{i+1}} \right)} \right) }
        \end{split}
    \end{equation}

    We will now use the concavity of the $\log$ function to bound the denominator.
    From \Cref{fact: ncm-log-concave},

    \begin{equation} \label{eqn: ncm-prod-log-psi}
        \begin{split}
            \left(\prod_{i=0}^{r-1} \log{(2\psi_i)}\right)^{1/r} &\leq \log{ \left( 2\left(\prod_{i=0}^{r-1} \psi_i \right)^{1/r} \right)}\\
            &\leq \log{ \left( 2 N^{1/r} \right)}\\
            &= \log{(2\eta)}.   
        \end{split}
        \end{equation}

    Similarly,
    
    \begin{equation}\label{eqn: ncm-prod-log-zi-deltai}
        \begin{split}
            \left(\prod_{i=0}^{r-1} \log{ \left(\frac{2 Z_i \Delta_i}{Z_{i+1} \Delta_{i+1}} \right) }\right)^{1/r} &\leq \log{ \left( 2\left(\prod_{i=0}^{r-1} \frac{Z_i \Delta_i}{Z_{i+1} \Delta_{i+1}} \right)^{1/r} \right)} \\
            &= \log{ \left( 2 \left(\frac{Z_0 \Delta_0}{Z_r \Delta_r} \right)^{1/r} \right)}\\
            &\leq \log{ \left( 2 N^{1/r} \right)}\\
            &= \log{(2\eta)}.
        \end{split}
    \end{equation}

    Plugging \Cref{{eqn: ncm-prod-log-zi-deltai},{eqn: ncm-prod-log-psi}} in \Cref{eqn: ncm-s-star-with-log-prod}, we get
    \begin{equation}\label{eqn: ncm-s-star-bound}
        \begin{split}
            |S^*| &\geq \frac{|S_r|}{2^{30r} \cdot \log^{4r}{(2\eta)} \cdot \log^{10r}{(2\eta)} \cdot \log^{3r}{(2\eta)} }\\
            &= \frac{|S_r|}{2^{30r} \cdot \log^{17r}{(2\eta)}}\\
            &\geq \frac{|S_r|}{N^{O \left(\frac{\log \log \eta}{\log \eta}\right)}}\\
            &\geq \frac{\optlis(S)}{\eta^3 N^{O \left(\frac{\log \log \eta}{\log \eta}\right)}}.
        \end{split}
    \end{equation}

    For the second inequality, we have used the fact that $r = \floor{\frac{\log N}{\log \eta}}$ and hence,

    \[ \left( \log{(2\eta)}  \right)^{17r} \leq \left(\log{(2\eta)} \right)^{17 \frac{\log N}{\log \eta}} = N^{17 \frac{\log \log{(2\eta)}}{\log \eta}} = N^{O\left( \frac{\log \log \eta}{\log \eta} \right)}. \] 

    The last inequality in \Cref{eqn: ncm-s-star-bound} follows from \Cref{eqn: ncm-partition-s-r-bound} and the fact that $\eta \geq \log \eta$.
    This completes the proof of \Cref{lem: ncm-partition-lemma}.
\end{proof}

We will later use the following two simple facts about our partitioning lemma that we prove in \Cref{{prf-obs: ncm-level-j-to-i-count},{prf-obs: ncm-many-elements-before-median}}.

\begin{observation}\label{obs: ncm-level-j-to-i-count}
    Fix an ensemble $\Upsilon = \left(\Psi^*({N, \eta}), \Psi', \vectZ, \vectDelta, \vectMu\right)$ of length $1 + r = 1+r(N, \eta)$, where $\vectZ = (Z_0, \ldots, Z_r)$ and $\vectMu = (\mu_0, \ldots, \mu_r)$.
    Consider an $\Upsilon$-canonical subsequence $S^*$ of $S$, a level $0 \leq i < r$, and a level-$i$ yes-pair $(B, B')$ for $S^*$.
    Then for each level $j > i$ of $\Upsilon$, fewer than $\frac{Z_i}{4Z_j}$ level-$j$ descendant yes-pairs $(\hat B, \hat B')$ of $(B, B')$ have $|\hat B \cap B'| > 8 Z_j \mu_i$.
\end{observation}

\begin{observation} \label{obs: ncm-many-elements-before-median}
    Fix an ensemble $\Upsilon = \left(\Psi^*({N, \eta}), \Psi', \vectZ, \vectDelta, \vectMu\right)$ of length $1 + r = 1+r(N, \eta)$, where $\vectZ = (Z_0, \ldots, Z_r)$ and $\vectMu = (\mu_0, \ldots, \mu_r)$.
    Consider an $\Upsilon$-canonical subsequence $S^*$ of $S$, a level $0 \leq i \leq r$, and a level-$i$ yes-pair $(B, B')$ for $S^*$.
    Then, at least $\floor{\frac{Z_i}{2}}\mu_i$ elements of $B \cap B'$ appear before the $\left(\floor{\frac{Z_i}{2}} + 1 \right)^{th}$ element of $S^*$ in $B$.
\end{observation}

\subsection{Efficiently Enumerating Ensembles} \label{subsec: ncm-enumerating-ensembles}

So far in this section we have demonstrated that,
given a sequence $S$ of $N$ elements from the range $H^* = \set{1, \ldots, N}$,
there exists an ensemble $\Upsilon$, that additionally satisfies some properties, such that there is an $\Upsilon$-canonical increasing subsequence $S^*$ of $S$ of size comparable to $\optlis(S)$.
However, in our application, we are not given an ensemble $\Upsilon$ a priori and must instead obtain it before processing the input sequence $S$.
We accomplish this by showing that there are only a small number of potential ensembles that meet the necessary conditions, and we can efficiently guess the correct ensemble prior to processing $S$.

\begin{corollary} \label{cor: ncm-partition-lemma}
    For each pair of parameters $N$ and $1 < \eta < N$, both integral powers of $2$, there is a collection $\lset(N, \eta)$ of $N^{O\left(\frac{\log \log \eta}{\log \eta}\right)}$ sequences of integrals powers of $2$, such that the following holds.
    For each permutation $S$ of the range $H^* = (1, \ldots, N)$, there are sequences $\Psi', \vectZ, \vectDelta, \vectMu \in \lset(N, \eta)$ and an increasing subsequence $S^*$ of $S$ such that:
    (i) for the resulting ensemble $\Upsilon = \left(\Psi^*({N, \eta}), \Psi', \vectZ, \vectDelta, \vectMu \right)$, the sequence $S^*$ is an $\Upsilon$-canonical; and 
    (ii) the length of $S^*$ is at least $\frac{\optlis(S)}{\eta^3 N^{O \left(\frac{\log \log \eta}{\log \eta} \right)}}$.
\end{corollary}
\begin{proof}
    As before, we denote by $r = r(N,\eta) = \floor{\log N / \log \eta}$.
    Consider some permutation $S$ of the range $H^* = (1, \ldots, N)$ and let $\optlis(S)$ be the cardinality of the longest increasing subsequence in $S$.
    From \Cref{lem: ncm-partition-lemma}, there is an ensemble  $\Upsilon_S$ and an $\Upsilon_S$-canonical increasing subsequence $S^*$ in $S$ of length at least $\frac{\optlis(S)}{\eta^3 N^{O \left(\frac{\log \log \eta}{\log \eta} \right)}}$.
    We let  $\Upsilon_S = \left(\Psi, \Psi'_S, \vectZ_S, \vectDelta_S, \vectMu_S \right)$, where  $\Psi = \Psi^*(N, \eta) = (\eta, \ldots, \eta)$ is a sequence of exactly $r = r(N, \eta) = \floor{\frac{\log N}{\log \eta}}$ integers, all with value $\eta$.
    We will now show that there is a small enough collection $\lset(N, \eta)$ of sequences of integrals powers of $2$, such that $\Psi'_S, \vectZ_S, \vectDelta_S, \vectMu_S \in \lset(N, \eta)$.
    We will use the following combinatorial fact.

    \begin{fact}\label{fact: ncm-sample-replacement-bound}
        There are $\binom{n+r-1}{r} \leq \left(\frac{e(n+r-1)}{r}\right)^r$ ways of choosing $r$ elements with replacement from a universe of $n$ elements.
    \end{fact}

    Recall that $\vectZ_S$, $\vectDelta_S$, and $\vectMu_S$ are all non-decreasing sequences of consisting of exactly $1+r$ elements, all  integral powers of $2$ between $1$ and $N$, inclusive.
    We let $\lset_1(N, \eta)$ be the collection of all such possible sequences.
    Thus, $\vectZ_S, \vectDelta_S, \vectMu_S \in \lset_1(N, \eta)$ and using  \Cref{fact: ncm-sample-replacement-bound} we can bound its cardinality as follows.
    
    \begin{align*}
        \left| \lset_1(N, \eta) \right| \leq \binom{(1 + \log N) + (1+r) - 1}{(1+r)} &= \binom{1 + \log N + r}{1+r}\\
        &\leq \left( \frac{e (1 + \log N + r)}{r} \right)^{O(r)}\\
        &\leq \left( \frac{O(\log N)}{r} \right)^{O(r)}\\
        &\leq \left( \log \eta \right)^{O(r)}\\
        &\leq N^{O \left( \frac{\log \log \eta}{\log \eta} \right)}.
    \end{align*}

    On the other hand, let $\Psi'_S = (\psi_1, \ldots, \psi_r)$, where each $\psi_i$ is an integral power of $2$ and $\prod_i \psi_i \leq N$.
    For each level $0 \leq i \leq r$, we let $\ell_i := N / (\Pi_{j=1}^i \psi_j)$ be the length of level-$i$ blocks of $\bset_{\Psi'_S}(H^*)$.
    It is immediate to see that $\set{\ell_i}_{0 \leq i \leq r}$ is a sequence of non-decreasing integral powers of $2$ from $\set{1, 2, \ldots, N}$.
    Moreover, fixing $\set{\ell_i}_{0 \leq i \leq r}$ uniquely determines $\Psi'_S$.
    As before, from \Cref{fact: ncm-sample-replacement-bound}, the number of possible choices for $\set{\ell_i}_{0 \leq i \leq r}$ is bounded by $N^{O \left(\frac{\log \log \eta}{\log \eta} \right)}$, and we now conclude that the number of choices of $\Psi'_S$ is also bounded by $N^{O \left(\frac{\log \log \eta}{\log \eta} \right)}$.
    We let $\lset_2(N, \eta)$ be a collection of $N^{O \left(\frac{\log \log \eta}{\log \eta} \right)}$ such sequences, one for each potential choice for $\Psi'_S$.
    The corollary now follows by choosing $\lset(N, \eta) = \lset_1(N, \eta) \cup \lset_2(N, \eta)$.
\end{proof}

    \section{From \NCM in Hybrid Model to \LIS in Streaming Model} \label{sec: ncm-ncm-hybrid-to-lis}
    \toggletrue{ncm-algo-to-lis}
\newcommand{\load}{\mathsf{load}}
\newcommand{\good}{\mathsf{good}}
\newcommand{\perfect}{\mathsf{perfect}}
\newcommand{\final}{\mathsf{final}}
\renewcommand{\active}{\mathsf{active}}
\newcommand{\rstarconstant}{4}
\newcommand{\rstarVal}{\floor{(1/2 - \rstarconstant \eps) r}}
\newcommand{\Tbound}{10}
\newcommand{\deltabound}{10^7}
\newcommand{\savable}{\mathsf{savable}}
\newcommand{\nearZBound}{5\eps}
\newcommand{\nearZBoundPlusOne}{6\eps}
\newcommand{\nearZBoundPlusTwo}{7\eps}
\newcommand{\nearZBoundPlusThree}{8\eps}
\newcommand{\nearZBoundPlusFour}{9\eps}
\renewcommand{\land}{\mathsf{land}} 
\newcommand{\rem}{\mathsf{rem}}
\newcommand{\query}{\mathsf{query}}
\newcommand{\missedChildBlocks}{32}
\newcommand{\Next}{\mathsf{next}}
\newcommand{\iseteq}{\iset^{\mathsf{eq}}}
\newcommand{\alggen}{\alg{\mathsf{Gen}}}
\newcommand{\susp}{\mathsf{susp}}
\newcommand{\prom}{\mathsf{prom}}
\newcommand{\algjump}[1]{{\alg}\mathsf{Jump}{(#1)}}
\newcommand{\alglevel}[1]{{\alg}\mathsf{Level}{(#1)}}
\newcommand{\Gtau}{G^{(\tau)}}
\newcommand{\Ltau}{L^{(\tau)}}
\newcommand{\Rtau}{R^{(\tau)}}
\newcommand{\Etau}{E^{(\tau)}}
\newcommand{\Mtau}{M^{(\tau)}}
\newcommand{\tildebsettau}{\tilde \bset^{(\tau)}}
\newcommand{\tildebsettauprime}{\tilde \bset^{(\tau)'}}
\newcommand{\DeltaLowerBound}{Z_i N^{5\eps}}
\newcommand{\isetnew}{\iset^*}
\newcommand{\knew}{k^*}
\newcommand{\Tval}{100}
\newcommand{\algSavableLevels}{\alg\mathsf{SavableLevels}}
\newcommand{\algSavable}[1]{\alg\mathsf{Savable}{(#1)}}
\newcommand{\pval}{2^7 \zeta  \cdot \frac{Z_\ell}{Z_j}}
\newcommand{\pprimeval}{{\frac{2^{16} \zeta}{Z_j \mu_{\ell}}}}
\newcommand{\estarellval}{2^6 p' Z_\ell \mu_{i}}
\newcommand{\estarjval}{2^6 p' Z_i \mu_{i}}
\newcommand{\algjumpabletuples}{\alg\mathsf{JumpableTuples}}
\newcommand{\eq}{\mathsf{eq}}
\newcommand{\savpval}{{\frac{2^{16} \zeta}{Z_j \mu_{j}}}}
\newcommand{\savellstarjval}{2^6 p Z_j \mu_{i}}
\newcommand{\tupwt}[1]{w_{\iset(#1)}}
\newcommand{\alphaval}{4}
\newcommand{\algbp}{\alg\mathsf{BlockProc}}
\newcommand{\ncminssize}{\max{\set{\eta^2, \psi_{i+1}}}}

The goal of this section is to prove \Cref{thm: LIS to NCM final}.
Recall that the input to the \LIS problem in the streaming model is a sequence $S = (a_1, \ldots, a_N)$, whose length $N$ is known to us in advance.
We further assume that $S$ is a permutation of the range $H^* = \set{1, \ldots, N}$.
To simplify the analysis, we assume that $N$ is an integral power of $2$.
This can be assumed w.l.o.g. by appending a decreasing sequence of large elements at the beginning of $S$.
We denote the length of the longest increasing subsequence of $S$ by $\optlis(S)$.
Our goal is to estimate $\optlis(S)$ within a factor of $N^{o(1)}$ using at most $N^{1/2 - \eps + o(1)}$ units of space, for some constant $\eps > 0$.

We are given an algorithm $\algncm$ for the \NCM problem in the hybrid model that satisfies the following guarantees.
Given an \NCM instance $G = (L, R, E_\advice, E)$ in the hybrid model and a parameter $\gamma$ with $|L| = |R|$, $\gamma \geq \ncmgammaboundasd$, and $d(G) \geq \ncmdboundasG$, it solves $(\gamma |L|, \gamma |L|/\alphancm(|G|))$-gap \NCM problem with per-vertex query complexity $\left( d(G) \right)^{1 - \delta}$, where $\alphancm(|G|) = |G|^{o(1)}$.
In this section, we fix $\eps = \delta/10^{12}$.
To prove \Cref{thm: LIS to NCM final}, we will show a randomized algorithm $\alglis$ for the \LIS problem, that achieves $N^{o(1)}$-approximation to $\optlis(S)$ using space $N^{1/2 - \eps + o(1)}$ in a single pass over the input sequence $S$.

Using standard techniques, it suffices to show the algorithm $\alglis$ for the \LIS problem, where, in addition to the input sequence $S$, we are given a `guess' $\tau^*$ for $\optlis(S)$.
Our goal is then to distinguish the case where $\optlis(S) \geq \tau^*$ from the case where $\optlis(S) < \tau^*/N^{o(1)}$.

Let $\eta = \eta(N)$ be an integral power of $2$ such that $\log \eta = \floor{\log \log \log N}$.
Notice that $\eta$ is a parameter growing sufficiently slowly with $N$, so that $2^{\poly(\eta)} = \left( \log N \right)^{o(1)}$.
We let $r = r(N, \eta) = \floor{\frac{\log N}{\log \eta}}$ and note that $r = o(\log N)$.
From Partition Lemma \ref{lem: ncm-partition-lemma}, there is an ensemble $\Upsilon_S$ of length $1+r$ such that there is a $\Upsilon_S$-canonical increasing subsequence $S^*$ of $S$ with length at least $\frac{\optlis(S)}{\eta^3 N^{O \left(\frac{\log \log \eta}{\log \eta} \right)}} = \frac{\optlis(S)}{\eta^3 N^{\left(\frac{c\log \log \eta}{\log \eta} \right)}}$ for some absolute constant $c$.
In the case where $\optlis(S) \geq \tau^*$, the size of this increasing subsequence $S^*$ is at least, 

\begin{equation}
    |S^*| \geq \frac{\optlis(S)}{\eta^3 N^{\left(\frac{c\log \log \eta}{\log \eta} \right)}} \geq \frac{\tau^*}{\eta^3 N^{\left(\frac{c\log \log \eta}{\log \eta} \right)}}.
\end{equation}

\paragraph{Case of small $\tau^*$.}
If $\tau^* \leq \eta^3 \cdot N^{\left(\frac{1}{2} - \eps + \frac{c\log \log \eta}{\log \eta} \right)}$, we can use the algorithm $\alg_1$ from \Cref{obs: ncm-det-opt-space-sound} to distinguish the case where $\optlis(S) \geq \tau^*$ from the case where $\optlis(S) < \tau^*$.
The space complexity of this algorithm is $O(\tau^*) \leq N^{1/2 - \eps + o(1)}$, since $\eta = \eta(N) = 2^{\floor{\log \log \log N}}$.

\paragraph{Case of large $\tau^*$.}
If $\tau^* \geq N^{1/2 + \eps}$, we can use the algorithm $\alg_2$ from \Cref{obs: ncm-det-n-by-opt-space-sound} to distinguish the case where $\optlis(S) \geq \tau^*$ from the case where $\optlis(S) < \tau^*/2$.
The space complexity of this algorithm is $\tilde O(N / \tau^*) = \tilde O(N^{1/2 - \eps}) \leq N^{1/2 - \eps + o(1)}$.

\paragraph{Case of intermediate $\tau^*$.}
We now focus on the case where $\eta^3 \cdot N^{\left(\frac{1}{2} - \eps + \frac{c\log \log \eta}{\log \eta} \right)} < \tau^* < N^{\left( \frac{1}{2} + \eps \right)}$.
From \Cref{cor: ncm-partition-lemma}, there is a collection $\lset = \lset_{N, \eta}$ of $N^{O\left(\frac{\log \log \eta}{\log \eta} \right)} = N^{o(1)}$ potential ensembles, such that $\Upsilon_S \in \lset$.
We can process all potential ensembles $\Upsilon \in \lset$ in parallel, incurring a multiplicative factor of $|\lset| = N^{o(1)}$ in our space complexity.
Using standard techniques, it suffices to show the following.
Assume that we are processing an ensemble $\Upsilon = (\Psi, \Psi',\vectZ, \vectDelta, \vectMu)$ from $\lset$, where $\Psi = \Psi^*(N, \eta) = (\eta, \ldots, \eta)$ is a sequence of $r$ integers, each with value $\eta$.
Let $\vectZ~=~(Z_0, \ldots, Z_r)$, and we are guaranteed that $\frac{\tau^*}{\eta^3 \cdot N^{\left(\frac{c\log \log \eta}{\log \eta} \right)}} \leq Z_0 \leq \tau^*$, which implies, $N^{1/2 - \eps} \leq Z_0 \leq N^{1/2 + \eps}$.
Our objective is then to distinguish between two cases: whether $S$ contains a $\Upsilon$-canonical increasing subsequence of length exactly $Z_0$ (\yi), or $\optlis(S) < Z_0/N^{o(1)}$ (\ni).
We formalize this goal as follows.

\paragraph{$\alpha$-canonical distinguisher algorithm.}
Consider an algorithm $\alg$ that is given an ensemble $\Upsilon = (\Psi, \Psi',\vectZ, \vectDelta, \vectMu)$ from $\lset$ as mentioned above and let $\vectZ~=~(Z_0, \ldots, Z_r)$.
It is then given access to the elements of the original input sequence $S$ in the streaming model.
For $\alpha \geq 1$, we say that $\alg$ is an \emph{$\alpha$-canonical distinguisher algorithm} if given a sequence $S$ in the streaming model and an ensemble $\Upsilon$ as mentioned above, it achieves the following guarantees:
\begin{itemize}
    \item \emph{Completeness Guarantee.} If $S$ has an $\Upsilon$-canonical increasing subsequence of length $Z_0$ (\yi), it reports yes with probability at least $3/4$; and
    \item \emph{Soundness Guarantee.} If $\optlis(S) < Z_0/\alpha$ (\ni), it reports no with probability at least $3/4$.
\end{itemize}

In the remainder of this section, our goal is to compute a $N^{o(1)}$-canonical distinguisher algorithm for the case where we are given an ensemble $\Upsilon = (\Psi, \Psi',\vectZ, \vectDelta, \vectMu)$ from $\lset$ as mentioned above with $\vectZ~=~(Z_0, \ldots, Z_r)$, where $N^{1/2 - \eps} \leq Z_0 \leq N^{1/2 + \eps}$.
We consider the hierarchical partition $\bset_{\Psi}(S)$ of the input sequence $S$ into stream-blocks and the hierarchical partition $\bset_{\Psi'}(H^*)$ of the range $H^*$ into range-blocks.
We let $\iset = \set{0, \ldots, r}$ be the set of levels of these hierarchical partitions.
For each such level $i \in \iset$, we denote by $X_i$ the size of the level-$i$ stream-blocks of $\bset_{\Psi}^{(i)}(S)$.
It is immediate to see that $X_i = N/\eta^i$.
We let $\Psi' = (\psi_1, \ldots, \psi_r)$, $\vectDelta = (\Delta_0, \ldots, \Delta_r)$, and $\vectMu = (\mu_0, \ldots, \mu_r)$.
We start with the following simple observation.

\begin{observation} \label{obs: ncm-decreasing-xi-by-zi}
    We can assume w.l.o.g. that
    (i) $N^{1/2 + \eps} \geq \frac{X_0}{Z_0} \geq \ldots \geq \frac{X_r}{Z_r}$;
    (ii) $\Delta_0 \geq \ldots \geq \Delta_r$;
    (iii) $\mu_0 \geq \ldots \geq \mu_r$; and
    (iv) for each level $0 \leq i < r$, $\mu_i \leq \Delta_i$.
\end{observation}
\begin{proof}
    As mentioned earlier, we have $Z_0  \geq N^{1/2 - \eps}$, and hence, $\frac{X_0}{Z_0} = \frac{N}{Z_0} < N^{1/2 + \eps}$.
    We claim that if either of the three guarantees of \Cref{obs: ncm-decreasing-xi-by-zi} do not hold, then we must have $\optlis(S) < Z_0$.
    We can then correctly report that $S$ is a \ni obtaining a $1$-distinguisher algorithm.
    Thus, assume from now that at least one of the guarantees of \Cref{obs: ncm-decreasing-xi-by-zi} does not hold, or in other words, there is a level $0 \leq i < r$ with either $\frac{X_i}{Z_i} < \frac{X_{i+1}}{Z_{i+1}}$ or $\Delta_i < \Delta_{i+1}$ or $\mu_i < \mu_{i+1}$ or $\mu_i > \Delta_i$.
    Further, assume for contradiction that $S$ is a \yi, or equivalently, has an $\Upsilon$-canonical increasing subsequence $S^*$ of length $Z_0$.

    We first deal with the case where there is a level $0 \leq i < r$ with $\frac{X_i}{Z_i} < \frac{X_{i+1}}{Z_{i+1}}$, or in other words, $Z_i > \frac{X_i}{X_{i+1}} Z_{i+1} = \eta Z_{i+1}$.
    Consider some level-$i$ yes-block $B \in \bset^{i}_\Psi(S)$ for $S^*$.
    Recall that each level-$(i+1)$ child-block of $B$ may contribute at most $Z_{i+1}$ elements to $S^*$ (see, Property \ref{prop: ncm-p1} of the ensemble $\Upsilon$).
    Since there are only $\eta$ child-blocks of $B$, their collective contribution to $S^*$ is at most $\eta Z_{i+1}$, contradicting the fact that $B$ contributes $Z_i > \eta Z_{i+1}$ elements to $S^*$.

    We now consider the second case where there is a level $0 \leq i < r$ with  $\Delta_i < \Delta_{i+1}$.
    Recall the Properties \ref{prop: ncm-p1} and \ref{prop: ncm-p2} of the ensemble $\Upsilon$.
    For each level-$i$ yes-pair $(B, B')$, we have $|B~\cap~B'|~<~2Z_i \Delta_i$.
    Moreover, $(B, B')$ has exactly $\frac{Z_i}{Z_{i+1}}$ level-$(i+1)$ child-pairs that are level-$(i+1)$ yes-pairs, where each such pair $(\hat B, \hat B')$ has $|\hat B \cap \hat B'| \geq Z_{i+1} \Delta_{i+1}$.
    Hence, 
        
        \[2 Z_i \Delta_{i} > |B \cap B'| \geq \frac{Z_i}{Z_{i+1}} \cdot Z_{i+1}\Delta_{i+1} = Z_i \Delta_{i+1}. \]
    
    We conclude that $\Delta_i < \Delta_{i+1} < 2 \Delta_i$, a contradiction to the fact that both, $\Delta_i$ and $\Delta_{i+1}$ are integral powers of $2$.

    We now consider the third case where there is a level $0 \leq i < r$ with  $\mu_i < \mu_{i+1}$.
    Notice that $i \leq r-2$ must hold since $\mu_{r-1} \geq 1 = \mu_r$.
    Proceeding as in the second case, consider an arbitrary level-$i$ yes-pair $(B, B')$ and an arbitrary level-$(i+1)$ descendant yes-pair $(\hat B, \hat B')$ of $(B^*, B^{*'})$.
    From Property \ref{prop: ncm-p3} of the ensemble $\Upsilon$, we have $|\hat B \cap B'| < 2 Z_{i+1} \mu_{i}$.
    Note that $(\hat B, \hat B')$ has $Z_{i+1}/Z_{i+2}$ level-$(i+2)$ child-pairs that are level-$(i+2)$ yes-pairs, where each such pair $(\tilde B, \tilde B')$ has $|\tilde B \cap \hat B'| \geq Z_{i+2} \mu_{i+1}$, implying, $|\hat B \cap \hat B'| \geq \frac{Z_{i+1}}{Z_{i+2}} \cdot Z_{i+2} \mu_{i+1} = Z_{i+1} \mu_{i+1}$.
    On the other hand, $\hat B'$ is a subblock of $B'$, and we obtain,
    
    \[ Z_{i+1} \mu_{i+1} \leq |\hat B \cap \hat B'| \leq |\hat B \cap B'| < 2 Z_{i+1} \mu_{i-1} \], 
    
    or equivalently, $\mu_i < \mu_{i+1} < 2 \mu_{i}$.
    This is a contradiction to the fact that both, $\mu_i$ and $\mu_{i+1}$ are integral powers of $2$.

    Finally, consider the last case where there is a level $0 \leq i < r$ with   $\mu_i > \Delta_i$.
    Proceeding as in the second case, consider an arbitrary level-$i$ yes-pair $(B, B')$.
    From Property \ref{prop: ncm-p2} of the ensemble $\Upsilon$, $Z_i \Delta_i \leq |B \cap B'| < 2Z_i \Delta_i$.
    Recall Properties \ref{prop: ncm-p1} and \ref{prop: ncm-p3} of the ensemble $\Upsilon$.
    Note that $(B, B')$ has $Z_i/Z_{i+1}$ level-$(i+1)$ child-pairs that are level-$(i+1)$ yes-pairs, where each such pair $(\hat B, \hat B')$ has $|\hat B \cap B'| \geq Z_{i+1} \mu_{i}$, implying, $|B \cap B'| \geq \frac{Z_i}{Z_{i+1}} \cdot Z_{i+1}\mu_{i} = Z_i \mu_{i}$.
    We now conclude that $Z_i \mu_{i} \leq |B \cap B'| < 2 Z_i \Delta_i$, or in other words, $\Delta_i < \mu_i < 2 \Delta_i$, a contradiction to the fact that both, $\mu_i$ and $\Delta_i$ are integral powers of $2$.
    This completes the proof of \Cref{obs: ncm-decreasing-xi-by-zi}.
\end{proof}

We thus assume from now on that (i) $N^{1/2 + \eps} \geq \frac{X_0}{Z_0} \geq \ldots \geq \frac{X_r}{Z_r}$;
(ii) $\Delta_0 \geq \ldots \geq \Delta_r$;
(iii) $\mu_0 \geq \ldots \geq \mu_r$; and
(iv) for each level $0 \leq i < r$, $\mu_i \leq \Delta_i$.

\newcommand{\algspl}[1]{{\mathsf{AlgCase}}_{\mathsf{#1}}}

\paragraph*{Plan.}
We consider a number of cases, depending on the values of the parameters of our ensemble $\Upsilon$.
In \Cref{subsec: ncm-small-xi-by-zi}, we consider special case $1$ where there is some level $0 \leq i < (1/2 - \eps)r$ such that $X_i/Z_i \leq N^{1/2 - \eps}$.
In this case, we present a $O(1)$-canonical distinguisher algorithm $\algspl{1}$ with space complexity $N^{1/2 - \eps + o(1)}$.
In \Cref{subsec: ncm-large-deltai-by-zi}, we analyze the second special case where there is some level $3\eps r \leq i < (1/2 - \eps)r$ such that $\Delta_i \geq Z_i N^{5\eps}$.
In this case, we show an $O(1)$-canonical distinguisher algorithm $\algspl{2}$ with space complexity $N^{1/2 - \eps + o(1)}$.
We also define more technical special cases, namely $3$ and $4$, in \Cref{subsec: ncm-processing-upsilon}, and prove that at least one of these four special cases must occur.
Assuming the occurrence of special case $3$, we demonstrate an $N^{o(1)}$-canonical distinguisher algorithm $\algspl{3}$ with space complexity $N^{1/2 - \eps + o(1)}$.
The description of this algorithm $\algspl{3}$ is present in \Cref{subsec: ncm-jumpable} with some details deferred to \Cref{{subsec: ncm-jumpable-alggen-desc},{subsec: ncm-jump-jump-i}}.
Finally, if the special case $4$ occurs, we utilize the algorithm $\algncm$ to show an $N^{o(1)}$-canonical distinguisher algorithm $\algspl{4}$ with space complexity $N^{1/2 - \eps + o(1)}$.
We describe this algorithm in \Cref{subsec: ncm-reduction-to-ncm} with some details present in \Cref{subsec: ncm-savale-savable}.

\subsection{Special Case $1$ :  Stream-blocks are Small} \label{subsec: ncm-small-xi-by-zi}
Recall that we have fixed parameters $N$ and $\eta$, both integral powers of $2$.
We are given an \LIS problem instance $S$ in the streaming model, where $S$ is a permutation of the range $H^* = (1, \ldots, N)$.
Additionally, we are given an ensemble $\Upsilon = (\Psi, \Psi', \vectZ, \vectDelta, \vectMu)$ of length $1+r$ 
where $r = r(N, \eta)$, $\Psi = \Psi^*(N, \eta) = (\eta, \ldots, \eta)$, and $\vectZ = (Z_0, \ldots, Z_r)$.
Recall that we denote the size of the level-$i$ stream-blocks of the hierarchical partition $\bset_{\Psi}^{i}(S)$ by $X_i$, for each level $i \in \iset$. 
In this special case, we assume that there is exists a level $0 \leq i < (1/2 - \eps)r$ such that $X_i/Z_i \leq N^{1/2 - \eps}$.
Our goal is to devise an $O(1)$-canonical distinguisher algorithm $\algspl{1}$ with low space complexity.
We will show the following lemma that will imply such a space-efficient algorithm for this special case.

\begin{lemma} \label{lem: ncm-lis-small-xi-by-zi}
    Suppose we are given a sequence $S$ of length $N$ taking distinct values in the range $H^* = (1, \ldots, N)$, where $N$ is an integral power of $2$.
    We are also given a level $i$ and level-$i$ partitions $\bset = \bset^i_\Psi(S)$ of stream $S$ and $\bset' = \bset^i_{\Psi'}(H^*)$ of range $H^*$ into stream-blocks and range-blocks respectively along with parameters $Z_0$ and $Z_i$.
    Then there is a $O(1)$-canonical distinguisher algorithm with space complexity $\tilde O\left(\frac{X_i + Z_0}{Z_i} \right)$, where $X_i$ is the length of level-$i$ stream-blocks.
\end{lemma}

We will prove the above lemma later, after proving the following corollary assuming it.

\begin{corollary}\label{cor: ncm-small-xi-by-zi}
    Suppose we are given parameters $N$ and $\eta$, both integral powers of $2$, along with an ensemble $\Upsilon = (\Psi, \Psi', \vectZ, \vectDelta, \vectMu)$ of length $1 + r$, where $\vectZ = (Z_0, \ldots, Z_r)$.
    Further, assume that there is some level  $0 \leq i < (1/2 - \eps)r$ with $X_i/Z_i \leq N^{1/2 - \eps}$.
    Then there is a $O(1)$-canonical distinguisher algorithm $\algspl{1}$ with space complexity $N^{1/2 - \eps + o(1)}$.
\end{corollary}
\begin{proof}
    Consider the input \LIS problem instance stream $S$ that is a permutation of the range $H^* = (1, \ldots, N)$.
    We consider the partition $\bset = \bset^i_\Psi(S)$ of $S$ into stream-blocks and the partition $\bset' = \bset^i_{\Psi'}(H^*)$ of $H^*$ into range-blocks.
    Our algorithm $\algspl{1}$ now runs the algorithm of \Cref{lem: ncm-lis-small-xi-by-zi} on the level-$i$ partitions $\bset$ and $\bset'$ along with parameters $Z_0$ and $Z_i$, and reports its outcome.
    Indeed, $\algspl{1}$ is a $O(1)$-canonical distinguisher algorithm and has space complexity $\tilde O \left(\frac{X_i}{Z_i} + \frac{Z_0}{Z_i} \right)$.

    We will now show that this space complexity is upper bounded by $N^{1/2 - \eps + o(1)}$.
    Indeed, in this special case, we have $\frac{X_i}{Z_i} \leq N^{1/2 - \eps}$.
    Moreover, from \Cref{obs: ncm-decreasing-xi-by-zi}, we can assume w.l.o.g. that $\frac{X_i}{Z_i} \leq \frac{X_0}{Z_0} = \frac{N}{Z_0}$, or in other words, $\frac{Z_0}{Z_i} \leq \frac{N}{X_i} = \eta^i \leq N^{1/2 - \eps}$.
    Here, we have used the fact that $i < (1/2 - \eps)r$ and $r = r(N, \eta) = \floor{\frac{\log N}{\log \eta}}$.
    Thus, the space complexity of $\algspl{1}$ is indeed bounded by $\tilde O(N^{1/2 - \eps}) \leq N^{1/2 - \eps + o(1)}$ and the corollary now follows.
\end{proof}

This completes the proof of the special case $1$ where $X_i/Z_i \leq N^{1/2 - \eps}$ for some level $0 \leq i < (1/2 - \eps)r$.
We now turn towards proving \Cref{lem: ncm-lis-small-xi-by-zi}.
Recall that we are given a sequence $S$ of length $N$ taking distinct values in the range $H^* = (1, \ldots, N)$ and level-$i$ partitions $\bset$ and $\bset'$ along with parameters $Z_0$ and $Z_i$.
The central ingredient of our proof, that we will also use in the future, is the notion of \emph{block-processing algorithm} that we describe next.

\paragraph{Block-Processing Algorithm.}
The input to the block-processing algorithm is a \LIS problem is a stream-block $B$ of the original input stream $S$ along with a partition $\bset'$ of the range $H^*$ into range-blocks.
The algorithm is given access to the elements of $B$ as they arrive as a part of the original input stream $S$.
We say that a range-block $B' \in \bset'$ is \emph{$(Z, \Delta)$-perfect for $B$} iff there is an increasing subsequence in $B \cap B'$ of length $Z$ that does not contain any of the first $\Delta Z$ elements of $B \cap B'$.
The goal of the block-processing algorithm is to output some subset $\bset^* \subseteq \bset'$ by reporting their respective indices.
For thresholds $Z$ and $\Delta$, and a parameter $\alpha \geq 1$, we say that a (probabilistic) block-processing algorithm is an $(Z, \Delta, \alpha)$-approximation algorithm iff the following guarantees hold:
\begin{itemize}
    \item \emph{Soundness Guarantee.} With probability at least $3/4$, \emph{every} range-block $B' \in \bset^*$ has $\optlis(B \cap B') \geq Z/\alpha$;
    \item \emph{Completeness Guarantee.} Each range-block $B' \in \bset'$ that is $(Z, \Delta)$-perfect for $B$ is present in $\bset^*$ with probability at least $3/4$.
\end{itemize}

\begin{claim} \label{clm: ncm-block-processing-alg}
    For each parameter $Z$ there is a block-processing algorithm with space complexity $\tilde O\left({|B|/{Z}} \right)$ that achieves $(Z, 1, 2)$-approximation, where $B$ is the input stream-block.
\end{claim}

We give a sketch of the proof of the above claim here, while the complete proof, using standard techniques, is deferred to \Cref{appn: ncm-small-xi-by-zi}.
The key idea is to sample the elements of $B$ at the rate of roughly $1/Z$ as they appear in the stream.
We say that a range-block $B' \in \bset'$ is `suspicious' if we sample an element whose value lies in $B'$.
We show that we will identify all range-blocks $B'$ containing a large number of values from $B$, which in turn are the only ones that can possibly induce a large enough increasing subsequence in $B$.
Moreover, we can also guarantee that we will identify suspicious range-blocks early on: there is still a large enough induced subsequence present in the subsequent elements of $B$.
We can now use algorithm $\alg_2$ from \Cref{obs: ncm-det-n-by-opt-space-sound} on the sampled blocks for these subsequent elements to find ones having large enough induced increasing subsequence.
Since we mark roughly $|B|/Z$ range-blocks as suspicious and for each suspicious block $B'$ we use roughly $O(|B \cap B'|/Z)$ space, our overall space complexity is also bounded by $O \left(|B|/Z \right) + O \left(\sum_{B'} \ceil{|B \cap B'|/Z} \right) = O \left({|B|/Z} \right)$ as required.
Here, the summation is over all suspicious range-blocks, which in turn is a subset of the partition $\bset'$.
We are now ready to prove \Cref{lem: ncm-lis-small-xi-by-zi} that will complete the analysis of this case.

\proofof{\Cref{lem: ncm-lis-small-xi-by-zi}}
    We are given a partition $\bset = \bset^i_\Psi(S) =  \set{B_1, \ldots, B_{m}}$ of $S$ into $m = \eta^i$ stream-blocks, each of size exactly $X_i = N/\eta^i$.
    Similarly, we are given a partition $\bset' = \bset^i_{\Psi'}(H^*) = \set{B'_1, \ldots, B'_{m'}}$ of the range $H^*$ into a number of range-blocks.

    For each stream-block $B_k \in \bset$, when the elements of $B_k$ arrive as a part of the stream $S$,
    we apply the block-processing algorithm $\algbp$ of \Cref{clm: ncm-block-processing-alg} with parameter $Z = Z_i/2$ and the set $\bset'$ of range-blocks.
    Let $\bset'_k \subseteq \set{1, \ldots, m'}$ be the set of indices of the range-blocks returned by $\algbp$.
    Consider now a stream $\sset$ that contains elements of the sets $\bset'_1, \ldots, \bset'_{k^*}$ in the reversed order of that elements.
    In other words, the stream $\sset$ can be divided into $k^*$ chunks, one chunk corresponding to each stream-block $B_k \in \bset$.
    The chunk corresponding to $B_k$ contains the elements of $\bset'_k$ in the decreasing order of their values.
    Notice that the range of this stream $\sset$ is $\hset^* = \set{1, \ldots, m'}$.
    We apply the algorithm $\alg_1$ of \Cref{obs: ncm-det-opt-space-sound} on the stream $\sset$ with parameter $\ceil{Z_0/10 Z_i}$ and report its answer.
    This completes the description of our algorithm, and we now turn to  analyze its properties.

    From the guarantees of $\alg_1$ and \Cref{clm: ncm-block-processing-alg}, the space complexity of our algorithm is indeed bounded by $\max_{B \in \bset} \left(\tilde O \left({\frac{|B|}{Z_i}}  \right)\right) + O\left( \frac{Z_0}{Z_i} \right) = \tilde O \left( \frac{X_i + Z_0}{Z_i} \right)$ as claimed.
    Furthermore, it is immediate to verify that if we report yes, $\optlis(S) \geq \frac{Z_0}{10 Z_i} \cdot \frac{Z_i}{2} = \frac{Z_0}{20}$ must hold with probability at least $3/4$.
    This completes the proof of soundness guarantee of our algorithm.

    We now show the completeness guarantee.
    Assume that there is a $\Upsilon$-canonical subsequence $S^*$ of $S$ of length $Z_0$.
    Our goal is to show that in this case, we report in affirmative with probability at least $3/4$.
    Consider a level=$i$ yes-pair $(B_j, B'_{j'})$ of $S^*$.
    Note that $B'_{j'}$ is a $(Z_i/2, 1)$-perfect range-block for $B_j$.
    Hence, from the correctness guarantee of $\algbp$ of \Cref{clm: ncm-block-processing-alg}, $j' \in \bset'_j$ with probability at least $3/4$.
    We consider an increasing subsequence $\sset^*$ of $\sset$ that is obtained as follows.
    For each yes-pair $(B_j, B'_{j'}) \in \bset \times \bset'$ of $S^*$, if $B' \in \bset'_{j'}$, we choose the element $j'$ present in the chunk corresponding to $B_j$ of $\sset$ to lie in $\sset^*$.
    It is immediate to verify that $\sset^*$ is indeed an increasing subsequence of $\sset$.
    Moreover, each yes-block $B_j \in \bset$ participates in a fresh execution of $\algbp$ and hence, the element corresponding to each level-$i$ yes-pair of $S^*$ is present in $\sset^*$ independently with probability at least $3/4$ each.
    We first consider the case where $Z_i \leq Z_0 \leq 10Z_i$.
    In this case, with probability at least $3/4$, we correctly identify at least one yes-pair and report yes.
    Thus, assume from now on that $Z_0 > 10 Z_i$.
    From the above discussion, $\expect{\optlis(\sset^*)} \geq \frac{3}{4} \cdot \frac{|S^*|}{Z_i}  = \frac{3Z_0}{4 Z_i}$.
    From Chernoff bound (\Cref{fact: ncm-chernoff-version}),
    
    \[ \prob{|\sset^*| \leq \frac{Z_0}{4Z_i}} \leq e^{-\left(\frac{2}{9} \expect{\optlis(\sset^*)} \right)} \leq e^{-\left(\frac{2}{9} \cdot \frac{3Z_0}{4 Z_i}\right)} = e^{-\left( \frac{1}{6} \cdot \frac{Z_0}{Z_i} \right)} \leq e^{-\frac{10}{6}} < \frac{1}{4}. \]

    Thus, with probability at least $3/4$, we have $\optlis(\sset^*) \geq |\sset^*| > \frac{Z_0}{4 Z_i} > \ceil{\frac{Z_0}{10 Z_1}}$ and we report yes.
    This completes the proof of \Cref{lem: ncm-lis-small-xi-by-zi}.
\endproofof

In this subsection, we have shown an $O(1)$-canonical distinguisher algorithm $\algspl{1}$ assuming that there is a level $0 \leq i < (1/2 - \eps)r$ such that $\frac{X_i}{Z_i} \leq N^{1/2 - \eps}$.
In the remainder of this section we can now assume:

\begin{assumption} \label{assm: large-xi-by-zi}
    For each level $0 \leq i < (1/2 - \eps)r$, $\frac{X_i}{Z_i} > N^{1/2 - \eps}$.
\end{assumption}

\subsection{Special Case 2:  Yes-pairs Share Many Common Elements} \label{subsec: ncm-large-deltai-by-zi}
Recall that we have fixed parameters $N$ and $\eta$, both integral powers of $2$, such that $\eta$ grows with $N$.
In this subsection, we are given an \LIS problem instance $S$ in the streaming model.
We assume that $S = (a_1, \ldots, a_N)$ is a permutation of the range $H^* = (1, \ldots, N)$.
Additionally, we are given an ensemble $\Upsilon = (\Psi, \Psi', \vectZ, \vectDelta, \vectMu)$ of length $1+r$ 
where $r = r(N, \eta) = \floor{\frac{\log N}{\log \eta}}$, $\Psi = \Psi^*(N, \eta) = (\eta, \ldots, \eta)$, $\Psi' = (\psi_1, \ldots, \psi_r)$, $\vectZ = (Z_0, \ldots, Z_r)$, $\vectDelta = (\Delta_0, \ldots, \Delta_r)$, and $\vectMu = (\mu_0, \ldots, \mu_r)$.
We let $\iset = \set{0, \ldots, r}$ be the set of the levels of the underlying hierarchical partitions $\bset_\Psi(S)$ of $S$ into stream-blocks and $\bset_{\Psi'}(H^*)$ of $H^*$ into range-blocks.
From \Cref{assm: large-xi-by-zi}, $X_i/Z_i > N^{1/2 - \eps}$ holds for each level $0 \leq i < (1/2 - \eps)r$.
In this subsection, we assume that there is some level $3\eps r \leq i < (1/2 - \eps)r$ with $\Delta_i \geq Z_i N^{5\eps}$.
We fix such a level $i$, and we will show an $O(1)$-canonical distinguisher algorithm $\algspl{2}$ with space complexity $N^{1/2 - \eps + o(1)}$.

As in \Cref{subsec: ncm-small-xi-by-zi}, the core of our algorithm is a block-processing algorithm exploiting the structure of $\Upsilon$: the fact that, $\Delta_i \geq Z_i N^{5\eps}$.
Unlike in \Cref{subsec: ncm-small-xi-by-zi} this block-processing algorithm operates on a carefully chosen intermediate level $0 \leq j < i$.
We use the following simple claim, whose proof is present in \Cref{prf: ncm-level-with-strictly-small-z}.

\begin{claim} \label{clm: ncm-level-with-strictly-small-z}
    There is a level $j \geq i - 3\eps r$ such that $2Z_i \leq Z_j \leq \eta Z_i$.
\end{claim}

We fix a level $j \geq  i - 3 \eps r \geq 0$ as guaranteed by \Cref{clm: ncm-level-with-strictly-small-z}.
We consider the collection $\bset = \bset_\Psi^j(S)$  of level-$j$ stream-blocks and the collection $\bset' = \bset_{\Psi'}^j(H^*)$ of level-$j$ range-blocks.
We are now ready to state our block-processing algorithm for this special case. 
The input to this algorithm is a level-$j$ stream-block $B \in \bset$ of the original input stream $S$ along with the level-$j$ partition $\bset'$ of $H^*$.
Recall that we say a range-block $B'$ is $(Z, \Delta)$-perfect for $B$ iff there is an increasing subsequence that does not use first $\Delta Z$ elements of $B \cap B'$.
Further, recall that we say that a block-processing algorithm achieves $(Z, \Delta, \alpha)$-approximation iff with probability at least $3/4$, every reported range-block $B'$ has $\optlis(B \cap B') \geq Z/\alpha$ and each range-block that is $(Z, \Delta)$-perfect for $B$ is reported independently with probability at least $3/4$.
In the following claim, we show a block-processing algorithm that is optimized for this special case.

\begin{claim} \label{clm: ncm-double-level-block-processing}
    There is a block-processing algorithm $\algbp$ that achieves $(Z_j/2, \Delta_i/8, 1)$-approximation with space complexity $N^{1/2 - \eps + o(1)}$,  where the input stream-block is a level-$j$ stream-block and the partition of $H^*$ is the level-$j$ partition $\bset^j_{\Psi'}(H^*)$.
\end{claim}

The proof of \Cref{clm: ncm-double-level-block-processing} follows closely that of \Cref{clm: ncm-block-processing-alg} and is deferred to \Cref{prf: ncm-double-level-block-processing},
We are now ready to describe the main algorithm $\algspl{2}$ of this subsection.

\begin{corollary}
    Suppose we are given parameters $N$ and $\eta$, both integral powers of $2$, along with an ensemble $\Upsilon = (\Psi, \Psi', \vectZ, \vectDelta, \vectMu)$ of length $1 + r$, where $\vectZ = (Z_0, \ldots, Z_r)$ and $\vectDelta = (\Delta_0, \ldots, \Delta_r)$.
    Further assume that there is a level $3\eps r \leq i \leq (1/2 - \eps)r$ with $\Delta_i \geq \DeltaLowerBound$.
    Then there is a $O(1)$-canonical distinguisher algorithm $\algspl{2}$ with space complexity $N^{1/2 - \eps + o(1)}$.
\end{corollary}
\begin{proof}
    We proceed exactly as in the proof of \Cref{lem: ncm-lis-small-xi-by-zi}.
    Recall that we have fixed a level $j$ and level-$j$ partitions $\bset$ and $\bset'$.
    We let $\bset = \set{B_1, \ldots, B_m}$ and $\bset' = \set{B'_{1}, \ldots, B'_{m'}}$ be their constituent blocks in their natural order.
    
    We initialize a new \LIS problem instance $\sset$ with range $\hset^* = \set{1, \ldots, m'}$.
    We now describe the procedure to generate elements of $\sset$.
    We will ensure that the sequence $\sset$ can be divided into $m$ chunks, each corresponding to a level-$j$ stream-block of $\bset$.
    These chunks appear in the natural order of these level-$j$ stream-blocks.
    Consider a stream-block $B_k \in \bset$.
    When its elements arrive as a part of the sequence $S$, we apply the block-processing algorithm $\algbp$ of \Cref{clm: ncm-double-level-block-processing} and let $\bset'_k$ be the indices of the range-blocks of $\bset'$ it reports.
    The chunk of $\sset$ corresponding to $B_k$ contains the elements of $\bset'_k$ in decreasing order of their indices.
    We run in parallel algorithm $\alg_1$ of \Cref{obs: ncm-det-opt-space-sound} on this stream $\sset$ with the parameter $\ceil{Z_0/10 Z_j}$.
    At the end of stream $S$, we terminate the stream $\sset$ and report the answer of $\alg_1$ on $\sset$.
    This completes the description of our algorithm $\algspl{2}$.

    By proceeding exactly as in \Cref{lem: ncm-lis-small-xi-by-zi}, we verify that $\algspl{2}$ is indeed the desired $O(1)$-canonical distinguisher algorithm for this case.
    Notice that there is at most $1$ execution of $\algbp$ at a time.
    Thus, the space complexity of $\algspl{2}$ is bounded by that of $\algbp$ and $\alg_1$.
    From \Cref{obs: ncm-det-opt-space-sound,clm: ncm-double-level-block-processing} this is bounded by,
    \[ N^{1/2 - \eps + o(1)} +  O\left(\ceil{\frac{Z_0}{10Z_j}} \right) \leq N^{1/2 - \eps + o(1)}.\]

    Here, the inequality follows from \Cref{obs: ncm-decreasing-xi-by-zi} since,

    \[\frac{Z_0}{Z_j} \leq \frac{X_0}{X_j} = \frac{N}{N/\eta^j} = \eta^j < \eta^i < \eta^{(1/2 - \eps)r} \leq N^{1/2 - \eps}. \]
\end{proof}

In this subsection, we have shown an $O(1)$-canonical distinguisher algorithm $\algspl{2}$ with space complexity $N^{1/2 - \eps + o(1)}$ assuming that there is a level $3 \eps r \leq i < (1/2 - \eps)r$ such that $\Delta_i \geq N^{5\eps}$.
In the remainder of this section, in addition to \Cref{assm: large-xi-by-zi}, we also assume:

\begin{assumption} \label{assm: small-deltai-by-zi}
    For each level $3\eps r \leq i < (1/2 - \eps)r$, $\Delta_i < Z_i N^{5\eps}$.
\end{assumption}

\subsection{\LIS Streaming Algorithm with Space Complexity $N^{1/2 + O(\eps)}$}\label{subsec: ncm-simpler-lis-stream}
\newcommand{\algsimp}{\alg_{\mathsf{simp}}}

Recall that we have fixed parameters $N$ and $\eta$, both integral powers of $2$, such that $\eta$ grows with $N$.
We are given an \LIS problem instance $S$ in the streaming model, where $S$ is a permutation of the range $H^* = (1, \ldots, N)$.
Additionally, we are given an ensemble $\Upsilon = (\Psi, \Psi', \vectZ, \vectDelta, \vectMu)$ of length $1+r$ 
where $r = r(N, \eta) = \floor{\frac{\log N}{\log \eta}}$, $\Psi = \Psi^*(N, \eta) = (\eta, \ldots, \eta)$, $\Psi' = (\psi_1, \ldots, \psi_r)$, $\vectZ = (Z_0, \ldots, Z_r)$ $\vectDelta = (\Delta_0, \ldots, \Delta_r)$, and $\vectMu = (\mu_0, \ldots, \mu_r)$.
We let $\iset = \set{0, \ldots, r}$ be the set of the levels of the underlying hierarchical partitions $\bset_\Psi(S)$ of $S$ into stream-blocks and $\bset_{\Psi'}(H^*)$ of $H^*$ into range-blocks.
Recall that we denote the size of the level-$i$ stream-blocks of the hierarchical partition $\bset_{\Psi}^{i}(S)$ by $X_i$, for each level $i \in \iset$. 
Also recall that we have fixed an optimal $\Upsilon$-canonical subsequence $S^*$ of $S$.
In this section, we will devise an $N^{o(1)}$-canonical distinguisher algorithm with space complexity $N^{1/2 + O(\eps)}$.
Even though this is worse bound than that achieved by algorithm $\alg_3$ of \Cref{obs: ncm-det-sqrt-n-space-sound}, this algorithm will serve as an important guiding principle in the analysis of special cases $3$ and $4$ that we will describe and analyze in the subsequent subsections.

We fix a threshold-level $r^* = \rstarVal$ and define the set $\isetnew = \set{i \> : \> 0 \leq i < r^* \text{ and } Z_i > Z_{i+1}}$ of special levels.
Denote $\isetnew = \set{i_0, \ldots, i_{\knew}}$ where $\knew := |\isetnew| - 1$ and the levels are indexed in their natural order so that $i_0 < i_1 < \ldots < i_{\knew}$.
We will need the following consequence of \Cref{assm: large-xi-by-zi}, whose proof is deferred to \Cref{prf: ncm-simpler-algo-k-star}.

\begin{claim} \label{clm: ncm-simpler-algo-k-star}
    $\knew \geq \left(\frac{1}{2} - 7\eps\right)r \geq \frac{r}{3}$.
\end{claim}

For each pair $i_k < i_{k'}$ of levels of $\isetnew$, we let $\zeta(i_k, i_{k'}) = \ln{ \left(\eta^{2(i_{k'} - i_k)} \cdot \psi_{i_{k + 1}} \ldots \psi_{i_{k'}} \right)}$.
In other words, $e^{\zeta(i_k, i_{k'})} = \eta^{2(i_{k'} - i_k)} \cdot \left(\prod_{k < k'' \leq k'} \psi_{i_{k''}} \right)$.
Using Property \ref{prop: ncm-p4} of the ensemble $\Upsilon$, we immediately obtain the following fact:

\begin{fact}\label{fact: ncm-exp-zeta-bound}
    Fix levels $i_k, i_{k'} \in \isetnew$ with $0 \leq k < k' \leq \knew$.
    Consider a level-$i_k$ pair $(B, B')$ and the sets $\bset$ and $\bset'$ of level-$i_{k'}$ descendant-blocks of $B$ and $B'$ respectively.
    Then $e^{\zeta(i_k, i_{k'})} \geq |\bset|^2 |\bset'|$.
\end{fact}

\paragraph{Algorithm for different levels.}
In this section, we will present $1 + \knew$ different algorithms, one for each level $i_k \in \isetnew$.
These algorithms will be parametrized by the relevant parameters of the respective levels, and hence, it will be convenient to describe them based on the levels at which they operate.
Consider some level $i \in \isetnew$.
The level-$i$ algorithm, that we denote by $\alglevel{i}$, will be responsible for `classifying' level-$i$ pairs.
Formally, $\alglevel{i}$ takes as an input a level-$i$ pair $(B^*, B^{*'})$ and a subblock $B^{**}$ of $B^*$.
Note that $B^*$ is a level-$i$ block in the level-$i$ partition $\bset^i_\Psi$ of the original input stream $S$ into stream-blocks.
Similarly, $B^{*'}$ is a level-$i$ block in the level-$i$ partition $\bset^i_{\Psi'}$ of the original range $H^*$ into range-blocks.
Thus, $B^*$ and $B^{*'}$ can be identified by their indices in $\bset^i_\Psi$ and $\bset^i_{\Psi'}$ respectively.
We assume that $\alglevel{i}$ is given these indices as input.
On the other hand, $B^{**}$ is an arbitrary subblock of $B^*$: it may not belong to any level of the hierarchical partition $\bset_\Psi$ of $S$.
The algorithm $\alglevel{i}$ is then given access to the elements of $B^{**}$ as they arrive as a part of the block $B^*$, which in turn is a part of the original input sequence $S$.

\paragraph*{$(\alpha_i, \alpha'_i)$-canonical distinguisher algorithm.}
For parameters $\alpha'_i \geq \alpha_i \geq 1$, we say that the level-$i$ algorithm $\alglevel{i}$ is an \emph{$(\alpha_i, \alpha'_i)$-canonical distinguisher} iff for each input level-$i$ pair  $(B^*, B^{*'})$ and the subblock $B^{**}$ of $B^*$ it achieves the following guarantees:
\begin{itemize}
    \item if $(B^*, B^{*'})$ is a level-$i$ yes-pair for $S^*$ and $|S^* \cap B^{**}| \geq Z_i/\alpha_i$, then it reports yes with probability at least $3/4$; and
    \item if $\optlis(B^{**} \cap B^{*'}) < Z_i/\alpha'_i$, then it reports no with probability at least $3/4$.
\end{itemize}

The main result of this subsection is the existence of a level-$i_{k}$ $(\alpha_{i_k}, \alpha'_{i_k})$-canonical distinguisher algorithm $\alglevel{i_{k}}$ with low space complexity where $\alpha_{i_k} = 4$ and $\alpha'_{i_k} = 2^{3(\knew - k) + 2}$, for each level $i_k \in \isetnew$.
We summarize this in the following lemma:

\begin{lemma}\label{lem: ncm-simple-lis-main}
    Consider a level $i_k \in \isetnew$.
    Further assume that if $k < \knew$, we are also given a level-$i_{k+1}$ $(\alpha_{i_{k+1}}, \alpha'_{i_{k+1}})$-canonical distinguisher algorithm $\alglevel{i_{k+1}}$.
    Then, there is a level-$i_k$ $(\alpha_{i_{k}}, \alpha'_{i_{k}})$-canonical distinguisher algorithm $\alglevel{i_k}$ that achieves the following guarantees:
    \begin{itemize}
        \item \textbf{Calls to $\alglevel{i_{k+1}}$:} if $k < \knew$, it performs at most $\zeta^2(i_k, i_{k+1}) \frac{\mu_{i_k}}{\mu_{i_{k+1}}}$  concurrent calls to $\alglevel{i_{k+1}}$; and
        
        \item \textbf{Space complexity:} its space complexity, excluding the space required by the calls to $\alglevel{i_{k+1}}$, is at most $N^{6 \eps}$.
    \end{itemize}
\end{lemma}

We complete the proof of \Cref{lem: ncm-simple-lis-main} after describing the main algorithm of this subsection assuming it.
Applying the above lemma recursively for levels of $\isetnew$ in their decreasing order, we obtain the level-$i_0$ $(\alpha_{i_0}, \alpha'_{i_0})$-canonical distinguisher algorithm $\alglevel{i_0}$.
In the following claim, we analyze its space complexity.

\begin{claim}\label{clm: ncm-simple-alglevel-0-space}
    The space complexity of $\alglevel{i_0}$ is $N^{1/2 + 7\eps + o(1)}$.
\end{claim}
\begin{proof}
    The level-$i_0$ algorithm $\alglevel{i_0}$ performs a number of calls to the lower-level algorithms $\set{\alglevel{i_k} \> | \> i_k \in \iset^* \text{ and } i_k > i_0}$ as subroutines.
    Consider some level $i \in \iset^*$ and a level-$i$ stream-block $B$.
    We denote $\load(B)$ as the maximum number of concurrent calls to the level-$i$ algorithm $\alglevel{i}$ in which $B$ participates.
    We define $\load(i) := \max_{B \in \bset^i_\Psi} \load(B)$ to be the maximum `load' across all level-$i$ stream-blocks $\bset^i_\Psi$.
    To bound the space complexity of $\alglevel{i_0}$, we will need the following two claims whose proofs are deferred to \Cref{{prf: ncm-lis-somple-load-i-k-bound},{clm: ncm-lis-simple-pi-zeta-bound}} respectively.
    
    \begin{claim}\label{clm: ncm-lis-somple-load-i-k-bound}
        For each level $i = i_k \in \iset^*$, $\load(i_k) \leq \frac{\mu_{i_0}}{\mu_{i_k}} \cdot \zeta^2(i_0, i_1) \cdot \ldots \cdot \zeta^2(i_{k-1}, i_k)$.
    \end{claim}

    \begin{claim}\label{clm: ncm-lis-simple-pi-zeta-bound}
        $\zeta(i_0, i_1) \cdot \ldots \cdot \zeta(i_{k^*-1}, i_{k^*})   \leq N^{o(1)}$. %
    \end{claim}

    We are now ready to provide an upper bound for the space complexity of $\alglevel{i_0}$.
    We consider its state when processing some element $a_t \in S$.
    For each level $i \in \iset^*$, we let $B_i \in \bset^i_\Psi$ denote the unique level-$i$ stream-block that contains $a_t$.
    Consider now some level $i \in \iset^*$ and observe that $B_i$ participates in all active calls to the level-$i$ algorithm $\alglevel{i}$.
    Since there are at most $\load(i)$ such calls and from  \Cref{lem: ncm-simple-lis-main}, the space used by each such run, excluding the space used by the lower level algorithms is bounded by $N^{6\eps}$, we conclude that the space complexity of $\alglevel{i_0}$ is at most,

    \begin{align*}
        \sum_{i \in \iset^*} \load(i) \cdot N^{6\eps} &\leq |\iset^*| \cdot \max_{i \in \isetnew}{\left( \load(i) \right)} \cdot N^{6\eps} \\
        &\leq r^* \cdot \frac{\mu_{i_0}}{\mu_{i_{k^*}}} \cdot \zeta^2(i_0, i_1) \cdot \ldots \cdot \zeta^2(i_{k^*-1}, i_{k^*}) \cdot N^{6\eps} \\
        &\leq N^{o(1)} \cdot \mu_{0} \cdot N^{6\eps + o(1)}\\
        &\leq N^{1/2 + 7\eps + o(1)}.
    \end{align*}

    Here, the third inequality follows from \Cref{clm: ncm-lis-simple-pi-zeta-bound} and the last inequality follows from the fact that $\mu_0 \leq N^{1/2 + \eps}$ (see, \Cref{obs: ncm-decreasing-xi-by-zi}).
    This completes the proof of \Cref{clm: ncm-simple-alglevel-0-space}.
\end{proof}

In \Cref{prf: ncm-alglevel-0-suffices}, we prove the following observation.

\begin{observation}\label{obs: ncm-alglevel-0-suffices}
    Suppose there is a level-$i_{0}$ $(\alpha, \alpha')$-canonical distinguisher algorithm $\alg(i_{0})$.
    Then there is a $\alpha'$-canonical distinguisher algorithm $\algsimp$,
    that performs at most $N^{o(1)}$ concurrent calls to $\alg(i_{0})$,
    and whose space complexity, excluding the space required by the calls to $\alg(i_{0})$, is at most $N^{o(1)}$.
\end{observation}

From \Cref{{clm: ncm-simple-alglevel-0-space},{obs: ncm-alglevel-0-suffices}}, and using the fact that $\alpha'_{i_0} = 2^{O(r^*)} = N^{o(1)}$, we immediately obtain the main algorithm of this subsection.

\begin{corollary}
    There is a $N^{o(1)}$-canonical distinguisher algorithm $\algsimp$ with space complexity $N^{1/2 + 7\eps + o(1)}$.
\end{corollary}

In the remainder of the subsection, we prove \Cref{lem: ncm-simple-lis-main}.
We fix an integer $0 \leq k \leq \knew$ and the corresponding level $i_k \in \iset^*$.
We first consider the case where $k \in \set{\knew - 1, \knew}$, that serves as the `base cases' of our algorithm.

\paragraph*{$\alglevel{i_k}$ with $k \in \set{\knew - 1, \knew}$.}
The input to the algorithm $\alglevel{i_k}$ is a level-$i_k$ pair $(B^* B^{*'})$ and a subblock $B^{**}$ of $B^*$.
It is given access to the elements of $B^{**}$ as they arrive as the part of the original input stream $S$.
On this input, the algorithm $\alglevel{i_k}$ runs the algorithm $\alg_1$ from \Cref{obs: ncm-det-opt-space-sound} for the input stream $B^{**} \cap B^{*'}$ with the parameter $Z = Z_{i_k}/\alphaval$.
If $\alg_1$ reports yes, it reports yes, and otherwise it reports no.

\paragraph*{$\alglevel{i_k}$ with $0 \leq k \leq k^*-2$.}
The input to the algorithm $\alglevel{i_k}$ is a level-$i_k$ pair $(B^* B^{*'})$ and a subblock $B^{**}$ of $B^*$.
We consider the level $i_{k+1}$ and for the sake of readability, we denote $i := i_k$ and $j := i_{k+1}$.
Note that $j > i$ and $j \in \iset^*$.
In this case, we have already defined the level-$j$ algorithm $\alglevel{j}$, and we will use it as our subroutine.

Let $\hset^*$ be the set of indices of the level-$j$ descendant range-blocks of $B^{*'}$ in $\bset^j_{\Psi'}$.
We initialize a new \LIS problem instance $\sset$ in the streaming model with the range $\hset^*$.
We populate the elements of $\sset$ while processing the subblock $B^{**}$ when its elements arrive as the part of the block $B^*$ of the original input sequence $S$, as follows.
We process the elements of $B^{**}$ in their natural order, discarding the elements that do not have their values in the range-block $B^{*'}$.
We sample the elements of $B^{**} \cap  B^{*'}$ independently at random with probability $p_j := \frac{2^{6}}{Z_{j} \mu_{j}}$ each as they arrive as the part of the original input stream.
For each element $a_{t}$ that is chosen, we proceed as follows.
Let $\hat B \in \bset^j_{\Psi}$ be the unique level-$j$ descendant stream-block of $B^*$ containing $a_t$.
Similarly, let $\hat B' \in \bset^j_{\Psi'}$ be the unique level-$j$ descendant range-block of $B^*$ containing its value.
We denote by $\hat B_{>t}$ the subblock of $\hat B$ consisting of its elements appearing after $a_t$.
We now execute $\floor{2^5 \zeta(i,j)}$ parallel instances of the level-$j$ algorithm $\alglevel{j}$ with input level-$j$ pair $(\hat B, \hat B')$ along with the subblock $\hat B_{>t}$ of $\hat B$.
If at any point, we sample more than $\frac{2^{10} \mu_{i}}{\mu_{j}}$ elements from $B^{*}$, we immediately halt our algorithm and report no.
At the end of the level-$j$ stream-block $\hat B$, all the calls to level-$j$ algorithms terminate.
Let $\set{r_1, \ldots, r_\ell}$ be the indices of the level-$j$ range-blocks in $\bset^j_{\Psi'}$ in their natural order, for whom at least $2^4 \zeta(i,j)$ corresponding executions of $\algjump{j}$ reported yes.
We now append the elements $\set{r_\ell, \ldots, r_1}$ to $\sset$ in this order.
At the end of the subblock $B^{**}$, we terminate the sequence $\sset$.

We run the algorithm $\alg_1$ of \Cref{obs: ncm-det-opt-space-sound} on this instance $\sset$ with the parameter $Z = \frac{Z_i}{8Z_j}$.
If $\alg_1$ on input $\sset$ answers yes, we report yes, and otherwise we report no.
This completes the description of our algorithm $\alglevel{i}$.
We now turn to analyze its properties, starting with correctness.

\paragraph{Soundness.}
Consider the execution of $\alglevel{i}$ with input level-$i$ pair $(B^*, B^{*'})$ and the subblock $B^{**}$ of $B^*$.
Assume that $\optlis(B^* \cap B^{*'}) < Z_i/\alpha'_i$.
Our goal is to show that we report no with probability at least $3/4$.

We proceed by induction.
We first consider the case where $k \in \set{k^*, k^*-1}$.
In this case, $\alpha'_i \geq 4$ and from the correctness guarantee of \Cref{obs: ncm-det-opt-space-sound}, if we report yes, $\optlis(B \cap B^{*'}) \geq Z_i/\alphaval \geq Z_i/\alpha'_i$ must hold.
Thus, assume from now on that $0 \leq k \leq k^* - 2$ and the soundness guarantee holds for each level $i_{k'} \in \iset^*$ with $k < k' \leq k^*$.
Let $j = i_{k+1}$ and let $\sset$ be the sequence as constructed by the run of $\alglevel{i}$.

\begin{claim}\label{clm: ncm-lis-simple-alg-soundness}
    With probability at least $3/4$, for \textbf{every} increasing subsequence $\sset'$ of $\sset$, there is an increasing sequence of size at least $|\sset'| \cdot \frac{Z_j}{\alpha'_j}$ in $B^{**} \cap B^{*'}$.
\end{claim}

The proof of \Cref{clm: ncm-lis-simple-alg-soundness} follows from a standard application of Chernoff bound and is deferred to \Cref{prf: ncm-lis-simple-alg-soundness}.
If we report yes, from the correctness guarantee of \Cref{obs: ncm-det-opt-space-sound}, $\optlis(\sset) \geq {Z_{i}}/{8 Z_j}$ must hold.
But then from \Cref{clm: ncm-lis-simple-alg-soundness}, with probability at least $3/4$,

\[ \optlis(B^{**} \cap B^{*'}) \geq \optlis(\sset) \cdot \frac{Z_j}{\alpha'_j} \geq \frac{Z_i}{8 Z_j} \cdot \frac{Z_j}{\alpha'_j} = \frac{Z_j}{8 \alpha'_j} = \frac{Z_i}{\alpha'_i}. \]

Here, the inequality follows since $j > i$ and $\alpha'_i \alpha'_{i_k} = 8\alpha'_{i_{k+1}} = 8 \alpha'_j$.
This completes the proof of soundness of $\alglevel{i_k}$, completing the induction step.

\paragraph{Completeness.}
Consider level $i = i_k$ and the execution of $\alglevel{i}$ with input level-$i$ pair $(B^*, B^{*'})$ and the subblock $B^{**}$ of $B^*$.
Assume that there is an $\Upsilon$-canonical increasing subsequence $S^*$ of $S$ of length $Z_0$.
Further assume that $(B^*, B^{*'})$ is a level-$i$ yes-pair for $S^*$ with $|B^{**} \cap S^*| \geq {Z_i}/\alpha_i$.
Our goal is to show that we report yes with probability at least $3/4$.

We proceed by induction.
We first consider the case where $k \in \set{k^*, k^*-1}$.
From the correctness guarantee of \Cref{obs: ncm-det-opt-space-sound}, if $\optlis(B \cap B^{*'}) \geq Z_i/\alpha_i = Z_i/\alphaval$, we indeed report yes.
Thus, assume from now on that $0 \leq k \leq k^* - 2$ and the completeness guarantee holds for each level $i_{k'} \in \iset^*$ with $k < k' \leq k^*$.
We fix level $j = i_{k+1}$ and let $\sset$ be the sequence as constructed by the execution of $\alglevel{i}$.
Recall that we execute the algorithm $\alg_1$ from \Cref{obs: ncm-det-opt-space-sound} with parameter $Z = \frac{Z_i}{8 Z_j}$.
From the correctness guarantee of \Cref{obs: ncm-det-opt-space-sound}, it suffices to show that, $\optlis(\sset) \geq \frac{Z_i}{8 Z_j}$ with probability at least $\frac{3}{4}$.

We let $\bset$ and $\bset'$ be the set of all level-$j$ descendant stream-blocks of $B^*$ and the set of all level-$j$ descendant range-blocks of $B^{*'}$ respectively.
Consider a level-$j$ descendant yes-pair $(B, B') \in \bset \times \bset'$ of $(B^*, B^{*'})$.
We say that $(B, B')$ is a \emph{happy pair} iff:
    (i) we sample some element $a_{t'} \in B \cap B'$;
    (ii) at least $2^4 \zeta(i,j)$ executions of the level-$j$ algorithm $\alglevel{j}$ with input level-$j$ pair $(B, B')$ and the subblock $B_{>t'}$ of $B$ reports in affirmative;
    and
    (iii) at the end of processing stream-block $B$, we sample at most $2^{10} \frac{\mu_{i}}{\mu_{j}}$ of elements of $B \cap B^{*'}$.
We use the following claim whose proof via standard techniques is deferred to \Cref{prf: ncm-large-eset}.

\begin{claim}\label{clm: ncm-large-eset}
    With probability at least $0.9$, there are at least $\frac{Z_i}{8 Z_{j}}$ happy pairs.
\end{claim}

We assume from now on that there is a set $\pset$ of at least $\ell \geq \frac{Z_i}{8 Z_j}$ happy pairs that we denote by $\pset = \set{(B_{r_1}, B'_{r'_1}), \ldots, (B_{r_\ell}, B'_{r'_\ell}) }$ in their natural order.
Here, $\set{r_1, \ldots, r_\ell}$ and $\set{r'_1, \ldots, r'_\ell}$ are the indices of the underlying level-$j$ stream-blocks and level-$j$ range-blocks in $\bset^j_\Psi$ and $\bset^j_{\Psi'}$ respectively, and $r_1 < r_2 < \ldots < r_\ell$ and $r'_1 < r'_2 < \ldots < r'_\ell$.
Notice that the corresponding elements in $\sset$ with values $\set{r'_1, \ldots, r'_\ell}$ appear in this order and form an increasing subsequence in $\sset$.
Hence, $\optlis(\sset) \geq \ell \geq \frac{Z_i}{8 Z_j}$.
This completes the proof of completeness of $\alglevel{i}$.

\paragraph{Calls to $\alglevel{i_{k+1}}$.}
Consider a level $i = i_k \in \isetnew$.
If $k \in \set{k^*, k^*-1}$, we do not perform any calls to lower level algorithms and there is nothing to show.
Assume from now on that $0 \leq k \leq k^*-2$ and fix the level $j := i_{k+1}$.
Recall that for each sampled element $a_t \in B^* \cap B^{*'}$, we perform either exactly $\floor{2^5 \zeta(i,j)}$ or $0$ calls to $\alglevel{j}$.
Also recall that if we sample more than $\frac{2^{10} \mu_{i}}{\mu_{j}}$ elements from $B^{*}$, we immediately halt our algorithm.
Thus, the number of concurrent calls to $\alglevel{j}$ is at most $\floor{2^5 \zeta(i,j)} \cdot \frac{2^{10} \mu_{i}}{\mu_{j}} \leq \zeta^2(i,j) \cdot \frac{\mu_i}{\mu_j}$.
Here, the inequality follows since $\zeta(i,j) \geq \log \eta \geq 2^{15}$ since $\eta$ is large enough.

\paragraph{Space complexity.}
Consider a level $i = i_k \in \isetnew$.
If $k \in \set{k^*, k^*-1}$, from the guarantee of \Cref{obs: ncm-det-opt-space-sound}, our space complexity is $O(Z_{i_k})$.
It is also easy to verify that $Z_{i_{k^*}} = Z_{r^*}$ and $Z_{i_{k^* - 1}} \leq \eta Z_{i_{k^*}} = \eta Z_{r^*} < \eta^2 N^{5\eps}$.
Here, the last inequality follows from \Cref{assm: large-xi-by-zi} and the facts that $r^* = \rstarVal$ and $r = r(N, \eta) = \floor{\log N / \log \eta}$, since,
\[ Z_{r^*} < \frac{X_{r^*}}{N^{1/2 - \eps}} = \frac{N/\eta^{r^*}}{N^{1/2 - \eps}} = \frac{N^{1/2 + \eps}}{\eta^{r^*}} \leq \eta \cdot \frac{N^{1/2 + \eps}}{N^{1/2 - 4\eps}} = \eta N^{5 \eps}. \]

Thus, the space complexity of a single execution of level-$i$ algorithms $\alglevel{i}$ is bounded by $O(Z_i) = O(\eta^2 N^{5\eps}) \leq N^{5\eps + o(1)}$, since $\eta = N^{o(1)}$ is small enough.

Assume from now on that $0 \leq k \leq k^*-2$, and we let $j := i_{k+1}$ be the corresponding level.
Let $\sset$ be the sequence as constructed by the execution of $\alglevel{i}$.
Recall that the range of $\sset$ is $\hset^*$: the indices of range-blocks in the level-$j$ partition $\bset_{\Psi'}^j$ of the original range $H^*$; and each element with value in $\hset^*$ can be stored in unit space.
From the guarantee of \Cref{obs: ncm-det-opt-space-sound}, we can execute a run of $\alg_1$ on the stream $\sset$ with values in $\hset^*$ with parameter $Z = Z_i/(8 Z_j)$ in space $O(Z_i/Z_j) = O(\eta)$.
We now analyze the space required to populate elements of $\sset$.
Consider some level-$j$ descendant stream-block $B$ of $B^{**}$.
When we are processing $B$, we perform a number of calls to the level-$j$ algorithm $\alglevel{j}$.
Depending on the outcome of these calls, we add elements to $\sset$.
We can charge the space used in computing these additional elements to the space freed after the respective calls to $\alglevel{j}$ terminate.
Thus, the space used by $\alglevel{i}$ in this case, excluding the space used by the calls to $\alglevel{j}$ is bounded by $O(\eta) \leq N^{o(1)}$.

This completes the analysis of $\alglevel{i}$ and \Cref{lem: ncm-simple-lis-main} now follows.

In this subsection, we presented an $N^{o(1)}$-canonical distinguisher algorithm with space complexity $N^{1/2 + O(\eps)}$.
To achieve this, we introduced the set $\isetnew$ of special levels and recursively defined algorithms $\set{\alglevel{i} \> | \> i \in \isetnew}$.
In \Cref{subsec: ncm-processing-upsilon} we establish that the underlying ensemble $\Upsilon$ must satisfy at least one of the two technical conditions that we define.
If it follows the first technical condition (special case $3$), we show in \Cref{subsec: ncm-jumpable} that there are `better' algorithms for certain levels.
We can then swap the algorithms of these respective levels in $\set{\alglevel{i}}$ and show that the resulting level-$i_0$ algorithm is an $N^{o(1)}$-canonical distinguisher algorithm with space complexity $N^{1/2 - \eps + o(1)}$.
Otherwise, if it follows the second technical condition (special case $4$), we show in \Cref{subsec: ncm-reduction-to-ncm} that exploiting the query-efficient algorithm $\algncm$ for the \NCM problem in hybrid model yields `better' algorithms for certain levels.
As before, we can then swap the algorithms of these respective levels in $\set{\alglevel{i}}$ and show that the resulting level-$i_0$ algorithm is an $N^{o(1)}$-canonical distinguisher algorithm with space complexity $N^{1/2 - \eps + o(1)}$.

\subsection{Processing the Ensemble} \label{subsec: ncm-processing-upsilon}
Recall that we have fixed parameters $N$ and $\eta$, both integral powers of $2$, such that $\eta$ grows with $N$.
We are given an \LIS problem instance $S$ in the streaming model, where $S$ is a permutation of the range $H^* = (1, \ldots, N)$.
We let $\iset = \set{0, \ldots, r}$ be the set of the levels of the underlying hierarchical partitions $\bset_\Psi(S)$ of $S$ into stream-blocks and $\bset_{\Psi'}(H^*)$ of $H^*$ into range-blocks.
We have also fixed an optimal $\Upsilon$-canonical subsequence $S^*$ of $S$.
From \Cref{assm: large-xi-by-zi,assm: small-deltai-by-zi}, we know that $X_i/Z_i > N^{1/2 - \eps}$ for all levels $0 \leq i < (1/2 - \eps) r$ and $\Delta_i < Z_i N^{5\eps}$ for all levels $3\eps r \leq i < (1/2  - \eps)r$.
Recall that in \Cref{subsec: ncm-simpler-lis-stream} we have fixed a threshold-level $r^* = \rstarVal$ and the set $\isetnew = \set{i_0, \ldots, i_{\knew}}$ of special levels.
The goal of this section is to describe the technical conditions that will serve as special cases $3$ and $4$.
The main result of this subsection is a dichotomy lemma that establishes that, at least one of these four special cases must hold.
Prior to stating this lemma, we require some further definitions.

\paragraph{Weight of levels and savable levels.}
For each level $0 \leq i_k < i_{\knew}$ of $\isetnew$, we define ${w_{i_k} := \log_{\eta}(\mu_{i_k} / \mu_{i_{k+1}})}$ to be its \emph{weight}.
For the level $i_{\knew}$ of $\isetnew$, we define $w_{i_{\knew}} := 0$.
For a subset $\iset' \subseteq \isetnew$ of levels, we say that $w_{\iset'}^{} := \sum_{i \in \iset'} w_i^{}$ is the \emph{weight of $\iset'$}.
We say that a level $i \in \isetnew$ is a \emph{savable level} if $0.9 < w_i < \Tval$.

\paragraph{Jumpable tuples.}
Consider a tuple $\tau = (i_k, j, \ell)$ of levels of $\isetnew$ where $j=i_{k'}$ and $\ell = i_{k''}$.
Consider the collection $\iset(\tau) := \set{i_k, i_{k+1}, \ldots, i_{k'-1}}$ of levels of $\iset^*$ corresponding to the tuple $\tau$.
We refer to $\tupwt{\tau} = \log_\eta{\left( \frac{\mu_{i_k}}{\mu_{j}} \right)}$ as the \emph{level-weight} of $\tau$.
We say that $\tau$ is a \emph{jumpable tuple} iff $0 \leq k < k' < k'' < \knew$ and $k'' = k' + \floor{0.3 \tupwt{\tau}}$.
We say that a jumpable tuple $\tau$ is a \emph{perfect} jumpable tuple if $\tupwt{\tau} \geq \Tval$, $k' \leq k + \frac{\tupwt{\tau}}{100}$, and $Z_{j} \geq {Z_{\ell}} \cdot { \eta^{0.2 \tupwt{\tau}}}$.

\paragraph{Independent collection of jumpable tuples.}
We say that a collection $\jset$ of jumpable tuples is \emph{independent} iff for each pair of jumpable tuples $\tau = (i, j, \ell)$ and $\tau' = (i', j', \ell')$ of $\jset$, either $\ell < i'$ or $\ell' < i$.
We let $\iset(\jset) = \bigcup_{\tau \in \jset} \iset(\tau)$ be the collection of levels of $\iset^*$ corresponding to the tuples present in $\jset$.
We denote by $\tupwt{\jset} = \sum_{\tau \in \jset} \tupwt{\tau}$ the \emph{level-weight of the collection $\jset$ of jumpable tuples}.

We are now ready to state the main dichotomy lemma of this subsection.

\begin{lemma} \label{lem: ncm-merged-levels-exists}
    Suppose we are given parameters $N$ and $\eta$, both integral powers of $2$, along with an ensemble $\Upsilon = (\Psi, \Psi', \vectZ, \vectDelta, \vectMu)$ of length $1 + r$ with the corresponding collection $\iset$ of levels that satisfies \Cref{assm: large-xi-by-zi,assm: small-deltai-by-zi} as mentioned earlier.
    Then either there is an independent collection $\jset^*$ of perfect jumpable tuples with level-weight $\tupwt{\jset^*} \geq \frac{r}{10^5}$ or there is a collection $\iset_\savable \subseteq \iset$ of savable levels with weight $w_{\iset_\savable} \geq \frac{r}{10^5}$.
\end{lemma}

\begin{proof}
    We are given an ensemble $\Upsilon = (\Psi = \Psi^*(N, \eta), \Psi', \vectZ, \vectDelta, \vectMu)$ with levels $\iset = \set{0, \ldots, r}$, where we let $\Psi' = (\psi_1, \ldots, \psi_r)$, $\vectZ = (Z_0, \ldots, Z_r)$, $\vectDelta = (\Delta_0, \ldots, \Delta_r)$, and $\vectMu = (\mu_0, \ldots, \mu_r)$.
    Recall that we have fixed a subset $\isetnew = \set{i_0, \ldots, i_{\knew}} \subseteq \iset$ of special levels.
    Let $\iset_\savable \subseteq \isetnew$ be the set of all the savable levels.
    We assume that $w_{\iset_\savable} < \frac{r}{10^5}$ since otherwise there is nothing to show.
    Our objective is to construct an independent collection $\jset^*$ of perfect jumpable tuples with level-weight $\tupwt{\jset^*} \geq \frac{r}{10^5}$.

    Our construction consists of two phases.
    In the first phase, we develop an algorithm that processes all the levels of $\isetnew$, while gradually building a collection of independent jumpable tuples $\jset$ with sufficiently large level-weight.
    Unfortunately, some of these jumpable tuples $\jset$ may not be perfect.
    To rectify this, in the second phase, we refine this set of jumpable tuples to obtain a subset $\jset^* \subseteq \jset$ of perfect jumpable tuples with large enough level-weight.
    Before we describe our algorithm of the first phase, we need some additional definitions.
    
    Consider a pair $0 \leq k < k' \leq \knew$ of integers and let $i_k$ and $i_{k'}$ be the corresponding levels of $\isetnew$.
    We define $\cover(k, k') := \min{\left( k' + \floor{0.3 \log_\eta{\left( \frac{\mu_{i_k}}{\mu_{i_{k'}}} \right)}}, \knew  \right)}$.

    Consider an integer $0 \leq k < \knew$ and the corresponding level $i_k \in \isetnew$.
    We define $\jump(k) > k$ to be the smallest integer $k'$ such that $\log_\eta{\left( \frac{\mu_{i_k}}{\mu_{\ell}} \right)} \leq 100 \log_\eta{\left( \frac{\mu_{i_k}}{\mu_{j}} \right)}$, where $j = i_{k'}$ and $\ell = i_{\cover(k, k')}$.
    We first claim that such an integer $\jump(k)$ exists and $\jump(k) \leq \knew$ holds.
    To this end, it suffices to show that the above-mentioned condition holds for $k' = \knew$.
    Indeed, in this case, $j = i_{\knew}$, $\cover(k, \knew) = \knew$, and $\ell = i_{\knew}$, implying that the above-mentioned condition trivially holds (see, \Cref{obs: ncm-decreasing-xi-by-zi}).
    We also define $\land(k) := \cover(k, \jump(k))$.

    We are now ready describe the first phase of our algorithm.
    We initialize the set $\jset = \emptyset$ and will ensure that $\jset$ remains a collection of independent jumpable tuples. 
    We start by processing levels of $\isetnew = \set{i_0, \ldots, i_{\knew}}$, starting at level $i_0$, in their natural increasing order.
    
    Consider a level $i_{k} \in \isetnew$.
    If $k = \knew$, we terminate our process and report our current collection of independent jumpable tuples $\jset$.
    In this case, we say that $i_k$ is the last level that we processed.
    Otherwise, if $w_{i_k} < \Tval$, we do not update $\jset$ and continue to process the subsequent level of $\isetnew$.
    Assume from now on that $w_{i_k} \geq \Tval$.
    Let $\jump(k)$ and $\land(k)$ be the integers as defined above.
    We consider the tuple $\tau(k) := (i_k, i_{\jump(k)}, i_{\land(k)})$.
    If $\land(k) < \knew$, $\tau(k)$ is indeed a jumpable tuple, and we update our collection of jumpable tuples $\jset \gets \jset \cup \set{\tau(k)}$.
    Also observer that $\jset$ remains an independent collection of jumpable tuples.
    We now proceed to process the level $i_{1 + \land(k)}$ of $\isetnew$.
    Otherwise, if $\land(k) = \knew$, we terminate our algorithm without updating our collection $\jset$.
    In this case, we say that $i_k$ is the last level that we processed.
    This completes the description of the first phase our algorithm.
    Before we describe our pruning step, we analyze the guarantees of this algorithm.

    Let $i_{k'} \in \isetnew$ with $0 \leq k' \leq \knew$ be the last level that we process before terminating our algorithm.
    We also let $\iset' := \set{i_0, \ldots, i_{k'-1}}$ be the set of levels of $\isetnew$ appearing before $i_{k'}$.
    Recall that we say that a level $i_k \in \isetnew$ is a savable level if $0.9 < w_i < \Tval$.
    Consider an integer $0 \leq k \leq \knew$ such that the corresponding level $i_k$ is not a savable level.
    We say that it is a \emph{featherweight} level if $w_i \leq 0.9$, and a \emph{heavyweight} level if $w_i \geq \Tval$.
    We will use the following claims whose proofs are deferred to \Cref{{prf-clm: ncm-i-star-sep},{prf-clm: ncm-large-w-star}} respectively.

    \begin{claim} \label{clm: ncm-i-star-sep}
        $i_{k'} \geq k' \geq  \frac{r}{6}$.
    \end{claim}

    \begin{claim}\label{clm: ncm-large-w-star}
        $w_{\iset'} > k' - \frac{r}{1000}$.
    \end{claim}

    We are now ready to show that the level-weight $w_{\jset}$ of our collection $\jset$ of independent jumpable levels is sufficiently large.

    \begin{claim} \label{clm: ncm-large-special}
        $\tupwt{\jset} \geq \frac{r}{10^4}$.
    \end{claim}
    \begin{proof}
        Before giving a lower bound on the level-weight of the set $\jset$ of jumpable levels, we first show that it is not an empty set.

        \begin{observation}\label{obs: ncm-jset-not-empty}
            $\jset \neq \emptyset$.
        \end{observation}
        \begin{proof}
            Assume otherwise for contradiction that $\jset = \emptyset$.
            We claim that there are no heavyweight levels in $\iset'$.
            Indeed, assume that there is some heavyweight level $i_k \in \iset'$.
            Then, we must have processed the level $i_k$ to obtain levels $\jump(k)$ and $\land(k)$.
            If $\land(k) < \knew$, we must have added a jumpable tuple $\tau(k)$ to our collection $\jset$, contradicting the fact that $\jset = \emptyset$.
            On the other hand, if $\land(k) = \knew$, we must have terminated our algorithm immediately, contradicting the fact that $i_{k}$ is not the last level that we processed.
            Thus, there are no heavyweight levels in $\iset'$, or in other words, each level of $\iset'$ is either a savable level or a featherweight level.
            
            Next, we claim that $w_{\iset'}\leq 0.9k' + \frac{r}{1000}$.
            To this end, note that $w_{\iset'} = \sum_{i \in \iset'} w_{i}$, and we bound the contribution of each level in $\iset'$ separately.
            Since there are only $k'$ such levels, the contribution due to featherweight levels, each of which has weight at most $0.9$, is trivially bounded by $0.9k'$.
            On the other hand, from our assumption, the weight of savable levels is less than $\frac{r}{10^5}$.
            Thus, $w_{\iset'} \leq \frac{9}{10}k' + \frac{r}{10^5}$.
            But from \Cref{clm: ncm-large-w-star}, we must have $w_{\iset'} > k' - \frac{r}{1000}$.
            Combining these two bounds, we obtain, $\frac{1}{10}k' < \frac{2r}{1000}$, or in other words, $k' < \frac{r}{50}$, a contradiction to \Cref{clm: ncm-i-star-sep}.
            This completes the proof of \Cref{obs: ncm-jset-not-empty}. 
        \end{proof}

        We now focus on showing that the level-weight $w_{\jset}$ is large.
        Consider a jumpable tuple $\tau(k) \in \jset$.
        Let $\tau(k) = (i_k, j, \ell)$, where $j = i_{\jump(k)}$, $\ell = i_{\land(k)}$, and $\land(k) = \cover(k, \jump(k))$.
        We let $\tilde \iset(k) := \set{i_k, \ldots,  i_{\land(k)-1}}$ and say that the jumpable tuple $\tau(k)$ is \emph{responsible} for these levels.
        Notice that,
        
        \begin{equation} \label{eqn: ncm-tilde-iset-jset-to-w-jset}
            w_{\tilde \iset(k)} = \log_{\eta}{\left( \frac{\mu_{i_k}}{\mu_\ell} \right)} \leq 100 \log_{\eta}{\left( \frac{\mu_{i_k}}{\mu_j} \right)} = 100 \tupwt{\tau}. 
        \end{equation}

        We let $\tilde \iset(\jset) := \bigcup_{\tau \in \jset} \tilde \iset(\tau)$ be the set of all levels of $\iset$ for which the jumpable tuples of $\jset$ are responsible.
        Since the jumpable tuples of $\jset$ are independent, for distinct jumpable tuples $\tau, \tau' \in \jset$, the levels of $\tilde \iset(\tau)$ and $\tilde \iset(\tau')$ are disjoint.
        Thus,
        
        \begin{equation} \label{eqn: ncm-tilde-iset-jset-to-jset}
            w_{\tilde \iset(\jset)} = \sum_{\tau \in \jset} w_{\tilde \iset(\tau)} \leq \sum_{\tau \in \jset} 100 \tupwt{\tau} = 100 \tupwt{\jset}.
        \end{equation}

        Here, the inequality follows from \Cref{eqn: ncm-tilde-iset-jset-to-w-jset}.
        We now conclude that $\tupwt{\jset} \geq w_{\tilde \iset(\jset)}/100$.
        Thus, to show that the level-weight $\tupwt{\jset}$ of our collection $\jset$ of jumpable tuples is large, it suffices to prove that $w_{\tilde \iset(\jset)}$ is large, which we show next.

        \begin{claim} \label{clm: ncm-w-tilde-iset-hat-jset-is-large}
            $w_{\tilde \iset(\jset)} \geq \frac{r}{100}$.
        \end{claim}
        \begin{proof}
            We first claim that $\tilde \iset(\jset) \subseteq \set{0, \ldots, i_{k'-1}}$, or in other words, each jumpable tuple $\tau(k) = (i_k, i_{\jump(k)}, i_{\land(k)}) \in \jset$ has $\land(k) \leq k'$.
            Indeed, consider the level $i_{k'}$.
            If $k' = \knew$, $\land(k) \leq \knew = k'$ by definition and there is nothing to show.
            Otherwise, if $i_{k'}$ is not a heavyweight level, for each jumpable tuple $\tau(k) \in \jset$, we must have $\land(k) < k'$ and there is nothing to show.
            Thus, assume that $i_{k'}$ is a heavyweight level and let $\jump(k')$ be the level as computed by our algorithm.
            Since $i_{k'}$ is the last level that we processed, we must have found that $\cover(k', \jump(k')) = \knew$.
            But then the jumpable tuple $\tau(k')$ is not added to $\jset$ and as a result, each jumpable tuple $\tau(k) \in \jset$ has $\land(k) < k'$ as claimed.

            We let $\iset_{\rem} := \iset' \backslash \tilde \iset(\jset)$ be the levels of $\iset'$ for which no jumpable tuple of $\jset$ is responsible.
            We claim that no level in $\iset_{\rem}$ is heavyweight.
            Indeed, assume for contradiction that there is a heavyweight level $i_k \in \iset_{\rem}$.
            But since $k < k'$, level $i_k$ is not the last level that we process.
            Hence, we must have added a jumpable tuple $\tau(k)$ to $\jset$, and we get $i_k \in \tilde \iset(\jset)$, a contradiction.
            Thus, each level in $\iset_\rem$ is either a featherweight level or a savable level.
            As in \Cref{obs: ncm-jset-not-empty}, we can now bound the weight of levels $\iset_\rem$ by, $w_{\iset_\rem} \leq 0.9k' + \frac{r}{10^5}$.
            We now conclude,

            \begin{align*}
                w_{\tilde \iset(\jset)} &= w_{\iset'} - w_{\iset_\rem}\\
                &> k' - \frac{r}{1000} - \left(\frac{9k'}{10} + \frac{r}{10^5} \right)\\
                &> \frac{k'}{10} - \frac{r}{500}\\
                &\geq \frac{r}{100}
            \end{align*}

            Here, the first inequality follows from \Cref{clm: ncm-large-w-star} and the last inequality follows from \Cref{clm: ncm-i-star-sep}.
            This completes the proof of \Cref{clm: ncm-w-tilde-iset-hat-jset-is-large}.
        \end{proof}
         
        From \Cref{{clm: ncm-w-tilde-iset-hat-jset-is-large},{eqn: ncm-tilde-iset-jset-to-jset}}, we have, $\tupwt{\jset} \geq \frac{1}{100} \cdot w_{\tilde \iset(\jset)} \geq \frac{r}{10^4}$.
        This completes the proof of \Cref{clm: ncm-large-special}.
    \end{proof}

    This completes the description of the first phase of our process.
    So far, we have shown that the level-weight $\tupwt{\jset}$ of our independent collection $\jset$ of jumpable tuples is large.
    However, as we previously noted, some of these jumpable tuples in $\jset$ may not be perfect. 
    In our second phase, we will show that level-weight of such imperfect jumpable tuples present in $\jset$ is relatively small.
    We can thus discard all imperfect jumpable tuples without significantly reducing the level-weight of the surviving jumpable tuples.
    We now formalize this approach.

    Consider a jumpable tuple $\tau(k) \in \jset$ and let $\tau(k) = (i_k, j, \ell)$, where $j = i_\jump(k)$, $\ell = i_{\land(k)}$, and ${\land(k) = {\cover(k, \jump(k))}}$.
    First we claim that $\tupwt{\tau(k)} \geq\Tval$.
    It is immediate to verify that $i_k$ is a heavyweight level and $\jump(k) > k$, implying  $\tupwt{\tau(k)} = \log_\eta{\left( \frac{\mu_{i_k}}{\mu_{j}} \right)} \geq w_{i_k} \geq \Tval$.
    We say that $\tau(k)$ is a \emph{bad jumpable tuple} if $\frac{Z_\ell}{Z_j} > {\eta^{- \frac{\tupwt{\tau(k)}}{5}}}$, or equivalently, ${\log_\eta{\left( \frac{Z_j}{Z_\ell} \right)} <  \frac{\tupwt{\tau(k)}}{5}}$.
    We say that it is a \emph{good jumpable tuple} otherwise.
    Let $\jset' \subseteq \jset$ be the set of all bad jumpable tuples.
    From the above discussion, the set $\jset^* = \jset \backslash \jset'$ is indeed a collection of perfect jumpable tuples of level-weight $\tupwt{\jset^*} = \tupwt{\jset} - \tupwt{\jset'}$, and it now suffices to upper bound the level-weight $\tupwt{\jset'}$.
    We prove the following claim after completing the proof of \Cref{lem: ncm-merged-levels-exists} assuming it.

    \begin{claim} \label{clm: ncm-small-bad}
        $\tupwt{\jset'} \leq 30 \eps r$.
    \end{claim}

    From \Cref{clm: ncm-large-special,clm: ncm-small-bad}, we have,
    
    \[ \tupwt{\jset^*} = \tupwt{\jset} - \tupwt{\jset'} \geq \frac{r}{10^4} - 30 \eps r \geq \frac{r}{10^5},  \]

    for all $\eps \leq 10^{-6}$.
    This completes the proof of \Cref{lem: ncm-merged-levels-exists} assuming \Cref{clm: ncm-small-bad} that we show next.
\end{proof}

\proofof{\Cref{clm: ncm-small-bad}}
    If $\jset' = \emptyset$, we have $\tupwt{\jset'} = 0$ and there is nothing to show.
    Thus, assume from now on that $\jset' \neq \emptyset$.
    We will use the following observation whose proof is deferred to \Cref{prf-obs: ncm-z-r-star-lower-bound}.

    \begin{observation} \label{obs: ncm-z-r-star-lower-bound}
        $\log_\eta{\left(\frac{Z_0}{Z_{i_{\knew}}}\right)} < {\knew - 0.09 \tupwt{\jset'}}$.
    \end{observation}

    From \Cref{assm: large-xi-by-zi}, $X_{i_{\knew}}/Z_{i_{\knew}} > N^{1/2 - \eps}$, and we obtain,

    \begin{align*}
        \frac{Z_0}{Z_{i_{\knew}}} > \frac{Z_0}{X_{i_{\knew}} / N^{1/2 - \eps}} = \frac{Z_0 \cdot N^{1/2 - \eps}}{X_{i_{\knew}}} \geq \frac{N^{1 - 2\eps}}{N / \eta^{i_\knew}} = \frac{\eta^{i_{\knew}}}{N^{2\eps}} \geq \frac{\eta^{\knew}}{N^{2\eps}},
    \end{align*}

    where the second inequality follows from the facts that $Z_0 \geq N^{1/2 - \eps}$ and $X_{i_{\knew}} = \frac{N}{\eta^{i_{\knew}}}$.
    Combined with \Cref{obs: ncm-z-r-star-lower-bound} we obtain, $ \frac{\eta^{\knew}}{N^{2\eps}} < \eta^{\knew - 0.09 w_{\jset'}}$,
    or equivalently, $\knew - 2\eps r < \knew - 0.09 \tupwt{\jset'}$.
    We now conclude $\tupwt{\jset'} < \frac{2}{0.09} \eps r < 30 \eps r$, completing the proof of \Cref{clm: ncm-small-bad}.
\endproofof

\subsection{Special Case 3: Jumpable Tuples have Large Level-Weight} \label{subsec: ncm-jumpable}
Recall that we have fixed parameters $N$ and $\eta$, both integral powers of $2$, such that $\eta$ grows with $N$.
We are given an \LIS problem instance $S$ in the streaming model, where $S$ is a permutation of the range $H^* = (1, \ldots, N)$.
Additionally, we are given an ensemble $\Upsilon = (\Psi, \Psi', \vectZ, \vectDelta, \vectMu)$ of length $1+r$
where $r = r(N, \eta) = \floor{\frac{\log N}{\log \eta}}$, $\Psi = \Psi^*(N, \eta) = (\eta, \ldots, \eta)$, $\Psi' = (\psi_1, \ldots, \psi_r)$, $\vectZ = (Z_0, \ldots, Z_r)$, $\vectDelta = (\Delta_0, \ldots, \Delta_r)$, and $\vectMu = (\mu_0, \ldots, \mu_r)$.
We let $\iset = \set{0, \ldots, r}$ be the set of the levels of the underlying hierarchical partitions $\bset_\Psi(S)$ of $S$ into stream-blocks and $\bset_{\Psi'}(H^*)$ of $H^*$ into range-blocks.
We have also fixed an optimal $\Upsilon$-canonical subsequence $S^*$ of $S$.
From \Cref{assm: large-xi-by-zi,assm: small-deltai-by-zi}, we know that $X_i/Z_i > N^{1/2 - \eps}$ for all levels $0 \leq i < (1/2 - \eps) r$ and $\Delta_i < Z_i N^{5\eps}$ for all levels $3\eps r \leq i < (1/2  - \eps)r$.
Recall that in \Cref{subsec: ncm-simpler-lis-stream} we have fixed a threshold-level $r^* = \rstarVal$ and the set $\isetnew = \set{i_0, \ldots, i_{\knew}} \subseteq \iset$ of special levels.

In this special case, we are given an independent collection $\jset^*$ of perfect jumpable tuples with level-weight $\tupwt{\jset^*} \geq \frac{r}{10^5}$.
Our goal is to devise an $N^{o(1)}$-canonical distinguisher algorithm $\algspl{3}$ for the \LIS problem with space complexity $N^{1/2 - \eps + o(1)}$.
Similar to \Cref{subsec: ncm-simpler-lis-stream}, we will create a collection of $1+\knew$ different algorithms, where each algorithm corresponds to a level $i_k \in \isetnew$.
These algorithms will be parameterized by the relevant parameters of the respective levels, and hence, it will be convenient to describe them based on the levels at which they operate. 
For each level $i \in \isetnew$ we denote the corresponding level-$i$ algorithm by $\algjump{i}$.
As in \Cref{subsec: ncm-simpler-lis-stream}, $\algjump{i}$ takes as an input a level-$i$ pair $(B^*, B^{*'})$ and a subblock $B^{**}$ of $B^*$.
It is given the description of $(B^*, B^{*'})$ by giving it the indices of these blocks $B^*$ and $B^{*'}$ in the level-$i$ partitions $\bset^i_\Psi$ and $\bset^i_{\Psi'}$ respectively.
Next, it has access to the elements of $B^{**}$ as they arrive as a part of the original input sequence $S$.
As in \Cref{subsec: ncm-simpler-lis-stream}, we ensure that $\algjump{i}$ is an $(\alpha_{i}, \alpha'_{i})$-canonical distinguisher algorithm for $\alpha_{i} = 4$ and $\alpha'_{i} = 2^{3(r^*-i)+2}$.
Additionally, recall that for each pair $i_k < i_{k'}$ of levels of $\isetnew$, we have fixed a parameter $\zeta(i_k, i_{k'}) = \ln{ \left(\eta^{2(i_{k'} - i_k)} \cdot \psi_{i_{k + 1}} \ldots \psi_{i_{k'}} \right)}$.

Consider a level $i_k \in \iset^*$ for some $0 \leq k \leq k^*$.
For readability, we denote $i := i_k$.
We are now ready to describe our level-$i$ algorithm $\algjump{i}$.

\paragraph{$\algjump{i}$ for the level $i \in \iset^*$ such that there is no jumpable tuple $\tau = (i, j, \ell)$ in $\jset^*$.}
Our algorithm $\algjump{i}$ is identical to the algorithm $\alglevel{i}$ of \Cref{{lem: ncm-simple-lis-main}} from \Cref{subsec: ncm-simpler-lis-stream}, with one exception.
Recall that if $k < \knew$, the algorithm $\alglevel{i}$ is allowed to use the level-$i_{k+1}$ algorithm $\alglevel{i_{k+1}}$ as subroutine.
In our new algorithm $\algjump{i}$, we replace each of these calls by the respective calls to our newly defined algorithm $\algjump{i_{k+1}}$.
\footnote{Note that the algorithm $\algjump{i_{\knew}}$ is identical to the algorithm $\alglevel{i_{\knew}}$.}
From \Cref{{lem: ncm-simple-lis-main}}, we immediately obtain the following observation:

\begin{observation}\label{obs: ncm-jump-no-jump}
    Assume that $k < \knew$ and for the level $j := i_{k+1}$ we are given a level-$j$ $(\alphaval, \alpha'_{j})$-canonical distinguisher algorithm $\algjump{j}$.
    Then, there is a level-$i$ $(\alphaval, \alpha'_i)$-canonical distinguisher algorithm $\algjump{i}$ that achieves the following guarantees:
    \begin{itemize}
        \item \textbf{Calls to $\algjump{j}$:} it performs at most $\zeta^2(i, j) \frac{\mu_{i}}{\mu_{j}}$  concurrent calls to the level-$j$ algorithm $\algjump{j}$; and
        
        \item \textbf{Space complexity:} its space complexity, excluding the space required by the calls to $\algjump{j}$ is at most $N^{6\eps}$.
    \end{itemize}
\end{observation}

This completes the description of our algorithm $\algjump{i}$ for this case.
We now focus on the remaining case.

\paragraph{$\algjump{i}$ for the level $i \in \iset^*$ such that there is a jumpable tuple $\tau = (i, j, \ell)$ in $\jset^*$.}
We fix such a jumpable tuple $\tau = \tau(k) = (i_k, j, \ell)$, with $j = i_{\jump(k)}$ and $\ell = i_\land(k)$.
In this case, we show the following algorithm, whose analysis is deferred to \Cref{subsec: ncm-jump-jump-i}.

\begin{lemma}\label{lem: ncm-jump-jump}
    Fix a tuple of levels $i_0 < i < j < \ell < \knew$ of $\isetnew$ such that $\frac{Z_j}{Z_\ell} \leq \sqrt{\frac{\mu_i}{\mu_j}}$.
    Assume that we are given a $(\alphaval, \alpha'_{j})$-canonical distinguisher level-$j$ algorithm $\algjump{j}$ and a $(\alphaval, \alpha'_{\ell})$-canonical distinguisher level-$\ell$ algorithm $\algjump{\ell}$.
    Then there is a $(\alphaval, \alpha'_i)$-canonical distinguisher level-$i$ algorithm $\algjump{i}$ that achieves the following guarantees:
    \begin{itemize}
        \item \textbf{Calls to $\algjump{j}$:} it performs at most $O \left( \zeta(i, \ell) \cdot \frac{Z_i}{Z_j} \cdot \eta^{\ell - j} \right)$  concurrent calls to $\algjump{j}$; 
        
        \item \textbf{Calls to $\algjump{\ell}$:} it performs at most $O \left( \zeta^2(i, \ell) \cdot \frac{Z_i \cdot Z_\ell^2}{Z_j^3} \cdot \frac{\mu_{i}}{\mu_{\ell}} \right)$  concurrent calls to $\algjump{\ell}$; and
        
        \item \textbf{Space complexity:} its space complexity, excluding the space required by the calls to $\algjump{j}$ and $\algjump{\ell}$, is at most $O \left( \zeta^3(i, \ell) \cdot \eta^{\ell - j} \cdot \frac{Z_i \cdot Z_\ell^2 }{Z_j^3} \cdot \frac{\mu_{i}}{\mu_{\ell}} \right)$.
    \end{itemize}
\end{lemma}

We are now ready to describe our algorithm $\algspl{3}$ for this special case.

\subsubsection{The algorithm $\algspl{3}$.}

Applying \Cref{obs: ncm-jump-no-jump} and \Cref{lem: ncm-jump-jump} recursively for the levels of $\isetnew$ in their decreasing order, we obtain a  $(\alphaval, \alpha'_0)$-canonical distinguisher level-$i_0$ algorithm $\algjump{i_0}$.
From \Cref{{obs: ncm-alglevel-0-suffices}}, there is a $\alpha'_0 = N^{o(1)}$-canonical distinguisher algorithm $\algspl{3}$ that performs at most $N^{o(1)}$ concurrent calls to $\algjump{i_0}$ and whose space complexity, excluding the space required by the calls to $\algjump{i_0}$, is at most $N^{o(1)}$.
We prove the following claim after describing the algorithm of this subsection assuming it.

\begin{claim}\label{clm: ncm-jumpable-level-0-space}
    The space complexity of $\alglevel{i_0}$ is $N^{1/2 -\eps}$.
\end{claim}

We immediately obtain the following corollary that is the main result of this subsection.

\begin{corollary}\label{cor: ncm-main-jumpable}
    Suppose we are given parameters $N$ and $\eta$, both integral powers of $2$, along with an ensemble $\Upsilon = (\Psi, \Psi', \vectZ, \vectDelta, \vectMu)$ of length $1 + r$ and an independent collection $\jset^*$ of perfect jumpable tuples with level-weight $\tupwt{\jset^*} \geq \frac{r}{10^5}$.
    Then there is a $N^{o(1)}$-canonical distinguisher algorithm $\algspl{3}$ with space complexity $N^{1/2 - \eps + o(1)}$.
\end{corollary}

Before proving \Cref{clm: ncm-jumpable-level-0-space}, we need the following simple fact whose proof is deferred to \Cref{prf-clm: zeta-convex}.

\begin{claim}\label{clm: zeta-convex}
    For each pair $0 \leq k_1 < k_2 \leq \knew$, $\zeta(i_{k_1}, i_{k_2})  \leq \prod_{k_1 \leq k' < k_2} \zeta(i_{k'}, i_{k'+1})$.
\end{claim}

\proofof{\Cref{clm: ncm-jumpable-level-0-space}}
    Consider a level $i \in \iset^*$ and a level-$i$ stream-block $B \in \bset_\Psi^{i}$.
    While executing $\algjump{i_0}$, we also execute a number of calls to the level-$i$ algorithm  $\algjump{i}$ in which $B$ participates.
    We denote by $\load(B)$ the maximum number of such concurrent calls to $\algjump{i}$ in which $B$ participates.
    We denote by ${\load(i) := \max_{B \in \bset_\Psi^{i}}{\left( \load(B) \right)}}$ the maximum `load' across all level-$i$ stream-blocks.
    We need the following additional parameters.

    For each pair $0 \leq i' < i'' \leq r$ of levels, we define $\eq(i', i'') := \left| \set{i', \ldots, i''} \backslash \isetnew \right|$.
    For each level $i_k \in \iset^*$, we have three parameters $q_1(i_k)$, $q_2(i_k)$, and $q_3(i_k)$ whose values are defined recursively as follows.
    We let $q_1(i_0) = q_2(i_0) = q_3(i_0) = 1$.
    Consider now a level $i_k \in \iset^*$ with $0 < k \leq k^*$.
    First consider the case where there is a jumpable tuple $\tau = (i, j, \ell) \in \jset^*$ with $j = i_k$.
    In this case, we let,
    $q_1(i_k) = q_1(i) \cdot \eta^{\eq(i, \ell)}$,
    $q_2(i_k) =  q_2(i) \cdot \eta^{0.39 \tupwt{\tau}}$, and
    $q_3(i_k) = q_3(i) \cdot \zeta^4(i, \ell)$.
    Otherwise, if there is no such jumpable tuple, we let $q_1(i_k) = q_1(i_{k-1})$, $q_2(i_k) = q_2(i_{k-1})$, and $q_3(i_k) = q_3(i_{k-1}) \cdot \zeta^4(i_{k-1}, i_k)$.
    For each level $i \in \iset^*$, we also define $L^*(i) = \frac{q_1(i) \cdot q_3(i)}{q_2(i)} \cdot \frac{\mu_{i_0}}{\mu_{i}}$.
    In \Cref{{prf-clm: ncm-jump-load-induction},{prf-clm: ncm-load-i-small}}, we prove the following two claims.

    \begin{claim} \label{clm: ncm-jump-load-induction}
        For each level $i_k \in \iset^*$, $\load(i_k) \leq L^*(i_k)$.
    \end{claim}
   
    \begin{claim} \label{clm: ncm-load-i-small}
        For each level $i_k \in \iset^*$, $L^*(i_k) \leq {N^{\frac{1}{2} + 6\eps}}/{\eta^{0.39 w_{\jset^*}}} \leq N^{\frac{1}{2} - \frac{2}{10^6}}$.
    \end{claim}

    We are now ready to analyze the space complexity of $\algjump{i_0}$.
    We consider its state when processing some element $a_t \in S$.
    Consider a level $i \in \iset^*$ and let $B_i \in \bset^i_\Psi$ be the unique level-$i$ stream-block containing $a_t$.
    We will show that the contribution to the space complexity of $\algjump{0}$ due to level-$i$ is at most $N^{1/2 - 2\eps}$.
    We analyze this contribution differently, depending on whether there is a jumpable tuple $\tau = (i, j, \ell) \in \jset^*$ for some levels $\ell > j > i$.

    We first consider the case where there is no jumpable tuple $\tau = (i, j, \ell) \in \jset^*$.
    From the guarantee of \Cref{obs: ncm-jump-no-jump}, the space complexity used by such an execution of $\algjump{i}$, excluding the space required by the calls to lower level algorithms, is at most $N^{6\eps}$.
    Observe that $B_i$ participates in all the active calls to the level-$i$ algorithm $\algjump{i}$.
    Hence, there are at most $\load(i)$ active calls to $\algjump{i}$.
    From \Cref{{clm: ncm-jump-load-induction},{clm: ncm-load-i-small}}, the contribution to the space complexity of $\algjump{0}$ due to the level $i$ is,

    \begin{equation} \label{eqn: ncm-jumpable-bound-one}
        \begin{split}
            \load(i) \cdot O (N^{6\eps}) &\leq L^*(i) \cdot O\left( N^{6\eps} \right)\\
            &\leq O \left( N^{\frac{1}{2} - \frac{2}{10^6}} \cdot N^{6\eps} \right) = O \left( N^{\frac{1}{2} - \frac{2}{10^6} + 6\eps} \right)\\
            &\leq N^{1/2 - 2\eps},
        \end{split}
    \end{equation}

    for all $\eps \leq 10^{-7}$.

    We now consider the remaining case and fix the jumpable tuple $\tau = (i, j, \ell) \in \jset^*$.
    Observe that $B_i$ participates in all the active calls to the level-$i$ algorithm $\algjump{i}$.
    Hence, there are at most $\load(i)$ active calls to $\algjump{i}$.
    From the guarantee of \Cref{lem: ncm-jump-jump}, the space complexity used by such an execution of $\algjump{i}$, excluding the space required by the calls to lower level algorithms, is at most,
    
    \[ O \left( \zeta^3(i, \ell) \cdot \eta^{\ell - j} \cdot \frac{Z_i \cdot Z_\ell^2 }{Z_j^3} \cdot \frac{\mu_{i}}{\mu_{\ell}} \right) \leq \frac{q_3(j)}{q_3(i)} \cdot \eta^{\ell - j} \cdot \frac{Z_i \cdot Z_\ell^2 }{Z_j^3} \cdot \frac{\mu_{i}}{\mu_{\ell}}. \]

    Since $\tau$ is a perfect jumpable tuple, we have $\ell - j = \floor{0.3 \tupwt{\tau}} + \eq(j, \ell)$, $\frac{Z_i}{Z_j} \leq \eta^{0.01 \tupwt{\tau}}$, and $Z_j \geq Z_\ell \cdot \eta^{0.2 \tupwt{\tau}}$.
    Using these relations, we can further bound the space used by,

    \begin{align*}
        \frac{q_3(j)}{q_3(i)} \cdot \eta^{\ell - j} \cdot \frac{Z_i \cdot Z_\ell^2 }{Z_j^3} \cdot \frac{\mu_{i}}{\mu_{\ell}}
        &\leq \frac{q_3(j)}{q_3(i)} \cdot  \eta^{0.3 \tupwt{\tau} + \eq(j, \ell)} \cdot \frac{Z_i}{Z_j} \cdot \frac{Z_\ell^2}{Z_j^2} \cdot \frac{\mu_{i}}{\mu_{\ell}}  \\
        &\leq \frac{q_1(j) q_3(j)}{q_1(i) q_3(i)} \cdot \eta^{0.3 \tupwt{\tau} + 0.01 \tupwt{\tau} - 0.4 \tupwt{\tau} } \cdot \frac{\mu_{i}}{\mu_{\ell}}  \\
        &= \frac{q_1(j) q_3(j)}{q_1(i) q_3(i)} \cdot \eta^{-0.09 \tupwt{\tau} } \cdot \frac{\mu_{i}}{\mu_{\ell}}.
    \end{align*}

    Using the above bound along with \Cref{{clm: ncm-jump-load-induction}}, we can bound the contribution to the space complexity of $\algjump{0}$ due to level $i$ by,

    \begin{align*}
        \load(i) \cdot \frac{q_1(j) q_3(j)}{q_1(i) q_3(i)} \cdot \eta^{-0.09 \tupwt{\tau} } \cdot \frac{\mu_{i}}{\mu_{\ell}}
        &\leq L^*(i) \cdot \frac{q_1(j) q_3(j)}{q_1(i) q_3(i)} \cdot \eta^{-0.09 \tupwt{\tau} } \cdot \frac{\mu_{i}}{\mu_{\ell}} \\
        &= \frac{q_1(j) \cdot q_3(j)}{q_2(i)} \cdot \eta^{-0.09 \tupwt{\tau} } \cdot \frac{\mu_{i_0}}{\mu_{\ell}}                   \\
        &= \frac{q_1(j) \cdot q_3(j)}{q_2(j)} \cdot \eta^{0.3 \tupwt{\tau} } \cdot \frac{\mu_{i_0}}{\mu_{\ell}}     \\
        &\leq \frac{q_1(\ell) \cdot q_3(\ell)}{q_2(\ell)} \cdot \eta^{0.3 \tupwt{\tau} } \cdot \frac{\mu_{i_0}}{\mu_{\ell}}\\
        &= L^*(\ell) \cdot \eta^{0.3 \tupwt{\tau} }\\
        &\leq L^*(\ell) \cdot \eta^{0.3 \tupwt{\jset^*} }.
    \end{align*}

    From \Cref{{clm: ncm-load-i-small}} we can further bound this contribution by,

    \begin{equation} \label{eqn: ncm-jumpable-bound-two}
        \begin{split}
            L^*(\ell) \cdot \eta^{0.3 \tupwt{\jset^*} } &\leq \frac{N^{\frac{1}{2} + 6\eps}}{\eta^{0.39 \tupwt{\jset^*}}} \cdot \eta^{0.3 \tupwt{\jset^*} } = \frac{N^{\frac{1}{2} + 6\eps}}{\eta^{0.09 \tupwt{\jset^*}}} \\
            &\leq \frac{N^{\frac{1}{2} + 6\eps}}{\eta^{\frac{0.09r}{10^6}}} \leq N^{\frac{1}{2} + 6\eps - 0.09 \frac{1}{10^6}} \leq N^{\frac{1}{2} - 2\eps}.
        \end{split}
    \end{equation}

    The second-last inequality follows from the fact that $\tupwt{\jset^*} \geq r/10^6$ and the last inequality holds for all $\eps \leq \frac{1}{10^8}$.
    
    From \Cref{{eqn: ncm-jumpable-bound-one},{eqn: ncm-jumpable-bound-two}}, for each level $i \in \iset^*$, the contribution of $\algjump{i}$ in the space complexity of $\algjump{i_0}$ is at most $N^{1/2 - 2\eps}$.
    We now conclude that the space complexity of $\algjump{i_0}$ is indeed bounded by,
        
        \[ |\iset^*| \cdot N^{1/2 - 2\eps} \leq r \cdot N^{1/2 - 2\eps} < N^{1/2 - \eps}. \]

    This completes the analysis of the space complexity of $\algjump{i_0}$ and \Cref{clm: ncm-jumpable-level-0-space} now follows.
\endproofof

\subsection{$\algjump{i}$ and its Properties --- Proof of Lemma \ref{lem: ncm-jump-jump}} \label{subsec: ncm-jump-jump-i}
The goal of this subsection is to prove \Cref{lem: ncm-jump-jump}.
We are given a tuple $(i, j, \ell)$ of levels of $\isetnew$ with $i_0 < i < j < \ell < i_\knew$.
We are also given an $(\alpha_j, \alpha'_{j})$-canonical distinguisher level-$j$ algorithm $\algjump{j}$ and an  $(\alpha_\ell, \alpha'_{\ell})$-canonical distinguisher level-$\ell$ algorithm $\algjump{\ell}$.
Our goal is to devise an $(\alpha_i, \alpha'_i)$-canonical distinguisher level-$i$  algorithm  $\algjump{i}$ with low space complexity that performs a small number of calls to $\algjump{j}$ and $\algjump{\ell}$.
Recall that the input to $\algjump{i}$ is a level-$i$ pair $(B^*, B^{*'})$ and a subblock $B^{**}$ of $B^*$.
We consider the collection $\bset$ of level-$j$ descendant stream-blocks of $B^*$ that are entirely contained in $B^{**}$.

Observe that if $|\bset| \leq 100$, we can immediately report no and there is nothing to show.
Indeed, if that is the case, $B^{**}$ may contain at most $(2 + |\bset|)Z_j \leq 102 Z_j < Z_i/\alpha_i$ elements of an $\Upsilon$-canonical subsequence.
Here, the last inequality follows from the fact that $\alpha_i = \alphaval$ and Property \ref{prop: ncm-p4} of $\Upsilon$-canonical subsequences.
We assume from now on that $|\bset| > 100$.
Consider the collection $\bset'$ of level-$j$ descendant range-blocks of $B^{*'}$.
We denote $\tilde \bset_j := \bset^{j}_{\Psi'}(H^*)$ the partition of the range $H^*$ into level-$j$ range-blocks.
Before describing our algorithm $\algjump{i}$, we need a notion of \emph{generalized subblock processing algorithm}, that is analogous to the block processing algorithm of \Cref{subsec: ncm-small-xi-by-zi}.
We will later show using the standard dynamic-programming approach that it suffices to show an `efficient' {generalized subblock processing algorithm}.

\paragraph*{Region of range-blocks.}
Note that the collection $\bset'$ of level-$j$ descendant-blocks of $B^{*'}$ is a partition of $B^{*'}$.
We view this collection $\bset'$, ordered in the natural increasing order of the range-block indices, as a sequence with values in the range $\tilde \bset_j $: the set of all level-$j$ range-blocks.
We refer to a block $\rset$ of this sequence $\bset$ as a \emph{region}.
It is worth noting that a region $\rset$ consists of $|\rset|$ contiguous range-blocks, and it can be uniquely identified by the indices of its first and last blocks of $\tilde \bset_j$; and as a consequence, stored in $O(1)$ units of space.
Consider a collection $\tilde \rset$ of such regions.
If $\tilde \rset$ constitute a partition of the sequence $\bset'$, we refer to it as  \emph{partition of $\bset'$ into regions}.

\paragraph*{Generalized Sunblock Processing Algorithm}
The generalized subblock processing algorithm, that we refer to as $\alggen$, is a streaming algorithm.
It is initialized with a region $\rset \subseteq \bset'$ of level-$j$ range-blocks determined by the indices of its first and last range-blocks in $\tilde \bset_j$.
It is then provided access to the elements of $B$ as they arrive as a part of the stream $B^{**}$ of the original input stream $S$.
At the end of the stream-block $B$, $\alggen$ reports a (possibly empty) region $\rset'$ of range-blocks of $\rset$ by outputting the interval of their respective indices in $\bset'_j$.
We allow $\alggen$ to use a number of calls to the level-$j$ algorithm $\algjump{j}$ and the level-$\ell$ algorithm $\algjump{\ell}$ as subroutine.

Recall that in \Cref{subsec: ncm-simpler-lis-stream}, we have defined, for each pair $i_k < i_{k'}$ of levels of $\isetnew$ a parameter $\zeta(i_k, i_{k'}) = \ln{ \left(\eta^{2(i_{k'} - i_k)} \cdot \psi_{i_{k + 1}} \cdot \ldots \cdot \psi_{i_{k'}} \right)}$.
We use the following lemma:

\begin{lemma}\label{lem: ncm-alggen-exists}
    Consider the collections $\bset$ and $\bset'$ of level-$j$ stream-blocks and level-$j$ range-blocks respectively as mentioned earlier.
    Assume that we are given an $(\alpha_j, \alpha'_{j})$-canonical distinguisher level-$j$ algorithm $\algjump{j}$ and an $(\alpha_\ell, \alpha'_{\ell})$-canonical distinguisher level-$\ell$ algorithm  $\algjump{\ell}$.
    Then there is a generalized subblock processing algorithm $\alggen$ that, on input $B \in \bset$, a region $\rset$ of $\bset'$, and the set $\hat \bset$ of level-$\ell$ descendants of $B$, achieves the following guarantees:
    \begin{itemize}
        \item \textbf{Completeness guarantee:} if there is a range-block $B' \in \rset$ such that $(B,B')$ is a yes-pair and $|B \cap B^{*'}| \leq 8 Z_j \mu_i$, then with probability at least $1 - e^{-\zeta(i, \ell)}$ over the randomness used by $\alggen$, the reported region $\rset'$ is non-empty and it does not contain any range-blocks of $\bset'$ appearing after $B'$;
        
        \item \textbf{Soundness guarantee:} the probability, over the randomness used by $\alggen$, that it reports a non-empty region $\rset'$ such that the longest increasing subsequence of $B$ using values collectively from the range-blocks of $\rset'$ has length less than $\frac{Z_j}{\alpha'_j}$, is at most $e^{-\zeta(i, \ell)}$;
        
        \item \textbf{Calls to $\algjump{j}$:} it performs at most $O \left( \zeta(i, \ell) \cdot \eta^{\ell - j} \right)$ concurrent calls to $\algjump{j}$;
        
        \item \textbf{Calls to $\algjump{\ell}$:} it marks each level-$\ell$ stream-block $\hat B \in \hat \bset$ independently randomly with probability $O \left( \zeta(i, \ell) \cdot \frac{Z_\ell}{Z_j} \right)$ each, and for each marked block $\hat B$ it performs at most $O \left( \zeta(i, \ell) \cdot \frac{Z_\ell \mu_{i}}{Z_j \mu_{\ell}} \right)$ calls to the level-$\ell$ algorithm $\algjump{\ell}$ in which $\hat B$ participates; and
        
        \item \textbf{Space complexity:} its space complexity, excluding the space required by the calls to $\algjump{j}$ and $\algjump{\ell}$ is at most $O \left( \zeta^3(i, \ell) \cdot \eta^{\ell - j} \cdot \frac{Z_\ell^2 }{Z_j^2} \cdot \frac{\mu_{i}}{\mu_{\ell}} \right)$.
    \end{itemize}
\end{lemma}

We complete the proof of \Cref{lem: ncm-alggen-exists} in \Cref{subsec: ncm-jumpable-alggen-desc} after we prove \Cref{lem: ncm-jump-jump} assuming it.
We are now ready to describe our level-$i$ algorithm $\algjump{i}$.
It is an adaptation of the algorithm $\alg_1$ of \Cref{obs: ncm-det-opt-space-sound}.

\paragraph{Description of $\algjump{i}$.}
Recall that the input to the level-$i$ algorithm $\algjump{i}$ is a level-$i$ pair $(B^{*}, B^{*'})$ and a subblock $B^{**}$ of $B^*$.
We have also fixed the sets $\bset$ and $\bset'$ of level-$j$ descendants of $B^*$ and $B^{*'}$ respectively.
We fix the parameter $\beta := \frac{Z_i}{8 Z_j}$.
We will process level-$j$ stream-blocks of $\bset$ as they arrive as a part of the subsequence $B^{**}$ of the original input sequence $S$.
While processing these elements, we will maintain an ordered collection $\tilde \rset$ of at most $\beta$ disjoint non-empty regions of $B^{*'}$.
Let $\leftover(\tilde \rset) := \bset' \backslash \bigcup_{\rset \in \tilde \rset} \rset$ be the set of range-blocks not appearing in the regions of $\tilde \rset$.
We will ensure that $\leftover(\tilde \rset)$ is a (possibly empty) region and the non-empty regions of $\tilde \rset \cup \set{\leftover(\tilde \rset)}$ constitute a partition of $\bset'$.
We initialize this collection by $\tilde \rset \gets \emptyset$.

Consider a level-$j$ stream-block $B \in \bset$ and let $\tilde \rset$ be our collection of the regions of $B^{*'}$ just before processing the first element of $B$.
We fix the special region $\leftover(\tilde \rset) = \bset' \backslash \bigcup_{\rset \in \tilde \rset} \rset$.
For each region $\rset \in \tilde \rset \cup \set{\leftover(\tilde \rset)}$, we run in parallel the generalized subblock processing algorithm $\alggen$ of \Cref{lem: ncm-alggen-exists} with input level-$j$ stream-block $B$ and the region $\rset$ of level-$j$ blocks of $\bset'$.
At the end of stream-block $B$, all the calls to the generalized subblock processing algorithm $\alggen$ terminate and we are ready to update our collection $\tilde \rset$.
Let $\rset'$ be the region reported by the call to $\alggen$ with input $B$ and the special region $\leftover(\tilde \rset)$.
If $\tilde \rset = \emptyset$ and $\rset' = \emptyset$, we do not update the regions and let $\tilde \rset = \emptyset$.
Thus, assume from now on that either $\rset' \neq \emptyset$ or $\tilde \rset \neq \emptyset$ (or both).

\paragraph*{Case $1$: $\rset' = \emptyset$ and $\tilde \rset \neq \emptyset$.}
In this case, the regions in $\tilde \rset$ constitute a partition of $\bset'$.
We let $\tilde \rset = (\rset_1, \ldots, \rset_k)$ be these regions in their natural order, where $k = |\tilde \rset| \leq \beta$.
In this case, we will ensure that after the update, the resulting collection $\tilde \bset$ contains exactly $k$ regions.
For each $1 \leq k' \leq k$, let $\rset'_{k'} \subseteq \rset_{k'}$ be the region reported by the respective call to $\alggen$ with input stream-block $B$ and region $\rset_{k'}$.
If $\rset'_{1} = \emptyset$, we do not update the first region: $\rset^{(\new)}_1 \gets \rset_1$.
Otherwise, we set $\rset^{(\new)}_1 \gets \rset'_1$.
For each successive $2 \leq k' \leq k$, we obtain the new region $\rset^{(\new)}_{k'}$ as follows.
If $\rset'_{k'} \neq \emptyset$, we let the new region $\rset^{(\new)}_{k'}$ contain all the range-blocks that appear after the region  $\rset^{(\new)}_{k'-1}$ until the last range-block of $\rset'_{k'}$.
Otherwise, if $\rset'_{k'} = \emptyset$, we let the new region $\rset^{(\new)}_{k'}$ contain all the range-blocks that appear after the region  $\rset^{(\new)}_{k'-1}$ until the last range-block of $\rset_{k'}$.
Finally, we update $\tilde \rset \gets (\rset^{(\new)}_{1}, \ldots, \rset^{(\new)}_{k})$.
Let $\leftover(\tilde \rset) \gets \bset' \backslash \bigcup_{\rset \in \tilde \rset} \rset$ be the special set of range-blocks not present in the regions of $\tilde \rset$.
From our construction, it is immediate to verify that $\leftover(\tilde \rset)$ is a region and the non-empty regions of $\tilde \rset \cup \set{\leftover(\tilde \rset)}$ indeed constitute a partition of $\bset'$.

\paragraph*{Case $2$: $\rset' \neq \emptyset$.}
Note that in this case, we must have $\leftover(\tilde \rset) \neq \emptyset$.
Since all the regions of $\tilde \rset \cup \set{\leftover(\tilde \rset)}$ are non-empty, they constitute a partition of $\bset'$.
We append $\tilde \rset \gets \tilde \rset \cup \set{\leftover(\tilde \rset)}$ and let $\tilde \rset = (\rset_1, \ldots, \rset_k)$ be these regions in their natural order, for $k = |\tilde \rset|$.
Notice that $1 \leq k \leq \beta + 1$.
In this case, we will ensure that after the update, $\tilde \bset$ contains exactly $k^* := \min{(k, \beta)}$ regions.
We proceed as in case $1$.
For each $1 \leq k' \leq k$, let $\rset'_{k'} \subseteq \rset_{k'}$ be the region reported by the respective call to $\alggen$ with input stream-block $B$ and region $\rset_{k'}$.
If $\rset'_{1} = \emptyset$, we do not update the first region: $\rset^{(\new)}_1 \gets \rset_1$.
Otherwise, we set $\rset^{(\new)}_1 \gets \rset'_1$.
For each successive $2 \leq k' \leq k$, we obtain the new region $\rset^{(\new)}_{k'}$ as follows.
If $\rset'_{k'} \neq \emptyset$, we let the new region $\rset^{(\new)}_{k'}$ contain all the range-blocks that appear after the region  $\rset^{(\new)}_{k'-1}$ until the last range-block of $\rset'_{k'}$.
Otherwise, if $\rset'_{k'} = \emptyset$, we let the new region $\rset^{(\new)}_{k'}$ contain all the range-blocks that appear after the region  $\rset^{(\new)}_{k'-1}$ until the last range-block of $\rset_{k'}$.
Finally, we update $\tilde \rset \gets (\rset^{(\new)}_{1}, \ldots, \rset^{(\new)}_{k^*})$.
Let $\leftover(\tilde \rset) \gets \bset' \backslash \bigcup_{\rset \in \tilde \rset} \rset$ be the set of range-blocks not present in the regions of $\tilde \rset$.
From our construction, it is immediate to verify that $\leftover(\tilde \rset)$ is a region and the non-empty regions of $\tilde \rset \cup \set{\leftover(\tilde \rset)}$ indeed constitute a partition of $\bset'$.
This completes the description of case $2$, and as a result, the description of our algorithm for processing the level-$j$ stream-block $B$.

We now proceed to process the subsequent level-$j$ stream-blocks of $\bset$.
Recall that $\alggen$ performs a number of calls to the level-$j$ and level-$\ell$ algorithms $\algjump{j}$ and $\algjump{\ell}$ respectively.
We define $L^* := c \zeta^2(i, \ell) \cdot \frac{Z_i \cdot Z_\ell^2}{Z_j^3} \cdot \frac{\mu_{i}}{\mu_{\ell}}$, for some constant $c$ to be fixed later.
Consider a level-$\ell$ descendant stream-block $\hat B$ of $B^*$.
We denote by $\eset^*_{\bad}(\hat B)$ the event that $\alggen$ performs at least $L^*$ calls to $\algjump{\ell}$ (excluding the calls performed while executing level-$j$ algorithm $\algjump{j}$) in which $\hat B$ participates.
We let $\eset^*_{\bad}$ be the event that there is a some level-$\ell$ descendant stream-block $\hat B$ of $B^*$ for which the event $\eset^*_\bad(\hat B)$ occurs.
Throughout our algorithm $\algjump{i}$, if at any point the event $\eset^*_\bad$ occurs, we immediately stop and report no.
Once we process the last stream-block of $\bset$, we are ready to report our answer.
Let $\tilde \rset$ be our collection of regions of $B^{*'}$, just after processing the last element of the last stream-block of $\bset$.
If $\tilde \rset$ contains exactly $\beta$ regions, we report yes; otherwise, we report no. 
This completes the description of our level-$i$ algorithm $\algjump{i}$.
We now turn to analyze its properties, starting with an upper bound on the number of concurrent calls to $\alggen$.

\paragraph*{Calls to $\alggen$.}
Recall that we process the level-$j$ stream-blocks of $\bset$ in their natural sequential order.
Consider some block $B \in \bset$ and let $\tilde \rset$ be our collection of regions just before processing $B$.
We also fix the special region $\leftover(\tilde \rset) = \bset' \backslash \bigcup_{\rset \in \tilde \rset} \rset$ just before processing $B$.
Since we execute the generalized subblock processing algorithm $\alggen$ with input $B$ and $\rset$, for each non-empty region $\rset \in \tilde \rset \cup \set{\leftover(\tilde \rset)}$, we perform at most $1 + |\tilde \rset| \leq 1 + \beta \leq O\left( \frac{Z_i}{Z_j} \right)$ concurrent calls to $\alggen$.
In \Cref{prf-clm: ncm-jump-eset-star-bad-low-prob}, we prove the following claim.

\begin{claim} \label{clm: ncm-jump-eset-star-bad-low-prob}
    $\prob{\eset^*_\bad} \leq e^{-2\zeta(i,\ell)}$.
\end{claim}

We assume from now on that the event $\eset^*_\bad$ does not occur.
We are now ready to show the correctness guarantees of our algorithm $\algjump{i}$.

\paragraph*{Completeness.}
Recall that we have fixed an optimal $\Upsilon$-canonical increasing subsequence $S^*$ of $S$.
Assume that $(B^*, B^{*'})$ is a yes-pair for $S^*$ and the subblock $B^{**}$ contains at least $Z_i/\alpha_i = Z_i/\alphaval$ elements of $S^*$.
Our goal is to show that we report yes with probability at least $3/4$.
As mentioned earlier, we assume that the event $\eset^*_\bad$ does not occur.

We let $\bset_\yes$ be the set of yes-blocks of $\bset$ for $S^*$.
Since $B^{**}$ contains at least $\frac{Z_i}{\alphaval}$ elements of $S^*$ and $S^*$ is an $\Upsilon$-canonical increase subsequence, we must have $|\bset_\yes| \geq \frac{Z_i}{\alphaval Z_j} - 2 \geq \beta$.
Here, the last inequality follows since $i < j$ and $i,j \in \isetnew$ implying that $Z_i/Z_j > $ from Property \ref{prop: ncm-p4} of the $\Upsilon$-canonical subsequence.
We arbitrarily discard additional blocks $\bset_\yes$ to ensure that $|\bset_\yes| = \beta$ and let $\bset_\yes = \set{B(1), \ldots, B(\beta)}$ in their natural order.
We also let $\bset'_\yes = \set{B'(1), \ldots, B'(\beta)}$ be the corresponding range-blocks of $\bset'$ such that $(B(1), B'(1)), \ldots, (B(\beta), B'(\beta))$ are level-$j$ yes-pairs.
For each $1 \leq s \leq \beta$, we let $\tilde \rset^{(s)}$ be the collection of regions of $\bset'$ just after processing the stream-block $B(s)$.
The following claim, whose proof is deferred to \Cref{prf-clm: ncm-jump-dp-completeness}, now shows the completeness of $\algjump{i}$.

\begin{claim}\label{clm: ncm-jump-dp-completeness}
    With probability at least $9/10$, for each $1 \leq s \leq \beta$, the following holds:
    (i) $|\tilde \rset^{(s)}| \geq s$; and
    (ii) the $s^{th}$ region in $\tilde \rset^{(s)}$ does not contain any range-block appearing after $B'(s)$.
\end{claim}

Recall that we report yes if we have exactly $\beta$ regions in our collection after processing the last stream-block of $\bset$.
From \Cref{clm: ncm-jump-eset-star-bad-low-prob,clm: ncm-jump-dp-completeness}, with probability at least $0.9 - e^{-\zeta(i,\ell)} \geq 3/4$, the collection $\tilde \rset^{(\beta)}$ has $\beta$ regions, showing the completeness guarantee of $\algjump{i}$.

\paragraph*{Soundness.}
Assume that $\optlis(B^{**} \cap B^{*'}) < \frac{Z_i}{\alpha'_i}$.
Our goal is to show that we report no with probability at least $3/4$.
As mentioned earlier, assume that the event $\eset^*_\bad$ does not occur.

Consider a level-$j$ stream-block $B \in \bset$ and let $\tilde \rset$ be our collection of the regions of $B^{*'}$ just before processing the first element of $B$.
We also fix the special region $\leftover(\tilde \rset) = \bset' \backslash \bigcup_{\rset \in \tilde \rset} \rset$.
Consider some non-empty region $\rset \in \tilde \rset \cup \set{\leftover(\tilde \rset)}$ and the execution of $\alggen$ with input stream-block $B$ and region $\rset$.
Let $\rset'$ be the region that it reports.
We let $\eset_\bad(B, \rset)$ be the event that $\rset' \neq \emptyset$ and there is no increasing subsequence of $B$ of length $\frac{Z_j}{\alpha'_j}$ using elements with values in the range-blocks of $\rset'$.
From the correctness guarantee of $\alggen$ and \Cref{fact: ncm-exp-zeta-bound}, the probability that $\eset_\bad(B, \rset)$ occurs is at most $e^{-\zeta(i, \ell)} \leq \frac{1}{10 |\bset| |\bset'|}$.
We let $\eset_\bad(B)$ be the event $\eset_\bad(B, \rset)$ occurs for some non-empty region $\rset \in \tilde \rset \cup \set{\leftover(\tilde \rset)}$.
From union bound over at most $|\bset'|$ such non-empty regions, the probability that $\eset_\bad(B)$ occurs is at most $\frac{1}{10 |\bset|}$.
We let $\eset_\bad$ be the bad event $\eset_\bad(B)$ occurs for some level-$j$ block $B \in \bset$.
From union bound over $|\bset|$ such stream-blocks, the probability that $\eset_\bad$ occurs is at most $\frac{1}{10}$.
We assume from now on that the event $\eset_\bad$ does not occur.

We let $\bset = \set{B(1), \ldots, B(|\bset|)}$ in their natural order.
For each $1 \leq s \leq |\bset|$, we let $\tilde \rset^{(s)}$ be the collection of regions of $\bset'$ just after processing the stream-block $B^{(s)}$.
In \Cref{prf-clm: ncm-jump-dp-soundness} we prove the following claim using techniques similar those used in proving \Cref{clm: ncm-jump-dp-completeness}.

\begin{claim}\label{clm: ncm-jump-dp-soundness}
    For each $1 \leq s \leq |\bset|$ and $1 \leq s' \leq |\tilde \rset^{(s)}|$, there is an increasing subsequence of length at least $\frac{s' Z_j}{\alpha'_j}$ using elements in $B(1) \cup \ldots \cup B(s)$ with values in the range-blocks of the first $s'$ regions of $\tilde \rset^{(s)}$.
\end{claim}

Recall that $\algjump{i}$ reports yes only if there are exactly $\beta$ regions in our collection at the end of $B^{**}$, or in other words, $|\tilde \rset^{|\bset|}| = \beta$.
But then from the above claim, there must be an increasing subsequence of $B^{**}$ of length at least $\frac{\beta Z_j}{\alpha'_j} = \frac{Z_i}{8 Z_j} \cdot \frac{Z_j}{\alpha'_j} \geq \frac{Z_i}{\alpha'_i}$ with values in $B^{*'}$.
This is in contradiction with our assumption that $\optlis(B^{**} \cap B^{*'}) < \frac{Z_i}{\alpha'_i}$, completing the proof of soundness.

\paragraph*{Calls to $\algjump{j}$.}
Consider some some level-$j$ stream-block $B \in \bset$.
Notice that all the calls to the level-$j$ algorithm $\algjump{j}$ are performed through the respective calls to $\alggen$ in which $B$ participates.
Consider one such call to $\alggen$ in which $B$ participates.
From \Cref{lem: ncm-alggen-exists}, each such execution of $\alggen$ performs at most $O \left(\zeta(i, \ell) \cdot \eta^{\ell - j} \right)$ calls to $\algjump{j}$.
Since there are at most $O(Z_i/Z_j)$ concurrent calls to $\alggen$, the number of concurrent calls to $\algjump{j}$ in which $B$ participates is bounded by,

\begin{align*}
    O\left( \frac{Z_i}{Z_j} \right) \cdot O \left(\zeta(i, \ell) \cdot \eta^{\ell - j} \right) &= O \left( \zeta(i, \ell) \cdot \frac{Z_i}{Z_j} \cdot \eta^{\ell - j} \right)
\end{align*}

\paragraph*{Calls to $\algjump{\ell}$.}
Recall that we immediately terminate our algorithm if the event $\eset^*_\bad$ occurs.
Thus, for each level-$\ell$ descendant-block $\hat B$ of $B^*$, the number of concurrent calls to the level-$\ell$ algorithm $\algjump{\ell}$ in which $\hat B$ participates is at most 
$L^* = O \left( \zeta^2(i, \ell) \cdot \frac{Z_i \cdot Z_\ell^2}{Z_j^3} \cdot \frac{\mu_{i}}{\mu_{\ell}} \right)$.

\paragraph*{Space complexity.}
We first analyze the space complexity of $\algjump{i}$, excluding the space used by the calls to $\alggen$.
Consider some level-$j$ stream-block $B \in \bset$ and let $\tilde \rset$ be our collection of regions just before processing $B$.
The space used by $\algjump{i}$ can be divided into two parts:
(i) the space required to store these regions; and
(ii) the space required to update these regions.
Recall that we can store a region $\rset \in \tilde \rset$ by storing the indices of its first and last blocks in the level-$j$ partition $\tilde \bset_j$ of range $H^*$.
Since there are at most $\beta$ regions in $\tilde \rset$, the space used to store them is at most $O(\beta)$.
Similarly, we can store the special region $\leftover(\tilde \rset) = \bset' \backslash \bigcup_{\rset \in \tilde \rset} \rset$ in $O(1)$ units of space.
For each region $\rset \in \tilde \rset \cup \set{\leftover(\tilde \rset)}$, we execute $\alggen$ with input $B$ and $\rset$.
Let $\rset'$ be the reported region, which we can store in $O(1)$ units of space.
It is immediate to verify that we can update our collection $\tilde \rset$ using $O(\beta)$ additional space.
We are now ready to analyze the space complexity of $\algjump{i}$, excluding the space used by the calls to $\algjump{j}$ and $\algjump{\ell}$.
Recall that we perform at most $O(Z_i/Z_j)$ concurrent calls to $\alggen$.
From \Cref{lem: ncm-alggen-exists}, the space complexity of each such call, excluding the space required by the calls to $\algjump{j}$ and $\algjump{\ell}$ is $O \left( \zeta^3(i, \ell) \cdot \eta^{\ell - j} \cdot \frac{Z_\ell^2 }{Z_j^2} \cdot \frac{\mu_{i}}{\mu_{\ell}} \right)$.
Hence, the overall space complexity of $\algjump{i}$, excluding the space required by the calls to $\algjump{j}$ and $\algjump{\ell}$, is

\begin{align*}
    O \left( \beta \right) + O \left( \frac{Z_i}{Z_j} \right) \cdot O \left( \zeta^3(i, \ell) \cdot \eta^{\ell - j} \cdot \frac{Z_\ell^2 }{Z_j^2} \cdot \frac{\mu_{i}}{\mu_{\ell}} \right) &= O \left( \zeta^3(i, \ell) \cdot \eta^{\ell - j} \cdot \frac{Z_i \cdot Z_\ell^2 }{Z_j^3} \cdot \frac{\mu_{i}}{\mu_{\ell}} \right).
\end{align*}

This completes the analysis of the properties of $\algjump{i}$ and \Cref{{lem: ncm-jump-jump}} now follows.
It now remains to prove \Cref{lem: ncm-alggen-exists}, that we show next.

\subsection{$\alggen$ and its Properties --- Proof of Lemma \ref{lem: ncm-alggen-exists}}\label{subsec: ncm-jumpable-alggen-desc}
We now focus on describing our generalized subblock processing algorithm $\alggen$ and completing the proof of \Cref{lem: ncm-alggen-exists}.
We are given a tuple $(i, j, \ell)$ of levels of $\isetnew$ with ${i_0 < i < j < \ell < i_\knew}$ with $\left(\frac{Z_j}{Z_\ell} \right)^2 \leq \frac{\mu_i}{\mu_j}$.
We are also given an $(\alpha_j, \alpha'_{j})$-canonical distinguisher level-$j$ algorithm $\algjump{j}$ and an  $(\alpha_\ell, \alpha'_{\ell})$-canonical distinguisher level-$\ell$ algorithm $\algjump{\ell}$.
We consider the collection $\bset$ of level-$j$ descendant stream-blocks of $B^*$ that are entirely contained in $B^{**}$.
We also consider the collection $\bset'$ of level-$j$ descendant range-blocks of $B^{*'}$.
We denote $\tilde \bset_j := \bset^{j}_{\Psi'}(H^*)$ the partition of the range $H^*$ into level-$j$ range-blocks.
Our algorithm $\alggen$ is given as input a level-$j$ stream-block $B \in \bset$ and a region $\rset \subseteq \bset'$.
If $\rset = \emptyset$, we report $\rset' = \emptyset$ and there is nothing to show.
Thus, we assume from now on that $|\rset| \geq 1$.
Recall that we are given a parameter $\zeta(i,\ell)$, that we denote by $\zeta$ for readability.
Let $\hat \bset$ be the set of all level-$\ell$ descendant-blocks of $B$.
Similarly, let $\hat \bset'$ be the set of all level-$\ell$ descendant-blocks of range-blocks in $\rset$.
We will need the following definition.

\paragraph*{Promising pairs and suspicious blocks.}
We say that a level-$\ell$ pair $(\hat B, \hat B') \in \hat \bset \times \hat \bset'$ is a \emph{promising pair} iff $\optlis(\hat B \cap \hat B') \geq \frac{Z_\ell}{\alpha'_\ell}$.
In this case, we also say that $\hat B$ participates in a promising pair with $\hat B'$.
We say that a level-$j$ range-block $B' \in \bset'$ is a \emph{suspicious block for $\hat \bset$} iff there are at least $\frac{Z_j}{Z_\ell}$ stream-blocks in $\hat \bset$ that participate in promising pairs with level-$\ell$ descendant range-blocks of $B'$.

Consider a level-$j$ yes-pair $(B, B')$ with $B' \in \rset$.
We claim that $B'$ is a suspicious block for $\hat \bset$.
Indeed, if $(B, B')$ is a yes-pair, there are $\frac{Z_j}{Z_\ell}$ descendant level-$\ell$ yes-pairs of $(B, B')$.
Thus, each of the $\frac{Z_j}{Z_\ell}$ yes-blocks of $\hat \bset$ participate in a promising pair with some level-$\ell$ descendant range-block of $B'$.
For the reverse direction, we claim that if there are sufficiently many suspicious blocks, they together contribute to a large increasing subsequence in $B^{**}$.
We summarize it in the following observation, that follows by `random shift' argument.

\begin{observation} \label{obs: ncm-more-suspicious-blocks-is-good-easy}
    Suppose there is a collection $\bset'' \subseteq \bset'$ of at least $3|\hat \bset|$ level-$j$ range-blocks that are suspicious for $\hat \bset$.
    Then there is an increasing subsequence of $B$  of length more than $\frac{Z_j}{\alpha'_j}$ with values in the range-blocks of $\bset''$.
\end{observation}
\begin{proof}
    We denote $m = |\hat \bset|$.
    We let $\hat \bset = \set{\hat B_1, \ldots, \hat B_m}$ be the stream-blocks in their natural order.
    We discard additional range-blocks of $\bset''$ and let $\bset'' = \set{B'_1, \ldots, B'_{3m}}$ in their natural order.
    We consider a bipartite graph $\hset$ with vertices $V(\hat \bset)$ on one side and $V(\bset'')$ on the other side.
    For each stream-block $\hat B_{k'} \in \hat \bset$, there is a unique vertex $u({k'}) \in V(\hat \bset)$.
    Similarly, for each range-block $B'_{k''} \in \bset_\good$, there is a unique vertex $v({k''}) \in V(\bset'')$.
    Consider a pair $(\hat B_{k'}, B'_{k''})$ of blocks with $\hat B_{k'} \in \hat \bset$ and $B'_{k''} \in \bset''$.
    If there is a level-$\ell$ descendant range-block $\hat B'_{k''}$ of $B'_{k'}$ such that $(\hat B_{k'}, \hat B_{k''})$ is a promising pair, we add an edge $(u(k'), v(k''))$ in the graph $\hset$.
    Let $\fset \subseteq V(\hat \bset) \times V(\bset'')$ be the resulting set of edges.
    Notice that since each range-block $B'_{k''} \in \bset''$ is a suspicious block for $\hat \bset$, each vertex $v(k'') \in V(\bset'')$ has at least $\frac{Z_j}{Z_\ell}$ edges incident on it.
    We now conclude that the number of edges in $\hset$ is

    \[ |\fset| \geq |\bset''| \cdot \frac{Z_j}{Z_\ell}  = 3m\frac{Z_j}{Z_\ell}.\]

    We divide this set of edges $\fset$ into $2m+1$ equivalence classes, where for each $-m \leq k \leq m$, the class $\fset_k$ consists of all the edges $(u(k'), v({k''}))$ of $\fset$ with $k' - k'' = k$.
    It is immediate to verify that for each such class $\fset_k$, there is an increasing subsequence of $B$ with values in the range-blocks of $\rset$ of length at least $|\fset_k| \cdot \frac{Z_\ell}{\alpha'_\ell}$.
    It now remains to show that there is an equivalence class containing a large number of edges.
    Indeed, from the pigeonhole principle, there is some class $\fset_{k^*}$ that contains at least $\frac{|\fset|}{2m+1} > \frac{|\fset|}{3m} \geq \frac{Z_j}{Z_\ell}$ edges.
    Thus, we obtain an increasing subsequence of size at least,
    
    \[ |\fset_{k^*}| \cdot \frac{Z_\ell}{\alpha'_\ell} > \frac{Z_j}{Z_\ell}\cdot \frac{Z_\ell}{\alpha'_\ell} = \frac{Z_j}{\alpha'_\ell} \geq \frac{Z_j}{\alpha'_j}\]
    
    with values in range-blocks of $\bset''$.
    This completes the proof of \Cref{obs: ncm-more-suspicious-blocks-is-good-easy}.
\end{proof}

In our generalized subblock processing algorithm  $\alggen$, we will use the level-$j$ algorithm $\algjump{j}$ as subroutine a number of times.
The input to each such call of $\algjump{j}$ will be some appropriately chosen level-$j$ pair $(B, B')$ along with a subblock $\tilde B$ of $B$, where $B' \in \rset$ is a level-$j$ range-block.
Before describing our algorithm, we explore the usefulness of this notion of suspicious blocks with the help of a thought experiment.

\paragraph{$\alggen$ -- A wishful thinking.}
For simplicity, we assume that each call to $\algjump{j}$ reports correct answer with probability $1$.
In other words, consider some level-$j$ pair $(B, B')$ along with a subblock $\tilde B$ of $B$.
If $(B, B')$ is a yes-pair and $\tilde B$ contains at least $Z_j/2$ elements of $S^*$, we assume that $\algjump{j}$ must answer this query in affirmative.
On the other hand, if this query is indeed answered in affirmative, $\optlis(\hat B \cap B') > Z_j /\alpha'_j$ must hold.
Ideally, $\alggen$ should work as follows.

Assume that we have processed elements $\set{, \ldots, a_t}$ of $B$ and somehow came up with a collection $\bset'' \subseteq \rset$ of suspicious blocks.
We discard all but lowest $\min{(|\bset''|, 3|\hat \bset|)}$ range-blocks from $\bset''$ and denote by $\bset'''$ the subset of surviving suspicious blocks.
For each surviving suspicious range-block $B'' \in \bset''$, we perform an independent call to the level-$j$ algorithm $\algjump{j}$ with input pair $(B, B'')$ and the subblock $B_{>t}$\footnote{Recall that $B_{>t}$ is the subblock of $B$ containing all elements appearing after $a_t$ in it.} of $B$.
First consider the case where at least one of these calls is answered in affirmative.
Let $B'_{\min} \in \bset'''$ be the lowest range-block such that the corresponding call to $\algjump{j}$ reports yes.
In this case, we report the region $\rset' = \set{B'_{\min{{}}}}$ containing that single range-block.
Now consider the case where $|\bset'''| = 3|\hat \bset|$ and none of the calls to $\algjump{j}$ returned in affirmative.
We then report the smallest region $\rset' \subseteq \rset$ that encompasses all the range-blocks of $\bset'''$.
Finally, if $|\bset'''| < 3|\hat \bset|$ and none of the calls to $\algjump{j}$ returned in affirmative, we report the empty region $\rset' = \emptyset$.

We claim that if we report a non-empty region $\rset'$, there must be a large enough increasing subsequence of $B$ taking values from the range-blocks of $\rset'$.
Indeed, if there is a range-block $B'' \in \bset'''$ such that the corresponding call to $\algjump{j}$ returned yes, then from our correctness assumption, $\optlis(B \cap B'') \geq \optlis(B_{>t} \cap B') > \frac{Z_j}{\alpha'_j}$ must hold.
On the other hand, if $|\bset'''| = 3|\hat \bset|$, then from \Cref{obs: ncm-more-suspicious-blocks-is-good-easy} there is an increasing subsequence of $B$ of length more than $\frac{Z_j}{\alpha'_j}$ taking values from the range-blocks of $\bset'''$, which are in turn contained in $\rset'$.

We now examine the behavior of our fictional algorithm in the case where $(B, B')$ is a yes-pair with $B' \in \rset$. 
In this case, we would like to ensure that with high probability, $\rset'$ is non-empty and does not contain any range-block appearing after $B'$.
Suppose, we are guaranteed that if such $B' \in \bset'$ exists, then $B' \in \bset''$ and the (yet-unseen) elements $B_{>t}$ contain at least $\frac{Z_j}{2}$ elements of $S^*$.
Under this assumption we claim that $\alggen$ reports the desired region $\rset' \subseteq \bset'$.
Indeed, if $B' \in \bset'''$, the level-$j$ algorithm $\algjump{j}$ with input pair $(B, B')$ and the subblock $B_{>t}$ of $B$ will return yes.
In this case, we indeed report a region consisting of a single range-block that is either $B'$ or some other one appearing before $B'$ in $\rset$.
Otherwise, if $B' \in \bset''$ is not chosen to be in $\bset'''$, we must have had $|\bset''| > 3|\hat \bset|$ and $B'$ appears after all $|\hat \bset|$ range-blocks of $\bset'''$.
In this case, we report the non-empty region $\rset'$ encompassing $\bset'''$, which in turn does not contain $B'$ and any other range-block of $\bset'$ appearing after $B'$. 

From the above discussion, it suffices to show an algorithm that correctly identifies the set of such suspicious blocks $\bset''$ without processing too many elements of $B \cap S^*$.
Recall that $\alggen$ is allowed to use lower level algorithms $\algjump{j}$ and $\algjump{\ell}$ as subroutines.
We will then also need to ensure that we do not call such algorithms too often, as these `trickle-down' calls contribute to space complexity of our algorithm.
Unfortunately, we cannot naively find such suspicious blocks and will have to make certain accommodations that we describe next.

\subsubsection{Description of $\alggen$ -- Pseudo-suspicious Blocks and How to Find Them?}
We are now ready to describe the main result of this subsection: a generalized subblock processing algorithm $\alggen$ for processing a level-$j$ stream-block $B \in \bset$ and a region $\rset \subseteq \bset'$ of contiguous level-$j$ range-blocks.
As mentioned earlier, we cannot naively find the suspicious blocks with desired guarantee.
To get around this hurdle, we will define `pseudo-suspicious' blocks that will serve as a proxy for suspicious blocks.
Informally, our algorithm works as follows.
We process elements of $B$ as they arrive as a part of the original input stream $S$.
As we process these elements, we mark certain level-$\ell$ descendant-pairs of $\set{(B, B') \> | \> B' \in \rset}$ as pseudo-promising.
Once we determine that a level-$j$ range-block $B' \in \rset$ has large number of level-$\ell$ descendant pseudo-promising pairs that `span across a large number of level-$\ell$ stream-blocks', we mark it as pseudo-suspicious block.
When a range-block $B'$ is marked pseudo-suspicious, we run in parallel a level-$j$ algorithm $\algjump{j}$ with input level-$j$ pair $(B, B')$ and the subblock $\tilde B$, where $\tilde B$ is the yet-unseen subblock of $B$.
We might also decide to discard a pseudo-suspicious range-block $B'$, in which case, we also discard all the computations of $\algjump{j}$ in which $B'$ participates.
We now describe our algorithm in detail.

Before starting to process the elements of $B$, we initialize the \emph{multiset} $\bset_{\prom} \gets \emptyset$ of \emph{pseudo-promising} range-blocks and the set $\bset_{\susp} \gets \emptyset$ of \emph{pseudo-suspicious} range-blocks.
We sample each level-$\ell$ subblock of $B$ independently with probability $p := \min{\left(\pval, 1 \right)}$ when it arrives as a part of $B$; or discard it otherwise.
If at any point we sample more than $12 \zeta p|\hat \bset|$ stream-blocks of $\hat \bset$, we immediately terminate $\alggen$ and report the empty region $\rset' = \emptyset$ as our output.
Consider a sampled level-$\ell$ subblock $\hat B$ of $B$.
We run the following algorithm while we process its elements arriving as a part of $B$.

We sample elements of $\hat B$ with values in the range-blocks of $\rset$ independently with probability $p' = \min{ \left( \pprimeval, 1 \right)}$ each.
Consider an element $a_t \in \hat B$ that is sampled and let $\hat B'$ be the unique level-$\ell$ range-block containing its value.
If no element of $\hat B \cap \hat B'$ was previously sampled, we run in parallel the level-$\ell$ algorithm $\algjump{\ell}$ with input level-$\ell$ pair $(\hat B, \hat B')$ along with the subblock $\hat B_{>t}$ of $\hat B$.
If at any point, we sample more than  $E^*_\ell := 2^6 p' Z_\ell \mu_{i}$ elements of $\hat B$, we discard the computation due to $\hat B$ and proceed to process the subsequent level-$\ell$ stream-block of $\hat \bset$.
We now examine the state of our algorithm at the end of the stream-block $\hat B$.
At this point, all the calls to level-$\ell$ algorithms in which $\hat B$ participates terminate.
We consider the subset $\rset_{\yes} \subseteq \rset$ of level-$\ell$ range-blocks for which the corresponding run of $\algjump{\ell}$ returned yes.
Let $\rset^{(j)}_\yes$ be the set of level-$j$ ancestor-blocks of these range-blocks in $\rset_\yes$.
Note that $\rset^{(j)}_\yes$ is a set and each level-$j$ range-block appears at most once in it.
We mark these level-$j$ ancestors as \emph{pseudo-promising} by appending them to the multiset $\bset_{\prom} \gets \bset_\prom \cup \rset^{(j)}_\yes$.
Note that $\bset_{\prom}$ is a multiset and a level-$j$ range-block may appear multiple times in it.
If some range-block appears at least $2^5 \zeta$ times in $\bset_{\prom}$, we mark it as \emph{pseudo-suspicious} by moving it to $\bset_{\susp}$.
Consider the time when some level-$j$ range-block $B' \in \rset$ is marked pseudo-suspicious, say, after processing the element $a_{t'}$ of the stream-block $B$.
We then run in parallel $\floor{2^6 \zeta}$ independent instances of the level-$j$ subblock processing algorithm $\algjump{j}$ with input level-$j$ pair $(B, B')$ and the subblock $B_{>t'}$ of $B$.
If at any point there are more than $|\hat \bset|$ pseudo-suspicious blocks, we discard the ones with the highest indices and their respective computations by $\algjump{j}$.
We refer to surviving pseudo-suspicious blocks as \emph{active} pseudo-suspicious blocks.

Finally, at the end of the stream-block $B$, we are ready to compute our output region $\rset' \subseteq \rset$.
At this point, all our calls to the level-$j$ algorithm $\algjump{j}$ terminate. 
If there is an active pseudo-suspicious range-block for which at least $2^5 \zeta$ of the corresponding runs of $\algjump{j}$ returned in affirmative, let $B'_{\min{}}$ be such a range-block with the lowest index.
We then report the region $\rset' = \set{B'_{\min{}}}$ consisting of this single range-block.
If there is no such range-block and there are exactly $|\hat \bset|$ active pseudo-suspicious blocks, we report the smallest region $\rset'$ that encompasses all of these active pseudo-suspicious range-blocks.
Otherwise, we report the empty region $\rset' = \emptyset$.
This completes the description of our generalized subblock processing algorithm $\alggen$.
We now turn to analyze its properties, starting with completeness.

\paragraph{Completeness.}
Assume that there is a range-block $B' \in \rset$ such that $(B, B')$ is a level-$j$ yes-pair and $|B \cap B^{*'}| \leq 8 Z_j \mu_{i}$.
Our goal is to show that with probability at least $1 - e^{-\zeta}$, we report a non-empty region $\rset'$ that does not contain any range-block appearing after $B'$.
We start with the following simple observation whose proof is present in \Cref{prf-obs: ncm-jumpable-sample-few}.

\begin{observation}\label{obs: ncm-jumpable-sample-few}
    With probability $1 - e^{-2 \zeta}$, we sample at most $12 \zeta p|\hat \bset|$ stream-blocks of $\hat \bset$.
\end{observation}

We assume from now on that this event indeed occurs and hence, we terminate $\alggen$ only after processing all the elements of $B$.
We will use the following observation that is proved in \Cref{prf-obs: ncm-b-prime-marked-pseudo-suspicious-early-on}.

\begin{observation}\label{obs: ncm-b-prime-marked-pseudo-suspicious-early-on}
    With probability at least $1 - e^{-2\zeta}$, the range-block $B'$ is marked as pseudo-suspicious while processing the first $\frac{Z_j}{2}$ elements of $S^*$ in $B$.
\end{observation}

We assume that the event of the above observation indeed occur.
Consider the time when $B'$ is marked pseudo-suspicious, say, after processing the element $a_t$ of $B$.
Recall that we execute multiple parallel runs of the level-$j$ algorithm $\algjump{j}$ with input level-$j$ pair $(B, B')$ and the subblock $B_{>t}$ of $t$.
Since the subblock $B_{>t}$ contains at least $\frac{Z_j}{2}$ elements of $S^*$, if $B'$ remains active pseudo-suspicious block throughout the run of our algorithm, each such execution of $\algjump{j}$ returns yes with probability at least $3/4$.
We now consider the state of our algorithm $\alggen$ at the end of the stream-block $B$ and distinguish between the following two cases.

\begin{itemize}
    \item \textbf{Case $1$: $B'$ remains an active pseudo-suspicious block throughout the processing of $B$.}
    In this case, we have executed $\floor{2^6 \zeta}$ independent instances of the level-$j$ algorithm $\algjump{j}$ with input level-$j$ pair $(B, B')$ and the subblock $B_{>t}$ of $B$.
    Since $B'$ remains an active pseudo-suspicious block throughout the processing of $B$, from the correctness guarantee of $\algjump{j}$, each of this instances report yes with probability at least $3/4$.
    From Chernoff bound (see, \Cref{fact: ncm-chernoff-version}), with probability at least $1 - e^{-2\zeta}$, more than $2^5 \zeta$ of these calls return yes.
    We assume from now on that this event indeed occurs.
    Let $B'_{\min{}}$ be the lowest range-block for which at least corresponding $2^5 \zeta$ runs of $\algjump{j}$ returns in affirmative.
    Recall that, if such $B'_{\min}$ exists, we report the region $\rset' = \set{B'_{\min{}}}$.
    We claim that such a range-block $B'_{\min{}}$ exist, and it does not appear after $B'$.
    Indeed, $B'$ is a candidate for being $B'_{\min{}}$ and the claim follows. 

    \item \textbf{Case $2$: $B'$ ceases to be an active pseudo-suspicious during the processing of $B$.}
    Since $B'$ ceases to be an active pseudo-suspicious  block, we must have had at least $|\hat \bset|$ active pseudo-suspicious blocks.
    Moreover, each such active pseudo-suspicious block appear before $B''$.
    Recall that in this case we report the smallest region $\rset' \subseteq \rset$ encompassing all active pseudo-suspicious blocks.
    In particular, this non-empty region $\rset'$ does not contain any range-block appearing after $B'$.
\end{itemize}

Thus, with probability at least $1 - e^{-\zeta}$, we report the region $\rset'$ with claimed properties.
This completes the proof of completeness of our generalized subblock processing algorithm.

\paragraph{Soundness.}
We denote by the event $\eset^*_{\bad}$ the event that the reported region $\rset'$ satisfies the following two properties:
(i) $\rset' \neq \emptyset$; and
(ii) the longest increasing subsequence of $B$ with values in the range-blocks of $\rset'$ has size less than $Z_j/\alpha'_j$.
Our goal is to show that $\prob{\eset^*_\bad} \leq e^{-\zeta}$, where the probability is over the randomness used by $\alggen$ and the executions of $\algjump{j}$ and $\algjump{\ell}$ that it performs.

We will need the following definition.
Recall that we say that a level-$\ell$ pair $(\hat B, \hat B')$ is a promising pair iff $\optlis(\hat B \cap \hat B') \geq \frac{Z_\ell}{\alpha'_\ell}$.
We say that a range-block $B' \in \rset$ is a \emph{good block} if there are at least $3\frac{Z_j \alpha'_\ell}{Z_\ell \alpha'_j}$ stream-blocks of $\hat \bset$ that participate in promising pairs with the level-$\ell$ descendants of $B'$.
We say that it is a \emph{bad block} otherwise.
We let $\hat \eset$ be the event that \emph{every} block marked pseudo-suspicious by our algorithm is also a good block.

\begin{claim}
    $\prob{\hat \eset} \geq 1 - e^{-2\zeta}$.
\end{claim}
\begin{proof}
    We will use the following claim whose proof is deferred to \Cref{prf-clm: ncm-bad-block-unlikely-to-be-pseudo-suspicious}.

    \begin{claim}\label{clm: ncm-bad-block-unlikely-to-be-pseudo-suspicious}
        Fix a range-block $B' \in \rset$ that is a bad block.
        Then $B'$ is marked pseudo-suspicious with probability at most $e^{-3\zeta}$.
    \end{claim}

    Since there are most $|\rset| \leq |\bset'|$ bad blocks, from \Cref{clm: ncm-bad-block-unlikely-to-be-pseudo-suspicious}, the expected number of bad blocks that are marked suspicious is at most $|\bset'| e^{-3\zeta} < e^{-2\zeta}$.
    Here, the last inequality follows size $\zeta = \zeta(i, \ell) \geq \ln{\left(|\bset'| \right)}$.
    From Markov's inequality, the probability that there exists a bad block that is marked pseudo-suspicious is at most $e^{-2\zeta}$.
    In other words, $\prob{\hat \eset} \geq 1 - e^{-2\zeta}$ and the claim follows.
\end{proof}

To show that $\prob{\eset^*_\bad} \leq e^{-\zeta}$, it now suffices to show that for each non-empty region $\rset'$ that is contained in $\rset$ such that the longest increasing subsequence of $B$ with values in the range-blocks of $\rset'$ has size less than $Z_j/\alpha'_j$, the probability that we report $\rset'$ is small.
The claim will then follow by a union bound over $|\rset| + \binom{|\rset|}{2}$ such possible regions.
To this end, fix a region $\rset' \subseteq \rset$.
We let $\eset^*_\bad(\rset')$ be the event that we report the region $\rset'$ and the event $\eset^*_\bad$ occurs.
Note that for us to potentially report the region $\rset'$, either $|\rset'| = 1$ or $|\rset'| \geq |\hat \bset|$ must hold.
We analyze both these cases separately.

\paragraph{Case $1$: $|\rset'| = 1$.}
Let $B' \in \rset$ be the level-$j$ range-block such that $\rset' = \set{B'}$.
For us to report the region $\rset'$, the range-block $B'$ must remain an active pseudo-suspicious block throughout the algorithm and at least $2^5 \zeta$ corresponding calls to $\algjump{j}$ must have returned yes.
Furthermore, for the event $\eset^*_\bad(\rset')$ to occur, $\optlis(B \cap B') < Z_j/\alpha'_j$ must hold.
But then from the correctness guarantee of $\algjump{j}$, each call to $\algjump{j}$ in which the level-$j$ pair $(B, B')$ participates, return yes with probability at most $1/4$.
From Chernoff bound (\Cref{fact: ncm-chernoff-version}), the probability that at least $2^5 \zeta$ out of $\floor{2^6 \zeta}$ such calls return yes is at most $e^{-3\zeta}$.
We now conclude that $\prob{\eset^*_\bad(\rset') \> | \> |\rset'| = 1} \leq e^{-3 \zeta}$.

\paragraph{Case $2$: $|\rset'| \geq |\hat \bset|$.}
We assume that the event $\hat \eset$ indeed occurs.
For the event $\eset^*_\bad(\rset')$ to occur in this case, we must have found a set $\bset''' \subseteq \rset' \subseteq \rset$ of $|\hat \bset|$ active pseudo-suspicious range-blocks.
Fix such a set $\bset'''$ of active pseudo-suspicious range-blocks, each of which is also a good block.
We will use the following observation, whose proof follows that of \Cref{obs: ncm-more-suspicious-blocks-is-good-easy} and is deferred to \Cref{prf-obs: ncm-large-is-in-good-blocks}.

\begin{observation}\label{obs: ncm-large-is-in-good-blocks}
    For every set $\bset_{\good} \subseteq \rset$ of $|\hat \bset|$ good blocks, there is an increasing subsequence of $B$ of length more than $\frac{Z_j}{\alpha'_j}$ using values in range-blocks of $\bset_{\good}$.
\end{observation}

We now conclude that $\prob{\eset^*_\bad(\rset') \> | \>  \left( \hat \eset \text{ and } |\rset'| \geq |\hat \bset|  \right) } = 0$.
We are now ready to complete the proof of soundness of $\alggen$ by union bound over $|\rset| + \binom{|\rset|}{2}$ possible regions $\rset'$ that are contained in $\rset$.
From the above discussion,

\begin{align*}
    \prob{\eset^*_\bad} &= \prob{\bigcup_{\substack{\rset' \text{ is a region in } \rset}} \eset^*_\bad(\rset')}\\
    &\leq \left( \sum_{\substack{\rset' \text{ is a region in } \rset \\ |\rset'| = 1}} \prob{\eset^*_\bad(\rset')} \right) + \prob{\neg \hat \eset} + \left( \sum_{\substack{\rset' \text{ is a region in } \rset \\ |\rset'| \geq |\hat \bset|}} \prob{\eset^*_\bad(\rset') \> | \> \hat \eset} \right) \\
    &\leq |\rset| \cdot e^{-3\zeta} + e^{-2 \zeta} +  \binom{|\rset|}{2} \cdot 0\\
    &\leq e^{-2\zeta} + e^{-2 \zeta}\\
    &\leq e^{-\zeta}.
\end{align*}

Here, the third inequality follows since $|\rset| \leq |\bset'| \leq e^\zeta$.
This completes the proof of soundness, and as a result, the proof of correctness of $\alggen$.
We now turn to analyze the space complexity of $\alggen$, excluding the space used by the calls to algorithms $\algjump{j}$ and $\algjump{\ell}$.

\paragraph{Space Complexity.}
Recall that our algorithm $\alggen$ is given as input a level-$j$ stream-block $B \in \bset$ and a region $\rset \subseteq \bset'$.
We assume that we are given $\rset$ by providing us with the indices of the first and last range-blocks in $\rset$ in the level-$j$ partition $\tilde \bset_j$ of the range $H^*$.
Similarly, we are given the index of $B$ in the level-$j$ partition $\bset_j$ of the original input stream $S$.
Thus, we have stored the descriptions of $B$ and $\rset$ in $O(1)$ units of space.
We perform a number of calls to level-$j$ and level-$\ell$ algorithms $\algjump{j}$ and $\algjump{\ell}$ respectively.
Our goal is to analyze the space complexity of $\alggen$, excluding the space used by such calls to these algorithms.

Before processing the elements of $B$, we initialize the multiset $\bset_\prom \gets \emptyset$ of pseudo-promising range-blocks and the set $\bset_\susp \gets \emptyset$ of pseudo-suspicious range-blocks.
As we process the elements of the stream-block $B$, we mark certain range-blocks of $\rset$ pseudo-promising by appending them to the multiset $\bset_\prom$.
Similarly, we also mark certain range-blocks of $\rset$ pseudo-suspicious by appending them to the set $\bset_\susp$.
As mentioned earlier, the space used in maintaining $\bset_\prom$ and $\bset_\susp$ is trivially upper bounded by $|\bset_\prom| + |\bset_\susp|$, which again is bounded by the number of elements that we sample in $B$ with values in the range-blocks of $\rset$.
Indeed, a range-block $B' \in \rset$ may be marked pseudo-promising only if an element in $B \cap B'$ was sampled.
Moreover, each entry of such a pseudo-promising range-block in the multiset $\bset_\prom$ can be charged to a unique sampled element of $B \cap B'$.
Similarly, each pseudo-suspicious range-block in $\bset_\susp$ can also be charged to a unique sampled element of $B \cap B'$.
Recall that at any point, if we sample more $12 \zeta p |\hat \bset|$ stream-blocks of $\hat \bset$, we immediately terminate the algorithm $\alggen$.
Furthermore, for each sampled stream-block $\hat B \in \hat \bset$, we sample at most $E^*_\ell$ of its elements.
Thus, the space used in storing $\bset_\prom$ and $\bset_\susp$ is $O(|\bset_\prom| + |\bset_\susp|) = O(\zeta p E^*_\ell |\hat \bset|)$.

We now turn to analyze the space used in updating $\bset_\prom$ and $\bset_\susp$ at the end of each level-$\ell$ stream-block.
Notice that the space used in maintaining $\bset_\prom$ can be charged to the space freed after the respective calls to level-$\ell$ algorithm $\algjump{\ell}$ terminate.
Thus, we do not need any additional space for this step.
It is also immediate to verify that we need only $O(|\bset_\prom|)$ additional space to maintain the set $\bset_\susp$ of suspicious blocks.
Finally, the space used to compute the output region $\rset'$ can be charged to the space freed after the respective calls to level-$j$ algorithm $\algjump{j}$ terminate.
Thus, the total space used by $\alggen$, excluding the space used by the calls to $\algjump{j}$ and $\algjump{\ell}$ is at most 

\begin{align*}
    O \left( \zeta p E^*_\ell |\hat \bset| \right) &= O \left( \zeta |\hat \bset|  \cdot \pval \cdot \estarellval \right)   \\
    &\leq O \left( \zeta^3 \cdot \eta^{\ell - j} \cdot \frac{Z_\ell^2 }{Z_j^2} \cdot \frac{\mu_{i}}{\mu_{\ell}} \right).
\end{align*}

Here, the last inequality follows since $|\hat \bset| = \eta^{\ell - j}$.

\paragraph*{Concurrent calls to level-$j$ algorithm $\algjump{j}$.}
We claim that throughout the run of $\alggen$, the number of active calls to the level-$j$ algorithm $\algjump{j}$ is at most $O \left( \zeta \cdot \eta^{\ell - j} \right)$.
Indeed, consider some element $a_t \in B$.
Let $\bset_\active$ be the set of at most $|\hat \bset|$ active pseudo-suspicious range-blocks just after processing $a_t$.
The number of active calls to $\algjump{j}$ is then  at most $|\bset_\active| \cdot {2^6 \zeta} \leq 2^6 \zeta |\hat \bset| = O \left( \zeta \cdot \eta^{\ell - j} \right)$, as claimed.

\paragraph*{Concurrent calls to level-$\ell$ algorithm $\algjump{\ell}$.}
Recall that we do not count the contribution due to the level-$j$ algorithm $\algjump{j}$ to the number of concurrent calls to the level-$\ell$ algorithm $\algjump{\ell}$.
Consider a level-$\ell$ stream-block $\hat B \in \hat \bset$.
If it is not marked, we do not perform any calls to $\algjump{\ell}$ in which $\hat B$ participates.
If it is marked, we process the elements of $\hat B$ with values in the range-blocks of $\rset$, marking each such element independently at random with probability $p'$ each.
If some element $a_t$ of $\hat B$ with value in the level-$\ell$ range-block $\hat B'$ is marked, we perform a call to the level-$\ell$ algorithm $\algjump{\ell}$ with input level-$\ell$ pair $(\hat B, \hat B')$ and the subblock $\hat B_{>t}$ of $\hat B$.
Recall that if we end up marking more than $E^*_\ell$ elements of $\hat B$, we terminate all such calls to $\algjump{\ell}$ in which $\hat B$ participates and discard their computation.
Thus, we perform at most

\begin{align*}
    E^*_\ell = \estarellval &= O \left( Z_\ell \mu_{i} \right) \cdot \pprimeval =  O \left( \frac{\zeta Z_\ell \mu_{i}}{Z_j \mu_{\ell}} \right)
\end{align*}

calls to $\algjump{\ell}$ in which $\hat B$ participates.
This completes the analysis of $\alggen$ and \Cref{lem: ncm-alggen-exists} now follows.

\subsection{Special Case 4: Savable Levels have Large Weight} \label{subsec: ncm-reduction-to-ncm}

Recall that we have fixed parameters $N$ and $\eta$, both integral powers of $2$, such that $\eta$ grows sufficiently slowly with $N$.
We are given an \LIS problem instance $S$ in the streaming model, where $S$ is a permutation of the range $H^* = (1, \ldots, N)$.
Additionally, we are given an ensemble $\Upsilon = (\Psi, \Psi', \vectZ, \vectDelta, \vectMu)$ of length $1+r$
where $r = r(N, \eta) = \floor{\frac{\log N}{\log \eta}}$, $\Psi = \Psi^*(N, \eta) = (\eta, \ldots, \eta)$, $\Psi' = (\psi_1, \ldots, \psi_r)$, $\vectZ = (Z_0, \ldots, Z_r)$, $\vectDelta = (\Delta_0, \ldots, \Delta_r)$, and $\vectMu = (\mu_0, \ldots, \mu_r)$.
We let $\iset = \set{0, \ldots, r}$ be the set of the levels of the underlying hierarchical partitions $\bset_\Psi(S)$ of $S$ into stream-blocks and $\bset_{\Psi'}(H^*)$ of $H^*$ into range-blocks.
We have also fixed an optimal $\Upsilon$-canonical subsequence $S^*$ of $S$.
Note that we do not explicitly know $S^*$ before processing the sequence $S$.
If $S$ is indeed a \YI, then $|S^*| = Z_0$.
On the other hand, if $S$ is a \NI, then there is no $\Upsilon$-canonical subsequence $S^*$ of $S$, and hence, $S^* = \emptyset$.

We have also assumed the existence of the algorithm $\algncm$ for the \NCM problem in the hybrid model with the following guarantees.
Given an \NCM instance $G = (L, R, E_\advice, E)$ in the hybrid model and a parameter $\gamma$ with $|L| = |R|$, $\gamma \geq \left(d(G)\right)^{10^{-3}}$, and $d(G) \geq |G|^{10^{-9}}$, it solves $(\gamma |L|, \gamma |L|/\alphancm(|G|))$-gap \NCM problem with per-vertex query complexity $\left( d(G) \right)^{1 - \delta}$, where $\alphancm(|G|) = |G|^{o(1)}$.
In this subsection, we will use this algorithm $\algncm$ as a subroutine on a number of carefully crafted \NCM instances in the hybrid model.
We fix ${\alpha^* := \alphancm \left(\eta^{10^{10}} \right) = \eta^{o(1)}}$.
We will ensure that each such instance $G = (L, R, E_\advice, E)$ has $|L|, |R|, d(G) \leq \eta^{10^9}$.
Hence, for each such instance $G$, we have $\alphancm(|G|) \leq \alpha^*$.

Recall that we have fixed $\eps = \delta/10^{12}$.
From \Cref{assm: large-xi-by-zi,assm: small-deltai-by-zi}, we assume w.l.o.g. that $X_i/Z_i > N^{1/2 - \eps}$ for all levels $0 \leq i < (1/2 - \eps) r$ and $\Delta_i < Z_i N^{5\eps}$ for all levels $3\eps r \leq i < (1/2  - \eps)r$.
Recall that in \Cref{subsec: ncm-simpler-lis-stream} we have fixed a threshold-level $r^* = \rstarVal$ and the set $\isetnew = \set{i_0, \ldots, i_{\knew}} \subseteq \iset$ of special levels.
In the special case that we deal in this subsection, we are given a set $\iset_\savable \subseteq \isetnew$ of savable levels with weight $w_{\iset_\savable} \geq r/10^5$.
Our goal is to obtain an $N^{o(1)}$-canonical distinguisher algorithm $\algspl{4}$ with space complexity $N^{1/2 - \eps}$.
This subsection is dedicated to the proof of the following lemma.

\begin{lemma}\label{lem: ncm-ncm-lis-reduction-main}
    Suppose we are given parameters $N$ and $\eta$, both integral powers of $2$, along with an ensemble $\Upsilon = (\Psi, \Psi', \vectZ, \vectDelta, \vectMu)$ of length $1 + r$ along with a set $\iset_\savable$ of its savable levels with weight $w_{\iset_\savable} \geq r/10^5$ as mentioned above.
    Further assume that we are given an algorithm $\algncm$ that satisfies the guarantees of \Cref{thm: LIS to NCM final}.
    Then there is a $N^{o(1)}$-canonical distinguisher algorithm $\algspl{4}$ with space complexity $N^{1/2 - \eps}$.
\end{lemma}

Similar to \Cref{subsec: ncm-simpler-lis-stream}, we will create a collection of $1+\knew$ different algorithms, where each algorithm corresponds to a level $i_k \in \isetnew$.
These algorithms will be parameterized by the relevant parameters of the respective levels, and hence, it will be convenient to describe them based on the levels at which they operate. 
Consider some level $i \in \isetnew$, and we denote the corresponding level level-$i$ algorithm by $\algSavable{i}$.
As in \Cref{subsec: ncm-simpler-lis-stream}, $\algSavable{i}$ takes as an input a level-$i$ pair $(B^*, B^{*'})$ and a subblock $B^{**}$ of $B^*$.
It is given the description of $(B^*, B^{*'})$ by giving it the indices of these blocks $B^*$ and $B^{*'}$ in the level-$i$ partitions $\bset^i_\Psi$ and $\bset^i_{\Psi'}$ respectively.
Next, it has access to the elements of $B^{**}$ as they arrive as a part of the original input sequence $S$.
We will ensure that $\algSavable{i}$ is an $(\alpha_{i}, \alpha'_{i})$-canonical distinguisher algorithm for $\alpha_{i} = 4$ and $\alpha'_{i} = \left(16 \alpha^* \log \eta \right)^{(r^*-i)+1}$.
Additionally, we recall that for each pair $i_k < i_{k'}$ of levels of $\isetnew$, we have fixed a parameter $\zeta(i_k, i_{k'}) = \ln{ \left(\eta^{2(i_{k'} - i_k)} \cdot \psi_{i_{k + 1}} \ldots \psi_{i_{k'}} \right)}$.
We introduce the following additional definition.

\paragraph{Perfectly savable levels.}
We say that a savable level $i = i_k \in \iset_\savable$ is \emph{perfectly savable} iff it satisfies the following properties:
(i) $k \leq \knew-2$;
(ii) $i_{k+1} = i+1$ and $i_{k+2} = i+2$;
(iii) $\frac{Z_i}{Z_{i+1}} \geq \eta^{3/4}$ and $\frac{Z_{i+1}}{Z_{i+2}} \geq \eta^{3/4}$; and
(iv) $\psi_{i+1} \leq \eta^{10^{8}}$;

\begin{claim}\label{clm: ncm-perfectly-savable}
    There is a set $\iset'_\savable$ of perfectly savable levels with weight $w_{\iset'_\savable} > r/10^6$.
\end{claim}

The proof of this claim follows from standard techniques is deferred to \Cref{prf-clm: ncm-perfectly-savable}.
Next, we discard from $\iset_\savable$ all levels that are not perfectly savable, but still refer to the resulting set as $\iset_\savable$.
We are now ready to provide the description of our level-$i$ algorithm $\algSavable{i}$.

\paragraph{$\algSavable{i}$ for levels $i \in \iset^* \backslash \iset_\savable$.}
In this case, our algorithm $\algSavable{i}$ is identical to the algorithm $\alglevel{i}$ from \Cref{{lem: ncm-simple-lis-main}} discussed in \Cref{subsec: ncm-simpler-lis-stream}, with one exception.
Recall that if $k < \knew$, the algorithm $\alglevel{i}$ is allowed to the level-$i_{k+1}$ algorithm $\alglevel{i_{k+1}}$ as a subroutine.
In our updated algorithm $\algSavable{i}$, we substitute each of these calls with the corresponding calls to $\algSavable{i_{k+1}}$. 
From \Cref{{lem: ncm-simple-lis-main}}, we immediately deduce the following observation, that is similar to \Cref{obs: ncm-jump-no-jump}:

\begin{observation}\label{obs: ncm-savable-no-savable}
    Assume that $k < \knew$ and for the level $j := i_{k+1}$ we are given a {level-$j$} $(\alphaval, \alpha'_{j})$-canonical distinguisher algorithm $\algSavable{j}$.
    Then, there is a level-$i$ $(\alpha_i, \alpha'_i)$-canonical distinguisher algorithm $\algSavable{i}$ that achieves the following guarantees:
    \begin{itemize}
        \item \textbf{Calls to $\algSavable{j}$:} it performs at most $\zeta^2(i, j) \frac{\mu_{i}}{\mu_{j}}$  concurrent calls to the level-$j$ algorithm $\algSavable{j}$; and
        
        \item \textbf{Space complexity:} its space complexity, excluding the space required by the calls to $\algSavable{j}$ is at most $N^{6\eps}$.
    \end{itemize}
\end{observation}

This completes the description of our algorithm $\algSavable{i}$ for the levels of $\isetnew$ that are not perfectly savable.
We now focus on the perfectly savable levels.

\paragraph{$\algSavable{i}$ for levels $i \in \iset_\savable$.}
In this case, we show the following algorithm, whose analysis is present in \Cref{subsec: ncm-savale-savable} after we complete the description of $\algspl{4}$ assuming it.

\begin{lemma}\label{lem: ncm-savable-savable}
    Fix a perfectly savable level $i := i_k \in \iset_\savable$ and $j := i_{k+1}$.
    Assume that we are given a level-$j$ $(\alpha_j, \alpha'_{j})$-canonical distinguisher algorithm $\algSavable{j}$.
    Further assume that we are given an algorithm $\algncm$ that satisfies the guarantees of \Cref{thm: LIS to NCM final}.
    Then, there is a level-$i$ $(\alpha_i, \alpha'_i)$-canonical distinguisher algorithm $\algSavable{i}$ that achieves the following guarantees:
    \begin{itemize}
        \item \textbf{Calls to $\algSavable{j}$:} it performs at most $\frac{\zeta^6}{\eta^{0.9\delta}} \cdot \frac{\mu_i}{\mu_j}$  concurrent calls to $\algSavable{j}$; and
        
        \item \textbf{Space complexity:} its space complexity, excluding the space required by the calls to $\algSavable{j}$, is at most $N^{o(1)}$.
    \end{itemize}
\end{lemma}

We are now ready to describe our \LIS algorithm for this subsection.
Applying \Cref{obs: ncm-savable-no-savable} and \Cref{lem: ncm-savable-savable} for the levels of $\isetnew$ in their decreasing order, we obtain a level-$i_0$ $(\alpha_0, \alpha'_0)$-canonical distinguisher algorithm $\algSavable{i_0}$.
We prove the following claim after describing $\algspl{4}$ assuming it.

\begin{claim}\label{clm: ncm-savable-level-0-space}
    The space complexity of $\alglevel{i_0}$ is $N^{1/2 -\eps}$.
\end{claim}

From \Cref{{obs: ncm-alglevel-0-suffices}}, there is a $\alpha'_0$-canonical distinguisher algorithm $\algspl{4}$ that performs at most $N^{o(1)}$ concurrent calls to $\algSavable{i_0}$ and whose space complexity, excluding the space required by the calls to $\algSavable{i_0}$, is at most $N^{o(1)}$.
Notice that

\[\alpha'_0 = \left(16 \alpha^* \log \eta \right)^{r^* + 1}  \leq \left(16 \alpha^* \log \eta \right)^{\frac{\log N}{\log \eta}} \leq N^{ \frac{\log \alpha^*}{\log \eta} + o(1)} \leq N^{o(1)}.\]

This completes the description of the algorithm $\algspl{4}$ for this special case and \Cref{{lem: ncm-ncm-lis-reduction-main}} follows assuming \Cref{{lem: ncm-savable-savable},{clm: ncm-savable-level-0-space}}.
This also completes the proof of \Cref{thm: LIS to NCM final}.
In the remainder of this subsection, we prove \Cref{clm: ncm-savable-level-0-space}, while the proof of \Cref{lem: ncm-ncm-lis-reduction-main} is deferred to \Cref{subsec: ncm-savale-savable}.

\proofof{\Cref{clm: ncm-savable-level-0-space}}
    We proceed as in \Cref{{subsec: ncm-simpler-lis-stream},{subsec: ncm-jumpable}}.
    Consider a level $i \in \iset^*$ and fix a level-$i$ stream-block $B$ in the level-$i$ partition $\bset_\Psi^{i}$.
    While executing $\algSavable{i_0}$, we also execute a number of calls to the level-$i$ algorithm  $\algSavable{i}$ in which $B$ participates.
    We denote by $\load(B)$ the maximum number of such concurrent calls to $\algSavable{i}$ in which $B$ participates.
    We denote by $\load(i) := \max_{B \in \bset_\Psi^{i}}{\left( \load(B) \right)}$ the maximum `load' across all level-$i$ stream-blocks.
    To analyze the load, we need the following additional parameters.

    For each level $i_k \in \iset^*$, we have two parameters $q_1(i_k)$ and $q_2(i_k)$, whose values are defined recursively as follows.
    We let $q_1(i_0) = 1$ and $q_2(i_0) = 1$.
    Consider now a level $i_k \in \iset^*$ with $0 < k \leq k^*$.
    We let $q_1(i_k) = q_1(i_{k-1}) \cdot \zeta^6(i_{k-1}, i_{k})$.
    If $i_{k-1} \in \iset_\savable$, we let $q_2(i_k) = \frac{q_2(i_{k-1})}{\eta^{0.9 \delta}}$ and $q_2(i_k) = q_2(i_{k-1})$ otherwise.
    For each level $i_k \in \iset^*$, we also define $L^*(i_k) = \frac{q_1(i_k)}{q_2(i_k)} \cdot \frac{\mu_{i_0}}{\mu_{i_k}}$.
    We use the following two claims, whose proofs follow those of \Cref{{clm: ncm-jump-load-induction},{{clm: ncm-load-i-small}}}, and are deferred to \Cref{{prf-clm: ncm-savable-load-induction},{prf-clm: ncm-savable-load-i-small}} respectively.

    \begin{claim} \label{clm: ncm-savable-load-induction}
        For each level $i_k \in \iset^*$, $\load(i_k) \leq L^*(i_k)$.
    \end{claim}

    \begin{claim} \label{clm: ncm-savable-load-i-small}
        For each level $i_k \in \iset^*$, $\L^*(i_k) \leq N^{\frac{1}{2} - \frac{\delta}{10^9}}$.
    \end{claim}

    We are now ready to analyze the space complexity of $\algSavable{i_0}$.
    We consider its state when processing some element $a_t \in S$.
    Consider a level $i = i_k \in \iset^*$ and let $B_i \in \bset^i_\Psi$ be the unique level-$i$ stream-block that contains $a_t$.
    Observe that $B_i$ participates in all the active calls to the level-$i$ algorithm $\algSavable{i}$.
    Hence, there are at most $\load(i)$ active calls to $\algSavable{i}$.

    We first consider the case where $i_{k} \not \in \iset_\savable$.
    From the guarantee of \Cref{obs: ncm-savable-no-savable}, the space complexity used by such an execution of $\algSavable{i}$, excluding the space required by the calls to lower level algorithms, is at most $N^{6\eps}$.
    From \Cref{{clm: ncm-savable-load-induction},{clm: ncm-savable-load-i-small}}, the contribution to the space complexity of $\algSavable{i_0}$ due to this level is,

    \begin{equation} \label{eqn: ncm-savable-bound-one}
        \begin{split}
            \load(i) \cdot N^{6\eps} &\leq L^*(i) \cdot N^{6\eps} \\
            &\leq N^{\frac{1}{2} - \frac{\delta}{10^9}} \cdot N^{6\eps} = N^{\frac{1}{2} - \frac{\delta}{10^9} + 6\eps} \leq N^{\frac{1}{2} - 2\eps},
        \end{split}
    \end{equation}

    since $\eps = \delta/10^{12}$.
    We now consider the remaining case where $i \in \iset_\savable$.
    From the guarantee of \Cref{lem: ncm-savable-savable}, the space complexity used by such an execution of $\algSavable{i}$, excluding the space required by the calls to lower level algorithms, is at most $N^{o(1)}$.
    As before, we can now bound the contribution due to level $i$ to the space complexity of $\algSavable{i_0}$ by,

    \begin{equation} \label{eqn: ncm-savable-bound-two}
        \begin{split}
            \load(i) \cdot N^{o(1)} &\leq L^*(i) \cdot N^{o(1)}\\
            &\leq N^{\frac{1}{2} - \frac{\delta}{10^9}} \cdot N^{o(1)} = N^{\frac{1}{2} - \frac{\delta}{10^9} + o(1)} \leq N^{\frac{1}{2} - 2\eps},
        \end{split}
    \end{equation}

    since $\eps = \delta/10^{12}$.
    From \Cref{{eqn: ncm-savable-bound-one},{eqn: ncm-savable-bound-two}}, we now conclude that the space complexity of $\algSavable{i_0}$ is indeed bounded by,
        
        \[ |\iset^*| \cdot N^{1/2 - 2\eps} < r \cdot N^{1/2 - 2\eps} < N^{1/2 - \eps}. \]

    Here, the last inequality follows since $r < \log N$ and $\eps$ is a constant.
    This completes the analysis of the space complexity of $\algSavable{i_0}$, completing the proof of  \Cref{clm: ncm-savable-level-0-space}.
\endproofof

\subsection{$\algSavable{i}$ and its Properties --- Proof of Lemma \ref{lem: ncm-savable-savable}} \label{subsec: ncm-savale-savable}

We are given a perfectly savable level $i := i_k \in \iset_\savable$.
We are also given a level $j := i_{k+1} = i+1$ of $\isetnew$ along with a level-$j$ $(\alpha_j, \alpha'_j)$-canonical distinguisher algorithm $\algSavable{j}$.
Our goal is to prove \Cref{lem: ncm-savable-savable} by devising a level-$i$ $(\alpha_i, \alpha'_i)$-canonical distinguisher algorithm $\algSavable{i}$ with low space complexity that performs a small number of calls to $\algSavable{j}$.
Recall that the input to $\algSavable{i}$ is a level-$i$ pair $(B^*, B^{*'})$ and a subblock $B^{**}$ of $B^*$.

We consider the collection $\bset$ of level-$j$ descendant stream-blocks of $B^*$ that are entirely contained in $B^{**}$.
We also consider the collection $\bset'$ of level-$j$ descendant range-blocks of $B^{*'}$.
We first claim that if $\bset < \frac{Z_i}{2\alpha_i Z_j}$, we can report no.
Indeed, if that is the case, $B^{**}$ shares elements with fewer than $\frac{Z_i}{2\alpha_i Z_j} + 2 \leq \frac{Z_i}{\alpha_i Z_j}$ level-$j$ blocks.
Here, the inequality follows from the fact that $\alpha_i = \alphaval$ and Property \ref{prop: ncm-p4} of the ensemble $\Upsilon$.
But then, $B^{**}$ must contain fewer than $\frac{Z_i}{\alpha_i Z_j} \cdot Z_j = \frac{Z_i}{\alpha_i}$ elements of each $\Upsilon$-canonical subsequence, and in particular, of $S^*$.
Thus, assume from now on that $|\bset| \geq \frac{Z_i}{2\alpha_i Z_j} \geq \frac{\eta^{3/4}}{8}$.
Here, the last inequality follows since $\alpha_i = \alphaval$ and $\frac{Z_i}{Z_j} = \frac{Z_i}{Z_{i+1}} \geq \eta^{3/4}$ since $i$ is a perfectly savable level.
On the other hand, we have $|\bset'| = \psi_{j} = \psi_{i+1} \leq \eta^{\ncminssize}$.
Recall that we have defined $\zeta(i_k, i_{k+1}) = \zeta(i,j) = \ln{ \left(\eta^{2} \cdot \psi_{i+1} \right)}$.
For readability, we denote $\zeta := \zeta(i, j)$ from now on.
Notice that,

\begin{equation} \label{eqn: ncm-savable-zeta-bound}
    2 \ln \eta \leq \zeta \leq \ln{\left(\eta^{2 + 10^8}\right)} \leq \left(2 \cdot 10^8 \right) \ln \eta.
\end{equation}

Before describing our algorithm $\algSavable{i}$, we define a number of instances of the \NCM problem in the hybrid model.
We will use the algorithm $\algncm$ as an oracle on these carefully crafted instances.
We consider $\log \eta$ different \NCM problem instances: for each  $\tau \in \set{2^0, \ldots, 2^{\log{(\eta)}-1}}$, there is an instance $\Gtau = (\Ltau, \Rtau, \Etau_\advice, \Etau)$ of the \NCM problem in the hybrid model that is probabilistically obtained as follows.

\paragraph{The set $\Ltau$ of left-vertices.}
For each level-$j$ stream-block $B \in \bset$, there is a collection $\Ltau(B)$ of $\tau$ vertices obtained as follows.
Consider the partition $\bset^{(\tau)}(B)$ of $B$ into $\tau$ subblocks, that we denote by $\bset^{(\tau)}(B) = \set{\tilde B_1, \ldots, \tilde B_\tau}$ in their natural order.
For each subblock ${\tilde B \in \bset^{(\tau)}(B)}$, there is a unique vertex $v{(\tilde B)}$ in $\Ltau(B)$.
These vertices appear in their natural order: $\Ltau(B) = \left( v(\tilde B_1), \ldots, v(\tilde B_\tau) \right)$.
Note that $\left| \bigcup_{B \in \bset} \Ltau(B) \right| = |\bset| \tau < \eta^2$.
The set $\Ltau$ now contains all the vertices $\Ltau(B)$ for each level-$j$ stream-block $B \in \bset$ in their natural order followed by ${\ncminssize} - |\bset|\tau$ special vertices.
We will ensure that no edge of $\Etau_\advice$, and hence, of $\Etau$ is incident on these special vertices.
This completes the description of our set $\Ltau$ of exactly ${\ncminssize}$ vertices.

\paragraph{The set $\Rtau$ of right-vertices.}
For each level-$j$ range-block $B' \in \bset'$, there is a unique vertex $v(B')$ in $\Rtau$.
Note that $\left| \bigcup_{B' \in \bset'} \set{v(B')} \right| = |\bset'| = \psi_{i+1}$.
The set $\Rtau$ now contains all such vertices in the natural order of their underlying range-blocks followed by ${\ncminssize} - |\bset'|$ special vertices.
As before, will ensure that no edge of $\Etau_\advice$, and hence, of $\Etau$ is incident on these special vertices.
This completes the description of our set $\Rtau$ of exactly ${|\Ltau| = {\ncminssize}}$ vertices.

\paragraph{The set $\Etau_\advice$ of advice-edges.}
The set $\Etau_\advice$ does not contain any edge-slot incident on special vertices of $\Ltau \cup \Rtau$.
Consider now some vertex $v(\tilde B) \in \Ltau$ that is not special and let $B \in \bset$ be the unique level-$j$ block that contains $\tilde B$.
The set of advice-edges $\Etau_\advice$ incident to $v(\tilde B)$ is obtained as follows.
Let $(\tilde B^{(1)}, \tilde B^{(2)})$ be the partition of $\tilde B$ into two equal subblocks.
We sample the elements of $\tilde B^{(1)} \cap B^{*'}$ independently with probability $p = \savpval$ each.
For each level-$j$ range-block $B' \in \bset'$ whose value is sampled, we add the corresponding edge $(v(\tilde B), v(B'))$ to the set of advice-edges incident to $v(\tilde B)$.
If we sample more than $d := \savellstarjval$ elements from $\tilde B^{(1)}$, we discard all the advice-edges incident to $v(\tilde B)$ found so far, and report that there are no advice-edges incident to it.
This completes the description of set of advice-edges incident to the vertex $v(\tilde B)$.
The final set of $\Etau_\advice$ is the union of all such advice-edges.

\paragraph{The set $\Etau$ of edges.}
The set $\Etau$ does not contain any edge absent from $\Etau_\advice$.
Consider some vertex $v(\tilde B) \in \Ltau$ that is not special and let $B \in \bset$ be the unique level-$j$ block that contains $\tilde B$.
Also consider the partition $(\tilde B^{(1)}, \tilde B^{(2)})$ of $\tilde B$ into two subblocks.
The set of ${\Etau \subseteq \Etau_\advice}$ edges incident to $v(\tilde B)$ is obtained as follows.
For each level-$j$ range-block $B' \in \bset^{*'}$ such that there is an advice-edge $(v(\tilde B), v(B')) \in \Etau_\advice$, we run $\floor{2^6 \zeta^2}$ independent instances of the level-$j$ algorithm $\algSavable{j}$ with input level-$j$ pair $(B, B')$ along with the subblock $\tilde B^{(2)}$ of $B$.
We say that $(v(\tilde B), v(B'))$ is an edge in $\Etau$ iff at least $2^5 \zeta^2$ such executions of $\algSavable{j}$ return yes.

This completes the description of our \NCM instance $\Gtau = (\Ltau, \Rtau, \Etau_\advice, \Etau)$ for each  $\tau \in \set{2^0, \ldots, 2^{\log{(\eta)}-1}}$.
Before describing our algorithm $\algSavable{i}$, we analyze the properties of these \NCM instances.

\paragraph{Properties of $\Gtau$.}
Fix some $\tau \in \set{2^0, \ldots, 2^{\log{(\eta)}-1}}$.
We note that the vertex-sets $\Ltau$ and $\Rtau$ are completely determined by the input level-$i$ pair $(B^*, B^{*'})$ and the subblock $B^{**}$ of $B^*$.
Consider now the corresponding \NCM problem instance $\Gtau = (\Ltau, \Rtau, \Etau_\advice, \Etau)$.
The advice-edge set $\Etau_\advice$ of $\Gtau$ is a random variable that depends solely on the randomness used in sampling the corresponding elements of $B^{**}$.
On the other hand, the edge set $\Etau$ of $\Gtau$ is a random  variable that depends on both, on the randomness used in sampling the corresponding elements of $B^{**}$ and the randomness used by the corresponding calls to $\algSavable{j}$. 
It is worthwhile to note that $\Gtau$ itself is a random variable independent of the randomness used by (potentially multiple) executions of $\algncm$ with $\Gtau$ as input.
Note that the maximum vertex degree  of $L$ for $E_\advice$ in $\Gtau$ is at most $d$.
We fix the parameter $\gamma := \frac{Z_i }{16 Z_j \log \eta}$ and will subsequently use $\algncm$ with input $\Gtau$ to distinguish between $\optncm(\Gtau) \geq \gamma$ from the case where $\optncm(\Gtau) < \gamma / \alpha(\Gtau)$.
To ensure that $\algncm$ succeeds on this decision problem, we need to ensure that $\gamma$ is large enough, as a function of $d$.
From our choice of $d = \savellstarjval$ and $p = \savpval$, we immediately obtain,

\begin{equation}\label{eqn: ncm-savable-d-eq-bound}
    d = \savellstarjval = 2^6 \cdot \frac{2^{16} \zeta}{Z_j \mu_j} \cdot Z_j \mu_i = 2^{22} \zeta \frac{\mu_i}{\mu_j} = 2^{22} \zeta \eta^{w_i}.
\end{equation}

From  \Cref{eqn: ncm-savable-zeta-bound} and the fact that $w_i < 100$ since level-$i$ is a savable level, we can upper bound $d$ by,

\begin{equation}\label{eqn: ncm-savable-d-upper-bound}
    d < O(\zeta) \cdot \eta^{w_i} < O(\ln \eta) \cdot \eta^{100}.
\end{equation}

Recall that $w_i > 0.9$ and ${|G| = |L| \cdot |R| = \left(  {\ncminssize} \right)^2 \leq \eta^{2 \cdot 10^8}}$, since $i$ is a perfectly savable level.
Plugging this inequalities in \Cref{{eqn: ncm-savable-d-eq-bound}} to obtain a lower bound on $d$,

\begin{equation}\label{eqn: ncm-savable-d-lower-bound}
    d > \eta^{w_i} > \eta^{0.9} > |G|^{0.9/(2 \cdot 10^8)} > |G|^{10^{-9}}.
\end{equation}

Plugging \Cref{eqn: ncm-savable-d-upper-bound} in our choice of $\gamma = \frac{Z_i}{16 Z_j \log \eta}$, we now obtain,

\begin{equation} \label{eqn: ncm-savable-gamma-lower-bound}
    \gamma = \frac{Z_i}{16 Z_j \log \eta} \geq \frac{\eta^{3/4}}{16 \log \eta} > d^{10^{-3}}.
\end{equation}

Here, the first inequality follows since $\frac{Z_i}{Z_j} \geq \eta^{3/4}$ as $i$ is a perfectly savable level and the second inequality follows since $\eta = \eta(N) = \omega(1)$ is an increasing function of $N$.
From \Cref{{eqn: ncm-savable-d-lower-bound},{eqn: ncm-savable-gamma-lower-bound}} along with our construction of $\Gtau$, we immediately obtain the following observation.

\begin{observation} \label{obs: ncm-algncm-works-on-gtau}
    Fix an instance $\Gtau = (\Ltau, \Rtau, \Etau_\advice, \Etau)$ of the \NCM problem as detailed above.
    Then the algorithm $\algncm$ assumed in \Cref{lem: ncm-savable-savable} with input instance $\Gtau$ in the hybrid model distinguishes the case where $\optncm(\Gtau) \geq \gamma$ from the case where $\optncm(\Gtau) < \gamma/\alpha^*$, with per-vertex query complexity at most $d^{\left(1 - \delta \right)}$, succeeding with probability at least $2/3$.
    Here, the probability is solely over the randomness used by $\algncm$ and is independent of the same used in the generation of the instance $\Gtau$.
\end{observation}

This completes the analysis of the generated \NCM problem instances.
We are now ready to describe our algorithm $\algSavable{i}$ for processing the input level-$i$ pair $(B^*, B^{*'})$ and the subblock $B^{**}$ of $B^*$.

\paragraph{Description of $\algSavable{i}$.}
For each $\tau \in \set{2^0, \ldots, 2^{\log{(\eta)} - 1}}$, we run $\floor{2^6 \zeta}$ independent instances of the algorithm $\algncm$ on the \NCM problem instance $\Gtau$ to distinguish between the case where  $\optncm(\Gtau) \geq \gamma$ from the case where $\optncm(\Gtau) < \frac{\gamma}{\alpha(\Gtau)}$.
We report yes iff there is a $\tau \in \set{2^0, \ldots, 2^{\log{(\eta)} - 1}}$ such that at least $2^5 \zeta$ corresponding executions of $\algncm$ return yes.
This completes the description of our level-$i$ algorithm $\algSavable{i}$.

It is immediate to see that the run of algorithm $\algncm$ on $\Gtau$ in the hybrid model can be simulated in the streaming model while processing the elements of $B^{**}$.
Indeed, we know the vertices $\Ltau$ and $\Rtau$ before processing the elements of $B^{**}$.
We run the algorithm $\algncm$ on the instance $\Gtau$ in the hybrid model as follows.
Consider the time just before $\algncm$ starts to process some vertex $v(\tilde B) \in \Ltau$.
Let $B \in \bset$ be the unique level-$j$ stream-block that contains $\tilde B$.
We will simulate the processing of $\algncm$ on $v(\tilde B)$ while processing the elements of $\tilde B$ as they arrive as the part of the sequence $S$.
Let $(\tilde B^{(1)}, \tilde B^{(2)})$ be the partition of $\tilde B$ into two subblocks, that we can compute before the arrival of elements of $\tilde B$.
We now compute the set of advice-edges $\Etau_\advice$ incident on $v(\tilde B)$ while processing the elements of $\tilde B^{(1)}$.
We reveal this set of advice-edges to $\algncm$ and wait for the subset of advice-edges that it choses to query.
For each level-$j$ range-block $B'$ such that $\algncm$ chose to query the advice-edge $(v(\tilde B), v(B') )$, we run the algorithm $\algSavable{j}$ with input level-$j$ pair $(B, B')$ and the subblock $\tilde B^{(2)}$ of $B$ while processing the elements of $\tilde B^{(2)}$ as they arrive as a part of $B^{**}$.
At the end of the subblock $\tilde B^{(2)}$, we know which runs of $\alg^{(j)}$ returned yes.
We report the corresponding edges to $\algncm$ as edges of $\Etau \cap \Etau_\advice$ and proceed to process subsequent vertices of $\Ltau$.
Thus, we can indeed simulate the run of algorithm $\algncm$ on $\Gtau$ in the hybrid model while processing the elements of $B^{**}$ in the streaming model.

We now turn to analyze the properties of $\algSavable{i}$, starting with correctness guarantee.

\paragraph{Soundness guarantee.}
Assume that $\optlis(B^{**} \cap B^{*'}) < Z_i/\alpha'_i$.
Our goal is to show that we report yes with probability less than $1/4$.
For each $\tau \in \set{2^0, \ldots, 2^{\log{(\eta)} - 1}}$, we fix the corresponding \NCM problem instance $\Gtau$.
Recall that $\Gtau$ is a random variable and is independent of the randomness used by the executions of $\algncm$.
We let $\eset_\bad(\tau)$ be the event that $\optncm(\Gtau) \geq \frac{Z_i/\alpha'_i}{Z_j/\alpha'_j} = \frac{\gamma}{\alpha^*}$.
We also let $\eset_\bad := \bigcup_{\tau \in \set{2^0, \ldots, 2^{\log{(\eta)} - 1}}} \eset_\bad(\tau)$ be the event that at least one of these events occur.
We first bound the probability that the event $\eset_\bad$ occurs in the following claim whose proof is deferred to \Cref{prf-clm: ncm-soundness-eset-bad}.

\begin{claim}\label{clm: ncm-soundness-eset-bad}
    $\prob{\eset_\bad} \leq 1/8$.
\end{claim}

We assume from now on that the event $\eset_\bad$ does not occur, or in other words, for each $\tau \in \set{2^0, \ldots, 2^{\log{(\eta)} - 1}}$, the corresponding \NCM problem instance $\Gtau$ has ${\optncm(\Gtau) < \frac{\gamma}{\alpha^*} \leq \frac{\gamma}{\alpha(\Gtau)}}$.
Consider some  $\tau \in \set{2^0, \ldots, 2^{\log{(\eta)} - 1}}$.
We let $\eset'_\bad(\tau)$ be the event that at least $2^5 \zeta$  executions of $\algncm$ with input $\Gtau$ out of  $\floor{2^6 \zeta}$ independent ones  return yes.
From \Cref{obs: ncm-algncm-works-on-gtau}, for each such $\tau$ and the corresponding instance $\Gtau$, an execution of $\algncm$ returns yes with probability at most $1/3$.
Thus, from Chernoff bound (\Cref{fact: ncm-chernoff-version}), ${\prob{\eset'_\bad(\tau) \> | \> \neg \eset_\bad} \leq e^{-3 \zeta}}$.
We denote by $\eset'_\bad$ the event that there is a  $\tau \in \set{2^0, \ldots, 2^{\log{(\eta)} - 1}}$ such that the event $\eset'_\bad(\tau)$ occurs.
From union bound, the probability that the event $\eset'_\bad$ occurs conditioned that the event $\eset_\bad$ does not occur is at most ${\prob{\eset'_\bad \> | \> \neg \eset_\bad} \leq \eta \cdot e^{-3 \zeta} \leq \eta \cdot \eta^{-6} \leq 1/8}$.
Hence, ${\prob{\eset'_\bad} \leq \prob{\eset_\bad} + \prob{\eset'_\bad \> | \> \neg \eset_\bad}  \leq 1/4}$.
Recall that we report yes iff the event $\eset'_\bad$ occurs and the soundness guarantee of $\algSavable{i}$ now follows.

\paragraph{Completeness guarantee.}
Recall that we have fixed an optimal $\Upsilon$-canonical increasing subsequence $S^*$ of $S$.
Assume that $(B^*, B^{*'})$ is a yes-pair for $S^*$ and the subblock $B^{**}$ of $B^*$ contains at least $\frac{Z_i}{2}$ elements of $S^*$.
Our goal is to show that we report yes with probability at least $3/4$.
We use the following claim whose proof follows standard techniques and is present in \Cref{prf-clm: ncm-g-tau-large-ncm}.

\begin{claim}\label{clm: ncm-g-tau-large-ncm}
    With probability at least $0.9$,  there is some $\tau \in \set{2^0, \ldots, 2^{\log{(\eta)} - 1}}$ such that  $\optncm(\Gtau) \geq \frac{Z_i}{16 Z_j \log \eta} = \gamma$. 
\end{claim}

Assume from now on that this event indeed occurs.
We fix such $\tau$ and corresponding \NCM problem instance $\Gtau$.
From \Cref{obs: ncm-algncm-works-on-gtau}, each execution of $\algncm$ with input $\Gtau$ returns yes with probability at least $2/3$, where the probability is over the randomness used by $\algncm$.
From Chernoff bound (\Cref{fact: ncm-chernoff-version}), it is immediate to verify that with probability at least $0.9$, at least $2^5 \zeta^2$ executions of $\algncm$ on input $\Gtau$ report yes.
Thus, we report yes with probability at least $0.8$ and the completeness guarantee of $\algSavable{i}$ follows.

\paragraph*{Calls to $\algjump{j}$.}
Consider a $\tau \in \set{2^0, \ldots, 2^{\log(\eta) - 1}}$ and the corresponding \NCM problem instance $\Gtau = (\Ltau, \Rtau, \Etau_\advice, \Etau)$.
We examine one out of $\floor{2^6 \zeta}$ parallel runs of $\algncm$ that we execute with input $\Gtau$.
Consider a level-$j$ stream-block $B \in \bset$ and a partition $\bset^{(\tau)}(B)$ of $B$ into $\tau$ subblocks.
Finally, consider a block $\tilde B \in \bset^{(\tau)}(B)$ and its partition $\left(\tilde B^{(1)}, \tilde B^{(2)} \right)$ into $2$ subblocks.
Notice that when we are processing the block $\tilde B$, the subblock $\tilde B^{(2)}$ participates in all the ongoing calls to $\algSavable{j}$.
Specifically, for each advice-edge incident to $v(\tilde B)$ that is queried by $\algncm$, we perform $\floor{2^6 \zeta^2}$ calls to $\algSavable{j}$ in which $\tilde B^{(2)}$ participates.
Recall that $\algncm$ performs at most $d^{1 - \delta}$ such queries.
Hence, the number of concurrent calls to $\algjump{j}$ is bounded by,

\begin{align*}
    \log{(\eta)} \cdot 2^6 \zeta \cdot d^{1 - \delta} \cdot 2^6 \zeta^2 &\leq O(\log \eta) \cdot \zeta^3 \cdot d^{1-\delta} \\
    &\leq O(\log \eta) \cdot \zeta^4 \cdot \eta^{(1-\delta)w_i}\\
    &\leq O(\zeta^5) \cdot \frac{\mu_i}{\mu_j} \cdot \frac{1}{\eta^{\delta w_i}}\\
    &\leq O(\zeta^5) \cdot \frac{\mu_i}{\mu_j} \cdot \frac{1}{\eta^{0.9\delta}}.
\end{align*}

Here, the second inequality follows from \Cref{eqn: ncm-savable-d-eq-bound} and the last inequality follows since $i$ is a savable level and hence, $w_i > 0.9$.

\paragraph{Space complexity.}
We first analyze the space complexity of $\algSavable{i}$ excluding the space used by the calls to $\algncm$ and $\algSavable{j}$.
Consider a $\tau \in \set{2^0, \ldots, 2^{\log(\eta) - 1}}$ and the corresponding \NCM problem instance $\Gtau = (\Ltau, \Rtau, \Etau_\advice, \Etau)$.
Consider one out of $\floor{2^6 \zeta}$ parallel instances of $\algncm$ that we execute with input $\Gtau$.
Consider a vertex $v(\tilde B) \in \Ltau$ and the corresponding block $\tilde B$.
Let $B \in \bset$ be the level-$j$ stream-block containing $\tilde B$.
It is immediate to verify that we can compute the subset of advice edges incident to $v(\tilde B)$ in space $O(d)$.
For each advice-edge that $\algncm$ chose to query, we execute a number of parallel instances of $\algSavable{j}$.
It is also easy to see that we can charge the space used in computing the edges to report to $\algncm$ to the space freed after the respective calls to $\algSavable{j}$ terminate.
Thus, the space used by $\algSavable{i}$, excluding the space used by the calls to $\algncm$ and $\algSavable{j}$, is bounded by,

\begin{align*}
    \log{(\eta)} \cdot 2^6 \zeta \cdot O(d) &\leq O(\zeta^2) \cdot d = O(\zeta^2) \cdot p Z_j \mu_i = O(\zeta^3) \cdot \frac{\mu_i}{\mu_j} \leq O(\zeta^3) \cdot \eta^{100} < N^{o(1)}.
\end{align*}

Here, second-last inequality follows since $i$ is a savable level with weight $w_i = \log_\eta{\left(\frac{\mu_i}{\mu_j} \right)} < 100$.
The last inequality follows since $\zeta \leq \poly \log \eta$ and $\eta = O(\log \log N)$ are small enough.

We now bound the space used by calls to $\algncm$.
Consider an execution of $\algncm$ with input \NCM instance $\Gtau = (\Ltau, \Rtau, \Etau_\advice, \Etau)$
It is immediate to verify that the space complexity of this execution is at most $2^{\poly(|\Gtau|)} = 2^{\poly(|\Ltau| \cdot |\Rtau|)} = 2^{\poly \left(\eta^{O(1)} \right)} = 2^{\eta^{O(1)}}$.
But from our choice of $\eta = O(\log \log N)$, we can further bound this space complexity by ${2^{\eta^{O(1)}} = 2^{\poly \log \log N} = 2^{o(\log N)} =  N^{o(1)}}$.
Since we perform at most $\log(\eta) \cdot 2^6 \zeta \leq N^{o(1)}$ concurrent calls to $\algncm$, the overall space used the calls to $\algncm$ is bounded by $N^{o(1)}$.
We now conclude that the space complexity of $\algSavable{i}$, excluding the space used by the calls $\algSavable{j}$ is bounded by $N^{o(1)}$.

This completes the analysis of $\algSavable{i}$ and \Cref{{lem: ncm-savable-savable}} now follows.

    \section{Randomized \LIS Algorithm in Streaming Model with $\alpha$-approximation} \label{sec: ncm-alpha-approx}
    \toggletrue{alpha-approx-lis}
\newcommand{\alphalislogfactor}{9}

The goal of this section is to prove \Cref{thm: sqrt n by alpha randomized lis algo}.
Consider an input stream $S = (a_1, \ldots, a_N)$ for the \LIS problem in streaming model consisting of elements with values in the universe $\uset = \set{1, \ldots, M}$.
We assume that each element of $\uset$ can be stored in unit space.
As earlier, we also assume that the length $N$ of the stream  is an integral power of $2$, and is known to us in advance.
Given a parameter $\alpha \leq N^{1/4}$ such that $\alpha \geq 1 + \eps$ for some constant $\eps$, our goal is to estimate $\optlis(S)$ within a factor of $\alpha$ using at most $\tilde O(\sqrt{N}/\alpha)$ units of space. 
We assume that $N$ is large enough, so that, $\log N \geq 2^{32}$.

Note that in the case where $\alpha \leq \log^{12} N$, the (deterministic) algorithm $\alg_3$ from \Cref{{obs: ncm-det-sqrt-n-space-sound}} already achieves factor-$(1+\eps)$ approximation in space $O(\sqrt{N})$.
Thus, we assume from now on that $\alpha > \log^{12} N$.
Using the standard techniques, it suffices to show the algorithm $\alglis$ for the \LIS problem, where, in addition to the input sequence $S$, we are given a `guess' $\tau^*$ for $\optlis(S)$.
Our goal is then to distinguish the case where $\optlis(S) \geq \tau^*$ from the case where $\optlis(S) < 2\tau^*/\alpha$.
We thus assume that we are given such a guess $\tau^*$.
If $\tau^* \leq \sqrt{N}/\alpha$, we can use the algorithm $\alg_1$ from \Cref{obs: ncm-det-opt-space-sound} to distinguish the case where $\optlis(S) \geq \tau^*$ from the case where $\optlis(S) < \tau^*$ in space $O(\tau^*) = O(\sqrt{N}/\alpha)$.
On the other hand, if $\tau^* \geq \alpha \sqrt{N}$, we can use the algorithm $\alg_2$ from \Cref{obs: ncm-det-n-by-opt-space-sound} to distinguish the case where $\optlis(S) \geq \tau^*$ from the case where $\optlis(S) < \tau^*/2$.
The space complexity of the algorithm $\alg_2$ is $\tilde O(N / \tau^*) \leq \tilde O(\sqrt{N}/\alpha)$ as required.
We assume from now on that $\tau^*$ satisfies,

\begin{equation}\label{eqn: ncm-alpha-approx-tau-star}
    \sqrt{N}/\alpha < \tau^* < \alpha \sqrt{N}.    
\end{equation}

We set parameters $\frac{\tau^*}{2 \alpha} < Z_1 \leq \frac{\tau^*}{\alpha}$ and $\frac{Z_1}{\alpha} < Z_2 \leq \frac{2Z_1}{\alpha}$, such that both, $Z_1$ and $Z_2$ are integral powers of $2$.
We also let $\frac{\tau^*}{2\log^\alphalislogfactor N} < Z_0 \leq \frac{\tau^*}{\log^\alphalislogfactor N}$ such that $Z_0$ is an integral power of $2$.
It is worthwhile to note that 

\begin{equation} \label{eqn: ncm-alpha-approx-z0-by-z1}
    \frac{Z_0}{Z_1} \geq \frac{\tau^*/\log^\alphalislogfactor N}{\tau^*/\alpha} =  \frac{\alpha}{\log^\alphalislogfactor N}   
\end{equation}

We need the following observation that is analogous to the Partition \Cref{lem: ncm-partition-lemma} but optimized for two levels of hierarchical partition instead of an arbitrary increasing number of levels.

\begin{observation}\label{obs: ncm-alpha-lis-part}
    There are $\psi_1$ and $\psi_2$, both integral powers $2$, such that for $\vectZ = (Z_1, Z_2)$ and $\Psi = (\psi_1, \psi_2)$, the resulting hierarchical partition $\bset_\Psi(S)$ of $S$ into $2$ levels of stream-blocks achieves the following guarantee.
    There is an increasing subsequence $S^*$ of $S$ of size at least ${\optlis(S)}/{\log^\alphalislogfactor N}$ that is $\vectZ$-canonical w.r.t. the hierarchical partition $\bset_\Psi(S)$.
\end{observation}

The proof of this observation is similar to that of \Cref{{lem: ncm-partition-lemma},{lem: ncm-vanilla-partition}} and is present in \Cref{prf-obs: ncm-alpha-lis-part}.
We are now ready to describe our algorithm of this section, that is similar to the one of \Cref{subsec: ncm-jump-jump-i}.

We fix the parameters $\psi^*_1$ and $\psi^*_2$ such that the resulting hierarchical partition $\bset_{\Psi^*}(S)$, where $\Psi^* = (\psi^*_1, \psi^*_2)$, achieves the claimed guarantee of \Cref{obs: ncm-alpha-lis-part}.
Note that there are only $O(\log^2 N)$ potential choices of these parameters  $\psi^*_1$ and $\psi^*_2$.
Hence, at the cost of multiplicative overhead of $O(\log^2)$ in our space complexity, we run in the following algorithm in parallel for all pairs of $\psi_1$ and $\psi_2$, that are integral powers of $2$ in $\set{1, \ldots, N}$.
Let $\bset^1 = \bset^1_\Psi(S)$ and $\bset^2 = \bset^2_\Psi(S)$ be the partitions of the resulting hierarchical partition $\bset_\Psi(S)$ of $S$ into $2$ levels of stream-blocks.
We also fix an optimal $\vectZ = (Z_1, Z_2)$-canonical-increasing subsequence $S^*$ of $S$.
Note that if $\psi_1 = \psi^*_1$ and $\psi_2 = \psi^*_2$, the length of $S^*$ is at least $\optlis(S)/\log^\alphalislogfactor N$.
Furthermore, if $S$ is a \yi, or in other words, $\optlis(S) \geq \tau^*$, we have,

\[ |S^*| \geq \frac{\optlis(S)}{\log^\alphalislogfactor N} \geq \frac{\tau^*}{\log^\alphalislogfactor N} \geq Z_0. \]

It now suffices to show an algorithm that achieves the following guarantee.
It is given access to the original input sequence $S$ in the streaming model along with parameters $\psi_1$, $\psi_2$, $\tau^*$, $Z_0$, $Z_1$, and  $Z_2$.
If $\psi_1 = \psi^*_1$, $\psi_2 = \psi^*_2$, and $S$ is a \yi, or in other words, there is an increasing subsequence $S^*$ of size at least $Z_0$ that is $\vectZ$-canonical w.r.t. the partition $\bset_{\Psi}(S)$, where $\Psi = (\psi_1, \psi_2)$ of $S$; then it must report yes with probability at least $1 - 1/N$.
On the other hand, the probability that it report yes even though $\optlis(S) < Z_0/\alpha$, is at most $1/N$ over the randomness used by our algorithm.
We are now ready to describe our algorithm.

We let $\Psi~=~(\psi_1, \psi_2)$ and consider the hierarchical partition $\bset_{\Psi}(S) = (\bset^1, \bset^2)$ of the stream $S$ into stream-blocks.
We will process stream-blocks of $\bset^1$ as they arrive as a part of the original input sequence $S$.
While processing these elements, we will maintain an ordered collection $\tilde \rset$ of at most $Z_0/Z_1$ disjoint non-empty blocks of $H^*$.
Let $\leftover(\tilde \rset) := H^* \backslash \bigcup_{R \in \tilde \rset} R$ be the set of elements not appearing in the blocks of $\tilde \rset$.
We will ensure that $\leftover(\tilde \rset)$ is also a (possibly empty) range-block and the non-empty blocks of $\tilde \rset \cup \set{\leftover(\tilde \rset)}$ constitute a partition of the range $H^*$.
We initialize this collection by $\tilde \rset \gets \emptyset$ and hence, $\leftover(\tilde \rset) = H^*$.

Consider a stream-block $B \in \bset^1$ and let $\tilde \rset$ be our collection of the blocks of $H^*$ just before processing the first element of $B$.
We also fix the special range-block $\leftover(\tilde \rset)$.
For each non-empty range-block $R_i \in \tilde \rset \cup \set{\leftover(\tilde \rset)}$, we run the following procedure in parallel.

We mark each descendant stream-block $B'$ of $B$ in $\bset^2$ independently at random with probability $p = \frac{Z_2}{Z_1} \log^{11} N = \frac{\log^{11} N}{\alpha}$ each.
Consider a stream-block $B' \in \bset^2$ that is marked.
We run the algorithm $\alg_1$ from \Cref{obs: ncm-det-opt-space-sound} with input $B' \cap R_i$ and the threshold $Z = Z_2$.
If it reports yes, let $v_i(B') \in R_i$ be the element that it additionally reports.
At the end of the stream-block $B$, if at least one such execution returned yes, we let $v_i(B)$ be the smallest of the corresponding reported elements.
Otherwise, we let $v_i(B)$ remain undefined.
This completes the description of our algorithm for processing the stream-sub-block $B'$ and the range-block $R_i$.

At the end of stream-block $B$, we are ready to update our collection $\tilde \rset$.
Let $v_\leftover(B)$ be as computed by our algorithm when processing the special region $\leftover(\tilde \rset)$.
If $\tilde \rset = \emptyset$ and $v_\leftover(B)$ is left undefined, we do not update the regions and let $\tilde \rset = \emptyset$.
Assume now on that either $v_\leftover(B)$ is defined or $\tilde \rset \neq \emptyset$ (or both).

\paragraph*{Case $1$: $v_\leftover(B)$ is left undefined and $\tilde \rset \neq \emptyset$.}
Let $\tilde \rset = (R_1, \ldots, R_k)$ be the range-blocks of our collection in their natural order, where $k = |\tilde \rset|$.
Since we ensure that $|\tilde \rset| \leq Z_0/Z_1$, we have $1 \leq k \leq Z_0/Z_1$.
In this case, we will ensure that after the update, the resulting collection $\tilde \rset$ contains exactly $k$ range-blocks.
For each $1 \leq k' \leq k$, let $v_{k'}(B)$ be as computed while processing the stream-block $B$ and the range-block $R_{k'}$.
If $v_1(B)$ is undefined, we do not update the first block: $R^{(\new)}_1 \gets R_1$.
Otherwise, we set $R^{(\new)}_1$ to be the range-block containing all elements until $v_1(B')$ of $H^*$ beginning from the first one.
For each successive $2 \leq k' \leq k$, we obtain the new block $R^{(\new)}_{k'}$ as follows.
If $v_{k'}(B)$ is defined, we let the new block $R^{(\new)}_{k'}$ contain all the elements that appear after the block  $R^{(\new)}_{k'-1}$ until the element $v_{k'}(B)$ of $H^*$.
Otherwise, if $v_{k'}(B)$ is undefined, we let the new block $R^{(\new)}_{k'}$ contain all the elements that appear after the block  $R^{(\new)}_{k'-1}$ until the last element of $R_{k'}$.
Finally, we update $\tilde \rset \gets (R^{(\new)}_{1}, \ldots, R^{(\new)}_{k})$.
Let $\leftover(\tilde \rset) := H^*  \backslash \bigcup_{R_i \in \tilde \rset} R_i$ be the set of elements not present in the blocks of $\tilde \rset$.
From our construction, it is immediate to verify that $\leftover(\tilde \rset)$ is indeed a (possibly empty) range-block and the non-empty blocks of $\tilde \rset \cup \set{\leftover(\tilde \rset)}$ constitute a partition of $H^*$.

\paragraph*{Case $2$: $v_\leftover(B)$ is defined.}
In this case,  $v_\leftover(B) \in \leftover(\tilde \rset)$ and $\tilde \rset \cup \set{\leftover(\tilde \rset)}$ is a partition of $H^*$.
We append $\tilde \rset \gets \tilde \rset \cup \set{\leftover(\tilde \rset)}$ for convenience, and let ${\tilde \rset = (\rset_1, \ldots, )}$ be these non-empty range-blocks in their natural order.
We let $|\tilde \rset| = k$ and notice that ${1 \leq k \leq \frac{Z_0}{Z_1} + 1}$ holds in this case.
We will ensure that after the update, our collection $\tilde \bset$ contains exactly $k^* := \min{(k, \frac{Z_0}{Z_1})}$ range-blocks.
We proceed as in case $1$.
For each $1 \leq k' \leq k$, let $v_{k'}(B)$ be as computed while processing stream-block $B$ and the range-block $R_{k'}$.
If $v_1(B)$ is undefined, we do not update the first block: $R^{(\new)}_1 \gets R_1$.
Otherwise, we set $R^{(\new)}_1 \gets R'_1$.
For each successive $2 \leq k' \leq k$, we obtain the new block $R^{(\new)}_{k'}$ as follows.
If $v_{k'}(B) \in R_i$, we let the new block $R^{(\new)}_{k'}$ contain all the elements that appear after the block  $R^{(\new)}_{k'-1}$ until the element $v_{k'}(B)$.
Otherwise, if $v_{k'}(B)$ is undefined, we let the new block $R^{(\new)}_{k'}$ contain all the elements that appear after the block  $R^{(\new)}_{k'-1}$ until the last element of $R_{k'}$.
Finally, we update $\tilde \rset \gets (R^{(\new)}_{1}, \ldots, R^{(\new)}_{k^*})$.
Let $\leftover(\tilde \rset) := H^* \backslash \bigcup_{R_i \in \tilde \rset} R_i$ be the set of elements not present in the blocks of $\tilde \rset$.
From our construction, it is immediate to verify that $\leftover(\tilde \rset)$ is indeed a range-block and the non-empty blocks of $\tilde \rset \cup \set{\leftover(\tilde \rset)}$  constitute a partition of $H^*$.
This completes the description of case $2$, and as a result, the description of our algorithm for processing the stream-block $B$.

We define $L^* := 32pZ_0/Z_1$.
For a stream-block $B' \in \bset^2$, we denote by $\eset^*_{\bad}(B')$ the event that the block $B'$ is marked at least $L^*$ times while processing its ancestor-block $B \in \bset^1$.
We let $\eset^*_{\bad}$ be the event that there is a some block $B' \in \bset^2$ for which the event $\eset^*_\bad(B')$ occurs.
Throughout our algorithm, if at any point the event $\eset^*_\bad$ occurs, we immediately stop our algorithm and report no.
Once we process the last stream-block of $\bset^1$, we are ready to report our answer.
Let $\tilde \rset$ be our collection of blocks, just after processing the last element of $S$.
If $\tilde \rset$ contains exactly $\frac{Z_0}{Z_1}$ regions, we report yes; otherwise, we report no. 
This completes the description of our algorithm $\alglis$.
We now turn to analyze its properties, starting with the following simple observation whose proof is included in \Cref{prf-obs: ncm-alpha-approx-eset-star-bad}.

\begin{claim}\label{obs: ncm-alpha-approx-eset-star-bad}
    $\prob{\eset^*_\bad} \leq 1/N^2$.
\end{claim}

\paragraph*{Completeness.}
Assume that there is a $\vectZ$-canonical increasing subsequence $S^*$ of $S$ with length at least $Z_0$, or in other words, $|\bset^1_\yes| \geq Z_0/Z_1$.
Our goal is to show that we report yes with probability at least $3/4$.

We let $\bset^1_\yes$ be the set of yes-blocks of $\bset^1$ for $S^*$.
We discard the additional yes-blocks to ensure that $|\bset^1_\yes| = Z_0/Z_1$ holds, and we let $\bset_\yes = \set{B(1), \ldots, B(Z_0/Z_1)}$ in their natural order.
Notice that each such yes-block $B \in \bset^1_\yes$ has $\frac{Z_1}{Z_2}$ descendant yes-blocks in $\bset^2$.
For each block $B \in \bset^1_\yes$, we let $\eset^{**}_\bad(B)$ be the event that we do not mark any of its descendant yes-blocks.
From Chernoff bound, it is immediate to verify that $\prob{\eset^{**}_\bad(B)} \leq 1/N^{2}$.
We let $\eset^{**}_\bad$ be the event that $\eset^{**}_\bad(B)$ occurs for some yes-block $B \in \bset^1_\yes$.
From union bound over at most $N$ yes-blocks, we obtain that $\prob{\eset^{**}_\bad} \leq 1/N^{2}$.
Combining this with \Cref{obs: ncm-alpha-approx-eset-star-bad}, we conclude that $\prob{\eset^{*}_\bad \cup \eset^{**}_\bad} \leq 2/N^2 < 1/N$.

For each integer $1 \leq s \leq Z_0/Z_1$ and the corresponding yes-block $B(s)$ of $\bset^1$, let $\tilde \rset^{(s)}$ be our collection of range-blocks just after processing $B(s)$.
Using techniques similar to \Cref{{clm: ncm-jump-dp-completeness}}, we obtain the following claim showing the completeness guarantee.

\begin{claim}\label{clm: ncm-lis-alpha-jump-dp-completeness}
    If the events $\eset^*_\bad$ and $\eset^{**}_\bad$ does not occur, then for \textbf{each} $1 \leq s \leq \frac{Z_0}{Z_1}$, $|\tilde \rset^{(s)}| \geq s$.
\end{claim}

We defer the proof of \Cref{clm: ncm-lis-alpha-jump-dp-completeness} to \Cref{prf-clm: ncm-lis-alpha-jump-dp-completeness}.

\paragraph*{Soundness.}
We let $\bset^1 = \set{B(1), \ldots, B(|\bset^1|)}$ be the stream-blocks in their natural order.
For each integer $1 \leq s \leq |\bset^1|$ and the corresponding stream-block $B(s) \in \bset^1$, let $\tilde \rset^{(s)}$ be our collection of range-blocks of $H^*$ just after processing $B(s)$.
Using techniques similar to \Cref{clm: ncm-jump-dp-completeness}, it is immediate to verify the following claim.

\begin{claim}\label{clm: ncm-lis-alpha-jump-dp-soundness}
    If the events $\eset^*_\bad$ and $\eset^{**}_\bad$ does not occur, then for each $1 \leq s \leq |\bset^1|$ and $1 \leq s' \leq |\tilde \rset^{(s)}|$, there is an increasing subsequence of length at least $s' Z_2$ using the elements in $B(1) \cup \ldots \cup B(s)$ with values in the range-blocks $\tilde \rset^{(s)}$.
\end{claim}

We include the proof of \Cref{clm: ncm-lis-alpha-jump-dp-soundness} in \Cref{prf-clm: ncm-lis-alpha-jump-dp-soundness} for the sake of completeness.
Recall that we report yes only if there are exactly $Z_0/Z_1$ blocks in our collection at the end of $S$, or in other words, $|\tilde \rset^{(s)}| = Z_0/Z_1$, where $s = |\bset^1|$.
But then from \Cref{clm: ncm-lis-alpha-jump-dp-soundness}, there must be an increasing subsequence of length at least $\frac{Z_0}{Z_1} \cdot Z_2 \geq \frac{Z_0}{\alpha}$.
We can now conclude that the probability that $\optlis(S) < Z_0/\alpha$ and we report yes, is at most $\prob{\eset^*_\bad \cup \eset^{**}_\bad} \leq 2/N^2 < 1/N$ as required.

\paragraph{Space Complexity.}
Consider some stream-block $B \in \bset^1$ and let $\tilde \rset$ be our collection of range-blocks just before processing $B$.
The space used by $\alglis$ can be divided into three parts:
(i) the space required to store the range-blocks $\tilde \rset$;
(ii) the space used by executions of $\alg_1$ from \Cref{obs: ncm-det-opt-space-sound} and
(iii) the space required to update the range-blocks $\tilde \rset$.

Recall that we can store a block $R \in \tilde \rset$ by storing the indices of its first and last elements in $H^*$.
Since there are at most $Z_0/Z_1$ regions in $\tilde \rset$, the space used to store them is at most $O(Z_0/Z_1)$.
Similarly, we can store the special block $R_\leftover = H^* \backslash \bigcup_{R \in \tilde \rset} R$ in $O(1)$ units of space.
Thus, the space used in part (i) is bounded by $O(Z_0/Z_1)$.
Recall that we halt our algorithm immediately if the event $\eset^*_\bad$ occurs.
Thus, throughout the execution of our algorithm, each descendant stream-block $B'$ of $B$ is marked at most $L^*$ times and the space required by $\alg_1$ of \Cref{obs: ncm-det-opt-space-sound} can be bounded by $O(L^* \cdot Z_2)$.
It is also immediate to verify that we can update our collection $\tilde \rset$ using $O(Z_0/Z_1)$ additional space.
Hence, the overall space complexity of $\alglis$ is bounded by,

\begin{align*}
    O\left(\frac{Z_0}{Z_1} \right) + L^* \cdot O(Z_2) &= O\left(\frac{Z_0}{Z_1}  + p \cdot \frac{Z_0}{Z_1} \cdot Z_2\right)\\
    &\leq \tilde O \left( \frac{Z_0}{Z_1} + \frac{Z_0}{\alpha^2} \right)\\
    &\leq \tilde O \left( \alpha + \frac{\sqrt{N}}{\alpha} \right)  = \tilde O\left( \frac{\sqrt{N}}{\alpha} \right).
\end{align*}

Here, the first equality follows from our choice of $L^* = 32pZ_0/Z_1$.
The inequalities follow from \Cref{{eqn: ncm-alpha-approx-z0-by-z1},{eqn: ncm-alpha-approx-tau-star}}.
This completes the analysis of the properties of $\alglis$ and \Cref{thm: sqrt n by alpha randomized lis algo} now follows.

\appendix
    \addtocontents{toc}{\protect\setcounter{tocdepth}{1 }}
\setcounter{secnumdepth}{3}
\begin{appendices}
    
            \chapter{Proofs Omitted from Chapter \ref{chap: ndp}}  \label{appn-chap: ndp}
            \iftoggle{ndp}{
                \iftoggle{ndp-algo}{
                    \section{Proofs Omitted from \Cref{sec: ndp-algo}} \label{appn-sec: proofs-of-ndp-algo}

\subsubsection{Proof of \Cref{clm: ndp-algo-hierarchical system of squares}}

Before we define a hierarchical partition of $\tG$ into squares, we need to define a hierarchical system of intervals.
\begin{definition} Given an integer $1\leq \rho'\leq \rho$, a \emph{$\rho'$-hierarchical system of intervals} is a sequence $\hset=(\iset_1,\iset_2,\ldots,\iset_{\rho'})$ of sets of intervals, such that:

\begin{itemize}
    \item for all $1\leq h\leq \rho'$, $\iset_h$ is a $d_h$-canonical family of intervals; and
    \item for all $1<h\leq \rho'$, for every interval $I\in \iset_h$, there is an interval $I'\in \iset_{h-1}$, such that $I\subseteq I'$.
\end{itemize}

We let $U(\hset)=\bigcup_{I\in \iset_{\rho'}}I$, and we say that the integers in $U(\hset)$ belong to the system $\hset$.
\end{definition}

We use the following simple observation.

\begin{observation}\label{obs: ndp-algo-canonical intervals}
 There is an efficient algorithm that constructs a collection  $\hset_1,\ldots,\hset_{2^{\rho}}$ of $2^{\rho}$ $\rho$-hierarchical systems  of intervals, such that every integer in $[\ell']$ belongs to exactly one such system.
\end{observation}

\begin{proof}
 It is enough to prove that there is an efficient algorithm, that, given an integer $1\leq \rho'\leq \rho$, constructs a collection  $\hset_1,\ldots,\hset_{2^{\rho'}}$ of $2^{\rho'}$ $\rho'$-hierarchical systems  of intervals, such that every integer in $[\ell']$ belongs to exactly one such system.
The proof is by induction on $\rho'$. The base case is when $\rho'=1$. We partition $[\ell']$ into consecutive intervals, where every interval contains exactly $d_1$ integers (recall that $\ell'$ is an integral multiple of $d_1=\eta^{\rho+2}$). Let $(I_1,I_2,\ldots,I_r)$ be the resulting sequence of intervals, where we assume that the intervals appear in the sequence in their natural order. Let $\iset_1$ be the set of all odd-indexed intervals and $\iset_1'$ the set of all even-indexed intervals in the sequence. Clearly, each of $\iset_1$ and $\iset_1'$ is a $d_1$-canonical set of intervals. We define two $1$-hierarchical systems, the first one containing only the set $\iset_1$, and the second one containing only the set $\iset_1'$. Note that every integer in $[\ell']$ belongs to exactly one resulting system.

We assume now that the statement holds for all integers between $1$ and $(\rho'-1)$, for some $\rho'> 1$, and we prove it for $\rho'$. We assume that we are given a collection of $2^{\rho'-1}$ $(\rho'-1)$-hierarchical systems of intervals, such that every integer in $[\ell']$ belongs to exactly one system. Let $\hset$ be one such $(\rho'-1)$-hierarchical system of intervals. We will construct two $\rho'$-systems, $\hset'$ and $\hset''$, such that every integer that belongs to $\hset$ will belong to exactly one of the two systems. This is enough in order to complete the proof of the observation.

Assume that $\hset=(\iset_1,\iset_2,\ldots,\iset_{\rho'-1})$. For simplicity, denote $\iset_{\rho'-1}$ by $\iset$. We now construct two new sets $\iset',\iset''$ of intervals, as follows. Start with $\iset'=\iset''=\emptyset$, and process every interval $I\in \iset$ one-by-one. Consider some interval $I\in \iset$. We partition $I$ into consecutive intervals containing exactly $d_{\rho'}$ integers each. Let $\set{I_1,\ldots,I_r}$ be the resulting partition, where we assume that the intervals are indexed in their natural order. We add to $\iset'$ all resulting odd-indexed intervals, and to $\iset''$ all resulting even-indexed intervals. Once every interval $I\in \iset$ is processed in this manner, we obtain our final sets $\iset',\iset''$ of intervals. It is immediate to verify that each set $\iset',\iset''$ is $d_{\rho'}$-canonical; that $\bigcup_{I'\in \iset'\cup \iset''}I'=\bigcup_{I\in \iset}I$; and that every integer of $\bigcup_{I\in \iset}I$ belongs to exactly one interval of $\iset'\cup \iset''$. We then set $\hset'=(\iset_1,\iset_2,\ldots,\iset_{\rho'-1},\iset')$, and $\hset''=(\iset_1,\iset_2,\ldots,\iset_{\rho'-1},\iset'')$.
\end{proof}

From Observation~\ref{obs: ndp-algo-canonical intervals}, we can construct $2^{\rho}$ hierarchical $\rho$-systems $\hset_1,\hset_2,\ldots,\hset_{2^{\rho}}$ of intervals of $[\ell']$. For every pair $1\leq i,j\leq 2^{\rho}$ of integers, we construct a single hierarchical family $\thset_{i,j}$ of squares, such that $V(\thset_{i,j})=\set{v_{x,y}\mid x\in U(\hset_i),y\in U(\hset_j)}$. Since every integer in $[\ell']$ belongs to exactly one set $U(\hset_z)$ for $1\leq z\leq 2^{\rho}$, it is immediate to verify that every vertex of $G'$ belongs to exactly one resulting hierarchical family $\thset_{i,j}$ of squares.

We now define the construction of the system $\thset_{i,j}=(\qset^{i,j}_1,\qset^{i,j}_2,\ldots,\qset^{i,j}_{\rho})$. 
Denote $\hset_i=(I_1,\ldots,I_{\rho})$ and $\hset_j=(I'_1,\ldots,I'_{\rho})$
The construction is simple: for all $1\leq r\leq \rho$, we let $\qset^{i,j}_r=\qset(\iset_r,\iset'_r)$. From the above discussion, since sets $\iset_r,\iset'_r$ are $d_r$-canonical, so is set $\qset^{i,j}_r$. It is also easy to verify that the set of vertices contained in the squares of $\qset^{i,j}_r$ is exactly $\set{v(x,y)\mid x\in \bigcup_{I\in \iset_r}I, y\in \bigcup_{I'\in \iset_r'}I'}$, and that $\thset_{i,j}$ is indeed a hierarchical system of squares of $G$.
This completes the proof of \Cref{clm: ndp-algo-hierarchical system of squares}.

\subsubsection{Proof of \Cref{obs: ndp-algo-boosting shadow}}
Assume that $\hmset=\set{(s_1,t_1),\ldots,(s_z,t_z)}$, where the source vertices $s_1,\ldots,s_z$ appear in this left-to-right order on $R^*$. We then let $\hmset'=\set{(s_i,t_i)\mid i\equiv 1\mod 2\ceil{\beta_2/\beta_1}}$. Clearly, $|\hmset'|\geq  \floor{\frac{|\hmset|}{2\ceil{\beta_2/\beta_1}}}\geq \floor{\frac{\beta_1|\hmset|}{4\beta_2}}$.

We claim that every square $Q\in \qset$ has the $\beta_2$-shadow property with respect to $\hmset'$. Indeed, let $Q\in \qset$ be any such square, and assume that its dimensions are $(d\times d)$. Then $J_{\hmset'}(Q)\subseteq J_{\hmset}(Q)$, and $J_{\hmset}(Q)$ contained at most $\beta_2 d$ source vertices of the demand pairs of $\hmset$. From our construction of $\hmset'$, it contains at most $\ceil{\frac{\beta_2 d}{2\ceil{\beta_2/\beta_1}}}\leq \beta_1 d$ demand pairs of $\hmset'$ and \Cref{obs: ndp-algo-boosting shadow} follows.

\subsubsection{Proof of \Cref{clm: ndp-algo-partition the forest}}
We compute a partition $\yset(\tau)$ for every tree $\tau\in F$ separately. The partition is computed in iterations, where in the $j$th iteration we compute the set $Y_j(\tau)\subseteq V(\tau)$ of vertices, together with the corresponding collection $\pset_j(\tau)$ of paths. For the first iteration, if $\tau$ contains a single vertex $v$, then we add this vertex to $Y_1(\tau)$ and terminate the algorithm. Otherwise, for every leaf $v$ of $\tau$, let $P(v)$ be the longest directed path of $\tau$, starting at $v$, that only contains degree-1 and degree-2 vertices, and does not contain the root of $\tau$. We then add the vertices of $P(v)$ to $Y_1(\tau)$, and the path $P(v)$ to $\pset_1(\tau)$. Once we process all leaf vertices of $\tau$, the first iteration terminates. It is easy to see that all resulting vertices in $Y_1(\tau)$ induce a collection $\pset_1(\tau)$ of disjoint paths in $\tau$, and moreover if $v,v'\in Y_1(\tau)$, and there is a path from $v$ to $v'$ in $\tau$, then $v,v'$ lie on the same path in $\pset_1(\tau)$. We then delete all vertices of $Y_1(\tau)$ from $\tau$.

The subsequent iterations are executed similarly, except that the tree $\tau$ becomes smaller, since we delete all vertices that have been added to the sets $Y_j(\tau)$ from the tree.

It is now enough to show that this process terminates after $\ceil{\log n}$ iterations. In order to do so, we can describe each iteration slightly differently. Before each iteration starts, we gradually contract every edge $e$ of the current tree, such that at least one endpoint of $e$ has degree $2$ in the tree, and $e$ is not incident on the root of $\tau$. We then obtain a tree in which every inner vertex (except possibly the root) has degree at least $3$, and delete all leaves from this tree. The number of vertices remaining in the contracted tree after each such iteration therefore decreases by at least factor $2$. It is easy to see that the number of iteration in this procedure is the same as the number of iterations in our algorithm, and is bounded by $\ceil{\log n}$.
For each $1\leq j\leq \ceil{\log n}$, we then set $Y_j=\bigcup_{\tau\in F}Y_j(\tau)$.
This completes the proof of \Cref{clm: ndp-algo-partition the forest}.

\subsubsection{Proof of \Cref{clm: ndp-algo- modified instance preserves solutions}}

Let $\pset^*$ be the optimal solution to instance $(G,\mset)$. If $|\pset^*|>d$, then we discard paths from $\pset^*$ arbitrarily, until $|\pset^*|=d$ holds. 
Recall that $G'$ is a sub-grid of $G$ spanned by some subset $\wset'$ of its columns and some subset $\rset'$ of its rows. Recall also that we have defined a sub-grid $G''\subseteq G'$, obtained from $G'$ by deleting $4d$ of its leftmost columns, $4d$ of its rightmost columns, and $|\wset'|-4d$ of its topmost rows.

Let $\mset^*\subseteq \mset'$ be the set of the demand pairs routed by $\pset^*$. For every path $P\in \pset^*$, we define a collection $\Sigma(P)$ of sub-paths of $P$, as follows. Assume that $P$ routes some demand pair $(s,t)$. Let $x_1,x_2,\ldots,x_r$ be all vertices of $\Gamma(G'')$ that appear on $P$, and assume that they appear on $P$ in this order (where we view $P$ as directed from $s$ to $t$). Denote $x_0=s$ and $x_{r+1}=t$. We then let $\Sigma(P)$ contain, for each $0\leq i\leq r$, the sub-path of $P$ from $x_i$ to $x_{i+1}$. We say that a segment $\sigma\in \Sigma(P)$ is of type $1$ if one of its endpoints is the source $s$; we say that it is of type $2$ if both its endpoints belong to $\Gamma(G'')$, and $\sigma$ is internally disjoint from $G''$; otherwise we say that it is of type $3$. We let $\Sigma_1$ contain all type-1 segments in all sets $\Sigma(P)$ for $P\in \pset^*$, and we define $\Sigma_2$ and $\Sigma_3$ similarly for all type-2 and type-3 segments.

Notice that all type-3 segments are contained in $G''$. Let $\tilde{\mset_2}$ be the set of all pairs $(u,v)$ of vertices, such that some segment $\sigma\in \Sigma_2$ connects $u$ to $v$. We also define a set $\tilde{\mset_1}$ of pairs of vertices, corresponding to the segments of $\Sigma_1$ as follows. Let $\sigma\in \Sigma_1$ be any type-$1$ segment, and assume that its endpoints are $s$ and $v$, with $s\in S(\mset')$ and $v\in \Gamma(G'')$. Let $s'\in \Gamma(G')$ be the new source vertex to which $s$ was mapped. Then we add $(s',v)$ to $\tilde{\mset_1}$. In order to complete the proof of the claim, it is now enough to prove the following observation.

\begin{observation}
There is a set $\pset$ of node-disjoint paths that routes all pairs in $\tmset_1\cup \tmset_2$ in graph $G'$, so that the paths in $\pset$ are internally disjoint from $G''$.
\end{observation}

Indeed, combining the paths in $\pset$ with the segments in $\Sigma_3$ provides a set of node-disjoint paths in graph $G'$ that routes $|\mset^*|$ demand pairs of $\mset''$ -- the demand pairs corresponding to the pairs in $\mset^*$. 

The proof of the above observation is straightforward; we only provide its sketch here. Let $J$ be the sub-path of $\Gamma(G'')$, obtained by deleting all vertices lying on the bottom boundary of $G''$ from it, excluding the two bottom corners of $G''$. For every pair $(u,v)\in \tmset_2$, we can think of the corresponding sub-path of $J$ between $u$ and $v$ as an interval $I(u,v)$. It is immediate to verify that all intervals defined by the pairs in $\tmset_2$ are nested: that is, if $(u,v),(u',v')\in \tmset_2$,  then either intervals $I(u,v)$ and $I(u',v')$ are disjoint, or one of them is contained in the other. 

Let $\iset=\set{I(u,v)\mid (u,v)\in \tmset_2}$ be the corresponding set of intervals.
Notice that no interval in $\iset$ may contain a destination vertex $v$ of any demand pair $(s,v)\in \tmset_1$.
Let $\iset^0\subseteq\iset$ be the set of all intervals containing both the top left and the top right corners of $G''$. 
Let $\iset^1\subseteq \iset\setminus \iset^0$ be the set of all intervals containing the top left corner of $G''$, and similarly, let $\iset^2\subseteq \iset\setminus \iset^0$ be the set of all intervals containing the top right corner of $G''$. 
Observe that for all $0\leq i\leq 2$, for every pair $I,I'\in \iset^i$ of intervals, one of the intervals is contained in the other.
Let $\iset'=\iset\setminus\left(\bigcup_{i=0}^2\iset^i\right )$ be the set of all remaining intervals.

We partition the intervals of $\iset'$ into levels, as follows. We say that $I(u,v)$ is a level-$1$ interval, iff no other interval of $\iset'$ is contained in it. Let $\iset_1$ denote the set of all level-$1$ intervals. Assume now that we have defined the sets $\iset_1,\ldots,\iset_{i}$ of intervals of levels $1,\ldots, i$, respectively. We say that an interval $I\in \iset'$ belongs to level $(i+1)$ iff it does not contain any interval from set $\iset'\setminus\left(\bigcup_{i'=1}^i\iset_{i'}\right)$.

We say that an interval $I(u,v)\in \iset'$ is a left interval, iff $u$ and $v$ belong to the left boundary edge of $G''$. Similarly, we say that $I(u,v)\in\iset'$ is a right interval, iff $u$ and $v$ belong to the right boundary edge of $G''$. Otherwise, we say that it is a top interval. In that case, both $u$ and $v$ must belong to the top boundary of $G''$. It is easy to verify that, if $I$ is a left or a right interval, then it must belong to levels $1,\ldots,2d$, as the height of the grid $G''$ is $4d$. Moreover, if $h$ is the largest level to which a left interval belongs, then $h+|\iset_0|+|\iset_1|\leq 4d$, and if $h'$ is the largest level to which a right interval belongs, then $h+|\iset_0|+|\iset_2|\leq 4d$.

Let $W'_1,W'_2,\ldots,W'_{4d}$ denote the $4d$ columns of $G'$ that lie to the left of $G''$, and assume that they are indexed in the right-to-left order. For each level $1\leq r\leq 4d$, for every level-$r$ interval $I(u,v)$, we route the demand pair $(u,v)$ via the column $W'_r$ in a straightforward manner: we connect $u$ and $v$ to $W'_r$ by horizontal paths, and then complete the routing using the corresponding sub-path of $W'_r$. 

The routing of the right intervals is performed similarly. The top intervals are routed similarly by exploiting the rows of $G'$ that lie above $G''$ -- recall that there are $|\wset''|+4d$ such rows, where $|\wset''|$ is the width of $G''$. We then add paths routing the pairs corresponding to the intervals in $\iset^1,\iset^2$ and $\iset^0$ in a straightforward manner (see \Cref{fig: modified routing}). 

\begin{figure}[h]
\scalebox{0.5}{\includegraphics{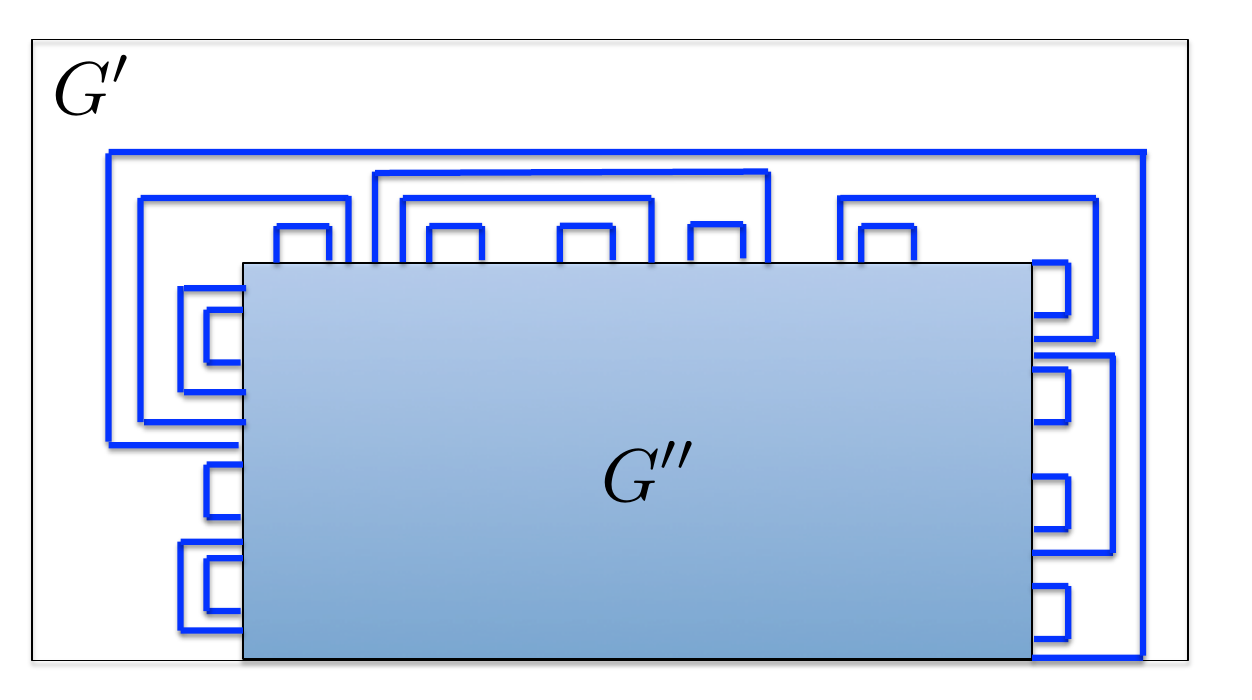}}
\caption{Routing in the modified graph.\label{fig: modified routing}}
\end{figure}

Let $H$ be the graph obtained from $G'$, after we delete all non-boundary vertices of $G''$ from it, and all vertices that participate in the routing of the pairs in $\tmset_2$ that we just defined. It is easy to verify that all demand pairs in $\tmset_1$ can be routed in $H$. In order to do so, we set up a flow network, where we start from the graph $H$, and add two special vertices $s$ and $t$ to it. We connect $s$ to every vertex in $S(\tmset_1)$, and we connect $t$ to every vertex in $T(\tmset_1)$, setting the capacity of every vertex of $H$ to be $1$. It is easy to verify that there is an $s$--$t$ flow of value $|\tmset_1|$ in this network (since every cut separating $s$ from $t$ must contain at least $|\tmset_1|$ vertices). From the integrality of flow, and due to the way in which the mapping between the vertices of $S(\mset')$ and $S(\mset'')$ was defined, we can obtain an integral routing of all demand pairs in $\tmset_1$ via node-disjoint paths in $H$.
This completes the proof of \Cref{clm: ndp-algo- modified instance preserves solutions}.

                }{}
                \iftoggle{ndp-hard}{
                    \section{Proof Omitted from Section \ref{subsec: ndp-hard-prelims}}\label{appn-sec: ndp-hard-prelims}
\subsubsection{Proof of \Cref{thm: ndp-hard-yi-partition-of-answers}}\label{appdx: ndp-hard-proof of yi-partition-of-answers thm}

Suppose $G$ is a \yi, and let $\chi$ be a valid coloring of $V(G)$. Let $\pi_1,\ldots,\pi_6$ be $6$ different permutations of $\set{r,g,b}$. 
For each $1\leq i\leq 6$, permutation $\pi_i$ defines a valid coloring $\chi_i$ of $G$: for every vertex $v\in V(G)$, if $v$ is assigned a color $c\in \set\cset$ by $\chi$, then 
$\chi_i$ assigns the color $\pi_i(c)$ to $v$. Notice that for each vertex $v$ and for each color $c\in \cset$, there are exactly two indices $i\in \set{1,\ldots, 6}$, such that $\chi_i$ assigns the color $c$ to $v$. Notice also that for each edge $(u,v)$, if $c,c'\in \cset$ is any pair of distinct colors, then there is exactly one index $i\in \set{1,\ldots,6}$, such that $u$ is assigned the color $c$ and $v$ is assigned the color $c'$ by $\chi_i$.

Let $B$ be the set of all vectors of length $\ell$, whose entries belong to $\set{1,\ldots,6}$, so that $|B|=6^{\ell}$. For each such vector $b\in B$, we define a perfect global assignment $f_b$ of answers to the queries, as follows. Let $Q\in \qset^E$ be a query to the edge-player, and assume that $Q=(e_1,\ldots,e_{\ell})$. Fix some index $1\leq j\leq \ell$, and assume that $e_j=(v_j,u_j)$. Assume that $b_j=z$, for some $1\leq z\leq 6$. We assign to $v_j$ the color $\chi_z(v_j)$, and we assign to $u_j$ the color $\chi_z(u_j)$. Since $\chi_z$ is a valid coloring of $V(G)$, the two colors are distinct. This defines an answer $A\in \aset^E$ to the query $Q$, that determines $f_b(Q)$.

Consider now some query $Q'\in \qset^V$ to the vertex-player, and assume that $Q'=(v_1,\ldots,v_{\ell})$. Fix some index $1\leq j\leq \ell$, and assume that $b_j=z$, for some $1\leq z\leq 6$. We assign to $v_j$ the color $\chi_z(v_j)$. This defines an answer $A'\in \aset^V$ to the query $Q'$, that determines $f_b(Q')$. Notice that for each $1\leq j\leq \ell$, the answers that we choose for the $j$th coordinate of each query are consistent with the valid coloring $\chi_{b_j}$ of $G$. Therefore, it is immediate to verify that for each $b\in B$, $f_b$ is a perfect global assignment.

We now fix some query $Q\in \qset^E$ of the edge-prover, and some answer $A\in \aset^E$ to it. Assume that $Q=(e_1,\ldots,e_{\ell})$, where for $1\leq j\leq \ell$, $e_j=(v_j,u_j)$. Let $c_j,c'_j$ are the assignments to $v_j$ and $u_j$ given by the $j$th coordinate of $A$, so that $c_j\neq c'_j$. Recall that there is exactly one index $z_j\in \set{1,\ldots,6}$, such that $\chi_{z_j}$ assigns the color $c_j$ to $v_j$ and the color $c'_j$ to $u_j$. Let $b^*\in B$ be the vector, where for $1\leq j\leq \ell$, $b^*_j=z_j$. Then $f_{b^*}(Q)=A$, and for all $b\neq b^*$, $f_b(Q)\neq A$.

Finally, fix some query $Q'\in \qset^V$ of the vertex-prover, and some answer $A'\in \aset^V$ to it. Let $Q'=(v_1,\ldots,v_{\ell})$. Assume that for each $1\leq j\leq \ell$, the $j$th coordinate of $A'$ contains the color $c_j$. Recall that there are exactly two indices $z\in \set{1,\ldots,6}$, such that $\chi_z$ assigns the color $c_j$ to $v_j$. Denote this set of two indices by $Z_j\subseteq\set{1,\ldots,6}$. Consider now some vector $b\in B$. If, for all $1\leq j\leq \ell$, $b_j\in Z_j$, then $f_b(Q')=A'$; otherwise, $f_b(Q')\neq A'$. Therefore, the total number of vectors $b\in B$, for which $f_b(Q')=A'$ is exactly $2^{\ell}$.

\section{Proofs Omitted from Section \ref{subsec: ndp-hard-from WGP to NDP}}\label{appdx: ndp-hard-from WGP to NDP}
\subsection{Proofs of Auxillary Lemmas \ref{lem: ndp-hard-random ordering} and \ref{lem: ndp-hard-random ordering2}} \label{appdx-subsec: ndp-hard-auxiliarly lemmas}
The goal of this subsection is to prove \Cref{lem: ndp-hard-random ordering,lem: ndp-hard-random ordering2}.
We will use the following simple observation.

\begin{observation} \label{obs: ndp-hard-decreasing ratios}
    For any two positive integers $a$ and $b$, $\frac{b-1}{a+b-1} < \frac{b}{a+b}$.
\end{observation}

Recall that we are given a set $U$ of $n$ items, such that  $P$ of the items are pink, and the remaining $Y=n-P$ items are yellow. We consider a random permutation $\pi$ of these items. Given a set $S\subseteq\set{1,\ldots,n}$ of $\delta$ indices, we let $\event(S)$ be the event that for all $i\in S$, the item of $\pi$ located at the $i$th position is yellow.

\begin{claim}\label{clm: ndp-hard-event delta}
    $\prob{\event(S)}\leq \left(\frac{Y}{n}\right )^{|S|}.$
\end{claim}
\begin{proof}
    As every subset of $\delta$ items of $U$ is equally likely to appear at the indices of $S$, we get that: 

    \[\prob{\event(S)} = \frac{\binom{Y}{\delta}}{\binom{P+Y}{\delta}}=
    \frac{Y \cdot (Y-1)\cdots (Y-\delta+1)}{(P+Y) \cdot (P+Y-1)\cdots (P+Y-\delta+1)}\leq  \left (\frac{Y}{P+Y}\right )^\delta=\left(\frac{Y}{n}\right )^{|S|}.\]

    (the last inequality follows from \Cref{obs: ndp-hard-decreasing ratios}).
\end{proof}

We now turn to prove \Cref{lem: ndp-hard-random ordering}. 

\begin{lemma}
    [Restatement of Lemma \Cref{lem: ndp-hard-random ordering}.] For any $\log n\leq \mu\leq Y$, the probability that there is a sequence of $\ceil{4n\mu/P}$ consecutive items in $\pi$ that are all yellow, is at most $n/e^{\mu}$.
\end{lemma}
\begin{proof}
    Let $\delta=\ceil{\frac{4n\mu}{P}}$. Consider a set $S$ of $\delta$ consecutive indices of $\set{1,\ldots,n}$. 
    From \Cref{clm: ndp-hard-event delta}, the probability that all items located at the indices of $S$ are yellow is at most:

    \[\left (\frac{Y}{n}\right )^\delta =  \left (1-\frac{P}{n}\right )^{\ceil{4n\mu/P}}\leq \left (1-\frac{P}{n}\right )^{4n\mu/P}\leq e^{-\mu}.\]
    
    Since there are at most $n$ possible choices of a set $S$ of $\delta$ consecutive indices, from the Union Bound, the probability that any such set only contains yellow items is bounded by $n/e^{\mu}$.
\end{proof}

\begin{lemma}
    [Restatement of \Cref{lem: ndp-hard-random ordering2}.] For any $\log n\leq \mu\leq P$, the probability that there is a set $S$ of $\floor{\frac{n\mu}{P}}$ consecutive items in $\pi$, such that more than $4\mu$ of the items are pink, is at most $n/4^{\mu}$.
\end{lemma}

\begin{proof}
    Let $x=\floor{\frac{n\mu}{P}}$, and let $S$ be any set of $x$ consecutive indices of $\set{1,\ldots,n}$. Denote $\delta=\ceil{4\mu}$, and let $S'\subseteq S$ be any subset of $\delta$ indices from $S$. From \Cref{clm: ndp-hard-event delta} (by reversing the roles of the pink and the yellow items), the probability that all items located at the indices of $S'$ are pink is at most $\left (\frac{P}{n}\right )^\delta$. 
    Since there are  ${x\choose \delta}$ ways to choose the subset $S'$ of $S$, from the Union Bound,  the probability that at least $\delta$ items of $S$ are pink is at most:

    \[\begin{split}
        \left (\frac{P}{n}\right )^\delta\cdot {x\choose \delta}&\leq \left (\frac{P}{n}\right )^\delta\cdot \left(\frac{ex}{\delta}\right )^\delta\\
        &\leq  \left (\frac{P}{n}\right )^\delta\cdot \left(\frac{e\cdot n\mu}{P \delta}\right )^\delta\\
        &= \left (\frac{e\cdot \mu}{ \ceil{4\mu}}\right )^{\ceil{4\mu}}\\
        &\leq \left(\frac{e}{4}\right)^{4 \mu}<\frac{1}{4^{\mu}}.
    \end{split}\]

    Since there are at most $n$ sets $S$ of $x$ consecutive items, taking the union bound over all such sets completes the proof.
\end{proof}

\subsection{\proofof{Observation \ref{obs: ndp-hard-h is large}}} \label{prf-obs: ndp-hard-h is large}
From our assumption that $|\mset^0|>c\log^3M$, $E^2\neq \emptyset$. Therefore, there must be an index $1\leq i\leq r$ with $E_i\cap E^2\neq \emptyset$. Fix any such index $i$. Then there is a vertex $v'_j\in W_i\cap V_2$, such that at least one edge of $\delta(v'_j)$ belongs to $E^2$. But then, from the definition of $E^2$, at least $2^{q-1}$ edges of $\delta(v'_j)$ belong to $E^2$. Assume without loss of generality that these edges connect $v'_j$ to vertices $v_1,\ldots,v_{2^{q-1}}\in V_1$. All these vertices must also belong to $W_i$, and  for each $1\leq x\leq 2^{q-1}$, vertex $v_x$ has at least one edge in $\delta(v_x)\cap E^2$. From our definition of $E^2$ and $E^1$, at least $2^{p-1}$ edges of $\delta(v_x)$ belonged to $E^1$. Therefore, $|E^1\cap E_i|\geq 2^{q-1}\cdot 2^{p-1}$. But $|E_i|\leq h$, and so $h\geq 2^p\cdot 2^q/4$.
\endproofof

\subsection{\proofof{Observation \ref{obs: ndp-hard-no heavy path exist}}} \label{prf-obs: ndp-hard-no heavy path exist}
    Since $|V_1|\leq M$, from the union bound, it is enough to prove that for a fixed vertex $v_i\in V_1$, the probability that a heavy sub-path $Q\subseteq \block_i$ exists is at most $1/(100M)$. We now fix some vertex $v_i\in V_1$. Observe that $\block_i$ may only contain a heavy sub-path if $\beta(v_i)=|X_i|\geq 16\log M$. We call the vertices of 
    $X'_i$ pink, and the remaining vertices of $X_i$ yellow. Let $P$ denote the number of the pink vertices. Then $2^{p-1}\leq P<2^p$. Let $\mu=4\log M$, and let $\event_i$ be the bad event that there is a set of $\floor{|X_i|\mu/P}$ consecutive vertices of $X_i$, such that at least $4\mu$ of them are pink. 

    Observe that the selection of the pink vertices only depends on the solution to the \WGPwB problem, and is independent of our construction of the \NDPgrid instance. 
    The ordering of the vertices in $X_i$ is determined by the permutation $\rho'$ of $\uset_2$, and is completely random.
    Therefore, from \Cref{lem: ndp-hard-random ordering2}, the probability of $\event_i$ is at most $|X_i|/4^{\mu}\leq M/4^{4\log M}\leq 1/M^7$. 

    Let $\event'_i$ be the event that some sub-path $Q$ of $\block_i$ is heavy. We claim that $\event'_i$ may only happen if event $\event_i$ happens. Indeed, consider some sub-path $Q$ of $\block_i$, and assume that it is heavy. Recall that $Q$ contains $\floor{\frac{512 h\log^2M}{2^p}}$ vertices. Since every pair of vertices in $X_i$ is separated by at least $512\ceil{\frac{h\log M}{\beta(v_i)}}$ vertices, we get that:

    \[ |V(Q)\cap X_i| \leq  \frac{\floor{512 h\log^2M/2^p}}{512\ceil {h\log M/\beta(v_i)}}+1
    \leq \frac{\beta(v_i)\log M}{2^p}+1
    \leq \floor{\frac{|X_i|\mu} P},
    \]

    as $2^{p-1}\leq P<2^p$. 
    Since $Q$ is heavy, at least $4\mu=16\log M$ of the vertices of $V(Q)\cap X_i$ belong to $X'_i$, that is, they are pink. Therefore, there is a set of  $\floor{|X_i|\mu/P}$ consecutive vertices of $X_i$, out of which $4\mu$ are pink, and $\event_i$ happens. We conclude that $\prob{\event'_i}\leq \prob{\event_i}\leq 1/M^7$, and overall, since we have assumed that $M>2^{50}$, the probability that a heavy path exists in any block $\block_i$ is bounded by $0.99$ as required.
\endproofof

\subsection{\proofof{Observation \ref{obs: ndp-hard-nset is small}}} \label{prf-obs: ndp-hard-nset is small}
    Since we have started with a perfect solution to $\iset$, for each group $U \in \uset'$, there is exactly one vertex of $U$ in $W_i$.
    Due to Step 1 of regularization, each such vertex contributed at least $2^{q-1}$ edges to $E^1\cap E_i$, while $|E^1\cap E_i|\leq h$.
    Therefore, $n^*\leq h/2^{q-1}$.
\endproofof

\subsection{Routing via Spaced-out Paths - Proof of Claim \ref{clm: ndp-hard-the paths}}\label{appdx: ndp-hard-routing in yi}
The goal of this subsection is to prove \Cref{clm: ndp-hard-the paths}.
We show an efficient algorithm to construct a set $\pset^1=\set{P_1,\ldots,P_{M'}}$ of spaced-out paths, that originate at the vertices of $X$ on $R$, and traverse the boxes $\hat \block_j$ in a snake-like fashion (see \Cref{fig: ndp-hard-routing}).
We will ensure that for each $1 \leq i \leq M'$ and $1\leq j\leq N_1$, the intersection of the path $P_i$ with the box $\hat\block_j$ is the $i$th column of $\wset_j$, and that $P_i$ contains the vertex $x_i$.

Fix some index $1\leq j\leq N_1$, and consider the box $\hat \block_j$. Let $I'_j$ and $I''_j$ denote the top and the bottom boundaries of $\hat \block_j$, respectively. 
For each $1 \leq i \leq M'$, let $W_j^i$ denote the $i$th column of $\wset_j$, and let $x'(j,i)$ and $x''(j,i)$ denote the topmost and the bottommost vertices of $W_j^i$, respectively.

For convenience, we also denote  $I''_0 = R$, and, for each $1\leq i\leq M'$, $x''(0,i) = x_i$.
The following claim is central to our proof.

\begin{claim}\label{clm: ndp-hard-sub-paths}
    There is an efficient algorithm to construct, for each $1\leq j\leq N_1$, a set $\pset_j=\set{P_j^1,\ldots,P_j^{M'}}$ of paths, such that:

    \begin{itemize}
        \item For each $1\leq j\leq N_1$ and $1\leq i\leq M'$, path $P_j^i$ connects $x''(j-1,i)$ to $x'(j,i)$; 
        \item The set $\bigcup_{j=1}^{N_1}\pset_j$ of paths is spaced-out; and
        \item All paths in $\bigcup_{j=1}^{N_1}\pset_j$ are contained in $G^t$, and are internally disjoint from $R$.
    \end{itemize}
\end{claim}

Notice that by combining the paths in $\bigcup_{j=1}^{N_1}\pset_j$ with the set $\bigcup_{j=1}^{N_1}\wset_j$ of columns of the boxes, we obtain the desired set $\pset^1$ of paths, completing the proof of \Cref{clm: ndp-hard-the paths}. In the remainder of this section we focus on proving \Cref{clm: ndp-hard-sub-paths}.

Recall that each box $\hat \block_j$ is separated by at least $2M$ columns from every other such box, and from the left and right boundaries of $\hat G$. It is also separated by at least $4M$ rows from the top boundary of $\hat G$ and from the row $R$.
We exploit this spacing in order to construct the paths, by utilizing a special structure called a \emph{snake}, that we define next.

Given a set $\lset$ of consecutive rows of $\hat G$ and a set $\wset$ of consecutive columns of $\hat G$, we denote by $\Y(\lset, \wset)$ the subgraph of $\hat G$ spanned by the rows in $\lset$ and the columns in $\wset$; we refer to such a graph as a \emph{corridor}.
Let $\Y=\Y(\lset,\wset)$ be any such corridor.
Let $L'$ and $L''$ be the top and the bottom row of $\lset$ respectively, and let $W'$ and $W''$ be the first and the last column of $\wset$ respectively.
The four paths $\Y\cap L',\Y\cap L'',\Y\cap W'$ and $\Y\cap W''$ are called the top, bottom, left and right boundaries of $\Y$ respectively, and their union is called the \emph{boundary} of $\Y$. The width of the corridor $\Y$ is $w(\Y)=\min\set{|\lset|,|\wset|}$.
We say that two corridors $\Y,\Y'$ are \emph{internally disjoint}, iff every vertex $v\in \Y\cap \Y'$ belongs to the boundaries of both corridors.
We say that two internally disjoint corridors $\Y,\Y'$ are \emph{neighbors} iff $\Y\cap \Y'\neq \emptyset$.
We are now ready to define snakes.

A snake $\yset$ of length $z$ is a sequence $(\Y_1,\Y_2,\ldots,\Y_{z})$ of $z$ corridors that are pairwise internally disjoint, such that for all $1\leq z',z'' \leq z$, $\Y_{z'}$ is a neighbor of $\Y_{z''}$ iff $|z'-z''|=1$. The width of the snake is defined to be the minimum of two quantities: (i) $\min_{1\leq z'<z}\set{|\Y_{z'}\cap \Y_{z'+1}|}$; and (ii) $\min_{1\leq z'\leq z}\set{w(\Y_{z'})}$.
Notice that, given a snake $\yset$, there is a unique simple cycle $\sigma(\yset)$ contained in $\bigcup_{z'=1}^z \Y_{z'}$, such that, if $D$ denotes the disc on the plane whose boundary is $\sigma(\yset)$, then every vertex of $\bigcup_{z'=1}^z \Y_{z'}$ lies in $D$, while every other vertex of $\hat G$ lies outside $D$.
We call $\sigma(\yset)$ the \emph{boundary} of $\yset$.
We say that a vertex $u$ belongs to a snake $\yset$, and denote $u \in \yset$, iff $u \in \bigcup_{z'=1}^z \Y_{z'}$.
 We use the following simple claim, whose proof can be found, e.g. in~\cite{NDP-hard-old}.

\begin{claim}\label{clm: ndp-hard-routing in a snake}
Let $\yset=(\Y_1,\ldots,\Y_{z})$ be a snake of width $w$, and let $A$ and $A'$ be two sets of vertices with $|A|=|A'|\leq w-2$, such that the vertices of $A$ lie on a single boundary edge of $\Y_1$, and the vertices of $A'$ lie on a single boundary edge of $\Y_{z}$. There is an efficient algorithm, that, given the snake $\yset$, and the sets $A$ and $A'$ of vertices as above, computes a set $\qset$ of node-disjoint paths contained in $\yset$, that  connect every vertex of $A$ to a distinct vertex of $A'$. 
\end{claim}

 Let $L, L'$ be any pair of rows of $G$.
 Let $\qset$ be a set of node-disjoint paths connecting some set of vertices $B \subseteq L$ to $B' \subseteq L'$.
 We say that the paths in $\qset$ are \emph{order-preserving} iff the left-to-right ordering of their endpoints on $L$ is same as that of their endpoints on $L'$.

 \begin{corollary} \label{cor: ndp-hard-spaced out snakes}
Let $\yset=(\Y_1,\ldots,\Y_{z})$ be a snake of width $w$, and let $B$ and $B'$ be two sets of $r \leq \floor{w/2}-1$ vertices each, such that the vertices of $B$ lie on the bottom boundary edge of $\Y_1$, the vertices of $B'$ lie on the top boundary edge of $\Y_{z}$ and for every pair $v,v'\in B\cup B'$ of vertices, $d_{\hat G}(v,v')\geq 2$.
There is an efficient algorithm, that, given the snake $\yset$, and the sets $B$ and $B'$ of vertices as above, computes a set $\hat \qset$ of spaced-out order-preserving paths contained in $\yset$.
 \end{corollary}

\begin{proof}
Let $B=\set{b_1, b_2, \ldots, b_r}$ and $B'=\set{b'_1, b'_2, \ldots, b'_r}$. Assume that the vertices in both sets are indexed according to their left-to-right ordering on their corresponding rows of the grid.
Since set $B$ does not contain a pair of neighboring vertices, we can augment it to a larger set $A$, by adding a vertex between every consecutive pair of vertices of $B$. In other words, we obtain $A=\set{a_1, \ldots, a_{2r-1}}$, such that for  all $1 \leq i \leq r$, $a_{2i-1} = b_i$, and the vertices of $A$ are indexed according to their left-to-right ordering on the bottom boundary of $\Y_1$. Similarly, we can augment the set $B'$ to a set $A' = \set{a'_1, \ldots, a'_{2r-1}}$ of vertices, such that for all $1 \leq i \leq r$, $a'_{2i-1} = b'_i$, and the vertices of $A'$ are indexed according to their left-to-right ordering on the top boundary of $\Y_z$.

We apply \Cref{clm: ndp-hard-routing in a snake} to the sets $A,A'$ of vertices, obtaining a set $\qset$ of node-disjoint paths, that are contained in $\yset$, and connect every vertex of $A$ to a distinct vertex of $A'$.  For all $1\leq i\leq r$, let $Q_i$ be the path originating from $a_i$.  We claim that $\qset$ is an order-preserving set of paths.
Indeed, assume for contradiction that some path $Q_i \in \qset$ connects $a_i$ to $a'_{i'}$, for $i \neq i'$. Notice that the path $Q_i$ partitions the snake $\yset$ into two sub-graphs: one containing $(i-1)$ vertices of $A$ and $(i'-1)$ vertices of $A'$; and the other containing the remaining vertices of $A$ and $A'$ (excluding the endpoints of $Q_i$). Since $i \neq i'$, there must be a path $Q_{i''} \in \qset$ intersecting the path $Q_i$, a contradiction to the fact that $\qset$ is a  set of node-disjoint paths.

Similarly, it is easy to see that for all $1\leq i<r$, $d(Q_{2i-1},Q_{2i+1})\geq 2$. This is since the removal of the path $Q_{2i}$ partitions the snake $\yset$ into two disjoint sub-graphs, with path $Q_{2i-1}$ contained in one and path $Q_{2i+1}$ contained in the other.

Our final set of path is $\hat \qset = \set{Q_{2i-1} : 1 \leq i \leq r}$. From the above discussion, it is a spaced-out set of paths contained in $\yset$, and for each $1\leq i\leq r$, path $Q_{2i-1} \in \hat \qset$ connects $b_i$ to $b'_{i}$. 
\end{proof}

In order to complete the proof, we need the following easy observation.

\begin{observation}\label{obs: ndp-hard-snakes}
    There is an efficient algorithm that constructs, for each $1\leq j\leq N_1$, a snake $\yset_j$ of width at least $2M$ in $G^t$, such that all resulting snakes are mutually disjoint, and for each $1\leq j\leq N_1$:

    \begin{itemize} 
        \item the bottom boundary of the first corridor of $\yset_j$ contains $I''_{j-1}$; 
        \item the top boundary of the last corridor of $\yset_j$ contains $I'_j$; and
        \item all snakes are disjoint from $R$, except for $\yset_1$, that contains $I_0''\subseteq R$ as part of is boundary, and does not contain any other vertices of $R$.
    \end{itemize}
\end{observation}

The construction of the snakes is immediate and exploits the ample space between the boxes $\hat \block_j$;  (see \Cref{fig: ndp-hard-routing} for an illustration).
From \Cref{cor: ndp-hard-spaced out snakes}, for each $1 \leq j \leq N_1$, we obtain a set $\pset_j$ of spaced-out paths contained in $\yset_j$, such that for each $1 \leq i \leq M'$, there is a path $P^i_j \in \pset_j$ connecting $x''(j-1,i)$ to $x'(j,i)$.
For each $1 \leq i \leq M'$, let $P_i$ be the path obtained by concatenating the paths $\set{P^i_1, \wset^i_1, P^i_2, \ldots, P^i_{N_1}, \wset^i_{N_1}}$.
The final set of paths is $\pset^1 = \set{P_1, \ldots, P_{M'}}$.

\subsection{\proofof{Observation \ref{obs: ndp-hard-number of vertices in hat H}}} \label{prf-obs: ndp-hard-number of vertices in hat H}
    Clearly, $|V(\hat \bfH)|=\sum_{v\in V(\bfH)}d^2_v$. Observe that for every pair $a\geq b\geq 0$ of integers, $(a+1)^2+(b-1)^2\geq a^2+b^2$. Since maximum vertex degree in $\bfH$ is bounded by $d$, the sum is maximized when all but possibly one summand are equal to $d$, and, since $\sum_{v\in \bfH}d_v=2m$, there are at most $\ceil{2m/d}$ summands. Therefore, $|V(\hat \bfH)|\leq \floor{2m/d}\cdot d^2+d^2\leq (2m+d)d$.
\endproofof

\subsection{Observation \proofof{\ref{obs: ndp-hard-cuts in grids}}} \label{prf-obs: ndp-hard-cuts in grids}
It is easy to verify that for any bi-partition $(X',Y')$ of $U$ into two disjoint subsets, there is a set $\pset$ of $\min\set{|X'|,|Y'|}$ node-disjoint paths in $Q$ connecting vertices of $X'$ to vertices of $Y'$. The observation follows from the maximum flow -- minimum cut theorem.
\endproofof

\section{Proof Omitted from Section \ref{subsec: ndp-hard-from NDP to EDP}} \label{appdx: ndp-hard-from NDP to EDP}
\proofof{\Cref{clm: ndp-hard-NDPwall to EDP wall}}
The assertion that $\opt'\geq \opt$ is immediate, as any set $\pset$ of node-disjoint paths in the wall $G$ is also a set of edge-disjoint paths. 

Assume now that we are given a set $\pset'$ of edge-disjoint paths in $G$. We show an efficient algorithm to compute a subset $\pset\subseteq \pset'$ of $\Omega(|\pset'|)$ paths that are node-disjoint. Since the maximum vertex degree in $G$ is 3, the only way for two paths $P,P'\in\pset'$ to share a vertex $x$ is when $x$ is an endpoint of at least one of these two paths. If $x$ is an endpoint of $P$, and $x\in V(P')$, then we say that $P$ has a conflict with $P'$.

We construct a directed graph $H$, whose vertex set is $\set{v_P\mid P\in\pset'}$, and there is an edge $(v_P,v_{P'})$ iff $P$ has a conflict with $P'$. It is immediate to verify that the maximum out-degree of any vertex in $H$ is at most $4$, as each of the two endpoints of a path $P$ may be shared by at most two additional paths. Therefore, every sub-graph $H'\subseteq H$ of $H$ contains a vertex of total degree at most $8$. We construct a set $U$ of vertices, such that no two vertices of $U$ are connected by an edge, using a standard greedy algorithm: while $H\neq \emptyset$, select a vertex $v\in H$ with total degree at most $8$ and add it to $U$; remove $v$ and all its neighbors from $H$. It is easy to verify that at the end of the algorithm, $|U|=\Omega(|V(H)|)=\Omega(|\pset'|)$, and no pair of vertices in $U$ is connected by an edge. Let $\pset=\set{P\mid v_P\in U}$. Then the paths in $\pset$ are node-disjoint, and $|\pset|=\Omega(|\pset'|)$.
\endproofof
                }{}
            }{}

            \chapter{Proofs Omitted from Chapter \ref{chap: exp}}  \label{appn-chap: exp}
            \iftoggle{exp}{
                \section{Proofs Omitted from Sections \ref{sec: exp-overview} and \ref{sec: exp-prelims}} \label{appn-sec: exp-proofs of exp-intro}
\subsection{\proofof{Corollary \ref{cor: exp-random} }}\label{appn-subsec: exp-cor proof}

In this subsection we prove \Cref{cor: exp-random}.
    We use the following result of Krivelevich~\cite{exp_random}:
    \begin{theorem}[Corollary 1 of \cite{exp_random}]
        For every $\epsilon > 0$, there exists $\gamma > 0$, such that for every $n>0$, a random graph $G \sim \gset(n, \frac{1+\epsilon}{n})$ contains an induced bounded-degree $\gamma$-expander $\tilde G$ on at least $\gamma n$ vertices w.h.p.
    \end{theorem}

    Let $G \sim \gset(n,\frac{1+\epsilon}{n})$.
    From the above theorem, w.h.p., there is an induced bounded-degree $\gamma$-expander $\tilde G\subseteq G$ on at least $\gamma n$ vertices, for some $\gamma$ depending only on $\epsilon$.
    From \Cref{thm: exp-general main}, every graph $H$ of size at most $c_\epsilon n/\log n$ is a minor of $\tilde G$, where $c_\epsilon$ is some constant  depending on $\epsilon$ only.
    \Cref{cor: exp-random} now follows.
\endproofof

\newcommand{\expBound}{{20}}

\subsection{\proofof{Observation \ref{obs: exp-lower bound}}} \label{appn-subsec: exp-lower bound}

Recall that we are given a integer $s$ and a graph $G = (V,E)$ of size $s$.
Assume for now that $2 \leq s < 2^\expBound$.
Let $H_G$ be a graph with $s+1$ vertices and $0$ edges.
Notice that the number of vertices in $H_G$ is strictly more than that in $G$, and hence $H_G$ is not a minor of $G$.
The observation now follows since $\expBound s / \log s \geq s + 1$.
Thus from now on, we assume that $s \geq 2^\expBound$ and hence, $\expBound s/ \log s \geq 2^{\expBound}$.

We denote by $\mu(G) = |\set{H \> | \> H \text{ is a minor of } G}|$.
For an integer $r$, let $\fset_r$ be the set of all graphs of size at most $r$.
The following two observations now complete the proof of \Cref{obs: exp-lower bound}.

\begin{observation}
	$\mu(G) \leq 3^s$.
	\vspace{-1em}
\end{observation}
\begin{proof}
	From the definition of minors, every minor $H$ of $G$ can be identified by a subset $E^{\text{del}}_H \subseteq E$ of deleted edges, a subset $E^{\text{cont}}_H \subseteq E$ of contracted edges and a subset $V^{\text{del}}_H \subseteq V$ of deleted vertices.
	Thus,
	\[\mu(G) \leq 2^{|V|} \cdot 3^{|E|} \leq 3^{|V| + |E|} \leq 3^s.\]
\end{proof}

\begin{observation}
	For every even integer $r \geq 2^{10}$, $|\fset_r| \geq r^{r/10}$.
	\vspace{-1em}
\end{observation}
\begin{proof}
	Let $k = \floor{r^{0.9}}$.
	We lower-bound the number of graphs containing exactly $k$ vertices and exactly $r/2$ edges.
	Notice that, since $r \geq 2^{10}$, $k + r/2 \leq r$.
	For convenience, assume that the set $V^* = \set{1, \ldots, k}$ of vertices and their indices are fixed.	We will first lower-bound the number of vertex-labeled graphs with the set $V^*$ of vertices, that contain exactly $r/2$ edges.
	Since there are only $\binom{k}{2}$ `edge-slots', this number is at least:
	\[ \binom{\binom{k}{2}}{r/2} \geq \binom{r^{1.6}}{r/2} \geq \left( \frac{r^{1.6} - r/2}{r/2}\right)^{r/2} \geq \left(r^{0.6} \right)^{r/2} \geq r^{0.3 r}.\]

	Here, the inequalities hold for all $r \geq 2^{10}$.
	Notice that two graphs $G_1 = (V^*, E_1)$ and $G_2 = (V^*, E_2)$ with labeled vertices are isomorphic to each other iff there is a permutation $\psi$ of the vertices, mapping $E_1$ to $E_2$.
	Thus, the number of non-isomorphic graphs on $k$ vertices and $r/2$ edges is at least:
	\[ \frac{r^{0.3r}}{k!} \geq \frac{r^{0.3r}}{ \left(r^{0.9} \right)!} > \frac{r^{0.3r}}{r^{0.9r^{0.9}}} \geq r^{r^{0.9}(0.3r^{0.1} - 0.9)} \geq r^{r/10}.\]
\end{proof}

We are now ready to complete the proof of \Cref{obs: exp-lower bound}.
Assume for contradiction that $G$ contains every graph in the family $\fset^* = \fset_{( \expBound s/\log s)}$ as a minor.
Recall that $\expBound s/\log s \geq 2^\expBound$.
However, from the above two observations, $|\fset^*|\geq (\expBound s/\log s)^{\expBound s/(10 \log s)}$, while $\mu(G)\leq 3^s$. It is immediate to verify that $|\fset^*|>\mu(G)$, a contradiction.
\endproofof

\subsection{\proofof{Observation \ref{obs: exp-simple partition}} }\label{appn-subsec: exp-simple partition proof}
	We assume without loss of generality that $x_1\geq x_2\geq\cdots\geq x_r$, and process
	the integers in this order. When $x_i$ is processed, we add $i$ to
	$A$ if $\sum_{j\in A}x_j\leq \sum_{j\in B}x_j$, and we add it to $B$
	otherwise. We claim that at the end of this process, $\sum_{i\in
		A}x_i,\sum_{i\in B}x_i\geq N/4$ must hold. Indeed, 
	$1$ is always added to $A$. If $x_1\geq N/4$, then, since $x_1\leq 3N/4$, it is easy to see
	that both subsets of integers sum up to at least $N/4$.  Otherwise,
	$|\sum_{i\in A}x_i-\sum_{i\in B}x_i|\leq \max_i\set{x_i}\leq x_1\leq
	N/4$, and so $\sum_{i\in
		A}x_i,\sum_{i\in B}x_i\geq N/4$.
\endproofof

\subsection{\proofof{Claim \ref{clm: exp-large expanding subgraph-edges}}} \label{appn-subsec: exp-large expanding subgraph}

Our algorithm iteratively removes edges from $T\setminus E'$, until we obtain a connected component of the resulting graph that is an $\alpha/4$-expander.
We start with $T'=T\setminus E'$ (notice that $T'$ is not necessarily connected).  We also maintain a set $E''$ of edges that we remove from $T'$, initialized to $E''=\emptyset$. While $T'$ is not an $\alpha/4$-expander, let $(X,Y)$ be a cut of sparsity less than $\alpha/4$ in $T'$, that is $|E_{T'}(X,Y)| < \alpha \min{(|X|,|Y|)}/4$. Assume w.l.o.g. that $|X|\geq |Y|$. Update $T'$ to be $T'[X]$, add the edges of $E(X,Y)$ to $E''$, and continue to the next iteration. 

Assume that the algorithm performs $r$ iterations, and for each $1\leq i\leq r$, let $(X_i,Y_i)$ be the cut computed by the algorithm in iteration $i$. Since $|X_i|\geq |Y_i|$, $|Y_i|\leq |V(T')|/2$. At the same time, if we denote $E_i=E''\cap E(X_i,Y_i)$, then $|E_i|< \alpha |Y_i|/4$. Therefore:

\[|E''|=\sum_{i=1}^r|E_i|\leq \alpha\sum_{i=1}^r|Y_i|/4.\]

On the other hand, since $T$ is an expander, the total number of edges leaving each set $Y_i$ in $T$ is at least $\alpha|Y_i|$, and all such edges lie in $E'\cup E''$. Therefore:

\[|E'|+|E''|\geq \alpha\sum_{i=1}^r|Y_i|/2.\]

Combining both bounds, we get that $|E'|\geq \alpha\sum_{i=1}^r|Y_i|/4$.
We get that $\sum_{i=1}^r|Y_i|\leq \frac{4|E'|}{\alpha}$, and therefore $|V(T')|\geq |V(T)|-\frac{4|E'|}{\alpha}$.
\endproofof

\section{Proofs Omitted from Section \ref{sec: exp-embedding in poe}} \label{appn-sec: exp-proofs of exp-new main}
\subsection{\proofof{Observation \ref{obs: exp-decompose by spanning tree}}} \label{appn-subsec: exp-proof of decompose by spanning tree}
Let $\tau$ be any spanning tree of $\hG$, rooted at an arbitrary degree-$1$ vertex of $\tau$. We start with $\uset=\emptyset$. Our algorithm performs a number of iterations, where in each iteration we add one new set $U\subseteq V(\hG)$ of vertices to $\uset$, such that $\hG[U]$ is connected and $\floor{|R|/(d r)}\leq |U\cap R|\leq |R|/r$, and we remove the vertices of $U$ from $\tau$. We execute the iterations as long as $|V(\tau)\cap R|\geq \floor{|R|/(d r)}$, after which we terminate the algorithm, and return the current collection $\uset$ of vertex subsets. 

In order to execute an iteration, we let $v$ be the lowest vertex of $\tau$, such that the subtree $\tau_v$ of $\tau$ rooted at $v$ contains at least $\floor{|R|/(d r)}$ vertices of $R$.
Since the maximum vertex degree in $\hG$ is bounded by $d$, tree $\tau_v$ contains fewer than $d \cdot \floor{|R|/(d r)} \leq |R|/r$ vertices of $R$. We add a new set $U=V(\tau_v)$ of vertices to $\uset$, delete the vertices of $U$ from $\tau$, and continue to the next iteration.

Let $\uset$ be the final collection of vertex subsets obtained at the end of the algorithm. It is immediate to verify that for every set $U\in \uset$, $\hG[U]$ is connected and, from the above discussion, $\floor{|R|/(d r)}\leq  |U\cap R|\leq |R|/r$. Therefore, $|\uset|\geq r$.	
\endproofof

\subsection{\proofof{Claim \ref{clm: exp-two does}}} \label{appn-subsec: clm: exp-two does}
    From the definition of the \poefull System, for $3\leq j\leq 6$, the set $A_j\cup B_j$ of vertices is well-linked in $S_j$. Therefore, there is a set $\pset_j$ of $w$ node-disjoint paths in $S_j$, connecting $A_j$ to $B_j$. By concatenating the path sets $\pset_3,\pset_4,\pset_5,\pset_6$, and the edge sets $\mset_3,\mset_4,\mset_5$, we obtain a collection $\pset$ of $w$ node-disjoint paths in $G''_{\Pi}$, connecting $A_3$ to $B_6$. We partition $\pset$ into two subsets: set $\pset^{(1)}$ contains all paths originating at the vertices of $W_1\cup Y_1$, and set $\pset^{(2)}$ contains all paths originating at the vertices of $W_2\cup Y_2$. 

    We are now ready to define the two graphs $G^{(1)}$ and $G^{(2)}$. Graph $G^{(1)}$ is obtained from the union of the expanders $T_3$ and $T_4$, the paths of $\pset^{(1)}$, and the edges of $\mset'_3\cup \mset'_4$ that have an endpoint lying on the paths of $\pset^{(1)}$. Graph $G^{(2)}$ is defined similarly by using $T_5,T_6$, the paths of $\pset^{(2)}$, and the edges of $\mset'_5\cup\mset'_6$ that have an endpoint lying on the paths of $\pset^{(2)}$. It is immediate to verify that the graphs $G^{(1)}$ and $G^{(2)}$ are disjoint.

    It now remains to show that each of the resulting graphs contains a \doefull System as a minor, with the required properties. We show this for $G^{(1)}$; the proof for $G^{(2)}$ is symmetric. Our first step is to contract every path of $\pset^{(1)}$ into a single vertex. For each such path $P\in \pset^{(1)}$, let $w\in W_1\cup Y_1$ be the first vertex of $P$. We denote the new vertex obtained by contracting $P$ by $v(w)$.
    We let the backbone $X^{(1)}$ of the new \doefull System $\dset^{(1)}$ be $X^{(1)}=\set{v(w)\mid w\in W_1}$, so $|X^{(1)}|=w/4$. We map every vertex $w\in W_1$ to the corresponding vertex $v(w)$ in the model of $G_{\dset^{(1)}}$ that we are constructing in $G^{(1)}$; that is, we set $f^{(1)}(w)=v(w)$.
    We also map the two expanders $T^{(1)}_1,T^{(1)}_2$ of $\dset^{(1)}$ to $T_3$ and $T_4$, respectively, by setting $T^{(1)}_1=T_3$ and $T^{(1)}_2=T_4$.

    Consider some vertex $w\in W_1\cup Y_1$ and the path $P\in \pset^{(1)}$ originating from $w$. Let $w'$ be the unique vertex of $P$ that belongs to $B_3$, and let $w''$ be the unique vertex of $P$ that belongs to $B_4$, in the original \poefull System $\Pi$. Recall that there is an edge of $\mset'_3$, connecting $w'$ to some vertex $u_w\in C_3$, and there is an edge of $\mset'_4$, connecting $w''$ to some vertex $u'_{w}\in C_4$. Therefore, there are edges $(v(w),u_w)$ and $(v(w),u'(w))$ in the new contracted graph.

    We set $D_0^{(1)}=\set{u_{w}\mid w\in  W_1}$, and we let $\tmset^{(1)}=\set{(v(w),u_w)\mid w\in W_1}$. 
    We also set $D_1^{(1)}=\set{u_{y}\mid y\in \hat Y_1}$, and $D_1^{(2)}=\set{u'_{y}\mid y\in \hat Y_1}$. Observe that all three sets $D^{(1)}_0,D^{(1)}_1,D^{(1)}_2$ of vertices are disjoint, and they contain $w/4$ vertices each. It now remains to define the set $(\tmset')^{(1)}$ of edges, that connect vertices of $D_1^{(1)}$ and $D_2^{(1)}$. In order to do so, for every vertex $y\in Y_1$, we merge the two edges $(v(y),u_{y})$ and $(v(y),u'_{y})$ into a single edge, by contracting one of these two edges. The resulting edge is added to $(\tmset')^{(1)}$. It is easy to see that we have obtained a \doefull System $\dset^{(1)}$, whose width is $w/4$ and expansion $\alpha$. It is easy to verify that the maximum vertex degree in the corresponding graph $G_{\dset^{(1)}}$ is bounded by $d$.  Notice that for every vertex $w\in W_1$, there is a distinct vertex $v(w)\in X^{(1)}$, such that $w\in f^{(1)}(v(w))$.
    Thus, $G^{(1)}$ indeed contain a \doefull System $\dset^{(1)}$ with claimed properties as minor and we have computed its model $f^{(1)}$ in $G^{(1)}$.
    The proof for $G^{(2)}$ is analogous and \Cref{clm: exp-two does} now follows.
\endproofof

\subsection{\proofof{Claim \ref{clm: exp-short paths in expanders}}}\label{appn-subsec: exp-proof of short paths in expanders}
Consider the following sequence of vertex subsets. Let $S_0=Z$, and for all $i>0$, let $S_i$ contain all vertices of $S_{i-1}$, and all neighbors of vertices in $S_{i-1}$. Notice that, if $|S_{i-1}|\leq |V(T)|/2$, then, since $T$ is an $\alpha'$-expander, there are at least $\alpha' |S_{i-1}|$ edges leaving the set $S_{i-1}$, and, since the maximum vertex degree in $T$ is at most $d$, there are at least $\frac{\alpha'|S_{i-1}|}{d}$ vertices that do not belong to $S_{i-1}$, but are neighbors of vertices in $S_{i-1}$. Therefore, $|S_i|\geq |S_{i-1}|\left (1+\frac{\alpha'}{d}\right )$.
We claim that there must be an index $i^*\leq \frac{8d}{\alpha'}\log (n/z)$, such that $|S_{i^*}|> |V(T)|/2$. Indeed, otherwise, we get that for $i=\ceil{\frac{8d}{\alpha'}\log (n/z)}$:

\[|S_{i^*}|\geq |S_0|\left (1+\frac{\alpha'}{d}\right )^{i}\geq z\cdot e^{i\alpha'/(2d)}\geq z\cdot e^{4\log (n/z)}>n/2.\]

Here, the second inequality follows from the fact that $(1+1/x)^{2x}>e$ for all $x>1$.
We construct a similar sequence $S'_0,S'_1,\ldots,$ for $Z'$. Similarly, there is an index $i^{**}\leq  \frac{8d}{\alpha'}\log (n/z')$, such that $S'_{i^{**}}$ contains more than half the vertices of $T$. Therefore, there is a path connecting a vertex of $Z$ to a vertex of $Z'$, whose length is at most $\frac{8d}{\alpha'}(\log (n/z)+\log(n/z'))$. 
\endproofof

\subsection{\proofof{Claim \ref{clm: exp-happy index}}} \label{prf-clm: exp-happy index}
    We say that a vertex $v$ of $D_0\cap V(T_1')$ is \emph{happy} iff there is a path in $T_1'$, of length at most $(\gamma\log\log n)/4$,
    connecting $v$ to a vertex of $D_1'$. Assume for contradiction that the claim is false. Then for each good index $j$, either all vertices of $Y_{j}'$ are unhappy, or all vertices of $Y_{j+r}'$ are unhappy. Let $Z\subseteq D_0\cap V(T_1')$ be the set of all unhappy vertices. Since $|Y_{j}'|,|Y_{j+1}'|\geq \sigma/2$, and $|J'|\geq  \frac{r\log\log n}{2\log n}$, we get that:

    \begin{align*}
        |Z|&\geq \frac{r\log\log n}{2\log n}\cdot \frac{\sigma}{2}\\
        &\geq \frac{w\alpha^2(\log\log n)^3}{2d^3\log^4n}\cdot 2^{14}\cdot\floor{\frac{d^3n\log n}{w\alpha^2}}\\
        &\geq \frac{2^{12} n (\log\log n)^3}{\log^3n}.        
    \end{align*}

    Let $Z'=D_1'$, so $|Z'|\geq w/16$. From \Cref{clm: exp-short paths in expanders}, there is a path in $T_1'$, connecting a vertex of $Z$ to a vertex of $Z'$, of length at most: 
    
    \begin{align*}
        \frac{32d}{\alpha}\left( \log{\left(\frac{n}{|Z|} \right)}+\log{\left(\frac{n}{|Z'|}\right)} \right) &\leq \frac{32d}{\alpha} \left (\log\left (\frac{\log^3n}{2^{13}(\log\log n)^3}   \right )+\log{\left(\frac{16n}{w}\right)}\right )\\
        &\leq \frac{32d}{\alpha} \left( 3\log\log n+\log{\left(\frac{16n}{w} \right)} \right)\\
        &\leq \frac{\gamma \log\log n}{4},
    \end{align*}
    
    since $\gamma=512 n d^2/(w\alpha)$.
    This completes the proof of \Cref{clm: exp-happy index}.
\endproofof

\subsection{\proofof{Lemma \ref{lem: exp-can route large disjoint subsets in expanders}}
}\label{appn-subsec: exp-routing along short paths}

Recall that we are given a graph $G = (V,E)$, with  $|V|\leq n$ and maximum vertex degree at most $d$, and a parameter $0<\alpha<1$.
We are also given a collection $\set{C_1, \ldots, C_{2 r}}$ of disjoint subsets of $V$, each containing $q = \ceil{cd^2 \log^2n/\alpha^2}$ vertices, for some constant $c$ to be fixed later. 
Our goal is to either find a set $\qset = \set{Q_1, \ldots, Q_{r}}$ of disjoint paths, such that for each $1\leq j\leq r$, path $Q_j$ connects $C_j$ to $C_{j+r}$; or compute a cut $(S, S')$ in $G$ of sparsity less than $\alpha$.

We use a standard definition of multicommodity flow. 
A \emph{flow} $f$ consists of a collection $\pset$ of paths  in $G$, called \emph{flow-paths}, and, for each path $P\in \pset$, an associated flow value $f(P)>0$. The \emph{edge-congestion} of $f$ is the maximum amount of flow passing through any edge, that is, $\max_{e\in E}\set{\sum_{\stackrel{P\in \pset:}{e\in P}}f(P)}$. We say that the flow in $f$ causes \emph{no edge-congestion} iff the edge-congestion due to $f$ is at most $1$. Similarly, the \emph{vertex congestion} of $f$ is the maximum flow passing through any vertex, that is,  $\max_{v\in V}\set{\sum_{\stackrel{P\in \pset:}{v\in P}}f(P)}$. 
If a path $P$ does not lie in $\pset$, then we implicitly set $f(P)=0$. For any pair $s,t\in V$ of vertices, let $\pset(s,t)$ be the set of all paths connecting $s$ to $t$ in $G$. We say that $f$ \emph{transfers $z$ flow units between $s$ and $t$} iff $\sum_{P\in\pset(s,t)} f(P) \geq z$.

The following theorem is a consequence of Theorem 18 from~\cite{LeightonRao}
that we prove after completing the proof of \Cref{lem: exp-can route large disjoint subsets in expanders}.

\begin{theorem}\label{thm: exp-integral routing on expanders}
    There is an efficient randomized algorithm, that,
    given a graph $G=(V,E)$ with $|V|=n$ and maximum vertex degree at most $d$,
    and a parameter $0<\alpha<1$,
    together with a (possibly partial) matching $\mset$ over the vertices of $G$,
    computes one of the following:
    \begin{itemize}
        \item either a collection $\qset'=\set{Q(u,v)\mid (u,v)\in \mset}$ of paths, such that for all $(u,v)\in \mset$, path $Q(u,v)$ connects $u$ to $v$;
        the paths in $\qset'$ with high probability cause vertex-congestion at most $\eta= O(d \log n/\alpha)$,
        and the length of every path in $\qset$ is at most $L= O(d \log n/\alpha)$; or
        \item a cut $(S, S')$ in $G$ of sparsity less than $\alpha$.
    \end{itemize}
\end{theorem}

We are now ready to complete the proof of \Cref{lem: exp-can route large disjoint subsets in expanders}.
We construct a matching $\mset$ over the vertices of $V$, as follows.
For each $1\leq j\leq r$, we add an arbitrary matching $\mset_j$, containing $q$ edges, between the vertices of $C_{j}$ and the vertices $C_{j + r}$. We then set $\mset=\bigcup_{j=1}^r\mset_j$.
We apply the algorithm from \Cref{thm: exp-integral routing on expanders} to the graph $G$, parameter $\alpha$ and the matching $\mset$.
If the algorithm returns a cut of sparsity less than $\alpha$, we terminate the algorithm and return the cut. Therefore, we assume from now on that the algorithm returns a set $\qset'$ of paths with the following properties:
\begin{itemize}
    \item For each $j \in [r]$, there is a subset $\qset'_j\subseteq \qset'$ of $q$ paths connecting vertices of $C_j$ to vertices of $C_{j+r}$;
    \item All paths in $\qset'$ have length at most $L=O(d\log n/\alpha)$; and
    \item With high probability, every vertex of $G$ participates in at most $\eta=O(d \log n/\alpha)$  paths of $\qset'$.
\end{itemize}

If the vertex-congestion caused by the paths in $\qset'$ is greater than $\eta$, the algorithm terminates with a failure.
Therefore, we assume from now on that the paths in $\qset'$ cause vertex-congestion  at most $\eta$.
We use the constructive version of the Lovász Local Lemma by Moser and Tardos~\cite{Moser-Tardos} in order to select one path from each set $\qset_j'$, so that the resulting paths are node-disjoint  with high probability.
The next theorem summarizes the symmetric version of the result of~\cite{Moser-Tardos}.

\begin{theorem}[\cite{Moser-Tardos}]\label{thm: exp-constructive symmetric LLL}
Let  $X$ be a finite set of mutually independent random variables in some probability space. Let $\aset$  be a finite set of bad events determined by these variables. 
 For each event $A\in \aset$, let $\vbl(A)\sse X$ be the unique minimal subset of variables determining $A$, and let $\Gamma(A)\sse \aset$ be a subset of bad events $B$, such that $A\neq B$, but $\vbl(A)\cap \vbl(B)\neq \emptyset$. Assume further that for each $A\in \aset$, $|\Gamma(A)|\leq D$, $\prob{A}\leq p$, and $ep(D+1)\leq 1$. Then there is an efficient randomized algorithm that computes an assignment to the variables of $X$, such that with high probability none of the events in $\aset$ holds.
 \end{theorem}

For each $1\leq i\leq r$, we choose one of its paths $Q_i\in \qset_i$ independently at random. We let $z_i$ be the random variable indicating which path has been chosen. 
For every pair $Q,Q'\in \qset'$ of intersecting paths, such that $Q,Q'$ belong to distinct sets $\qset'_i,\qset'_j$ let $\event(Q,Q')$ be the bad event that both these paths were selected. Notice that the probability of $\event(Q,Q')$ is $1/q^2$. Notice also that $\vbl(\event(Q,Q'))=\set{z_i,z_j}$, where $Q\in \qset'_i,Q'\in \qset'_j$. There are at most $qL\eta$ events $\event(\hat Q,\hat Q')$, with $z_i\in \vbl(\event(Q,Q'))$: set $\qset'_i$ contains $q$ paths; each of these paths has length at most $L$, so there are at most $qL$ vertices that participate in the paths in $\qset'_i$. Each such vertex may be shared by at most $\eta$ other paths. Similarly, there are at most  $qL\eta$ events $\event(\hat Q,\hat Q')$, with $z_j\in \vbl(\event(Q,Q'))$. 
Therefore, $|\Gamma(\event(Q,Q'))|\leq 2qL\eta$. Let $D=2qL\eta$.
It now only remains to show that $(D+1)ep\leq 1$.
Indeed,
\[ (D+1)ep=\frac{O(qL\eta)}{q^2}=\frac{O(L\eta)}{q}=O\left (\frac{d^2\log^2n}{\alpha^2q}\right ).\] 

By choosing the constant $c$ in the definition of $q$ to be large enough, we can ensure that $(D+1)ep\leq 1$ holds.
Using the algorithm from \Cref{thm: exp-constructive symmetric LLL}, we obtain a collection $\qset = \set{Q_1, \ldots, Q_r}$ of paths in $G$,
where for each $j \in [r]$, path $Q_j$ connects a vertex of $C_j$ to a vertex of $C_{j+r}$, and with high probability the resulting paths are disjoint.
This completes the proof of \Cref{lem: exp-can route large disjoint subsets in expanders}, except for the proof of \Cref{thm: exp-integral routing on expanders} that we provide next.

\subsection{\proofof{Theorem \ref{thm: exp-integral routing on expanders}}}
    We use a slight adaptation of Theorem~18 from~\cite{LeightonRao}. 

    \begin{theorem}[Adaptation of Theorem 18 from~\cite{LeightonRao}]\label{thm: exp-Leighton-Rao}
        There is an efficient algorithm, that,
        given a $n$-vertex graph $G$ with maximum vertex degree at most $d$,
        together with a parameter $0 < \alpha <1$ computes one of the following:
        \begin{itemize}
            \item either a flow $f$ in $G$, with every pair of vertices in $G$ transferring
            $\frac{\alpha}{64 n\log n}$ flow units to each other with no edge-congestion,
            such that every flow-path has length at most $\frac{64 d \log n}{\alpha}$; or
            \item a cut $(S, S')$ in $G$ of sparsity less than $\alpha$.
        \end{itemize}
    \end{theorem}

We provide the proof of \Cref{thm: exp-Leighton-Rao} below, after completing the proof of \Cref{thm: exp-integral routing on expanders} using it.

    We apply \Cref{thm: exp-Leighton-Rao} to the graph $G$ and the parameter $\alpha$.
    If the algorithm returns a cut $(S, S')$ of sparsity less than $\alpha$, then we terminate the algorithm and return this cut. Therefore, we assume from now on that the algorithm returns the flow $f$. Let $f'$ be a flow obtained from $f$ by scaling it up by factor $64 \log n/\alpha$, so that every pair of vertices in $G$ now sends $1/n$ flow units to each other, with total edge-congestion at most $64 \log n/\alpha$. 

    We start by showing that there is a multi-commodity flow $f^*$, where every pair $(u,v)\in \mset$ of vertices sends one flow unit to each other simultaneously, on flow-paths of length at most $128 d\log n/\alpha$, with total vertex-congestion at most $128 d\log n/\alpha$.
    Let $(u,v)\in \mset$ be any pair of vertices.
    The new flow between $u$ and $v$ is defined as follows: $u$ sends $1/n$ flow units to every vertex of $G$, using the flow $f'$,
    and $v$ collects $1/n$ flow units from every vertex of $G$, using the flow $f'$.
    In other words, the flow $f^*$ between $u$ and $v$ is obtained by concatenating all flow-paths in $f'$ originating at $u$ with all flow-paths in $f'$ terminating at $v$.
    It is easy to see then that every flow-path in $f'$ is used at most twice: once by each of its endpoints; all flow-paths in $f^*$ have length at most $128 d\log n/\alpha$; and the total edge-congestion due to flow $f^*$ is at most $128 \log n/\alpha$.
    Since the maximum vertex degree in $G$ is at most $d$, flow $f^*$ causes vertex-congestion at most $128 d\log n/\alpha$.
    
    Next, for every pair $(u,v)\in \mset$, we select one path $Q(u,v)\in \pset(u,v)$ at random, where a path $P\in \pset(u,v)$ is selected with probability $f^*(P)$  -- the amount of flow sent on $P$ by $f^*$.
    We then let $\qset'=\set{Q(u,v)\mid (u,v)\in \mset}$.
    Notice that the length of every path in $\qset'$ is at most $128 d\log n/\alpha$.
    It remains to show that the total vertex-congestion due to paths in $\qset'$ is at most $O(d \log n / \alpha)$ with high probability.
    This is done by standard techniques.
    Consider some vertex $x\in V$.
    We say that the bad event $\event(x)$ happens if more than $8 \cdot 128 d\log n/\alpha$ paths of $\qset$ use the vertex $x$. %
    We use the following variation of the Chernoff bound (see~\cite{measure-concentration}):
    
    \begin{fact}
        Let $X_1,\ldots,X_n$ be independent random variables taking values in $[0,1]$, let $X=\sum_iX_i$, and
        let $\mu=\expect{X}$. Then for all $t>2e\mu$, 
        $\prob{X>t}\leq 2^{-t}$.
    \end{fact}

    It is easy to see that the expected number of paths in $\qset'$ that contain $x$ is at most $128 d\log n/\alpha$, and so the probability of $\event(x)$ is bounded by $1/n^4$.
    From the Union Bound, the probability that any such event happens for any vertex $x\in V$ is bounded by $1/n^3$.
    Therefore, with high probability, every vertex of $G$ belongs to $2^{10} d \log n/\alpha = O(d \log n / \alpha)$  paths in $\qset'$.
    This finishes the proof of \Cref{thm: exp-integral routing on expanders} except for the proof of \Cref{thm: exp-Leighton-Rao}, that we prove in the next sub-section.
\endproofof

\subsection{Proof of Theorem \ref{thm: exp-Leighton-Rao}}

   \newcommand{\LRconst}{64}

    The proof follows closely that of \cite{LeightonRao}; we provide it here for completeness.
    Recall that we are given a graph $G = (V,E)$ with maximum vertex-degree at most $d$, $|V|=n$ and a parameter $0 < \alpha < 1$.
    We let $L = \LRconst d \log n / \alpha$.
    For every pair $u, v$ of vertices in $V$, let $\pset^{\leq L}(u,v)$ be the set of all paths in $G$ between $u$ and $v$ that contain at most $L$ vertices.
    We employ standard linear program for uniform multicommodity flow:

\begin{eqnarray*}
\mbox{(LP-1)}&	\max\quad\quad f^*&\\
	\mbox{s.t.}&&\\
	&\sum_{P \in \pset^{\leq L}(u,v)}  f(P)\geq f^*& \forall u,v \in V \\
&	\sum_{u,v\in V}\sum_{\stackrel{P \in \pset^{\leq L}(u,v):}{e \in P}} f(P) \leq 1 & \forall e \in E \\
 & f(P) \geq 0 & \forall u,v \in V; \forall P \in \pset^{\leq L}(u,v)
\end{eqnarray*}

    In general, the dual of the standard relaxation of the uniform multicommodity flow problem is the problem of assigning lengths $\ell(e)$ to the edges $e\in E$, so as to minimize $\sum_e\ell(e)$, subject to the constraint that the total sum of all pairwise distances between pairs of vertices is at least $1$, where the distance between pairs of vertices is defined with respect to $\ell$. 
    
    In our setting, given lengths $\ell(e)$ on edges $e\in E$, we need to use $L$-hop bounded distances between vertices, defined as follows:
    for all $u,v \in V$, if  $\pset^{\leq L}(u,v) \neq \emptyset$, then we let:
    \[ D^{\leq L}_{\ell}(u,v) = \min_{P \in \pset^{\leq L}(u,v)} \set{\sum_{e \in P} \ell(e)}; \]
    
    otherwise, we set $D^{\leq L}_{\ell}(u,v) = \infty$.
    The dual of (LP-1) can now be written as follows:

\begin{eqnarray*}
	\mbox{(LP-2)}&\min\quad\quad \sum\limits_{e\in E}\ell(e)&\\
	\mbox{s.t.}&&\\
&	\displaystyle\sum\limits_{u,v \in V} D^{\leq L}_{\ell}(u,v) \geq 1  & \\
	& \ell(e) \geq 0 & \forall e \in E\\
	\end{eqnarray*}

    Even though Linear Programs (LP-1) and (LP-2) are of exponential size, they can be solved efficiently using standard techniques (that is, edge-based flow formulation).
    Let $f^*_{\opt}$ be the value of the optimal solution to (LP-1).
    We let $W^* = \frac{d}{nL} = \frac{\alpha}{\LRconst n \log n}$.
    If $f^*_{\opt} \geq W^*$, then we return the flow $f$ corresponding to the optimal solution of (LP-1); it is immediate to verify that it satisfies all requirements.
    Therefore, we assume from now on that $f^*_{\opt} < W^*$.
    We will provide an efficient algorithm to compute a cut $(S, S')$ in $G$ of sparsity less than $\alpha$.

    Given a length function $\ell : E \mapsto \reals_{\geq 0}$, we denote by $W(\ell) = \sum_{e \in E} \ell(e)$ the total `weight' of $\ell$.
   We need the following definition.
   
   \begin{definition}
   	Given an integer $r$ and a length function $\ell(e)$ on edges $e\in E$, the $r$-hop bounded diameter of $G$ is $\max_{u,v\in V}\set{D^{\leq r}_{\ell}(u,v)}$.
   	\end{definition}

    Consider the optimal solution $\ell_{\opt}:E\rightarrow \reals^+$ to (LP-2).
    Observe that, by the strong duality, the value of the solution $W(\ell_{\opt})  = f^*_{\opt}$, and so $W(\ell_\opt) < W^*$ holds.
    
    We define a new solution $\ell$ to (LP-2) as follows: for each edge $e$, we let $\ell(e)=\ell_{\opt}(e)\cdot \frac{W^*}{W(\ell_{\opt})}$. Since $W^* > W(\ell_{\opt})$, it immediate to verify that we obtain a valid solution to (LP-2), of value $W(\ell)=W^*$.
    Moreover, the constraint governing the sum of pairwise $L$-hop bounded distances is now satisfied with strict inequality:
    \begin{equation} \label{eqn: exp-strinct inequality}
        \sum_{u,v} D^{\leq L}_{\ell}(u,v) > 1.
    \end{equation}
    The lengths $\ell(e)$ on edges are fixed from now on, and we denote $D^{\leq L}_{\ell}$ by $D^{\leq L}$ from now on. We will also use 
    the distance function $D^{\leq L/4}_{\ell}$, that we denote by $D^{\leq L/4}$ from now on.
    
    We use the following lemma. 

    \begin{lemma}[Adaptation of Corollary 20 from \cite{LeightonRao}]\label{cor: exp-LR 20}
        There is an efficient algorithm, that,
        given a graph $G = (V,E)$, a parameter $0 < \alpha < 1$ and any edge length function $\ell : E \mapsto \reals_{\geq 0}$ of total weight $W(\ell) = \sum_{e \in E} \ell(e) \leq \frac{\alpha}{64 n \log n}$,
        returns one of the following:
        \begin{itemize}
            \item 
            either a subset $T \subseteq V$ of at least $\ceil{\frac{2|V|}{3}}$ vertices, such that, for $r=\frac{|E|}{2 n^2 W(\ell)}$, the $r$-hop bounded diameter of $G[T]$ is at most $\frac{1}{2n^2}$; or
            \item a cut $(S, S')$ in $G$ of sparsity less than $\alpha$.
        \end{itemize}
    \end{lemma}

    We complete the proof of \Cref{cor: exp-LR 20} later, after we complete the proof of \Cref{thm: exp-Leighton-Rao} using it.
    Recall that $W(\ell) = W^* = \frac{\alpha}{64 n \log n}$.
    We apply the algorithm from \Cref{cor: exp-LR 20} to graph $G$, with parameter $\alpha$ and distance function $\ell$.

    If the algorithm returns a cut $(S, S')$ of sparsity less than $\alpha$, we terminate the algorithm and return this cut. Therefore, we assume from now on that the algorithm from \Cref{cor: exp-LR 20} returns a subset $T \subseteq V$ of at least $2|V|/3$ vertices such that
    $G[T]$ has $r$-hop bounded diameter at most $\frac{1}{2n^2}$, where $r=\frac{|E|}{2 n^2 W(\ell)}$.
    Observe that for all $r'>r$, for every pair $u,v$ of vertices, $D^{\leq r'}(u,v)\leq D^{\leq r}(u,v)$. Observe also that: 
    
    \[r=\frac{|E|}{2 n^2 W(\ell)} \leq \frac{\frac{dn}{2}}{2n^2 \frac{d}{nL}} = \frac{L}{4}.\]

Therefore, the $L/4$-hop bounded diameter of $G[T]$ is at most $\frac 1{2n^2}$.

    For convenience, for a subset $S \subseteq V$ of vertices and a vertex $u \in V$, we denote by $D^{\leq L/4}(u, S) := \min_{v \in S} D^{\leq L/4}(u,v)$.
    We use the following lemma.

    \begin{lemma}[Adaptation of Lemma 21 from \cite{LeightonRao}]\label{lem: exp-LR 21}
        There is an efficient algorithm, that,
        given a graph $G = (V,E)$,
        a parameter $0 < \alpha < 1$,
        any edge length function $\ell : E \mapsto \reals_{\geq 0}$,
        a length parameter $L \geq \frac{2 d \ln n}{\alpha}$ and
        a subset $T \subseteq V$ of at least $\ceil{2|V|/3}$ vertices,
        such that $\sum_{v \in V} D^{\leq L}(v,T) > \frac{4 W(\ell)}{\alpha}$,
        returns a cut $(S, S')$ of $V$ with sparsity less than $\alpha$.
    \end{lemma}

We prove \Cref{lem: exp-LR 21} later, after we complete the proof of \Cref{thm: exp-Leighton-Rao} using it.

    First, we claim that $\sum_{v \in V} D^{\leq L/4}(v, T) > \frac{4 W^*}{\alpha}$.
    Indeed, assume for contradiction otherwise, that is:
    
    \[ \sum_{v \in V} D^{\leq L/4}(v, T) \leq \frac{4 W^*}{\alpha} =\frac{4}{\alpha}\cdot\frac{\alpha}{64n\log n} = \frac{1}{16 n \log n}.\]

    Recall that the $L/4$-hop bounded diameter of $G[T]$ is at most $\frac{1}{2n^2}$.
    From the triangle inequality, for any pair $u, v \in V$ of vertices:

    \[ D^{\leq L}(u,v) \leq D^{\leq L/4}(u,T) + D^{\leq L/4}(v,T) + \frac{1}{2n^2}. \]
    Hence,
    \[ \sum_{u,v \in V} D^{\leq L}(u,v) \leq \sum_{u,v \in V} \left( D^{\leq L/4}(u,T) + D^{\leq L/4}(v,T) + \frac{1}{2n^2} \right) \]
    \[ \indent \leq \frac{1}{2} + 2n \sum_{u \in V}D^{\leq L/4}(u,T)\]
    \[ \indent \leq \frac{1}{2} + 2n \frac{1}{16 n \log n}  = \frac{1}{2} + \frac{1}{8 \log n} < 1,\]

    contradicting the fact that $\ell$ is a valid solution to (LP-2). Therefore, $\sum_{u \in V} D^{L/4}(v, T) > \frac{4 W^*}{\alpha}$ must hold.
    Moreover, notice that $\frac{L}{4} = \frac{16 d \log n}{\alpha} \geq \frac{2 d \ln n}{\alpha}$ holds.
    We now apply the algorithm from \Cref{lem: exp-LR 21} to $G$, with parameters $\alpha$ and $L/4$, edge length function $\ell$ and the subset $T$ of vertices, to 
     obtain a cut $(S,S')$ of $V$ with sparsity less than $\alpha$.
    This completes the proof of \Cref{thm: exp-Leighton-Rao}, except for the proofs of \Cref{cor: exp-LR 20} and \Cref{lem: exp-LR 21} that we provide in the next subsection.

\subsection{Proof of Lemma \ref{cor: exp-LR 20}}

        We start with the following definition:
        \begin{definition}
            Given a graph $G = (V,E)$, a partition of $G$ into components is a collection $\gset = \set{G[V_1], \ldots, G[V_z]}$ of vertex-induced subgraphs such that $\bigcup_{i \in [z]} V_i = V$ and for every $i \neq j$, $V_i \cap V_j = \emptyset$.
        \end{definition}

        We use the following lemma, that we prove later for completeness after completing the proof of \Cref{cor: exp-LR 20} using it.
        \begin{lemma}[Adaptation of Lemma 19 from \cite{LeightonRao}] \label{lem: exp-LR 19}
            There is an efficient algorithm, that,
            given a graph $G = (V,E)$, a parameter $\Delta > 0$, and any edge length function $\ell : E \mapsto \reals_{\geq 0}$,
            partitions $G$ into components
            $\gset = \set{G[V_1], \ldots, G[V_z]}$ such that the following holds:
            \begin{itemize}
                \item For each $G[V_i] \in \gset$, the $r'$-hop bounded diameter of $G[V_i]$ is at most $\Delta$, for $r'=\Delta |E| / W(\ell)$; and
                \item $\sum_{i < j} |E(V_i, V_j)| < 8 W(\ell) \log n/ \Delta$.
            \end{itemize}
        \end{lemma}

        We use \Cref{lem: exp-LR 19} with $\Delta = \frac{1}{2n^2}$ and edge length function $\ell$ to obtain a collection $\gset = \set{G[V_1], \ldots, G[V_z]}$ of components.
        Notice that $r'
        =\frac{\Delta |E|}{W(\ell)} = \frac{|E|}{2n^2 W(\ell)}=r$, so the $r$-hop bounded diameter of each subgrpaph $G[V_i]$ is at most $1/n^2$.
        
        If, for some subgraph $G[V_{i^*}] \in \gset$, $|V_{i^*}| \geq \frac{2|V|}{3}$, then we return $V_{i^*}$.
        Otherwise, we use \Cref{obs: exp-simple partition}, to obtain a partition of the graphs in $\gset$ into two subsets, $\gset'$ and $\gset''$, such that, if we let $S=\bigcup_{G_i\in \gset'}V(G_i)$, and $S'=\bigcup_{G_i\in \gset'}V(G_i)$, then $|S|, |S'| \geq |V|/4$ and $|E(S,S')| < \frac{8 W(\ell) \log n}{\Delta} = 16 W(\ell) n^2 \log n$. 
        Therefore, the sparsity of the cut $(S, S')$ is less than:
        \[ \frac{16 W(\ell) n^2 \log n}{n/4} = 64 W(\ell) n \log n  \leq 64 n \log n \cdot \frac{\alpha}{64 n \log n} = \alpha. \]

        This completes the proof of \Cref{cor: exp-LR 20} except for the proof of \Cref{lem: exp-LR 19} that we provide next.

    \proofof{\Cref{lem: exp-LR 19}}
        If $\Delta < \frac{8 W(\ell) \log n}{|E|}$, we output $\gset = \set{G[\set{v}] \> | \> v \in V}$.
        Notice that for each $G[V_i]$, we have $G[V_i] = G[\set{v_i}]$ for some $v_i \in V$.
        Hence, the $r'$-hop bounded diameter of $G[V_i]$ is $0$, and we have $\sum_{i<j} |E(V_i, V_j)| = |E| < \frac{8 W(\ell) \log n}{\Delta}$ as required.
        Therefore, we assume from now on that $\Delta \geq \frac{8 W(\ell) \log n}{|E|} > \frac{8 W(\ell) \ln n}{|E|}$ holds.
        For convenience, we denote $\epsilon := \frac{2 W(\ell) \ln n}{\Delta |E|}$.
        Notice that $\epsilon \leq 1/4$ holds.

        Consider an auxiliary graph $G^{+} = (V^+, E^+)$ obtained from $G$ by replacing each edge $e$ with a path consisting of $\ceil{|E| \ell(e) / W(\ell)}$ edges.
        Notice that $|E^+| \leq 2|E|$.
        For simplicity, we identify the common vertices of $G$ and $G^+$.
        The following observation is now immediate:
        \begin{observation}\label{obs: exp-LR 19 path length}
            For any path of length $\gamma$ in $G^+$, the corresponding path in $G$ has length at most $\frac{W(\ell) \gamma}{|E|}$.
        \end{observation}

        Next, we iteratively partition vertices of $G^+$ into $V_0^+, V_1^+, \ldots,$ and the required partition of $G$ into components will be given by $G[V_1] = G[V_1^+ \cap V], G[V_2] = G[V_2^+ \cap V], \ldots, $.
        We start with $V^+_0 = \emptyset$ and then iterate.
        We now show how to compute $V_{i+1}^+$ given $V_0^+, \ldots, V_i^+$.

        We denote  $V_i^* := V^+ \backslash \bigcup_{j \leq i} V_j^+$.
        If $V \cap V_i^* = \emptyset$, we have computed the desired partition and the algorithm terminates.
        Thus, we assume from now on that there is a vertex $v_{i+1} \in V_i^*$.
        For every integer $j \geq 0$, we denote by $B^{i+1}_{j}$ the subset of vertices $u \in V_i^*$, such that there is some path of length at most $j$ connecting $v_{i+1}$ and $u$ in $G^+[V^*_i]$.

        We let $C_j := \frac{2|E|}{n} + |E_G[B^{i+1}_j]|$ for every integer $j \geq 0$.
        Let $j^*_{i+1}$ be the smallest $j \geq 0$ such that $C_{j+1} < (1+\epsilon)C_j$.
        Notice that some such $j^*_{i+1}$ must exist, since $\epsilon > 0$ and $C_{j+1} = C_j$ for $j \rightarrow \infty$.
        We set $V_{i+1}^+ = B^{i+1}_{j^*_{i+1}}$ and proceed to the next iteration.
        The following observation is now immediate:
        \begin{observation}\label{obs: exp-LR 19}
            For every index $i > 0$, $V_i^+ \cap V \neq \emptyset$ and $|E(V_i^+, V_i^*)| < \epsilon \left( \frac{2|E|}{n} + \left|  E[V_i^+] \right|\right)$.
        \end{observation}
        \begin{proof}
            Notice that for every index $i > 0$ and $j \geq 0$, we have $v_i \in B_j^i$.
            Thus, $v_i \in V_i^+ \cap V$, and hence $V_i^+ \cap V \neq \emptyset$.
            From our construction, we have
            \[\frac{2|E|}{n} + \left| E[B_{j_i^* + 1}^i] \right| < (1+\epsilon) \left( \frac{2|E|}{n} + \left| E[B_{j_i^*}^i] \right| \right). \]
            
            Equivalently:
            
            \[ \left|  E[B_{j_i^* + 1}^i] \right| - \left| E[B_{j_i^*}^i] \right| < \epsilon \left( \frac{2|E|}{n} + \left|  E[B_{j_i^*}^i] \right|\right).\]
            
            Therefore,
            \[  |E(V_i^+, V_i^*)| \leq \left|  E[B_{j_i^* + 1}^i] \right| - \left| E[B_{j_i^*}^i] \right| < \epsilon \left( \frac{2|E|}{n} + \left|  E[B_{j_i^*}^i] \right|\right) = \epsilon \left( \frac{2|E|}{n} + \left|  E[V_i^+] \right|\right).\]
        \end{proof}

        The following two claims will complete the proof of \Cref{lem: exp-LR 19}.

        \begin{claim}
            $\sum_{i<j} |E(V_i, V_j)| < \frac{8 W(\ell) \log n}{\Delta}$.
        \end{claim}
        \begin{proof}
            \[\sum_{i<j} |E(V_i, V_j)| = \sum_{i>0} \left| E \left( V_i, \bigcup_{j> i} V_j \right) \right| \leq \sum_{i>0} |E(V_i^+, V_i^*)| < \sum_{i > 0} \epsilon \left( \frac{2|E|}{n} + |E(|V_i^+|)| \right)\]
            \[ \indent \leq \epsilon \left( 2|E| + |E^+| \right) \leq 4|E|\epsilon  = \frac{8 W(\ell) \ln n}{\Delta} < \frac{8 W(\ell) \log n}{\Delta} .\]
            Here, the second inequality follows from \Cref{obs: exp-LR 19} and the penultimate inequality follows from the fact that $|E^+| \leq 2|E|$.
        \end{proof}

        \begin{claim}
            For each $G[V_i]$, the $r'$-hop bounded diameter of $G[V_i]$ is at most $\Delta$, for $r' = \frac{\Delta |E|}{W(\ell)}$.
        \end{claim}
        \begin{proof}
            We claim that it suffices to show that, for each $G[V_i]$, the diameter of $G^+[V^+_i]$ is at most $r'= \frac{\Delta |E|}{W(\ell)}$.
            Indeed, if this is the case, \Cref{obs: exp-LR 19 path length} implies that the $r'$-hop bounded diameter of $G[V_i]$ is at most $\frac{W(\ell) r'}{|E|} = \Delta$.
            Notice that, in order to show that the diameter of $G^+[V_i^+]$ is at most $r'$, it suffices to show that $j^*_i \leq \frac{r'}{2} = \frac{\Delta |E|}{2 W(\ell)}$.
            Fix any index $i$ and the corresponding graph $G^+[V_i^+]$. 
            If $j_i^* \neq 0$, we must have:
            \[ 2|E| \geq |E^+| \geq |E(V_i^+)| > (1+\epsilon)^{j^*_i} \frac{2|E|}{n}.\]
            
            Therefore, 
            $ (1+\epsilon)^{j^*_i} < n$ must hold, and so: 
            \[ j^*_i < \frac{\ln n}{\epsilon} = \frac{\Delta |E|}{2 W(\ell)}.\]
            (We have used the fact that $\epsilon < 1/4$).
        \end{proof}
    \endproofof

\subsection{Proof of Lemma \ref{lem: exp-LR 21}}    
        Similarly to the proof of \Cref{lem: exp-LR 21}, consider an auxiliary graph $G^{+} = (V^+, E^+)$ obtained from $G$ by replacing each edge $e$ with a path consisting of $\ceil{|E| \ell(e) / W(\ell)}$ edges.
        Notice that $|E^+| \leq 2|E|$.
        For simplicity, we identify the common vertices of $G$ and $G^+$.
        Given a subset $S \subseteq V(G^+)$ of vertices, we denote by $N(S)$ the set of all vertices $v\in V(G^+)$ such that $v\not\in S$, but $v$ has a neighbor in $S$.
        
        Next, we iteratively partition the vertices of $G^+$ into layers, $V^+_0,V^+_1,\ldots$, and for each $i\geq 0$, we define the corresponding graph $G^+_i=G^+[V^+_i]$, as follows. We start with $V^+_0=T$, $G^+_0 = G^+[T]$ and then iterate. We now show how to compute $V^+_{i+1}$ and $G^+_{i+1}$, given  $V^+_i$ and $G^+_i$, assuming that $V^+_i\neq V^+$ (otherwise, the algorithm terminates). 
        
        Let $E_i := \delta_{G^+}(V^+_i)$ and $C_i := |E_i|$. We partition $E_i$ into two subsets: set $E'_i$ containing all edges $(u,v)$ with $u\in V^+_i$, such that $v$ is a vertex of the original graph $G$; and set $E''_i$ containing all remaining edges. Let $C'_i=|E_i'|$, and let $C''_i=|E''_i|$.
        We distinguish between the following two cases:
        \begin{itemize}
            \item \textbf{Case 1: $C'_i \geq C_i/2$}.
                    In this case, we let $V^+_{i+1}$ contain all vertices of $V^+_i \cup N(V^+_i)$.
                    We also set $G^+_{i+1}=G^+[V^+_{i+1}]$.
                    Notice that in this case, $|E[G^+_{i+1}]\setminus E[G^+_i]|\geq C_i$.

            \item \textbf{Case 2: $C''_i > C_i/2$}.
                    In this case, we let $V^+_{i+1}$ only contain the vertices of $V^+_i$, and those vertices of $N(V^+_i)$ that do not lie in the original graph $G$, that is:
                    
                    \[V^+_{i+1}=V^+_i\cup (N(V^+_i)\setminus V(G)).\] 
                    
                    As before, we set $G^+_{i+1}=G^+[V^+_{i+1}]$. Notice that in this case, $E[G^+_{i+1}]\setminus E[G^+_i]$ contains all edges of $E''_i$, and so $|E[G^+_{i+1}]\setminus E[G^+_i]|\geq C''_i> C_i/2$.
        \end{itemize}

        From the above discussion we obtain the following observation:
        \begin{observation} \label{obs: exp-LR edges grow}
            For each level $i$, $|E(G^+_{i+1})\setminus E(G^+_i)|\geq \frac{C_i}{2}$, and in particular $\sum_iC_i\leq 2|E^+|$.
        \end{observation}

        For each level $i$, let $n_i = |V(G) \backslash V^+_i|$ -- the number of vertices of the original graph $G$ that do not lie in $V^+_i$.
        Recall that $|T| \geq \ceil{2|V|/3}$, and so for all $i$, $n_i \leq |V|/3 \leq |V|/2$.
        Moreover, $C_i = |\delta_{G^+}(V^+_i)| \geq |\delta_G(V \cap V^+_i)|$.
        
        If, for any level $i$, $C_i<\alpha n_i$, then we return the cut $(V \cap V^+_i, V \backslash V^+_i)$; it is immediate to see that its sparsity is less than $\alpha$. Therefore, we assume from now on, that for all $i$, $C_i\geq \alpha n_i$. 
        We will reach a contradiction by showing that $\sum_{v \in V} D^{\leq L}(v,T) \leq \frac{4 W(\ell)}{\alpha}$ must hold.
        In order to do so, we use the following two claims.
        
        \begin{claim}\label{clm: exp-bound Case 1}
        	The number of indices $i$ for which Case $1$ is invoked is at most $L$.
        	\end{claim}
        
        \begin{proof}
        	Let $i$ be an index for which Case $1$ is invoked, so $C'_i \geq C_i/2$. Recall that we have assumed that $C_i\geq\alpha n_i$. 
        Since the maximum vertex-degree of $G$ is bounded by $d$, the number of new vertices of $V$ that are added to $V^+_{i+1}$ is at least $\frac{C'_i}{d}\geq \frac{\alpha n_i}{2d} $. Therefore, $n_{i+1} \leq n_i(1- \frac{\alpha }{2d})$, and the total number of indices $i$ in which Case 1 is invoked must be bounded by  $\frac{2d \ln n}{\alpha} \leq L$.
        \end{proof}

        \begin{claim}\label{clm: exp-bound nis}
        	$\sum_i n_i\leq \frac{4|E|}{\alpha}$.
        \end{claim}
        \begin{proof}
            Recall that we have assumed $C_i \geq \alpha n_i$ for all $i$.
            Thus, 
            \[ \sum_i n_i \leq \sum_i \frac{C_i}{\alpha} = \frac{\sum_{i} C_i}{\alpha} \leq \frac{2 |E^+|}{\alpha} \leq \frac{4 |E|}{\alpha}.\]
            Here, the second-last inequality follows from \Cref{obs: exp-LR edges grow} and the last inequality follows from the fact that $|E^+| \leq 2|E|$.
        \end{proof}

        For each vertex $v \in V \backslash T$, let $i_v$ be the unique index, such that $v \in V(G^+_i)$ and $v \not \in V(G^+_{i-1})$.
        For the remaining vertices $v \in T$, we set $i_v = 0$. Notice that $v$ must be connected by an edge to a vertex $u$ with $i_u<i_v$. Therefore, we can construct a path $P^+_v=(v_0,v_1,\ldots,v_r)$ in $G^+$, where $v_0\in T$, $v_r=v$, and for all $1\leq j\leq r$, $i_{v_{j-1}}<i_{v_j}$. 
        
        Let $P_v$ be the path corresponding to $P^+_v$ in the original graph $G$. Since we invoke Case $1$ at most $L$ times, it is easy to verify that $P_v$ contains at most $L$ edges. 
        Moreover:

        \[D^{\leq L}(v, T) \leq \sum_{e \in P_v} \ell(e) \leq \sum_{e \in P_v} \frac{W(\ell)}{|E|} \ceil{\frac{|E| \ell(e)}{W(\ell)}} = \frac{W(\ell)}{|E|} |E(P^+_v)| \leq i_v \frac{W(\ell)}{|E|}.\]

        Altogether:

        \[\sum_v D^{\leq L}(v, T) \leq \frac{W(\ell)}{|E|} \sum_v i_v = \frac{W(\ell)}{|E|} \sum_i n_i \leq \frac{4W(\ell)}{\alpha}, \]
        where the last inequality follows from \Cref{clm: exp-bound nis}. This contradicts the assumption that $\sum_v D^{\leq L}(v, T) > \frac{4W(\ell)}{\alpha}$,

        completing the proof of \Cref{lem: exp-LR 21}.

\section{Proofs Omitted from Section \ref{sec: exp-exp to poe}}\label{appn-sec: proofs for sec 7}
\subsection{\proofof{Claim \ref{clm: exp-cheeger toy}}} \label{appn-subsec: clm: exp-cheeger toy}
We start with an arbitrary balanced cut $(U',U'')$ in $G$ with $|U'|\geq |U''|$, and perform a number of iterations. In every iteration, we will either establish that $G[U']$
is an $\Omega(\frac{\beta^2}{d})$-expander, or compute the desired partition $(S,T)$ of $V$, or find a new balanced cut $(J',J'')$ in $G$ with $|E(J',J'')|<|E(U',U'')|$. In the first two cases, we terminate the algorithm and return either $V'=U'$ (in the first case), or the cut $(S,T)$ (in the second case). In the last case, we replace $(U',U'')$ with $(J',J'')$, and continue to the next iteration.

We now describe the execution of an iteration. Recall that we are given a balanced cut $(U',U'')$ of $G$ with  $|U'|\geq |U''|$. If $|E(U',U'')|<\beta\cdot \min\set{|U'|,|U''|}$, then we return the cut $(S,T)=(U',U'')$ and terminate the algorithm. Therefore, we assume that $|E(U',U'')|\geq \beta\cdot \min\set{|U'|,|U''|}$. We apply the algorithm from \Cref{thm: exp-spectral} to graph $G[U']$, and consider the cut $(S,T)$ of $G[U']$ computed by the algorithm. We then consider two cases. First, if $|E(S,T)|\geq \frac{\beta}{4}\min\set{|S|,|T|}$, then from \Cref{thm: exp-spectral}, we are guaranteed that $G[U']$ is an $\Omega(\frac{\beta^2}{d})$-expander. We terminate the algorithm and return $V'=U'$. 

We assume that $|E(S,T)|<\frac{\beta}{4}\min\set{|S|,|T|}$ from now on, and we assume w.l.o.g. that $|T|\leq |S|$. We consider again two cases. First, if $|E(T,U'')|\leq \frac{\beta}{2} |T|$, we define a new cut $(S',T)$ in $G$, where $S'=S\cup U''$. We then get that $|T|\leq |S'|$, and moreover, $|E_G(S',T)|=|E_G(S,T)|+|E_G(U'',T)|<\beta |T|$. We return the cut $(S',T)$ and terminate the algorithm.

The final case is when $|E(T,U'')|>\frac{\beta}{2}|T|$. In this case, we are guaranteed that $|E(T,U'')|>|E(S,T)|$. Therefore, if we consider the cut $(J',J'')$, where $J'=S$ and $J''=T\cup U''$, then $(J',J'')$ is a balanced cut in $G$, and moreover:

\[|E(J',J'')|=|E(S,U'')|+|E(S,T)|<|E(S,U'')|+|E(T,U'')|=|E(U',U'')|.\]

We then replace $(U',U'')$ with the new cut $(J',J'')$, and continue to the next iteration.
It is easy to verify that every iteration can be executed in time $\poly(n)$. Since the number of the edges in the set $E(U',U'')$ decreases in every iteration, the number of iterations is also bounded by $\poly(n)$.
This completes the proof of \Cref{clm: exp-cheeger toy}.

\subsection{Corollary \ref{cor: exp-expander or balanced}} \label{appn-subsec: cor: exp-expander or balanced}
Throughout the algorithm, we maintain a set $E'$ of edges of $G$ that we remove from the graph, starting with $E'=\emptyset$, and a collection $\gset$ of disjoint induced subgraphs of $G\setminus E'$, starting with $\gset=\set{G}$. The algorithm continues as long as there is some graph $H\in \gset$, with $|V(H)|>3|V(G)|/4$. In every iteration, we select the unique graph $H\in \gset$ with  $|V(H)|>3|V(G)|/4$, and apply \Cref{clm: exp-cheeger toy} to it, with the parameter $\beta/4$. If the outcome is a subset $V'\subseteq V(H)$ of vertices, such that $|V(H)|/2\leq |V'|\leq 3|V(H)|/4$, and $H[V']$ is an $\Omega(\frac{\beta^2}{d})$-expander, then we return $V'$: it is easy to verify that $n/4\leq |V'|\leq 3n/4$, so $V'$ is a valid output. Otherwise, we obtain a partition $(S',T')$ of $V(H)$ with $|E(S',T')| < \frac{\beta}{4} \cdot \min\set{|S'|,|T'|}$. We add the edges of $E(S',T')$ to $E'$,  remove $H$ from $\gset$, and add $H[S']$ and $H[T']$ to $\gset$ instead. If $|S'|<|T'|$, then our algorithm will never attempt to process the graph $H[S']$ again, so we \emph{charge} the edges of $E(S',T')$ to the vertices of $S'$, where every vertex of $S'$ is charged fewer than $\beta/4$ units. The algorithm terminates when every graph $H\in \gset$ has $|V(H)|\leq 3n/4$ (unless it terminates earlier with an expander). Notice that from our charging scheme, at the end of the algorithm, $|E'|< n\beta/4$. Moreover, using \Cref{obs: exp-simple partition}, we can partition the final collection $\hset$ of graphs into two subsets, $\hset',\hset''$, such that $\sum_{H\in \hset'}|V(H)|,\sum_{H\in \hset''}|V(H)|\geq n/4$. Letting $S=\bigcup_{H\in \hset'}V(H)$ and $T=\bigcup_{H\in \hset''}V(H)$, we obtain a balanced partition $(S,T)$ of $V(G)$. Since $E(S,T)\subseteq E'$, we get that $|E(S,T)|< \frac{\beta n}{4}\leq \beta \cdot \min\set{|S|,|T|}$.

\subsection{\proofof{Claim \ref{clm: exp-large expanding subgraph cheeger}}}\label{appn-subsec: exp-efficient expander fixing}

The proof is very similar to the proof of \Cref{clm: exp-large expanding subgraph-edges}.
The algorithm iteratively removes edges from $G\setminus E'$, until we obtain a connected component of the resulting graph that is an $\Omega\left(\frac{\alpha^2}{d}\right)$-expander.
We start with $G'=G\setminus E'$ (notice that $G'$ is not necessarily connected).  We also maintain a set $E''$ of edges that we remove from $G'$, initialized to $E''=\emptyset$. 
We then perform a number of iterations. In every iteration, we apply \Cref{thm: exp-spectral} to $G'$, and obtain a cut $(X,Y)$ in $G'$. If $|E_{G'}(X,Y)| \geq \alpha \cdot\min{(|X|,|Y|)}/4$, then, from \Cref{thm: exp-spectral}, $G'$ is an $\Omega\left(\frac{\alpha^2}{d}\right)$-expander. We terminate the algorithm and return $G'$. We later show that $|V(G')|\geq |V| - \frac{4|E'|}{\alpha}$. Assume now that  $|E_{G'}(X,Y)| < \alpha \cdot \min{(|X|,|Y|)}/4$, and assume w.l.o.g. that $|X|\geq |Y|$. Update $G'$ to be $G'[X]$, add the edges of $E(X,Y)$ to $E''$, and continue to the next iteration. 
Clearly, at the end of the algorithm, we obtain a graph $G'$ that is an $\Omega\left(\frac{\alpha^2}{d}\right)$-expander. It only remains to show that $|V(G')|\geq |V| - \frac{4|E'|}{\alpha}$. The remainder of the analysis is identical to the analysis of \Cref{clm: exp-large expanding subgraph-edges}.

Assume that the algorithm performs $r$ iterations, and for each $1\leq i\leq r$, let $(X_i,Y_i)$ be the cut computed by the algorithm in iteration $i$. Since $|X_i|\geq |Y_i|$, $|Y_i|\leq |V(G)|/2$. At the same time, if we denote $E_i=E''\cap E(X_i,Y_i)$, then $|E_i|< \alpha |Y_i|/4$. Therefore:

\[|E''|=\sum_{i=1}^r|E_i|< \alpha\sum_{i=1}^r|Y_i|/4.\]

On the other hand, since $G$ is an $\alpha$-expander, the total number of edges leaving each set $Y_i$ in $G$ is at least $\alpha|Y_i|$, and all such edges lie in $E'\cup E''$. Therefore:

\[|E'|+|E''|\geq \alpha\sum_{i=1}^r|Y_i|/2.\]

Combining both bounds, we get that $|E'|\geq \alpha\sum_{i=1}^r|Y_i|/4$, and so $\sum_{i=1}^r|Y_i|\leq \frac{4|E'|}{\alpha}$. 
Therefore, $|V(G')|=|V|-\sum_{i=1}^r|Y_i|\geq |V|-\frac{4|E'|}{\alpha}$.
\endproofof

\subsection{\proofof{Claim \ref{clm: exp-flow in expander}}} \label{appn-subsec: exp-flow in expander}
We can compute the largest-cardinality set of disjoint paths connecting vertices of $A$ to vertices of $B$  in $G$ using standard maximum $s$--$t$ flow computation and the integrality of flow. Therefore, it is sufficient to show that there exists a set of $\ceil{\alpha z/d}$ disjoint paths connecting $A$ to $B$ in $G$.

Assume otherwise. Then, from Menger's theorem, there is a set $Z$ of fewer than $\alpha z/d$ vertices in $G$, such that $G\setminus Z$ contains no path from a vertex of $A\setminus Z$ to a vertex of $B\setminus Z$. Let $E'$ be the set of all edges of $G$ incident to the vertices of $Z$. Since the maximum vertex degree in $G$   is at most $d$, $|E'|<\alpha z$.
Therefore, graph $G\setminus E'$ contains no path connecting a vertex of $A$ to a vertex of $B$. Let $X$ be the union of all connected components of $G\setminus E'$ containing the vertices of $A$, and let $Y=V(G)\setminus X$. Then $|E(X,Y)|\leq |E'|<\alpha z\leq \alpha\cdot\min\set{|X|,|Y|}$, contradicting the fact that $G$ is an $\alpha$-expander.
\endproofof

\subsection{\proofof{Theorem \ref{thm: exp-exp to wl}}} \label{appn-subsec: exp-exp to wl}
The main tool that we use for the proof of \Cref{thm: exp-exp to wl} is the following theorem, whose proof appeared in~\cite{CC_gmt}; we include the proof here for completeness.

\begin{theorem}[Restatement of Theorem A.4 in~\cite{CC_gmt}]\label{thm: exp-path grouping}
  There is an efficient algorithm, that, given a graph $G$ with maximum vertex degree at most $d$, an integer $q\geq 1$, and a set $\pset$ of at least $16dq$ disjoint paths in $G$, computes a subset $\pset'\subseteq \pset$ of at least $|\pset|/2$ paths, and a collection $\cset$ of
  disjoint connected subgraphs of $G$, such that each path $P\in
  \pset'$ is completely contained in some subgraph $C\in \cset$, and
  each such subgraph contains at least $q$ and at most $4d q$
  paths in $\pset$.
\end{theorem}

\begin{proof}
Starting from $G$, we construct a new graph $H$, by contracting every path $P\in \pset$ into a super-node $u_P$. Let $U=\set{u_P\mid P\in \pset}$ be the resulting set of super-nodes. Let $\tau$ be any spanning tree of $H$, rooted at an arbitrary vertex $r$. Given a vertex $v\in V(\tau)$, let $\tau_v$ be
  the sub-tree of $\tau$ rooted at $v$. Let $J'_v\subseteq V(G)$ be the set of all vertices of
  $\tau_v$ that  belong to the original graph $G$ (that is, they are not super-nodes), and let $J''_v$ be the set of all vertices of $G$ that lie on paths $P\in \pset$ with $u_P\in \tau_v$. We then let $J_v=J'_v\cup J''_v$. We also denote $G_v=G[J_v]$; observe that it must be a connected graph. Over the course of the algorithm, we will delete some vertices from $\tau$. The notation $\tau_v$ and $G_v$ is always computed with respect to the most current tree
  $\tau$. We start with $\cset=\emptyset,\pset'=\emptyset$, and then iterate.

  Each iteration is performed as follows. If $q\leq |V(\tau)\cap U|\leq 4d q$, then we add the graph $G_r$ corresponding to the root $r$ of $\tau$ to $\cset$, and terminate
  the algorithm. If $|V(\tau)\cap U|<q$, then we also terminate the
  algorithm (we will show later that $|\pset'|\geq |\pset/2|$ at this point). Otherwise, let $v$ be the lowest
  vertex of $\tau$ with $|\tau_v\cap U|\geq q$. If $v\not \in U$, then,
  since the degree of every vertex in $G$ is at most $d$, $|\tau_v\cap
  U|\leq d q$. We add $G_v$ to $\cset$, and all paths in
  $\set{P\mid u_P\in \tau_v}$ to $\pset'$. We then delete all vertices of
  $\tau_v$ from $\tau$, and continue to the next iteration.

  Assume now that $v=u_P$ for some path $P\in \pset$. If $|\tau_v\cap
  U|\leq 4d q$, then we add $G_v$ to $\cset$, and all paths in
  $\set{P'\mid u_{P'}\in \tau_v}$ to $\pset'$ and continue to the next
  iteration. So we assume that $|\tau_v\cap U|> 4d q$.

  Let $v_1,\ldots,v_z$ be the children of $v$ in $\tau$.  Build a new
  tree $\tau'$ as follows. Start with the path $P$, and add the vertices
  $v_1,\ldots,v_z$ to $\tau'$. For each $1\leq i\leq z$, let
  $(x_i,y_i)\in E(G)$ be any edge connecting some vertex $x_i\in V(P)$ to
  some vertex $y_i\in V(G_{v_i})$; such an edge must exist from the definition
  of $G_{v_i}$ and $\tau$. Add the edge $(v_i,x_i)$ to $\tau'$. Therefore,
  $\tau'$ is the union of the path $P$, and a number of disjoint stars
  whose centers lie on the path $P$, and whose leaves are the vertices
  $v_1,\ldots,v_z$. The degree of every vertex of $P$ is at most
  $d$. We define the \emph{weight} of the vertex $v_i$ as the number
  of the paths in $\pset$ contained in $G_{v_i}$ (equivalently, it is $|U\cap \tau_{v_i}|$). Recall that the weight
  of each vertex $v_i$ is at most $q$, by the choice of $v$. For each
  vertex $x\in P$, the weight of $x$ is the total weight of its
  children in $\tau'$. Recall that the total weight of all vertices
  of $P$ is at least $4d q$, and the weight of every vertex is at
  most $d q$. We partition $P$ into a number of disjoint segments
  $\Sigma=(\sigma_1,\ldots,\sigma_{\ell})$ of weight at least $q$ and
  at most $4dq$ each, as follows. Start with $\Sigma=\emptyset$,
  and then iterate. If the total weight of the vertices of $P$ is at
  most $4d q$, we build a single segment, containing the whole
  path. Otherwise, find the shortest segment $\sigma$ starting from
  the first vertex of $P$, whose weight is at least $q$. Since the
  weight of every vertex is at most $d q$, the weight of $\sigma$
  is at most $2d q$. We then add $\sigma$ to $\Sigma$, delete it
  from $P$ and continue. Consider the final set $\Sigma$ of
  segments. For each segment $\sigma$, we add a new graph
  $C_{\sigma}$ to $\cset$. Graph $C_\sigma$ consists of the union of
  $\sigma$, the graphs $G_{v_i}$ for each $v_i$ that is connected to a
  vertex of $\sigma$ with an edge in $\tau'$, and the corresponding edge
  $(x_i,y_i)$. Clearly, $C_{\sigma}$ is a connected subgraph of
  $G$, containing at least $q$ and at most $4d q$ paths of
  $\pset$. We add all those paths to $\pset'$, delete all vertices of
  $\tau_v$ from $\tau$, and continue to the next iteration. We note that path $P$ itself is not added to $\pset'$, but all paths $P'$ with $u_{P'}\in V(\tau_v)$ are added to $\pset'$.

  At the end of this procedure, we obtain a collection $\pset'$ of
  paths, and a collection $\cset$ of disjoint connected subgraphs of
  $G$, such that each path $P\in \pset'$ is contained in some $C\in
  \cset$, and each $C\in \cset$ contains at least $q$ and at most
  $4d q$ paths from $\pset'$. It now remains to show that
  $|\pset'|\geq |\pset|/2$. We discard at most $q$ paths in the last
  iteration of the algorithm. Additionally, when $v=u_P$ is processed,
  if $|\tau_v\cap U|> 4d q$, then path $P$ is also discarded, but at
  least $4d q$ paths are added to $\pset'$. Therefore, overall,
  $|\pset'|\geq |\pset|-\frac{|\pset|}{4d q+1}-q\geq |\pset|/2$,
  since $|\pset|\geq 16dq$.
\end{proof}

We now turn to prove \Cref{thm: exp-exp to wl}. Recall that we are given an $\alpha$-\posexp System $\Sigma = (\sset, \mset, A_1, B_3)$ of width $w$ and length $3$, where $0<\alpha <1$, and the corresponding graph $G_{\Sigma}$ has maximum vertex degree at most $d$. Our goal is to compute subsets $\hat A_1\subseteq A_1,\hat B_3\subseteq B_3$ of $\Omega(\alpha^2 w/d^3)$ vertices each, such that $\hat A_1\cup \hat B_3$ is well-linked in $G_{\Sigma}$. Notice that we can assume w.l.o.g. that $w\geq 256 d^3/\alpha^2$, as otherwise it is sufficient that each set $\hat A_1,\hat B_3$ contains a single vertex, which is trivial to ensure.

We apply \Cref{clm: exp-flow in expander} to graph $S_1$, together with the sets $A_1,B_1$ of vertices, to compute a set $\pset_1$ of $\ceil{\alpha w/d}$ node-disjoint paths in $S_1$, connecting vertices of $A_1$ to vertices of $B_1$. We then set $q=\floor{16d/\alpha}$, and use \Cref{thm: exp-path grouping}, 
to compute a subset $\pset'_1\subseteq \pset_1$ of at least $|\pset_1|/2\geq \alpha w/(2d)$ paths, and a collection $\cset$ of
  disjoint connected subgraphs of $S_1$, such that each path $P\in
  \pset'_1$ is completely contained in some subgraph $C\in \cset$, and
  each such subgraph contains at least $q$ and at most $4d q$ paths of $\pset'_1$. 
  (Note that from our assumption that $w\geq 256 d^3/\alpha^2$, $|\pset_1|\geq 16dq$).
  Clearly, $|\cset|\geq \frac{|\pset_1'|}{4dq}\geq \frac{\alpha^2 w}{256d^3}$. We select one representative path $P\in \pset'_1$ from each subgraph $C\in \cset$, so that $P\subseteq C$, and we let $\pset^*_1\subseteq \pset'_1$ be the resulting set of paths. We are now ready to define the set $\hat A_1\subseteq A_1$ of vertices: set $\hat A_1$ contains, for every path $P\in \pset^*_1$, the endpoint of $P$ that lies in $A_1$. Note that $|\hat A_1|=|\pset^*_1|=|\cset|\geq  \frac{\alpha^2 w}{256d^3}$. For convenience, for every vertex $a\in \hat A_1$, we denote by $P_a\in \pset^*_1$ the unique path originating at $a$, and we denote by $C_a\in \cset$ the unique subgraph of $S_1$ containing $P_a$.

We select a subset $\hat B_3\subseteq B_3$ of at least $ \frac{\alpha^2 w}{256d^3}$ vertices similarly, by running the same algorithm in $S_3$.
The set of paths obtained as the outcome of \Cref{thm: exp-path grouping} is denoted by $\pset_3'$, and the set of connected subgraphs of $S_3$ by $\cset'$. We also denote by $\pset^*_3\subseteq \pset'_3$ the set of representative paths that we select from each subgraph of $\cset'$.
 For every vertex $b\in \hat B_3$, we denote by $P_b\in \pset^*_b$ the unique path originating at $b$, and we denote by $C_b\in \cset'$ the unique subgraph containing $P_b$.
 
 It remains to show that $\hat A_1\cup \hat B_3$ is well-linked in $G_{\Sigma}$. We show this using the same arguments as in ~\cite{CC_gmt}.
Let $X,Y\subseteq \hat A_1\cup \hat B_3$ be two equal-cardinality sets of vertices. We need to show that there is a set $\qset$ of $|X|=|Y|$ disjoint paths connecting them in $G_{\Sigma}$, such that the paths in $\qset$ are internally disjoint from $\hat A_1\cup \hat B_3$. We define a new subgraph $H\subseteq G_{\Sigma}$ as follows: graph $H$ is the union of the graph $S_2$ and the matchings $\mset_1$ and $\mset_2$; additionally, for every vertex $v\in X\cup Y$, we add the graph $C_v$ to $H$. It is now enough to show that  there is a set $\qset$ of $|X|=|Y|$ disjoint paths connecting $X$ to $Y$ in $H$; such paths are guaranteed to be internally disjoint from $\hat A_1\cup \hat B_3$. From the integrality of flow, it is sufficient to show a flow $F$ in $H$, where every vertex in $X$ sends one flow unit, every vertex in $Y$ receives one flow unit, and every vertex of $H$ carries at most one flow unit. We now construct such a flow.
This flow will be
a concatenation of three flows, $F_1,F_2,F_3$.

We start by defining the flows $F_1$ and $F_3$. 
Consider some vertex $v\in X\cup Y$, and assume w.l.o.g. that $v\in \hat A_1$. We select an arbitrary subset $U_v\subseteq B_1$ of $q=\floor {16d/\alpha}$ vertices that serve as endpoints of paths $P\in \pset_1'$ that are contained in $C_v$. Since $C_v$ is a connected graph, vertex $v$ can send $1/q$ flow units to every vertex in $U_v$ simultaneously, inside the graph $C_v$, so that the flow on every vertex is at most $1$. We denote the resulting flow by $F^v$.

We obtain the flow $F_1$ by taking the union of all flows $F^v$ for $v\in X$, and we obtain the flow $F_3$ by taking the union of all flows $F^v$ for $v\in Y$ (we reverse the direction of the flow $F^v$ in the latter case). 

Let $R_1=\bigcup_{v\in X}U_v$, and let $R_2=\bigcup_{v\in Y}U_v$. Note that  $R_1\cup R_2\subseteq B_1\cup A_3$. For every vertex $x\in R_1\cup R_2$ that lies in $B_1$, we let $x'$ be the vertex of $A_2$, that is connected to $x$ by an edge of $\mset_1$. Similarly,  for every vertex $x\in R_1\cup R_2$ that lies in $A_3$, we let $x'$ be the vertex of $B_2$, that connects to $x$ by an edge of $\mset_2$. Let $R_1'=\set{x'\mid x\in R_1}$ and $R_2'=\set{x'\mid x\in R_2}$. Note that $R_1',R_2'$ are disjoint sets of vertices in $S_2$. Since graph $S_2$ is an $\alpha$-expander, there is a flow $F_2'$ in $S_2$, where every vertex in $R_1'$ sends one flow unit, every vertex in $R_2'$ sends one flow unit, and every edge carries at most $1/\alpha$ flow units. Scaling this flow down by factor $q=\floor{16d/\alpha}$, we obtain a new flow $F_2$ in $S_2$, where every vertex of $R_1'$ sends $1/ {q}$ flow units, every vertex of $R_2'$ receives $1/ {q}$ flow units, and every vertex of $S_2$ carries at most one flow unit.

The final flow $F$ is obtained by concatenating the flows $F_1,F_2$ and $F_3$, and sending $1/{q}$ flow units on every edge of $\mset_1\cup \mset_2$ that is incident to a vertex of $R_1\cup R_2$. The flow in $F$ guarantees that every vertex of $X$ sends one flow unit, every vertex in $Y$ receives one flow unit, and every vertex of $G_{\Sigma}$ carries at most one flow unit.
            }{}

            \chapter{Proofs Omitted from Chapter \ref{chap: ncm}}  \label{appn-chap: ncm}

        \iftoggle{ncm}{

\iftoggle{partitioning}{
    \section{Proofs Omitted from Section \ref{sec: ncm-partition-lemma}} \label{appn: ncm-partition-lemma}

\subsubsection{\proofof{Lemma \ref{lem: ncm-vanilla-partition}}} \label{prf-lem: ncm-vanilla-partition}
    For each $0 \leq i \leq \log N$, we consider the partition $\bset^i := \bset_{2^i}(S)$ of sequence $S$ into exactly $2^i$ blocks, each of length exactly $N/2^i$.
    We associate a partition tree $T$ with the partitions $\bset^0, \ldots, \bset^{\log N}$ of $S$ into blocks, as follows.
    We start with the tree $T$ containing a single vertex $v(S)$, representing the unique block $S \in \bset^0$.
    We will view vertex $v(S)$ as the root of the tree $T$, and we say that it lies at level $0$ of $T$.
    For all $1\leq i\leq \log N$, the set of vertices at level $i$ of the tree $T$ is $\set{v(B) \mid B \in \bset^i}$,
    that is, we have one vertex for every block in the partition $\bset^i$.
    For each such vertex $v(B)$ with $B\in \bset^i$, we let $B'$ be the unique block of $\bset^{i-1}$ with $B\subseteq B'$,
    and we add an edge connecting $v(B')$ to $v(B)$ to the tree $T$, so that $v(B)$ becomes a child vertex of $v(B')$.
    This completes the construction of the partition tree $T$.
    Notice that each non-leaf vertex of $T$ has exactly two children,
    and each leaf vertex of $T$ represents a block of $\bset^{\log N}$, which contains exactly one element of $S$.
        
    We will say that block $B$ is a parent of block $B'$ iff $v(B)$ is a parent of $v(B')$ in the tree $T$.
    Similarly, we will say that block $B$ is an \emph{ancestor} of block $B'$ iff vertex $v(B)$ is an ancestor of $v(B')$
    (we assume that a vertex of $T$ may not be an ancestor of itself).
        
    For every block $B \in  \bigcup_{i=0}^{\log N}\bset^i$, we denote by $n(B)$ the number of elements of $S'$ lying in  $B$.
    Note that, if $P$ is any leaf-to-root path in the tree $T$, then the values of $n(B)$ of the blocks $B$ with $v(B)\in P$ are non-decreasing, as we traverse $P$ towards the root of $T$.
    We will now \emph{mark} some vertices of the tree $T$.
    Specifically, we mark a vertex $v(B)$ of $T$, iff
    (i) $n(B)\leq 2Z$; and
    (ii) for every ancestor block $B'$ of $B$, $n(B')>2Z$.
    If vertex $v(B)$ is marked, we will sometimes say that block $B$ is marked.
    Clearly, we never mark both a vertex and its ancestor.
    Moreover, every leaf-to-root path $P$ in tree $T$ has exactly one marked vertex: the last vertex $v(B)$ on$P$ with $n(B)\leq 2Z$.

    Let $\bset^*$ denote the set of all marked blocks $B \in  \bigcup_{i=0}^{\log N}\bset^i$.
    Then the blocks in $\bset^*$ define a partition of the sequence $S$. Moreover, since each marked block contains at most $2Z$ elements of $S'$, we get that $|\bset^*|\geq |S'|/(2Z)$.
    We let $x_1, \ldots, x_{1 + \log N}$ be such that for each $1 \leq i \leq 1 + \log N$, we have $x_i = |\bset^* \cap \bset^{i-1}|$.
    Notice that $\sum_i x_i = |\bset^*|$ and from \Cref{fact: ncm-pigeonhole}, there is an index $1 \leq j \leq \log N$ such that $x_j \geq \sum_i x_i / (2j^2) = |\bset^*|/(2j^2)$.
    We fix such $j$ and let $\eta = 2^{j-1}$ so that $\bset^{j-1} = \bset_\eta(S)$.
    Let $\bset^{**} = \bset^* \cap \bset_\eta(S)$ and we have,

    \begin{equation} \label{eqn: ncm-bset-star-star-bound}
        |\bset^{**}| = x_j \geq \frac{|\bset^*|}{2j^2} = \frac{|\bset^*|}{2 (1 + \log{\eta})^2}  \geq \frac{|S'|}{4Z \log^2{(2\eta)}}.
    \end{equation}

    We say that a block $B \in \bset^{**}$ is a \emph{bad block} iff its parent block $B'$ has $n(B') > 16Z \log^2{(2\eta)}$.
    We need the following observation.
        
    \begin{observation}\label{obs: ncm-num of bad blocks}
        The number of bad blocks in $\bset^{**}$ is at most $|\bset^{**}|/2$.
    \end{observation}
    \begin{proof}
        If $\eta = 1$, there are no bad blocks and there is nothing to show.
        Assume for contradiction that $\eta \geq 2$ and there are more than $|\bset^{**}|/2$ bad blocks in $\bset^{**}$.
        If $B \in \bset^{**}$ is a bad block, and $B'$ is its parent, then $n(B') > 16Z \log^2{(2\eta)}$.
        As $B'$ may be a parent of at most two bad blocks in $\bset^{**}$, we get that there is a set $\tilde \bset\subseteq \bset_{\eta/2}(S)$ of blocks (that are parent-blocks of bad blocks in $\bset^{**})$,
        such that $|\tilde \bset|\geq |\bset^{**}|/4$, and every block $B'\in \tilde \bset$ has $n(B') > 16Z \log^2{(2\eta)}$.
        Since all blocks in $\tilde \bset$ are disjoint, from \Cref{eqn: ncm-bset-star-star-bound}, they collectively contain more than
        
        \[ |\tilde \bset| \cdot 16Z \log^2{(2\eta)} \geq \frac{|\bset^{**}|}{4} \cdot 16Z \log^2{(2\eta)} \geq |S'| \]
        
        elements of $S'$, a contradiction.
    \end{proof}

    We discard all bad blocks from $\bset^{**}$.
    Notice that now $|\bset^{**}| \geq \frac{|S'|}{8Z \log^2{2\eta}}$ holds.
    Next, we construct a set $\hat \bset \subseteq \bset^{**}$ of blocks, as follows.
    We start with $\hat \bset = \emptyset$.
    We then iterate, as long as $\bset^{**} \neq \emptyset$.
    In every iteration, we consider some arbitrary block $B \in \bset^{**}$.
    We denote by $B'$ its sibling block and by $B''$ its parent block 
    (that is, $v(B)$ and $v(B')$ are both child vertices of $v(B'')$ in the tree $T$).
    Since block $B$ was marked, $n(B'') > 2Z$ must hold.
    Moreover, $B$ and $B'$ form a partition of the block $B''$, and hence, every element of $B'' \cap S'$ must lie in either $B$ or $B'$.
    Therefore, either $n(B) \geq Z$ or $n(B') \geq Z$ must hold.
    In the former case, we add $B$ to $\hat \bset$, and in the latter case, we add $B'$ to $\hat \bset$.
    Observe that in either case, the block that was added to $\hat \bset$ contains at most $16Z \log^2{2\eta}$ elements of $S'$, since the block $B$ is not a bad block.
    We then delete $B$ from $\bset^{**}$, and, if block $B'$ lies in $\bset^{**}$, then we delete it from $\bset^{**}$ as well.
    The algorithm terminates once $\bset^{**} = \emptyset$.
    It is immediate to verify that, at the end of the algorithm, the number of blocks in set $\hat \bset$ is at least half the number of blocks that originally lied in $\bset^{**}$.
    Moreover, every block of $\hat \bset$ contains at least $Z$ and at most $16Z \log^2{2\eta}$ elements of $S'$.
    Let $\bset' \subseteq \bset_{\eta}(S)$ be the set of all blocks $B_i\in \bset_{\eta}$, for which $Z \leq |B \cap S'| \leq 16Z \log^2{2\eta}$ holds.
    Since $\hat \bset \subseteq \bset'$,
    \[ \sum_{B_i\in \bset'}|B_i \cap S'|\geq |\hat \bset'|\cdot Z \geq \frac{|S'|}{16\log^2{2\eta}}. \]

    We now let $S'' := \bigcup_{B \in \bset'} B \cap S'$ and the lemma follows.
\endproofof

\subsubsection{\proofof{Claim \ref{clm: ncm-yset-3-all-or-none}}} \label{prf-clm: ncm-yset-3-all-or-none}
    For the sake of readability, we let $\alpha := \log^2{(2\psi_{i+1})}$.
    Let $\bset^{(2)}_i \subseteq \bset_i$ be the set of all stream-blocks that contribute yes-pairs to $\yset^{(2)}_{i+1}$.
    We say that a block $B \in \bset^{(2)}_i$ is a \emph{good block} iff it contributes at least $2^j/(32\alpha)$ pairs to $\yset^{(2)}_{i+1}$.
    Let $\bset^*_i \subseteq \bset^{(2)}_{i}$ be the set of all such good blocks.

    \begin{claim}\label{clm: ncm-many-good-stream-blocks}
        $|\bset^*_i| \geq \frac{|\ysup{2}_{i+1}|}{2^{j+7} \alpha}$.
    \end{claim}
    \begin{proof}
        Since $\yset^{(2)}_{i+1} \subseteq \yset^{(1)}_{i+1}$, if a stream-block $B \in \bset_i$ contributes pairs to $\yset^{(2)}_{i+1}$, it also contributes pairs to $\yset^{(1)}_{i+1}$.
        But from our construction, if a stream-block contributes pairs to $\yset^{(1)}_{i+1}$, it contributes at least $2^j$ of them.
        Thus, $|\yset^{(1)}_{i+1}| \geq |\bset^{(2)}_{i}|2^j$, or in other words,
        
        \begin{equation} \label{eqn: ncm-partition-appn-bset-2-i-upper-bound}
            |\bset^{(2)}_i| \leq \frac{|\ysup{1}_{i+1}|}{2^j} \leq \frac{16 \alpha |\ysup{2}_{i+1}|}{2^j}
        \end{equation}
        
        where the last inequality follows from \Cref{eqn: ncm-yset-2-to-yset-1}.
        Next we claim that $|\bset^*_i| \geq |\bset^{(2)}_{i}|/(64 \alpha)$.
        Indeed, assume otherwise for contradiction.
        Since each stream-block in $\bset_i$ contributes less than $2^{j+1}$ pairs to $\yset^{(1)}_{i+1}$, which in turn, is a superset of $\ysup{2}_{i+1}$, 
        the total contribution to pairs of $\ysup{2}_{i+1}$ due to good stream-blocks is at most $|\bset^*_i| \cdot 2^{j+1} < \frac{|\bset^{(2)}_i|}{64 \alpha} \cdot 2^{j+1} = \frac{|\bset^{(2)}_i| 2^j}{32 \alpha}$.
        On the other hand, each stream-block in $\bset^{(2)}_i \backslash \bset^*_i$ contributes less than $2^j/(32 \alpha)$ pairs to $\yset^{(2)}_{i+1}$.
        Thus, 
        \[|\yset^{(2)}_{i+1}| < \frac{|\bset^{(2)}_i|2^j}{32 \alpha} +  \frac{|\bset^{(2)}_i| 2^j}{32 \alpha} = \frac{|\bset^{(2)}_{i}|2^{j}}{16 \alpha} \leq |\yset^{(2)}_{i+1}|,\]

        a contradiction.
        Here, the last inequality follows from \Cref{eqn: ncm-partition-appn-bset-2-i-upper-bound}.
        So far, we have shown that $|\bset^*_i| \geq |\bset^{(2)}_{i}|/(64 \alpha) = |\bset^{(2)}_i|/(2^6 \alpha)$.

        We now claim that $|\bset^{(2)}_{i}| > |\yset^{(2)}_{i+1}|/2^{j+1}$.
        Indeed, consider some stream-block $B \in \bset^{(2)}_i$.
        Recall that $B$ contributes less than $2^{j+1}$ pairs to $\yset^{(1)}_{i+1}$, which in turn, is a superset of $\yset^{(2)}_{i+1}$.
        Thus, $|\yset^{(2)}_{i+1}| < |\bset^{(2)}_{i}| 2^{j+1}$, or in other words, $|\bset^{(2)}_{i}| > |\yset^{(2)}_{i+1}|/2^{j+1}$.
        We can now bound,
        
        \[ |\bset^*_i| \geq \frac{|\bset^{(2)}_{i}|}{64 \alpha} > \frac{|\ysup{2}_{i+1}|}{2^{j+1} \cdot 64\alpha} = \frac{|\ysup{2}_{i+1}|}{2^{j+7} \alpha}.\]
        
        This completes the proof of \Cref{clm: ncm-many-good-stream-blocks}.
    \end{proof}

    Consider some good stream-block $B \in \bset^{*}_{i}$.
    We claim that there are at most two range-blocks of $\tilde \bset_i$ that contribute to the pairs of $\yset^{(2)}_{i+1}$ belonging to $B$.
    Indeed, assume otherwise for contradiction that there are $3$ such range-blocks $B', B'', B''' \in \tilde \bset_i$ in the increasing order of their respective range-elements.
    Recall that if a range-block contributes pairs to $\yset^{(2)}_{i+1}$, it contributes at least $2^{j+1}$ of them.
    Moreover, all the pairs contributed by $B''$ must have their respective stream-blocks contained in $B$.
    Thus, $(B, B'')$ must contribute at least $2^{j+1}$ pairs to $\yset^{(2)}_{i+1}$, a contradiction to the fact that $B$ contributes less than $2^{j+1}$ pairs to $\yset^{(1)}_{i+1}$, which in turn, is a superset of $\ysup{2}_{i+1}$.

    We let $\pset_i \subseteq \bset^*_i \times \tilde \bset_i$ be the set of exactly $|\bset^*_i|$ pairs, obtained by choosing for each good stream-block $B \in \bset^*_i$, a unique range-block $B' \in \tilde \bset_i$ such that $(B, B')$ contribute at least $\ceil{\frac{1}{2} \cdot \frac{2^j}{32 \alpha}} \geq \frac{2^{j-6}}{\alpha}$ pairs to $\yset^{(2)}_{i+1}$.
    Consider some range-block $B' \in \tilde \bset_i$.
    Since it contributes at most $2^{j+5} \alpha$ pairs to $\ysup{2}_{i+1}$, it may appear in at most $\frac{2^{j+5}\alpha}{{2^{j-6}}/{\alpha}} = 2^{11} \alpha^2$ pairs of $\pset_i$.
    It is now immediate to see that there is a subset $\pset'_i \subseteq \pset_i$ of at least $\frac{|\pset_i|}{2^{11} \alpha^2}$ pairs such that each range-block $B' \in \tilde \bset_i$ appears in at most one pair.
    We let $\yset^{(3)}_{i+1} \subseteq \yset^{(2)}_{i+1}$ be the set of pairs obtained by choosing exactly $\ceil{2^{j-6}/\alpha}$ pairs belonging to each pair $(B, B') \in \pset'_i$.
    From our construction and \Cref{clm: ncm-many-good-stream-blocks},

    \begin{align*}
        |\yset^{(3)}_{i+1}| = |\pset'_i| \cdot \ceil{\frac{2^{j-6}}{\alpha}} &\geq |\pset_i| \cdot \frac{2^{j-17}}{\alpha^3}\\
        &= |\bset^*_i| \cdot \frac{2^{j-17}}{\alpha^3}\\
        & \geq \frac{|\yset^{(2)}_{i+1}|}{2^{j+7} \alpha} \cdot \frac{2^{j-17}}{\alpha^3} \\
        &= \frac{|\yset^{(2)}_{i+1}|}{2^{24} \alpha^4}.
    \end{align*}

    This completes the proof of \Cref{clm: ncm-yset-3-all-or-none}.
\endproofof

\subsubsection{\proofof{Claim \ref{clm: ncm-sigma-i}}} \label{prf-clm: ncm-sigma-i}
    Consider some pair $(\hat B, \hat B') \in \yset^{(3)}_{i+1}$ and its parent-pair $(B, B') \in \bset_i \times \tilde \bset_i$.
    Since ${(\hat B, \hat B') \in \yset_{i+1}}$, we have $|B \cap B'| \geq |\hat B \cap \hat B'| \geq Z_{i+1} \Delta_{i+1}$.
    We now partition pairs of $\yset^{(3)}_{i+1}$ into a number of classes, where a pair $(\hat B, \hat B')$ belongs to the class $\cset'_{j'}$ iff it is $\left(2^{j'-1} Z_{i+1} \Delta_{i+1}\right)$-friendly.
    For each positive integer $j'$, we let $x_{j'} := |\cset'_{j'}|$, the number of pairs assigned to the class $\cset'_{j'}$.
    Since $\sum_{j'} x_{j'} = |\yset^{(3)}_{i+1}|$, from \Cref{fact: ncm-pigeonhole}, there is an index $j''$ such that $x_{j''} \geq |\yset^{(3)}_{i+1}|/2(j'')^2$.
    We fix such an integer $j''$ and the corresponding class $\cset'_{j''}$ containing at least $\frac{|\yset^{3}_{i+1}|}{2(j'')^2}$ pairs of $\yset^{(3)}_i$.
    We choose $\sigma_i := 2^{j''-1} Z_{i+1} \Delta_{i+1}$ and let $\yset^{(4)}_{i+1}$ be the set of all pairs of $\yset^{(3)}_i$ belonging to the class $\cset'_{j''}$.
    From our choice of these pairs, each pair in $\ysup{4}_{i+1}$ is $\sigma_i$-friendly.
    Furthermore, from our choice of $\sigma_i$ such that $j'' = \log{\left(\frac{2 \sigma_i}{Z_{i+1} \Delta_{i+1}} \right)}$, we have,
    
    \[ |\ysup{4}_{i+1}| \geq \frac{|\yset^{3}_{i+1}|}{2(j'')^2} = \frac{\ysup{3}_{i+1}}{2\log^2{\left(\frac{2\sigma_i}{Z_{i+1} \Delta_{i+1}} \right)}}. \]

    The claim now follows.
\endproofof

\subsubsection{\proofof{\ref{clm: ncm-sigma-prime-heavy}}} \label{prf-clm: ncm-sigma-prime-heavy}
    Consider some pair $(\hat B, \hat B') \in \yset^{(4)}_{i+1}$ and let $B'$ be the parent-block of $\hat B'$ in $\tilde \bset_i$.
    Since ${(\hat B, \hat B') \in \yset_{i+1}}$, we have $|\hat B \cap B'| \geq |\hat B \cap \hat B'| \geq Z_{i+1} \Delta_{i+1}$.
    On the other hand, since $(\hat B, \hat B') \in \ysup{4}_{i+1}$ and $\hat B$ is contained in $B$, we have $|\hat B \cap B'| \leq |B \cap B'| < 2\sigma_i$.
    We now partition pairs of $\yset^{(4)}_{i+1}$ into $\log{ \left( \frac{2\sigma_i}{Z_{i+1} \Delta_{i+1}} \right)}$ classes, where a pair $(\hat B, \hat B')$ belongs to class $\cset'_{j'}$ iff ${2^{j'} Z_{i+1} \Delta_{i+1} \leq |\hat B \cap B'| < 2^{j'+1} Z_{i+1} \Delta_{i+1}}$.
    Note that if a pair $(\hat B, \hat B')$ belongs to class $\cset'_{j'}$, it is $(2^{j'} Z_{i+1} \Delta_{i+1})$-heavy.
    From the pigeonhole principle, there is a class $\cset'_{j''}$ containing at least $\frac{|\yset^{4}_{i+1}|}{\log{ \left( \frac{2\sigma_i}{Z_{i+1} \Delta_{i+1}} \right)}}$ pairs of $\yset^{(4)}_{i+1}$.
    We let $\yset^{(5)}_{i+1}$ be the set of such pairs.
    Note that all the pairs of $\yset^{(5)}_{i+1}$ are $\sigma'_i$-heavy, where $\sigma'_i = 2^{j''} Z_{i+1} \Delta_{i+1}$.
    This completes the proof of \Cref{{clm: ncm-sigma-prime-heavy}}.
\endproofof

\subsubsection{\proofof{\Cref{obs: ncm-level-j-to-i-count}}} \label{prf-obs: ncm-level-j-to-i-count}
    We first consider the case where $j = i+1$.
    Let $(\hat B, \hat B')$ be a level-$j$ descendant yes-pair of $(B, B')$.
    Since $S^*$ is an $\Upsilon$-canonical subsequence, $|\hat B \cap B'| < 2 Z_j \mu_i$.
    Thus, there are no level-$j$ descendant yes-pairs $(\hat B, \hat B')$ of $(B, B')$ with $|\hat B \cap B'| > 8 Z_j \mu_i$ and the observation follows.

    We now consider the case where $j > i+1$.
    Assume for contradiction that there are at least $\frac{Z_i}{4Z_j}$ such level-$j$ descendant yes-pairs $(\hat B, \hat B')$ of $(B, B')$ with $|\hat B \cap B'| > 8 Z_j \mu_i$.
    We let $\pset_j$ be the set of such yes-pairs.
    We consider the intermediate level $j' = i+1$.
    By pigeonhole principle, there is some level-$j'$ yes-pair $(\tilde B, \tilde B')$ that has at least $\frac{{Z_i}/\left({4 Z_j}\right)}{{Z_i}/{Z_{j'}}} = \frac{Z_{i+1}}{4Z_j}$ pairs of $\pset_j$ as its descendants.
    Since the stream-blocks of $\pset_j$ are disjoint, $|\tilde B \cap B'| > \frac{Z_{i+1}}{4 Z_j} \cdot 8 Z_j \mu_i = 2 Z_{i+1} \mu_i$, a contradiction to the fact that $(\tilde B, \tilde B')$ is a level-$(i+1)$ yes-pair.
\endproofof

\subsubsection{\proofof{\Cref{obs: ncm-many-elements-before-median}}} \label{prf-obs: ncm-many-elements-before-median}
    Notice that the observation is trivial when $Z_i = 1$.
    Hence, assume from now on that $Z_i > 1$ and let $j > i$ be the lowest level with $Z_j < Z_i$.
    Recall that each level-$j$ yes-block contributes exactly $Z_j$ elements to $S^*$.
    Thus, $\frac{Z_i}{2Z_j}$ such level-$j$ yes-blocks must appear before the $\left({\frac{Z_i}{2}} + 1 \right)^{th}$ element of $S^*$ in $B$.
    Moreover, from \Cref{obs: ncm-level-j-to-i-count} each such yes-block shares at least $Z_j \mu_i$ elements with $B'$.
    Thus, there are at least $Z_j \mu_i \cdot \frac{Z_i}{2 Z_j} = \frac{Z_i \mu_i}{2}$ elements of $B \cap B'$ appearing before the $\left({\frac{Z_i}{2}} + 1 \right)^{th}$ element of $S^*$ in $B$.
\endproofof

}{}

\iftoggle{ncm-algo-to-lis}{
    \section{Proofs Omitted from Section \ref{sec: ncm-ncm-hybrid-to-lis}} \label{appn: ncm-ncm-hybrid-to-lis}

\subsection{Proof Omitted from Section \ref{subsec: ncm-small-xi-by-zi}} \label{appn: ncm-small-xi-by-zi}

\proofof{\Cref{clm: ncm-block-processing-alg}}
    We assume that the block size $|B|$ is known to us in advance.
    We first analyze the trivial case when $Z \leq 100 \log N$.
    In this case, we store all the elements of $B$ and naively compute the desired collection $\bset^*$ of the range-blocks offline after $B$ terminates.
    Note that we can perform the offline computation using only $O(|B|) = \tilde O(|B|/Z)$ additional space and there is nothing to show.
    If $|B| < Z$, we report the empty collection $\bset^* = \emptyset$.
    Thus, assume from now on that $|B| \geq Z > 100 \log N$.

    We will process elements of $B$ as they arrive while maintaining a collection $\bset''$ of the indices of the range-blocks that we call \emph{suspicious blocks}.
    Initially, $\bset'' = \emptyset$.
    Whenever a new element $a_t$ of the stream $S$ arrives, we ignore it with probability $1 - \frac{10 \log N}{Z}$ and select it with the remaining probability.
    If $a_i$ is selected, we consider the range-block $B'_{j} \in \bset'$ in which the value of $a_t$ lies.
    If $B'_{j}$ is not already marked suspicious, we mark it thus and add its index $j$ to $\bset''$.
    Whenever an range-block $B'_j$ is marked suspicious (by adding the index $j$ to the set $\bset''$), we initialize a new algorithm $\algcheckblock(B'_j)$, that proceeds as follows. 
    Let $a_t \in B \cap B'_j$ be the element that caused the range-block $B'_j$ to marked suspicious.
    We run the algorithm $\alg_2$ from \Cref{obs: ncm-det-n-by-opt-space-sound} on the subsequent elements of $B \cap B'_j$ with parameter $Z$.
    At the end of processing the last element of $B$, if $\alg_2$ determines returns yes, we add the index $j$ to $\bset^*$.
    This completes the description of $\algcheckblock(B')$.
    If at any point in the execution $|\bset''|$ exceeds $\frac{20 |B|\log N}{Z}$, we immediately terminate our block-processing algorithm and report $\bset^* = \emptyset$.
    Otherwise, at the end of stream, we report $\bset^*$.
    This completes the description of our block-processing algorithm.
    We now analyze its properties, starting with its space complexity.

    Consider the execution of $\alg_2$ from \Cref{obs: ncm-det-n-by-opt-space-sound} on a subblock $\hat B$ of the stream-block $B$ along with the parameter $Z$.
    The space complexity of this run is at most ${\tilde O \left( \ceil{\frac{|\hat B|}{Z}} \right) = \ceil{\frac{|\hat B|}{Z}} \cdot \log^{O(1)} N}$.
    It is now immediate to see that the space complexity of our block-processing algorithm is,

    \begin{align*}
        O(|\bset''|) + \sum_{B' \in \bset''} \left( \ceil{\frac{|B \cap B'|}{Z}}  \cdot \log^{O(1)} N \right) &\leq \left( |\bset''| \cdot \log^{O(1)} N \right) + \sum_{B' \in \bset''} \left( {\frac{|B \cap B'|}{Z}}  \cdot \log^{O(1)} N \right)\\
        &\leq \left( |\bset''| \cdot \log^{O(1)} N \right) + \left( {\frac{|B|}{Z}}  \cdot \log^{O(1)} N \right)\\
        &\leq {\frac{|B|}{Z}}  \cdot \log^{O(1)} N.
    \end{align*}

    Here, the first inequality follows since for each $B' \in \bset''$, $\ceil{\frac{|B \cap B'|}{Z}} \leq 1 + \frac{|B \cap B'|}{Z}$ and the last inequality follows since we ensure that $|\bset''| \leq O \left( \frac{|B| \log N}{Z} \right)$ throughout our algorithm.

    Consider a range-block $B' \in \bset'$ with $\optlis(B \cap B') < Z/2$.
    From the correctness guarantee of \Cref{obs: ncm-det-n-by-opt-space-sound}, $\prob{B' \in \bset^*} \leq 1/N^2$, where the probability is over the randomness used by $\alg_2$.
    Since there are only $|\bset'| \leq N$ such range-blocks, with probability $1 - 1/N$, \emph{every} range-block $B' \in \bset^*$ has $\optlis(B \cap B') \geq Z/2$.
    
    We now show that each range-block $B' \in \bset'$ that is $(Z,1)$-perfect for $B$ is present in $\bset''$ with probability at least $3/4$.
    We denote by $\eset^*_\bad$ the event that $|\bset''| > \frac{20 |B|\log N}{Z}$ at some point in our algorithm.
    For each range-block $B' \in \bset'$ that is $(Z,1)$-perfect for $B$, we denote by $\eset_\bad(B')$ be the event that its index is not added to $\bset''$ before processing the $(Z+1)^{th}$ element of $B \cap B'$.
    It now suffices to show that
    (i) $\prob{\eset^*_\bad} \leq 1/N$; and
    (ii) for each range-block $B' \in \bset'$ that is $(Z,1)$-perfect for $B$, $\prob{\eset_\bad(B')} \leq 0.1$.

    To bound $\prob{\eset^*_\bad}$, it suffices to bound the probability of sampling more than $\frac{20 |B|\log N}{Z}$ elements.
    Since we sample each element independently at random with probability $\frac{10 \log N}{Z}$, the expected number of sampled elements is $\frac{10 |B| \log N}{Z}$.
    From Chernoff bound (see, \Cref{fact: ncm-chernoff-version}), the probability that we sample more than $\frac{20 |B|\log N}{Z}$ elements is at most $e^{-\frac{20 |B| \log N}{6Z}} \leq e^{-3 \log N} \leq \frac{1}{N^3}$.
    We assume from now on that the event $\eset^*_\bad$ does not occur.
    Consider now a range-block $B' \in \bset'$ that is $(Z,1)$-perfect for $B$. 
    We have, 
    \[ \prob{\eset_\bad(B') \> | \> \neg\>\eset^*_\bad} \leq \left( 1 - \frac{10 \log N}{Z} \right)^{Z} \leq \frac{1}{N}, \]

    and we conclude, $\prob{\eset_\bad(B')} \leq \prob{\eset^*_\bad} + \prob{\eset_\bad(B') \> | \> \neg\>\eset^*_\bad} \leq 2/N < 0.1$ as claimed.
    This completes the proof of correctness of our algorithm and \Cref{clm: ncm-block-processing-alg} now follows.

\subsection{Proofs Omitted from Section \ref{subsec: ncm-large-deltai-by-zi}} \label{appn: ncm-large-deltai-by-zi}

\subsubsection{\proofof{\Cref{clm: ncm-level-with-strictly-small-z}}} \label{prf: ncm-level-with-strictly-small-z}
    We start by claiming that there is a level $j \geq i - 3\eps r$ such that $Z_j > Z_i$.
    Indeed, assume for contradiction that the claim is false.
    Then for the level $k = \ceil{i - 3\eps r} = i - \floor{3\eps r} \geq 0$, $Z_k = Z_i$.
    Here, the inequality follows since $i \geq 3\eps r$.
    But then

    \[ \frac{X_i}{Z_i} = \frac{X_k / \eta^{i-k}}{Z_k}  = \frac{X_k/Z_k}{\eta^{\floor{3\eps r}}} \leq \frac{N^{1/2 + \eps + o(1)}}{\eta^{\floor{3\eps r}}} < N^{1/2 - \eps}, \]
    
    where the first inequality follows from \Cref{obs: ncm-decreasing-xi-by-zi}.
    The last inequality follows since $\eta^r~\geq~N/\eta$ and $\eps$ is a small enough constant.
    But this is a contradiction to \Cref{assm: large-xi-by-zi} that $\frac{X_i}{Z_i} > N^{1/2 - \eps}$.
    We now fix the largest level $i - 3\eps r \leq j < i$ such that $Z_j > Z_i$.
    Notice that $X_j = \eta X_{j+1}$ and from our choice of level $j$, we have $Z_{j+1} = Z_i$.
    From \Cref{obs: ncm-decreasing-xi-by-zi}, we now conclude that $\frac{X_{j}}{Z_j} \geq \frac{X_{j+1}}{Z_{j+1}}$, or in other words, $Z_j \leq \eta Z_i$.
    \Cref{clm: ncm-level-with-strictly-small-z} now follows by noting that both $Z_i$ and $Z_j$ are integral powers of $2$.
\endproofof

\subsubsection{\proofof{\Cref{clm: ncm-double-level-block-processing}}} \label{prf: ncm-double-level-block-processing}
We are given a level-$j$ stream-block $B$ and a partition $\bset' = \bset^j_{\Psi'}(H^*)$ of the range $H^*$ into range-blocks.
Our goal is to achieve $(Z_j/2, \Delta_i/8, 1)$-approximation in space $N^{1/2 - \eps + o(1)}$.
We proceed exactly as in the proof of \Cref{clm: ncm-block-processing-alg}.

We will maintain a collection $\bset'' \subseteq \bset'$ of range-blocks that we call \emph{suspicious blocks}.
Initially, $\bset'' = \emptyset$.
We are given access to the elements of the stream-block $B$ as they arrive as a part of the original input stream $S$.
Whenever a new element $a_k$ of $B$ arrives, we ignore it with probability $1 - \frac{2^5}{Z_i \Delta_i}$ and select it with remaining probability.
If an element $a_k$ is selected, we consider the range-block $B' \in \bset'$ in which the value of $a_k$ lies.
If $B'$ is not already marked suspicious, we mark it thus and add it to $\bset''$.
We also run the algorithm $\alg_1$ from \Cref{obs: ncm-det-opt-space-sound} on the subsequent elements of $B \cap B'$ with parameter $Z = Z_j/2$.
If at any point in the processing, $|\bset''|$ exceeds $2^{10} \frac{X_j}{Z_i \Delta_i}$, we immediately stop and report $\bset^* = \emptyset$.
At the end of the processing last element of $B$, we are ready to report our output.
We let $\bset^* \subseteq \bset''$ be the set of all range-blocks for which the respective execution of $\alg_1$ reported yes.
We report the indices of all range-blocks of $\bset^*$.
This completes the description of our algorithm.
We now analyze its properties.

\paragraph{Space complexity.}
Consider an element $a_t \in B$ and the set $\bset''$ of suspicious blocks just after processing $a_t$.
Since we ensure that $|\bset''| \leq 2^{10} \frac{X_j}{Z_i \Delta_i}$ throughout the execution of our algorithm, the space used by our algorithm at the end of processing the element $a_t$ is at most,

\begin{align*}
    O \left( |\bset''| \right) + \sum_{B' \in \bset''} O (Z_j) = O \left(|\bset''| \cdot Z_j \right) &\leq O \left( \frac{X_j}{Z_i \Delta_i} \cdot Z_j \right)\\
    &\leq O \left( \frac{X_j Z_j}{Z_i^2 N^{5\eps}} \right) \\
    &\leq O \left(\frac{\eta^{3\eps r+ 1}}{N^{5\eps}} \cdot \frac{X_i}{Z_i} \right) \leq \frac{N^{o(1)}}{N^{2\eps}} \cdot \frac{X_i}{Z_i}\\
    &\leq N^{1/2 - \eps + o(1)}
\end{align*}

The second inequality follows from the assumption in this special case,  that,  $\Delta_i \geq Z_i N^{5\eps}$.
The third inequality follows from \Cref{clm: ncm-level-with-strictly-small-z} and the fourth from our choices $r = \floor{\frac{\log N}{\log \eta}}$ and $\eta = O(\log \log N)$.
Finally, the last inequality follows from \Cref{obs: ncm-decreasing-xi-by-zi} since ${\frac{X_i}{Z_i} \leq N^{1/2 - \eps}}$.
Thus, the space used by our algorithm throughout its execution is indeed bounded by ${N^{1/2 - \eps + o(1)}}$.

\paragraph{Soundness guarantee.}
We will show that each block $B' \in \bset^*$ has $\optlis(B \cap B') \geq Z_j/2$.
Indeed, a range-block $B'$ is present in $\bset^*$ iff the respective execution of algorithm $\alg_1$ from \Cref{obs: ncm-det-n-by-opt-space-sound} with parameter $Z = Z_j/2$ reported yes.
But $\alg_1$ is a deterministic algorithm and if it reports yes, $\optlis(B \cap B') \geq Z_j/2$ must hold. 

\paragraph{Correctness guarantee.}
Our goal is to show that each $(Z_j/2, \Delta_i/8)$-perfect range-block $B'$ of $\bset'$ is present in $\bset^*$ with probability at least $3/4$.
Consider a range-block $B' \in \bset'$ that is $(Z_j/2, \Delta_i/8)$-perfect for $B$.
Since $B'$ is $\left(\frac{Z_j}{2}, \frac{\Delta_i}{8} \right)$-perfect for $B$, there is an increasing subsequence of $B \cap B'$ with size $\frac{Z_j}{2} \geq Z_i$ that does not use the first $\frac{Z_j}{2} \cdot \frac{\Delta_i}{8} = \frac{Z_j \Delta_i}{16}$ elements of $B \cap B'$.
We denote by the event $\eset^*_{\bad}$ the event that we do not sample at least one element out of the first $Z_j \Delta_i/16$ elements of $B \cap B'$.
We denote by $\eset^{**}_\bad$ the event that we terminate our algorithm because $|\bset''|$ exceeded $2^{10} \frac{X_j}{Z_i \Delta_i}$ during the execution of our algorithm.
From the correctness guarantee of \Cref{obs: ncm-det-n-by-opt-space-sound}, if the events $\eset^*_{\bad}$ and $\eset^{**}_{\bad}$ does not occur, the index of $B'$ is present in $\bset^*$.
We can now bound,

\begin{align*}
    \prob{\eset^*_\bad} = \left(1 - \frac{2^5}{Z_i \Delta_i} \right)^{\frac{Z_j \Delta_i}{2^4}} &\leq e^{-\frac{Z_j}{Z_i}} \leq \frac{1}{8}.
\end{align*}

Here, the last inequality follows from \Cref{prop: ncm-p4} of our ensemble $\Upsilon$.
We now upper bound the probability of the event $\eset^{**}_\bad$.
Recall that $|B| = X_j$ and we sample each element independently at random with probability $\frac{2^5}{Z_i \Delta_i}$.
Moreover, for the event $\eset^{**}_\bad$ to occur, a necessary condition is that we sample at least $2^{10} \frac{X_j}{Z_i \Delta_i}$ elements of $B$.
From Chernoff bound (see, \Cref{fact: ncm-chernoff-version}), this probability is at most $1/8$.
We now conclude that  $\prob{\eset^{**}_{\bad} \cup \eset^{**}_{\bad}} \leq 1/4$ and the soundness guarantee follows.
This completes the proof of correctness of our block processing algorithm and \Cref{clm: ncm-double-level-block-processing} now follows.
\endproofof

\subsection{Proof Omitted from Section \ref{subsec: ncm-simpler-lis-stream}} \label{appn: ncm-simpler-lis-stream}

\subsubsection{\proofof{\Cref{clm: ncm-simpler-algo-k-star}}} \label{prf: ncm-simpler-algo-k-star}
    Consider some integer $0 \leq k < \knew$ and the corresponding pair of levels $i_k$ and $i_{k+1}$.
    From our choice of the set $\isetnew$ of levels, we have $Z_{i'} = Z_{i_{k+1}}$ for all levels $i_k < i' \leq i_{k+1}$.
    It is also easy to see that $Z_{i'} = Z_{i_0}$ for all levels $0 \leq i' \leq i_0$ and $Z_{i''} = Z_{r^*}$ for all levels $i_{\knew} < i'' \leq r^*$.
    We also claim that $\eta^{r^*} \leq \frac{N^{1/2 - \rstarconstant \eps}}{\eta^2}$.
    Indeed, since $r^* = \rstarVal$ and $r = r(N, \eta) = \floor{\frac{\log N}{\log \eta}}$, we obtain,

    \[\eta^{r^*} = \eta^{\rstarVal} > \frac{\eta^{\left(1/2 - \rstarconstant \eps \right) r}}{\eta} > \frac{\eta^{\left(1/2 - \rstarconstant \eps \right) \cdot \left( \frac{\log N}{\log \eta} \right)}}{\eta^2} = \frac{N^{1/2 - \rstarconstant \eps}}{\eta^2}.\]

    We now show that $Z_{r^*}$ is small and $Z_0$ is large.
    Combined with the above two claims, we will be able to argue that the number of special levels $|\iset^*| = 1 + k^*$ is large enough.
    We start by showing that $Z_{r^*}$ is small.
    From \Cref{assm: large-xi-by-zi}, for each level $0 \leq i < (1/2 - \eps)r$, we have $\frac{X_i}{Z_i} > N^{1/2 - \eps}$.
    In particular, for the level $r^*$, we have $\frac{X_{r^*}}{Z_{r^*}} > N^{1/2 - \eps}$.
    In other words, 

    \begin{align*}
        Z_{r^*} < \frac{X_{r^*}}{N^{1/2 - \eps}} = \frac{N/\eta^{r^*}}{N^{1/2 - \eps}} &=  \frac{N^{1/2 + \eps}}{\eta^{r^*}} < \frac{N^{1/2 + \eps}}{N^{1/2 - 4\eps}/\eta^2} =  \eta^2 N^{5\eps} < N^{6\eps}.
    \end{align*}

    Here, last inequality follows since $\eta = O(\log \log N)$.
    Next, we show that $Z_0$ is large.
    From \Cref{obs: ncm-decreasing-xi-by-zi}, $Z_0 \geq N^{1/2 - \eps}$ and for each level $0 \leq i < r$, we have $Z_{i+1} \leq  Z_{i} \leq \eta Z_{i+1}$.
    It is now immediate to verify that,

    \begin{equation*}
        \begin{split}
            k^* \geq \log_\eta{\left( \frac{Z_0}{Z_{r^*}} \right)} &\geq \log_\eta{\left( \frac{N^{1/2 - \eps}}{N^{6\eps}} \right)} = \log_{\eta}{\left(N^{1/2 - 7 \eps} \right)} \geq \left(\frac{1}{2} - 7 \eps\right)r.
        \end{split}
    \end{equation*}

    Since $\eps \leq 1/100$, from the above inequality, we immediately obtain that $\knew \geq r/3$.
\endproofof

\subsubsection{\proofof{\Cref{clm: ncm-lis-somple-load-i-k-bound}}} \label{prf: ncm-lis-somple-load-i-k-bound}
    We proceed by induction.
    The base case is when $k = 0$ and the claim is trivial since $\load(0) = 1$.
    Consider now some $0 < k \leq k^*$ and the corresponding level $i_k \in \iset^*$.
    We assume that the induction hypothesis holds for the level-$i_{k-1}$, or in other words,

    \[ \load(i_{k-1}) \leq 2^{10(k-1)}\frac{\mu_{i_0}}{\mu_{i_{k-1}}} \cdot \zeta(i_0, i_1) \cdot \ldots \cdot \zeta(i_{k-2}, i_{k-1}).\]

    It is immediate to verify that each execution to $\alglevel{i_k}$ in which $B$ participates is called by a run of the level-$i_{k-1}$ algorithm $\alglevel{i_{k-1}}$ in which $B(i_{k-1})$ participates.    
    By our induction hypothesis, there are at most $\load(i_{k-1})$ concurrent calls to $\alglevel{i_{k-1}}$ in which $B^*$ participates. 
    Furthermore, from \Cref{lem: ncm-simple-lis-main}, each such execution of $\alglevel{i_{k-1}}$ performs at most $2^{10} \zeta(i_{k-1}, i_k) \frac{\mu_{i_{k-1}}}{\mu_{i_k}}$ concurrent calls to $\alglevel{i_k}$.
    Thus, the maximum number of concurrent calls to the level-$i$ algorithm $\alglevel{i}$ in which $B$ participates is,

    \begin{align*}
        \load(B) &\leq \load(B(i_{k-1})) \cdot 2^{10} \zeta(i_{k-1}, i_k) \frac{\mu_{i_{k-1}}}{\mu_{i_k}} \\
        &\leq 2^{10(k-1)} \frac{\mu_{i_0}}{\mu_{i_{k-1}}} \cdot 2^{10} \frac{\mu_{i_{k-1}}}{\mu_{i_k}} \cdot \zeta(i_0, i_1) \cdot \ldots \cdot \zeta(i_{k-1}, i_k)\\
        &= 2^{10k}\frac{\mu_{i_0}}{\mu_{i_k}} \cdot \zeta(i_0, i_1) \cdot \ldots \cdot \zeta(i_{k-1}, i_k).
    \end{align*}
\endproofof

\subsubsection{\proofof{\Cref{clm: ncm-lis-simple-pi-zeta-bound}}} \label{prf: ncm-lis-simple-pi-zeta-bound}
    For convenience, for each $0 \leq k < k^*$, we let $x_k := e^{\zeta(i_k, i_{k+1})} = \eta^{2 \left(i_{k+1} - i_k \right)} \cdot \psi_{i_{k+1}}$.
    Thus, for each $0 \leq k < k^*$ we have $\zeta(i_k, i_{k+1}) = \ln{x_k}$.
    From \Cref{fact: ncm-log-concave}, we can now bound,

    \begin{align*}
        \left( \zeta(i_0, i_1) \cdot \ldots \cdot \zeta(i_{k^*-1}, i_{k^*}) \right)^{1/k^*} &= \left( \Pi_{0 \leq k < k^*} \ln{x_k} \right)^{1/k^*} \\
        &\leq \ln{ \left( \left(\Pi_{0 \leq k < k^*} x_k \right)^{1/k^*} \right)} \\
        &\leq \ln{ \left( \eta^{\frac{2\left( i_{k^*} - i_0 \right)}{k^*}} \cdot N^{\frac{1}{k^*}} \right)} \\
        &\leq \ln{ \left( \eta^{\frac{2r}{k^*}} \cdot N^{\frac{1}{k^*}} \right)} \leq \ln{ \left( N^{\frac{3}{k^*}} \right) }\\
        &\leq \frac{3}{k^*} \log N \leq 9 \log \eta.
    \end{align*}

    Here, the second inequality holds since $\Pi_{1 \leq i < r} \Psi_i \leq N$ and the last inequality follows from \Cref{clm: ncm-simpler-algo-k-star} and fact that $r = \floor{\frac{\log N}{\log \eta}}$.
    We now conclude that,
    \begin{align*}
        \zeta(i_0, i_1) \cdot \ldots \cdot \zeta(i_{k^*-1}, i_{k^*}) \leq \left( 9 \log \eta \right)^{k^*} &\leq \left(9 \log \eta \right)^r \leq N^{\frac{\log{(9 \log \eta)}}{\log \eta}} \leq N^{o(1)}.
    \end{align*}
\endproofof

\subsubsection{\proofof{\Cref{obs: ncm-alglevel-0-suffices}}} \label{prf: ncm-alglevel-0-suffices}
    Since $Z_0 = \ldots = Z_{i_0}$, it is immediate to verify that $\bset^{i_0}_{\Psi'}(H^*) = \set{H^*}$, or in other words, there is a unique level-$i_0$ range-block $H^*$.
    For each level-$i_0$ stream-block $B \in \bset^{i_0}_\Psi(S)$, we run $2^6 \left(\zeta(0, i_0) + 1 \right)$ parallel executions of the algorithm $\alglevel{i_0}$ with input level-$i_0$ pair $(B, H^*)$ and the subblock $B$ of $B$.
    We report yes iff at least $2^5 \left(\zeta(0, i_0) + 1 \right)$ of these executions report yes.
    This completes the description of $\algsimp$.
    It is immediate to verify that the space complexity of $\algsimp$, excluding the space required by the calls to $\alglevel{i_{0}}$, is at most $O \left( 1 + \zeta(0, i_0) \right) = N^{o(1)}$.
    We now analyze its correctness.

    \paragraph*{Completeness.}
    Assume that $S$ is a \yi.
    Then, there is the level-$i_0$ yes-block $B \in \bset^{i_0}_\Psi(S)$, for which each corresponding execution of $\alglevel{i_0}$ returns yes with probability at least $3/4$.
    From Chernoff bound, the probability that more than $2^5 \left(\zeta(0, i_0) + 1 \right)$ executions of $\alglevel{i_0}$ report no is at most $e^{2(\zeta(0, i_0) + 1)} \leq \frac{1}{e^2}$.
    We now conclude that we report yes with probability at least $1 - \frac{1}{e^2} \geq \frac{3}{4}$.

    \paragraph*{Soundness.}
    Assume that $\optlis(S) < Z_0/\alpha'_0$.
    Consider a level-$i_0$ stream-block $B \in \bset^{i_0}_\Psi(S)$.
    Since $\optlis(B) \leq \optlis(S) < Z_0/\alpha'_0$, each run of $\alglevel{i_0}$ in which $B$ participates returns no with probability at least $3/4$.
    The probability that at least $2^5 \left(\zeta(0, i_0) + 1 \right)$ executions of $\alglevel{i_0}$ report yes is at most $e^{-2 - \zeta(0, i_0)} \leq \frac{e^{-2}}{\eta^{i_0}}$.
    The soundness now follows by union bound over at most $\eta^{i_0}$ level-$i_0$ stream blocks.
\endproofof

\subsubsection{\proofof{\Cref{clm: ncm-lis-simple-alg-soundness}}} \label{prf: ncm-lis-simple-alg-soundness}
    Recall that for each element $e \in \sset$, we have a level-$j$ pair $(\hat B(e), \hat B'(e))$ with the following guarantees:
    (i) $(\hat B(e), \hat B'(e))$ is a level-$j$ descendant-pair of $(B^*, B^{*'})$;
    (ii) out of $\floor{32 \zeta(i,j)}$ calls to $\alglevel{j}$ in which the pair $(\hat B(e), \hat B'(e))$ participates, at least $16 \zeta(i,j)$ of them returned yes.

    \begin{claim}
        Consider an element $e \in \eset$ and the corresponding level-$j$ pair $(\hat B(e), \hat B'(e))$.
        Then with probability at least $1 - \frac{1}{10 |\bset| |\bset'|}$, $\optlis(\hat B(e) \cap \hat B'(e)) \geq \frac{Z_j}{\alpha'_j}$.
    \end{claim}
    \begin{proof}
        Recall that we perform $\floor{32 \zeta(i,j)}$ calls to $\alglevel{j}$,  in which the level-$j$ pair $(\hat B(e), \hat B'(e))$ participates.
        Consider one such execution of $\alglevel{j}$.
        From the correctness guarantee of $\alglevel{j}$, if it returns yes, with probability at least $3/4$, we have ${\optlis(\hat B(e) \cap \hat B'(e)) \geq \frac{Z_j}{\alpha'_j}}$.
        From Chernoff bound, the probability that at least $16 \zeta(i,j)$ such calls return yes even though ${\optlis(\hat B(e) \cap \hat B'(e)) < \frac{Z_j}{\alpha'_j}}$ is at most $e^{-\zeta(i,j)} \leq \frac{1}{\eta^2|\bset||\bset'|} \leq \frac{1}{10 |\bset| |\bset'|}$.
    \end{proof}

    Since there are only at most $|\bset| \cdot |\bset'|$ elements in $\sset$, with probability at least $0.9$, for \textbf{each} $e \in \sset$ the corresponding level-$j$ pair $(\hat B(e), \hat B'(e))$ has $\optlis(\hat B(e) \cap \hat B'(e)) \geq \frac{Z_j}{\alpha'_j}$.
    Assume from now on that this indeed holds.
    We are now ready to complete the proof of \Cref{clm: ncm-lis-simple-alg-soundness}.
    
    Consider an increasing subsequence $\sset' = (e_1, \ldots, e_\ell)$ of $\sset$ with length $\ell$.
    It is immediate to verify that the corresponding level-$j$ stream-blocks $\hat B(e_1), \ldots, \hat B(e_\ell)$ are unique and appear in this order.
    Similarly, the corresponding level-$j$ stream-blocks $\hat B(e_1), \ldots, \hat B(e_\ell)$ are unique and appear in this order.
    We now conclude that there is an increasing subsequence of $B^{*} \cap B^{*'}$ in which each such level-$j$ pair contributes at least $\frac{Z_j}{\alpha'_j}$ elements.
    Thus, $\optlis(B^* \cap B^{*'}) \geq \ell \cdot \frac{Z_j}{\alpha'_j}$ and the claim follows.
\endproofof

\subsubsection{\proofof{\Cref{clm: ncm-large-eset}}} \label{prf: ncm-large-eset}
    Let $\pset$ be the set of all level-$j$ descendant yes-pairs $(B, B')$ of $(B^*, B^{*'})$ such $B$ is a subblock of $B^{**}$.
    We first show that $|\pset|$ is large enough.
    We will then show that each such pair $(B, B') \in \pset$ is a happy pair independently with good probability, and then complete the proof of \Cref{clm: ncm-large-eset} with a simple application of Chernoff bound.

    \begin{claim} \label{clm: ncm-pset-z-i-z-j}
        $|\pset| \geq \frac{Z_i}{2 Z_j}$.
    \end{claim}
    \begin{proof}
        From our assumption, $|B^{**} \cap S^*| \geq \frac{Z_i}{2}$.
        Since each level-$j$ yes-block contributes exactly $Z_j$ elements to $S^*$, there are at least $\frac{Z_i}{2Z_j}$ level-$j$ yes-blocks that are subblocks of $B^{**}$.
    \end{proof}

    \begin{claim} \label{clm: ncm-b-b-prime-happy}
        Each pair $(B, B') \in \pset$ is a happy pair independently with probability at least $0.9$.
    \end{claim}
    \begin{proof}
        Fix some pair $(B, B') \in \pset$.
        Notice that we sample elements of $B \cap B'$ independently at random with probability $p_j = \frac{2^{6}}{Z_{j} \mu_{j}}$ each.
        Consider the level $j' = j+1$ of $\isetnew$ and a level-$j'$ descendant stream-block $\hat B$ of $B$.
        Since $S^*$ is an $\Upsilon$-canonical subsequence, if $\hat B$ is a yes-block, it must contribute at least $|\hat B \cap B'| \geq Z_{j'} \mu_{j'-1} = Z_{j'} \mu_j$ elements to $B'$.
        Moreover, a level-$j'$ stream-block contributes elements to $S^*$ only if it is a yes-block, and if it contributes elements, it contributes exactly $Z_{j'}$ of them.
        Thus, before processing the $\left(\frac{Z_j}{2} + 1 \right)^{th}$ element of $S^*$ in $B$, we must have encountered at least $\frac{Z_j}{2 Z_{j'}}$ level-$j'$ yes-blocks, which in turn, contain at least $\frac{Z_j}{2Z_{j'}} \cdot  Z_{j'} \mu_{j} = \frac{Z_j \mu_{j}}{2}$ elements of $B \cap B'$.
        We denote by $\eset_1$ the event that we sample some element of $B \cap B'$ before processing the $\left(\frac{Z_j}{2} + 1 \right)^{th}$ element of $S^*$ in $B$.
        From the above discussion,

        \[\prob{\eset_1} \geq 1 - \left(1 - \frac{2^{6}}{Z_{j} \mu_{j}}\right)^{\frac{1}{2} Z_{j} \mu_{j}} \geq 1 - e^{-2^4} \geq 0.99. \]

        We assume from now on that this event $\eset_1$ indeed occurs.
        Let $a_{t'}$ be the element such that we call $\floor{32 \zeta(i,j)}$ parallel instances of the level-$j$ algorithm $\alglevel{j}$ with input level-$j$ pair $(B, B')$ and the subblock $B_{>t'}$ of $B$.
        We denote by $\eset_2$ the event that at least $16\zeta(i,j)$ such executions of $\alglevel{j}$ return yes.
        Since the event $\eset_1$ does occur, $B_{>t'} \cap B'$ contains at least $Z_j/2$ elements of $S^*$.
        Furthermore, each such execution of $\alglevel{j}$ reports in affirmative with probability at least $3/4$.
        From Chernoff bound, it is immediate to verify that $\prob{\eset_2 \> | \> \eset_1} \geq 0.99$.
        We assume from now on that the events $\eset_1$ and $\eset_2$ indeed occur.

        We denote by $\eset_3$ the event that we sample at most $\frac{2^{10} \mu_{i}}{\mu_{j}}$ elements of $B \cap B^*$.
        We first claim that $B^{*'}$ is in fact, the level-$(j-1)$ ancestor block of $B'$.
        Indeed, for each level $i < i' < j$, we have $\psi_{i'} = 1$ and hence, the partitions $\bset^{i}_{\Psi'}(H^*) = \ldots = \bset^{j-1}_{\Psi'}(H^*)$ of the range $H^*$ are identical.
        Thus, $B^{*'}$ is the level-$(j-1)$ ancestor block of $B'$.
        Since $S^*$ is an $\Upsilon$-canonical sequence, and $B$ is a level-$j$ yes-block, $Z_j \mu_{j-1} \leq |B \cap B^{*'}| < 2 Z_j \mu_{j-1}$.
        Recall that we sample the elements of $B \cap B^{*'}$ independently with probability $\frac{2^6}{Z_{j} \mu_{j}}$ each.
        Using Chernoff bound, it is now immediate to verify that $\prob{\eset_3 \> | \> \eset_1 \text{ and } \eset_2} > 0.99$.
        From union bound, $\prob{\eset_1 \text{ and } \eset_2 \text{ and } \eset_3} \geq 0.97$, and we conclude that $(B, B')$ is a happy pair with probability at least $0.97$.
    \end{proof}

    Using \Cref{clm: ncm-pset-z-i-z-j,clm: ncm-b-b-prime-happy} and applying Chernoff bound, with probability at least $0.9$, there are at least $\frac{Z_i}{8 Z_j}$ happy pairs.
    This completes the proof of \Cref{clm: ncm-large-eset}.
\endproofof

\subsection{Proofs Omitted from Section \ref{subsec: ncm-processing-upsilon}} \label{appn: ncm-processing-upsilon}

\subsubsection{\proofof{\Cref{clm: ncm-i-star-sep}}} \label{prf-clm: ncm-i-star-sep}
    We first consider the case where $w_{i_{k'}} < \Tval$.
    In this case, we must have $k' = \knew$ and from \Cref{clm: ncm-simpler-algo-k-star}, $\knew \geq r/3$.
    Assume from now on that $k' < \knew$ and $w_{i_{k'}} \geq \Tval$.
    We consider the execution of our algorithm on level $i_{k'}$.
    Since it is the last level that we process, we must have had $\land(k') = \cover(k', \jump(k')) = \knew$ and hence, $\jump(k') + \floor{0.3 \log_\eta{\left( \frac{\mu_{i_{k'}}}{\mu_{i_{\jump(k')}}} \right)}} \geq \knew$.
    But $\mu_{i_{\jump(k')}} \geq 1$ and we obtain
    
    \begin{equation} \label{eqn: ncm-jump-k-prime-lb}
        \jump(k') \geq \knew - 0.3 \log_\eta{\left( \mu_{i_{k'}} \right)}        
    \end{equation}

    We will use the following claim that we prove after completing the proof of \Cref{clm: ncm-i-star-sep} assuming it.

    \begin{claim}\label{clm: ncm-k-jump-start-small-new-new}
        $\jump(k') \leq k' + \frac{31}{100} \cdot \log_\eta{\left(\mu_{i_{k'}}\right)}$.
    \end{claim}

    From \Cref{eqn: ncm-jump-k-prime-lb,clm: ncm-k-jump-start-small-new-new}, $k' + \frac{31}{100} \cdot \log_\eta{\left(\mu_{i_{k'}}\right)} \geq \knew - \frac{3}{10} \log_\eta{\left( \mu_{i_{k'}} \right)}$, or equivalently, 

    \begin{align*}
        k' &\geq \knew - \frac{61}{100} \log_\eta{\left( \mu_{i_{k'}} \right)}\\
        &\geq \left( \frac{1}{2} - 7 \eps \right)r - \frac{61}{100} \left( \frac{1}{2} + \eps \right)r\\
        &> \left( \frac{39}{200} - 8 \eps \right)r > \frac{r}{6}.
    \end{align*}
    
    Here, the second inequality follows from \Cref{clm: ncm-simpler-algo-k-star} and the fact that $\mu_{i_{k'}} \leq \mu_0 \leq \Delta_0 \leq N^{1/2 + \eps}$ (see, \Cref{obs: ncm-decreasing-xi-by-zi}).
    The last inequality holds for all $\eps < 1/1000$.
    This completes the proof of \Cref{{clm: ncm-i-star-sep}} assuming \Cref{{clm: ncm-k-jump-start-small-new-new}} that we prove next.

    \proofof{\Cref{clm: ncm-k-jump-start-small-new-new}}
        Recall that $k' < \jump(k') \leq \knew$ is the smallest index such that $\log_\eta{\left( \frac{\mu_{i_{k'}}}{\mu_{\ell}} \right)} \leq 100 \log_\eta{\left( \frac{\mu_{i_{k'}}}{\mu_{j}} \right)}$, where $j = i_{\jump(k')}$, $\ell = i_{\land(k')}$, and $\land(k') = \cover(k', \jump(k'))$.
        We consider a sequence $\left(p(0), \ldots, \right)$ of integers  with values in the range $\set{k' + 1, \ldots, \knew}$ that is obtained as follows.
        We set $p(0) = 1 + k'$ and for each subsequent $u > 0$, we let $p({u}) = \min{\left( \cover(k', p(u-1)), \jump(k') \right)}$.
        It is immediate to verify that this sequence is non-decreasing.
        This sequence naturally corresponds to a sequence $\left( q(0), \ldots, \right)$ of levels of $\isetnew$, where for each $u \geq 0$, we let $q(u) := i_{p(u)}$.

        \begin{observation}\label{obs: ncm-tu-tu-plus-1-new}
            For each $u \geq 0$ such that $p(u+1) < \jump(k')$, $\log_\eta{\left( \frac{\mu_{i_{k'}}}{\mu_{q(u+1)}} \right)} > 100 \log_\eta{\left( \frac{\mu_{i_{k'}}}{\mu_{q(u)}} \right)}$.
        \end{observation}
        \begin{proof}
            Fix an integer $u \geq 0$ such that $p(u+1) = \min{\left( \cover(k', p(u)), \jump(k') \right)} < \jump(k')$, and hence, $p(u+1) = \cover(k', p(u))$.
            But $p(u) \leq p(u+1) < \jump(k')$, and we must have had $\log_\eta{\left( \frac{\mu_{i_{k'}}}{\mu_{\ell}} \right)} > 100 \log_\eta{\left( \frac{\mu_{i_{k'}}}{\mu_{q(u)}} \right)}$, where $\ell = i_{\cover(k', p(u))} = i_{p(u+1)} = q(u+1)$.
        \end{proof}

        Let $v \geq 0$ be the smallest integer such that $p(v) = \jump(k')$.
        We first claim that such an integer $v$ exists.
        Indeed, assume for contradiction that for all $u \geq 0$ we have $p(u) < \jump(k') \leq \knew$.
        But then,

        \begin{align*}
            p(u+1) &= \min{\left( \cover(k', p(u)), \jump(k') \right)} = \cover(k', p(u))\\
            &= \min{\left(p(u) + \floor{0.3 \log_\eta{\left( \frac{\mu_{i_{k'}}}{\mu_{q(u)}} \right)}}, \knew \right)}\\
            &= p(u) + \floor{0.3 \log_\eta{\left( \frac{\mu_{i_{k'}}}{\mu_{q(u)}} \right)}}\\
            &\geq p(u) + \floor{0.3 w_{i_{k'}}}\\
            &> p(u).
        \end{align*}

        Here, the equalities follow from that fact that $p(u) \leq p(u+1) < \jump(k') \leq \knew$.
        The first inequality follows since $p(u) \geq 1+k'$ and the last inequality follows from the fact that $w_{i_{k'}} \geq 100$.
        Thus, the sequence $(p(0), \ldots, )$ is simultaneously strictly increasing in perpetuity and also bounded by $\jump(k')$, a contradiction.
        We fix the smallest integer $v$ such that ${p(v) = \jump(k')}$.

        \begin{observation} \label{obs: ncm-pu-lt-v-upper-bound}
            For each $0 \leq u < v$, $p(u) \leq k' +  \frac{1}{100} \cdot \log_\eta{\left( \frac{\mu_{i_{k'}}}{\mu_{q(u)}} \right) } $.
        \end{observation}
        \begin{proof}
            We proceed by induction.
            The base case is when $u = 0$.
            In this case, $p(0) = k'+1$ and $q(0) = i_{p(0)} = i_{k'+1}$, implying,
            
            \[ p(0) = k' + 1 \leq k' + \frac{1}{100} \cdot w_{i_{k'}} = k' + \frac{1}{100} \cdot \log_\eta{\left( \frac{\mu_{i_{k'}}}{\mu_{q(0)}} \right) },\]
            
            and the assertion follows.
            Here, the inequality follows since $w_{i_{k'}} \geq 100$.
            Assume now that the induction hypothesis holds for $0 \leq u < v-1$, and we show it for $u+1$.
            Recall that $v$ is the smallest integer such that ${p(v) = \jump(k')}$.
            Thus, $p(u+1) < \jump(k')$ and hence, 

            \begin{align*}
                p(u+1) &= \cover(k', p(u))\\
                &\leq p(u) + {\frac{3}{10} \cdot \log_\eta{\left( \frac{\mu_{i_{k'}}}{\mu_{q(u)}} \right)}}\\
                &\leq k' + \frac{1}{100} \cdot \log_\eta{\left( \frac{\mu_{i_{k'}}}{\mu_{q(u)}} \right) } + {\frac{3}{10} \cdot \log_\eta{\left( \frac{\mu_{i_{k'}}}{\mu_{q(u)}} \right)}}\\
                &= k' + \frac{31}{100} \cdot \log_\eta{\left( \frac{\mu_{i_{k'}}}{\mu_{q(u)}} \right) }\\
                &\leq k' + \frac{31}{10000} \cdot \log_\eta{\left( \frac{\mu_{i_{k'}}}{\mu_{q(u+1)}} \right) } < k' + \frac{1}{100} \cdot \log_\eta{\left( \frac{\mu_{i_{k'}}}{\mu_{q(u+1)}} \right) }.
            \end{align*}
    
            Here, the second inequality follows from the induction hypothesis and the second-last inequality follows from \Cref{obs: ncm-tu-tu-plus-1-new}.
        \end{proof}

        If $v = 0$, we have $\jump(k') = p(0) = k' + 1 < k' + \frac{31}{100} \cdot \log_\eta{\left(\mu_{i_{k'}}\right)}$ and there is nothing to show.
        Thus, assume from now on that $v > 0$.
        Recall that $p(v) = \jump(k')$ and hence, $\cover(k', p(v-1)) \geq \jump(k')$.
        In other words,

        \begin{align*}
            \jump(k') \leq \cover(k', p(v-1)) &\leq p(v-1) + \floor{0.3 \log_\eta{\left( \frac{\mu_{i_{k'}}}{\mu_{q(v-1)}} \right)}}\\
            &\leq k' + \frac{1}{100} \log_\eta{\left( \frac{\mu_{i_{k'}}}{\mu_{q(v-1)}} \right)} + \frac{3}{10} \log_\eta{\left( \frac{\mu_{i_{k'}}}{\mu_{q(v-1)}} \right)}\\
            &= k' + \frac{31}{100} \log_\eta{\left( \frac{\mu_{i_{k'}}}{\mu_{q(v-1)}} \right)} \leq k' + \frac{31}{100} \log_\eta{\left(\mu_{i_{k'}}\right)}.
        \end{align*}

        Here, the second inequality follows from \Cref{obs: ncm-pu-lt-v-upper-bound}.
        This completes the proof of \Cref{clm: ncm-k-jump-start-small-new-new}.
    \endproofof
    This also completes the proof of  \Cref{{clm: ncm-i-star-sep}}.
\endproofof

\subsubsection{\proofof{\Cref{clm: ncm-large-w-star}}} \label{prf-clm: ncm-large-w-star}
    For convenience, we let $i' := i_{k'}$.
    Notice that,

    \begin{align*}
        w_{\iset'} = w_{\set{i_0, \ldots, i_{k' - 1}}} = \sum_{0 \leq k < k'} w_{i_{k}} &= \sum_{0 \leq k < k'} \log_\eta{\left( \frac{\mu_{i_k}}{\mu_{i_{k+1}}} \right)} = \log_\eta{\left( \frac{\mu_{i_0}}{\mu_{i'_{}}} \right)}.
    \end{align*}

    First, we claim that $\mu_{i_0} \geq N^{1/2 -\eps}$.
    Recall that $i_0$ is the smallest level $i$ such that $Z_i > Z_{i+1}$, and hence, $Z_{i_0} = Z_0$.
    But then $\mu_{i_0} = \frac{X_{i_0+1}}{Z_{i_0 + 1}} \geq N^{1/2 - \eps}$.
    Here, the equality follows since $\psi_0 = \ldots = \psi_{i_0} = 1$ and the inequality follows from \Cref{assm: large-xi-by-zi}.

    Next, we claim that $3 \eps r \leq i' < (1/2 - \eps)r$.
    Indeed, $i' \leq i_{\knew} \leq {r^*} < (1/2 - \eps)r$.
    On the other hand, from \Cref{clm: ncm-i-star-sep}, $i' \geq r/6 \geq 3 \eps r$, for all $\eps \leq 1/100$.

    We now claim that $\mu_{i'} < \frac{N^{1/2 + 6\eps}}{\eta^{i'}}$.
    Indeed, from \Cref{assm: large-xi-by-zi} we get, $Z_{i'} < \frac{X_{i'}}{N^{1/2 - \eps}} = \frac{N^{1/2 + \eps}}{\eta^{i'}}$.
    On the other hand, from \Cref{assm: small-deltai-by-zi}, $Z_{i'} > \Delta_{i'}/N^{5\eps}$, implying $\Delta_{i'} < Z_{i'} N^{5\eps} < \frac{N^{1/2 + 6\eps}}{\eta^{i'}}$.
    From \Cref{obs: ncm-decreasing-xi-by-zi}, we now obtain $\mu_{i'} \leq \Delta_{i'} < \frac{N^{1/2 + 6\eps}}{\eta^{i'}}$ as claimed.
    We now conclude,

    \begin{align*}
        w_{\iset'} = \log_\eta{\left( \frac{\mu_{i_0}}{\mu_{i'_{}}} \right)} \geq \log_\eta{\left( \frac{{\eta^{i'_{}}}}{N^{7\eps}} \right)} = i' - \log_\eta{\left(N^{7\eps}\right)} &> i' - 8 \eps r\\
        &\geq k' - 8 \eps r > k' - \frac{r}{1000},
    \end{align*}

    for all $\eps < 1/10^4$.
    Here, the second inequality follows since $r = \floor{\frac{\log N}{\log \eta}}$ and $\eta = o(\log N)$ is small enough while the third inequality follows from \Cref{clm: ncm-i-star-sep}.
    This completes the proof of \Cref{clm: ncm-large-w-star}.
\endproofof

\subsubsection{\proofof{\Cref{obs: ncm-z-r-star-lower-bound}}} \label{prf-obs: ncm-z-r-star-lower-bound}
    Consider some bad jumpable tuple $\tau(k) \in \jset'$ and let $\tau(k) = (i_k, j, \ell)$, where $j = i_\jump(k)$ and $\ell = i_{\land(k)}$, and $\land(k) = {\cover(k, \jump(k))}$.
    We let $\hat \iset(\tau(k)) := \set{i_{\jump(k)}, \ldots, i_{\land(k)-1}}$ and think of $\tau(k)$ as being responsible for $\hat \iset(\tau)$.
    Note that,

    \begin{equation} \label{eqn: ncm-hat-iset-tau-bound}
        \begin{split}
            |\hat \iset(\tau(k))| = \land(k) - \jump(k) &=  \cover(k, \jump(k)) - \jump(k)\\
            &= \floor{0.3 \tupwt{\tau(k)}} \geq 0.29 \tupwt{\tau(k)}.
        \end{split}
    \end{equation}

    Here, the last equality follows since we chose to add the jumpable tuple $\tau(k)$ to $\jset$, we must have had $\cover(k, \jump(k)) < \knew$ implying that $\cover(k, \jump(k)) = \jump(k) + \floor{0.3 \tupwt{\tau(k)}}$.
    The last inequality follows since $\tupwt{\tau(k)} \geq \Tval$.
    We let $\hat \iset(\jset') = \bigcup_{\tau \in \jset'} \hat \iset(\tau)$ be the set of all levels for which the jumpable tuples of $\jset'$ are responsible.
    Since the respective sets $\hat \iset(\tau)$ are disjoint for distinct jumpable tuples of $\jset$, we have

    \begin{equation} \label{eqn: ncm-iset-jset-prime-bound}
        \begin{split}
            |\hat \iset(\jset')| = \sum_{\tau \in \jset'} |\hat \iset(\tau)| \geq \sum_{\tau \in \jset'} 0.29 \tupwt{\tau} &= 0.29 \tupwt{\jset'}.
        \end{split}
    \end{equation}

    Consider now a bad jumpable tuple $\tau(k) = (i_k, j, \ell)$ of $\jset'$.
    Since it is a bad jumpable tuple, we have $\frac{Z_j}{Z_\ell} < \eta^{0.2 \tupwt{\tau(k)}}$, and hence,
    $\prod_{i' \in \hat \iset(\tau(k))} \frac{Z_{i'}}{Z_{i'+1}} = \frac{Z_j}{Z_\ell} < \eta^{0.2 \tupwt{\tau(k)}}$.
    Next, recall that for each level $i_k \in \isetnew$ with $0 \leq k < \knew$, we have $\frac{Z_{i_{k}}}{Z_{i_{k+1}}} \leq \eta$.
    Using these two facts, we bound $\frac{Z_0}{Z_{i_{\knew}}}$ by writing it as a telescopic product:

    \begin{align*}
        \frac{Z_0}{Z_{i_{\knew}}} = \frac{Z_{i_0}}{Z_{i_{\knew}}} = \prod_{0 \leq k < \knew} \frac{Z_{i_k}}{Z_{i_{k+1}}} &= \left( \prod_{\substack{0 \leq k < \knew \\ \text{ and } \\  i_k \not \in \hat \iset(\jset')}} \frac{Z_{i_k}}{Z_{i_{k+1}}} \right) \cdot \left( \prod_{\tau \in \jset'} \left( \prod_{i \in \hat \iset(\tau)}  \frac{Z_i}{Z_{i+1}} \right) \right)\\
        &< \left( \eta^{\knew - |\hat \iset(\jset')|} \right) \cdot \left( \prod_{\tau \in \jset'} \eta^{0.2 \tupwt{\tau}} \right)\\
        &\leq \eta^{\knew - 0.29 \tupwt{\jset'}} \cdot \eta^{0.2 \tupwt{\jset'}} = \eta^{\knew - 0.09 \tupwt{\jset'}}.
    \end{align*}

    Here, the inequalities follow from \Cref{eqn: ncm-hat-iset-tau-bound,eqn: ncm-iset-jset-prime-bound}.
    This completes the proof of \Cref{obs: ncm-z-r-star-lower-bound}.
\endproofof

\subsection{Proofs Omitted from Sections \ref{subsec: ncm-jumpable}, \ref{subsec: ncm-jump-jump-i}, and \ref{subsec: ncm-jumpable-alggen-desc}} \label{appn: ncm-jumpable}

\subsubsection{\proofof{\Cref{clm: zeta-convex}}} \label{prf-clm: zeta-convex}
    We will first show that $\zeta(i_{k_1}, i_{k_2}) = \sum_{k_1 \leq k' < k_2} \zeta(i_{k'}, i_{k'+1})$.
    Indeed,

    \begin{align*}
        e^{\zeta(i_{k_1}, i_{k_2})} &= \eta^{2(i_{k_2} - i_{k_1})} \cdot \psi_{i_{k_1+1}} \cdot \ldots \cdot \psi_{i_{k_2}} \\
        &= \prod_{k_1 \leq k' < k_2} \eta^{2(i_{k'+1} - i_{k'})} \cdot \psi_{i_{k'+1}}\\
        &= \prod_{k_1 \leq k' < k_2} e^{\zeta(i_{k'}, i_{k'+1})}    \\
        &= e^{ \sum_{k_1 \leq k' < k_2} \left( \zeta(i_{k'}, i_{k'+1}) \right)}.
    \end{align*}

    The claim now follows by noting that for each $0 \leq k < \knew$, we have $\zeta(i_k, i_{k+1}) \geq \ln{\eta} > 2$.
\endproofof

\subsubsection{\proofof{\Cref{clm: ncm-jump-load-induction}}} \label{prf-clm: ncm-jump-load-induction}
    We proceed by induction.
    The base case is when $k = 0$ and the assertion is trivial since we execute level-$i_0$ algorithm $\algjump{i_0}$ only once.
    Consider now some integer $0 < k \leq k^*$ and the corresponding level $i_k \in \iset^*$.
    We assume that the induction hypothesis holds for level $i_{k-1}$.
    We fix a level-$i_k$ stream-block $B$ and analyze $\load(B)$.

    We first consider the case where there is no jumpable tuple $\tau = (i_{k'}, i_{k''}, i_{k'''})$ in $\jset^*$ with $k \in \set{k'', k'''}$.
    Let $B^*$ be the unique level-$(i_{k-1})$ ancestor-block of $B$.
    It is immediate to verify that each execution to $\algjump{i_k}$ in which $B$ participates is called by a run of the level-$(i_{k-1})$ algorithm $\algjump{i_{k-1}}$ in which $B^*$ participates.
    Moreover, from \Cref{obs: ncm-jump-no-jump}, each such execution of $\algjump{i_{k-1}}$ performs at most $\zeta^2(i_{k-1}, i_k) \cdot \frac{\mu_{i_{k-1}}}{\mu_{i_k}} \leq \frac{q_3(i_k)}{q_3(i_{k-1})} \cdot \frac{\mu_{i_{k-1}}}{\mu_{i_k}}$ concurrent calls to $\algjump{i_k}$.
    Thus, from induction hypothesis, the maximum number of concurrent calls to the level-$i$ algorithm $\algjump{i}$ in which $B$ participates is indeed,

    \begin{align*}
        \load(B) & \leq \load(B^*) \cdot  \frac{q_3(i_k)}{q_3(i_{k-1})} \cdot \frac{\mu_{i_{k-1}}}{\mu_{i_k}}    \\
        & \leq \frac{q_1(i_{k-1}) \cdot q_3(i_{k-1})}{q_2(i_{k-1})} \cdot \frac{\mu_{i_0}}{\mu_{i_{k-1}}} \cdot  \frac{q_3(i_k)}{q_3(i_{k-1})} \cdot \frac{\mu_{i_{k-1}}}{\mu_{i_k}}\\
        & = \frac{q_1(i_{k}) \cdot q_3(i_{k})}{q_2(i_{k})} \cdot \frac{\mu_{i_0}}{\mu_{i_{k}}}.
    \end{align*}
    
    Next, we  consider the second case where there is a perfect jumpable tuple $\tau = (i_{k'}, i_{k''}, i_{k'''})$ in $\jset^*$ such that $k \in \set{k'', k'''}$.
    We analyze the subcases $k = k''$ and $k = k'''$ separately.

    \paragraph{Subcase 1: $k = k''$.}
    In this case, the jumpable tuple is $\tau = (i_{k'}, i_{k''}, i_{k'''}) = (i_{k'}, i_k, i_{k'''})$.
    Let $B^*$ be the level-$i_{k'}$ ancestor block of $B$.
    It is immediate to verify that each execution to $\algjump{i_k}$ in which $B$ participates is called by a run of the level-$i_{k'}$ algorithm $\algjump{i_k'}$ in which $B^*$ participates.
    From \Cref{lem: ncm-jump-jump} each such execution of $\algjump{i_{k'}}$ performs at most $O \left( \zeta(i_{k'}, i_k) \cdot \frac{Z_{i_{k'}}}{Z_{i_k}} \cdot \eta^{i_{k'''} - i_k} \right)$ concurrent calls to $\algjump{i_k}$.
    From \Cref{clm: zeta-convex} and using the fact that $\eta$ is large enough, we can bound this by $\frac{q_3(i_k)}{q_3(i_{k'})} \cdot \frac{Z_{i_{k'}}}{Z_{i_k}} \cdot \eta^{i_{k'''} - i_k}$.
    The number of concurrent calls to $\algjump{j}$ in which $B$ participates can now be bounded by,

    \begin{equation} \label{eqn: ncm-algjump-subcase-1-1}
        \begin{split}
            \load(B) & \leq \load(B^*) \cdot \frac{q_3(i_k)}{q_3(i_{k'})} \cdot \frac{Z_{i_{k'}}}{Z_{i_k}} \cdot \eta^{i_{k'''} - i_k} \\
        &= \frac{q_1(i_{k'}) \cdot q_3(i_k)}{q_2(i_{k'})} \cdot \frac{\mu_{i_0}}{\mu_{i_{k'}}}  \cdot \frac{Z_{i_{k'}}}{Z_{i_k}} \cdot \eta^{i_{k'''} - i_k} \\
        &= \frac{q_1(i_{k'}) \cdot q_3(i_k)}{q_2(i_{k'})} \cdot \frac{\mu_{i_0}}{\mu_{i_k}} \cdot \left( \frac{\mu_{i_k}}{\mu_{i_{k'}}}  \cdot \frac{Z_{i_{k'}}}{Z_{i_k}} \cdot \eta^{i_{k'''} - i_k} \right).
        \end{split}
    \end{equation}

    But,

    \begin{equation} \label{eqn: ncm-algjump-subcase-1-2}
        \begin{split}
            \frac{\mu_{i_k}}{\mu_{i_{k'}}}  \cdot \frac{Z_{i_{k'}}}{Z_{i_k}} \cdot \eta^{i_{k'''} - i_k} &\leq \eta^{-0.99 \tupwt{\tau} + i_{k'''} - i_k} \\
            &\leq \eta^{-0.69 \tupwt{\tau} + \eq(i_k, i_{k'''})}\\
            &= \frac{q_1(i_k)}{q_1(i_{k'})} \cdot \frac{q_2(i_{k'})}{q_2(i_k)}.
        \end{split}
    \end{equation}

    Here, the first inequality follows since the level-weight of the jumpable tuple $\tau$ is ${\tupwt{\tau} = \log_\eta{\left( \frac{\mu_{i_{k'}}}{\mu_{i_k}} \right)}}$ and from the fact that $\frac{Z_{i_{k'}}}{Z_{i_k}} \leq \eta^{k - k''} \leq \eta^{0.01 \tupwt{\tau}}$, as $\tau$ is a perfect jumpable tuple.
    The second inequality follows since $i_{k'''} - i_{k} = k''' - k + \eq(i_{k}, i_{k'''}) = \floor{0.3 \tupwt{\tau}} + \eq(i_{k}, i_{k'''})$.
    The last inequality follows since $q_1(i_k) = q_1(i_{k'}) \cdot \eta^{\eq(i, i_{k'''})}$ and $q_2(i_k) = q_2(i_{k'}) \cdot \eta^{0.39 \tupwt{\tau}}$.
    Plugging \Cref{eqn: ncm-algjump-subcase-1-2} in \Cref{eqn: ncm-algjump-subcase-1-1}, we obtain,

    \begin{equation*}
        \load(B) \leq \frac{q_1(i_{k'}) \cdot q_3(i_k)}{q_2(i_{k'})} \cdot \frac{\mu_{i_0}}{\mu_{i_k}} \cdot \left( \frac{q_1(i_k)}{q_1(i_{k'})} \cdot \frac{q_2(i_{k'})}{q_2(i_k)} \right) = \frac{q_1(i_{k}) \cdot q_3(i_k)}{q_2(i_{k})} \cdot \frac{\mu_{i_0}}{\mu_{i_k}},
    \end{equation*}

    as claimed.
    This completes the analysis of subcase $1$.

    \paragraph{Subcase 2: $k = k'''$.}
    In this subcase, the jumpable tuple is $\tau = (i_{k'}, i_{k''}, i_{k'''}) = (i_{k'}, i_{k''}, i_{k})$.
    As before, let $B^*$ be the level-$i_{k'}$ ancestor block of $B$.
    We also consider the level-$(i_{k-1})$ ancestor $\tilde B$ of $B$.
    It is immediate to verify that each execution to $\algjump{i_k}$ in which $B$ participates is either called by a run of the level-$i_{k'}$ algorithm $\algjump{i_{k'}}$ in which $B^*$ participates or by a run of the level-$(i_{k-1})$ algorithm $\algjump{i_{k-1}}$ in which $\tilde B$ participates.

    We first analyze the contribution of the executions of $\algjump{i_{k'}}$ to $\load(B)$.
    Consider a run of $\algjump{i_{k'}}$ in which $B^*$ participates.
    From \Cref{lem: ncm-jump-jump} each such execution of $\algjump{i_{k'}}$ performs at most $O \left( \zeta^2(i_{k'}, i_k) \cdot \frac{Z_{i_{k'}} \cdot Z_{i_k}^2}{Z_{i_{k''}}^3} \cdot \frac{\mu_{i_{k'}}}{\mu_{i_k}} \right)$ concurrent calls to $\algjump{i_k}$.
    As before, we can bound this number by $\frac{q_3(i_k)}{2q_3(i_{k'})} \cdot \frac{Z_{i_{k'}} \cdot Z_{i_k}^2}{Z_{i_{k''}}^3} \cdot \frac{\mu_{i_{k'}}}{\mu_{i_k}}$.
    Thus, the number of concurrent calls to $\algjump{i_k}$ in which $B$ participates due to all the executions of $\algjump{i_{k'}}$ in which $B^*$ participates is at most,

    \begin{equation} \label{eqn: ncm-algjump-subcase-2-1}
        \begin{split}
            \load(B^*) \cdot \frac{q_3(i_k)}{2q_3(i_{k'})} \cdot \frac{Z_{i_{k'}} \cdot Z_{i_k}^2}{Z_{i_{k''}}^3} \cdot \frac{\mu_{i_{k'}}}{\mu_{i_k}}
            &\leq \load(i_{k'}) \cdot \frac{q_3(i_k)}{2q_3(i_{k'})} \cdot \frac{Z_{i_{k'}} \cdot Z_{i_k}^2}{Z_{i_{k''}}^3} \cdot \frac{\mu_{i_{k'}}}{\mu_{i_k}}\\
            &\leq \frac{q_1(i_{k'}) q_3(i_k)}{2q_2(i_{k'})} \cdot \frac{\mu_{i_0}}{\mu_{i_k}} \cdot \left(\frac{Z_{i_{k'}} \cdot Z_{i_k}^2}{Z_{i_{k''}}^3} \right)\\
            &\leq \frac{q_1(i_{k'}) q_3(i_k)}{2q_2(i_{k'})} \cdot \frac{\mu_{i_0}}{\mu_{i_k}} \cdot \left( \eta^{-0.39 \tupwt{\tau}} \right)\\
            &\leq \frac{q_1(i_k) q_3(i_k)}{2q_2(i_k)} \cdot \frac{\mu_{i_0}}{\mu_{i_k}}.
        \end{split}
    \end{equation}

    Here, the third inequality follows since $\tau = (i_{k'}, i_{k''}, i_k)$ is a perfect jumpable tuple implying $\frac{Z_{i_k}}{Z_{i_{k''}}} \leq \eta^{-0.2 \tupwt{\tau}}$ and $\frac{Z_{i_{k'}}}{Z_{i_{k''}}} \leq \eta^{k'' - k'} \leq \eta^{0.01 \tupwt{\tau}}$.
    The last inequality follows since $k > k''$ and $q_2(i_k) = q_2(i_{k'}) \cdot \eta^{0.39 \tupwt{\tau}}$.
    This completes the analysis of the contribution of the executions of $\algjump{i_{k'}}$ to $\load(B)$.

    We now analyze the contribution of the executions of $\algjump{i_{k-1}}$ to $\load(B)$.
    From \Cref{obs: ncm-jump-no-jump}, each execution of $\algjump{i_{k-1}}$ in which $\tilde B$ participates, performs at most
    $O \left(\zeta(i_{k-1}, i_k) \cdot  \frac{\mu_{i_{k-1}}}{\mu_{i_k}} \right) \leq \frac{q_3(i_k)}{2 q_3(i_{k-1})} \cdot \frac{\mu_{i_{k-1}}}{\mu_{i_k}}$ concurrent calls to $\algjump{i_k}$.
    The number of concurrent calls to $\algjump{\ell}$ in which $B$ participates due to the execution of $\algjump{i_{k-1}}$ in which $B'$ participates can now be bounded by,

    \begin{equation} \label{eqn: ncm-algjump-subcase-2-2}
        \begin{split}
            \load(\tilde B) \cdot \frac{q_3(i_k)}{2 q_3(i_{k-1})} \cdot \frac{\mu_{i_{k-1}}}{\mu_{i_k}} &\leq \load(i_{k-1}) \cdot \frac{q_3(i_k)}{2 q_3(i_{k-1})} \cdot \frac{\mu_{i_{k-1}}}{\mu_{i_k}}\\
            &\leq \frac{q_1(i_{k-1}) q_3(i_k)}{2q_2(i_{k-1})} \cdot \frac{\mu_{i_0}}{\mu_{i_{k-1}}} \cdot \frac{\mu_{i_{k-1}}}{\mu_{i_k}}\\
            &=  \frac{q_1(i_k) q_3(i_k)}{2q_2(i_k)} \cdot \frac{\mu_{i_0}}{\mu_{i_k}}.
        \end{split}
    \end{equation}

    Here, the last inequality follows since $q_1(i) = q_1(i_{k-1})$ and $q_2(i) = q_2(i_{k-1})$.
    We now conclude that $\load(B)$ is indeed bounded by $\frac{q_1(i) q_3(i)}{q_2(i)}  \cdot \frac{\mu_{i_0}}{\mu_{i}}$.
    Thus, $\load(i) \leq \frac{q_1(i) q_3(i)}{q_2(i)}  \cdot \frac{\mu_{i_0}}{\mu_{i}}$ as claimed.
    This completes the analysis of the contribution of the executions of $\algjump{i_{k-1}}$ to $\load(B)$.

    From \Cref{{{eqn: ncm-algjump-subcase-2-1},{{eqn: ncm-algjump-subcase-2-2}}}}, we now conclude that the number of concurrent calls to $\algjump{i_k}$ in which $B$ participates is indeed bounded by,

    \[ \load(B) \leq 2 \cdot \frac{q_1(i_k) q_3(i_k)}{2 q_2(i_k)} \cdot \frac{\mu_{i_0}}{\mu_{i_k}} = \frac{q_1(i_k) q_3(i_k)}{q_2(i_k)} \cdot \frac{\mu_{i_0}}{\mu_{i_k}},\]

    as claimed.
    This completes the analysis of subcase $2$, and hence the induction step.
    This also completes the proof of \Cref{clm: ncm-jump-load-induction}.
\endproofof

\subsubsection{\proofof{\Cref{clm: ncm-load-i-small}}} \label{prf-clm: ncm-load-i-small}
    We claim that $\max_{i \in \isetnew} L^*(i) = L^*(i_{k^*})$.
    To show this, it suffices to show that for each jumpable tuple $\tau = (i, j, \ell) \in \jset^*$, $L^*(i) \leq L^*(j)$.
    Indeed, 

    \begin{align*}
        \frac{L^*(j)}{L^*(i)} &= \frac{q_1(j) \cdot q_3(j) \cdot q_2(i)}{q_2(j) \cdot q_1(i) \cdot q_3(i)} \cdot \frac{\mu_{i}}{\mu_{j}}\\
        &\geq \frac{q_2(i)}{q_2(j)} \cdot \eta^{\tupwt{\tau}}\\
        &= \eta^{\tupwt{\tau} - 0.39 \tupwt{\tau}} = \eta^{0.61 \tupwt{\tau}} > 1.
    \end{align*}

    It now remains to upper-bound $L^*(i_{\knew}) = \frac{q_1(i_{k^*}) \cdot q_3(i_{k^*})}{q_2(i_{k^*})} \cdot \frac{\mu_{i_0}}{\mu_{i_{k^*}}}$.
    Since $Z_{i_0} = Z_0 \geq N^{1/2 - \eps}$, we have,

    \begin{equation} \label{eqn: ncm-jump-loadi-small-1}
        \mu_{i_0} = \frac{X_{i_0}}{Z_{i_0}} = \frac{N/\eta^{i_0}}{Z_0} \leq \frac{N^{1/2 + \eps}}{\eta^{i_0}} \leq N^{1/2 + \eps}.  
    \end{equation}

    On the other hand, from \Cref{clm: ncm-simpler-algo-k-star} and the fact that $r^* = \rstarVal$, we obtain,
    
    \begin{equation}\label{eqn: ncm-jump-loadi-small-2}
        q_1(\knew) = \eta^{\eq(i_0, i_{\knew})} \leq \eta^{r^* - \knew} \leq N^{4\eps}.    
    \end{equation}

    Moreover,
    
    \[ \log_\eta{\left(q_2(i_{\knew})\right)} = \sum_{\tau \in \jset^*} 0.39 \tupwt{\tau} = 0.39 \tupwt{\jset^*} \geq \frac{39}{100} \cdot \frac{r}{10^5} \geq \frac{3r}{10^6},\]
    
    or in other words,
    
    \begin{equation} \label{eqn: ncm-jump-loadi-small-3}
        q_2(i_\knew) \geq \eta^{\frac{3r}{10^6}} \geq {N^{\frac{3}{10^6}}}/{\eta},
    \end{equation}
    
    as $r = \floor{\frac{\log N}{\log \eta}}$.
    Finally, from \Cref{{clm: ncm-jump-load-induction},{clm: ncm-lis-simple-pi-zeta-bound}}, we obtain,
    
    \begin{equation} \label{eqn: ncm-jump-loadi-small-4}
        q_3(i_{k^*}) \leq \prod_{0 \leq k' < k^*} \zeta^4(i_{k'}, i_{k'+1}) = N^{o(1)}.
    \end{equation}

    We are now ready to provide an upper bound for $L^*(i_{\knew}) = \frac{q_1(i_{k^*}) \cdot q_3(i_{k^*})}{q_2(i_{k^*})} \cdot \frac{\mu_{i_0}}{\mu_{i_{k^*}}}$.
    From \Cref{{eqn: ncm-jump-loadi-small-1},{eqn: ncm-jump-loadi-small-2},{eqn: ncm-jump-loadi-small-3},{eqn: ncm-jump-loadi-small-4}},

    \begin{align*}
        L^*(i_{\knew}) &= \frac{q_1(i_{k^*}) \cdot q_3(i_{k^*})}{q_2(i_{k^*})} \cdot \frac{\mu_{i_0}}{\mu_{i_{k^*}}}\\
        &\leq \frac{N^{4\eps + o(1)}}{N^{\frac{3}{10^6}}} \cdot N^{\frac{1}{2} + \eps} \leq N^{\frac{1}{2} + 6\eps - \frac{3}{10^6}} \leq N^{\frac{1}{2} - \frac{2}{10^6}},
    \end{align*}

    where the last inequality holds for all $\eps \leq 10^{-7}$.
    Here, the second inequality follows from the facts that $r = \floor{\frac{\log N}{\log \eta}}$ and $\eps$ is an absolute constant.    
\endproofof

\subsubsection{\proofof{\Cref{clm: ncm-jump-eset-star-bad-low-prob}}} \label{prf-clm: ncm-jump-eset-star-bad-low-prob}
    We claim that it suffices to show that $\prob{\eset^*_\bad}(\hat B) \leq e^{-3\zeta(i,\ell)}$ for each level-$\ell$ descendant-block $\hat B$ of $B^*$. 
    Indeed, if we could show this, then \Cref{clm: ncm-jump-eset-star-bad-low-prob} follows by union bound over at most $e^{\zeta(i,\ell)}$ possible level-$\ell$ descendant stream-blocks of $B^*$.

    Consider now a level-$\ell$ descendant-block $\hat B$ of $B^*$ and let $B$ be the level-$j$ ancestor-block of $\hat B$.
    Notice that all the calls to the level-$\ell$ algorithm $\algjump{\ell}$ in which $\hat B$ participates are performed through the calls to $\alggen$ in which $B$ participates.
    Consider one such call to $\alggen$.
    From \Cref{lem: ncm-alggen-exists}, this execution of $\alggen$ independently randomly marks $\hat B$ with probability $O \left( \zeta(i,\ell) \cdot \frac{Z_\ell}{Z_j} \right)$.
    If $\hat B$ is marked, it performs at most $O \left(\zeta(i,\ell) \cdot \frac{Z_\ell \mu_{i}}{Z_j \mu_{\ell}} \right)$ calls to $\algjump{\ell}$ in which $\hat B$ participates.
    Otherwise, if $\hat B$ is not marked, it does not perform any call to $\algjump{\ell}$ in which $\hat B$ participates.
    \footnote{Note that we are excluding the calls to $\algjump{\ell}$ performed by $\algjump{j}$.}
    Since there are at most $O \left( \frac{Z_i}{Z_j} \right)$ concurrent calls to $\alggen$, the expected number of concurrent calls to $\algjump{\ell}$ in which $\hat B$ participates is bounded by,

    \begin{align*}
        O\left( \frac{Z_i}{Z_j} \right) \cdot O \left( \frac{\zeta(i, \ell) Z_\ell}{Z_j} \right) \cdot O \left( \frac{\zeta(i, \ell) Z_\ell \mu_{i}}{Z_j \mu_{\ell}} \right)
        &= O \left(\zeta^2(i, \ell) \right) \frac{Z_i \cdot Z_\ell^2}{Z_j^3} \cdot \frac{\mu_i}{\mu_\ell}   \\
        &\leq c' \cdot \zeta^2(i, \ell) \cdot \frac{Z_i \cdot Z_\ell^2}{Z_j^3} \cdot \frac{\mu_i}{\mu_\ell} ,
    \end{align*}

    for some absolute constant $c'$.
    We now chose the constant $c = 18 \max{(c', 1)}$ in the definition of $L^*$ so that this expected number of concurrent calls to $\algjump{\ell}$ is bounded by $\frac{L^*}{6}$.
    But then,
    
    \begin{align*}
        L^* &\geq 18 \zeta^2(i,\ell) \cdot \frac{Z_i \cdot Z_\ell^2}{Z_j^3} \cdot  \frac{\mu_i}{\mu_\ell} \\
        &> 18 \zeta(i, \ell) \cdot \frac{Z_\ell^2}{Z_j^2} \cdot  \frac{\mu_i}{\mu_j} \\
        &\geq 18 \zeta(i, \ell).\\
    \end{align*}

    Here, the second inequality follows since $i< j < \ell$ and hence, $Z_i > Z_j$ and $\mu_{j} \geq \mu_{\ell}$.
    The last inequality follows from our assumption that $\left(\frac{Z_j}{Z_\ell} \right)^2 \leq \frac{\mu_i}{\mu_j}$.
    From Chernoff bound (\Cref{fact: ncm-chernoff-version}), the probability that we perform more than $L^*$ calls to $\algjump{\ell}$ in which $\hat B$ participates is bounded by $e^{\left( - \frac{L^*}{6} \right)} < e^{ - 3 \zeta(i, \ell)}$ as claimed.
\endproofof

\subsubsection{\proofof{\Cref{clm: ncm-jump-dp-completeness}}} \label{prf-clm: ncm-jump-dp-completeness}
    Consider a level-$j$ yes-pair $(B, B')$ with $B \in \bset_\yes$ and $B' \in \bset'_\yes$.
    Let $\tilde \rset$ be our collection of the regions of $B^{*'}$ just before processing the first element of $B$.
    We also fix the special region $\leftover(\tilde \rset) = \bset' \backslash \bigcup_{\rset \in \tilde \rset} \rset$.
    Since the non-empty regions of $\tilde \rset \cup \set{\leftover(\tilde \rset)}$ constitute a partition of $\bset'$, there is exactly one region $\rset \in \tilde \rset \cup \set{\leftover(\tilde \rset)}$ such that $B' \in \rset$.
    We fix such a region $\rset$ and consider the execution of $\alggen$ with input $B$ and $\rset$.
    We let $\rset'$ be the output of this execution.
    We let $\eset_\bad(B)$ be the event that either the output region $\rset' = \emptyset$ or $\rset'$ contains some range-block appearing after $B'$ in $\bset'$.
    From the correctness guarantee of $\alggen$, the probability that $\eset_\bad$ occurs is at most $e^{-\zeta(i, \ell)} \leq \frac{1}{10|\bset|}$.
    We let $\eset_\bad$ be the bad event $\eset_\bad(B)$ occurs for some level-$j$ yes-block $B \in \bset_\yes$.
    From union bound over at most $|\bset_\yes| \leq |\bset|$ yes-blocks of $\bset$, the probability that $\eset_\bad$ occurs is at most $0.1$.
    We assume from now on that the event $\eset_\bad$ does not occur.

    We will need the following simple observation.
    Consider some stream-block $B \in \bset$.
    Let $\tilde \rset^{\mathsf{pre}}$ be the collection of regions of $\bset'$ just before processing the stream-block $B$.
    Similarly, let $\tilde \rset^{\mathsf{post}}$ be the collection of regions of $\bset'$ just after processing $B$.
    Let $\tilde \rset^{\mathsf{pre}} = \set{\rset^{\mathsf{pre}}_1, \ldots, \rset^{\mathsf{pre}}_{k'}}$ and $\tilde \rset^{\mathsf{post}} = \set{\rset^{\mathsf{post}}_1, \ldots, \rset^{\mathsf{post}}_{k''}}$ in their natural order.
    It is immediate to see that $k'' \geq k'$ and for each $1 \leq s \leq k'$, the region $\rset^{\mathsf{post}}_s$ does not contain any range-block that appears after the region $\rset^{\mathsf{pre}}_{s}$.
    We are now ready to complete the proof of \Cref{clm: ncm-jump-dp-completeness} by induction.

    The base case is when $s = 1$, where we consider the first stream-block $B(1)$ of $\bset_\yes$.
    Let $\tilde \rset^\mathsf{pre}$ be the collection of regions of $\bset'$ just before processing $B^{(1)}$.
    We also let $\leftover(\tilde \rset^{\mathsf{pre}})$ be the special region consisting of range-blocks not contained in the regions of $\tilde \rset^{\mathsf{pre}}$.
    For convenience, we let $\tilde \rset = \tilde \rset^\mathsf{pre} \cup \set{\leftover(\tilde \rset^{\mathsf{pre}})}$ and let $\tilde \rset = \set{\rset_1, \ldots, \rset_k}$ be these regions in their natural order where $k = |\tilde \rset|$.
    Since the non-empty regions of $\tilde \rset$ form a partition of the level-$j$ blocks $\bset'$, there is a unique integer $1 \leq k' \leq k$ such that $B'(1) \in \rset_{k'}$.
    Recall that $\tilde \rset^{(1)}$ is the collection of regions of $\bset'$ just after processing $B^{(1)}$.
    We first consider the case where $k' > 1$.
    In this case, all the range-blocks of $\rset_1$ appear before $B'(1)$.
    But then all the range-blocks of the first region of $\rset^{(1)}$ appear before $B'(1)$ and the assertion follows.
    Thus, assume from now on that $k' = 1$, or in other words, $B' \in \rset_1$.
    Recall that we call the subblock processing algorithm $\alggen$ with input $B^{(1)}$ and the region $\rset_1$ and let $\rset'_1$ be the region it reports.
    Since the event $\eset_\bad(B^{(1)})$ does not occur, $\rset'_1 \neq \emptyset$, and it does not contain any range-block appearing after $B'(1)$.
    In this case, the first region of $\rset^{(1)}$ is $\rset'_1$ and the assertion follows for the base case where $s=1$.

    We now fix some $1 < s \leq \beta$.
    We assume that the assertion holds for $s-1$ and show it for $s$.
    Let $\tilde \rset^\mathsf{pre}$ be the collection of regions of $\bset'$ just before processing $B^{(s)}$.
    We also let $\leftover(\tilde \rset^{\mathsf{pre}})$ be the region consisting of range-blocks not contained in the regions of $\tilde \rset^{\mathsf{pre}}$.
    Recall that $\tilde \rset^{(s)}$ is the collection of regions of $\bset'$ just after processing $B^{(1)}$.
    As before, we let $\tilde \rset = \tilde \rset^\mathsf{pre} \cup \set{\leftover(\tilde \rset^{\mathsf{pre}})}$.
    Let $\tilde \rset = \set{\rset_1, \ldots, \rset_k}$ be these regions in their natural order where $k = |\tilde \rset|$.
    Since the non-empty regions of $\tilde \rset$ form a partition of the level-$j$ blocks $\bset'$, there is a unique integer $1 \leq k' \leq k$ such that $B'(s) \in \rset_{k'}$.
    If $k' > s$, there is nothing to show since all the range-blocks of $\rset_s$ appear before $B'(s)$.
    In this case, all the range-blocks of the $s^{th}$ region of $\rset^{(s)}$ appear before $B'(s)$ and the assertion follows.
    Thus, assume that $k' \leq s$.
    From our induction hypothesis, it is immediate to verify that the regions $\set{\rset_1, \ldots, \rset_{s-1}}$ does not contain any range-block that appear after $B'(s-1)$.
    Since the range-block $B'(s)$ appears after $B'(s-1)$, we must have $k' = s$, or in other words, $B'(s) \in \rset_s$.
    Recall that we execute the subblock processing algorithm $\alggen$ with input $B^{(1)}$ and the region $\rset_s$.
    Let $\rset'_s$ be the region it reports.
    Since the event $\eset_\bad(B^{(s)})$ does not occur, $\rset'_s \neq \emptyset$, and it does not contain any range-block appearing after $B'(s)$.
    In this case, the last range-block of $s^{th}$ region of $\rset^{(1)}$ is the last range-block of $\rset'_s$ and the assertion follows.

    This completes the induction step and \Cref{clm: ncm-jump-dp-completeness} follows.
\endproofof

\subsubsection{\proofof{\Cref{clm: ncm-jump-dp-soundness}}} \label{prf-clm: ncm-jump-dp-soundness}
    We proceed by induction.
    The base case is when $s = 1$.
    It is immediate to see that $|\tilde \rset^{(1)}| \leq 1$.
    If $\tilde \rset^{(1)} = \emptyset$, there is nothing to show.
    Thus, assume that $|\tilde \rset^{(1)}| = 1$ and let $\rset^{(1)}$ be the single region of $\tilde \rset^{(1)}$.
    Since the event $\eset_\bad(B(1))$ does not occur, we conclude that there is an increasing subsequence of $B(1)$ of length at least $\frac{Z_j}{\alpha'_j}$ with values in the range-blocks of $\rset^{(1)}$.
    The assertion now follows.

    We fix some $1 < s \leq |\bset|$ and assume that the induction hypothesis holds for $s-1$.
    We consider the collection $\tilde \rset^{(s-1)}$ of regions just before processing elements of $B(s)$.
    For readability, we drop the superscript and denote this collection by $\tilde \rset$.
    Let $\tilde \rset = \set{\rset(1), \ldots, \rset(|\tilde \rset|)}$ be these regions in their natural order.
    From our induction hypothesis, for each $1 \leq s' \leq |\tilde \rset|$, there is an increasing subsequence of $B(1) \cup \ldots \cup B(s-1)$ of length at least $\frac{s' Z_j}{\alpha'_j}$ with values in the range-blocks of $\rset(1) \cup \ldots \cup \rset(s')$.
    We are now ready to show the assertion for $s$, starting with the case where $1 \leq s' \leq |\tilde \rset|$.

    Consider the execution of $\alggen$ with the input stream-block $B(s)$ and the $s'{th}$ region ${\rset(s') \in \tilde \rset}$.
    Let $\rset'(s')$ be the region it reports.
    We first consider the case where $\rset'(s') \neq \emptyset$.
    Since the event $\eset_\bad(B(s), \rset(s'))$ did not occur, there must be an increasing subsequence of $B(s)$ of length at least $\frac{Z_j}{\alpha'_j}$ with values in the range-blocks of $\rset'(s')$.
    Moreover, $\rset'(s')$ does not contain any range-block of $\rset(1) \cup \ldots \rset(s'-1)$.
    But from our induction hypothesis, there is an increasing subsequence of  $B(1) \cup \ldots \cup B(s-1)$ of length at least $\frac{(s'-1) Z_j}{\alpha'_j}$ with values in the range-blocks of $\rset(1) \cup \ldots \cup \rset(s'-1)$.
    Thus, there is an increasing subsequence of  $B(1) \cup \ldots \cup B(s)$ of length at least $\frac{s' Z_j}{\alpha'_j}$ with values in the range-blocks of $\rset(1) \cup \ldots \cup \rset(s')$.
    It is now immediate to verify that all these range-blocks $\rset(1) \cup \ldots \cup \rset(s')$ are contained in the first $s'$ regions of $\tilde \rset^{(s)}$ and the claim follows.
    We now consider the remaining case where $\rset'(s') = \emptyset$.
    In this case, it is immediate to verify that the range-blocks contained in the first $s'$ regions of $\tilde \rset^{(s)}$ are precisely the ones in the first $s'$ regions of $\tilde \rset$.
    The assertion now follows from our induction hypothesis that there is an increasing subsequence of length at least $\frac{s' Z_j}{\alpha'_j}$ using elements in $B(1) \cup \ldots \cup B(s-1)$ with values in the range-blocks of $\rset(1) \cup \ldots \cup \rset(s')$.

    It now remains to show the assertion for the case where $|\tilde \rset| < s' \leq |\tilde \rset^{(s)}|$.
    Since $|\tilde \rset^{(s)}| \leq 1 + |\tilde \rset|$, we must have $s' = |\tilde \rset^{(s)}| = 1 + |\tilde \rset|$.
    We let $\leftover(\tilde \rset) = \bset' \backslash \bigcup_{\rset \in \tilde \rset} \rset$ be the special region before processing $B(s)$.
    We consider the run of $\alggen$ with the input stream-block $B(s)$ and the region $\leftover(\tilde \rset)$.
    Let $\rset'$ be the region it reports.
    For $|\tilde \rset^{(s)}| = 1 + |\tilde \rset|$ to occur, we must have had $\rset' \neq \emptyset$.
    Moreover, since $\eset_\bad(B(s))$ does not occur, there is an increasing subsequence of $B(s)$ of length at least $\frac{Z_j}{\alpha'_j}$ with values in the range-blocks of $\rset'$.
    As before, it is immediate to verify that there is an increasing subsequence of $B(1) \cup \ldots \cup B(s)$ of length at least $\frac{s' Z_j}{\alpha'_j}$ with values in the range-blocks of $\rset(1) \cup \ldots \cup \rset(s'-1) \cup \rset'$, which are precisely the range-blocks contained in the regions of $\tilde \rset^{(s)}$.
    
    We now conclude that the induction hypothesis indeed holds for $s$, completing the induction step.
    This completes the proof of \Cref{clm: ncm-jump-dp-soundness}.
\endproofof

\subsubsection{\proofof{\Cref{obs: ncm-jumpable-sample-few}}} \label{prf-obs: ncm-jumpable-sample-few}
    Recall that we sample each stream-block of $\hat \bset$ independently randomly with probability $p$ each.
    If $p = 1$, we sample $|\hat \bset|$ stream-blocks and there is nothing to show.
    Otherwise, on expectation, we sample $p|\hat \bset| = 2^7 \zeta \cdot \frac{Z_\ell}{Z_j} \cdot |\hat \bset| \geq 2^7 \zeta$ blocks.
    Here, the last inequality follows since $|\hat \bset|$ is the set of level-$\ell$ descendant-blocks of $B$, implying, $|\hat \bset| \geq Z_j/Z_\ell$.
    From Chernoff bound (\Cref{fact: ncm-chernoff-version}), the probability that we sample more than $12 \zeta p|\hat \bset|$ such blocks is at most $e^{-2\zeta}$.
\endproofof

\subsubsection{\proofof{\Cref{obs: ncm-b-prime-marked-pseudo-suspicious-early-on}}} \label{prf-obs: ncm-b-prime-marked-pseudo-suspicious-early-on}
    Recall that we have $j > \ell > r^*$, and hence, $Z_j > Z_\ell > Z_{r^*}$.
    From \Cref{obs: ncm-many-elements-before-median}, at least $\frac{Z_j/Z_\ell}{2} = \frac{Z_j}{2 Z_\ell}$ level-$\ell$ yes-blocks appear before $\left(\frac{Z_j}{2} + 1\right)^{th}$ element of $B \cap S^*$.
    Let $\tilde \bset$ be the set of such level-$\ell$ yes-blocks.
    We claim that at most $\frac{Z_j}{4 Z_\ell}$  stream-blocks in $\tilde \bset$ share at least $2^5 Z_\ell \mu_{i}$ elements with $B^{*'}$.
    Indeed, assume otherwise for contradiction.
    Since each of these stream-blocks are disjoint and are descendants of $B$, we have
        
        \[ |B \cap B^{*'}| > \frac{Z_j}{4 Z_\ell} \cdot 2^5 Z_\ell \mu_{i} = 8 Z_j \mu_{i},\]

    a contradiction to the fact that $B \cap B^{*'} \leq 8 Z_j \mu_{i}$.
    We now discard from $\tilde \bset$ all stream-blocks $\hat B$ that have $|\hat B \cap B^{*'}| > 2^5 Z_\ell \mu_{i}$ and still denote by $\tilde \bset$ the set of at least $\frac{Z_j}{4 Z_\ell}$ surviving stream-blocks.

    Recall that we sample each level-$\ell$ stream-block of $\hat \bset$ independently with probability  $p$ each.
    We denote by $\eset_1$ the bad event that we sample fewer than $\frac{p Z_j}{8 Z_\ell}$ blocks of $\tilde \bset$.
    Recall that $p~=~\min{\set{\pval, 1}}$.
    If $p = 1$, $\prob{\eset_1 \> | \> p = 1} = 0$.
    Thus, assume that $p = \pval < 1$ and from Chernoff bound,
    
    \[ \prob{\eset_1 \> | \> p < 1} \leq e^{ - \frac{p |\tilde B|}{8}} \leq e^{- \frac{p Z_j}{2^5 Z_\ell}} = e^{- 4 \zeta}. \]

    We assume from now on that the event $\eset_1$ does not occur.
    Consider some level-$\ell$ stream-block $\hat B \in \tilde \bset$ that we have sampled.
    Recall that we then sample element of $\hat B$ with value in the range-blocks $\rset$ independently randomly with probability $p'$ each.
    We say that $\hat B$ is an \emph{oversampled} block if we sample more than $E^*_\ell = \estarellval$ of its elements.
    Recall that $\hat B \in \tilde \bset$ and hence, $|\hat B \cap B^{*'}| \leq 2^5 Z_\ell \mu_{i}$.
    Also recall that $p' = \min{\set{\pprimeval, 1}}$.
    Hence, if $p' = 1$, $\prob{\hat B \text{ is oversampled } \> | \> p' = 1} = 0$.
    Otherwise, $p' = \pprimeval$ and from Chernoff bound (\Cref{fact: ncm-chernoff-version}),

    \begin{align*}
        \prob{\hat B \text{ is oversampled } \> | \> p' < 1} \leq e^{-\frac{E^*_\ell}{6}} &= e^{-\frac{\estarellval}{6}}\\
        &\leq e^{- 8 Z_\ell \mu_i \cdot \frac{2^{16} \zeta}{Z_j \mu_\ell}}\\
        &= e^{- 2^{19} \zeta \cdot \frac{Z_\ell \mu_i}{Z_j \mu_\ell}}\\
        &\leq e^{- 2^{19} \zeta}.
    \end{align*}

    Here, the last inequality follows from the guarantee of \Cref{lem: ncm-jump-jump} that $\frac{Z_j}{Z_\ell} \leq \sqrt{\frac{\mu_i}{\mu_j}}$ holds, implying,  $\frac{Z_\ell \mu_i}{Z_j \mu_\ell} \geq \frac{\mu_i}{\mu_\ell} \cdot \sqrt{\frac{\mu_j}{\mu_i}} = \frac{\sqrt{\mu_i \mu_j}}{\mu_\ell} \geq \frac{\mu_j}{\mu_\ell} \geq 1$.
    We denote by $\eset_2$ the bad event that at least $1$ block of $\tilde \bset$ is oversampled.
    Note that $\prob{\eset_2 \> | \> p' = 1} = 0$.
    On the other hand, from union bound,
    
    \[ \prob{\eset_2 \> | \> p' < 1}  \leq |\tilde \bset| \cdot e^{- 2^{19} \zeta} \leq \frac{Z_j}{Z_\ell} \cdot e^{- 2^{19} \zeta} \leq e^{-2^{18} \zeta}.\]

    Here, the last inequality follows since $e^{\zeta} = e^{\zeta(i, \ell)} \geq e^{\ln{\left( \eta^{\ell - i} \right)}} = \eta^{\ell - i} \geq |\tilde \bset|$.
    We assume from now on that the event $\eset_2$ does not occur, and hence, we are left with a set $\tilde \bset$ of at least $\frac{Z_j}{4 Z_\ell}$ stream-blocks that are not over-sampled.
    
    Consider some level-$\ell$ yes-pair $(\hat B, \hat B')$ with $\hat B \in \tilde \bset$.
    We say that $\hat B$ is a \emph{happy block} if we sample some element of $\hat B \cap \hat B'$ before processing $\left( \frac{Z_\ell}{2} + 1\right)^{th}$ element of $\hat B \cap S^*$.
    We say that it is \emph{sad block} otherwise.
    From \Cref{obs: ncm-many-elements-before-median}, at least $\frac{Z_\ell \mu_{\ell}}{2}$ elements of $\hat B \cap \hat B'$ appear before $\left( \frac{Z_\ell}{2} + 1\right)^{th}$ element of $\hat B \cap S^*$.
    Since we sample each such element independently with probability $p'$ each, the probability that $\hat B$ is a sad block is at most,

    \begin{align*}
        \prob{\hat B \text{ is sad}} \leq \left(1 - p' \right)^{\frac{Z_\ell \mu_\ell}{2}} &\leq e^{ - p' \cdot \frac{Z_\ell \mu_{\ell}}{2} } \\
        &\leq e^{- \pprimeval \cdot \frac{Z_\ell \mu_{\ell}}{2} }\\
        &= e^{- \frac{2^{15} \zeta Z_\ell}{Z_j}}\\
        &\leq 1 -  \frac{2^{14} \zeta Z_\ell}{Z_j}.
    \end{align*}

    In other words, $\prob{\hat B \text{ is happy}} \geq  \frac{2^{14} \zeta Z_\ell}{Z_j}$.
    If $\hat B$ is a happy block, we execute the level-$\ell$ algorithm $\algjump{\ell}$ with input level-$\ell$ pair $(\hat B, \hat B')$ along with subblock $\tilde B$ that contains at least $\frac{Z_\ell}{2} \geq \frac{Z_\ell}{\alpha_\ell}$ elements of $S^*$.
    Since $\hat B$ is not oversampled, each such execution of $\algjump{\ell}$ is not preemptively terminated.
    From the correctness guarantee of $\algjump{\ell}$, it reports yes with probability at least $3/4$, and we mark the level-$j$ ancestor range-block $B'$ as pseudo-promising.
    Thus, the probability that a $\hat B \in \tilde \bset$ marks $B'$ as pseudo-promising is at least $\frac{3}{4} \cdot \frac{2^{14} \zeta Z_\ell}{Z_j} \geq \frac{2^{13} \zeta Z_\ell}{Z_j}$.
    In other words, for each level-$\ell$ pair $(\hat B, \hat B')$ with $\hat B \in \tilde \bset$, we mark $B'$ as pseudo-promising independently randomly with probability at least $\frac{2^{13} \zeta Z_\ell}{Z_j}$.

    We are now ready to bound the probability that it $B'$ is not marked pseudo-suspicious at the end of stream-blocks of $\tilde \bset$.
    Recall that we mark $B'$ as pseudo-suspicious after it has been marked pseudo-promising $2^5 \zeta$ times.
    On expectation, we mark it pseudo-promising at least 
    
    \[ \frac{2^{13} \zeta Z_\ell}{Z_j}  \cdot |\tilde \bset| \geq \frac{2^{13} \zeta Z_\ell}{Z_j}  \cdot \frac{Z_j}{4 Z_\ell} = 2^{11} \zeta\]
    
    times.
    From Chernoff bound, the probability that we mark $B'$ as pseudo-promising fewer than $2^5 \zeta$ times is at most $e^{\left(-4 \zeta \right)}$.
    
    By union bound over all the above discussed bad events, we now conclude that $B'$ is marked pseudo-suspicious before processing all the elements present in the stream-blocks $\tilde \bset$ with probability at least $1 - e^{-2 \zeta}$.
    This completes the proof of \Cref{obs: ncm-b-prime-marked-pseudo-suspicious-early-on}.
\endproofof

\subsubsection{\proofof{\Cref{clm: ncm-bad-block-unlikely-to-be-pseudo-suspicious}}} \label{prf-clm: ncm-bad-block-unlikely-to-be-pseudo-suspicious}
    Fix a bad block $B' \in \rset$.
    Let $\tilde \bset \subseteq \hat \bset$ be the set of stream-blocks that participate in promising pairs with the level-$\ell$ descendants of $B'$.
    From the definition of bad blocks, ${|\tilde \bset| < 3\frac{Z_j \alpha'_\ell}{Z_\ell \alpha'_j} \leq \frac{1}{8}  \cdot \frac{Z_j}{Z_\ell}}$.
    Notice that the range-block $B'$ may be marked pseudo-suspicious only if we mark it pseudo-promising at least $2^5 \zeta$ times.
    For that to occur, we must have sampled at least $2^5 \zeta$ stream-blocks of $\tilde \bset$.
    Since we sample each such block independently randomly with probability $p$ each, the expected number of sampled blocks of $\tilde \bset$ is,
    
    \[ p |\tilde \bset| \leq \pval \cdot \frac{Z_j}{8 Z_\ell} = 8 \zeta.\]
    
    From Chernoff bound (\Cref{fact: ncm-chernoff-version}), the probability that we sample more than $2^5 \zeta$ such blocks is at most $e^{-2^5 \zeta/6} \leq e^{-3 \zeta}$.
\endproofof

\subsubsection{\proofof{\Cref{obs: ncm-large-is-in-good-blocks}}} \label{prf-obs: ncm-large-is-in-good-blocks}
    We proceed as in the proof of \Cref{obs: ncm-more-suspicious-blocks-is-good-easy}.
    Recall that we have fixed the set $\hat \bset$ of level-$\ell$ descendant stream-blocks of the level-$j$ stream-block $B$.
    We denote $m = |\hat \bset|$.
    We let $\hat \bset = \set{\hat B_1, \ldots, \hat B_m}$ be the stream-blocks in their natural order.
    Recall that each good block $B' \in \bset_\good$ is a level-$j$ range-block of $\rset$.
    We also let $\bset_{\good} = \set{B'_1, \ldots, B'_m}$ be these good blocks in their natural order.

    We consider a bipartite graph $\hset$ with vertices $V(\hat \bset)$ on one side and $V(\bset_\good)$ on the other side.
    For each stream-block $\hat B_{k'} \in \hat \bset$, there is a unique vertex $u({k'}) \in V(\hat \bset)$.
    Similarly, for each range-block $B'_{k''} \in \bset_\good$, there is a unique vertex $v({k''}) \in V(\bset_\good)$.
    Consider a pair $(\hat B_{k'}, B'_{k''})$ of blocks with $\hat B_{k'} \in \hat \bset$ and $B'_{k''} \in \bset_\good$.
    If there is a level-$\ell$ descendant range-block $\hat B'_{k''}$ of $B'_{k'}$ such that $(\hat B_{k'}, \hat B_{k''})$ is a promising pair, we add an edge $(u(k'), v(k''))$ in the graph $\hset$.
    Let $\fset \subseteq V(\hat \bset) \times V(\bset_\good)$ be the resulting set of edges.
    Recall that each range-block $B'_{k''} \in \bset_\good$ is a good block.
    Thus, each vertex $v(k'') \in V(\bset_\good)$ has at least $3\frac{Z_j \alpha'_\ell}{Z_\ell \alpha'_j}$ edges incident on it.
    We now conclude that the number of edges in $\hset$ is,

    \[ |\fset| \geq |\bset_\good| \cdot 3\frac{Z_j \alpha'_\ell}{Z_\ell \alpha'_j}  = 3m\frac{Z_j \alpha'_\ell}{Z_\ell \alpha'_j}.\]

    We divide this set of edges $\fset$ into $2m+1$ equivalence classes, where for each $-m \leq k \leq m$, the class $\fset_k$ consists of all the edges $(u(k'), v({k''}))$ of $\fset$ with $k' - k'' = k$.
    It is immediate to verify that for each such class $\fset_k$, there is an increasing subsequence of $B$ with values in the range-blocks of $\rset$ of length at least $|\fset_k| \cdot \frac{Z_\ell}{\alpha'_\ell}$.
    It now remains to show that there is an equivalence class containing a large number of edges.
    Indeed, from the pigeonhole principle, there is some class $\fset_{k^*}$ that contains at least $\frac{|\fset|}{2m+1} > \frac{|\fset|}{3m} \geq \frac{Z_j \alpha'_\ell}{Z_\ell \alpha'_j}$ edges.
    Thus, we obtain an increasing subsequence of size at least,
    
    \[ |\fset_{k^*}| \cdot \frac{Z_\ell}{\alpha'_\ell} > \frac{Z_j \alpha'_\ell}{Z_\ell \alpha'_j} \cdot \frac{Z_\ell}{\alpha'_\ell} = \frac{Z_j}{\alpha'_j}\]
    
    with values in range-blocks of $\bset_\good$.
    This completes the proof of \Cref{obs: ncm-large-is-in-good-blocks}.
\endproofof

\subsection{Proofs Omitted from Sections \ref{subsec: ncm-reduction-to-ncm} and \ref{subsec: ncm-savale-savable}} \label{appn: ncm-reduction-to-ncm}

\subsubsection{\proofof{\Cref{clm: ncm-perfectly-savable}}} \label{prf-clm: ncm-perfectly-savable}
    We start with the set $\iset_\savable$ of savable levels and refine it in multiple steps.
    We claim that there are only a small number of savable levels that do not satisfy properties (i), (ii), (iii), and (iv) of perfectly savable levels.
    Since the weight of each savable level is relative small, the contribution of the levels that are not perfectly savable is also small.
    We can then argue that the weight of perfectly savable levels is large enough.

    Let $\iset^*_1 \subseteq \isetnew$ be the set of levels $0 \leq i < r$ such that $\psi_{i+1} > \eta^{10^{8}}$.
    We claim that there are at most $2 r/10^8$ such levels.
    Indeed, assume otherwise for contradiction.
    But then,
    
    \[ \prod_{i \in \iset^*_1} \psi_{i+1} > \left(\eta^{10^8} \right)^{2r/10^8} = \eta^{2r} > N. \]

    This is a contradiction to the fact that $\prod_{1 \leq i \leq r} \psi_i \leq N$ since $\bset_\Psi(H^*)$, where $\Psi = (\psi_0, \ldots, \psi_r)$, is a hierarchical decomposition of the range $H^*$ consisting of $N$ elements.

    Let $\iset^*_2 \subseteq \isetnew \backslash \set{i_{\knew}}$ be the set of levels $i_k$ such that $i_{k+1} \neq i_k+1$, or in other words, $i_{k+1} \geq i+2$.
    We claim that $|\iset^*_2| \leq 3\eps r$.
    Indeed, each level of $\iset^*_2$ can be charged to a unique level of $\set{0, \ldots, r^*} \backslash \set{\isetnew}$.
    But from \Cref{{clm: ncm-simpler-algo-k-star}}, there are at most $3\eps r$ such levels and the claim follows.

    Let $\iset^*_3 \subseteq \iset^*$ be the set of levels $i_k$ with either $k = \knew$ or $Z_{i_k} < \eta^{3/4} Z_{i_{k+1}}$.
    As before, we write $\frac{Z_0}{Z_{i_{k^*}}} = \frac{Z_{i_0}}{Z_{i_{k^*}}}$ as telescopic product:

    \begin{align*}
        \frac{Z_{i_0}}{Z_{i_{k^*}}} = \prod_{0 \leq k < \knew} \frac{Z_{i_k}}{Z_{i_{k+1}}} &= \left( \prod_{i_k \in \isetnew \backslash \iset^*_3} \frac{Z_{i_k}}{Z_{i_{k+1}}}  \right) \cdot \left(\prod_{i_k \in \iset^*_3} \frac{Z_{i_k}}{Z_{i_{k+1}}} \right)   \\
        &\leq \eta^{\left| \isetnew \backslash \iset^*_3 \right|} \cdot \eta^{\frac{3}{4} |\iset^*_3| } \\
        &= \eta^{\knew - \frac{1}{4} \left| \iset^*_3 \right|}.
    \end{align*}

    Thus, $Z_{i_\knew} \geq \frac{Z_{0}}{\eta^{\knew - |\iset^*_3|/4}} \geq \frac{N^{1/2 - \eps}}{\eta^{\knew - |\iset^*_3|/4}}$.
    But from \Cref{assm: large-xi-by-zi} and \Cref{clm: ncm-simpler-algo-k-star}, we have,

    \begin{align*}
        Z_{i_{\knew}} &\leq \frac{X_{i_\knew}}{N^{1/2 - \eps}} = \frac{N^{1/2 + \eps}}{\eta^{i_\knew}} \leq \frac{N^{1/2 + \eps}}{\eta^{\knew}}.
    \end{align*}

    Combining these two inequalities, we obtain $\frac{N^{1/2 - \eps}}{\eta^{\knew - |\iset^*_3|/4}} \leq Z_{i_\knew} \leq \frac{N^{1/2 + \eps}}{\eta^{\knew}}$.
    We now conclude that $\eta^{|\iset^*_3|/4} \leq N^{2\eps}$, or equivalently, $|\iset^*_3| \leq 8\eps r$.

    We let $\iset^{**} := \iset^*_1 \cup \iset^*_2 \cup \iset^*_3$ and note that $|\iset^{**}| \leq  \frac{2 r}{10^8} + 11 \eps r$.
    Consider a level $i = i_k \in \iset_\savable$ that is not perfectly savable.
    It is immediate to verify that either $i_k \in \iset^{**}$ or $i_{k+1} \in \iset^{**}$.
    Thus, there are at most $\frac{4 r}{10^8} + 22 \eps r$ levels in $\iset_\savable$ that is not perfectly savable.
    We let $\iset'_\savable \subseteq \iset_\savable$ be the set of all perfectly savable levels.
    Recall that each savable level $i \in \iset_\savable$ has weight $0.9 < w_i < 100$.
    It is now immediate to verify that the weight of these perfectly savable levels is at lest,

    \begin{align*}
        w_{\iset'_\savable} \geq w_{\iset_\savable} - 100\left( \frac{4 r}{10^8} + 22 \eps r \right) &= w_{\iset_\savable} - \frac{4 r}{10^6} - 2200 \eps r\\
        &\geq \frac{r}{10^5} - \frac{4 r}{10^6} - 2200 \eps r\\
        &\geq \frac{r}{10^6}.
    \end{align*}
        
    Here, the last inequality follows since $\eps = \delta/10^{12} \leq 10^{-12}$.
\endproofof

\subsubsection{\proofof{\Cref{clm: ncm-savable-load-induction}}} \label{prf-clm: ncm-savable-load-induction}
We proceed by induction.
The base case is when $k = 0$ and the assertion is trivial since we execute level-$i_0$ algorithm $\algSavable{i_0}$ only once.
Consider now some $0 < k \leq k^*$ and the corresponding level $i = i_k \in \iset^*$.
We assume that the induction hypothesis holds for level $i_{k-1}$.
We fix a level-$i_k$ stream-block $B$ and analyze $\load(B)$.
Let $B^*$ be its unique level-$i_{k-1}$ ancestor-block.
It is immediate to verify that each execution to $\algSavable{i_k}$ in which $B$ participates is called by a run of the level-$i_{k-1}$ algorithm $\algSavable{i_{k-1}}$ in which $B^*$ participates.

We first consider the case where $i_{k-1} \not \in \iset_\savable$.
From \Cref{obs: ncm-savable-no-savable}, each such execution of $\algSavable{i_{k-1}}$ performs at most $2^{10} \cdot \zeta(i_{k-1}, i_k) \cdot \frac{\mu_{i_{k-1}}}{\mu_{i_k}} \leq \zeta^2(i_{k-1}, i_k) \cdot \frac{\mu_{i_{k-1}}}{\mu_{i_k}}$ concurrent calls to $\algSavable{i_k}$.
Thus, from induction hypothesis, the maximum number of concurrent calls to the level-$i$ algorithm $\algSavable{i}$ in which $B$ participates is,

\begin{align*}
    \load(B) & \leq \load(B^*) \cdot  \zeta^2(i_{k-1}, i_k) \cdot \frac{\mu_{i_{k-1}}}{\mu_{i_k}}    \\
    & \leq \frac{q_1(i_{k-1})}{q_2(i_{k-1})} \cdot \frac{\mu_{i_0}}{\mu_{i_{k-1}}} \cdot  \zeta^2(i_{k-1}, i_k) \cdot \frac{\mu_{i_{k-1}}}{\mu_{i_k}} \\
    & \leq \frac{q_1(i_{k})}{q_2(i_{k-1})} \cdot \frac{\mu_{i_0}}{\mu_{i_{k}}} = \frac{q_1(i_{k})}{q_2(i_{k})} \cdot \frac{\mu_{i_0}}{\mu_{i_{k}}},
\end{align*}

as claimed.
Assume now that $i_{k-1} \in \iset_\savable$.
From \Cref{lem: ncm-savable-savable}, each such execution of $\algSavable{i_{k-1}}$ performs at most $\frac{\zeta^6(i_{k-1}, i_{k})}{\eta^{0.9 \delta}} \cdot \frac{\mu_{i_{k-1}}}{\mu_{i_{k}}}$ concurrent calls to $\algSavable{i_k}$.
Thus, from induction hypothesis, the maximum number of concurrent calls to the level-$i$ algorithm $\algSavable{i}$ in which $B$ participates is,

\begin{align*}
    \load(B) & \leq \load(B^*) \cdot \zeta^6(i_{k-1}, i_{k}) \cdot  \frac{1}{\eta^{0.9\delta}} \cdot \frac{\mu_{i_{k-1}}}{\mu_{i_{k}}}   \\
    & \leq \frac{q_1(i_{k-1})}{q_2(i_{k-1})} \cdot \frac{\mu_{i_0}}{\mu_{i_{k-1}}} \cdot  \zeta^6(i_{k-1}, i_{k}) \cdot  \frac{1}{\eta^{0.9 \delta}} \cdot \frac{\mu_{i_{k-1}}}{\mu_{i_{k}}} \\
    & = \frac{q_1(i_{k})}{q_2(i_{k})} \cdot \frac{\mu_{i_0}}{\mu_{i_{k}}},
\end{align*}

as claimed.
This completes the analysis of the case where $i_{k-1} \in \iset_\savable$ and hence the induction step.
\Cref{clm: ncm-savable-load-induction} now follows.
\endproofof

\subsubsection{\proofof{\Cref{clm: ncm-savable-load-i-small}}} \label{prf-clm: ncm-savable-load-i-small}
    We start with the following claim.

    \begin{claim}
        For each $0 \leq k < \knew$, $L^*(i_k) \leq L^*(i_{k+1})$.
    \end{claim}
    \begin{proof}
        To show this claim, it suffices to show that $\frac{q_1(i_k)}{q_2(i_k)} \cdot \frac{\mu_{i_0}}{\mu_{i_k}} \leq \frac{q_1(i_{k+1})}{q_2(i_{k+1})} \cdot \frac{\mu_{i_0}}{\mu_{i_{k+1}}}$.
        But since $q_1(i_{k+1}) > q_1(i_{k})$, it suffices to show that $\frac{q_2(i_{k+1})}{q_2(i_{k})} \leq \frac{\mu_{i_k}}{\mu_{i_{k+1}}}$.
        Note that this assertion is trivial when $i_k \not \in \iset_\savable$ since in that case, $q_2(i_{k+1}) = q_2(i_k)$ and $\mu_{i_k} \geq \mu_{i_{k+1}}$.
        Assume now that $i_k \in \iset_\savable$.
        In this case, $\frac{q_2(i_{k+1})}{q_2(i_{k})} = \eta^{0.9} \leq \eta^{0.9} < \frac{\mu_{i_k}}{\mu_{i_{k+1}}}$ since $\delta \leq 1$ and the claim follows.
    \end{proof}

    Thus, to show \cref{clm: ncm-savable-load-i-small}, it now suffices to show that $L^*(i_{\knew}) = \frac{q_1(i_{\knew})}{q_2(i_{\knew})} \cdot \frac{\mu_{i_0}}{\mu_{i_{\knew}}} \leq N^{\frac{1}{2} - \frac{\delta}{10^9}}$.
    To this end, we first show bounds on $q_1(i_\knew)$, $q_2(i_\knew)$, and $q_3(i_\knew)$.
    Notice that
    
    \[ q_1(i_\knew) = \prod_{0 < k \leq \knew} \zeta^6(i_{k-1}, i_k) = \left( \zeta(i_{0}, i_1) \cdot \ldots \cdot \zeta(i_{\knew-1}, i_\knew) \right)^6 <  N^{o(1)},\]

    where, the inequality follows form \Cref{clm: ncm-lis-simple-pi-zeta-bound}.
    Recall that we are given a set $\iset_\savable$ of perfectly savable levels with level-weight $\tupwt{\iset_\savable} \geq r/10^6$.
    Since the weight of each savable level $i \in \iset_\savable$ is at most $w_i < 100$, the total number of perfectly savable levels is at least $|\iset_\savable| \geq \frac{w_{\iset_\savable}}{100} \geq \frac{r}{10^8}$.
    We now obtain,
    
    \[ q_2(i_{\knew}) = \eta^{0.9 \delta |\iset_\savable| } \geq \eta^{\frac{9 \delta r}{10^9}} > N^{\frac{8 \delta}{10^9}}. \]

    Here, the last inequality follows since $r = \floor{\frac{\log N}{\log \eta}}$.
    We are now ready to bound $L^*(i_{\knew})$.

    \begin{align*}
        L^*(i_{\knew}) = \frac{q_1(i_{\knew})}{q_2(i_{\knew})} \cdot \frac{\mu_{i_0}}{\mu_{i_{\knew}}} &\leq \frac{N^{o(1)}}{N^{\frac{8 \delta}{10^9}}} \cdot N^{1/2 + \eps}  \\
        &\leq N^{\frac{1}{2} - \frac{8 \delta}{10^9} + \eps + o(1)}   \\
        &\leq N^{\frac{1}{2} - \frac{\delta}{10^9}}.
    \end{align*}

    Here, the first inequality follows from the fact that $\mu_{i_0} = \frac{X_{i_0}}{Z_{i_0}} \leq \frac{N}{Z_0} \leq {N^{1/2 + \eps}}$ and the last inequality follows since $\eps = \delta/10^{12}$.
    This completes the proof of \Cref{clm: ncm-savable-load-i-small}.
\endproofof

\subsubsection{\proofof{\Cref{clm: ncm-soundness-eset-bad}}} \label{prf-clm: ncm-soundness-eset-bad}
    From union bound, $\prob{\eset_\bad} \leq \sum_{\tau} \prob{\eset_\bad(\tau)}$ and we show that for each ${\tau \in \set{2^0, \ldots, 2^{\log{(\eta)} - 1}}}$, $\prob{\eset_\bad(\tau)} \leq e^{-4\zeta}$.
    From now on, we fix some integral power of $2$ $\tau \in \set{2^0, \ldots, 2^{\log{(\eta)} - 1}}$ and let ${\Gtau = (\Ltau, \Rtau, \Etau_\advice, \Etau)}$ be the corresponding (random) instance of the \NCM problem constructed by $\algSavable{i}$.
    For convenience, we discard special vertices from $\Ltau \cup \Rtau$ and still denote the surviving vertices by $\Ltau$ and $\Rtau$ respectively.
    Consider an edge-slot $e \in \Ltau \times \Rtau$ of $\Gtau$.
    Let $\tilde B$ and $B'$ be the respective blocks such that $e = \left(v(\tilde B), v(B') \right)$.
    Let $B \in \bset$ be the level-$j$ stream-block that contains $\tilde B$.
    We also consider the partition $\left( \tilde B^{(1)}, \tilde B^{(2)} \right)$ of $\tilde B$ into $2$ subblocks.
    We denote by $\eset''_\bad(e)$ to be the event that $\optlis(\tilde B^{(2)} \cap B') < \frac{Z_j}{\alpha'_j}$ and at least $2^5 \zeta^2$ independent executions of $\algSavable{j}$ with input level-$j$ pair $(B, B')$ and the subblock $\tilde B^{(2)}$ of $\tilde B$ returns yes.
    From the correctness guarantee of $\algSavable{j}$ and Chernoff bound (\Cref{fact: ncm-chernoff-version}), it is immediate to verify that $\prob{\eset''_\bad(e)} \leq e^{-2\zeta^2}$.

    We denote by $\eset''_\bad$ be the event that for some edge-slot $e$, the event $\eset''_\bad(e)$ occurs.
    Since there are at most $|\Ltau| \cdot |\Rtau| \leq \left(\max{\set{\eta^2, \psi_{i+1}}} \right)^2 < \left( \eta^2 \psi_{i+1} \right)^2 = e^{2 \zeta}$ such edge-slots, from union bound, $\prob{\eset''_\bad} \leq e^{2 \zeta} \cdot e^{-2\zeta^2} < 1/8$.

    \begin{claim}\label{clm: ncm-redn-to-lis-soundness}
        If the event $\eset''_\bad$ does not occur, $\optncm(\Gtau) \leq \frac{Z_i/\alpha'_i}{Z_j/\alpha'_j}$.
    \end{claim}
    \begin{proof}
        Consider an edge $e \in \Etau$ of $\Gtau$ and let $\tilde B(e)$ and $B'(e)$ be the respective blocks such that $e = \left(v(\tilde B(e)), v(B'(e)) \right)$.
        Since the event $\eset''_\bad$ does not occur, there is an increasing subsequence $S^*(e)$ of $\tilde B(e)$ of length $\optlis(\tilde B(e) \cap B'(e)) \geq \frac{Z_j}{\alpha'_j}$ with values in $B'(e)$.

        Consider an optimum non-crossing matching $\Mtau \subseteq \Etau$ of $\Gtau$.
        We let $\Mtau = (e_1, \ldots, e_m)$ in the natural order, where $m = \optncm(\Gtau)$.
        Since $\Mtau$ is a non-crossing matching, the stream-blocks $\tilde B(e_1), \ldots, \tilde B(e_m)$ are disjoint and appear in this order as a part of the sequence $S^*$.
        Similarly, the range-blocks $B'(e_1), \ldots, B'(e_m)$ are disjoint and appear in this order in the range $H^*$.
        We consider the sequence $S^*(\Mtau) := S^*(e_1) \cup \ldots \cup S^*(e_m)$.
        It is now immediate to verify that $S^*(\Mtau)$  is an increasing subsequence of $B^{**}$ with values in the range-block $B^{*'}$.
        Moreover, the cardinality of $S^*(\Mtau)$ is at least $m \cdot \frac{Z_{j}}{\alpha'_{j}} = \optncm(\Gtau) \cdot \frac{Z_{j}}{\alpha'_{j}}$.
        The \Cref{clm: ncm-redn-to-lis-soundness} now follows by recalling that $\optlis(B^{**} \cap B^{*'}) < \frac{Z_i}{\alpha'_i}$.
    \end{proof}

    From the above claim, the event $\eset_\bad(\tau)$ may occur only if the event $\eset''_\bad$ occurs.
    We now conclude that $\prob{\eset_\bad(\tau)} \leq \prob{\eset''_\bad} < 1/8$ and \Cref{clm: ncm-soundness-eset-bad} follows.
\endproofof

\newcommand{\ratio}{4}

\subsubsection{\proofof{\Cref{{clm: ncm-g-tau-large-ncm}}}} \label{prf-clm: ncm-g-tau-large-ncm}
    We consider the set $\pset \subseteq \bset \times \bset'$ of all level-$j$ yes-pairs whose corresponding stream-blocks are present in $\bset$ and range-blocks are present in $\bset'$.
    Recall that we have assumed that $(B^*, B^{*'})$ is a level-$i$ yes-pair and the subblock $B^{**}$ of $B^{*}$ contains at least $Z_i/2$ elements of $S^*$.
    Thus, there are at least $\frac{Z_i}{2 Z_j}$ level-$j$ yes-pairs in $\pset$.
    We discard additional pairs from $\pset$ and let still denote by $\pset$ the resulting set of $\frac{Z_i}{2 Z_j}$ yes-pairs.

    Consider some $\tau \in \set{2^0, \ldots, 2^{\log{(\eta)} - 1}}$.
    Recall that we denote by $\bset_\tau(B)$ the partition of $B$ into exactly $\tau$ subblocks.
    Consider a subblock $\tilde B \in \bset_\tau(B)$ and the partition $(\tilde B^{(1)}, \tilde B^{(2)})$ of $\tilde B$ into two subblocks.
    We say that $\tilde B$ is a \emph{balanced} block iff $\tilde B^{(1)}$ and $\tilde B^{(2)}$ both contain at least $\frac{Z_j}{\ratio}$ elements of $S^*$.
    If there is such a balanced block $\tilde B \in \bset_\tau(B)$, we say that the yes-pair $(B, B')$ is \emph{$\tau$-good}.

    \begin{claim} \label{clm: ncm-each-b-is-tau-balanced}
        Each pair $(B, B') \in \pset$ is $\tau(B, B')$-good for some $\tau(B, B') \in \set{2^0, \ldots, 2^{\log{(\eta)} - 1}}$.
    \end{claim}
    \begin{proof}
        We start with $k = 0$ and $\tilde B^{(0)} = B$.
        We will ensure that at the end of every step $k$, we have a subblock $\tilde B^{(k)} \in \bset_{2^k}(B)$ with $|\tilde B^{(k)} \cap S^*| \geq \left(1 - \frac{1}{\ratio} \right)^{k} Z_j$.
        Notice that $k < \log \eta$ must hold.
        Indeed, assume otherwise for contradiction.
        Then, for $k = \log \eta$, we have some subblock $\tilde B^{(k)} \in \bset_{2^k}(B) = \bset_\eta(B)$ with 
        
        \[ |\tilde B^{(k)} \cap S^*| \geq \left(1 - \frac{1}{\ratio} \right)^{\log \eta} Z_j = \left(\frac{3}{4} \right)^{\log \eta} Z_j > \frac{Z_j}{\sqrt{\eta}}.\]

        In other words, for some level-$(j+1)$ stream-block $\hat B$, we have $|\hat B \cap S^*| > \frac{Z_{j}}{\eta^{1/2}}$, a contradiction to the fact that each level-$(j+1)$ stream-block may contribute at most $Z_{j+1} \leq \frac{Z_j}{\eta^{3/4}}$ elements to $S^*$.
        We now assume that throughout our algorithm, $k < \log \eta$ must hold.
        
        Assume now that for some $k \geq 0$, we are given a block $\tilde B^{(k)} \in \bset_{2^k}(B)$ with ${|\tilde B^{(k)} \cap S^*| \geq \left(1 - \frac{1}{\ratio} \right)^{k} Z_j}$.
        If $\tilde B^{(k)}$ is a balanced subblock of $B$, we report $\tau(B, B') = 2^k$ and halt.
        Since $0 \leq k < \log \eta$, we indeed have $\tau(B, B') \in \set{2^0, \ldots, 2^{\log{(\eta)} - 1}}$ as claimed.
        Thus, assume from now on that $\tilde B^{(k)}$ is not balanced and consider the partition $\bset_2(\tilde B^{(k)})$ of $\tilde B^{(k)}$ into two subblocks.
        Since $\tilde B^{(k)}$ is not balanced, there is some subblock $\tilde B^{(k+1)} \in \bset_2(\tilde B^{(k)})  \subseteq \bset_{2^{k+1}}(B)$ with
        
        \[ |\tilde B^{(k+1)} \cap S^*| > \left(1 - \frac{1}{\ratio} \right) |\tilde B^{(k)} \cap S^*| \geq \left(1 - \frac{1}{\ratio} \right)^{k+1} Z_j. \]

        Thus, $\tilde B^{(k+1)}$ is indeed the guaranteed subblock for $k+1$.
        We now update $k \gets k+1$ and continue the process.
        The claim now follows since we must have terminated our process before reaching $k = \log \eta$.
    \end{proof}

    From the pigeonhole principle, there is some $\tau \in \set{2^0, \ldots, 2^{\log{(\eta)} - 1}}$ such that at least ${\frac{|\pset|}{\log \eta} \geq \frac{Z_i}{4 Z_j \log \eta}}$ pairs of $\pset$ are $\tau$-good.
    We fix such $\tau$ and the set $\pset^{*} \subseteq \pset$ of at least $\frac{Z_i}{4 Z_j \log \eta}$ $\tau$-good pairs. 
    Let $\Gtau = (\Ltau, \Rtau, \Etau_\advice, \Etau)$ be the corresponding (random) instance of the \NCM problem.
    We now focus on showing that $\optncm(\Gtau)$ is large with probability at least $0.9$.

    Consider a $\tau$-good pair $(B, B') \in \pset^{*}$.
    Let $\tilde B \in \bset_\tau(B)$ be an arbitrary balanced subblock of $B$.
    We denote by $e(B, B') :=  (v(\tilde B), v(B'))$ the corresponding edge-slot of $\Gtau$.
    Notice that the edges $\set{e(B, B') \> | \> (B, B') \in \pset^{*}}$, if exist, form a non-crossing matching.
    Indeed, the endpoint of these edges are disjoint and appear in the natural order of the underlying stream-blocks in $S$ and the range-blocks in $H^*$.
    Thus, our goal reduces to showing that a large number of such edges are present in $\Gtau$.

    \begin{claim} \label{clm: ncm-e-present-in-e-tau}
        For each pair $(B, B') \in \pset^{*}$ the corresponding edge $e(B, B')$ is present in $\Etau$ independently randomly with probability at least $0.6$.
    \end{claim}
    \begin{proof}
        Fix a $\tau$-good pair $(B, B')$ and a balanced subblock $\tilde B \in \bset_\tau(B)$ along with the corresponding edge $e(B, B') = (v(\tilde B), v(B'))$.
        Also consider the partition $\bset_2(\tilde B) = (\tilde B^{(1)}, \tilde B^{(2)})$ of $\tilde B$ into two subblocks.
        Since $\tilde B$ is balanced, both $\tilde B^{(1)}$ and $\tilde B^{(2)}$ contain at least $\frac{Z_j}{\ratio}$ elements of $S^*$.
        It is easy to see that each of them contain at least $\frac{Z_j}{4 Z_{j+1}} - 2 \geq \frac{Z_j}{5 Z_{j+1}}$ level-$(j+1)$ yes-blocks.
        But $S^*$ is an $\Upsilon$-canonical subsequence and hence, each of these level-$(j+1)$ yes-block contains at least $Z_{j+1} \mu_j$ elements with values in $B'$.
        Thus, each of them, and in particular, the subblock $\tilde B^{(1)}$ contains at least $\frac{Z_j}{5 Z_{j+1}} \cdot Z_{j+1} \mu_j = \frac{Z_j \mu_j}{5}$ elements of $B'$.

        First, we claim that we mark the edge $e(B, B')$ as an advice-edge with probability at least $0.9$.
        Recall that we sample elements of $\tilde B^{(1)} \cap B^{*'}$ with probability $p = \savpval$ each.
        We denote by $\eset_1$ the bad event that we do not sample any element of $\tilde B^{(1)} \cap B'$, and hence, $e(B, B') \not \in \Etau_\advice$.
        From the above claim, 
            
        \[ \prob{\eset_1} = \left(1 - p \right)^{|\tilde B^{(1)} \cap B'|} =  \left(1 - \savpval \right)^{\frac{Z_j \mu_j}{5}}  < 0.1.\] 
            
        Next we denote by $\eset_2$ the bad event that $e(B, B') \not \in \Etau$, or in other words, our level-$j$ subblock processing algorithm $\alg^{(j)}$ for $(\tilde B^{(2)}, B')$ reports no.
        Recall that $\tilde B$ is a balanced block and hence, $\tilde B^{(2)} \cap B'$ contains at least $\frac{Z_j}{4}$ elements of $S^*$.
        From the correctness guarantee of $\alg^{(j)}$, we can bound $\prob{\eset_2 \> | \> \eset_1} \leq 1/4$.
        Finally, we conclude that $e \in \Etau$ with probability at least $1 - \prob{\eset_1 \cup \eset_2} \geq 0.6$.
        This completes the proof of \Cref{clm: ncm-e-present-in-e-tau}.
    \end{proof}

    We are now ready to show that $\optncm(\Gtau)$ is large.
    From \Cref{clm: ncm-e-present-in-e-tau},
    
    \[ \expect{\optncm(\Gtau)} \geq \frac{6}{10} \cdot |\pset^{*}| \geq  \frac{Z_i}{8 Z_j \log \eta}.\]

    From Chernoff bound (\Cref{fact: ncm-chernoff-version}), $\prob{\optncm(\Gtau) < \frac{Z_i}{16 Z_j \log \eta}} < 0.1$.
    This completes the proof of \Cref{clm: ncm-g-tau-large-ncm}.
\endproofof
}{}

\iftoggle{alpha-approx-lis}{
    \section{Proofs Omitted from Section \ref{sec: ncm-alpha-approx}} \label{appn: ncm-alpha-approx}

\subsubsection{\proofof{\Cref{obs: ncm-alpha-lis-part}}} \label{prf-obs: ncm-alpha-lis-part}
    We fix an optimal increasing subsequence $S'$ of $S$ of size $\optlis(S)$.
    From \Cref{cor: ncm-partition-lemma}, there is $\psi'_2$, an integral power of $2$, such that there is a subsequence $S''$ of $S'$ of size at least $|S'|/256 \log^4 N$ that is $Z_2$-canonical w.r.t. the partition $\bset_{\psi'_2}(S)$ of $S$.
    Let $\bset^{\yes}$ be the set of yes-blocks of $\bset_{\psi'_2}(S)$ w.r.t. the subsequence $S''$.

    We now consider a sequence $\sset$ of size exactly $\psi'_2$ obtained as follows.
    Each element of $\sset$ is a stream-block of $\bset_{\psi'_2}(S)$ appearing in its natural order.
    Note that $\bset^\yes$ form a subsequence of $\sset$.
    Another application of \Cref{cor: ncm-partition-lemma} yields $\psi_1$, an integral power of $2$, such that there is a subsequence $\hat \bset^\yes$ of $\bset^\yes$ of length at least $|\bset^\yes|/256 \log^4 N$ that is $Z_1/Z_2$-canonical w.r.t. the partition $\bset_{\psi_1}(\sset)$ of $\sset$.
    We fix such $\psi_1$ and the subset $\hat \bset^\yes$ of stream-blocks.

    Finally, consider the subsequence $S^*$ of $S''$ that is obtained as follows.
    For each stream-block $\hat B \in \hat \bset^\yes$, the subsequence $S^*$ contains all elements of $S''$ appearing in $\hat B$.
    It is immediate to verify that the cardinality of $S^*$ is at least 
    
    \[ |\hat \bset^\yes| \cdot Z_2 \geq \frac{|\bset^\yes|}{2^8 \log^4 N} \cdot Z_2 = \frac{|S''|}{Z_2 \cdot 2^8 \log^4 N} \cdot Z_2 \geq \frac{|S'|}{2^{16} \log^8 N} = \frac{\optlis(S)}{2^{16} \log^8 N}. \] 

    It is now immediate to verify that the sequence $S^*$ is a $\vectZ = (Z_1, Z_2)$-canonical subsequence of $S$ w.r.t. the hierarchical partition $\bset_\Psi(S)$, where $\Psi = (\psi_1, \psi_2)$ and $\psi_2 = \psi'_2/\psi_1$.
\endproofof

\subsubsection{\proofof{\Cref{obs: ncm-alpha-approx-eset-star-bad}}} \label{prf-obs: ncm-alpha-approx-eset-star-bad}
    We will show that for each block $B' \in \bset^2$, $\prob{\eset^*_\bad(B')} \leq 1/N^3$.
    The claim then follows by taking a union bound over all such $|\bset^2| \leq N$ blocks.
    Consider now a block $B' \in \bset^2$ and let $B \in \bset^1$ be its ancestor-block.
    Let $\tilde \rset$ be the set of blocks just before processing elements of $B$ and let $R_\leftover \gets H^* \backslash \bigcup_{R_ \in \tilde \rset} R$ be the corresponding special region.
    Recall that $|\tilde \rset| \leq Z_0/Z_1$.
    For each block $R \in \tilde \rset \cup \set{R_\leftover}$, we mark $B'$ independently at random with probability $p$.
    Thus, on expectation, we mark it at most $p \cdot (1 + Z_0/Z_1) \leq 2p Z_0/Z_1$ times.
    From Chernoff bound (\Cref{fact: ncm-chernoff-version}), the probability that we mark it at least $L^* = 32pZ_0/Z_1$ times is at most $e^{-L^*/6} \leq N^{-4}$.
    Here, the inequality follows, since,

    \[L^* = 32 p \frac{Z_0}{Z_1} \geq 32 \cdot \frac{\log^{11} N}{\alpha} \cdot \frac{\alpha}{\log^\alphalislogfactor N} = 32 \log^2 N. \]
\endproofof

\subsubsection{\proofof{\Cref{clm: ncm-lis-alpha-jump-dp-completeness}}} \label{prf-clm: ncm-lis-alpha-jump-dp-completeness}
    We will show a stronger guarantee.
    For each yes-block $B(i) \in \bset^1_\yes$, we let $e(i)$ be the value of the last element of $S^*$ in $B(i)$.
    We will show that for each $1 \leq s \leq Z_0/Z_1$, $|\tilde \rset^{(s)}| \geq s$ and the $s^{th}$ range-block of $\tilde \rset^{(s)}$ does not contain any element that appears after $e(s)$.

    We proceed by induction.
    The base case is when $s = 1$, where we consider the first stream-block $B(1)$ of $\bset^1_\yes$.
    Let $e(1) \in H^*$ be 
    Let $\tilde \rset^{\mathsf{pre}}$ be the collection of range-blocks just before processing $B^{(1)}$.
    We also let $R_\leftover$ be the special block consisting of elements not contained in the blocks of $\tilde \rset^{{\mathsf{pre}}}$.
    For convenience, we let $\tilde \rset = \tilde \rset^{\mathsf{pre}} \cup \set{R_\leftover}$.
    Let $\tilde \rset = \set{R_1, \ldots, R_k}$ in their natural order where $k = |\tilde \rset| = 1 + |\tilde \rset^{\mathsf{pre}}|$.
    Since $\tilde \rset$ form a partition of $H^*$, there is a unique $1 \leq k' \leq k$ such that $e(1) \in R_{k'}$.
    Recall that $\tilde \rset^{(1)}$ is the collection of range-blocks just after processing $B^{(1)}$, where the blocks of $\tilde \rset^{(1)}$ appear in their natural order.
    We first consider the case where $k' > 1$.
    In this case, it is immediate to verify that $|\tilde \rset^{(1)}| \geq 1$ and the first block of $\tilde \rset^{(1)}$ does not contain the element $e(1)$, and as a consequence, does not contain any element that appears after it.
    Thus, assume from now on that $k' = 1$.
    Since the events $\eset^{*}_\bad$ and $\eset^{**}_\bad$ does not occur, $|\tilde \rset^{(1)}| \geq 1$ and the first block of $\tilde \rset^{(1)}$ does not contain any element that appears after $e(1)$.
    This completes the analysis of the base case.

    We now fix some $1 < s \leq Z_0/Z_1$.
    We assume that the assertion holds for $s-1$ and show it for $s$.
    Let $\tilde \rset^{\mathsf{pre}}$ be the collection of range-blocks just before processing $B^{(s)}$.
    We also let $R_\leftover$ be the range-block consisting of elements not contained in the blocks of $\tilde \rset^{{\mathsf{pre}}}$.
    Recall that $\tilde \rset^{(s)}$ is the collection of range-blocks just after processing $B^{(1)}$.
    For convenience, we let $\tilde \rset = \tilde \rset^{\mathsf{pre}} \cup \set{R_\leftover}$.
    Let $\tilde \rset = \set{\rset_1, \ldots, \rset_k}$ in their natural order where $k = |\tilde \rset| = 1+|\tilde \rset^{\mathsf{pre}}|$.
    Since $\tilde \rset$ form a partition of $H^*$, there is a unique $1 \leq k' \leq k$ such that $e(s) \in \rset_{k'}$.
    If $k' > s$, as in the base case, there is nothing to show.
    Indeed, $|\tilde \rset^{(s)}| \geq s$ and the first $s$ range-blocks does not contain the element $e(s)$.
    Thus, assume from now on that $k' \leq s$.
    From our induction hypothesis, the range-blocks $\set{R_1, \ldots, R_{s-1}}$ does not contain any element that appear after $e(s-1)$.
    Since $e(s)$ appears after $e(s-1)$, we must have $k' = s$, or in other words, $\set{e(s-1)+1, \ldots, e(s)} \subseteq R_s$.
    Using the facts that the events $\eset^*_\bad$ and $\eset^{**}_\bad$ did not occur and the correctness guarantee of $\alg'_1$, $|\tilde \rset^{(s)}| \geq s$, and the $s^{th}$ range-block of $\tilde \rset^{(s)}$ does not contain any element that appears after $e(s)$.
    This completes the induction step and \Cref{clm: ncm-lis-alpha-jump-dp-completeness} now follows.
\endproofof

\subsubsection{\proofof{\Cref{clm: ncm-lis-alpha-jump-dp-soundness}}} \label{prf-clm: ncm-lis-alpha-jump-dp-soundness}
    We proceed by induction.
    The base case is when $s = 1$.
    It is immediate to see that $|\tilde \rset^{(1)}| \leq 1$.
    If $\tilde \rset^{(1)} = \emptyset$, there is nothing to show.
    Thus, assume that $|\tilde \rset^{(1)}| = 1$ and let $R^{(1)}$ be the unique range-block of $\tilde \rset^{(1)}$.
    Since the event $\eset^{**}_\bad$ does not occur and from the correctness of $\alg'_1$, we conclude that there is an increasing subsequence of $B(1) \cap R^{(1)}$ of length at least $Z_2$.
    The assertion now follows.

    We now fix some $1 < s \leq |\bset|$ and assume that the induction hypothesis holds for $s-1$.
    Our goal is to show it for $s$.
    We consider the collection $\tilde \rset^{(s-1)}$ of range-blocks just before processing elements of $B(s)$.
    For readability, we drop the superscript and denote this collection by $\tilde \rset$.
    Let $\tilde \rset = \set{R(1), \ldots, R|\tilde \rset|}$ be these blocks in their natural order.
    From our induction hypothesis, for each $1 \leq s' \leq |\tilde \rset|$, there is an increasing subsequence of $B(1) \cup \ldots \cup B(s-1)$ of length at least $s'Z_2$ with values in the range-blocks $R(1) \cup \ldots \cup R(s')$.
    We are now ready to show the assertion for $s$, starting with the case where $1 \leq s' \leq |\tilde \rset|$.

    Consider the $s'{th}$ range-block $R_{s'}$ of $\tilde \rset$ and let $v_{s'}(B)$ be as we compute.
    If $v_{s'}(B)$ is defined, from the correctness of $\alg'_1$, there must be an increasing subsequence of $B(s) \cap R_{s'}$ of length at least $Z_2$ using elements not appearing after $v_{s'}(B)$.
    But from induction hypothesis, there is an increasing subsequence of  $B(1) \cup \ldots \cup B(s-1)$ of length at least $(s'-1)Z_2$ with values in the range-blocks $R(1) \cup \ldots \cup R(s'-1)$.
    Thus, there is an increasing subsequence of  $B(1) \cup \ldots \cup B(s)$ of length at least $s'Z_2$ with values in the range-blocks $R(1) \cup \ldots \cup R(s')$.
    We now consider the remaining case where $v_{s'}(B)$ is undefined.
    In this case, it is immediate to verify that the elements contained in the first $s'$ range-blocks of $\tilde \rset^{(s)}$ are precisely the ones in the first $s'$ blocks of $\tilde \rset$.
    The assertion now follows from our induction hypothesis that there is an increasing subsequence of length at least $s'Z_2$ using elements in $B(1) \cup \ldots \cup B(s-1)$ with values in the range-blocks $R(1) \cup \ldots \cup R(s')$.

    It now remains to show the assertion for the case where $s' = |\tilde \rset^{(s)}|$ and $|\tilde \rset^{(s)}| = 1 + |\tilde \rset|$.
    We let $R_\leftover = \bset' \backslash \bigcup_{\rset \in \tilde \rset} \rset$ be the special range-block before processing $B(s)$.
    Let $v_\leftover(B(s))$ as computed by our algorithm.
    For this case to occur, we must have had $v_\leftover(B(s)) \in R_\leftover$.
    Thus,  there is an increasing subsequence of $B(s) \cap R_\leftover$ of length at least $Z_2$ using elements that do not appear after $v_\leftover(B(s))$.
    As before, it is immediate to verify that there is an increasing subsequence of $B(1) \cup \ldots \cup B(s)$ of length at least $s'Z_2$ with values in the range-blocks of $\tilde \rset^{(s)}$.
    
    We now conclude that the induction hypothesis indeed holds for $s$, completing the induction step.
    This completes the proof of \Cref{clm: ncm-lis-alpha-jump-dp-soundness}.
\endproofof

}{}
        }{}

\end{appendices}
\addtocontents{toc}{\protect\setcounter{tocdepth}{2}}

\bibliographystyle{Alpha}
\addcontentsline{toc}{chapter}{References}
\bibliography{main}

\newcommand{\etalchar}[1]{$^{#1}$}
\begin{thebibliography}{KMNFT20}

\bibitem[AAM{\etalchar{+}}11]{dks_average_hardness}
Noga Alon, Sanjeev Arora, Rajsekar Manokaran, Dana Moshkovitz, and Omri Weinstein.
\newblock Inapproximabilty of densest k-subgraph from average case hardness, 2011.

\bibitem[ACG{\etalchar{+}}10]{ACGKTZ}
Matthew Andrews, Julia Chuzhoy, Venkatesan Guruswami, Sanjeev Khanna, Kunal Talwar, and Lisa Zhang.
\newblock Inapproximability of edge-disjoint paths and low congestion routing on undirected graphs.
\newblock {\em Combinatorica}, 30(5):485--520, 2010.

\bibitem[AGLR94]{AwerbuchGLR94}
B.~Awerbuch, R.~Gawlick, T.~Leighton, and Y.~Rabani.
\newblock On-line admission control and circuit routing for high performance computing and communication.
\newblock In {\em Proceedings 35th Annual Symposium on Foundations of Computer Science}, pages 412--423, Nov 1994.

\bibitem[AKW00]{AKW}
Alok Aggarwal, Jon Kleinberg, and David~P. Williamson.
\newblock Node-disjoint paths on the mesh and a new trade-off in {VLSI} layout.
\newblock {\em SIAM J. Comput.}, 29(4):1321--1333, February 2000.

\bibitem[Alo86]{Cheeger2}
Noga Alon.
\newblock Eigenvalues and expanders.
\newblock {\em Combinatorica}, 6(2):83--96, 1986.

\bibitem[Alo96]{alon1996explicit}
Noga Alon.
\newblock Explicit expanders of every degree and size.
\newblock {\em Proceedings of the 28th Annual ACM Symposium on Theory of Computing}, pages 686--690, 1996.

\bibitem[Alo98]{Alon}
Noga Alon.
\newblock Spectral techniques in graph algorithms.
\newblock In {\em Latin American Symposium on Theoretical Informatics}, pages 206--215. Springer, 1998.

\bibitem[AM84]{Cheeger1}
Noga Alon and Vitali~D Milman.
\newblock Eigenvalues, expanders and superconcentrators.
\newblock In {\em Foundations of Computer Science, 1984. 25th Annual Symposium on}, pages 320--322. IEEE, 1984.

\bibitem[And10]{Andrews}
Matthew Andrews.
\newblock Approximation algorithms for the edge-disjoint paths problem via {Raecke} decompositions.
\newblock In {\em Proceedings of IEEE FOCS}, pages 277--286, 2010.

\bibitem[ANSS22]{ANSS22}
Alexandr Andoni, Negev~Shekel Nosatzki, Sandip Sinha, and Clifford Stein.
\newblock Estimating the longest increasing subsequence in nearly optimal time.
\newblock In {\em 2022 IEEE 63rd Annual Symposium on Foundations of Computer Science (FOCS)}, pages 708--719, 2022.

\bibitem[AR95]{grids1}
Yonatan Aumann and Yuval Rabani.
\newblock Improved bounds for all optical routing.
\newblock In {\em Proceedings of the sixth annual ACM-SIAM symposium on Discrete algorithms}, SODA '95, pages 567--576, Philadelphia, PA, USA, 1995. Society for Industrial and Applied Mathematics.

\bibitem[AST94]{planar-separator-theorem3}
Noga Alon, Paul Seymour, and Robin Thomas.
\newblock Planar separators.
\newblock {\em SIAM Journal on Discrete Mathematics}, 7(2):184--193, 1994.

\bibitem[AZ05]{AZ-undir-EDP}
Matthew Andrews and Lisa Zhang.
\newblock Hardness of the undirected edge-disjoint paths problem.
\newblock In {\em STOC}, pages 276--283. ACM, 2005.

\bibitem[BCC{\etalchar{+}}10]{dks10}
Aditya Bhaskara, Moses Charikar, Eden Chlamtac, Uriel Feige, and Aravindan Vijayaraghavan.
\newblock Detecting high log-densities: an \emph{O}(\emph{n}\({}^{\mbox{1/4}}\)) approximation for densest \emph{k}-subgraph.
\newblock In {\em Proceedings of the 42nd {ACM} Symposium on Theory of Computing, {STOC} 2010, Cambridge, Massachusetts, USA, 5-8 June 2010}, pages 201--210, 2010.

\bibitem[BCE80]{hadwiger_erdos}
B\'{e}la Bollob\'{a}s, Paul Catlin, and Paul Erd\H{o}s.
\newblock Hadwiger's conjecture is true for almost every graph.
\newblock {\em European Journal of Combinatorics}, 1(3):195 -- 199, 1980.

\bibitem[Beh22]{Behnezhad21}
Soheil Behnezhad.
\newblock Time-optimal sublinear algorithms for matching and vertex cover.
\newblock In {\em 2021 IEEE 62nd Annual Symposium on Foundations of Computer Science (FOCS)}, pages 873--884, 2022.

\bibitem[BFSU98]{BroderFSU94}
Andrei~Z. Broder, Alan~M. Frieze, Stephen Suen, and Eli Upfal.
\newblock Optimal construction of edge-disjoint paths in random graphs.
\newblock {\em SIAM Journal on Computing}, 28(2):541--573, 1998.

\bibitem[BFU92]{BFU92}
Andrei~Z. Broder, Alan~M. Frieze, and Eli Upfal.
\newblock Existence and construction of edge disjoint paths on expander graphs.
\newblock In {\em Proceedings of the Twenty-fourth Annual ACM Symposium on Theory of Computing}, STOC '92, pages 140--149, New York, NY, USA, 1992. ACM.

\bibitem[BFU94]{BFU}
Andrei. Broder, Alan. Frieze, and Eli Upfal.
\newblock Existence and construction of edge-disjoint paths on expander graphs.
\newblock {\em SIAM Journal on Computing}, 23(5):976--989, 1994.

\bibitem[BYJK{\etalchar{+}}02]{Bar-Yossefstream}
Ziv Bar-Yossef, T.~S. Jayram, Ravi Kumar, D.~Sivakumar, and Luca Trevisan.
\newblock Counting distinct elements in a data stream.
\newblock In Jos{\'e} D.~P. Rolim and Salil Vadhan, editors, {\em Randomization and Approximation Techniques in Computer Science}, pages 1--10, Berlin, Heidelberg, 2002. Springer Berlin Heidelberg.

\bibitem[Car88]{thomassen1988presence}
Thomassen Carsten.
\newblock On the presence of disjoint subgraphs of a specified type.
\newblock {\em Journal of Graph Theory}, 12(1):101--111, 1988.

\bibitem[CC16a]{NDPwC2}
Chandra Chekuri and Julia Chuzhoy.
\newblock Half-integral all-or-nothing flow, 2016.
\newblock Personal Communication.

\bibitem[CC16b]{CC_gmt}
Chandra Chekuri and Julia Chuzhoy.
\newblock Polynomial bounds for the grid-minor theorem.
\newblock {\em J. ACM}, 63(5):40:1--40:65, December 2016.

\bibitem[CE13]{ChekuriE13}
Chandra Chekuri and Alina Ene.
\newblock Poly-logarithmic approximation for maximum node disjoint paths with constant congestion.
\newblock In {\em Proc.\ of ACM-SIAM SODA}, 2013.

\bibitem[Cha12]{Chakrabarti-polylog-communication}
Amit Chakrabarti.
\newblock A note on randomized streaming space bounds for the longest increasing subsequence problem.
\newblock {\em Inf. Process. Lett.}, 112:261--263, 03 2012.

\bibitem[Chu15]{C_gmt}
Julia Chuzhoy.
\newblock Excluded grid theorem: Improved and simplified.
\newblock In {\em Proceedings of the Forty-seventh Annual ACM Symposium on Theory of Computing}, STOC '15, pages 645--654, New York, NY, USA, 2015. ACM.

\bibitem[Chu16a]{gmt_julia_arxiv}
Julia Chuzhoy.
\newblock {Improved Bounds for the Excluded Grid Theorem}.
\newblock {\em ArXiv e-prints}, February 2016.

\bibitem[Chu16b]{Chuzhoy11}
Julia Chuzhoy.
\newblock Routing in undirected graphs with constant congestion.
\newblock {\em {SIAM} J. Comput.}, 45(4):1490--1532, 2016.

\bibitem[CK15]{NDP-grids}
Julia Chuzhoy and David H.~K. Kim.
\newblock On approximating node-disjoint paths in grids.
\newblock In Naveen Garg, Klaus Jansen, Anup Rao, and Jos{\'{e}} D.~P. Rolim, editors, {\em Approximation, Randomization, and Combinatorial Optimization. Algorithms and Techniques, {APPROX/RANDOM} 2015, August 24-26, 2015, Princeton, NJ, {USA}}, volume~40 of {\em LIPIcs}, pages 187--211. Schloss Dagstuhl - Leibniz-Zentrum fuer Informatik, 2015.

\bibitem[CKL16]{NDP-planar}
Julia Chuzhoy, David H.~K. Kim, and Shi Li.
\newblock Improved approximation for node-disjoint paths in planar graphs.
\newblock In {\em Proceedings of the 48th Annual ACM SIGACT Symposium on Theory of Computing}, STOC 2016, pages 556--569, New York, NY, USA, 2016. ACM.

\bibitem[CKN17]{NDP-hard-old}
Julia Chuzhoy, David H.~K. Kim, and Rachit Nimavat.
\newblock New hardness results for routing on disjoint paths.
\newblock In {\em Proceedings of the 49th Annual ACM SIGACT Symposium on Theory of Computing}, STOC 2017, page 86–99, New York, NY, USA, 2017. Association for Computing Machinery.

\bibitem[CKN18a]{NDP-hard-new}
Julia Chuzhoy, David H.~K. Kim, and Rachit Nimavat.
\newblock Almost polynomial hardness of node-disjoint paths in grids.
\newblock In {\em Proceedings of the 50th Annual ACM SIGACT Symposium on Theory of Computing}, STOC 2018, page 1220–1233, New York, NY, USA, 2018. Association for Computing Machinery.

\bibitem[CKN18b]{NDP-algo}
Julia Chuzhoy, David H.~K. Kim, and Rachit Nimavat.
\newblock {Improved Approximation for Node-Disjoint Paths in Grids with Sources on the Boundary}.
\newblock In Ioannis Chatzigiannakis, Christos Kaklamanis, D{\'a}niel Marx, and Donald Sannella, editors, {\em 45th International Colloquium on Automata, Languages, and Programming (ICALP 2018)}, volume 107 of {\em Leibniz International Proceedings in Informatics (LIPIcs)}, pages 38:1--38:14, Dagstuhl, Germany, 2018. Schloss Dagstuhl--Leibniz-Zentrum fuer Informatik.

\bibitem[CKS05]{CKS}
Chandra Chekuri, Sanjeev Khanna, and F.~Bruce Shepherd.
\newblock Multicommodity flow, well-linked terminals, and routing problems.
\newblock In {\em Proc.\ of ACM STOC}, pages 183--192, 2005.

\bibitem[CKS06]{EDP-alg}
Chandra Chekuri, Sanjeev Khanna, and F.~Bruce Shepherd.
\newblock An ${O}(\sqrt n)$ approximation and integrality gap for disjoint paths and unsplittable flow.
\newblock {\em Theory of Computing}, 2(1):137--146, 2006.

\bibitem[CKS09]{EDP-planar-constant-cong}
Chandra Chekuri, Sanjeev Khanna, and F~Bruce Shepherd.
\newblock Edge-disjoint paths in planar graphs with constant congestion.
\newblock {\em SIAM Journal on Computing}, 39(1):281--301, 2009.

\bibitem[CL16]{ChuzhoyL12}
Julia Chuzhoy and Shi Li.
\newblock A polylogarithmic approximation algorithm for edge-disjoint paths with congestion 2.
\newblock {\em J. {ACM}}, 63(5):45:1--45:51, 2016.

\bibitem[CMS07]{ChekuriMS07}
Chandra Chekuri, Marcelo Mydlarz, and F.~Bruce Shepherd.
\newblock Multicommodity demand flow in a tree and packing integer programs.
\newblock {\em ACM Trans. Algorithms}, 3(3), August 2007.

\bibitem[CN19]{large-minors-in-expanders}
Julia Chuzhoy and Rachit Nimavat.
\newblock Large minors in expanders, 2019.

\bibitem[CS78]{Cutler-Shiloach}
M.~Cutler and Y.~Shiloach.
\newblock Permutation layout.
\newblock {\em Networks}, 8:253--278, 1978.

\bibitem[CT19]{CT18}
Julia Chuzhoy and Zihan Tan.
\newblock Towards tight(er) bounds for the excluded grid theorem.
\newblock In {\em SODA}, 2019.

\bibitem[CV20]{comparison-based-cite}
Graham Cormode and Pavel Vesel\'{y}.
\newblock A tight lower bound for comparison-based quantile summaries.
\newblock In {\em Proceedings of the 39th ACM SIGMOD-SIGACT-SIGAI Symposium on Principles of Database Systems}, PODS'20, page 81–93, New York, NY, USA, 2020. Association for Computing Machinery.

\bibitem[DH07]{DemaineH07}
Erik~D Demaine and MohammadTaghi Hajiaghayi.
\newblock {Quickly deciding minor-closed parameters in general graphs}.
\newblock {\em European Journal of Combinatorics}, 28(1):311--314, January 2007.

\bibitem[DH08]{bidimensionality}
Erik~D. Demaine and MohammadTaghi Hajiaghayi.
\newblock The bidimensionality theory and its algorithmic applications.
\newblock {\em The Computer Journal}, 51(3):292--302, 2008.

\bibitem[DP09]{measure-concentration}
Devdatt Dubhashi and Alessandro Panconesi.
\newblock {\em Concentration of Measure for the Analysis of Randomized Algorithms}.
\newblock Cambridge University Press, New York, NY, USA, 1st edition, 2009.

\bibitem[EHL{\etalchar{+}}18]{matching-streaming-det-lowerbound}
Hossein Esfandiari, Mohammadtaghi Hajiaghayi, Vahid Liaghat, Morteza Monemizadeh, and Krzysztof Onak.
\newblock Streaming algorithms for estimating the matching size in planar graphs and beyond.
\newblock {\em ACM Trans. Algorithms}, 14(4), aug 2018.

\bibitem[EIS76]{EDP-hardness}
Shimon Even, Alon Itai, and Adi Shamir.
\newblock On the complexity of timetable and multicommodity flow problems.
\newblock {\em {SIAM} J. Comput.}, 5(4):691--703, 1976.

\bibitem[EJ08]{ErgunJ08}
Funda Erg{\"{u}}n and Hossein Jowhari.
\newblock On distance to monotonicity and longest increasing subsequence of a data stream.
\newblock In Shang{-}Hua Teng, editor, {\em Proceedings of the Nineteenth Annual {ACM-SIAM} Symposium on Discrete Algorithms, {SODA} 2008, San Francisco, California, USA, January 20-22, 2008}, pages 730--736. {SIAM}, 2008.

\bibitem[Fei02]{Feige02}
Uriel Feige.
\newblock Relations between average case complexity and approximation complexity.
\newblock In {\em Proceedings of the Thiry-fourth Annual ACM Symposium on Theory of Computing}, STOC '02, pages 534--543, New York, NY, USA, 2002. ACM.

\bibitem[FHKS03]{Feige3COL5}
Uriel Feige, Magn\'{u}s~M. Halld\'{o}rsson, Guy Kortsarz, and Aravind Srinivasan.
\newblock Approximating the domatic number.
\newblock {\em SIAM J. Comput.}, 32(1):172--195, January 2003.

\bibitem[Fie73]{cheeger-alg}
Miroslav Fiedler.
\newblock Algebraic connectivity of graphs.
\newblock {\em Czechoslovak mathematical journal}, 23(2):298--305, 1973.

\bibitem[FKO09]{ccl_random}
Nikolaos Fountoulakis, Daniela K\"{u}hn, and Deryk Osthus.
\newblock The order of the largest complete minor in a random graph.
\newblock {\em Random Structures \& Algorithms}, 33(2):127--141, 2009.

\bibitem[FMS16]{fleszar_et_al}
Krzysztof Fleszar, Matthias Mnich, and Joachim Spoerhase.
\newblock {New Algorithms for Maximum Disjoint Paths Based on Tree-Likeness}.
\newblock In Piotr Sankowski and Christos Zaroliagis, editors, {\em 24th Annual European Symposium on Algorithms (ESA 2016)}, volume~57 of {\em Leibniz International Proceedings in Informatics (LIPIcs)}, pages 42:1--42:17, Dagstuhl, Germany, 2016. Schloss Dagstuhl--Leibniz-Zentrum fuer Informatik.

\bibitem[Fre75]{FREDMAN197529}
Michael~L. Fredman.
\newblock On computing the length of longest increasing subsequences.
\newblock {\em Discrete Mathematics}, 11(1):29--35, 1975.

\bibitem[Fri01]{journal-Frieze}
Alan~M Frieze.
\newblock Edge-disjoint paths in expander graphs.
\newblock {\em SIAM Journal on Computing}, 30(6):1790--1801, 2001.

\bibitem[FST11]{FominST11}
Fedor~V. Fomin, Saket Saurabh, and Dimitrios~M. Thilikos.
\newblock Strengthening {Erdos-P\'osa} property for minor-closed graph classes.
\newblock {\em Journal of Graph Theory}, 66(3):235--240, 2011.

\bibitem[GG07]{GalG07}
Anna G{\'{a}}l and Parikshit Gopalan.
\newblock Lower bounds on streaming algorithms for approximating the length of the longest increasing subsequence.
\newblock In {\em 48th Annual {IEEE} Symposium on Foundations of Computer Science {(FOCS} 2007), October 20-23, 2007, Providence, RI, USA, Proceedings}, pages 294--304. {IEEE} Computer Society, 2007.

\bibitem[GJKK07]{sqrt-n-det-lis-soda}
Parikshit Gopalan, T.~S. Jayram, Robert Krauthgamer, and Ravi Kumar.
\newblock Estimating the sortedness of a data stream.
\newblock In {\em Proceedings of the Eighteenth Annual ACM-SIAM Symposium on Discrete Algorithms}, SODA '07, page 318–327, USA, 2007. Society for Industrial and Applied Mathematics.

\bibitem[GVY97]{apx_tree}
N.~Garg, V.V. Vazirani, and M.~Yannakakis.
\newblock Primal-dual approximation algorithms for integral flow and multicut in trees.
\newblock {\em Algorithmica}, 18(1):3--20, 1997.

\bibitem[Had43]{hadwiger-original}
Hugo Hadwiger.
\newblock {\"U}ber eine klassifikation der streckenkomplexe.
\newblock {\em Vierteljschr. Naturforsch. Ges. Z{\"u}rich}, 88(2):133--142, 1943.

\bibitem[HLW06]{avi_survey}
Shlomo Hoory, Nathan Linial, and Avi Wigderson.
\newblock Expander graphs and their applications.
\newblock {\em Bull. Amer. Math. Soc. (N.S.)}, 43(4):439--561, 2006.

\bibitem[HMM{\etalchar{+}}21]{edp-fully-planar-constant-approx}
Chien-Chung Huang, Mathieu Mari, Claire Mathieu, Kevin Schewior, and Jens Vygen.
\newblock An approximation algorithm for fully planar edge-disjoint paths.
\newblock {\em SIAM Journal on Discrete Mathematics}, 35(2):752--769, 2021.

\bibitem[Hol07]{HolensteinParallelRep}
Thomas Holenstein.
\newblock Parallel repetition: Simplifications and the no-signaling case.
\newblock In {\em Proceedings of the Thirty-ninth Annual ACM Symposium on Theory of Computing}, STOC '07, pages 411--419, New York, NY, USA, 2007. ACM.

\bibitem[Kar75]{Karp-NDP-hardness}
R.~Karp.
\newblock On the complexity of combinatorial problems.
\newblock {\em Networks}, 5:45--68, 1975.

\bibitem[Kho04]{Khot04}
Subhash Khot.
\newblock Ruling out ptas for graph min-bisection, densest subgraph and bipartite clique.
\newblock In {\em Proceedings of the 45th Annual IEEE Symposium on Foundations of Computer Science}, FOCS '04, pages 136--145, Washington, DC, USA, 2004. IEEE Computer Society.

\bibitem[KK12]{KK_gmt}
Ken-ichi Kawarabayashi and Yusuke Kobayashi.
\newblock {Linear min-max relation between the treewidth of H-minor-free graphs and its largest grid}.
\newblock In {\em 29th International Symposium on Theoretical Aspects of Computer Science (STACS 2012)}, volume~14 of {\em Leibniz International Proceedings in Informatics (LIPIcs)}, pages 278--289, Dagstuhl, Germany, 2012.

\bibitem[KK13]{KK-planar}
Ken-Ichi Kawarabayashi and Yusuke Kobayashi.
\newblock An {O(log n)}-approximation algorithm for the edge-disjoint paths problem in {Eulerian} planar graphs.
\newblock {\em ACM Trans. Algorithms}, 9(2):16:1--16:13, March 2013.

\bibitem[Kle05]{Kleinberg-planar}
Jon Kleinberg.
\newblock An approximation algorithm for the disjoint paths problem in even-degree planar graphs.
\newblock In {\em Proceedings of the 46th Annual IEEE Symposium on Foundations of Computer Science}, FOCS '05, pages 627--636, Washington, DC, USA, 2005. IEEE Computer Society.

\bibitem[KMNFT20]{kapralov2020space}
Michael Kapralov, Slobodan Mitrovi{\'c}, Ashkan Norouzi-Fard, and Jakab Tardos.
\newblock Space efficient approximation to maximum matching size from uniform edge samples.
\newblock In {\em Proceedings of the Fourteenth Annual ACM-SIAM Symposium on Discrete Algorithms}, pages 1753--1772. SIAM, 2020.

\bibitem[KN18]{KN2018}
Michael Krivelevich and Rajko Nenadov.
\newblock Complete minors in graphs without sparse cuts.
\newblock {\em arXiv preprint arXiv:1812.01961}, 2018.

\bibitem[KR96]{KR}
Jon Kleinberg and Ronitt Rubinfeld.
\newblock Short paths in expander graphs.
\newblock In {\em Proceedings of the 37th Annual Symposium on Foundations of Computer Science}, FOCS '96, pages 86--, Washington, DC, USA, 1996. IEEE Computer Society.

\bibitem[KR10]{clique_or_sep_opt}
Ken-ichi Kawarabayashi and Bruce Reed.
\newblock A separator theorem in minor-closed classes.
\newblock In {\em 2010 IEEE 51st Annual Symposium on Foundations of Computer Science}, pages 153--162, Oct 2010.

\bibitem[Kri18]{exp_random}
Michael Krivelevich.
\newblock Finding and using expanders in locally sparse graphs.
\newblock {\em SIAM Journal on Discrete Mathematics}, 32(1):611--623, 2018.

\bibitem[Kri19]{expander-minor}
Michael Krivelevich.
\newblock Expanders-how to find them, and what to find in them.
\newblock In Allan Lo and et~al., editors, {\em Surveys in Combinatorics 2019}, pages 115--142. London Mathematical Society Lecture Notes, 2019.

\bibitem[KS04]{KolliopoulosS}
Stavros~G. Kolliopoulos and Clifford Stein.
\newblock Approximating disjoint-path problems using packing integer programs.
\newblock {\em Mathematical Programming}, 99:63--87, 2004.

\bibitem[KT95]{grids4}
Jon~M. Kleinberg and {\'E}va Tardos.
\newblock Disjoint paths in densely embedded graphs.
\newblock In {\em Proceedings of the 36th Annual Symposium on Foundations of Computer Science}, pages 52--61, 1995.

\bibitem[KT98]{grids3}
Jon~M. Kleinberg and {\'E}va Tardos.
\newblock Approximations for the disjoint paths problem in high-diameter planar networks.
\newblock {\em J. Comput. Syst. Sci.}, 57(1):61--73, 1998.

\bibitem[KvL84]{npc_grid}
MR~Kramer and Jan van Leeuwen.
\newblock The complexity of wire-routing and finding minimum area layouts for arbitrary vlsi circuits.
\newblock {\em Advances in computing research}, 2:129--146, 1984.

\bibitem[LR99]{LeightonRao}
Tom Leighton and Satish Rao.
\newblock Multicommodity max-flow min-cut theorems and their use in designing approximation algorithms.
\newblock {\em J. ACM}, 46(6):787--832, November 1999.

\bibitem[LS15]{leaf_gmt}
Alexander Leaf and Paul Seymour.
\newblock Tree-width and planar minors.
\newblock {\em Journal of Combinatorial Theory, Series B}, 111:38 -- 53, 2015.

\bibitem[LT79]{planar-separator-theorem1}
Richard~J Lipton and Robert~Endre Tarjan.
\newblock A separator theorem for planar graphs.
\newblock {\em SIAM Journal on Applied Mathematics}, 36(2):177--189, 1979.

\bibitem[LVZ06]{Liben-NowellVZ06}
David Liben{-}Nowell, Erik Vee, and An~Zhu.
\newblock Finding longest increasing and common subsequences in streaming data.
\newblock {\em J. Comb. Optim.}, 11(2):155--175, 2006.

\bibitem[Lyn75]{npc_planar}
James~F. Lynch.
\newblock The equivalence of theorem proving and the interconnection problem.
\newblock {\em SIGDA Newsl.}, 5(3):31--36, September 1975.

\bibitem[Man17]{Manurangsi16}
Pasin Manurangsi.
\newblock Almost-polynomial ratio eth-hardness of approximating densest k-subgraph.
\newblock In {\em Proceedings of the 49th Annual {ACM} {SIGACT} Symposium on Theory of Computing, {STOC} 2017, Montreal, QC, Canada, June 19-23, 2017}, pages 954--961, 2017.

\bibitem[MOP93]{ncm-first}
Federico Malucelli, Thomas Ottmann, and Daniele Pretolani.
\newblock Efficient labelling algorithms for the maximum noncrossing matching problem.
\newblock {\em Discrete Applied Mathematics}, 47(2):175--179, 1993.

\bibitem[MS21]{MS21}
Michael Mitzenmacher and Saeed Seddighin.
\newblock Improved sublinear time algorithm for longest increasing subsequence.
\newblock In {\em Proceedings of the Thirty-Second Annual ACM-SIAM Symposium on Discrete Algorithms}, SODA '21, page 1934–1947, USA, 2021. Society for Industrial and Applied Mathematics.

\bibitem[MT10]{Moser-Tardos}
Robin~A. Moser and G\'{a}bor Tardos.
\newblock A constructive proof of the general lov\'{a}sz local lemma.
\newblock {\em J. ACM}, 57:11:1--11:15, February 2010.

\bibitem[Mut05]{muthukrishnan2005data}
S.~Muthukrishnan.
\newblock {\em Data Streams: Algorithms and Applications}.
\newblock Now Publishers Inc., 2005.

\bibitem[NO08]{NguyenK08}
Huy~N. Nguyen and Krzysztof Onak.
\newblock Constant-time approximation algorithms via local improvements.
\newblock FOCS '08, page 327–336, USA, 2008. IEEE Computer Society.

\bibitem[NV21]{NV21}
Ilan Newman and Nithin Varma.
\newblock {New Sublinear Algorithms and Lower Bounds for LIS Estimation}.
\newblock In Nikhil Bansal, Emanuela Merelli, and James Worrell, editors, {\em 48th International Colloquium on Automata, Languages, and Programming (ICALP 2021)}, volume 198 of {\em Leibniz International Proceedings in Informatics (LIPIcs)}, pages 100:1--100:20, Dagstuhl, Germany, 2021. Schloss Dagstuhl -- Leibniz-Zentrum f{\"u}r Informatik.

\bibitem[ORRR12]{onak2012near}
Krzysztof Onak, Dana Ron, Michal Rosen, and Ronitt Rubinfeld.
\newblock A near-optimal sublinear-time algorithm for approximating the minimum vertex cover size.
\newblock In {\em Proceedings of the twenty-third annual ACM-SIAM symposium on Discrete Algorithms}, pages 1123--1131. Society for Industrial and Applied Mathematics, 2012.

\bibitem[R{\"a}c02]{Raecke}
Harald R{\"a}cke.
\newblock Minimizing congestion in general networks.
\newblock In {\em Proc.\ of IEEE FOCS}, pages 43--52, 2002.

\bibitem[Rao08]{RaoParallelRep}
Anup Rao.
\newblock Parallel repetition in projection games and a concentration bound.
\newblock In {\em Proceedings of the Fortieth Annual ACM Symposium on Theory of Computing}, STOC '08, pages 1--10, New York, NY, USA, 2008. ACM.

\bibitem[Raz98]{RazParallelRep}
Ran Raz.
\newblock A parallel repetition theorem.
\newblock {\em SIAM J. Comput.}, 27(3):763--803, June 1998.

\bibitem[Ree97]{Reed-chapter}
Bruce Reed.
\newblock {\em Surveys in Combinatorics}, chapter Treewidth and Tangles: A New Connectivity Measure and Some Applications.
\newblock London Mathematical Society Lecture Note Series. Cambridge University Press, 1997.

\bibitem[RS86]{gmt_5}
Neil Robertson and Paul Seymour.
\newblock Graph minors. v. excluding a planar graph.
\newblock {\em J. Comb. Theory Ser. B}, 41(1):92--114, August 1986.

\bibitem[RS88]{NDP-surface}
Neil Robertson and Paul~D. Seymour.
\newblock Graph minors. {VII.} disjoint paths on a surface.
\newblock {\em J. Comb. Theory, Ser. {B}}, 45(2):212--254, 1988.

\bibitem[RS90]{RobertsonS}
N.~Robertson and P.~D. Seymour.
\newblock Outline of a disjoint paths algorithm.
\newblock In {\em Paths, Flows and VLSI-Layout}. Springer-Verlag, 1990.

\bibitem[RS95]{flat-wall-RS}
Neil Robertson and Paul~D Seymour.
\newblock Graph minors. {XIII}. the disjoint paths problem.
\newblock {\em Journal of Combinatorial Theory, Series B}, 63(1):65--110, 1995.

\bibitem[RS04]{RS}
Neil Robertson and Paul Seymour.
\newblock Graph minors. xx. wagner's conjecture.
\newblock {\em Journal of Combinatorial Theory, Series B}, 92(2):325 -- 357, 2004.
\newblock Special Issue Dedicated to Professor W.T. Tutte.

\bibitem[RS10]{RaghavendraSteurer10}
Prasad Raghavendra and David Steurer.
\newblock Graph expansion and the unique games conjecture.
\newblock In {\em Proceedings of the Forty-second ACM Symposium on Theory of Computing}, STOC '10, pages 755--764, New York, NY, USA, 2010. ACM.

\bibitem[RSSS19]{RSSS19}
Aviad Rubinstein, Saeed Seddighin, Zhao Song, and Xiaorui Sun.
\newblock Approximation algorithms for lcs and lis with truly improved running times.
\newblock In {\em 2019 IEEE 60th Annual Symposium on Foundations of Computer Science (FOCS)}, pages 1121--1145, 2019.

\bibitem[RST94]{RST_exclude_planar}
Neil Robertson, Paul Seymour, and Robin Thomas.
\newblock Quickly excluding a planar graph.
\newblock {\em J. Comb. Theory, Ser. {B}}, 62(2):323--348, 1994.

\bibitem[RT87]{RaghavanT}
Prabhakar Raghavan and Clark~D. Tompson.
\newblock Randomized rounding: a technique for provably good algorithms and algorithmic proofs.
\newblock {\em Combinatorica}, 7:365--374, December 1987.

\bibitem[RZ10]{RaoZhou}
Satish Rao and Shuheng Zhou.
\newblock Edge disjoint paths in moderately connected graphs.
\newblock {\em SIAM J. Comput.}, 39(5):1856--1887, 2010.

\bibitem[SCS11]{EDP-planar-c2}
Lo\"{\i}c Seguin-Charbonneau and F.~Bruce Shepherd.
\newblock Maximum edge-disjoint paths in planar graphs with congestion 2.
\newblock In {\em Proceedings of the 2011 IEEE 52Nd Annual Symposium on Foundations of Computer Science}, FOCS '11, pages 200--209, Washington, DC, USA, 2011. IEEE Computer Society.

\bibitem[Sey16]{hadwiger-survey}
Paul Seymour.
\newblock {\em Hadwiger{\textquoteright}s conjecture}, pages 417--437.
\newblock Springer International Publishing, January 2016.
\newblock Publisher Copyright: {\textcopyright} Springer International Publishing Switzerland 2016.

\bibitem[SS13]{SaksS13}
Michael~E. Saks and C.~Seshadhri.
\newblock Space efficient streaming algorithms for the distance to monotonicity and asymmetric edit distance.
\newblock In Sanjeev Khanna, editor, {\em Proceedings of the Twenty-Fourth Annual {ACM-SIAM} Symposium on Discrete Algorithms, {SODA} 2013, New Orleans, Louisiana, USA, January 6-8, 2013}, pages 1698--1709. {SIAM}, 2013.

\bibitem[SS17]{SS10-estimate-lis-polylog}
M.~Saks and C.~Seshadhri.
\newblock Estimating the longest increasing sequence in polylogarithmic time.
\newblock {\em SIAM Journal on Computing}, 46(2):774--823, 2017.

\bibitem[SW07]{SunW07}
Xiaoming Sun and David~P. Woodruff.
\newblock The communication and streaming complexity of computing the longest common and increasing subsequences.
\newblock In Nikhil Bansal, Kirk Pruhs, and Clifford Stein, editors, {\em Proceedings of the Eighteenth Annual {ACM-SIAM} Symposium on Discrete Algorithms, {SODA} 2007, New Orleans, Louisiana, USA, January 7-9, 2007}, pages 336--345. {SIAM}, 2007.

\bibitem[Ung51]{planar-separator-theorem2}
Peter Ungar.
\newblock A theorem on planar graphs.
\newblock {\em Journal of the London Mathematical Society}, 1(4):256--262, 1951.

\bibitem[Wag37]{Wagner1937}
K.~Wagner.
\newblock Über eine eigenschaft der ebenen komplexe.
\newblock {\em Mathematische Annalen}, 114:570--590, 1937.

\bibitem[YYI09]{YoshidaYI09}
Yuichi Yoshida, Masaki Yamamoto, and Hiro Ito.
\newblock An improved constant-time approximation algorithm for maximum~matchings.
\newblock STOC '09, page 225–234, New York, NY, USA, 2009. Association for Computing Machinery.

\end{thebibliography}
\end{document}